\documentclass[10pt]{article}
\usepackage[lmargin=1in,rmargin=1in,tmargin=1in,bmargin=1in]{geometry}
\usepackage{amsmath,amsthm,amsfonts,amssymb,comment,slashed,mathtools}
\mathtoolsset{showonlyrefs=true}

\usepackage[T1]{fontenc}
\usepackage[english]{babel}
\usepackage{xcolor}

\usepackage[scr=rsfso]{mathalpha}

\usepackage[protrusion=true]{microtype}
\UseMicrotypeSet[protrusion]{basicmath}

\usepackage{csquotes}
\usepackage{bm}
\usepackage[shortlabels]{enumitem}
\usepackage[affil-it]{authblk}

\usepackage[section]{placeins}
\usepackage{soul} 

\usepackage{mathrsfs} 
\usepackage{graphicx}
\usepackage{stmaryrd}
\usepackage{epigraph}

\graphicspath{{figures/}}
\usepackage{accents}

\usepackage{etoolbox}
\robustify{\accentset}

\usepackage[backend=biber,style=alphabetic,giveninits=true,doi=false,isbn=false,url=false,sorting=nyt,maxbibnames=99,maxnames=99,maxalphanames=99,date=year]{biblatex}
\renewbibmacro{in:}{}
\usepackage{chngcntr}

\usepackage{imakeidx}
\makeindex[title = Index of Notation, columns = 2, intoc]

\usepackage{varioref} 
\usepackage[%
bookmarks=true,
colorlinks=true,
linkcolor=black,
urlcolor=black,
citecolor=black,
plainpages=false,
pdfpagelabels,
final]{hyperref}
\usepackage{aliascnt}
\usepackage[capitalise,nameinlink]{cleveref}

\renewcommand{\bf}{\textbf}

\numberwithin{equation}{section}

\theoremstyle{plain}
\newtheorem{thm}{Theorem}[section]
\crefname{thm}{Theorem}{Theorems}

\newaliascnt{prop}{thm}
\newtheorem{prop}[prop]{Proposition}
\aliascntresetthe{prop}
\crefname{prop}{Proposition}{Propositions}

\newaliascnt{lem}{thm}
\newtheorem{lem}[lem]{Lemma}
\aliascntresetthe{lem}
\crefname{lem}{Lemma}{Lemmas}

\newtheorem*{thm*}{Theorem}

\newtheorem{thmrough}{Theorem}

\crefname{thmrough}{Theorem}{Theorems} 

\newtheorem{ingredient}{Ingredient}

\crefname{ingredient}{Ingredient}{Ingredients}

\newaliascnt{cor}{thm}
\newtheorem{cor}[cor]{Corollary}
\aliascntresetthe{cor}
\crefname{cor}{Corollary}{Corollaries}

\newtheorem*{conj*}{Conjecture}

\newaliascnt{conj}{thm}
\newtheorem{conj}[conj]{Conjecture}
\aliascntresetthe{conj}
\crefname{conj}{Conjecture}{Conjectures}

\newaliascnt{question}{thm}
\newtheorem{question}[question]{Question}
\aliascntresetthe{question}
\crefname{question}{Question}{Questions}

\theoremstyle{definition}

\newaliascnt{defn}{thm}
\newtheorem{defn}[defn]{Definition}
\aliascntresetthe{defn}
\crefname{defn}{Definition}{Definitions} 

\newaliascnt{ass}{thm}
\newtheorem{ass}[ass]{Assumption}
\aliascntresetthe{ass}
\crefname{ass}{Assumption}{Assumptions}

\newaliascnt{ex}{thm}

\aliascntresetthe{ex}
\crefname{ex}{Example}{Examples}

\theoremstyle{remark}

\newaliascnt{rk}{thm}
\newtheorem{rk}[rk]{Remark}
\aliascntresetthe{rk}
\crefname{rk}{Remark}{Remarks} 

\renewcommand{\Bbb}{\mathbb}
\newcommand{\ve}{\varepsilon}
\newcommand{\les}{\lesssim}

\newcommand{\id}{\operatorname{id}}

\newcommand{\dom}{\operatorname{dom}}

\newcommand{\tr}{\operatorname{tr}}

\newcommand{\Ric}{\mathrm{Ric}}
\newcommand{\out}{\mathrm{out}}
\newcommand{\ing}{\mathrm{in}}

\newcommand{\loc}{\mathrm{loc}}

\newcommand{\Int}{\operatorname{int}}

\newcommand{\R}{\Bbb R}

\renewcommand{\Im}{\mathrm{Im}}

\newcommand{\stab}{\mathrm{stab}}

\renewcommand{\u}{\underline}

\DeclareMathOperator{\cyl}{cyl}

\DeclareFontFamily{U}{mathx}{}
\DeclareFontShape{U}{mathx}{m}{n}{<-> mathx10}{}
\DeclareSymbolFont{mathx}{U}{mathx}{m}{n}
\DeclareMathAccent{\widehat}{0}{mathx}{"70}
\DeclareMathAccent{\widecheck}{0}{mathx}{"71}

\newcommand{\qd}[1]{
  \accentset{ \vphantom{\mbox{\scalebox{0.5}{$\scriptscriptstyle (2)$}}}
    \smash[t]{\mbox{\scalebox{0.6}{$\scriptscriptstyle (2)$}}}
  }{#1}}

\newcommand{\lin}[1]{
  \accentset{  \vphantom{\mbox{\scalebox{0.5}{$\scriptscriptstyle (1)$}}}
    \smash[t]{\mbox{\scalebox{0.6}{$\scriptscriptstyle (1)$}}}
  }{#1}}
  
  \newcommand{\lint}[1]{
  \accentset{\vphantom{\mbox{\scalebox{0.5}{$\scriptscriptstyle (1)_{\mathrm{T}}$}}}
    \smash[t]{\mbox{\scalebox{0.6}{$\scriptscriptstyle (\mathrm{T})$}}}
  }{#1}}

\makeatletter
\renewcommand{\paragraph}{%
  \@startsection{paragraph}{4}%
  {\z@}{1.25ex \@plus 1ex \@minus .2ex}{-1em}%
  {\normalfont\normalsize\bfseries}%
}
\makeatother

\begin{document}

 \title{\makebox[0pt][c]{The moduli space of dynamical spherically symmetric black hole spacetimes}\\
 and the extremal threshold}

\author[1]{Yannis Angelopoulos\thanks{yannis@bimsa.cn}}
\author[2]{Christoph~Kehle\thanks{kehle@mit.edu}}
\author[3]{Ryan Unger\thanks{runger@berkeley.edu}}
\affil[1]{\small Beijing Institute of Mathematical Sciences and Applications,

No.~544, Hefangkou Village, Huairou District, 101408 Beijing, China \vskip.1pc \ 
}
\affil[2]{\small  Massachusetts Institute of Technology, Department of Mathematics,

Building 2, 77 Massachusetts Avenue, Cambridge, MA 02139, United States of America \vskip.1pc \ 
}

 \affil[3]{\small  University of California, Berkeley,  Department of Mathematics,
	
Evans Hall, Berkeley, CA 94720, United States of America \vskip.1pc \  
	}

\date{March 11, 2026}

\maketitle

\begin{abstract}
In this paper, we give a complete description of the black hole threshold, locally near the Reissner--Nordstr\"om family, in the infinite-dimensional moduli space $\mathfrak M$ of dynamical spherically symmetric solutions to the Einstein--Maxwell-neutral scalar field system. 

In a neighborhood of the full Reissner--Nordstr\"om family in $\mathfrak M$, we prove the following: (i) Any solution that forms a black hole eventually decays to a Reissner--Nordstr\"om black hole. (ii) Any solution that fails to collapse into a black hole eventually becomes superextremal along null infinity and exists globally in the domain of dependence of the bifurcate characteristic initial data. (iii) The subset of this neighborhood consisting of black hole solutions admits a $C^1$ foliation by hypersurfaces of constant final charge-to-mass ratio, up to and including extremality. (iv) The mutual boundary between the set of black hole solutions and noncollapsing solutions, i.e., the black hole threshold, is the extremal leaf of the foliation. Black holes which are not on the threshold are asymptotically subextremal.

Our quantitative control of near-threshold solutions allows us to prove ``universal'' scaling laws for the location of the event horizon and its final area and temperature (surface gravity), with critical exponent~$\frac 12$. Moreover, we show that the celebrated Aretakis instability is activated for an open and dense set of threshold solutions and that generic near-threshold subextremal black holes experience a transient horizon instability on the timescale of their inverse final temperature.
\end{abstract}

\thispagestyle{empty}
\newpage 
\setcounter{tocdepth}{4}
\setcounter{secnumdepth}{4}
\tableofcontents

\newpage
\section{Introduction} 

The \emph{Reissner--Nordstr\"om} family of metrics \cite{reissner1916eigengravitation, nordstrom1918energy},
\begin{equation*}
  g_{M,e}\doteq -\left(1-\frac{2M}{r}+\frac{e^2}{r^2}\right)dt^2+\left(1-\frac{2M}{r}+\frac{e^2}{r^2}\right)^{-1}dr^2+r^2(d\vartheta^2+\sin^2\vartheta\,d\varphi^2),
\end{equation*}
indexed by \emph{mass} $M>0$ and \emph{charge} $e\in\Bbb R$, is one of the simplest families of explicit solutions to the Einstein field equations (in fact, the coupled Einstein--Maxwell system). When $|e|\le M$, $g_{M,e}$ describes a \emph{black hole} spacetime. The case $|e|<M$ is known as \emph{subextremal}---which includes the celebrated \emph{Schwarzschild} solution at $e=0$---and $|e|=M$ is known as \emph{extremal}. Extremal Reissner--Nordstr\"om is the simplest example of an extremal black hole, that is, a stationary black hole spacetime with vanishing surface gravity. On the other hand, in the \emph{superextremal} case $|e|>M$, $g_{M,e}$ does not describe a black hole. 

Already in this simple explicit example, we can observe the transition between black hole formation and failure to form a black hole.  Understanding the \emph{black hole formation threshold} is a fundamental problem in classical general relativity that has been extensively studied numerically, starting with the influential work of Choptuik \cite{choptuik1993universality} on the spherically symmetric Einstein-scalar field model, in a regime where the critical solutions (i.e., the spacetimes lying on the threshold) are believed to be naked singularities. We refer to \cite{crit-review-update} for a survey of this problem, which is known as \emph{critical collapse}.  A key conjectured feature of critical collapse is the presence of certain \emph{universal scaling laws} of various quantities across the black hole threshold (for example, the mass of the ``first'' trapped surface in \cite{choptuik1993universality}), which are believed to be analogous to scaling laws associated with critical phenomena in statistical physics. 

By the incompleteness theorem \cite{Penrose} and Cauchy stability, if a critical solution contains a black hole, then it must be free of trapped surfaces and hence extremal. However, the exact Reissner--Nordstr\"om black holes are eternal and arise from two-ended Cauchy data, while the superextremal variants contain an eternal ``naked singularity'' that has historically caused much confusion. In fact, it was long thought that extremal black holes were forbidden from forming dynamically due to the \emph{third law of black hole thermodynamics} \cite{BCH,Israel-third-law}. Were this true, it would rule out extremal Reissner--Nordstr\"om as a critical solution.

In \cite{KU22}, the second- and third-named authors \emph{disproved} the third law in the Einstein--Maxwell-charged scalar field model and showed that an exactly extremal Reissner--Nordstr\"om event horizon can indeed form in gravitational collapse. Moreover, in \cite{KU24}, the same authors constructed one-parameter families of smooth, spherically symmetric solutions to the Einstein--Maxwell--charged Vlasov system (with regular centers!) which interpolate between dispersion and collapse and for which the critical solution contains an exact extremal Reissner--Nordstr\"om black hole at late times. This work showed that extremal Reissner--Nordstr\"om can rightly be considered a critical solution and clarified the role (or lack thereof) of the naked singularity of superextremal Reissner--Nordstr\"om in dynamical questions. 

The examples of \emph{extremal critical collapse}, that is, critical collapse involving extremal black holes, in \cite{KU24} consist of fine-tuned one-parameter families. As is believed to be true for other types of critical collapse, we conjecture that these examples are stable: a small perturbation of such a fine-tuned one-parameter family should result in a family which still crosses the black hole formation threshold and which forms an \emph{asymptotically} (as opposed to \emph{exact}) extremal Reissner--Nordstr\"om black hole at the critical value of the parameter. In other words, we conjecture that the set of asymptotically extremal Reissner--Nordstr\"om black holes is a codimension-one ``submanifold'' in the moduli space of solutions of the Einstein equations and (at least some portion of it) is a subset of the black hole formation threshold. For the precise form of this conjecture, we refer the reader to Section 1.5 of \cite{KU24} and \cref{sec:ECC} below.

\subsection{The main theorems}

\subsubsection{The setting and first version of the main theorem}

As a first step towards understanding the extremal regime of the black hole formation threshold in the gravitational collapse setting, we study the behavior of perturbations of the full family of Reissner--Nordstr\"om black holes in a neighborhood of the event horizon.

In this paper, we work in the spherically symmetric Einstein--Maxwell-neutral scalar field (EMSF) model, which consists of a spherically symmetric spacetime $(\mathcal M^{3+1},g)$, a spherically symmetric electromagnetic field~$\mathbf F$, together with a spherically symmetric massless scalar field $\phi:\mathcal M\to\Bbb R$ satisfying the system of equations
\begin{gather*}
    R_{\mu\nu}-\tfrac 12 Rg_{\mu\nu}= 2\mathbf F_{\mu\alpha}\mathbf F_\nu{}^\alpha-\tfrac 12 g_{\mu\nu}\mathbf F_{\alpha\beta} \mathbf F^{\alpha\beta}+2\partial_\mu\phi\partial_\nu\phi-g_{\mu\nu}\partial_\alpha\phi\partial^\alpha\phi,\\
  \nabla_\mu \mathbf F^{\mu\nu}=0, \qquad \Box_g\phi =0.
\end{gather*}
 This is one of the simplest self-gravitating models in which one can entertain dynamical nonlinear perturbations of Reissner--Nordstr\"om, and has been used in several important works in recent years \cite{Dafermos-thesis,dafermos2005interior,Price-law,Reall-numerical,LukOhI,luk2019strong,Gautam}. See already \cref{sec:the-model} for the precise definitions and equations. 

\begin{figure}
\centering{
\def\svgwidth{28pc}
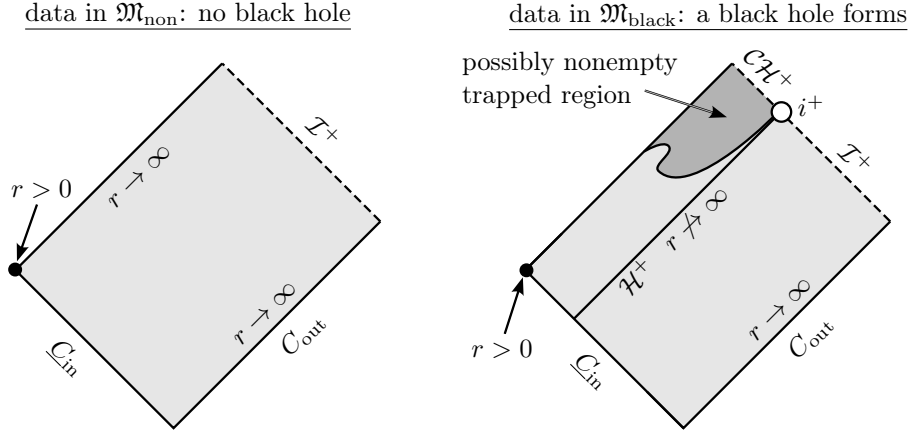}
\caption{Penrose diagrams depicting evolutions of characteristic data in the sets $\mathfrak M_\mathrm{non}$ and $\mathfrak M_\mathrm{black}$. Spacetimes arising from $\mathfrak M_\mathrm{non}$ have an incomplete null infinity $\mathcal I^+$ because the ingoing cone $\underline C{}_\ing$ is incomplete and no black hole has formed. However, $r\to \infty$ along every outgoing cone and the solution remains regular up to and including the final outgoing cone. Every solution arising from data in $\mathfrak M$ has one of these two Penrose diagrams (for more information on the possibilities in the black hole interior, we refer to \cite{Kommemi13}).} \label{fig:dichotomy}
\end{figure}

Consider the moduli space $\mathfrak M$\index{M@$\mathfrak M$, the moduli space} of characteristic data for the spherically symmetric EMSF system posed on a bifurcate null hypersurface $\mathcal C=C_\mathrm{out}\cup \u C{}_\ing$, as depicted in \vref{fig:dichotomy}. Note that the outgoing cone $C_\out$ goes to \emph{null infinity} $\mathcal I^+$, where $r=\infty$, but the ingoing cone $\u C{}_\ing$ stops at (and includes) a non-central symmetry sphere with $r>0$. Note that spherically symmetric EMSF solutions with nonvanishing charge $Q$ cannot have a regular center, so stopping before $r=0$ on $\u C{}_\ing$ is a natural condition. For the precise definition of $\mathfrak M$, see already \cref{sec:intro-mod-space,sec:AF-seed-data}.

By classical work of Dafermos \cite{dafermos-trapped-surface}, $\mathfrak M$ can be written as a disjoint union  $\mathfrak M_\mathrm{black}\sqcup\mathfrak M_\mathrm{non}$, where $\mathfrak M_\mathrm{black}$\index{M@$\mathfrak M_\mathrm{black}$, data leading to the formation of a black hole} is closed and consists of data leading to the formation of a black hole, and $\mathfrak M_\mathrm{non}$ \index{M@$\mathfrak M_\mathrm{non}$, noncollapsing data} is open and gives rise to solutions failing to form a black hole in the domain of dependence of $\mathcal C$ (``non'' for \emph{noncollapse}). The set $\mathfrak M_\mathrm{non}$ is characterized by $r\to \infty$ along every outgoing cone, which means that $\mathcal I^+$ extends ``as far as the data allows.'' (And yet, $\mathcal I^+$ is incomplete in the sense of Christodoulou \cite{christodoulou1999global} in this case!) Note that in models which allow for a regular center (such as charged scalar field or charged Vlasov), one could entertain \emph{dispersion} (decay to Minkowski space as opposed to just noncollapse). We will discuss this aspect of the problem in \cref{sec:ECC} below. We emphasize that incompleteness of $\mathcal I^+$ in $\mathfrak M_\mathrm{non}$ does not signal the presence of a naked singularity, but simply that one has ``run out of data'' in the setup.

The \emph{black hole threshold} in this setting is given by the topological boundary of $\mathfrak M_\mathrm{black}$ in $\mathfrak M$, $\partial \mathfrak M_\mathrm{black}$. Clearly, extremal Reissner--Nordstr\"om black holes belong to $\partial \mathfrak M_\mathrm{black}$. The goal of this paper is to describe the structure of $\partial \mathfrak M_\mathrm{black}$ in a neighborhood of Reissner--Nordstr\"om and to understand the behavior of solutions on and near it. We now state the first (informal) version of our main theorem:

\begin{thm*} In a neighborhood of extremal Reissner--Nordstr\"om in the moduli space of solutions to the spherically symmetric Einstein--Maxwell-neutral scalar field model,  $\partial\mathfrak M_\mathrm{black}$ is a regular hypersurface (codimension-one submanifold) consisting of black holes which decay to extremal Reissner--Nordstr\"om. 
\end{thm*}

This theorem can be viewed as resolving the spherically symmetric analogue of a conjecture by Dafermos--Holzegel--Rodnianski--Taylor for extremal Reissner--Nordstr\"om/very slowly rotating Kerr--Newman outside of symmetry (see \cite[Section IV.2]{DHRT}). We refer also to the essay \cite{dafermos2025stability} by Dafermos and to \cref{sec:EHC,sec:ECC} of this paper for discussion and further related conjectures. Moreover, this theorem and the associated results about transient behavior for near-extremal subextremal black holes expand on, and fully confirm, the pioneering numerical study of Murata--Reall--Tanahashi \cite{Reall-numerical}.

We will now expand on this by stating four more detailed theorems and making various remarks. The proof of the theorem is an extension of the black hole stability problem, and requires simultaneously understanding the \emph{entire range of black hole parameters}. We present this uniform stability statement in \cref{sec:intro-refined} below. In \cref{sec:intro-foliation}, we present the \emph{isologous foliation} of $\mathfrak M_\mathrm{black}$ near extremal Reissner--Nordstr\"om, consisting of hypersurfaces with constant final parameter ratio. The leaves of this foliation which correspond to extremality are precisely the hypersurfaces constituting the black hole threshold. Finally, we describe fine properties of dynamical extremal and near-extremal black holes in \cref{sec:intro-scaling,sec:intro-instability} (``universal'' scaling laws and horizon instabilities, respectively). We will outline the proofs of our main results in \cref{sec:intro-proof-overview}.

 \subsubsection{The refined dichotomy and uniform asymptotic stability}\label{sec:intro-refined}
 
 \begin{figure}
\centering{
\def\svgwidth{28pc}
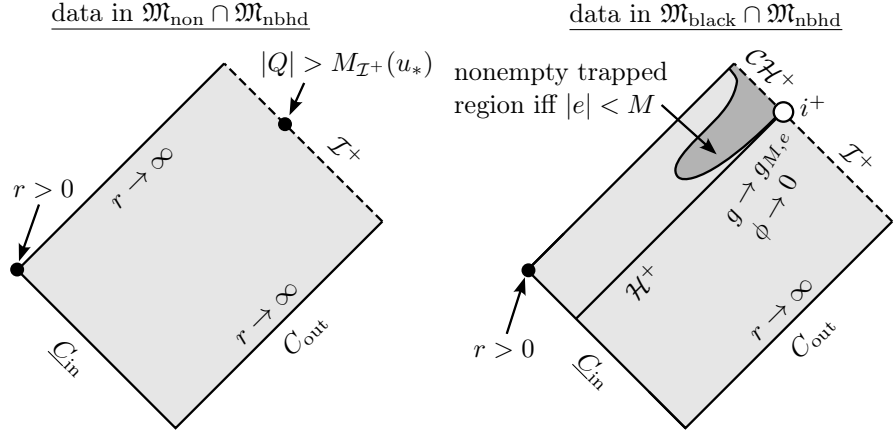}
\caption{Penrose diagrams depicting evolutions of characteristic data in the sets $\mathfrak M_\mathrm{non}\cap\mathfrak M_\mathrm{nbhd}$ and $\mathfrak M_\mathrm{black}\cap\mathfrak M_\mathrm{nbhd}$. In the first case, $|Q|>M_{\mathcal I^+}$ somewhere along $\mathcal I^+$. In the second case, the solution converges to a Reissner--Nordstr\"om black hole. The statement about (existence and absence of) trapped surfaces follows from \cite{C-91,dafermos2005interior,AKU24}.}  
\label{fig:refined-dichotomy}
\end{figure}
 
 Dafermos' dichotomy $\mathfrak M =\mathfrak M_\mathrm{non}\sqcup \mathfrak M_\mathrm{black}$ does not directly give much information about the kinds of black hole spacetimes generated from data in $\mathfrak M_\mathrm{black}$ or about why solutions fail to collapse in $\mathfrak M_\mathrm{non}$. In this model, it is reasonable to expect that all black holes are asymptotically Reissner--Nordstr\"om (see already \cref{sec:large-data}). In the perturbative (small scalar field) regime, we show that this is indeed the case and show that ``eventual superextremality'' is the cause of noncollapse.
 
 In the following, we denote data sets in $\mathfrak M$ by the letter $\Psi$, and the associated maximal future globally hyperbolic development by $\mathcal S[\Psi]$. The topology of $\mathfrak M$ includes a norm for the characteristic initial data of $\phi$, so when we speak about a neighborhood of an electrovacuum solution (Reissner--Nordstr\"om), we are implicitly indicating that the scalar field is (at least initially) small.
 
\begin{thmrough}[Refined dichotomy and uniform stability]\label{thm:intro-uniform-stability} There exists a neighborhood $\mathfrak M_\mathrm{nbhd}$\index{M@$\mathfrak M_\mathrm{nbhd}\subset\mathfrak M$, neighborhood of the Reissner--Nordstr\"om family} of the full family of Reissner--Nordstr\"om solutions in $\mathfrak M$ such that for any $\Psi\in \mathfrak M_\mathrm{nbhd}$, the following holds.
\begin{enumerate} 
\item \ul{Noncollapse criterion}: $\Psi\in \mathfrak M_\mathrm{non}$ if and only if $|Q|>M_{\mathcal I^+}$ somewhere on $\mathcal I^+$, where $Q$\index{Q@$Q$, electromagnetic charge} is the conserved charge of $\mathcal S[\Psi]$ and $M_{\mathcal I^+}$ is the \ul{Bondi mass} function of $\mathcal S[\Psi]$. In other words, becoming eventually superextremal at null infinity is precisely the obstruction to black hole formation in this setting.

\item \ul{Universality of the Reissner--Nordstr\"om family}: If $\Psi\in \mathfrak M_\mathrm{black}$, then the black hole exterior of $\mathcal S[\Psi]$, up to and including the event horizon $\mathcal H^+$, decays to a member of the Reissner--Nordstr\"om family of black holes relative to a teleologically normalized double null gauge in the following sense: There exist parameters $(M,e)$ with $|e|\le M$ such that the metric $g$ decays in $C^0$ to the Reissner--Nordstr\"om metric $g_{M,e}$ (written in Eddington--Finkelstein double null coordinates), the renormalized Hawking mass $\varpi$ decays uniformly to $M$, and the scalar field $\phi$ decays uniformly to zero pointwise and with respect to a hierarchy of energies. Moreover, each quantity that decays does so at least as fast as \ul{uniform rate}, i.e., the overall constants and polynomial decay upper bounds do not depend on $\Psi$ or $(M,e)$. 
\end{enumerate}
\end{thmrough}

For the precise statement, see already \cref{thm:uniform-stability-RN,thm:dichotomy-revisited}. The energies used to measure the decay of $\phi$ depend in a nontrivial way on the parameters $(M,e)$; see already \cref{sec:intro-unifying} for the definitions. For discussion about the structure of the black hole interior in $\mathfrak M_\mathrm{nbhd}$, we refer the reader to \cref{sec:interior} below. 

Stability of subextremal Reissner--Nordstr\"om in this model has been understood for many years, due to the work of Dafermos--Rodnianski \cite{Price-law} and later improvements by Luk--Oh \cite{luk2019strong} (see also \cite{Gautam}). In our previous paper \cite{AKU24}, we addressed the \emph{codimension-one stability} of the extremal case, which is more delicate than the subextremal case because it lacks the stabilizing mechanism of the celebrated horizon redshift effect. (Codimension one is sharp in view of the fact that the extremal case is codimension one in the full family of exact Reissner--Nordstr\"om solutions.) Part 2.~of \cref{thm:intro-uniform-stability} unifies the subextremal and extremal stories from \cite{luk2019strong} and \cite{AKU24}. 

In order to prove this refined dichotomy, we develop a new modulation scheme based on measurements of the charge-to-mass ratio on dyadic time steps \emph{at null infinity $\mathcal I^+$}. This allows us to take advantage of monotonicities of the system in a new way. The heart of the stability argument lies in unified energy estimates for $\phi$ in the full range of Reissner--Nordstr\"om parameters (including superextremal in a sense that will become clear later). These estimates specialize in the extremal case to the ones in \cite{AKU24}, but our interpolation with the redshift estimate in the subextremal case is novel.

\begin{rk}[Comparison with \cite{AKU24} I]
The teleological double null gauge referred to in \cref{thm:intro-uniform-stability} is morally the same as the one employed in \cite{AKU24}, but differs in an important way in its construction (see already \cref{sec:intro-gauge}) because of our new modulation scheme at $\mathcal I^+$. (The eschatological gauges differ only in the $\u v$-component, both $\u u$-coordinates are Bondi normalized.) It is easy to see that asymptotic stability in the ``new'' gauge implies asymptotic stability with respect to the ``old'' gauge.
\end{rk}

\begin{rk} The refined dichotomy presented in  \cref{thm:intro-uniform-stability} can only be true in a certain well-chosen neighborhood $\mathfrak M_\mathrm{nbhd}$: in general, there exist data in $\mathfrak M_\mathrm{non}$ which are nowhere superextremal on $\mathcal I^+$. For instance, even in full vacuum (zero charge), if the Bondi mass $M$ is small compared to the area-radius of the final solid point in $\u C{}_\ing$, then $\Psi$ fails to collapse. Indeed, $\mathcal S[\Psi]$ would be isometric to a Schwarzschild solution, but with $r>2M$ on the final cone, and hence there is no event horizon in the piece of Schwarzschild under consideration. Examples of this type are eliminated by imposing a relation between $r$ of the bifurcation sphere and the mass in the definition of $\mathfrak M_\mathrm{nbhd}$.
\end{rk}

\begin{rk}\label{rk:disp-whole-slab}
When $\Psi\in\mathfrak M_\mathrm{non}$, the final outgoing cone is at a finite Bondi time and the scalar field does not actually decay to zero on the final outgoing cone. (The estimates very near this cone are very general and are essentially contained already in \cite{dafermos-trapped-surface}.) It is an interesting problem to obtain more precise estimates for the geometry and scalar field in $\mathfrak M_\mathrm{non}\cap\mathfrak M_\mathrm{nbhd}$, see already \cref{rk:dispersion-long-time}.
\end{rk}

\begin{rk}
We emphasize that the decay rates in \cref{thm:intro-uniform-stability} are uniform, not sharp. For any given subextremal parameter ratio, the results in \cite{Price-law,luk2019strong,Gautam} give (at late times) stronger decay estimates, but with overall constants depending nontrivially on $|e|/M$. It is a very interesting problem to obtain asymptotics in $\mathfrak M_\mathrm{black}\cap\mathfrak M_\mathrm{nbhd}$ in a unified manner. 
\end{rk}

\subsubsection{The isologous foliation and the threshold property}\label{sec:intro-foliation} 

We now seek to precisely characterize the threshold $\partial\mathfrak M_\mathrm{black}\cap \mathfrak M_\mathrm{nbhd}$, which we do by means of the \emph{isologous foliation}. By \cref{thm:intro-uniform-stability}, any $\Psi\in\mathfrak M_\mathrm{black}\cap\mathfrak M_\mathrm{nbhd}$ converges to a Reissner--Nordstr\"om black hole with some parameters $(M,e)$; we may therefore define its 
\emph{(signed) final parameter ratio} by \index{P@$\mathscr P_\infty$, (signed) final parameter ratio}
 \[\mathscr P_\infty(\Psi)\doteq \frac{e}{M}\in [-1,1].\]
For \index{sigma@$\sigma$, level value of $\mathscr P_\infty$} $\sigma\in[-1,1]$, we then define \index{M@$\mathfrak M_\mathrm{stab}^\sigma$, stable manifold with final parameter ratio $\sigma$, isologous hypersurface}
\[\mathfrak M_\mathrm{stab}^\sigma\doteq \{\Psi\in\mathfrak M_\mathrm{black}\cap\mathfrak M_\mathrm{nbhd}:\mathscr P_\infty(\Psi)=\sigma \},\]
i.e., the set of all black holes in $\mathfrak M_\mathrm{black}\cap\mathfrak M_\mathrm{nbhd}$ with final parameter ratio $\sigma$ (hence, \emph{isologous}).

\begin{thmrough}[Isologous foliation and threshold property] \label{thm:intro-foliation} 
For $\mathfrak M_\mathrm{nbhd}$ chosen sufficiently small, the following holds: The isologous sets $\{\mathfrak M_\stab^\sigma\}_{\sigma\in[-1,1]}$ are $C^1$ hypersurfaces (codimension-one submanifolds) in $\mathfrak M_\mathrm{nbhd}$ which form a $C^1$ foliation of $\mathfrak M_\mathrm{black}\cap\mathfrak M_\mathrm{nbhd}$. Moreover, the local black hole threshold is characterized as
\[\partial\mathfrak M_\mathrm{black}\cap \mathfrak M_\mathrm{nbhd}=\mathfrak M_\stab^{+1}\sqcup\mathfrak M_\stab^{-1},\] the set of asymptotically extremal black holes. Each of the hypersurfaces $\mathfrak M_\stab^{+1}$ and $\mathfrak M_\stab^{-1}$ locally separate $\mathfrak M$ into $\mathfrak M_\mathrm{black}$ and $\mathfrak M_\mathrm{non}$. 
\end{thmrough}

\begin{figure}
\centering{
\def\svgwidth{18pc}
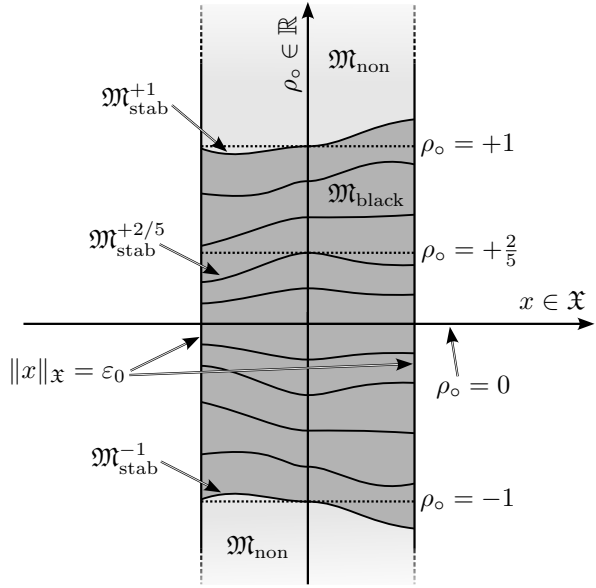}
\caption{A schematic depiction of the set $\mathfrak M_\mathrm{nbhd}$ and the isologous foliation from \cref{thm:intro-foliation}. The moduli space $\mathfrak M$ has a natural product structure, where the ordinate is $\rho_\circ$, the charge-to-mass ratio of the bifurcation sphere of $\mathcal C$. The abscissa $x$ lives in a Banach space $\mathfrak X$ and accounts for the remaining degrees of freedom (such as the initial data for $\phi$). Therefore, the horizontal axis represents an infinite-dimensional space in this picture and the stable manifolds $\mathfrak M_\mathrm{stab}^{\sigma}$ are infinite-dimensional hypersurfaces. Essentially, $\mathfrak M_\mathrm{nbhd}=\{(x,\rho_\circ):\|x\|_\mathfrak{X}<\ve_0\},$ where $\ve_0$ is a small parameter. The dashed horizontal lines are the constant-$\rho_\circ$ coordinate hyperplanes. The black hole portion of $\mathfrak M_\mathrm{nbhd}$ is shaded dark gray and the noncollapse portion is shaded light gray. For the definition of $\mathfrak M$ and its topology, see already \cref{sec:intro-mod-space}.}
\label{fig:foliation-intro-1}
\end{figure}

For the precise statement, see already \cref{thm:main}. See \vref{fig:foliation-intro-1} for a schematic picture of $\mathfrak M_\mathrm{nbhd}$ and the isologous foliation $\{\mathfrak M_\stab^\sigma\}_{\sigma\in[-1,1]}$. We set $\mathfrak M_\mathrm{stab}^{+1,-1}\doteq \mathfrak M_\stab^{+1}\sqcup\mathfrak M_\stab^{-1}$. For discussion of possible higher regularity of the foliation and the extremal threshold, see already \cref{sec:higher}.

In our previous work \cite{AKU24}, we identified the extremal isologous sets $\mathfrak M^{\pm 1}_\mathrm{stab}\subset\mathfrak M_\mathrm{nbhd}$ by a shooting-type argument (inspired by \cite{DHRT}) and proved decay to extremal Reissner--Nordstr\"om consistent with \cref{thm:intro-uniform-stability} above. In that setting, ``codimension one'' is interpreted in an \emph{effective} sense, namely, that in order to land on $\mathfrak M^{\pm 1}_\mathrm{stab}$ one only has to modulate one parameter of the data.\footnote{Note that in the case of nonlinear stability of Schwarzschild outside of vacuum considered in \cite{DHRT}, the stable ``submanifold'' of Schwarzschild has codimension three in this sense (one has to tune three parameters). This is because the Kerr family is correctly parametrized by \emph{four} parameters (a mass and three components of specific angular momentum).} This type of ``shooting argument'' fundamentally gives no information about the regularity or structure of $\mathfrak M_\mathrm{stab}^{\pm1}$.

Proving regularity of such stable manifolds/isologous hypersurfaces is technically challenging because the black hole stability analysis involves a choice of teleological gauge, which makes comparing nearby solutions (and hence proving ``difference estimates'') subtle. Moreover, even on the isologous hypersurface, the final masses of solutions are in general different, so that once differences are actually defined, they do not even necessarily decay. These issues have been handled in the subextremal case by Luk--Oh in \cite{luk2019strong} (one can extract Lipschitz stable manifolds in the subextremal regime from their work), but reaching extremality requires several new ideas.

In this paper, we develop machinery allowing us to take the Fr\'echet derivative of the final parameter ratio $\mathscr P_\infty$ with respect to the natural Banach space structure of the moduli space $\mathfrak M$. In fact, we show that $\mathscr P_\infty$ is a $C^1$ \emph{submersion}, though interpreting this is rather delicate since the domain $\mathfrak M_\mathrm{black}\cap\mathfrak M_\mathrm{nbhd}$ is not known to be a manifold a priori. Nevertheless, the regularity of $\mathfrak M_\mathrm{stab}^\sigma$ and the foliation property follow from the implicit function theorem once the submersion property has been established. The proof that $\mathfrak M_\mathrm{stab}^{+1,-1}$ constitutes the threshold uses part 1.~of \cref{thm:intro-uniform-stability}, because we show that if $\Psi$ lies ``beyond'' $\mathfrak M_\mathrm{stab}^{+1,-1}$, then it is superextremal somewhere along $\mathcal I^+$.

Computing the Fr\'echet derivative $\mathscr P_\infty'$ lies at the heart of this paper. We show that Gateaux derivatives of $\mathscr P_\infty$ can be computed using linearized gravity on \emph{dynamical} backgrounds, and continuity (in $\Psi$) of these derivatives is inferred through a delicate limiting argument. Since the domain of $\mathscr P_\infty$ is not open, this requires considerable care and the procedure will be outlined in \cref{sec:overview-II} below.

\begin{rk}[Physical interpretation of the isologous foliation] The regularity of the isologous foliation implies that forming an asymptotically extremal black hole in $\mathfrak M_\mathrm{nbhd}$ is no more difficult than forming a black hole of any other fixed charge to mass ratio: the leaves of the foliation do not ``bunch up'' or become ``harder'' to tune to as one approaches the threshold. In particular, \cref{thm:intro-foliation} disproves in this context the ``generalized'' third law of black hole thermodynamics formulated by Dafermos \cite{dafermos2025stability}. 
\end{rk}

\begin{rk}[No trapped surfaces on the threshold]\label{rk:no-trapped-surfaces}
In \cite{AKU24}, using an argument of Kommemi, we showed that an asymptotically extremal black hole in the spherically symmetric EMSF model does not contain trapped surfaces behind the event horizon. In particular, this result applies to $\mathfrak M_\mathrm{stab}^{+1,-1}$. \cref{thm:intro-foliation} provides a new proof of this fact, since having trapped surfaces is a stable property (having a trapped surface is an open condition in $\mathfrak M$) and solutions in $\mathfrak M_\mathrm{non}$ are manifestly free of trapped surfaces. 
\end{rk}

\begin{rk}[Asymptotically subextremal black holes]\label{rk:intro-subextremal}
Let {\index{M@$\mathfrak M_\mathrm{sub}$, subextremal black hole data}} $\mathfrak M_\mathrm{sub}$  denote the subset of $\mathfrak M_\mathrm{black}$ with asymptotically subextremal final parameter ratio. By the seminal work of Dafermos \cite{dafermos2005interior}, any such solution has trapped surfaces behind the horizon. As we just recalled, this is an open condition, and it follows that $\mathfrak M_\mathrm{sub}$ is open. \cref{thm:intro-foliation} provides a new proof that $\mathfrak M_\mathrm{sub}\cap \mathfrak M_\mathrm{nbhd}$ is open, since $\mathscr P_\infty$ is continuous on $\mathfrak M_\mathrm{black}\cap\mathfrak M_\mathrm{nbhd}$ and
\[\mathfrak M_\mathrm{sub}\cap \mathfrak M_\mathrm{nbhd}= \{|\mathscr P_\infty|<1\}\cap\mathfrak M_\mathrm{nbhd}.\]
\end{rk}

\begin{rk}[Comparison with \cite{AKU24} II]
For both fundamental technical reasons and convenience, the moduli space $\mathfrak M$ in the present paper is defined slightly differently than the moduli space $\mathfrak M'$ in \cite{AKU24}. Most importantly, $\mathfrak M$ involves a weighted $C^2$ norm of the scalar field $\phi$, whereas $\mathfrak M'$ only involved a weighted $C^1$ norm. If we strengthen the norm on $\phi$ in $\mathfrak M'$ to the one used here, then $\mathfrak M$ and $\mathfrak M'$ are diffeomorphic as will be explained in \cref{sec:intro-mod-space} below, and the isologous hypersurfaces $\mathfrak M_\mathrm{stab}^{\pm 1}$ in the present paper do correspond to the stable ``manifold'' constructed in \cite{AKU24}. In particular, the methods of the present paper prove uniqueness of the modulation parameter $\alpha_\star$ in \cite{AKU24} (recall \cite[Remark 8.7]{AKU24}).
\end{rk}

\begin{rk}\label{rk:dispersion-long-time}
For $\Psi\in\mathfrak M_\mathrm{non}$  close to the extremal threshold $\mathfrak M_\mathrm{stab}^{+1,-1}$, we show that the geometry of $\mathcal S[\Psi]$ is quantitatively close to superextremal Reissner--Nordstr\"om up to a large dyadic Bondi time depending on how close $\Psi$ is to the threshold. As already mentioned in \cref{rk:disp-whole-slab}, it is an interesting problem to understand the behavior of $\mathcal S[\Psi]$ beyond this specific dyadic time. \end{rk}

\subsubsection{``Universal'' scaling laws}\label{sec:intro-scaling}

Our proof of \cref{thm:intro-foliation} gives detailed information on the behavior of solutions on and near the extremal threshold. In particular, we can show certain natural universal \emph{scaling laws} for generic one-parameter families of solutions crossing the threshold (\emph{interpolating families} in terminology of \cite{KU24}). Scaling laws of this type are commonly studied numerically in other examples of critical phenomena in gravitational collapse (see the survey \cite{crit-review-update}).

Let $\Psi\in \mathfrak M_\mathrm{black}$. As before, we can associate to $\Psi$ its final parameter ratio $\mathscr P_\infty(\Psi)$. We now define the \emph{final event horizon area} $\mathscr A(\Psi)$, the \emph{location of the event horizon} $\mathscr U(\Psi)$ (that is, the retarded time coordinate of the event horizon $\mathcal H^+$ in the initial data gauge), and the \emph{final event horizon temperature} $\mathscr T(\Psi)$, which is defined by \index{A@$\mathscr A$, final horizon area function} \index{U@$\mathscr U$, final horizon location function} \index{T@$\mathscr T$, final horizon temperature function}
\[\mathscr T(\Psi) = \frac{\bm\kappa(M,e)}{2\pi},\]
where $\bm\kappa(M,e)$ is the surface gravity of the event horizon of the Reissner--Nordstr\"om black hole that $\mathcal S[\Psi]$ decays to.

\begin{thmrough}[Scaling laws] \label{thm:intro-scaling} 
For $\mathfrak M_\mathrm{nbhd}$ chosen sufficiently small, the following holds:
 Let $\{ \Psi_p\}_{p\in[0,1]}$ be a smooth one-parameter family of data sets in $\mathfrak M_\mathrm{nbhd}$ which crosses $\mathfrak M_\mathrm{stab}^{+1,-1}$ transversally such that $\Psi_p\in \mathfrak M_\mathrm{black}$ for $p\in [0,p_*]$ and $\Psi_{p_*}\in \mathfrak M_\mathrm{stab}^{+1,-1}$. Then it holds that
 \begin{align}
\label{eq:intro-P-scaling} |\mathscr P_\infty(\Psi_p)-\mathscr P_\infty(\Psi_{p_*})|&\sim |p-p_*|,\\
 \label{eq:intro-A-scaling}  |\mathscr A(\Psi_p)-\mathscr A(\Psi_{p_*})|&\sim |p-p_*|^{1/2},\\  \label{eq:intro-U-scaling}  |\mathscr U(\Psi_p)-\mathscr U(\Psi_{p_*})|&\sim |p-p_*|^{1/2},\\  
 \label{eq:intro-T-scaling}  |\mathscr T(\Psi_p)-\mathscr T(\Psi_{p_*})|&\sim |p-p_*|^{1/2}
 \end{align}
 as $p\nearrow p^*$. The implicit constants in these relations may depend on the specific curve, but the powers are \ul{universal}.\end{thmrough} \index{p@$p$, parameter in the context of critical phenomena}
 
 For the precise statement, see already \cref{thm:scaling}. The scaling of the final temperature presented here was first observed numerically by Murata--Reall--Tanahashi in \cite{Reall-numerical}.
 
\begin{rk}
The transversality condition in \cref{thm:intro-scaling} is equivalent to 
\begin{equation}
    \left.\frac{d}{dp}\right|_{p=p_*}\mathscr P_\infty(\Psi_p)\ne 0,\label{eq:intro-transversality}
\end{equation}
which is an open and dense condition on the space of smooth curves because $\mathscr P_\infty$ is $C^1$. 
\end{rk}
 
 The scaling law for $\mathscr P_\infty$ follows immediately from the fact that $\mathscr P_\infty$ is a $C^1$ submersion as was stated in the previous section. The fractional scaling laws for $\mathscr A$ and $\mathscr T$ then follow immediately from the formulas \index{r@$r_+\doteq M+\sqrt{M^2-e^2}$, RN outer horizon area-radius} \index{kappa@$\bm\kappa$, RN surface gravity}
 \[r_+(M,e)=M+\sqrt{M^2-e^2},\quad \bm\kappa(M,e) = \frac{\sqrt{M^2-e^2}}{(M+\sqrt{M^2-e^2})^2},\] where $r_+$ is the area-radius of the event horizon. The power $\frac 12$ is simply because of the square root in these formulas. The square root scaling for $\mathscr U$ is also natural: the initial data gauge has $\partial_ur=-1$ along $\u C{}_\ing$ (the initial ingoing cone), which links $\mathscr U$ to $r_+$ in the Reissner--Nordstr\"om family. The fact that the nonlinear error terms are subleading in powers of $p-p_*$ requires a very nontrivial calculation in linearized gravity which is more subtle than the proof of \cref{thm:intro-foliation}.
 
 In fact, we show that the functions
 \[\mathscr A,\mathscr U,\mathscr T:\mathfrak M_\mathrm{black}\to \Bbb R\]
 are exactly $C^{1/2}$ near $\mathfrak M_\mathrm{stab}^{+1,-1}$ and not $C^\alpha$ for any $\alpha>\frac 12$. However, when restricted to the set of asymptotically subextremal black holes $\mathfrak M_\mathrm{sub}\cap\mathfrak M_\mathrm{nbhd}$, we prove that these functions are all $C^1$. Again, this is because the function $x\mapsto \sqrt x$ is nonsmooth only at $x=0$. Combined with methods from \cite{Price-law}, it should not be difficult to prove that $\mathscr A$, $\mathscr U$, and $\mathscr T$ are $C^1$ everywhere in $\mathfrak M_\mathrm{sub}$, not just restricted to $\mathfrak M_\mathrm{nbhd}$.

\subsubsection{Horizon instabilities on and near the threshold}\label{sec:intro-instability}

Let \index{Y@$Y\doteq(\partial_u r)^{-1}\partial_u$, nondegenerate ingoing null vector field} \[Y \doteq \frac{1}{\partial_ur}\partial_u\] denote the ingoing gauge-invariant null derivative which is transverse to the event horizon $\mathcal H^+$ in the black hole case. In a remarkable work \cite{Aretakis-instability-1,Aretakis-instability-2,Aretakis-instability-3}, Aretakis showed that if $\phi$ (spherically symmetric) solves the linear wave equation on extremal Reissner--Nordstr\"om, then $Y\psi$ is conserved along $\mathcal H^+$, where $\psi\doteq r\phi$.\index{ps@$\psi\doteq r \phi$, rescaled scalar field} In particular, if $Y\psi|_{\mathcal H^+}$ is initially nonvanishing (a generic condition), then $Y\psi$ does not decay, in sharp contrast to the subextremal case. This constant is denoted by $H_0[\phi]$ and is called the \emph{Aretakis charge} of $\phi$. Moreover, if $H_0[\phi]\ne 0$, $Y^2\psi$ grows linearly in $v$ along $\mathcal H^+$, and higher derivatives grow at correspondingly higher polynomial rates. This instability for the linear wave equation is known as the \emph{Aretakis instability}. 

By Cauchy stability, one immediately expects to find a remnant of the Aretakis instability in the near-extremal case, which we call a \emph{transient horizon instability}. This is precisely the behavior observed in the numerical study of Murata--Reall--Tanahashi \cite{Reall-numerical}, who find that the instability along the event horizon persists for timescales of order $\bm\kappa^{-1}$, where $\bm\kappa$ is the horizon redshift. It is easy to heuristically identify this timescale: the exponential factor in the redshift calculation at the horizon (see for instance \cite{Price-law}) is $\sim e^{-2\bm\kappa v}$. Therefore, as long as $v\les \beta$,  where $\beta \doteq \frac{1}{\mathscr{T}}= \frac{2\pi}{\bm\kappa}$ is the inverse temperature,  \index{beta@$\beta$, inverse temperature, redshift time, transient time scale} the subextremal case should behave like the extremal case, where $\beta =\infty$.

In our previous work \cite{AKU24}, we showed that the Aretakis instability holds on $\mathfrak M_\mathrm{stab}^{+1,-1}$, which is now complicated by the nonlinearity of the EMSF system. For any $\Psi\in \mathfrak M_\mathrm{stab}^{+1,-1}$, we showed that the \emph{asymptotic Aretakis charge} \index{H1@$H_0[\phi]$, asymptotic Aretakis charge}
\begin{equation*}
    H_0[\phi]\doteq \lim_{v\to\infty}Y\psi|_{\mathcal H^+}
\end{equation*}  
exists and that $Y\psi$ is ``almost conserved'' along $\mathcal H^+$ in the sense that 
\begin{equation}\label{eq:intro-Aretakis-1}
    Y\psi|_{\mathcal H^+} = H_0[\phi] + O\big(\ve^3 v^{-1+\delta}\big),
\end{equation}
where $\ve$ is an \emph{upper bound} for the initial data of $\phi$. For instance, if $Y\psi=\ve$ on $\u C{}_\ing$ exactly, then $|H_0[\phi]|\sim \ve\ne 0$ if $\ve$ is sufficiently small. While this shows that the nonlinear Aretakis instability is activated for some data (it is easy to arrange open sets of data for which $|Y\psi|_{\u C{}_\ing}|\sim \ve$), \eqref{eq:intro-Aretakis-1} alone cannot prove true genericity. If $H_0[\phi]\ne 0$, we further showed that $Y^2\psi$ grows linearly in $v$ along $\mathcal H^+$. This confirmed numerical results of \cite{Reall-numerical}.

Through the coupling with the Einstein equations, the Aretakis instability for $\phi$ implies an instability for the Ricci tensor. The curvature quantities $\Ric(Y,Y)$ and $\nabla_Y\Ric(Y,Y)$ vanish on any Reissner--Nordstr\"om background, but if $H_0[\phi]\ne 0$, then $\Ric(Y,Y)$ is ``almost conserved'' and $\nabla_Y\Ric(Y,Y)$ grows linearly along $\mathcal H^+$ in the coupled problem.  

In this paper, we can use our detailed understanding of the threshold $\mathfrak M_\mathrm{stab}^{+1,-1}$ to show that the nonlinear Aretakis instability is activated on a truly generic subset of $\mathfrak M_\mathrm{stab}^{+1,-1}$. In fact, we show that the sets in $\mathfrak M_\mathrm{stab}^{+1,-1}$ with constant asymptotic Aretakis charge form a $C^1$ foliation of $\mathfrak M_\mathrm{stab}^{+1,-1}$ and therefore the Aretakis instability occurs away from a codimension-one submanifold of $\mathfrak M_\mathrm{stab}^{+1,-1}$ (which is of course codimension two in $\mathfrak M$). Moreover, we quantify precisely the transient instability in the near-extremal regime in the framework of interpolating families as in \cref{thm:intro-scaling} and prove that the transient instability is active up to the timescale $v\sim \beta$. In particular, this is well beyond the timescale $\sim \log\beta$ that follows from Cauchy stability considerations.

\begin{thmrough}[Horizon instabilities] \label{thm:intro-instability} For $\mathfrak M_\mathrm{nbhd}$ chosen sufficiently small, the following holds:
\begin{enumerate}
\item \label{thm:intro-instab-1} For every $h\in \Bbb R$, the set\index{H5@$\mathfrak H_h\doteq \{\Psi\in \mathfrak M_\mathrm{ext}:H_0[\phi]=h\}$, level set in extremal manifold}
\begin{equation*}
\mathfrak H_h\doteq \{\Psi\in \mathfrak M_\mathrm{ext}:H_0[\phi]=h\}
\end{equation*}
is a $C^1$ hypersurface inside of $\mathfrak M_\mathrm{stab}^{+1,-1}$ (and hence a $C^1$ submanifold of codimension two inside of $\mathfrak M_\mathrm{nbhd}$). In particular, $\mathfrak M_\mathrm{stab}^{+1,-1}\setminus\mathfrak H_0$ is open and dense in $ \mathfrak M_\mathrm{stab}^{+1,-1}$. These hypersurfaces assemble into a $C^1$ foliation of $\mathfrak M_\mathrm{stab}^{+1,-1}$. Combined with results of \cite{AKU24}, we obtain that if $\Psi\in\mathfrak H_h$, then 
\[\lim_{v\to\infty}Y\psi|_{\mathcal H^+}  = h,\quad \lim_{v\to\infty}\Ric(Y,Y)|_{\mathcal H^+} =2M^{-2}h^2,\]
\[\lim_{v\to\infty}v^{-1}\cdot Y^2\psi|_{\mathcal H^+} = 2 M^{-2} h ,\quad \lim_{v\to\infty} v^{-1}\cdot\nabla_Y\Ric(Y,Y)|_{\mathcal H^+}  = 8 M^{-4} h^2. \]

\item \label{thm:intro-instab-2}
  Let $\{ \Psi_p\}_{p\in[0,1]}$ be a smooth one-parameter family of data sets in $\mathfrak M_\mathrm{nbhd}$ which crosses $\mathfrak M_\mathrm{stab}^{+1,-1}$ transversally such that $\Psi_p\in \mathfrak M_\mathrm{sub}$ for $p\in [0,p_*)$ and $\Psi_{p_*}\in \mathfrak M_\mathrm{stab}^{+1,-1}\cap \mathfrak H_h$. 
      Denote by $\beta= \beta(\Psi_p) = 1/\mathscr T(\Psi_p)$ the inverse temperature. Then, we have
         \begin{align}\label{eq:transient-1}
           Y\psi|_{\mathcal H^+} (v)& = e^{-4 \pi v /\beta  }  h  +O(\ve_0 v^{-3\delta/2}),\\ \label{eq:transient-2}
             |Y^2\psi|_{\mathcal H^+} (v)| & \gtrsim  |h|v  + O(\ve_0 v^{1-3\delta/2}) \quad \text{for } v \leq \beta,    \\  \label{eq:transient-3}
               \Ric(Y,Y)|_{\mathcal H^+}(v)& =2r_+^{-2} e^{-8 \pi v /\beta  }  h^2+O( \ve_0^2 v^{-3\delta/2}),\\  \label{eq:transient-4}
        |\nabla_Y\Ric(Y,Y)|_{\mathcal H^+}(v)| & \gtrsim h^2+O( \ve_0^2 v^{1-3\delta/2})\quad\text{ for } v \leq \beta,
          \end{align}
for $p<p_*$ sufficiently close to $p_*$. The implicit constants in the $O$-error terms may depend on the specific curve.
\end{enumerate}
\end{thmrough}

For the precise statement, see already \cref{thm:instabilities}. 

 \begin{figure}
\centering{
\def\svgwidth{12pc}
\begingroup%
  \makeatletter%
  \providecommand\color[2][]{%
    \errmessage{(Inkscape) Color is used for the text in Inkscape, but the package 'color.sty' is not loaded}%
    \renewcommand\color[2][]{}%
  }%
  \providecommand\transparent[1]{%
    \errmessage{(Inkscape) Transparency is used (non-zero) for the text in Inkscape, but the package 'transparent.sty' is not loaded}%
    \renewcommand\transparent[1]{}%
  }%
  \providecommand\rotatebox[2]{#2}%
  \newcommand*\fsize{\dimexpr\f@size pt\relax}%
  \newcommand*\lineheight[1]{\fontsize{\fsize}{#1\fsize}\selectfont}%
  \ifx\svgwidth\undefined%
    \setlength{\unitlength}{83.20865962bp}%
    \ifx\svgscale\undefined%
      \relax%
    \else%
      \setlength{\unitlength}{\unitlength * \real{\svgscale}}%
    \fi%
  \else%
    \setlength{\unitlength}{\svgwidth}%
  \fi%
  \global\let\svgwidth\undefined%
  \global\let\svgscale\undefined%
  \makeatother%
  \begin{picture}(1,0.99961345)%
    \lineheight{1}%
    \setlength\tabcolsep{0pt}%
    \put(0,0){\includegraphics[width=\unitlength,page=1]{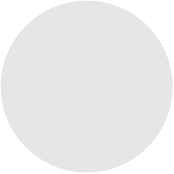}}%
    \put(0.62440032,0.58960751){\color[rgb]{0,0,0}\makebox(0,0)[lt]{\lineheight{1.25}\smash{\begin{tabular}[t]{l}$\phi_\circ=0$\end{tabular}}}}%
    \put(0.18817888,0.5324469){\color[rgb]{0,0,0}\makebox(0,0)[lt]{\lineheight{1.25}\smash{\begin{tabular}[t]{l}$T_\mathrm{ERN}\mathfrak H_0$\end{tabular}}}}%
    \put(0.40157096,0.26385703){\color[rgb]{0,0,0}\makebox(0,0)[lt]{\lineheight{1.25}\smash{\begin{tabular}[t]{l}$\mathfrak H_h$, $h<0$\end{tabular}}}}%
    \put(0.54301364,0.70532341){\color[rgb]{0,0,0}\makebox(0,0)[lt]{\lineheight{1.25}\smash{\begin{tabular}[t]{l}$\mathfrak H_h$, $h>0$\end{tabular}}}}%
    \put(0,0){\includegraphics[width=\unitlength,page=2]{Aretakis.pdf}}%
    \put(0.23591834,0.37456865){\color[rgb]{0,0,0}\makebox(0,0)[lt]{\lineheight{1.25}\smash{\begin{tabular}[t]{l}$\mathfrak H_0$\end{tabular}}}}%
    \put(0,0){\includegraphics[width=\unitlength,page=3]{Aretakis.pdf}}%
  \end{picture}%
\endgroup%
}
\caption{A schematic depiction of the level sets $\mathfrak H_h$ of the asymptotic Aretakis charge from \cref{thm:intro-instability}. One should think of this disk as the top or bottom face of the ``cylinder'' depicted in \cref{fig:foliation-intro-1}, i.e., it is a hypersurface in the moduli space $\mathfrak M$ parametrized by initial data for the scalar field, $\phi_\circ$. (We have suppressed infinitely many degrees of freedom in this picture, including $r_\circ$ and $\varpi_\circ$; see already \cref{sec:intro-mod-space}.) The sets $\mathfrak H_h$ are hypersurfaces in the faces, and hence codimension-two in $\mathfrak M$. The set $\mathfrak H_0$ corresponds to those initial data which evolve to extremal Reissner--Nordstr\"om but do not experience the Aretakis instability at arbitrarily large times. The tangent space of $\mathfrak H_0$ at exact extremal Reissner--Nordstr\"om, $T_\mathrm{ERN}\mathfrak H_0$, consists of those initial data for the wave equation with vanishing Aretakis charge on exact extremal Reissner--Nordstr\"om. However, because of the nonlinearity of the coupled system, the actual set $\mathfrak H_0$ is a (non-explicit) quadratic graph over $T_\mathrm{ERN}\mathfrak H_0$.}
\label{fig:foliation-Aretakis}
\end{figure}

\begin{rk}
    In the uncoupled linear problem (scalar wave equation on extremal Reissner--Nordstr\"om), we can use the exact conservation law to compute exactly 
    \begin{equation*}
        \mathfrak H_h^\mathrm{linear} = \{\phi_\circ:\partial_r(r\phi_\circ)|_{r=M}=-h\},
    \end{equation*}
    where $\phi_\circ$ denotes the characteristic data for the wave equation. It is clear that each $\mathfrak H_h^\mathrm{linear}$ is a codimension-one affine hyperplane in the space of initial data. In particular, $\mathfrak H_0^\mathrm{linear}$ is immediately seen to be nongeneric (and can be identified with $T_\mathrm{ERN}\mathfrak H_0$). Because the nonlinearity of the coupled EMSF system breaks Aretakis' conservation law, the level sets $\mathfrak H_h$ are curved and can only be determined teleologically. For an illustration, see \vref{fig:foliation-Aretakis}.
\end{rk}

\subsection{Discussion and outlook} We end the introduction with some discussion and possible extensions of the main theorems (\cref{sec:interior,sec:higher,sec:large-data}). We also formalize general conjectural versions of the main theorems for other matter models and outside of spherical symmetry in \cref{sec:EHC}. In \cref{sec:ECC}, we formulate a general ``extremal critical collapse'' conjecture that generalizes conjectures in \cite{KU24}. We also include in \cref{sec:dispersive-PDE} some comparisons with the extensive literature of threshold results for dispersive PDEs. 

In this section, we will repeatedly reference the \emph{proofs} of \cref{thm:intro-uniform-stability,thm:intro-foliation,thm:intro-scaling,thm:intro-instability}. Therefore, before continuing here, the reader may wish to consult \cref{sec:intro-proof-overview}.

\subsubsection{The black hole interior on approach to the extremal threshold}\label{sec:interior}

As explained in the caption to \cref{fig:refined-dichotomy} and \cref{rk:no-trapped-surfaces,rk:intro-subextremal}, the black hole interior in the spherically symmetric EMSF system contains strictly trapped surfaces if and only if the black hole parameters are asymptotically subextremal along $\mathcal H^+$. For any $\Psi\in\mathfrak M_\mathrm{black}\cap \mathfrak M_\mathrm{nbhd}$, the maximal development is bounded to the future by a smooth outgoing null hypersurface (a trivial, smooth Cauchy horizon) emanating from the final sphere of $\underline C{}_\ing$ (the solid point in \cref{fig:refined-dichotomy}). When $Q\ne 0$, the interior contains a potentially singular ingoing Cauchy horizon $\mathcal{CH}^+$\index{CH@$\mathcal{CH}^+$, Cauchy horizon} emanating from $i^+$ by \cite{Dafermos-thesis,dafermos2005interior,gajic-luk}, across which the metric may be extended as a continuous Lorentzian metric. If $\Psi\in\mathfrak M_\mathrm{black}\cap \mathfrak M_\mathrm{nbhd}$ and $Q=0$, then there is a spacelike singularity emanating from $i^+$ by \cite{C-91}. We refer the reader to \cite{Kommemi13} for the definitions of these boundary components $i^+$ and $\mathcal{CH}^+$ of the Penrose diagrams in question.  

Penrose's \emph{strong cosmic censorship conjecture} \cite{penrose1974gravitational} contends that, for generic initial data, the spacetime is \emph{inextendible} past the Cauchy horizon $\mathcal{CH}^+$. One can entertain several different formulations of this conjecture, and we refer the reader to the introduction of \cite{DL17} for discussion and to \cite{van2025strong} for a recent review article. As already mentioned above, the $C^0$-formulation is known to be false. 

In the asymptotically subextremal case, there is a well-known instability mechanism caused by the blueshift effect along $\mathcal{CH}^+$ \cite{Penrose:1968ar}, which can cause the Hawking mass $m$ to blow up identically along $\mathcal{CH}^+$. This phenomenon is called \emph{mass inflation} \cite{poisson1989inner}. Finishing the program initiated by Dafermos in \cite{Dafermos-thesis,dafermos2005interior}, Gautam has recently shown that mass inflation occurs for generic subextremal black holes in the spherically symmetric EMSF system \cite{Gautam}
(see also \cite{LukOhI,luk2019strong,sbierski2022holonomy,LOSR23}).\footnote{In fact, the genericity statement proved in \cite{Gautam} is stronger than what is required for genericity in our moduli space $\mathfrak M$, in the sense that we allow for weakly decaying data for which it is easier to prove the generic lower bounds required to apply the intermediate result \cite{LOSR23}. Note that our results in this paper hold under any \emph{stronger} assumptions on the scalar field, so that \cref{cor:mass-inflation} actually holds in whichever more restrictive class of data one wishes to consider.} 

These results require the teleological assumption of subextremality of the final black hole parameters. Therefore, the genericity statement must be further conditioned on the expectation that extremal black holes are nongeneric, which is precisely settled by our \cref{thm:intro-foliation}. It should be noted, however, that Kommemi's argument ruling out trapped surfaces in $\mathfrak M_\mathrm{stab}^{+1,-1}$ in our setting also implies that asymptotically extremal black holes cannot form in the two-ended case considered in \cite{LukOhI}. We conclude: 

\begin{cor}[Unconditional generic mass inflation]\label{cor:mass-inflation}
    For an open and dense subset of $\mathfrak M_\mathrm{black}\cap\mathfrak M_\mathrm{nbhd}$, the Hawking mass blows up identically along the Cauchy horizon $\mathcal{CH}^+$ emanating from timelike infinity $i^+$. The $C^{0,1}$-formulation of strong cosmic censorship holds in a full neighborhood of the Reissner--Nordstr\"om family in $\mathfrak M$. 
\end{cor}

If $\Psi\in \mathfrak M_\mathrm{stab}^{+1,-1}$, there are no strictly trapped surfaces in the black hole interior. This is equivalent to $1-\frac{2m}{r}\ge 0$ and hence $m$ is bounded. It follows that there is \emph{no mass inflation in the threshold case}. For a more general result, valid even for (weakly) charged scalar fields, see \cite{gajic-luk}.  

The exact generic regularity of the Cauchy horizon in the asymptotically extremal case remains open. Spherically symmetric solutions to the wave equation on exact extremal Reissner--Nordstr\"om are in fact $C^2$-extendible across $\mathcal{CH}^+$ \cite{gajic1,gajic2}.\footnote{Miethke has extended this result to higher $\ell$-modes outside of spherical symmetry \cite{Miethke}.} By \cite{gajic-luk}, solutions in $\mathfrak M_\mathrm{stab}^{+1,-1}$ can be extended in $C^{0,1/2}_\mathrm{loc}\cap H_{\mathrm{loc}}^1$ across $\mathcal{CH}^+$. This should be compared with the subextremal case, where solutions are generically inextendible in $C^{0,1}_\mathrm{loc}$ and are believed to be inextendible in $C^{0,1/2}_\mathrm{loc}\cap H^1_\mathrm{loc}$ \cite{LukOhI,sbierski2022holonomy}. In the extremal case, the existence of Cauchy data leading to any type of singularity at $\mathcal{CH}^+$ remains open, even for the linear wave equation. 
 
\subsubsection{Higher regularity of the extremal threshold}\label{sec:higher}

It is an interesting technical question to ask how regular the extremal threshold $\mathfrak M_\mathrm{stab}^{+1,-1}$ is. In this paper, we show that it is $C^1$, but it is unclear what is sharp. For starters, since the proof of \cref{thm:intro-foliation} seems to require a teleological normalization of the gauge (see already \cref{sec:intro-gauge}), which depends nontrivially on the solution itself, the regularity of  $\mathfrak M_\mathrm{stab}^{+1,-1}$ could very well depend on the regularity of the initial data in $\mathfrak M$. In this paper, we require $C^2$ data in $\mathfrak M$ to prove a $C^1$ estimate for $\mathfrak M_\mathrm{stab}^{+1,-1}$. Using our strategy, one would need $C^{k+1}$ data in $\mathfrak M$ to attempt to prove a $C^k$ estimate for $\mathfrak M_\mathrm{stab}^{+1,-1}$. To emphasize: it is not clear to us to what extent the regularity of the data affect the regularity of the threshold, but our methods certainly require more regular data to entertain higher regularity of the threshold, or of any of the manifolds $\mathfrak M^\sigma_\mathrm{stab}$. 

In fact, by adapting the methods of this paper, one can readily show that $\mathfrak M_\mathrm{stab}^\sigma$ is $C^k$ for every $k$ when $|\sigma|<1$. This is because the linearized (and second-order, third-order, etc.) estimates in the subextremal regime do not get much worse as we apply more derivatives in $\mathfrak M$.\footnote{There is actually a logarithmic loss, see already \cref{sec:intro-linear-estimates}, but it is harmless in this context.} As we will explain in \cref{sec:overview-II}, we have to handle \emph{growing} estimates towards $\mathcal H^+$ and along $\mathcal I^+$ in the extremal case. Taking higher derivatives of the Reissner--Nordstr\"om metric with respect to the parameters causes this growth to become worse. The methods presented in this paper do not seem to suffice to prove that $\mathfrak M_\mathrm{stab}^\sigma$ is $C^4$, regardless of how regular the data in $\mathfrak M_\mathrm{stab}^{+1,-1}$ are. Moreover, it is not clear if the foliation itself is even $C^2$ up to extremality, even though it is as regular as the leaves when $|\sigma|<1$.

\begin{question}
  If $\mathfrak M$ is modified to consist of arbitrarily regular data satisfying some natural falloff conditions, is the extremal threshold $\mathfrak M_\mathrm{stab}^{+1,-1}$ also arbitrarily regular? What about the regularity of the foliation $\{\mathfrak M_\mathrm{stab}^\sigma\}_{\sigma\in[-1,1]}$?
\end{question}

\subsubsection{``Large data'' versions of the main theorems}\label{sec:large-data}

In this paper, we have restricted attention to the perturbative small scalar field regime: in the main theorems, we assume that a weighted $C^2$ norm of the scalar field is initially small. For any black hole solution in the spherically symmetric EMSF model, one can define asymptotic parameters $r_{\mathcal H^+}>0$ (area radius) and $\varpi_{\mathcal H^+}>0$ (mass, see \eqref{eq:varpi-defn}) along the event horizon by monotonicity. It is then natural to define the sets of \emph{weakly subextremal} and \emph{weakly extremal} black holes
\begin{equation*}
   \mathfrak M_\mathrm{weak\,sub}\doteq \{\Psi\in\mathfrak M_\mathrm{black}:r_{\mathcal H^+}>|Q|\} ,\qquad  \mathfrak M_\mathrm{weak\,ext}\doteq \{\Psi\in\mathfrak M_\mathrm{black}:r_{\mathcal H^+}=|Q|\}.
\end{equation*}
It holds that $\mathfrak M_\mathrm{black}=\mathfrak M_\mathrm{weak\,sub}\sqcup \mathfrak M_\mathrm{weak\,ext}$ (see already \cref{cor:final-parameters}).

By the remarkable work of Dafermos--Rodnianski \cite{Price-law} (see also \cite{luk2019strong,Gautam}), every solution in $\mathfrak M_\mathrm{weak\,sub}$ decays in a strong sense to a Reissner--Nordstr\"om black hole with subextremal parameters $(\varpi_{\mathcal H^+},Q)$. This result requires \emph{no initial smallness assumption on the scalar field}. The key observation of \cite{Price-law} is that by monotonicities of the EMSF system, one can upgrade the (teleological!) assumption $r_{\mathcal H^+}>|Q|$ to $\varkappa\ge c$ on and near $\mathcal H^+$ at late times, where $\varkappa$ is the dynamical redshift factor defined in \eqref{eq:varkappa} below and $c>0$ is a constant. Therefore, before proving any decay to subextremal Reissner--Nordstr\"om, one can use the horizon redshift effect to control the scalar field. Using the method of characteristics, the redshift implies that $Y\phi$ is bounded up to and including $\mathcal H^+$. The proof of decay in \cite{Price-law} is substantially more involved, but this initial boundedness statement is crucial. 

On the other hand in the weakly extremal case $r_{\mathcal H^+}=|Q|$, simple monotonicity considerations show that $\varkappa\le 0$ along $\mathcal H^+$! Generically, conditioned on extremality, one expects $\varkappa <0$. Of course, in exact extremal Reissner--Nordstr\"om, $\varkappa =0$ at $\mathcal H^+$. In \cite{AKU24}, one of the key observations, in the context of a bootstrap argument in the small-data regime, is that $\varkappa$ is integrable along $\mathcal H^+$ despite the bad sign, and therefore $Y\phi$ remains bounded. Without some control over $Y\phi$, it is unclear how to even prove the basic estimate $\kappa\sim 1$ on the black hole exterior, where $\kappa$ is defined in \eqref{eq:greek-letters} below.

\begin{question}\label{quest:large-data-1}
    Do all solutions in $\mathfrak M_\mathrm{weak\,ext}$ eventually decay to an extremal Reissner--Nordstr\"om black hole in the sense of \cref{thm:intro-uniform-stability}? If not, is there at least a natural statement of orbital stability?
\end{question}

Regarding the ``full'' black hole threshold in the EMSF model, i.e., not restricted to the small data setting, we show the following theorem in \cref{sec:threshold-a-priori} using \cite{Dafermos-thesis,dafermos2005interior,Price-law}:
 
\begin{thm} For the spherically symmetric Einstein--Maxwell-neutral scalar field model, it holds that
    \begin{equation*}
         \partial \mathfrak M_\mathrm{black}\subset\mathfrak M_\mathrm{weak\,ext}\cup \mathfrak M_\mathrm{marg\,black},
    \end{equation*}
    where $\mathfrak M_\mathrm{marg\,black}$ consists of ``marginal'' black holes, i.e., those whose event horizon emanates from the final sphere on the initial ingoing cone $\u C{}_\ing$.
\end{thm}

This suggests the following extension of \cref{quest:large-data-1}:

\begin{question}\label{quest:large-data-2}
    Is it the case that $\mathfrak M_\mathrm{weak\,ext}\subset\partial\mathfrak M_\mathrm{black}$? Is $\mathfrak M_\mathrm{weak\,ext}$ a regular (locally Lipschitz or better) hypersurface in general?
\end{question}

The inclusion $\mathfrak M_\mathrm{weak\,ext}\subset\partial\mathfrak M_\mathrm{black}$ is false for the charged scalar field and charged Vlasov models as demonstrated in \cite{KU22,KU24}. Understanding this problem in the neutral scalar field model seems to be a very difficult problem. 

We note here the numerical work of Donninger--Schlag on the boundary of the forward scattering region of phase space for focusing cubic nonlinear Klein--Gordon equation \cite{donninger2011numerical}. In the ``large data'' regime for that problem, the threshold appears to have highly nontrivial structure, although the authors are unable to deduce any soft (ir)regularity statement from the numerics. Connections between the questions studied in this paper and those studied in the dispersive PDE literature will be presented in \cref{sec:dispersive-PDE} below.

\begin{rk}
    We expect $\partial \mathfrak M_\mathrm{black}$ to fail to be regular at the intersection $\mathfrak M_\mathrm{weak\,ext}\cap \mathfrak M_\mathrm{marg\,black}$ (which is nonempty, see already \cref{rk:meeting}).
\end{rk}

\subsubsection{The extremal event horizon threshold conjecture}\label{sec:EHC}

We now move away from the spherically symmetric Einstein--Maxwell-\emph{neutral} scalar field model and wish to consider more general matter models in spherical symmetry, as well as to leave symmetry altogether. When leaving symmetry, we must enlarge the class of extremal black holes under consideration. 

Recall the family of \emph{Kerr--Newman} metrics, given by
\begin{equation*}
    g_{M,a,e} \doteq  -\frac{\Delta}{|q|^2}\big(dt-a\sin^2\vartheta\,d\varphi\big)^2+\frac{|q|^2}{\Delta}dr^2+|q|^2d\vartheta^2 + \frac{\sin^2\vartheta}{|q|^2}\big(a\,dt-(r^2+a^2)\,d\varphi\big)^2,
\end{equation*} in Boyer--Lindquist coordinates $(t,r,\vartheta,\varphi)$, where $q\doteq r+ia\cos\vartheta$, $\Delta\doteq (r-r_-)(r-r_+)$, and $r_\pm \doteq M\pm \sqrt{M^2-a^2-e^2}$ \cite{newman1965note,newman1965metric}. The parameter $a\in\Bbb R$ is the \emph{specific angular momentum} (angular momentum per unit mass). These metrics solve the coupled Einstein--Maxwell system and include the Reissner--Nordstr\"om family $g_{M,e}$ as $a=0$. The other special family, with $e=0$, is known as the \emph{Kerr} family $g_{M,a}$ \cite{kerr1963gravitational}. When $a\ne 0$, $g_{M,a,e}$ is axisymmetric but not spherically symmetric. For Kerr--Newman, the parameter ranges are: subextremal $a^2+e^2<M^2$, extremal $a^2+e^2=M^2$, and superextremal $a^2+e^2 > M^2$. The presence of trapped surfaces behind the event horizon in the subextremal case, and absence of trapped surfaces behind the event horizon in the extremal case, is valid for the Kerr--Newman family. For work on linear stability of \emph{subextremal} Kerr--Newman for the Einstein--Maxwell system, we refer the reader to \cite{civinPhD,Elena-small-Q,Elena-linear-stability,Giorgi-KN-1,Giorgi-KN-2,Giorgi-KN-3,Giorgi-KN-4}.

Fix now a matter model for the Einstein equations and a compatible symmetry class. We have made \cref{conj:EVHT} below with a choice of one of the following in mind (although one can certainly entertain these questions for other matter, such as Yang--Mills):
\begin{itemize}[noitemsep]
    \item Einstein--Maxwell-neutral scalar field in spherical symmetry (as in this paper),
    \item Einstein--Maxwell-charged scalar field in spherical symmetry (e.g., \cite{Kommemi13}),
    \item Einstein--Maxwell--neutral massless\footnote{The black hole stability story is significantly more complicated in the presence of \emph{massive} Vlasov matter due to the presence of bound states. See \cite[Section 1.4]{KU24} for discussion.} Vlasov in spherical symmetry (e.g., \cite{KU24} with $\mathfrak e=0$),
    \item Einstein--Maxwell--charged massless Vlasov in spherical symmetry (e.g., \cite{KU24}),
    \item Einstein vacuum without symmetry,
    \item Einstein--Maxwell without symmetry.
\end{itemize}

Let $\mathcal C=C_\out\cup \u C{}_\ing$ denote a bifurcate null hypersurface as in the setting of this paper, with moduli space $\mathfrak M=\mathfrak M(\mathcal C)$ corresponding to a matter model-symmetry class combination which admits at least one extremal Kerr--Newman black hole. In particular, we will assume that the ``final sphere'' of $\u C{}_\ing$ has ``$r>0$'' in a suitable sense. In the above examples, the natural moduli space $\mathfrak M$ contains extremal Reissner--Nordstr\"om for the first four examples, extremal Kerr for the fifth, and the full Kerr--Newman family for the sixth. In this space, one can again make sense of the set of solutions which form black holes, $\mathfrak M_\mathrm{black}$. The noncollapsed set $\mathfrak M_\mathrm{non}$ is defined to consist of solutions which remain regular and for which $\mathcal C\subset J^-(\mathcal I^+)$. Outside of symmetry, it might not be the case that $\mathfrak M=\mathfrak M_\mathrm{black}\sqcup \mathfrak M_\mathrm{non}$. It is unknown if naked singularities can form from data in such an $\mathfrak M$; these data would not be included in either set.\footnote{Note that the examples of naked singularities for the Einstein vacuum equations constructed by Rodnianski--Shlapentokh-Rothman \cite{SR22,RSR} are \emph{not} included in the type of moduli space considered here.}

We now make the following conjecture, which generalizes and unifies \cite[Conjecture IV.2]{DHRT}, \cite[Conjecture 6.1]{dafermos2025stability}, \cite[Conjectures 1 and 3]{KU24}, and \cref{thm:intro-uniform-stability,thm:intro-foliation,thm:intro-scaling,thm:intro-instability} of the present paper. 

\begin{conj}[Extremal event horizon threshold conjecture]\label{conj:EVHT} Let $\mathcal C$ and $\mathfrak M$ be as above such that the extremal Kerr--Newman black hole $g_{M_0,a_0,e_0}$ arises from an initial data set $\Psi_0\in\mathfrak M$. There exists a neighborhood $\mathfrak M_\mathrm{nbhd}$ of $\Psi_0$ with the following properties:
\begin{enumerate}
    \item \ul{Universality of the Kerr--Newman family}: If $\Psi\in \mathfrak M_\mathrm{black}\cap\mathfrak M_\mathrm{nbhd}$, then $\mathcal S[\Psi]$ converges to a member of the Kerr--Newman family in an appropriate sense. Moreover, $\mathfrak M_\mathrm{nbhd}=(\mathfrak M_\mathrm{black}\sqcup \mathfrak M_\mathrm{non})\cap\mathfrak M_\mathrm{nbhd}$, i.e., no naked singularities arise from initial data in $\mathfrak M_\mathrm{nbhd}$. 

    \item \ul{Isologous foliation and threshold property}: There exists a $C^1$ foliation of $\mathfrak M_\mathrm{black}\cap\mathfrak M_\mathrm{nbhd}$ by hypersurfaces $\mathfrak M_\mathrm{stab}^\sigma\subset\mathfrak M_\mathrm{nbhd}$, $\sigma\in[0,1]$, such that if $\Psi\in \mathfrak M_\mathrm{stab}^\sigma$, then $\mathcal S[\Psi]$ converges to a Kerr--Newman black hole with parameters $(M,a,e)$ satisfying $\sigma = \sqrt{a^2+e^2}/M$. The extremal hypersurface $\mathfrak M^1_\mathrm{stab}$ is the local black hole threshold, $\partial\mathfrak M_\mathrm{black}\cap\mathfrak M_\mathrm{nbhd}=\mathfrak M^1_\mathrm{stab}$, and locally separates $\mathfrak M_\mathrm{black}$ and $\mathfrak M_\mathrm{non}$.\footnote{Here, unlike \cref{thm:intro-foliation}, $\sigma$ is not a signed quantity and it is implicit that $\mathfrak M_\mathrm{stab}^\sigma$ can be empty for $\sigma$ not sufficiently close to $1$, as the hypersurfaces can simply exit $\mathfrak M_\mathrm{nbhd}$.} For any $\Psi\in\mathfrak M^1_\mathrm{stab}$, the maximal development does not contain any strictly trapped surfaces.  

    \item \ul{Scaling laws}: For a generic one-parameter family of black hole initial data $\{\Psi_p\}_{p\in[0,1]}$ crossing the extremal threshold, the final parameter ratio, area of the event horizon, and temperature scale as in \cref{thm:intro-scaling}. In symmetry, the location of the event horizon scales in the same way as well.\footnote{It is not clear to us how to appropriately generalize this statement outside of symmetry.}

    \item \ul{Horizon instabilities}: A generic solution lying on $\mathfrak M_\mathrm{stab}^1$ will experience some form of growing hair on its event horizon for all time (see below on what the nature of these instabilities will be in various settings). A generic near-threshold solution will experience a transient instability on the timescale of its inverse final temperature. 
\end{enumerate}
\end{conj}

To summarize: the conjecture contends that the general picture of extremality is conceptually the same as is proved in this paper for the spherically symmetric Einstein--Maxwell-neutral scalar field system, though this is by far the simplest setting in which one can ask these questions. Note that the absence of trapped surfaces behind the event horizon in point 2.~is a necessary condition for lying on the threshold. For discussion of the possibility of what might happen if this conjecture is false outside of symmetry, and for the connection between this conjecture and the overcharging/overspinning problem, see \cite{dafermos2025stability}.

We now discuss the status of \cref{conj:EVHT} in the settings listed above. We will focus on the question of codimension-one stability of extremal Kerr--Newman (i.e., the analogue of \cite[Theorem 1]{AKU24}), as this is clearly a prerequisite for proving any part of \cref{conj:EVHT}. 

\paragraph{Charged scalar field in spherical symmetry.} The spherically symmetric Einstein--Maxwell-charged scalar field (EMCSF) system is a natural generalization of the neutral scalar field system considered in this paper and plays an especially important role in the study of strong cosmic censorship, as it allows for solutions with nontrivial charge and a regular center. 

In this model, the real scalar field is upgraded to a complex scalar field $\phi:\mathcal M\to\Bbb C$ satisfying the \emph{charged wave equation}
\begin{equation}\label{eq:CSF-wave}
    \mathrm D_\mu \mathrm D^\mu\phi=0,
\end{equation}
where $\mathrm D_\mu \doteq\nabla_\mu -iqA_\mu$ denotes the \emph{gauge covariant derivative} and $q\ne 0$ is the \emph{coupling constant}. The scalar field couples to the electromagnetic field via Maxwell's equation
\begin{equation}\label{eq:CSF-Maxwell}
    \nabla^\mu \mathbf F_{\mu\nu} =  q \Im(\phi\overline{\mathrm D_\nu\phi}). 
\end{equation}
For the full set of equations and the fundamental local theory of this model, see \cite{Kommemi13}.

Even ignoring \eqref{eq:CSF-Maxwell} and the Einstein equations, the analysis of \eqref{eq:CSF-wave} on a Reissner--Nordstr\"om black hole $g_{M,e}$ is significantly more complicated than the real scalar wave equation.\footnote{We are taking the gauge potential $A$ in $\mathrm D$ to be the one associated to the background Reissner--Nordstr\"om electromagnetic field with charge $e$, i.e., $A=-\frac{e}{r}dt$.} In particular, \eqref{eq:CSF-wave} contains an inverse square potential term $\sim q e r^{-2}\phi$, where $qe$ is dimensionless. By the heuristic work of Hod--Piran \cite{hod1998lateI,hod1998lateII,hod1998lateIII}, it is known that the decay properties of solutions to \eqref{eq:CSF-wave} depend on the magnitude of $qe$, even in the subextremal case. See the upcoming series \cite{Dejan-CSF-I,Dejan-CSF-II} by Gajic for a detailed study of \eqref{eq:CSF-wave} (for not necessarily spherically symmetric data) in the near-extremal regime. For $|qe|$ which is not very small, the techniques required to analyze \eqref{eq:CSF-wave} are related to those for extremal Kerr with fixed azimuthal number $m$; see the discussion below.

On extremal Reissner--Nordstr\"om, the conservation laws exploited by Aretakis in the case of the uncharged wave equation fail for \eqref{eq:CSF-wave}. In fact, the instability is \emph{enhanced} by the zeroth order term in the equation. In \cite{Dejan-CSF-II} (see also the earlier heuristic work of Zimmerman \cite{zimmerman2017horizon}), it is shown that for generic solutions to \eqref{eq:CSF-wave}, $|Y\psi|_{\mathcal H^+}|\sim v^{\frac 12(1-\sqrt{1-4q^2e^2})}$ when $q^2e^2<\frac 14$ and  $|Y\psi|_{\mathcal H^+}|\sim v^\frac{1}{2}$ when $q^2e^2>\frac 14$ (the case $q^2e^2=\frac 14$ is exceptional).

For small $|qe|$, the spherically symmetric coupled system \eqref{eq:CSF-wave}--\eqref{eq:CSF-Maxwell} has been studied on subextremal Reissner--Nordstr\"om black hole backgrounds by Van de Moortel \cite{van2022decay} (see also \cite{van2025extension}). (Note that the analysis of \cite{van2022decay} does not apply to \eqref{eq:CSF-wave} on its own, as the coupling with Maxwell and spherical symmetry is used in a crucial manner. Remarkably, Van de Moortel is able to show stronger estimates for the coupled nonlinear system than seem possible for the charged wave equation on its own!)

In a forthcoming note \cite{AKU26b}, we begin to generalize the results of this paper to the much more complicated spherically symmetric EMCSF system, under the assumption that $|qe|$ is small. To accurately describe the result of \cite{AKU26b}, we will use some of the notation established later in \cref{sec:intro-proof-overview}. In particular, we refer the reader to the definitions of the Banach spaces $\mathfrak F$, $\mathfrak X$, $\mathfrak Z$, and the moduli space $\mathfrak M$ in \cref{sec:intro-mod-space} and the moduli space cylinders $\cyl(\ve,\ell)$ in \cref{sec:intro-broad-strokes}.

For the coupled EMCSF system, we can immediately complexify $\mathfrak F$, denoting the result by $\mathfrak F^\Bbb C$, to allow for complex-valued $\phi_\circ$. We then define $\mathfrak X^\Bbb C\doteq \mathfrak F^\Bbb C\times\Bbb R^2$ and $\mathfrak Z^\Bbb C\doteq\mathfrak X^\Bbb C\times\Bbb R$. This gives a natural definition of $\mathfrak M^{\Bbb C,q}$, which depends on $q$ due to the asymptotic flatness condition in \cref{defn:mod-space}, as one must solve the equations along $C_\out$.\footnote{One must also fix the electromagnetic gauge to get a well-defined solution map on the moduli space. We make the essentially arbitrary choice of $A_v=0$ globally and $A_u=0$ on $\u C{}_\ing$.} For $M_0>0$, we set $x_0=(0,100M_0,M_0)\in\mathfrak X^\Bbb C$ as in the neutral case, which allows us to define the cylinders $\cyl^\mathbb C(\ve,\ell)$ in the natural manner using $\mathfrak X^\Bbb C$.

The following theorem establishes codimension-one stability of Reissner--Nordstr\"om for every charge-to-mass ratio when the coupling constant is small: 

\begin{thm}[\cite{AKU26b}, forthcoming] \label{thm:CSF}
   There exists an $\alpha_0>0$ such that for any $q\in\Bbb R$ and $M_0>0$ satisfying $|qM_0|\le\alpha_0$, the following holds. There exists an $\ve_0>0$, depending perhaps on $M_0$ but not on $q$, and a function $\mathscr W_\infty^q:\cyl^\Bbb C(\ve_0,1)\to \Bbb R$ such that for every $(x,\sigma)\in \cyl^\Bbb C(\ve_0,1)$, the solution to the EMCSF system with coupling constant $q$ arising from $\Psi=(x,\mathscr W^q_\infty(x,\sigma))$ converges to a Reissner--Nordstr\"om black hole with charge-to-mass ratio $\sigma$. In particular, extremal Reissner--Nordstr\"om is nonlinearly codimension-one stable for the EMCSF system with small coupling constant.  
\end{thm}

The functions $\mathscr W_\infty^q$ should be compared with $\mathscr W_\infty$ in \cref{sec:overview-II}, but we do not prove that they are $C^1$ (or even continuous). Therefore, the notion of codimension stability given by this theorem, proved via a shooting argument, is the same as in \cite{DHRT,AKU24}. Note that the stable ``hypersurfaces'' \[\mathfrak M_\mathrm{stab}^{\sigma,q}\doteq \{(x,\mathscr W^q_\infty(x,\sigma)):x\in B_{\ve_0}^{\mathfrak X^\Bbb C}(x_0)\}\] depend nontrivially on $q$. By generalizing arguments used to prove \cref{thm:intro-foliation}, it should be possible to prove that $\mathscr W^q_\infty$ is $C^1$, though this model presents interesting new difficulties even when $|qM_0|$ is small. 

The two main difficulties in the proof of \cref{thm:CSF} are (i) the inverse square potential in the wave equation \eqref{eq:CSF-wave}, which is an obstacle to proving decay estimates for $\phi$, and (ii) the (now dynamical) charge $Q$ which decays slowly to the background near $\mathcal H^+$ (strictly slower than $v^{-2}$). 

Difficulty (i) was resolved in the subextremal case in \cite{van2022decay,van2025extension}.\footnote{Again, under the hypothesis $|qe|\ll 1$. In fact, Van de Moortel gives the explicit estimate $|qe|\le 0.08267$ for the method to work. We have not attempted any such optimization ourselves. Moreover, it seems relatively straightforward to generalize \cite{van2022decay} to the fully coupled EMCSF system.} However, the proof of the Morawetz estimate in \cite{van2022decay} uses the redshift effect in a fundamental way, so it is not clear if the proof generalizes to the extremal case. Motivated by \cite{gajic2023late}, we instead couple the Morawetz estimate with the $r^p$- and $h_p$-hierarchies (see already \cref{sec:intro-unifying}) for $p=\delta\ll 1$. This obviously shortens the overall length of the hierarchy we can achieve, but only by the inconsequential small parameter $\delta$. Secondly, we replace the ``square root Gr\"onwall'' argument in \cite{van2022decay,van2025extension}, which unlocks the range $p\ge 2$, by an elementary dyadic decomposition and bootstrap argument that allows us to absorb the scale-invariant error terms into the $r^p$- and $h_p$-energies directly (with a little polynomial growth at the top of the $p$-range). These methods crucially exploit the smallness of $\alpha_0$ and thus seem specific to that regime. 

The source of difficulty (ii) is the following component of \eqref{eq:CSF-Maxwell}:
\begin{equation}\label{eq:CSF-Qu}
    \partial_uQ = - qr^2\Im(\phi\overline{\mathrm D_u\phi}).
\end{equation}
By this equation, the behavior of $Q$ is governed by the $p=1$ energy near $\mathcal H^+$ (see already \cref{def-of-energies}) and hence $Q$ decays to the background value strictly slower than $v^{-2}$ (this is sharp, as can be seen from the asymptotics of \cite{Dejan-CSF-II}). In the EMCSF system, the renormalized Hawking mass $\varpi$ also decays slower because of the interaction with the charge. Indeed, 
\begin{equation}\label{eq:CSF-varpiu}
    \partial_u\varpi = \frac{r^2}{2\gamma}|\mathrm D_u\phi|^2-qQ r\Im(\phi\overline{\mathrm D_u\phi}),
\end{equation}
compare with \eqref{eq:varpi-u}. Using \eqref{eq:CSF-Qu} and \eqref{eq:CSF-varpiu}, the (quasilocal) \emph{charge-to-mass ratio}
\begin{equation*}
    \rho \doteq\frac{Q}{\varpi}
\end{equation*}
satisfies 
\begin{equation}
    \partial_u\rho  = -\frac{\rho r^2}{2\varpi\gamma}|\mathrm D_u\phi|^2- q \frac{r^4\varkappa}{2\varpi^2}\Im(\phi\overline{\mathrm D_u\phi}),
\end{equation}
where $\varkappa$ is the dynamical redshift factor \eqref{eq:varkappa}. Therefore, $\rho$ is controlled by the $p=0$ energy and hence decays faster! Moreover, once we treat $\varpi$ and $\rho$ as the fundamental dependent variables, simple computations reveal that $\varpi$ always comes paired with redshift factors in the nonlinearity, and that this combination also enjoys better decay. With these observations, the geometric estimates in the proof of \cref{thm:CSF} proceed in essentially the same way as \cref{sec:K-1} of the present paper. We expect these properties of $\rho$ and $\varpi$ to play an important role for any value of $|qM_0|$. 

\begin{rk}
    We will also use the charge-to-mass ratio $\rho$ in this paper, not because of its better decay properties, but because the good structure of the linearized Reissner--Nordstr\"om family is more apparent when linearizing with respect to $\rho$. 
\end{rk} 

\cref{thm:CSF} does not address the threshold property of $\mathfrak M^{+1,q}_\mathrm{stab}\sqcup \mathfrak M^{-1,q}_\mathrm{stab}$, nor does it address the presence of enhanced horizon instabilities (or actually any instability at all) due to the charge. In particular, the starting point for our understanding of the threshold in this paper, \cref{ingredient-noncollapse} in \cref{sec:intro-proof-uniform}, fails spectacularly in the presence of any amount of dynamical charge. These issues, as well as the extension of \cref{thm:CSF} to large $|qM_0|$, will be the subject of future investigations.

Finally, we note the series of numerical works \cite{gelles2025accumulation,gellesII} by Gelles and Pretorius, studying first the spherically symmetric coupled system \eqref{eq:CSF-wave}--\eqref{eq:CSF-Maxwell} on an extremal background, and second the fully coupled EMCSF system, generalizing the work of Murata--Reall--Tanahashi \cite{Reall-numerical}. In particular, in \cite{gellesII} the authors study several aspects of \cref{conj:EVHT} for this model, including insights on the noncollapsing regime, scaling laws, and the ramifications of the enhanced instability for the Maxwell field. 

\paragraph{Massless Vlasov in spherical symmetry.}

Another case of interest, especially in view of the work of the second- and third-named authors \cite{KU24} on ``extremal critical collapse'' (to be discussed further in the following subsection) is the spherically symmetric Einstein--Maxwell--massless Vlasov system. Massless Vlasov matter represents a collisionless gas of particles with electric charge $q$ which obey the \emph{Lorentz force equation}
\begin{equation}\label{eq:Lorentz-force}
    \frac{Dp^\mu}{ds} = q \bf F^\mu{}_\nu p^\nu,
\end{equation}
where $p^\mu = dx^\mu/ds$ is the (null) $4$-momentum. The particle distribution is modeled by a function $f(x,p)\ge 0$ on the future-directed null shell $\mathcal P\to\mathcal M$ and $f$ is transported along the trajectories \eqref{eq:Lorentz-force}. The electromagnetic current and energy-momentum tensor of the Vlasov matter are given by
\begin{equation*}
   J^\mu_\mathrm{Vl}\doteq q\int_{\mathcal P_x}p^\mu f(x,p)\,d\mu_x(p) \quad\text{and}\quad T^{\mu\nu}_\mathrm{Vl}\doteq \int_{\mathcal P_x}p^\mu p^\nu f(x,p)\,d\mu_x(p),
\end{equation*}
respectively, where $d\mu_x$ is the canonical measure on the fibers $\mathcal P_x$ of $\mathcal P$. For the full set of equations and the fundamental local theory of this model, see \cite{KU24}.

When $q=0$, the neutral case, moments of solutions of the linear massless Vlasov equation decay exponentially on subextremal Reissner--Nordstr\"om black holes \cite{Bigorgne-Schw,W23}. Moreover, Velozo Ruiz has proved nonlinear asymptotic stability of Schwarzschild for the spherically symmetric Einstein--massless Vlasov system \cite{VelozoPhD}. In the extremal case, Weissenbacher has shown that moments of $f$ decay pointwise in general only like $\tau^{-2}$, and that this is sharp generically. Moreover, the first derivative of $T_\mathrm{Vl}$ generically does not decay along $\mathcal H^+$, which is the manifestation of the horizon instability in this setting. (Polynomial growth of higher derivatives of $T_\mathrm{Vl}$ is conjectured, but not proved, in \cite{W23}.) 

There is hope to use the kinds of phase space volume estimates discovered in \cite{W23} to establish $(r-M)^{-p}$-weighted integral estimates for $T_\mathrm{Vl}$ and hence prove codimension-one stability of extremal Reissner--Nordstr\"om (again, the analogue of \cite[Theorem 1]{AKU24}) in the $q=0$ case. Note that the charge-to-mass ratio in the neutral Vlasov model enjoys the same monotonicity properties at $\mathcal I^+$ as in the neutral scalar field model. This suggests that the basic strategy for proving \cref{thm:intro-foliation} could also work in this case, although the details and general strategy of the linear analysis would be necessarily quite different.\footnote{It is worth noting that the moduli space in Vlasov models does not have a natural underlying Banach space structure, as $f\ge 0$ is not a linear constraint. This would seem to also necessarily complicate the setup of the problem.} Curiously, however, even for neutral Vlasov the special monotonicities used to rule out trapped surfaces behind the event horizon of asymptotically extremal black holes in \cite{AKU24} break down. Understanding the black hole interior at extremality in this model is an interesting question. 

To our knowledge, the stability analysis for $q\ne 0$ is completely open, even in the subextremal case.

\paragraph{Electrovacuum outside of symmetry.} Outside of symmetry, \cref{conj:EVHT} is much more ambitious already because of the fundamental lack of monotonicities, explosion of number of degrees of freedom, and general difficulty of proving black hole stability statements outside of symmetry. Many of these issues are already present in the subextremal story and we refer the reader to \cite{DHR19,Elena-linear-stability,DHRT,GKS} and references therein. We comment here on developments for extremal Reissner--Nordstr\"om and extremal Kerr, as to our knowledge nothing yet is known about extremal Kerr--Newman with both $a\ne 0$ and $e\ne 0$. 

The linear wave equation on extremal Reissner--Nordstr\"om (without any symmetry assumptions on the initial data) is by now well understood, starting with the seminal work of Aretakis \cite{Aretakis-instability-1,Aretakis-instability-2}.  Sharp asymptotics have been obtained by the first-named author, Aretakis, and Gajic in \cite{angelopoulos2020late}. Lucietti--Murata--Reall--Tanahashi showed that linearized gravitational and electromagnetic perturbations on extremal Reissner--Nordstr\"om have conservation laws, just like the wave equation, that lead to a horizon instability \cite{lucietti2013horizon}. Decay and non-decay for these perturbations was rigorously established by Apetroaie in \cite{apetroaie}, who also discovered a worse instability for the curvature component $\u\alpha$. As of now, however, no stability or instability statements for the gauge-dependent sector of linearized gravity are known. 

We also have a fairly good understanding of small data global existence for semilinear wave equations satisfying the null condition on extremal Reissner--Nordstr\"om, starting with the work of the first-named author in symmetry \cite{A16}, and later extended outside of symmetry with Aretakis, Gajic, and the third-named author \cite{AAG20,angelopoulos2025semilinear}. By the elegant work of Aretakis \cite{aretakis2013nonlinear}, it is known that some form of the null condition is necessary even at the event horizon to prevent finite-time blowup (in sharp contrast to the subextremal case \cite{luk2013null,DHRT22,DHRT24}). Capturing this structure in the full Einstein--Maxwell system will be an important part of proving stability of extremal Reissner--Nordstr\"om outside of symmetry. 

The linear wave equation on extremal Kerr with \emph{axisymmetric} initial data behaves rather similarly to extremal Reissner--Nordstr\"om \cite{Aretakis-Kerr,Aretakis-instability-3}. See also \cite{giorgi2024physical} for physical space estimates and \cite{lucietti2012gravitational} for gravitational perturbations. In general, however, the wave equation on extremal Kerr is much more complicated than extremal Reissner--Nordstr\"om (as mentioned earlier, for fixed azimuthal number $m$, it has many similarities with the charged scalar field equation). Mode stability for the scalar wave equation on extremal Kerr has been shown by Teixeira da Costa \cite{Rita-mode-stability}. Casals--Gralla--Zimmerman predicted a generic growth rate of $|Y\psi_m|\sim v^{1/2}$ along $\mathcal H^+$ in \cite{cgz-exkerr}, which was confirmed by Gajic in \cite{Gajic23} (conditional on something even worse not happening, though it is fully expected that this technical assumption can be removed). For fixed $m$, \cite{Gajic23} also gives some basic boundedness and decay results for energies, but summing over $m$ is a significant challenge because of the complicated interplay between trapping and superradiance in the $m\to\infty$ limit. As of now, even pointwise boundedness for general solutions to the wave equation on extremal Kerr remains a fundamental open problem. 

\begin{rk}[The $(2+1)$-dimensional case I]\label{rk:2+1-I}
  One can entertain \cref{conj:EVHT} in $2+1$ dimensions with a negative cosmological constant. Indeed, the two-parameter family of \emph{BTZ black holes} \cite{banados1992black,carlip19952+} admit a regular extremal limit, and one can ask if \cref{conj:EVHT} holds near extremal BTZ under perturbations by a real, massless scalar field (without matter, solutions to the $(2+1)$-dimensional Einstein equations have constant curvature). For more discussion of the $(2+1)$-dimensional case, see already \cref{rk:2+1-II}.
\end{rk}

\begin{rk}[The $(4+1)$-dimensional case I]\label{rk:4+1-I}
 One can also entertain \cref{conj:EVHT} in $4+1$ dimensions (with zero cosmological constant). Indeed, the three-parameter family of \emph{Myers--Perry black holes} \cite{myers1986black} admit a regular extremal limit, and one can ask if \cref{conj:EVHT} holds near extremal Myers--Perry in vacuum. Remarkably, the $(4+1)$-dimensional vacuum equations admit a nontrivial $\mathrm{SU}(2)\times \Bbb Z_4$-symmetry reduction which still allows for a (special type of) extremal Myers--Perry black hole \cite{3rd-law-vac}. In this symmetry class, the vacuum equations look like a much more complicated version of the spherically symmetric EMCSF system, but now there is no dimensionful parameter like $q$. One can hope to address this model after the EMCSF picture has been resolved for large $q$. 
\end{rk}

\subsubsection{Extremal critical collapse}\label{sec:ECC}

In \cref{sec:EHC}, we considered a moduli space $\mathfrak M$ consisting of characteristic data on a bifurcate null hypersurface $\mathcal C$, where $\u C{}_\ing$ ends on a non-central sphere. Now, however, we want to consider a different moduli space $\widehat{\mathfrak M}$ consisting of \emph{Cauchy data} on $\Bbb R^3$, so that we can actually observe gravitational collapse from data in $\widehat{\mathfrak M}$.\footnote{One could also consider characteristic data on an outgoing null cone starting at a regular center and going to $\mathcal I^+$.} 

As before, we have the sets $\widehat{\mathfrak M}_\mathrm{black}$ and  $\widehat{\mathfrak M}_\mathrm{non}$, consisting of solutions leading to black hole formation and noncollapsing \emph{future causally geodesically complete} solutions, respectively. We also highlight the special class $\widehat{\mathfrak M}_\mathrm{disp}\subset\widehat{\mathfrak M}_\mathrm{non}$ of initial data with \emph{dispersive} developments, i.e, those solutions whose geometry asymptotically converges to Minkowski space in the far future and matter fields decay suitably. Nontrivial stationary states, if they exist, lie in $\widehat{\mathfrak M}_\mathrm{non}\setminus\widehat{\mathfrak M}_\mathrm{disp}$ since they do not decay. The question of \emph{critical collapse} is concerned with the study of phase transitions between $\widehat{\mathfrak M}_\mathrm{non}$ and $\widehat{\mathfrak M}_\mathrm{black}$. We refer the reader to \cite{crit-review-update} (and also the introduction of \cite{KU24}) for an introduction to this subject. 

\begin{defn} An \emph{interpolating family} is a continuous one-parameter family $\{\Psi_p\}_{p\in[0,1]}\subset\widehat{\mathfrak M}$ such that $\Psi_{0}\in\widehat{\mathfrak M}_\mathrm{non}$ and $\Psi_1\in\widehat{\mathfrak M}_\mathrm{black}$. Given such a family, we may define the \emph{critical parameter} $p_*$ and the \emph{critical solution} $\mathcal S[\Psi_{p_*}]$ by $p_*\doteq \sup\{p\in[0,1]: \Psi_p\in\widehat{\mathfrak M}_\mathrm{non}\}$.
\end{defn}

Here, we are interested in the special case of extremal black holes forming on the threshold:

\begin{defn}\label{def:ECC}
    An interpolating family $\{\Psi_p\}_{p\in[0,1]}$ exhibits \emph{extremal critical collapse} if the critical solution $\mathcal S[\Psi_{p_*}]$ asymptotically settles down to an extremal black hole.
\end{defn}

In the spirit of \cref{conj:EVHT}, generalizing \cite[Conjecture 3]{KU24}, we formulate:

\begin{conj}[Extremal critical collapse conjecture]\label{conj:ECC}
Let $\widehat{\mathfrak M}$ be as above, with a matter model and symmetry combination such that:
\begin{enumerate}
    \item[0.]\label{ECC:0} \ul{Extremal Kerr--Newman can form}: An exact extremal Kerr--Newman event horizon can form in finite time from an initial data set $\Psi\in\widehat{\mathfrak M}$.
\end{enumerate}
Then the following holds: 
\begin{enumerate}
\item\label{ECC:1} \ul{The third law is false}: There exist data $\Psi\in\widehat{\mathfrak M}$, whose evolution $\mathcal S[\Psi]$ describes a stationary subextremal apparent horizon forming in finite time, which at a later finite time evolves into an exactly extremal Kerr--Newman event horizon. Moreover, this can be achieved in true gravitational collapse, i.e., the event horizon $\mathcal H^+$ does not intersect the Cauchy data hypersurface.

\item\label{ECC:2} \ul{Extremal critical collapse can occur}: There exist interpolating families of initial data $\{\Psi_p\}_{p\in[0,1]}$ for which the critical solution $\mathcal S[\Psi_{p_*}]$ forms an asymptotically extremal Kerr--Newman black hole.

\item\label{ECC:3} \ul{Structure of the threshold}: Let $\Psi_{p_*}$ be critical as in the previous point. Then there exists a neighborhood $\widehat{\mathfrak M}_\mathrm{nbhd}$ of $\Psi_{p_*}$ such that all $\Psi\in \widehat{\mathfrak M}_\mathrm{nbhd}$ either (a) collapse to form an asymptotically Kerr--Newman black hole or (b) lie in $\widehat{\mathfrak M}_\mathrm{non}$. Moreover, there exists a $C^1$ isologous foliation by hypersurfaces $\widehat{\mathfrak M}_\mathrm{stab}^\sigma\subset\widehat{\mathfrak M}_\mathrm{nbhd}$, $\sigma\in [0,1]$, such that $\partial\widehat{\mathfrak M}_\mathrm{black}\cap\widehat{\mathfrak M}_\mathrm{nbhd}=\widehat{\mathfrak M}^1_\mathrm{stab}$.

\item\label{ECC:4} \underline{Scaling laws}: For a generic one-parameter family of black hole initial data $\{\Psi_p\}_{p\in[0,1]}$ crossing the extremal threshold, the final parameter ratio, area of the event horizon, and temperature scale as in \cref{thm:intro-scaling}. In symmetry, the location of the event horizon scales in the same way as well. Moreover, when $\Psi_p\in\widehat{\mathfrak M}_\mathrm{non}$, the timescale on which $\mathcal S[\Psi_p]$ resembles extremal Kerr--Newman is $\sim |p-p_*|^{-1/2}$.

\item\label{ECC:5} \underline{Horizon instabilities}: A generic solution lying on $\mathfrak M_\mathrm{stab}^1$ will experience some form of growing hair on its event horizon for all time (see below on what the nature of these instabilities will be in various settings). A generic near-threshold solution will experience a transient instability on the timescale of its inverse final temperature. 
\end{enumerate}
\end{conj}

\begin{rk}
    Note that in \hyperref[ECC:2]{part 2.}, we do not demand that the critical solution is exactly extremal Kerr--Newman after finite time, but one could also entertain this stronger form of the conjecture (in particular, note that the examples in \cite{KU24,East2025} are of this type). This means that \hyperref[ECC:0]{points 0.}~and \hyperref[ECC:1]{1.}~are not strictly necessary for \hyperref[ECC:2]{points 2.}--\hyperref[ECC:5]{5.}~to make sense, but we have included these anyway because they are of independent interest and it does not seem likely to us that one can rigorously construct examples of extremal critical collapse without understanding a finite time formation mechanism. 
\end{rk}

Note crucially that \cref{conj:ECC} does \emph{not} claim that all (asymptotically or exactly in finite time) extremal black holes arising in gravitational collapse lie on the threshold of black hole formation. Indeed, any counterexample to the third law contains a trapped surface and hence cannot possibly be a threshold solution because the existence of trapped surfaces is stable under perturbation of the initial data. Instead, this conjecture asserts that there exist extremal black holes on the threshold and that, near such solutions, the threshold consists of extremal black holes and that some ``global'' version of \cref{conj:EVHT} holds. 

However, \cref{conj:EVHT} does imply something nontrivial about gravitational collapse! Suppose $\Psi\in\widehat{\mathfrak M}_\mathrm{black}$ forms an extremal black hole which is either exactly extremal Kerr--Newman in finite time or settles down sufficiently well to extremal Kerr--Newman so that the induced data on a $\mathcal H^+$-penetrating bifurcate null hypersurface $\mathcal C$ enters the neighborhood $\mathfrak M_\mathrm{nbhd}\subset\mathfrak M(\mathcal C)$ from \cref{conj:EVHT}. Then \cref{conj:EVHT} implies that there exists a (locally defined) $C^1$ hypersurface $\widehat{\mathfrak M}_*\subset\widehat{\mathfrak M}$, passing through $\Psi$, consisting of data sets which evolve into asymptotically extremal black holes. On one side of $\widehat{\mathfrak M}_*$ (and near $\Psi$), every data set forms an asymptotically subextremal black hole. On the other side of $\widehat{\mathfrak M}_*$, black holes may or may not form. If $\mathcal S[\Psi]$ contains a trapped surface, then black holes will form on this side of $\widehat{\mathfrak M}_*$, but \cref{conj:EVHT} implies that the event horizon will not form in the domain of dependence of $\mathcal C$! In this case, as we cross $\widehat{\mathfrak M}_*$, we can observe the location of $\mathcal H^+$ \emph{jumping}. For an explicit example of this phenomenon, we refer the reader to \cite[Theorem 3]{KU24}.

We now briefly discuss the status of \cref{conj:ECC} in the settings from \cref{sec:EHC} for which the conjecture makes sense.

\paragraph{Charged scalar field in spherical symmetry.}

For the spherically symmetric EMCSF model, \hyperref[ECC:1]{parts 0.} and \hyperref[ECC:1]{1.}~of \cref{conj:ECC} were confirmed by the second two authors of the present paper in \cite{KU22}, using a characteristic gluing construction inspired by the constructions of Aretakis--Czimek--Rodnianski \cite{ACR1}. See \cite{third-law-numerical} for more examples of extremal black holes constructed (numerically) via characteristic gluing in the EMCSF model.

In \cite[Corollary 3]{KU22}, an example of a dynamically extremal black hole (arising from Cauchy data on $\Bbb R^3$, but not in true gravitational collapse) without any trapped surfaces is presented. The black hole is exactly extremal Reissner--Nordstr\"om along the event horizon. It is natural to conjecture that this example lies on the threshold, but it is not clear to us how to prove that statement. Thus, \hyperref[ECC:2]{part 2.}~remains open, and \hyperref[ECC:2]{parts 3.}--\hyperref[ECC:2]{5.}~hinge on this and \cref{conj:EVHT}. 

A critical solution $\mathcal S[\Psi_{p_*}]$ as in \cref{conj:ECC} cannot contain trapped surfaces behind its event horizon by the incompleteness theorem \cite{Penrose}. However, for the EMCSF model we can make a much more specific conjecture about the structure of the black hole interior. We refer the reader to \cite{Kommemi13} for the general structure of the Penrose diagram of a solution to the EMCSF system. Assuming mass inflation to be generic and locally naked singularities (null singularities emanating from the center but clothed by an event horizon) to be nongeneric, Van de Moortel has shown that a generic asymptotically \emph{subextremal} black hole arising from Cauchy data on $\Bbb R^3$ in this model must have a spacelike singularity in its interior \cite{van2023breakdown,maxime-coexistence}; the Cauchy horizon breaks down due to mass inflation. In contrast, under some assumptions, it is known that mass inflation does not occur for asymptotically extremal black holes in the EMCSF model \cite{gajic-luk}. 

\begin{conj}\label{conj:closing}
For \ul{generic} $\mathcal S[\Psi_{p_*}]$, conditioned on being critical in the sense of \cref{def:ECC} for the EMCSF model, the Cauchy horizon $\mathcal{CH}^+$ arising from future timelike infinity $i^+$ smoothly closes off the spacetime. In other words, Figure 1 of \cite{van2023breakdown} holds true, in sharp contrast to the generic subextremal case.
\end{conj}

We note that the example of \cite[Corollary 3]{KU22} satisfies the conclusion of this conjecture. 

\begin{rk}\label{rk:generics}
    It is important to compare and contrast the genericity assumptions in the results of Van de Moortel \cite{van2023breakdown,maxime-coexistence} and our \cref{conj:closing}.
    
    Of course, \cite{van2023breakdown,maxime-coexistence} assumes that the black holes in question are asymptotically subextremal, which, while expected to be generic, is a fact that has only been established in this paper for the neutral model (and not in the setting of gravitational collapse). Next, \cite{van2023breakdown,maxime-coexistence} assumes that mass inflation occurs generically, which is still an open problem for EMCSF, unlike the neutral case \cite{LOSR23,Gautam}. Finally, \cite{van2023breakdown,maxime-coexistence} assumes that a generic black hole does not have a locally naked singularity emanating from the center. This fact was proved (in a relatively coarse topology) by Christodoulou in \cite{C-WCC} for the spherically symmetric Einstein-scalar field model.\footnote{Recall that the EMSF model in this paper does not allow for a regular center when the globally conserved charge $Q\ne 0$.} 
    
     By contrast, the assumption in \cref{conj:closing} that $\Psi_{p_*}$ lie on the threshold $\partial\widehat{\mathfrak M}_\mathrm{black}$ is manifestly nongeneric in $\mathfrak M$ by \cref{conj:ECC}. \cref{conj:closing} also requires locally naked singularities to be nongeneric, at least conditioned on lying on $\partial\widehat{\mathfrak M}_\mathrm{black}$! The instability mechanism for locally naked singularities in \cite{C-WCC} involves trapped surface formation, which is however clearly inappropriate for showing instability conditioned on lying in $\partial\widehat{\mathfrak M}_\mathrm{black}$. One would need a mechanism related to ``local dispersion'' as in the numerical work \cite{choptuik1993universality} (see also \cref{rk:2+1-II} below). This is an extremely interesting open problem.
\end{rk}

\paragraph{Charged massless Vlasov in spherical symmetry.}

The concept of extremal critical collapse was introduced by the second two authors of the present paper in the context of the spherically symmetric charged Vlasov model in \cite{KU24}, where \hyperref[ECC:0]{parts 0.}--\hyperref[ECC:2]{2.}~of \cref{conj:ECC} were confirmed. Recently, East has constructed new numerical examples of interpolating families \cite{East2025} which are completely distinct from the examples in \cite{KU24}. (The examples in \cite{East2025} are related to stationary solutions constructed in \cite{andreasson2009numerical} and the examples in \cite{KU24} are based on desingularizing bouncing charged null dust, a model introduced by Ori in \cite{Ori91}.) In all of these examples, extremal Reissner--Nordstr\"om is the model critical solution. East's numerics lend some credence to \hyperref[ECC:4]{part 4.}~of \cref{conj:ECC}, but his examples are manifestly nongeneric because all of the matter falls into the horizon in finite advanced time (like in \cite{KU24}). Therefore, we still have no real understanding of horizon instabilities in this model beyond \cite{W23}. Making further progress on \cref{conj:ECC} requires first completely understanding \cref{conj:EVHT} for Vlasov matter. 

\paragraph{Vacuum outside of symmetry.} In $(3+1)$-dimensional vacuum gravity, no part of \cref{conj:ECC} is understood in any sense of the word. In \cite{KU23}, the second two authors of the present paper constructed examples of vacuum gravitational collapse which are isometric to Kerr black holes with prescribed $M$ and $a$ satisfying $0\le |a|/M\le \mathfrak a_0$ for some small positive constant $\mathfrak a_0$. Like in \cite{KU22}, the proof is by characteristic gluing and uses the perturbative and obstruction-free gluing results of Aretakis--Czimek--Rodnianski \cite{ACR1} and Czimek--Rodnianski \cite{Czimek2022-cl} as a black box (one can also use the result of Mao--Oh--Tao \cite{Mao2023-cm} to obtain solutions with arbitrarily high regularity). The proof does not work for large values of $a$ and whether extremal Kerr black holes can form in gravitational collapse remains open.

\begin{rk}[Negative results]
   Reall \cite{reall2024third} and McSharry--Reall \cite{mcsharry2025supersymmetric} have developed techniques for showing that \hyperref[ECC:0]{parts 0.}~and \hyperref[ECC:1]{1.}~of \cref{conj:ECC} cannot happen in certain moduli spaces. For instance, if we endow the charged scalar field with a Klein--Gordon mass $m$, then extremal Reissner--Nordstr\"om cannot form starting from Cauchy data on $\Bbb R^3$ if $m\ge |q|$, where $q$ is the charge coupling constant. These works do not address the question of asymptotically extremal black holes forming, and it would be interesting to obtain a result of that type in any nontrivial setting. 
\end{rk}

\begin{rk}[The $(2+1)$-dimensional case II]\label{rk:2+1-II} In view of the existence of extremal BTZ black holes in $2+1$ dimensions, recall \cref{rk:2+1-I}, it is reasonable to ask if they can be models of critical solutions. As mentioned before, $2+1$ dimensions requires nontrivial matter to have a dynamical theory. In forthcoming work, McSharry--Reall adapt techniques from \cite{mcsharry2025supersymmetric} to prove that extremal BTZ cannot form in collapse (in finite time) if the matter satisfies the dominant energy condition. Therefore, \hyperref[ECC:0]{parts 0.}~and \hyperref[ECC:1]{1.}~of \cref{conj:ECC} are false for a massless scalar field. The case of a conformally coupled scalar field is left open, however. 
  
We note that there has been recent mathematical progress on the non-extremal front of critical collapse, in $2+1$ dimensions, by Cicortas and Rodnianski \cite{2D-critical}. The authors construct a countable collection of interpolating families in $(2+1)$-dimensional gravity with a negative cosmological constant and a massless scalar field in hypersurface-orthogonal axisymmetry, for which the critical solutions are naked singularities,\footnote{Technically, collapsed light cone singularities, see \cite{Kommemi13}.} with a distinct critical solution for each family. These families form black holes on one side of the critical solution and fail to collapse (in a local sense, similar to our notion in this paper) on the other side. This problem was previously studied numerically by Pretorius and Choptuik \cite{pretorius2000gravitational} (see also \cite{garfinkle2001exact,jalmuzna2015scalar}). One of the main observations used in \cite{2D-critical} (and also the proof of weak cosmic censorship \cite{2D-WCC}) is the existence of a \emph{mass gap} which serves as an extension principle at the center. This is in sharp contrast to the case of a spherically symmetric scalar field in $(3+1)$-dimensions \cite{C-WCC,choptuik1993universality}, for which the critical collapse picture remains completely open. It would be extremely interesting to study the threshold hypersurface in the setting of \cite{2D-critical}, but the techniques of this paper do not seem tailored to that problem.
\end{rk}

\begin{rk}[The $(4+1)$-dimensional case II]
   Recently, Crump, Gadioux, Reall, and Santos have provided numerical evidence of a gluing construction of extremal Myers--Perry forming in collapse for $(4+1)$-dimensional vacuum gravity under $\mathrm{SU}(2)\times\Bbb Z_4$-symmetry \cite{3rd-law-vac} (recall \cref{rk:4+1-I}). Assuming this result can be made rigorous (e.g., by interval arithmetic and a perturbation argument), \hyperref[ECC:0]{parts 0.}~and \hyperref[ECC:1]{1.}~of \cref{conj:ECC} are true in this setting. 
\end{rk}

\subsubsection{Connections with threshold phenomena for dispersive equations}\label{sec:dispersive-PDE}

In the context of the Einstein equations, it does not really make sense to think of a ``phase space'' of solutions, as there is no canonical splitting into ``space'' and ``time'' directions, and so one cannot reasonably think of the Einstein evolution as defining a flow in some phase space. We have therefore deliberately used the terminology ``moduli space'' for the object that parametrizes solutions. The ``stable manifolds'' $\mathfrak M_\mathrm{stab}^\sigma$ in \cref{thm:intro-foliation} are therefore not invariant manifolds of the flow in the usual sense of dynamical systems, nor are they attractors in any sense. Note as well that in the (asymptotically flat and/or extremal) black hole stability problem, there is no exponential decay towards equilibrium at the linear level.

Nevertheless, the subject of this paper is thematically very related to the construction of invariant manifolds for nonlinear dispersive PDEs. This subject has a long history and a direct connection with the classical theory of hyperbolic dynamics for ODEs. In the context of focusing semilinear dispersive PDEs which admit unstable soliton solutions, there are several results that identify a transition phenomenon in phase space from solutions which exist globally and scatter to a free wave, to solutions which blow up in finite time. The interface is a codimension-one submanifold passing through the ground state. 

In \cite{bates-jones}, Bates and Jones construct the center, stable, and unstable manifolds for ODEs in Banach spaces using a method of Hadamard \cite{hadamard-graph}, and then apply this to the focusing energy subcritical semilinear Klein--Gordon equation with a power nonlinearity in spherical symmetry, near its ground state. The associated center-stable manifold is shown to be Lipschitz. The method has been extended outside of symmetry and for more nonlinearities in \cite{nakanishi-schlag_nlkg}. The Lyapunov--Perron method has also been used, and involves analyzing the linearized flow around the ground state. The associated center-stable manifold is then constructed through energy methods. This has been used in the context of the nonlinear Schr\"{o}dinger equation in \cite{schlag-nls}, \cite{krieger-schlag}, and \cite{beceanu} (in the last work the codimension-one manifold of data that asymptote to modulated copies of the ground state is shown to be real analytic!). For a more detailed picture we refer to the book of Nakanishi and Schlag \cite{nakanishi-schlag_book}. 

In the results discussed in \cite{nakanishi-schlag_book}, the soliton is exponentially unstable because the linearized operator possesses a negative eigenvalue. We also mention the works \cite{merle-raphael-szeftel_bw} and \cite{martel-merle-nakanishi-raphael} for similar results where the soliton is not exponentially unstable. Finally, let us also mention the works \cite{duyckaerts_merle-gafa}, \cite{duyckaerts_merle-imrn}, \cite{jendrej_lawrie-inv} on energy-critical dispersive semilinear equations, where threshold dynamics are studied establishing at that level the existence of so-called ``bubble decomposition'' (solutions that settle to a sum of rescaled solitons), that are a very useful tool towards an attempt to prove the ``soliton resolution conjecture'' in these contexts.

In the context of quasilinear wave equations, we are not aware of any construction of regular stable manifolds before this work besides \cite{donninger2016codimension}, where the soliton-like solution considered is the catenoid. The authors prove codimension-one stability of the catenoid under the timelike minimal surface equation in Minkowski space. However, they only prove Lipschitz regularity of the stable manifold and neither teleological gauge normalization nor modulation of the final state are involved. The work \cite{donninger2016codimension} takes place in axisymmetry on a $(2+1)$-dimensional manifold, although the codimension-one stability of the catenoid in the same context has been investigated outside symmetry in higher dimensions, see \cite{luhrmann-oh-shahshahani}, \cite{oh-shahshahani}, and \cite{ning}. Again, the instability in this problem is driven by a negative eigenvalue of the linearized operator. See \cite{donninger2016codimension} for a conjecture on what happens off of the center-stable manifold in this context. We also mention the recent work \cite{shatah} concerning invariant manifolds for certain quasilinear equations including irrotational water waves with surface tension and mean curvature flow. 

\subsection*{Acknowledgments}

The authors would like to thank Mihalis Dafermos, Dejan Gajic, Gustav Holzegel, Jonathan Luk, Sung-Jin Oh, Harvey Reall, and Maxime Van de Moortel for helpful conversations. R.U. acknowledges support from the NSF grant DMS-2401966 and a Miller Fellowship.

\section{Overview of the proofs}\label{sec:intro-proof-overview}

In this section, we give an overview of the main ideas of the paper. We begin in \cref{sec:EMSF-system} with the definition of the spherically symmetric Einstein--Maxwell-neutral scalar field system and fixing our basic notation. In \cref{sec:overview-I}, we introduce schematically the moduli space, the general setup of gauges, and describe the proof of the refined dichotomy of \cref{thm:intro-uniform-stability}. In \cref{sec:overview-II}, we describe the proof of the main theorem of the paper, \cref{thm:intro-foliation}. Finally, in \cref{sec:overview-III-IV}, we outline the proofs of \cref{thm:intro-scaling,thm:intro-instability}, which are essentially corollaries of the proof of \cref{thm:intro-foliation}.

\subsection{The Einstein--Maxwell-neutral scalar field system in spherical symmetry}\label{sec:EMSF-system}

\subsubsection{Spherically symmetric spacetimes in double null gauge}\label{sec:double-null-gauge}

Let \index{M@$(\mathcal M^{3+1},g)$, spacetime manifold} $(\mathcal M,g)$ be a connected, time-oriented, four-dimensional Lorentzian manifold. We say that $(\mathcal M,g)$ is \emph{spherically symmetric} if $\mathcal M $ splits diffeomorphically as \index{Q@$\mathcal Q$, $(1+1)$-dimensional quotient manifold} $\mathcal Q\times S^2$ with metric \index{r@$r$, area radius} \index{o@$\Omega^2=-2g_{uv}$, null lapse} 
\begin{align*}
    g = g_\mathcal{Q} + r^2g_{S^2},
\end{align*}
where $(\mathcal Q,g_\mathcal Q)$ is a (1+1)-dimensional Lorentzian spacetime with boundary (corresponding to the initial data hypersurface\footnote{In this paper, our spacetimes do not include a center of symmetry $\subset\{r=0\}$.}), $g_{S^2}$ is the round metric on the unit sphere, and $r:\mathcal Q\to\Bbb R_{>0}$ can be geometrically interpreted as the area-radius of the orbits of the isometric $\mathrm{SO}(3)$ action on  $(\mathcal M,g)$. We assume that $(\mathcal Q,g_\mathcal Q)$ admits a \emph{global double-null coordinate system} $(u,v)$ such that $\partial_u$ and $\partial_v$ are future-directed and
\begin{equation}
g = -\Omega^2\,dudv + r^2  g_{S^2},\label{eq:dn}
\end{equation}
  where $\Omega^2:\mathcal Q\to\Bbb R_{>0}$ is known as the \emph{null lapse}. The constant $u$ and $v$ curves are null in $(\mathcal Q,g_\mathcal Q)$ and correspond to null hypersurfaces ``upstairs'' in the full spacetime $(\mathcal M,g)$. We will often refer interchangeably to  $(\mathcal M,g)$ and the reduced spacetime $(\mathcal Q,r,\Omega^2)$.

The double null coordinates $(u,v)$ are not uniquely defined. For any strictly increasing smooth functions $U,V: \R \to \R$, $(\tilde u,\tilde v)=(U(u),V(v))$ defines a double null coordinate system on $\mathcal Q$ for which $g= -\tilde \Omega^2 \,d \tilde u d \tilde v + \tilde r^2g_{S^2}$, where $\tilde \Omega^2(\tilde u , \tilde v) = (U' V')^{-1} \Omega^2(U^{-1}(\tilde u), V^{-1}(\tilde v)) $ and $\tilde r (\tilde u, \tilde v) = r(U^{-1} (\tilde u), V^{-1} (\tilde v))$. \index{u0@$u$, initial data normalized retarded time}\index{v0@$v$, initial data normalized advanced time}

Recall the \emph{Hawking mass} $m:\mathcal M  \to\mathbb R$, which is defined by the relation $ 1-\frac{2m}{r} \doteq g^{-1}(d r, dr)$ and can be expressed as \index{m@$m$, Hawking mass}
\begin{equation}\label{eq:Hawking-mass}
    m = \frac r2 \left( 1+  \frac{4\partial_ur\partial_vr}{\Omega^2}\right)
\end{equation} in double null gauge. 

We will consider spherically symmetric spacetimes equipped with spherically symmetric electromagnetic fields with constant Coulomb charge. The field strength tensor takes the simple form \index{F@$\mathbf F$, EM field strength tensor}
\begin{equation}\label{eq:F-sph-sym}
   \mathbf  F = -\frac{\Omega^2Q}{2r^2}du\wedge dv,
\end{equation} where $Q\in\Bbb R$\index{Q@$Q$, electromagnetic charge} is a global constant. 

To see best the good structure of the spherically symmetric Einstein equations, it is very helpful to introduce some shorthand notation. We recall the \emph{renormalized Hawking mass} \index{pi0@$\varpi$, renormalized Hawking mass}
\begin{equation}
    \varpi\doteq m+\frac{Q^2}{2r},\label{eq:varpi-defn}
\end{equation}
 the \emph{mass aspect function} \index{mu@$\mu$, mass aspect function}
\begin{equation}
    \mu\doteq \frac{2m}{r}= \frac{2\varpi}{r}-\frac{Q^2}{r^2},\label{eq:mu-defn}
\end{equation} the traditional Greek notation \index{n@$\nu$, ingoing expansion $\partial_ur$}\index{l@$\lambda$, outgoing expansion $\partial_vr$}\index{kappa@$\kappa$, renormalized outgoing expansion $\lambda/(1-\mu)$}\index{g@$\gamma$, renormalized ingoing expansion $\nu/(1-\mu)$} \begin{equation}
    \nu\doteq\partial_ur,\quad\lambda\doteq \partial_vr, \quad\kappa\doteq-\frac{\Omega^2}{4\partial_ur} = \frac{\lambda}{1-\mu},\quad \gamma\doteq - \frac{\Omega^2}{4\partial_vr} =\frac{\nu}{1-\mu},\label{eq:greek-letters}
\end{equation}
the \index{kappa@$\varkappa$, dynamical redshift factor} \emph{dynamical redshift factor}
\begin{equation}
 \label{eq:varkappa}   \varkappa\doteq \frac{2}{r^2}\left(\varpi-\frac{Q^2}{r}\right),
\end{equation}
and the \emph{signed charge-to-mass ratio} \index{rho@$\rho$, quasilocal charge-to-mass ratio $Q/\varpi$}
\begin{equation} \label{eq:rho}  
    \rho\doteq\frac{Q}{\varpi},
\end{equation}
which is defined whenever $\varpi\ne 0$ (as will always be the case in this paper). 

The Reissner--Nordstr\"om metric written in Eddington--Finkelstein double null coordinates, \index{g1@$g_{M,e}$, RN metric} \index{g2@$g_{M,\varrho}$, also RN metric} \index{M@$M$, RN mass parameter} \index{e@$e$, RN charge parameter} \index{rho1@$\varrho$, RN charge-to-mass ratio} \index{D@$D(r_\diamond)\doteq 1-\frac{2M}{r_\diamond}+\frac{Q^2}{r_\diamond^2}$}
\begin{equation}\label{eq:RN-intro-1}
    g_{M,e} = -4Ddu dv+r^2g_{S^2},\qquad D\doteq 1 - \frac{2M}{r}+\frac{e^2}{r^2},
\end{equation}
is spherically symmetric with\index{D0@$D'(r_\diamond) \doteq \frac{dD}{dr}(r_\diamond)$}
\begin{equation*}
    \varpi=M,\quad Q=e,\quad \lambda=-\nu=D,\quad \kappa = -\gamma = 1,\quad \varkappa = D'\doteq \frac{dD}{dr}=\frac{2}{r^2}\left(M-\frac{e^2}{r}\right),\quad \rho=\frac{e}{M}.
\end{equation*}
We denote the constant value of $\rho$ on Reissner--Nordstr\"om by the cursive $\varrho$, and we will often parametrize the family by $(M,\varrho)$ instead of $(M,e)$. We will recall more facts about the Reissner--Nordstr\"om geometry in \cref{sec:geometry-RN} below.

\begin{rk}
    The definition of $\varkappa$ here differs from that of \cite{AKU24} by a factor of 2, as this greatly reduces the number of factors of 2 in the paper. However, the ``usual'' horizon redshift (recall \cite[Remark 2.3]{AKU24}) is now $\bm\kappa=\frac 12\varkappa|_{\mathcal H^+}$ in Reissner--Nordstr\"om.
\end{rk}

\subsubsection{The system of equations} \label{sec:the-model}

\begin{defn}
    The \emph{Einstein--Maxwell-(neutral/real) scalar field (EMSF) system} consists of a spacetime $(\mathcal M,g)$ equipped with an electromagnetic field $\bf F\in\Omega^2(\mathcal M)$ and a scalar field $\phi:\mathcal M\to\Bbb R$ satisfying the equations \index{ph0@$\phi$, (real) scalar field}
    \begin{equation}
    \Ric(g)-\tfrac 12 R(g)g = 2(T^\mathrm{EM}+T^\mathrm{SF}),\label{eq:EFE}
\end{equation}
\begin{equation}
    d\bf F=0,\quad d\star\bf  F=0,\label{eq:Maxwell}
\end{equation}
\begin{equation}
    \Box_g \phi =0,\label{eq:phi-wave-general}
\end{equation}
where the energy-momentum tensors are defined by 
\begin{equation*}
 \nonumber   T^\mathrm{EM}_{\mu\nu}\doteq \mathbf F_{\mu\alpha}\mathbf F_\nu{}^\alpha-\tfrac 14g_{\mu\nu}\mathbf F_{\alpha\beta}\mathbf F^{\alpha\beta},\quad
   T^\mathrm{SF}_{\mu\nu}\doteq \partial_\mu\phi\partial_\nu\phi-\tfrac 12 g_{\mu\nu}\partial_\alpha\phi\partial^\alpha\phi.
\end{equation*}
\end{defn}

We say that $(\mathcal M,g,\bf F,\phi)$ is \emph{spherically symmetric} if $(\mathcal M,g)$ is spherically symmetric as defined in \cref{sec:double-null-gauge}, $F$ has the form \eqref{eq:F-sph-sym}, and $\phi$ is independent of the $S^2$ factor. In this case, the field equations \eqref{eq:EFE} reduce to the wave equations
\begin{align}
\label{eq:r-wave}    \partial_u\partial_vr&=-\frac{\Omega^2}{4r}-\frac{\partial_ur\partial_vr}{r}+\frac{\Omega^2Q^2}{4r^3},\\
    \label{eq:Omega-wave}\partial_u\partial_v{\log\Omega^2}&=\frac{\Omega^2}{2r^2}+\frac{2\partial_ur\partial_vr}{r^2}-\frac{\Omega^2Q^2}{r^4} - 2 \partial_u \phi \partial_v \phi ,
\end{align}
and Raychaudhuri's equations
\begin{align}
   \label{eq:Ray-u} \partial_u\left(\frac{\partial_ur}{\Omega^2}\right)&=-\frac{r}{\Omega^2}(\partial_u\phi)^2,\\
  \label{eq:Ray-v}\partial_v\left(\frac{\partial_vr}{\Omega^2}\right)&=-\frac{r}{\Omega^2}(\partial_v\phi)^2.
\end{align}
Maxwell's equation \eqref{eq:Maxwell} is automatically satisfied since $Q$ is constant. Finally, the wave equation \eqref{eq:phi-wave-general} is equivalent to 
\begin{equation}
     \partial_u\partial_v\phi= - \frac{\partial_vr\partial_u\phi}{r}-\frac{\partial_ur\partial_v\phi}{r}.\label{eq:phi-wave-1}
\end{equation}
We consider a ``solution'' to the spherically symmetric Einstein--Maxwell-scalar field system to consist of the tuple $\mathcal S=(\mathcal Q,r,\Omega^2,\phi,Q)$. 

We will not actually work with the equations \eqref{eq:r-wave}--\eqref{eq:Ray-v} as written here and will instead use the quantities \eqref{eq:varpi-defn}--\eqref{eq:varkappa} which satisfy more helpful equations. First, the wave equation \eqref{eq:r-wave} can be written in the compact form 
\begin{equation}
    \label{eq:nu-v}\partial_u\lambda=\partial_v\nu=\kappa\nu\varkappa,
\end{equation}
which allows us to write \eqref{eq:phi-wave-1} as an inhomogeneous wave equation for $\psi\doteq r\phi$,\index{ps@$\psi\doteq r \phi$, rescaled scalar field}
\begin{equation}\label{eq:wave-equation-psi}
    \partial_u\partial_v\psi=\kappa\nu\varkappa\phi.
\end{equation}
Using \eqref{eq:r-wave}--\eqref{eq:Ray-v}, we derive the fundamental relations \begin{gather}\label{eq:m-u}
\partial_um=\frac{r^2}{2\gamma}(\partial_u\phi)^2 +\frac{Q^2}{2r^2}\nu, \qquad
\partial_vm=\frac{r^2}{2\kappa}(\partial_v\phi)^2 +\frac{Q^2}{2r^2}\lambda,\\
  \label{eq:varpi-u}  \partial_u\varpi = \frac{r^2}{2\gamma}(\partial_u\phi)^2,\qquad
   \partial_v\varpi = \frac{r^2}{2\kappa}(\partial_v\phi)^2,\\
\label{eq:kappa-u}    \partial_u\kappa =\frac{r\kappa}{\nu}(\partial_u\phi)^2, \qquad
       \partial_v\gamma=\frac{r\gamma}{\lambda}(\partial_v\phi)^2.
\end{gather}

The system \eqref{eq:r-wave}--\eqref{eq:kappa-u} is \emph{gauge covariant}
 in the following sense. Let $(\tilde u,\tilde v)=(U(u),V(v))$ be a double null gauge transformation as described in \cref{sec:double-null-gauge}. Then the quantities 
 {\mathtoolsset{showonlyrefs=false} 
 \begin{equation}\label{eq:COV-1}
     \tilde r(\tilde u,\tilde v)\doteq r(U^{-1}(\tilde u),V^{-1}(\tilde v)),\quad \tilde\varpi(\tilde u,\tilde v)\doteq \varpi(U^{-1}(\tilde u),V^{-1}(\tilde v)), \quad \tilde\phi(\tilde u,\tilde v)\doteq \phi(U^{-1}(\tilde u),V^{-1}(\tilde v)),
 \end{equation}
 \begin{equation}\label{eq:COV-2}
     \tilde\nu(\tilde u,\tilde v)\doteq \frac{\nu(U^{-1}(\tilde u),V^{-1}(\tilde v))}{U'(U^{-1}(\tilde u))},\quad \tilde\lambda(\tilde u,\tilde v)\doteq \frac{\lambda(U^{-1}(\tilde u),V^{-1}(\tilde v))}{V'(V^{-1}(\tilde v))},
 \end{equation}
 \begin{equation}\label{eq:COV-3}
        \tilde\gamma(\tilde u,\tilde v)\doteq \frac{\gamma(U^{-1}(\tilde u),V^{-1}(\tilde v))}{U'(U^{-1}(\tilde u))},\quad \tilde\kappa(\tilde u,\tilde v)\doteq \frac{\kappa(U^{-1}(\tilde u),V^{-1}(\tilde v))}{V'(V^{-1}(\tilde v))}
 \end{equation}}
 satisfy \eqref{eq:r-wave}--\eqref{eq:kappa-u} as well (with $\partial_u$ and $\partial_v$ replaced by $ \partial_{\tilde u}$ and $\partial_{\tilde v}$, respectively).
 
\subsection{Overview of the proof of \texorpdfstring{\cref{thm:intro-uniform-stability}}{Theorem I}}\label{sec:overview-I}

The proof of this theorem has many similarities with \cite{AKU24}, but has important differences because of the noncollapsing aspect and estimates which are uniform in the Reissner--Nordstr\"om parameters. For use later in the proof of \cref{thm:intro-foliation}, the setup is also necessarily much more complicated than in \cite{AKU24}. 

\subsubsection{Definition of the moduli space \texorpdfstring{$\mathfrak M$}{M} and its topology}\label{sec:intro-mod-space} 

We begin by sketching the definition of the moduli space $\mathfrak M$. While \cref{thm:intro-uniform-stability} is not so sensitive to this definition (it is much more important for \cref{thm:intro-foliation,thm:intro-scaling,thm:intro-instability}), it is useful to fix the notation now. 

We refer back to the Penrose diagram of the setup of our main results, \vref{fig:dichotomy}. \emph{Bifurcate characteristic seed data} for the spherically symmetric EMSF model on $\mathcal C\doteq C_\out\cup\underline C{}_\ing$ consists of 
\[\Psi\doteq (\phi_\circ,r_\circ,\varpi_\circ,\rho_\circ),\]
where $\phi_\circ\doteq \phi|_{\mathcal C}$, and $r_\circ$, $\varpi_\circ$, and $\rho_\circ$ denote the values of $r$, $\varpi$, and $\rho$ (respectively) at $C_\out\cap\u C{}_\ing$. The conserved charge $Q$ of the solution is recovered via the formula $Q=\rho_\circ\varpi_\circ$. We parametrize $C_\out$ by an advanced time coordinate $v\in [1,\infty)$ and $\u C{}_\ing$ by a retarded time coordinate $u\in[1,U_*]$, where $(U_*,1)$ is the solid black point depicted in \cref{fig:dichotomy}. After imposing the gauge conditions $\lambda = 1$ on $C_\out$ and $\nu = -1$ on $\u C{}_\ing$, we can define the maximal globally hyperbolic development $\mathcal S_\mathrm{max}[\Psi]=(\mathcal Q_\mathrm{max},r,\Omega^2,\phi)$ of $\Psi$, where $\mathcal Q_\mathrm{max}\subset\Bbb R^2_{u,v}$ is the maximal domain of the solution. 

We take $\phi_\circ$ to belong to a Banach space $(\mathfrak F,\|\cdot\|_\mathfrak{F})$, where $\|\cdot\|_\mathfrak{F}$ is a weighted $C^2$-norm on $\mathcal C$ (see already \eqref{eq:norm-defn} for the definition), and $\mathfrak F\doteq\{\phi_\circ\in C^2(\mathcal C):\|\phi_\circ\|_\mathfrak{F}<\infty\}$. We also set
\begin{equation*}
    \mathfrak X\doteq \mathfrak F\times \Bbb R^2\quad\text{with elements}\quad x=(\phi_\circ,r_\circ,\varpi_\circ),
\end{equation*}
\begin{equation*}
       \mathfrak Z\doteq \mathfrak X\times \Bbb R\quad\text{with elements}\quad \Psi=(x,\rho_\circ).
\end{equation*}
The \emph{moduli space} of seed data $\mathfrak M$ is then given by an open set of $\mathfrak Z$ satisfying natural asymptotic flatness and compatibility conditions (for instance, we require $r_\circ>0$). 

Note that the seed data set $(0,r_\circ,\varpi_\circ,\rho_\circ)$ evolves into (an appropriate piece of) a Reissner--Nordstr\"om solution with mass $M=\varpi_\circ$ and charge $e=\rho_\circ\varpi_\circ$. 

See already \cref{sec:seed-data,sec:AF-defns} for the precise definitions of seed data, $\mathfrak M$, etc. 

\begin{rk}
    In \cite{AKU24}, the moduli space of seed data on $\mathcal C$ was parametrized by quadruples $(\phi_\circ,r_\circ,\varpi_\circ,Q)$ (after changing notation to match that of this paper). These spaces are diffeomorphic, away from $\{\varpi_\circ =0\}$, via the map $(\phi_\circ,r_\circ,\varpi_\circ,\rho_\circ)\mapsto (\phi_\circ,r_\circ,\varpi_\circ,\rho_\circ\varpi_\circ)$.
\end{rk}

\subsubsection{Logic of the proof of \texorpdfstring{\cref{thm:intro-uniform-stability}}{Theorem I}}\label{sec:intro-proof-uniform}

Let $\Psi\in \mathfrak M$ with maximal development $\mathcal S_\mathrm{max}[\Psi]$ and null infinity $\mathcal I^+$. Recall the Bondi mass function $M_{\mathcal I^+}$ and let \[P_{\mathcal I^+}\doteq\frac{Q}{M_{\mathcal I^+}}.\] The proof begins with the following key observation:

\begin{ingredient}[Noncollapse criterion]\label{ingredient-noncollapse}
    If $|P_{\mathcal I^+}(u)|>1$ for some $(u,\infty)\in\mathcal I^+$, then $\Psi\in\mathfrak M_\mathrm{non}$.
\end{ingredient}

This result is soft and is not restricted to the small data regime; see already \cref{sec:superextremal} below for the proof. It relies crucially on the constancy of $Q$ in the EMSF model and is false in its generality for models with dynamical charge (counterexamples can be constructed using the methods of \cite{KU22,KU24}). 

Roughly, \cref{thm:intro-uniform-stability} is proved by showing that as long as $P_{\mathcal I^+}$ remains non-superextremal, then the spacetime is decaying to a non-superextremal Reissner--Nordstr\"om. (Note that $u\mapsto |P_{\mathcal I^+}(u)|$ is nondecreasing, so that solutions can become dynamically superextremal along $\mathcal I^+$.) To start making this precise, we define a Bondi time coordinate $\u u$ such that
\begin{equation}\label{eq:intro-Bondi}
    \partial_{\u u}r|_{\mathcal I^+} = -1.
\end{equation}
Assuming $\mathcal S_\mathrm{max}[\Psi]$ exists until at least $\u u = \u u{}_f$, we measure
\begin{equation}
    M=\u M{}_{\mathcal I^+}(\u u{}_f),\quad \varrho = \u P{}_{\mathcal I^+}(\u u{}_f),\label{eq:intro-RN-anchoring-1}
\end{equation}
where $\u M{}_{\mathcal I^+}$ and $\u P_{\mathcal I^+}$ are $M_{\mathcal I^+}$ and $P_{\mathcal I^+}$ written in Bondi time, respectively. After defining an appropriate teleologically normalized advanced time coordinate $\u v{}_{\u u{}_f}$ (which depends nontrivially on $\u u{}_f$, see already \cref{sec:intro-gauge}), we compare $\u{\mathcal S}{}_{\u u{}_f}[\Psi]$ (the solution written in this $\u u{}_f$-normalized gauge $(\u u, \u v{}_{\u u{}_f})$) to an appropriately anchored Reissner--Nordstr\"om solution $(r_\diamond,\Omega^2_\diamond)$ with mass $M$ and parameter ratio $\varrho$ (written in Eddington--Finkelstein double null gauge \eqref{eq:RN-intro-1}). 

\begin{rk}
    In some works in double null gauge, $\u u$ is used to denote the advanced time coordinate. We emphasize that this symbol is used to denote the Bondi-normalized retarded time coordinate in this paper. 
\end{rk}

We say that
\[\Psi\in\mathcal K(\ve,M,\varrho,\u u{}_f)\]
if the data satisfies $\|\phi_\circ\|_\mathfrak{F}\le\ve$, the geometry of $\u{\mathcal S}{}_{\u u{}_f}[\Psi]$ decays in an appropriate sense to  $(r_\diamond,\Omega^2_\diamond)$ up to $\u u= \u u{}_f$, and the scalar field $\phi$ decays in an appropriate sense (with respect to a hierarchy of weighted energies) up to $\u u= \u u{}_f$. This notion allows for $\varrho$ to be superextremal, which we will take advantage of for understanding noncollapsing solutions near the threshold. We introduce the dyadic time scales $L_i\doteq2^i$\index{L@$L_i\doteq 2^i$, dyadic time scale}. 

\begin{ingredient}[Dyadic iteration and modulation]\label{ingredient:iteration}
    For any $\ve$ sufficiently small, if $\Psi\in \mathcal K(\ve,M,\varrho,L_i)$ with $|\varrho|\le 1+\alpha L_i^{-2}$, then $\Psi\in \mathcal K(\ve,M',\varrho',L_{i+1})$, where the modulated parameters satisfy
    \begin{equation}\label{eq:parameter-change} |M-M'|+|\varrho-\varrho'|\les \ve^2L_i^{-3+\delta}.\end{equation}
    Here, $\alpha>0$ is a constant tied to the geometry of superextremal Reissner--Nordstr\"om and $\delta>0$ is an arbitrary (fixed) small parameter. 
\end{ingredient} 

 The assumption $|\varrho|\le 1+\alpha L_i^{-2}$ should be understood as allowing $\varrho$ to be superextremal, but not ``too'' superextremal, depending on the dyadic timescale $L_i$. \cref{thm:intro-uniform-stability} is now proved via the following \emph{dyadic induction} algorithm. 

\begin{enumerate}
    \item On the level of initial data: If $|P_{\mathcal I^+}(1)|>1$, then $\Psi\in\mathfrak M_\mathrm{non}$. Otherwise, for $\ve$ sufficiently small, we solve up to Bondi time $\u u=L_1$ and show that $\Psi\in \mathcal K(\ve,M_1,\varrho_1,L_1)$ for some parameters $(M_1,\varrho_1)$ (this step is essentially semiglobal Cauchy stability of the Reissner--Nordstr\"om family).  

    \item $i\ge 1$: If $\Psi\in\mathcal K(\ve,M_i,\varrho_i,L_i)$ and $|\varrho_i|>1$, then $\Psi\in\mathfrak M_\mathrm{non}$. Otherwise, $|\varrho_i|\le 1$ and we may apply \cref{ingredient:iteration} to find parameters $(M_{i+1},\varrho_{i+1})$ such that $\Psi\in\mathcal K(\ve,M_{i+1},\varrho_{i+1},L_{i+1})$.

    \item Step 2. can be repeated until $|\varrho_i|>1$ or indefinitely if $|\varrho_i|\le 1$ for all $i$. 

    \item If Step 2.~is repeated infinitely often, then the parameters $(M_{i},\varrho_i)$ converge to some $(M_\infty,\varrho_\infty)$ with $|\varrho_\infty|\le 1$ by \eqref{eq:parameter-change}. One then shows that $\mathcal S_\mathrm{max}[\Psi]$ contains a black hole which converges to Reissner--Nordstr\"om with parameters $(M_\infty,\varrho_\infty)$.
\end{enumerate}

\cref{ingredient:iteration} combines \emph{uniform stability estimates} for the geometry and scalar field, in the full range of parameters $|\varrho|\le 1+\alpha L_i^{-2}$, with the dyadic modulation technique from \cite{AKU24}, but now performed \emph{at $\mathcal I^+$}. The use of dyadic iteration and modulation was inspired by \cite{DHRT22,DHRT24}.  

\begin{rk}
    The proof of \cref{thm:intro-uniform-stability} only requires $|\varrho|\le 1$ in \cref{ingredient:iteration}. However, the uniform stability estimates in the range $1<|\varrho|\le 1+\alpha L_i^{-2}$ are crucial for understanding the threshold. 
\end{rk}

\subsubsection{Uniform energy estimates for the Reissner--Nordstr\"om family}\label{sec:intro-unifying}

A major component of the proof of \cref{ingredient:iteration} is to understand the decay of solutions to the wave equation
\begin{equation}\label{eq:intro-h-wave}
    \Box_{g_{M,\varrho}}\phi =0,
\end{equation}
where $g_{M,\varrho}$ is a Reissner--Nordstr\"om metric, up to a dyadic time $L_i$, for parameter ratios in the range $|\varrho|\le 1+\alpha L_i^{-2}$. For simplicity, we consider in this section only homogeneous waves on the exact Reissner--Nordstr\"om background, and will explain how the analysis is extended to dynamical backgrounds and inhomogeneous solutions in \cref{sec:intro-energy-estimates} below.

\begin{figure}
\centering{
\def\svgwidth{24pc}
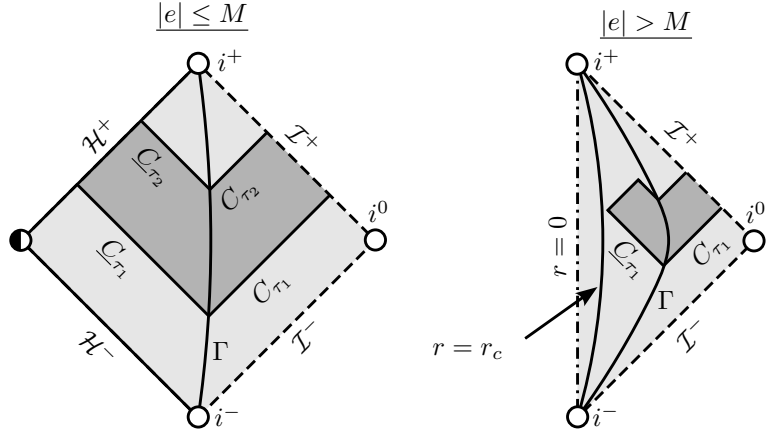}
\caption{A Penrose diagram of Reissner--Nordstr\"om depicting the foliations $C(\tau)$ and $\underline C(\tau)$ used in the estimates \eqref{eq:intro-T}--\eqref{eq:intro-hp}. The region of integration in the double integrals is shaded darker. In the left figure, the bifurcation sphere of $\mathcal H^+$ and $\mathcal H^-$ is included in the spacetime when $|e|<M$ (solid point) but not when $|e|=M$ (open point).}
\label{fig:RN-intro}
\end{figure}

The general strategy to prove energy decay statements for solutions to \eqref{eq:intro-h-wave} on Reissner--Nordstr\"om backgrounds $g_{M,\varrho}$ consists of deriving a hierarchy of weighted energy boundedness inequalities and time-integrated energy decay estimates. For subextremal Reissner--Nordstr\"om, this hierarchy takes the form
\begin{gather}
   \label{eq:intro-T}   \int_{C(\tau_2)}r^2(\partial_v\phi)^2\, dv+ \int_{\underline C(\tau_2)} (\partial_u\phi)^2\,du \les \int_{C(\tau_1)}r^2(\partial_v\phi)^2\, dv+ \int_{\underline C(\tau_1)}(\partial_u\phi)^2\,du, \\
 \label{eq:intro-N}   \int_{\u C(\tau_2)}D^{-1}(\partial_u\psi)^2\,du +\bm\kappa\int_{\tau_1}^{\tau_2}\int_{\u C(\tau_2)}D^{-1}(\partial_u\psi)^2\,du \les  \int_{\u C(\tau_1)}D^{-1}(\partial_u\psi)^2\,du+\cdots\\
  \label{eq:intro-rp}    \int_{C(\tau_2)}r^p(\partial_v\psi)^2\, dv+ \int_{\tau_1}^{\tau_2}\int_{C(\tau)} r^{p-1}(\partial_v\psi)^2\, dvd\tau\les \int_{C(\tau_1)}r^p(\partial_v\psi)^2\, dv+\cdots,
\end{gather}
where $(u,v)$ denote Eddington--Finkelstein double null coordinates on the domain of outer communication, $\tau$ is proportional to proper time along a timelike curve $\Gamma$ with constant area-radius, $\tau_1\le\tau_2$, $p\in (0,3)$, ``$\cdots$'' denote extra terms which are unimportant for now, the foliations $C(\tau)$ and $\underline C(\tau)$ are defined pictorially in \vref{fig:RN-intro}, $\psi\doteq r\phi$, $D$ is as in \eqref{eq:RN-intro-1}, and $\bm\kappa= \frac 12 D'(r_+)\sim \sqrt{1-\varrho}$ is the surface gravity.

 \index{g@$\Gamma=\{\u u=\u v\}$}

\begin{rk}
    The expression $D^{-1}(\partial_u\psi)^2\,du$ represents a \emph{non-degenerate} one-form along $\underline C(\tau)$. Indeed, written in ingoing Eddington--Finkelstein coordinates $(v,r)$, it corresponds to $(\partial_r\psi)^2\,dr$ along $\underline C(\tau)$, which is manifestly nondegenerate. 
\end{rk}

The first inequality, \eqref{eq:intro-T}, follows immediately from the usual conservation law associated to the time-translation Killing vector field $T=\partial_t$ in Reissner--Nordstr\"om. The inequality \eqref{eq:intro-N} is the celebrated \emph{horizon redshift estimate} of Dafermos--Rodnianski \cite{dafermos2009red,dafermos2013lectures} and follows from the divergence theorem applied to the energy current with multiplier vector field $Y\doteq D^{-1}\partial_u$.\footnote{In the original \cite{dafermos2009red,dafermos2013lectures}, the $Y$ vector field is timelike (and called $N$ in \cite{dafermos2013lectures}), not null, but this explicit choice is simpler in this context.} Note that the spacetime integral in the redshift estimate degenerates in the extremal limit $|\varrho|\to 1$. Finally, the inequality \eqref{eq:intro-rp} is the celebrated \emph{$r^p$-weighted estimate} of Dafermos and Rodnianski \cite{dafermos2010new} and relies only on the asymptotic flatness of the Reissner--Nordstr\"om geometry.

In the extremal case $|\varrho|=1$, the estimate \eqref{eq:intro-N} degenerates, and is replaced a hierarchy of estimates
\begin{equation}\label{eq:r-Mp-intro}  
        \int_{\u C(\tau_2)}(r-M)^{-p}(\partial_u\psi)^2\,du + \int_{\tau_1}^{\tau_2}\int_{\u C(\tau)}(r-M)^{-p+1}(\partial_u\psi)^2\,dud\tau \les   \int_{\u C(\tau_1)}(r-M)^{-p}(\partial_u\psi)^2\,du +\cdots
\end{equation} for $p\in (0,3)$.
This can be thought of as a horizon analogue of the $r^p$ hierarchy at $\mathcal I^+$: the event horizon $\mathcal H^+$ of extremal Reissner--Nordstr\"om is located at $r=M$ and hence $r-M$ is a degenerate weight on $\mathcal H^+$, just as $r^{-1}$ is a degenerate weight on $\mathcal I^+$ in \eqref{eq:intro-rp}.  This duality between $r$ and $(r-M)^{-1}$ is related to the so-called \emph{Couch--Torrence conformal isometry} \cite{couch1984conformal} that exchanges $\mathcal H^+$ and $\mathcal I^+$ in extremal Reissner--Nordstr\"om.  Special cases of \eqref{eq:r-Mp-intro} were introduced by Aretakis in \cite{Aretakis-instability-1} and the full hierarchy (in fact, even more elaborate extended versions) was derived by the first-named author of the present paper, Aretakis, and Gajic in \cite{angelopoulos2020late}. Note that the $p=2$ case of \eqref{eq:r-Mp-intro} gives the boundedness of the nondegenerate energy, since $D\sim  (r-M)^2$ near the horizon of extremal Reissner--Nordstr\"om. 
 
The horizon hierarchy \eqref{eq:r-Mp-intro} is derived using the family of multipliers $(r-M)^{-p}\partial_u\sim D^{-p/2}\partial_u$. In order to unify this with the redshift estimate \eqref{eq:intro-N}, we consider the family of multipliers $h_p\partial_u$ for $p\in [0,3)$, where\index{hp@$h_p$, horizon multiplier}
\begin{equation}\label{eq:hp-intro-defn}
    h_p(r)\sim  D'D^{-p/2-1/2}
\end{equation} for $p>0$ and $h_0\sim 1$. We also introduce the critical radius\index{r@$r_c\doteq M+ \sqrt{\lvert M^2-e^2\rvert} = M + \mathfrak t$}
\begin{equation*}
    r_c\doteq M+\sqrt{|M^2-e^2|},
\end{equation*}
which equals $r_+$ in the black hole case, but defines a timelike curve in the superextremal case; see \vref{fig:RN-intro}. One of the important properties of $r_c$ is that if $r\ge r_c$, then $D'(r)\ge 0$.  If the spacetime region of integration lies in the region $\{r\ge r_c\}$, then we obtain the estimates 
\begin{equation}\label{eq:intro-hp}
    \int_{\u C(\tau_2)}h_p(\partial_u\psi)^2\,du + \int_{\tau_1}^{\tau_2}\int_{\u C(\tau)}D'h_p(\partial_u\psi)^2\,dud\tau \les   \int_{\u C(\tau_1)}h_p(\partial_u\psi)^2\,du +\cdots
\end{equation}
for $p\in (0,3)$, where the implicit constant is \emph{independent of $\varrho$}. In the extremal case, $h_p$ and $(r-M)^{-p}$ are comparable, so this reproduces \eqref{eq:r-Mp-intro}. In the subextremal case, $h_1\sim \bm\kappa D^{-1}$ near $\mathcal H^+$, so \eqref{eq:intro-N} is encoded as the $p=1$ estimate in \eqref{eq:intro-hp}. In general, we have the hierarchy property
\begin{equation*}
    D'h_p\gtrsim h_{p-1},
\end{equation*} which gives the hierarchy of estimates
\begin{equation}\label{eq:intro-hp'}
    \int_{\u C(\tau_2)}h_p(\partial_u\psi)^2\,du + \int_{\tau_1}^{\tau_2}\int_{\u C(\tau)}h_{p-1}(\partial_u\psi)^2\,dud\tau \les   \int_{\u C(\tau_1)}h_p(\partial_u\psi)^2\,du +\cdots.
\end{equation}

Using the pigeonhole principle as in \cite{dafermos2010new}, \eqref{eq:intro-T}, \eqref{eq:intro-rp}, and \eqref{eq:intro-hp'} can be used to prove the decay estimate
\begin{equation}\label{eq:ED}
    \int_{C(\tau)} r^p(\partial_v\psi)^2\, dv+\int_{\underline C(\tau)}h_p (\partial_u\psi)^2\,du \le C_\star\tau^{-3+\delta+p},
\end{equation} for every $p\in[0,3-\delta]$ and $\tau\ge\tau_0$, where $C_\star$ is a constant depending on $\delta$ and the data at $\tau=\tau_0$, but not on $\varrho$, as long as the foliation remains in the region $\{r\ge r_c\}$. The condition $|\varrho|\le 1+\alpha L_i^{-2}$ in \cref{ingredient:iteration} is designed to ensure that a slab of Bondi length $L_i$ in superextremal Reissner--Nordstr\"om $(M,\varrho)$ with initial bifurcation sphere at $r\approx 100M$ remains in the region $\{r\ge r_c\}$.

\subsubsection{The gauge conditions and background anchoring procedure}\label{sec:intro-gauge}

 \begin{figure}
\centering{
\def\svgwidth{15pc}
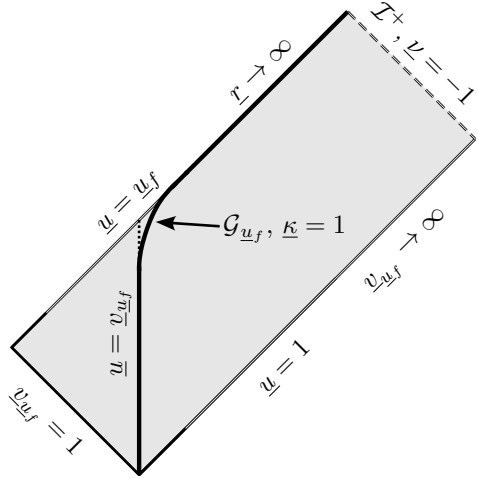}
\caption{Penrose diagram summarizing the $\u u{}_f$-normalized teleological gauge.}
\label{fig:intro-gauge}
\end{figure}

As mentioned several times already, we utilize a teleologically normalized double null gauge in this paper. The ``initial data'' coordinates are denoted $(u,v)$ and the Bondi time coordinate $\u u$, which correctly counts time at $\mathcal I^+$, is determined by the condition \eqref{eq:intro-Bondi}. 

We now explain how the advanced time coordinate $\u v{}_{\u u{}_f}$ is chosen and how the background Reissner--Nordstr\"om solution $(r_\diamond,\Omega^2_\diamond)$ in the definition of $\mathcal K(\ve,M,\varrho,\u u {}_f)$ from \cref{sec:intro-proof-uniform} is determined. We will go into some detail here which will pay off later in \cref{sec:intro-linear-estimates} when we describe the proof of \cref{thm:intro-foliation}.

We normalize $\u u$ so that $C_\out = \{\u u =1\}$. For $2\le \u u{}_f\le\infty$, we construct $\u v{}_{\u u{}_f}=\u v{}_{\u u{}_f}(v)$, which is normalized by $\u v{}_{\u u{}_f}=1$ on $\u C{}_\ing$, so that the quantity $\kappa$, when written in the coordinates $(\u u,\u v{}_{\u u{}_f})$, equals 1 along a causal curve $\u v{}_{\u u{}_f}\mapsto\mathcal G_{\u u{}_f}(\u v{}_{\u u{}_f})$ which has the following properties: for $\u v{}_{\u u{}_f}\le \u u{}_f -1$, $\mathcal G_{\u u{}_f}$ is given by $\u u=\u v{}_{\u u{}_f}$, for $\u v{}_{\u u{}_f}\ge \u u{}_f +1$, $\mathcal G_{\u u{}_f}$ is given by $\u u=\u u{}_f$, and interpolates suitably for $\u u{}_f-1\le\u v{}_{\u u{}_f}\le \u u{}_f+1$. See \vref{fig:intro-gauge}.

The original solution $\mathcal S_\mathrm{max}[\Psi]$ determined by $\Psi\in\mathfrak M$ consists of a quadruple $(\mathcal Q,r,\Omega^2,\phi)$, where $\mathcal Q\subset \Bbb R^2_{u,v}$ is the domain of the functions $r$, $\Omega^2$, and $\phi$. Relative to the coordinates $(\u u,\u v{}_{\u u{}_f})$, we obtain now a solution $\u{\mathcal S}{}_{\u u{}_f}[\Psi]=(\mathcal R_{\u u{}_f},\u r{}_{\u u{}_f},\u\Omega{}_{\u u{}_f}^2,\u \phi{}_{\u u{}_f})$, where
\[\mathcal R_{\u u{}_f}\doteq [1,\u u{}_f]\times[1,\infty).\]
The gauge normalization conditions for $\u{\mathcal S}{}_{\u u{}_f}[\Psi]$ can be succinctly summarized by
\begin{gather}
  \label{eq:intro-v-gauge}  \u u|_{C_\out} = 1 ,\qquad  \u \gamma{}_{\u u{}_f}|_{\mathcal I^+} = -1,\\
  \label{eq:intro-u-gauge}   \u v{}_{\u u{}_f}|_{\u C{}_\ing} = 1 ,\qquad  \u \kappa{}_{\u u{}_f}|_{\mathcal G_{\u u{}_f}} = 1.
\end{gather}

This setup is inherently different from that of \cite{AKU24} (which was inspired by \cite{luk2019strong}). In order to take advantage of \cref{ingredient-noncollapse}, which requires modulation \emph{at $\mathcal I^+$}, we are forced to use a semiglobal construction from the outset. So, rather than working with an approximate Bondi time on a ``far out'' incoming cone, we directly work in Bondi gauge. The curve $\mathcal G_{\u u{}_f}$\index{g@$\mathcal G_{\u u{}_f}$, curve where $\u \kappa_{\u u{}_f} \equiv 1$}, when $\u v{}_{\u u{}_f}\le \u u{}_f-1$, is a stand-in for the constant-$r$ curve $\Gamma$ in \cite{luk2019strong,AKU24}. The motivation for normalizing $\kappa$ on a (in this case only approximately) constant-$r$ curve is the same as in \cite{AKU24} and will be explained further in the following section. In order to ``complete'' the choice of $v$-coordinate, we force $\mathcal G_{\u u{}_f}$ to be null for $\u v{}_{\u u{}_f}\ge \u u{}_f+1$, and the interpolation in between can be done in an essentially arbitrary way.

\begin{rk}
    Because of the shape of the curves $\mathcal G_{\u u{}_f}$, which are all translates of each other, this gauge only makes sense for $\u u{}_f\ge 2$. Since teleological normalization is only important for large $\u u{}_f$, this is not an issue in the logic of the paper. 
\end{rk}

The background solution $(r_\diamond,\Omega^2_\diamond)$ is now determined as follows: We consider the unique Reissner--Nordstr\"om metric written in Eddington--Finkelstein double null coordinates
\begin{equation*}
    g_{M,\varrho} = -4D(r_\diamond)d\u u d\u v{}_{\u u{}_f}+r_\diamond^2 g_{S^2},
\end{equation*} where $r_\diamond$ satisfies the anchoring condition
\begin{equation}
    r_\diamond(\u u{}_f,\u u{}_f) = \u r{}_{\u u{}_f}(\u u{}_f,\u u{}_f),
\end{equation}
and the parameters $(M,\varrho)$ are given by \eqref{eq:intro-RN-anchoring-1}.

As will become clear in \cref{sec:intro-linear-estimates} below, it is important that the domain $\mathcal R_{\u u{}_f}$ and the distinguished curve $\mathcal G_{\u u{}_f}$ depend only on $\u u{}_f$, and not on $\Psi$. Of course, the coordinate transformation $(u,v)\mapsto (\u u,\u v{}_{\u u{}_f})$ does depend on $\Psi$ in a nontrivial manner. For continuity arguments, which now become subtle because of the semiglobal nature of the domains $\mathcal R_{\u u{}_f}$, it is important that $\u{\mathcal S}{}_{\u u{}_f}[\Psi]$ depends continuously on $\Psi$ and $\u u{}_f$ in a precise manner. This uses the structure of asymptotically flat solutions of the spherically symmetric EMSF system and is established in
 \cref{sec:semiglobal-1}.

\begin{rk}
    Note that $\u{\mathcal S}{}_{\u u{}_f}[\Psi]$ in general only corresponds to a subset of $\mathcal S_\mathrm{max}[\Psi]$. For instance, if $\mathcal S_\mathrm{max}[\Psi]$ contains a black hole, then for any finite $\u u{}_f$, $\u{\mathcal S}{}_{\u u{}_f}[\Psi]$ describes a strict subset of the black hole exterior, and $\u{\mathcal S}{}_{\infty}[\Psi]$ only ``contains'' the event horizon $\mathcal H^+$ as $\{\u u=\infty\}$.
\end{rk}

\subsubsection{Uniform stability in the class \texorpdfstring{$\mathcal K$}{K}} 

We now complete the definition of the decay class $\mathcal K(\ve,M,\varrho,\u u{}_f)$. Let $(r_\diamond,\Omega_\diamond^2)$ be the anchored background Reissner--Nordstr\"om metric from the previous section. Given a function $\zeta$ on $\mathcal R_{\u u{}_f}$, we define the following energy norms, motivated by the uncoupled case:
\begin{align}
\nonumber  \mathcal E_p^{(M,\varrho,\u u{}_f)}[\zeta](\tau)  &\doteq \int_{C_\tau}\u r^p(\partial_{\u v}(\u r\zeta))^2\, d\u v+\cdots ,\\
   \underline{\mathcal E}{}_p^{(M,\varrho,\u u{}_f)}[\zeta](\tau) & \doteq\int_{\u C{}_\tau\cap \mathcal R_{\u u{}_f}} h_p(r_\diamond)(\partial_{\u u}(\u r\zeta))^2\,d\u u+\cdots,\label{eq:E-bar-p-outline}
\end{align}
where we have only written the most important terms for now and the cones $C_{\u u}$ and $\u C{}_{\u v}$ are depicted in \vref{fig:intro-gauge}. Here the underlined quantities are expressed in the $\u u{}_f$-normalized teleological gauge of \cref{sec:intro-gauge}. We have omitted the $\u u{}_f$-subscript to reduce visual clutter.

\begin{rk}
    In the definition of $\mathcal E_p$, we use the ``dynamical'' $r$ and not the background $r_\diamond$. This is because the $r^p$ estimates in spherical symmetry do not generate nonlinear errors and $r$ works just as well as $r_\diamond$ in the far region. The use of $r_\diamond$ in $\underline{\mathcal E}{}_p$ is crucial, however. 
\end{rk}

For $\Psi$ to lie in $\mathcal K(\ve,M,\varrho,\u u{}_f)$, we demand that the ``bootstrap assumptions''
\begin{equation}\label{eq:boot-geo-intro}
    \left|\frac{\u\nu}{-D(r_\diamond)}-1\right|\le A^2\ve^{3/2}\tau^{-1+\delta},\quad |\u r-r_\diamond|\le A^2\ve^{3/2}\tau^{-2+\delta},\quad |\u\varpi-M|\le A \ve^{3/2}\tau^{-3+\delta}
\end{equation}
for the geometry hold on $\mathcal R_{\u u{}_f}$, where $\tau(\u u,\u v)\doteq \min\{\u u,\u v\}$, and 
\begin{equation}\label{eq:boot-sf-intro}
    \mathcal E_p^{(M,\varrho,\u u{}_f)}[\u\phi](\tau)+ \underline{\mathcal E}{}_p^{(M,\varrho,\u u{}_f)}[\u\phi](\tau)\le A \ve^2\tau^{-3+\delta+p}
\end{equation}
for $\tau\in[1,\u u{}_f]$ for the scalar field; compare with \eqref{eq:ED}.\footnote{We also have bootstrap assumptions for energy fluxes along outgoing cones in the near region and along ingoing cones in the far region, but suppress these at this level of discussion. We have also suppressed a logarithmic correction in the $\u r$ bootstrap coming from the final outgoing cone.} 

In the above inequalities, $A$ is a large constant that will be chosen according to the following result (and we will not track it carefully in this proof outline). The main analytic content of \cref{ingredient:iteration} consists of recovering the ``bootstrap assumptions'' \eqref{eq:boot-geo-intro} and \eqref{eq:boot-sf-intro}:

\begin{ingredient}[Recovering the ``bootstrap assumptions'']\label{ingredient:recovering}
    If $A$ is sufficiently large, $\ve$ is sufficiently small, and $\Psi\in \mathcal K(\ve,M,\varrho,\u u{}_f)$ with $|\varrho|\le 1+\alpha \u u{}_f^{-2}$, then the estimates \eqref{eq:boot-geo-intro} and \eqref{eq:boot-sf-intro} hold with constant $\frac 12$ on the right-hand side. Moreover, all other differences decay at appropriate polynomial rates independent of $\varrho$. 
\end{ingredient}

In the following, we use a dagger subscript to denote the difference from the background. For instance,
\begin{equation}\label{eq:outline-dagger-1}
    \u r{}_\dagger\doteq \u r- r_\diamond,\quad \u\kappa{}_\dagger\doteq \u\kappa-1,\quad \u\varpi{}_\dagger\doteq \u\varpi-M.
\end{equation}

\subsubsection{Decay estimates for the geometry} \label{sec:outline-geometry}

These estimates are mostly the same as in \cite{AKU24}, but now uniform in the full range of allowed parameters and with some shortcuts and refinements.
      
      We begin with the quantities $\u\kappa$ and $\u\gamma$, which satisfy good evolution equations \eqref{eq:kappa-u} and are normalized to $+1$ and $-1$, respectively, in the background solution. Recall the gauge conditions \eqref{eq:intro-v-gauge} and \eqref{eq:intro-u-gauge}. The equation for $\u\gamma$ sees the flux $\mathcal E_0[\u\phi]$ and hence gives us that $\u\gamma{}_\dagger = O(\ve^2\tau^{-3+\delta})$ in the far region, $\u v\ge \u u$, which is the best that can be expected. On the other hand, the $u$-equation for $\kappa$ sees $\underline{\mathcal E}{}_2[\u\phi]$, and hence decays much slower in the near region, $\u u\ge \u v$. However, the quantities $D'\u\kappa{}_\dagger$ and $D\u\kappa{}_\dagger$ obey evolution equations that see the fluxes $\u{\mathcal E}{}_1[\u\phi]$ and $\u{\mathcal E}{}_0[\u\phi]$, \emph{when integrating in the direction of increasing} $\u u$. Hence, because of the normalization condition \eqref{eq:intro-u-gauge}, we can estimate $\u\kappa$ in the near region by doing precisely this, and hence obtain the hierarchy of decay rates
      \begin{equation*}
      \u\kappa{}_\dagger = O(\ve^2\tau^{-1+\delta}),\quad     D'  \u\kappa{}_\dagger= O(\ve^2\tau^{-2+\delta}),\quad     D  \u\kappa{}_\dagger = O(\ve^2\tau^{-3+\delta}).
      \end{equation*}
      This replaces the $(r_\diamond-M)^{2-p}\u\kappa{}_\dagger$-hierarchy from \cite{AKU24} and uses the fact that the $h_p$-weight contains a power of $D'$ for $p>0$. In contrast to \cite{luk2019strong,AKU24}, we also prove $\u r$-decay for $\u\kappa{}_\dagger$ in the far region, which turns out to be quite useful for shortening the proof of the energy estimates later on.
      
      In \cite{AKU24}, it was important to track carefully powers of $r_\diamond-M$, which is comparable to both $D'$ and $D^{1/2}$ in the extremal case. In the full range of parameters, we have to track separately powers of $D'$ and $D$ to close the estimates, a theme which will appear over and over in this work. The ``Taylor expansions'' of $1-\u\mu$ and $\u\lambda$ in \cite{AKU24} are now replaced by 
     \begin{align} 
  \label{eq:Taylor-1}1-\u\mu&= D+ D'\u r{}_\dagger+O(\ve^{3/2}\u r^{-1}\tau^{-3+\delta}),\\
 \label{eq:Taylor-2} \u\lambda &=D+ \kappa D'\u r{}_\dagger+ O(\ve^{3/2}\min\{\{\tau^{-3+\delta},\u r^{-1}\tau^{-2+\delta}\}),
\end{align} 
      which serve a similar purpose. Note the $\u r$-decay of the error term in the $\u\lambda$ expansion over the corresponding estimate in \cite{AKU24}. 
      
The coefficients of the terms which are linear in $\u r{}_\dagger$ and the strong decay of the error terms in \eqref{eq:Taylor-1} and \eqref{eq:Taylor-2} are important. For example, $1-\u \mu$ appears naturally in the $\u u$-equation for $\u\varpi$ \eqref{eq:varpi-u}, and proving $\tau^{-3+\delta}$ decay for $\u\varpi{}_\dagger$ requires taking advantage of the linear term in \eqref{eq:Taylor-1}. In the subextremal case, the redshift effect implies one can prove $\tau^{-3+\delta}$ decay for $\u\varpi{}_\dagger$ while completely ignoring the $1-\u\mu$ weight, but the implicit constant obtained by this argument would depend (poorly) on the distance to extremality. In order to get uniform estimates in the full range, we need to use the $D'$-factor in the $h_p$-weight. Indeed, using \eqref{eq:hp-intro-defn}, \eqref{eq:boot-geo-intro}, and \eqref{eq:Taylor-1}, we have
\begin{equation*}
    |\u\nu^{-1}(1-\u\mu)|\les D^{-1} |1-\u\mu|\les 1 + D^{-1}D'\u r{}_\dagger+ \ve^{3/2} D^{-1} \tau^{-3+\delta}\les h_0 + \ve^{3/2}\tau^{-2+\delta}h_1+\ve^{3/2}\tau^{-3+\delta}h_2
\end{equation*}
in the near region, which is consistent with the desired decay for $\u\varpi{}_\dagger$. 

Note also that the sign of the $\u r{}_\dagger$ term in \eqref{eq:Taylor-2} is crucial: it is positive in the domain of outer communication, which reflects the \emph{global redshift effect} on the full Reissner--Nordstr\"om family under the assumption $|\varrho|\le 1+\alpha \u u{}_f^{-2}$. However, unlike the strictly subextremal case, where the redshift leads to an exponentially decaying integrating factor for the analogue of \eqref{eq:Taylor-2} (compare \cite[Lemma 8.19]{luk2019strong}), we have to track the degeneration of $D'$ and the rapidly decaying error term becomes important, since integrating \eqref{eq:Taylor-2} in $\u v$ now loses a power of $\tau$. 

The geometric estimates are proved in \cref{sec:K-1}.

\subsubsection{Energy estimates in the class \texorpdfstring{$\mathcal K$}{K}} \label{sec:intro-energy-estimates}

To recover \eqref{eq:boot-sf-intro}, we need to prove energy estimates for the wave equation \eqref{eq:phi-wave-1}. We consider the general inhomogeneous equation
\begin{equation}\label{eq:outline-wave}
    \partial_{\u u}\partial_{\u v}\zeta + \frac{\u\nu}{\u r}\partial_{\u v}\zeta+\frac{\u\lambda}{\u r}\partial_{\u u}\zeta = F,
\end{equation}
where $\zeta$ and $F$ are functions on $\mathcal R_{\u u{}_f}$ and $\u r$, $\u\nu$, and $\u\lambda$ arise from $\Psi\in\mathcal K(\ve,M,\varrho,\u u{}_f)$, with parameters in the range where the estimates from \cref{sec:intro-unifying} apply. To prove \cref{thm:intro-uniform-stability}, we prove estimates for \eqref{eq:outline-wave} with $\zeta=\u\phi$ and $F=0$, but for \cref{thm:intro-foliation} we take $\zeta=\lin{\u\phi}$, the linearization of $\u\phi$, and $F$ will in general be nontrivial.  

For solutions of \eqref{eq:outline-wave}, we obtain the boundedness estimates
\begin{equation}\label{eq:outline-wave-1}
    \mathcal E_p[\zeta](\tau_2)+\u{\mathcal E}{}_p[\zeta](\tau_2)\les \mathcal E_p[\zeta](\tau_1)+\u{\mathcal E}{}_p[\zeta](\tau_1)+ \mathbb E^\mathrm{nlin}_p[\zeta](\tau_1,\tau_2) + \mathbb E^\mathrm{inhom}_p[\zeta,F](\tau_1,\tau_2),
\end{equation} for $p\in \{0\}\cup[\delta,3-\delta]$ and $1\le\tau_1\le\tau_2\le \u u{}_f$, where $\mathbb E^\mathrm{nlin}_p[\zeta](\tau_1,\tau_2)$ are nonlinear error terms coming from the difference with the Reissner--Nordstr\"om background and $\mathbb E^\mathrm{inhom}_p[\zeta,F](\tau_1,\tau_2)$ are inhomogeneous error terms linear in $F$.\footnote{We do not use the notations $\mathbb E^\mathrm{nlin}_p[\zeta](\tau_1,\tau_2)$ and $\mathbb E^\mathrm{inhom}_p[\zeta,F](\tau_1,\tau_2)$ in the main body of the paper, but it is useful to introduce it here for brevity.} As usual, these error terms are spacetime integrals between $\tau=\tau_1$ and $\tau=\tau_2$. We also have the integrated decay estimates 
\begin{equation}\label{eq:outline-wave-2}
   \int_{\tau_1}^{\tau_2}\big(\mathcal E_{p-1}[\zeta](\tau)+\u{\mathcal E}{}_{p-1}[\zeta](\tau)\big)\,d\tau\les \mathcal E_p[\zeta](\tau_1)+\u{\mathcal E}{}_p[\zeta](\tau_1)+ \mathbb E^\mathrm{nlin}_p[\zeta](\tau_1,\tau_2) + \mathbb E^\mathrm{inhom}_p[\zeta,F](\tau_1,\tau_2),
\end{equation}
for $p\ge 1$. Since the decoupled linear estimates \eqref{eq:intro-rp} and  \eqref{eq:intro-hp} for \eqref{eq:intro-h-wave} are proved by direct integration by parts arguments, we repeat the proofs for the nonlinear, inhomogeneous equation \eqref{eq:outline-wave} and use the geometric estimates obtained in \cref{sec:K-1} to obtain \eqref{eq:outline-wave-1} and \eqref{eq:outline-wave-2}. These arguments are carried out in \cref{sec:energy-1}. 

We now discuss the structure of the error terms $\mathbb E^\mathrm{nlin}_p[\zeta](\tau_1,\tau_2)$ and $\mathbb E^\mathrm{inhom}_p[\zeta,F](\tau_1,\tau_2)$. We have
\begin{equation}\label{eq:E-nlin-str}
    \mathbb E^\mathrm{nlin}_p[\zeta](\tau_1,\tau_2) = \iint_{\mathcal D^{\le}} \ve \tau^{-6+2\delta}h_{1+\delta}(r_\diamond)(\partial_{\u u}\zeta)^2\,d\u u d\u v + \iint_{\mathcal D^{\le}} \ve\tau^{-4+2\delta}h_p(r_\diamond)(\partial_{\u u}(\u r\zeta))^2\, d\u u d\u v,
\end{equation}\vspace{-5mm}
\begin{multline}\label{eq:E-inhom-str}
      \mathbb E^\mathrm{inhom}_p[\zeta,F](\tau_1,\tau_2) = \iint_{\mathcal D}  r^2\big(|\partial_{\u u}\zeta |+|\partial_{\u v}\zeta | + \u r^{-1}D(r_\diamond)|\zeta|\big)|F|\,d\u u d\u v\\ + \iint_{\mathcal D^{\le}}h_p(r_\diamond)|\partial_{\u u}(\u r\zeta)||F|\,d\u u d\u v+ \iint_{\mathcal D^\ge}\u r^{p+1}|\partial_{\u v}(\u r\zeta)||F|\,d\u u d\u v,
\end{multline}
where $\mathcal D\doteq \mathcal R_{\u u{}_f}\cap\{\tau_1\le\tau\le\tau_2\}$, $\mathcal D^{\le}\doteq\mathcal D\cap\{\u v\le \u u\}$, and $\mathcal D^{\ge}\doteq \mathcal D \cap\{\u v\ge \u u\}$. Note that \eqref{eq:E-nlin-str} does not contain any contribution from the far region: this is because of the new $\u r$-weighted estimate for  $\u\kappa{}_\dagger$ mentioned in \cref{sec:outline-geometry}. The first term in \eqref{eq:E-nlin-str} comes from the Morawetz estimate and the second from the $h_p$-estimate. We have omitted some less important terms in \eqref{eq:E-nlin-str} which are essentially lower order than the first term. In \eqref{eq:E-inhom-str}, the first term comes from the Morawetz estimate, the second term from the $h_p$-estimate, and the third term from the $r^p$-estimate.

Note that the decaying time weights in \eqref{eq:E-nlin-str} are integrable and hence the nonlinear errors can be absorbed into the fluxes during a dyadic pigeonholing argument (as in \cite{dafermos2010new}) as long as we take $p_\mathrm{max}\ge 1+\delta$. More precisely, given a dynamical background arising from $\Psi\in\mathcal K(\ve,M,\varrho,\u u{}_f)$, we can prove the estimates \eqref{eq:outline-wave-1} and \eqref{eq:outline-wave-2} for any $\delta\le p\le 3-\delta$, but the estimates cannot ``close'' unless one works with $p\in \{0\}\cup[\delta,p_\mathrm{max}]$, where $p_\mathrm{max}\ge 1+\delta$. This observation will be important in the proof of \cref{thm:intro-foliation}.

\subsubsection{Recovering the energy estimates for the scalar field.} 

To improve \eqref{eq:boot-sf-intro}, we take $\zeta= \u\phi$ and $F=0$ in \eqref{eq:outline-wave}. Then \eqref{eq:boot-sf-intro} and \eqref{eq:outline-wave-1}--\eqref{eq:E-nlin-str} immediately give
\begin{align*}
    \mathcal E_p[\u\phi](\tau_2)+ \underline{\mathcal E}{}_p[\u\phi](\tau_2) &\les  \mathcal E_p[\u\phi](\tau_1)+ \underline{\mathcal E}{}_p[\u\phi](\tau_1)+ \ve^3(\text{quickly decaying error}),\\
    \int_{\tau_1}^{\tau_2}\big(\mathcal E_{p-1}[\u\phi](\tau)+\u{\mathcal E}{}_{p-1}[\u\phi](\tau)\big)\,d\tau &\les  \mathcal E_p[\u\phi](\tau_1)+ \underline{\mathcal E}{}_p[\u\phi](\tau_1)+\ve^3(\text{quickly decaying error}).
\end{align*}
From this hierarchy, decay (with an improved constant over \eqref{eq:boot-sf-intro}, using the higher power of $\ve$ on the error term on the right-hand side) is inferred by a standard application of the pigeonhole principle on dyadic time intervals as in \cite{dafermos2010new}. This argument is carried out in \cref{sec:improved-energy} and is essentially the same as in \cite{AKU24}. 

This completes the outline of the proof of \cref{thm:intro-uniform-stability}.

\subsection{Overview of the proof of \texorpdfstring{\cref{thm:intro-foliation}}{Theorem II}}\label{sec:overview-II}

We now outline the proof of the main result of the paper, \cref{thm:intro-foliation}. The sections of the paper explicitly dedicated to its proof, \cref{sec:proof-of-main-thm,sec:estimates-linear-perturbations}, take up a deceptively small portion of the paper---in reality the proof relies on the dichotomy and detailed stability estimates from \cref{thm:intro-uniform-stability}, the full semiglobal theory of \cref{sec:semiglobal-1}, the Reissner--Nordstr\"om calculations of \cref{sec:RN}, and the inhomogeneous energy estimates of \cref{sec:energy-1}.

\subsubsection{Logic of the proof of \texorpdfstring{\cref{thm:intro-foliation}}{Theorem II}}\label{sec:intro-broad-strokes}

We now outline the construction of the isologous foliation $\{\mathfrak M^\sigma_\mathrm{stab}\}_{\sigma\in [-1,1]}$, which consists of $C^1_b$ hypersurfaces in $\mathfrak M$ with constant final parameter ratio $\sigma$, as well as the proof that $\mathfrak M^{+1,-1}_\mathrm{stab}=\mathfrak M^{+1}_\mathrm{stab}\sqcup \mathfrak M^{-1}_\mathrm{stab}$ constitutes the black hole threshold near Reissner--Nordstr\"om.  

Per the discussion in \cref{sec:intro-foliation}, the goal is to show that the final parameter ratio map, $\mathscr P_\infty:\mathfrak M_\mathrm{black}\to \Bbb R$, is a $C^1_b$ submersion (has surjective derivative at every point), at least for $\|\phi_\circ\|_\mathfrak{X}\le \ve_0$. An immediate issue is that $\mathfrak M_\mathrm{black}$ is closed, not open, and we have no a priori information on $\partial\mathfrak M_\mathrm{black}$, so interpreting the statement ``$\mathscr P_\infty\in C^1_b(\mathfrak M_\mathrm{black})$'' requires care. We solve this issue and resolve the threshold property simultaneously by approximating $\mathscr P_\infty$ by a sequence of maps defined at \emph{finite Bondi time}. 

Let $L_i= 2^i$ be a dyadic time step and define\footnote{The $+1$ in the definition of $\mathfrak M_i$ is because $\u u$ starts at $1$ on $C_\out$.} 
\begin{equation*}
    \mathfrak M_i\doteq \{\Psi\in\mathfrak M: \text{range of Bondi time $\u u$ is at least $L_i+1$}\},
\end{equation*}
which is \emph{open} in $\mathfrak M$. Clearly $\mathfrak M_{i+1}\subset\mathfrak M_i$ and it follows from Dafermos' dichotomy that
\begin{equation*}
   \mathfrak M_\mathrm{black} = \bigcap_{i\ge 0}\mathfrak M_i.
\end{equation*} 
We now define the \emph{parameter ratio map at $\u u=L_i$}\index{P@$\mathscr P_i$, (signed) parameter ratio map at time $\u u = L_i$}
,
\begin{align}\nonumber
    \mathscr P_i : \mathfrak M_i &\to \Bbb R,\\
    \Psi &\mapsto \u P{}_{\mathcal I^+}(L_i).\label{eq:intro-Pi-defn}
\end{align}
We prove in \cref{sec:semiglobal-1} that $\mathscr P_i\in C^1(\mathfrak M_i)$ for every $i$, in the usual sense of Fr\'echet derivatives, which makes sense since $\mathfrak M_i$ is an open set of the Banach space $\mathfrak Z$.  

Recall the splitting $\mathfrak Z=\mathfrak X\times\Bbb R$ described in \cref{sec:intro-mod-space}. Fix $M_0>0$ and set $x_0\doteq (0,100M_0,M_0)\in\mathfrak X$. As $\rho_\circ$ ranges over $\Bbb R$, the seed data sets $(x_0,\rho_\circ)$ parametrize the Reissner--Nordstr\"om family for fixed mass $M=M_0$. In order to describe a neighborhood of Reissner--Nordstr\"om, we introduce the open ball $  B^\mathfrak{X}_{\ve_0}(x_0)$ and the \emph{cylinder of width $\ve_0$ and length $\ell$}\index{cyl@$ \cyl(\ve,\ell)$, cylinder in moduli space of radius $\ve$ and length $\ell$} 
\begin{equation*}
    \cyl(\ve_0,\ell)\doteq B^\mathfrak{X}_{\ve_0}(x_0)\times[-\ell,\ell].
\end{equation*}
Given a function $f:U\to \Bbb R$, $U\subset \mathfrak Z=\mathfrak X\times \Bbb R$, we define the \emph{augmented map}\index{*@$\check{\cdot}$, augmentation of a map} $\check f:U\to\mathfrak Z,(x,t)\mapsto(x,f(x,t))$.

\begin{ingredient}[Existence, regularity, and convergence of $\mathscr W_i$]\label{ingredient:Wi-existence} Let $\alpha>0$ be as in \cref{ingredient:iteration}. There exists an $\ve_0>0$ depending only on $M_0$ and a sequence of $C^1_b$ functions $\mathscr W_i:\cyl(\ve_0,1+\alpha L_i^{-2})\to \Bbb R$\index{W@$\mathscr W_i$, implicit function for $\mathscr P_i$}
 with the following properties:
    \begin{enumerate}
        \item $\mathscr W_i$ is an implicit function for $\mathscr P_i$ in the sense that 
    \begin{equation}\label{eq:intro-Wi-1}
        \mathscr P_i(x,\mathscr W_i(x,\sigma))=\sigma
    \end{equation}
    for every $(x,\sigma)\in \cyl(\ve_0,1+\alpha L_i^{-2})$. In fact,
    \begin{equation}\label{eq:intro-Wi-2}
        (x,\mathscr W_i(x,\sigma))\in \mathcal K(\ve_0,M,\sigma,L_i)
    \end{equation} for some $M$ which is $O(\ve_0^2)$-close to $M_0$. 

        \item The functions $\mathscr W_i$ satisfy the estimates
         \begin{align}
\label{eq:intro-Wi-est-1}   \|\mathscr W_i\|_{C^1(\cyl(\ve_0,1+\alpha L_i^{-2}))} &\les 1,\\
  \label{eq:intro-Wi-est-2} \sup_{\cyl(\ve_0,1+\alpha L_i^{-2})}|\partial_\sigma\mathscr W_i-1|  &\les \ve_0.
    \end{align}   
        Therefore, the augmentation $\check{\mathscr W}_i:\cyl(\ve_0,1+\alpha L_i^{-2})\to\mathfrak M_i$ is a diffeomorphism onto its image.
        
        \item   As $i\to\infty$, the functions $\mathscr W_i$ converge uniformly to a $C^1_b$ function $\mathscr W_\infty:\cyl(\ve_0,1)\to\Bbb R$ for which $\check{\mathscr W}_\infty:\cyl(\ve_0,1)\to \mathfrak M_\mathrm{black}$ is a diffeomorphism onto its image and
    \begin{equation}\label{eq:intro-W-infty}
        \mathscr P_\infty(x,\mathscr W_\infty(x,\sigma))=\sigma
    \end{equation}
    for every $(x,\sigma)\in \cyl(\ve_0,1)$.
    \end{enumerate}
\end{ingredient}

\begin{rk}
    The sets $\cyl(\ve_0,\ell)$ are not open, but the boundary is nice, and we show directly that $\mathscr W_i$ is $C^1$ on a neighborhood of $\cyl(\ve_0,1+\alpha L_i^{-2})$ for every $i\in \Bbb N\cup\{\infty\}$. 
\end{rk}

\begin{rk} 
   Note that the convergence $\mathscr W_i\to\mathscr W_\infty$ is only stated to be in the uniform topology, not in the $C^1_b$ topology. Nevertheless, $\mathscr W_\infty$ is a $C^1_b$ function and we do have pointwise convergence $\mathscr W_i'\to \mathscr W_\infty'$ of the Fr\'echet derivatives.
\end{rk}

The stable manifolds $\mathfrak M_\mathrm{stab}^\sigma$ in \cref{thm:intro-foliation} are given by the graphs of $\mathscr W_\infty|_\sigma\doteq\mathscr W_\infty(\cdot,\sigma)$\index{W@$\mathscr W_\infty$, implicit function for $\mathscr P_\infty$ whose graph defines the stable manifolds}:
\begin{equation*}
    \mathfrak M_\mathrm{stab}^\sigma \doteq \check{\mathscr W}_\infty|_\sigma\big(B^\mathfrak{X}_{\ve_0}(x_0)\big),
\end{equation*}
and these assemble into a $C^1_b$ foliation since $\check{\mathscr W}_\infty$ is a $C^1_b$ diffeomorphism. We will describe the construction of $\mathscr W_i$ in some detail below, which uses estimates for the Fr\'echet derivative $\mathscr P_i'$ and involves a continuity argument in the $x$-variable, using $\mathscr W_i(0,\sigma)=\sigma$ as a starting point (which follows from \eqref{eq:intro-Wi-1}). The convergence statement uses estimates for the iterative differences $\mathscr P_{i+1}-\mathscr P_{i}$ and $\mathscr P_{i+1}'-\mathscr P_{i}'$.

Note that \cref{ingredient:Wi-existence} does not yet show that $\mathfrak M_\mathrm{stab}^{+1,-1}$ constitutes the black hole threshold near Reissner--Nordstr\"om, which we capture as follows:

\begin{ingredient}[No gaps property]\label{ingredient:no-gaps}
    Let $\alpha>0$ be as in \cref{ingredient:iteration}. Then,
    \begin{align*}
    \mathscr W_i(x,1)&< \mathscr W_{i+1}(x,1+\alpha L_{i+1}^{-2}),\\
   \mathscr W_i(x,-1)&> \mathscr W_{i+1}(x,-1-\alpha L_{i+1}^{-2})
\end{align*}
for every $i\ge 1$ and $x\in B^\mathfrak{X}_{\ve_0}(x_0)$.
\end{ingredient}

By \cref{ingredient-noncollapse}, a data set of the form $\check{\mathscr W}_i(x,\sigma)$ with $\sigma\in (1,1+\alpha L_i^{-2}]$ or $\sigma\in [-1-\alpha L_i^{-2},-1)$ lies in $\mathfrak M_\mathrm{non}$. The no gaps property implies that the finite-time foliations cover a full neighborhood of $\bigcup_{\sigma\in[-1,1]}\mathfrak M^\sigma_\mathrm{stab}$ and we can therefore rule out any black hole solutions existing outside of this set. \cref{ingredient:no-gaps} is a straightforward consequence of \cref{ingredient:Wi-existence} and its proof. 

\cref{ingredient:Wi-existence} is proved in \cref{sec:proof-of-main-thm} but relies on estimates proved in \cref{sec:estimates-linear-perturbations}. \cref{ingredient:no-gaps} is proved in \cref{sec:no-gaps}. The proof of \cref{thm:intro-foliation} is completed in \cref{sec:thm-2-proof}.

 This completes the basic outline of the logic of the proof of \cref{thm:intro-foliation}. In the following section, we sketch the proof of \cref{ingredient:Wi-existence}.

\subsubsection{Building the maps \texorpdfstring{$\mathscr W_i$}{Wi}: the role of linearized gravity} \label{sec:outline-role-linearized}

As one might expect from the definition \eqref{eq:intro-Wi-1}, the construction of $\mathscr W_i$ uses the implicit function theorem applied to $\mathscr P_i$, which we know is a $C^1$ function on the open set $\mathfrak M_i\subset\mathfrak Z$ from general considerations. For $\mathscr W_i$ to exist locally, we need to show that
\begin{equation*}
    \partial_{\rho_\circ} \mathscr P_i\ne 0.
\end{equation*}
In order to construct $\mathscr W_i$ on a cylinder of uniform width $\ve_0$, and then prove convergence of the implicit functions $\mathscr W_i$, we require quantitative estimates for the Fr\'echet derivatives $\mathscr P_i'$ and $\mathscr P_{i+1}'-\mathscr P_i'$.  

Since $\mathscr P_i$ is $C^1$, we may compute $\mathscr P_i'$ using the Gateaux derivative. Let $\lin\Psi=(\lin\phi_\circ,\lin r_\circ,\lin\varpi_\circ,\lin\rho_\circ)\in \mathfrak Z$ be a tangent vector at $\Psi\in\mathfrak M_i$. For $z\in\Bbb R$ small, $\Psi(z)\doteq \Psi+z\lin\Psi\in \mathfrak M_i$ (since $\mathfrak M_i$ is open) and 
\begin{equation}\lin{\mathscr P}_i\doteq \left.\frac{d}{dz}\right|_{z=0}\mathscr P_i(\Psi(z)),\label{eq:lin-Pi-intro-1}\end{equation}
the Gateaux derivative of $\mathscr P_i$ at $\Psi$ in the direction $\lin \Psi$, equals $\mathscr P_i'(\Psi)\in\mathfrak Z^*$ applied to $\lin\Psi$. We may assume that $\|\lin\Psi\|_\mathfrak Z=1$ without loss of generality, since $\lin{\mathscr P}_i$ is linear in $\lin\Psi$. 

\begin{ingredient}[Estimates for the linearized parameter ratio]\label{ingredient:parameter-ratio} If $\ve_0$ is sufficiently small depending on $M_0$, $\Psi\in\mathcal K(\ve_0,M,\varrho,L_i)$ with $|\varrho|\le 1+\alpha L_i^{-2}$, $\lin\Psi\in\mathfrak Z$, and $\lin{\mathscr P}_i$ is defined by \eqref{eq:lin-Pi-intro-1}, then the following are true:
    \begin{enumerate}
        \item \ul{Uniform boundedness}: It holds that
        \begin{equation}
            |\lin{\mathscr P}_i|\les 1.\label{eq:intro-Pi-bound}
        \end{equation}
        \item \ul{Submersion property}: It holds that
        \begin{equation}\label{eq:intro-Pi-surj}
            |\lin{\mathscr P}_i-\lin\rho_\circ|\les \ve_0 .
        \end{equation}
        Therefore, if $\lin\Psi=(0,0,0,1)$, then 
        \begin{equation*}
            \lin{\mathscr P}_i = \partial_{\rho_\circ}\mathscr P_i = 1 + O(\ve_0),
        \end{equation*}
        and hence $\mathscr P_i$ is a submersion at $\Psi$. 
        \item \ul{Cauchy property}: If $\Psi$ is also an element of $\mathfrak M_{i+1}$, then $\lin{\mathscr P}_{i+1}$ is defined and it holds that
        \begin{equation}\label{eq:intro-Pi-Cauchy}
            |\lin{\mathscr P}_{i+1}-\lin{\mathscr P}_i|\les \ve_0 L_i^{-1+\delta/2},
        \end{equation}
        where $\delta>0$ is the same parameter as in \cref{ingredient:iteration}.
    \end{enumerate}
\end{ingredient}

At least formally, the submersion property of $\lin{\mathscr P}_i$ allows us to prove \cref{ingredient:Wi-existence} using the implicit function theorem. We will explain this argument more carefully in \cref{sec:intro-linear-estimates-II} below. In reality, the two ingredients have to be proved simultaneously with a continuity argument.

We compute $\lin{\mathscr P}_i$ using linearized gravity on $\u{\mathcal S}{}_{L_i}[\Psi]$ as follows. By the definition \eqref{eq:intro-Pi-defn}, $\mathscr P_i=\u{\rho}(L_i,\infty)$, where $\u\rho=\u{\rho}{}_{L_i}$ is the $L_i$-teleologically normalized charge-to-mass ratio $\rho=Q/\varpi$. We now consider $\u\rho$ of the family $\Psi(z)$ for small $z$: this is a function of the variables $(\u u,\u v,z)$ and we define the linearization 
\begin{equation}\label{eq:intro-rho-lin}
    \lin{\u\rho}(\u u,\u v) \doteq \left.\frac{\partial}{\partial z}\right|_{z=0}\u\rho(\u u,\u v, z),
\end{equation}
where $\partial_z$ is taken at \emph{fixed $(\u u,\u v)$}. We compute immediately from the definitions
\begin{equation*}
    \lin{\u\rho} = \frac{\varpi_\circ}{\u\varpi}\lin\rho_\circ+ \rho_\circ\frac{\lin\varpi_\circ}{\u\varpi}-\u\rho\frac{\lin{\u\varpi}}{\u\varpi},
\end{equation*}
where $\lin{\u\varpi}$ is defined similarly to \eqref{eq:intro-rho-lin}. By \cref{ingredient:recovering}, $\u\varpi=\varpi_\circ+O(\ve_0^2)$ and $\u\rho=\rho_\circ+O(\ve_0^2)$, so that
\begin{equation}\label{eq:outline-lin-u-rho}
    \lin{\u\rho}=\lin\rho_\circ+\frac{\rho_\circ}{\varpi_\circ}(\lin\varpi_\circ-\lin{\u\varpi}) +O(\ve_0^2|\lin{\u\varpi}|) + O(\ve_0^2\|\lin\Psi\|_\mathfrak{Z}).
\end{equation}
The goal is now to show that
\begin{equation}\label{eq:lin-varpi-intro-3}
    |\lin{\u\varpi}(L_i,\infty)-\lin\varpi_\circ|\les \ve_0,
\end{equation}
which would prove \eqref{eq:intro-Pi-bound} and \eqref{eq:intro-Pi-surj}.

In order to estimate $\lin{\u\varpi}$, we linearize the evolution equations for $\u\varpi$ about $\u{\mathcal S}{}_{L_i}[\Psi]$. For instance, we linearize $\partial_{\u u}\u\varpi= \frac 12 \u r^2\u\gamma^{-1}(\partial_{\u u}\u\phi)^2$ to obtain
\begin{equation}\label{eq:lin-varpi-intro-1}
    \partial_{\u u}\lin{\u\varpi}= \frac{\u r}{\u\gamma}(\partial_{\u u}\u\phi)^2 \lin{\u r} - \frac{\u r^2}{2\u\gamma^2}(\partial_{\u u}\u\phi)^2\lin{\u\gamma} + \frac{\u r^2}{\u\gamma}\partial_{\u u}\u\phi\partial_{\u u}\lin{\u\phi},
\end{equation}
where $\u r$, $\u\gamma$, and $\u\phi$ are taken from $\u{\mathcal S}{}_{L_i}[\Psi]$ and $\lin{\u r}$, $\lin{\u\gamma}$, and $\lin{\u\phi}$ are determined by their respective (linearized) evolution equations and the gauge conditions. We denote the collection of all linearized quantities by $\lin{\u{\mathcal S}}{}_{L_i}[\Psi,\lin\Psi]$. By integrating \eqref{eq:lin-varpi-intro-1} (and the analogous formula for $\partial_{\u v}\lin{\u\varpi}$), we obtain
\begin{equation}\label{eq:lin-varpi-intro-2}
    \lin{\u\varpi}(L_i,\infty) = \lin\varpi_\circ - \int_1^{L_i}\u r^2\partial_{\u u}\u\phi\partial_{\u u}\lin{\u\phi}\big|_{\mathcal I^+}\,d\u u + \cdots,
\end{equation}
where $\cdots$ denote other (for now less important) terms which are $O(\ve_0)$. It turns out that $\lin{\u\phi}$ is quite poorly behaved at extremality, and in the following sections we will explain how to prove the uniform \emph{growing} dyadic estimate 
\begin{equation}\label{eq:intro-dyadic-growing}
    \int_{L_{j-1}}^{L_j}\u r^2(\partial_{\u u}\lin{\u\phi})^2\big|_{\mathcal I^+}\,d\u u \les L_j
\end{equation}
for $j\le i$. Using the decay for $\partial_{\u u}\u\phi$ proved as a part of \cref{ingredient:recovering}, we then obtain
\begin{align*}
    \int_1^{L_i}\u r^2|\partial_{\u u}\u\phi||\partial_{\u u}\lin{\u\phi}|\big|_{\mathcal I^+}\,d\u u&\le \sum_{j=1}^i \left(\int_{L_{j-1}}^{L_j}\u r^2(\partial_{\u u}\u\phi)^2\big|_{\mathcal I^+}\,d\u u \right)^{1/2}\left(\int_{L_{j-1}}^{L_j}\u r^2(\partial_{\u u}\lin{\u\phi})^2\big|_{\mathcal I^+}\,d\u u \right)^{1/2}\\&\les\sum_{j=1}^i (\ve_0^2L_j^{-3+\delta})^{1/2} L_j^{1/2}\les \ve_0 ,
\end{align*}
which together with \eqref{eq:lin-varpi-intro-2}, proves \eqref{eq:lin-varpi-intro-3}. The estimate \eqref{eq:intro-Pi-Cauchy} is proved in the same way, since each piece of the dyadic sum contributes $\les\ve_0 L_j^{-1+\delta/2}$. 

\begin{rk}\label{rk:comparison-with-DHR}
   The linearization procedure here (and hence the notation used) is reminiscent of the seminal work of Dafermos--Holzegel--Rodnianski \cite{DHR19}, but differs in important ways. First, we make the (double null) gauge rigid \emph{before} linearizing. This means we do not need to add a ``pure gauge solution'' to obtain ``decay'' for the perturbation (we put decay in quotes since not everything can actually decay; we will elaborate on this in detail in the following sections). Secondly, the linearization occurs around a \emph{dynamical solution}, not Reissner--Nordstr\"om. For instance, in general all three terms on the right-hand side of \eqref{eq:lin-varpi-intro-1} will be nonzero, breaking conservation of mass at the linear level. This also means that the linearized system remains highly coupled, like the nonlinear system, and we have to employ a bootstrap argument (in $x$!) to prove estimates.    
\end{rk}

\begin{rk}\label{rk:Luk-Oh-differences}
   In fact, the teleologically normalized linearized quantities such as $\lin{\u\phi}$ and $\lin{\u r}$ are closely related to the difference quantities such as $\tilde\phi$ and $\tilde r$ in the work of Luk--Oh \cite{luk2019strong}. In that work, the authors consider two black hole solutions to the spherically symmetric EMSF system, say $(r,\Omega^2,\phi)$ and $(\bar r,\bar \Omega^2,\bar\phi)$, adapt the gauge to the first solution (based on a finite bootstrap domain as we used in \cite{AKU24}), and then consider differences $(\tilde r,\tilde\Omega^2,\tilde\phi)$. The structure of the resulting equations is of course very similar to our linearized system, but the estimates are complicated by the fact that the gauge is only asymptotically adapted to the second solution, which introduces subtlety in the bootstrap argument. In some sense, we completely avoid this issue by synchronizing the gauges at $\mathcal I^+$ and along $\mathcal G_{\u u{}_f}$ before linearizing. Away from extremality, our linear estimates recover (and in some cases simplify and improve) the difference estimates of Luk--Oh. 
\end{rk}

\begin{rk}\label{rk:overview-second-order}
    In this overview, we will only discuss the linearization of the teleologically normalized solutions. However, in \cref{sec:semiglobal-1}, we consider the linearization of general asymptotically flat solutions in the initial data gauge $(u,v)$. This is important for conducting continuity arguments in the teleological gauges, as well as for establishing regularity of $\mathcal I^+$-based quantities such as $M_{\mathcal I^+}$, since the domains involved are not compact. In fact, in order to perform the continuity argument in \cref{sec:intro-linear-estimates-II} below, we need to prove certain semiglobal estimates at \emph{second order in perturbation theory}, in the initial data gauge. 
\end{rk}

\subsubsection{The linearized Reissner--Nordstr\"om contribution}\label{sec:intro-linear-estimates}

We now outline our strategy for estimating $\lin{\u{\mathcal S}}{}_{L_i}[\Psi,\lin\Psi]$ (henceforth denoted $\lin{\u{\mathcal S}}$ since $\Psi$, $\lin\Psi$, and $i$ are fixed), and in particular, how to prove the key estimate \eqref{eq:intro-dyadic-growing}. 

As in \cite{DHR19}, the first step is to identify the ``linearized Reissner--Nordstr\"om'' part of $\lin{\u{\mathcal S}}$. As mentioned in \cref{rk:comparison-with-DHR}, mass is not linearly conserved if $\Psi$ is not vacuum initial data. Therefore, the ``linearized Reissner--Nordstr\"om'' part of $\lin{\u{\mathcal S}}$ cannot be read off from the data, and we obtain it as follows: By the procedure of \cref{sec:intro-gauge}, we obtain a one-parameter family $z\mapsto r_\diamond(\u u,\u v,z)$ of teleologically normalized background solutions. We then set
\begin{equation}\label{eq:intro-lin-r-diamond}
    \lin r_\diamond(\u u,\u v) \doteq \left.\frac{\partial}{\partial z}\right|_{z=0}r_\diamond(\u u,\u v, z).
\end{equation} From this procedure, we also obtain $\lin\nu_\diamond =-\lin \lambda_\diamond= \partial_{\u u}\lin r_\diamond$ and 
\begin{equation*}
   \lin\varpi_\diamond  = \lin{\mathscr M}_i,\quad 
   \lin\rho_\diamond  = \lin{\mathscr P}_i,
\end{equation*}
where \index{M@$\mathscr M_i$, Bondi mass function at time $\underline{u} = L_i$} $\mathscr M_i:\mathfrak M_i\to \Bbb R$, that is the $\u u=L_i$ Bondi mass function. Note that the corresponding quantities $\lin\gamma_\diamond$ and $\lin\kappa_\diamond$ vanish identically since each $ r_\diamond(\cdot,z)$ is written in Eddington--Finkelstein double null coordinates. 

\begin{rk}\label{rk:DHR-lin-S} Our notion of ``linearized Schwarzschild'' (linearized Reissner--Nordstr\"om with $\varrho=\lin{\mathscr P}_i=0$) differs slightly from \cite[Section 6.2.1]{DHR19}. Indeed, our notion is related to theirs by a ``pure gauge solution'' because they include an $M$-dependent rescaling of the double null coordinates before linearizing (see also \cite[Remark 6.4]{DHR19}).
\end{rk}

We study these linear perturbations of Reissner--Nordstr\"om in \cref{sec:linearized-RN}. It turns out that $\lin r_\diamond$ has a lot of interesting and important structure. In the near region $\u u\ge \u v$, we have the expansion
\begin{equation}\label{eq:outline-lin-r-diamond}
    \lin r_\diamond = \frac{r_c}{M}\lin{\mathscr M}_i + O(D'^{-1}|\lin{\mathscr P}_i|)+O(D').
\end{equation}
The coefficient of the $\lin{\mathscr P}_i$ term is sharp: in extremal Reissner--Nordstr\"om, $\lin r_\diamond$ blows up towards $\mathcal H^+$ like $D'^{-1}\sim D^{-1/2}$ for a generic linear perturbation. We also compute
\begin{equation}\label{eq:outline-lin-nu-diamond}
    |\lin\nu_\diamond|=|\lin\lambda_\diamond|\les D\big(1+|{\log(D'^{-2}D)}|\big)(D'^{-2}|\lin{\mathscr P}_i|+1)
\end{equation}
in the near region. Note that, already in subextremal Reissner--Nordstr\"om, $\lin\nu_\diamond$ receives a logarithmic correction: it does not behave exactly as $D$ towards the horizon. This behavior was noticed already in \cite[(8.110)]{luk2019strong}, and \eqref{eq:outline-lin-nu-diamond} gives the sharp  behavior in the full Reissner--Nordstr\"om family. (Recall \cref{rk:Luk-Oh-differences} for the analogy between our $\lin\cdot$ procedure and Luk--Oh's $\tilde\cdot$ procedure.)

\begin{rk}
  Note that there is no logarithmic correction to $\lin{(\Omega\tr\u\chi)}$ for linearized Schwarzschild in \cite{DHR19} (in fact, it vanishes). Consistency is restored by computing the $\lin{(\Omega\tr\u\chi)}$ contribution from the pure gauge transformation $f_1= \frac{u}{2M}$, $f_2 = \frac{v}{2M}$ from \cite[Remark 6.4]{DHR19} and using $e^{\frac{v-u}{2M}}=(\frac{r}{2M}-1)e^{\frac{r}{2M}}$.
\end{rk}

In order to control the linearized solution $\lin{\u{\mathcal S}}$, we want to estimate the difference from the linearized background. As in \eqref{eq:outline-dagger-1}, we define 
\begin{equation*}
    \lin{\u r}{}_\dagger\doteq \lin{\u r}- \lin r_\diamond,\quad \lin{\u\nu}{}_\dagger\doteq \lin{\u\nu}-\lin\nu_\diamond,\quad \lin{\u\varpi}{}_\dagger\doteq \lin{\u\varpi}-\lin{\mathscr M}_i,
\end{equation*}
etc. The goal will be to show that all dagger quantities (which includes $\lin{\u\phi}$ since the linearized Reissner--Nordstr\"om solution has no scalar field) decay in $\tau$, but are allowed to have degenerating $D'$- and $D$-weights, as in \eqref{eq:outline-lin-r-diamond} and \eqref{eq:outline-lin-nu-diamond}. 

\subsubsection{The linearized scalar field}\label{sec:outline-interlude} At this point, we can already assess the expected behavior for the linearized scalar field $\lin{\u\phi}$. Linearizing \eqref{eq:phi-wave-1}, we find the inhomogeneous wave equation\footnote{At this point, the reader will hopefully understand our decision to remove the underlines from teleologically normalized variables in parts of the paper (\cref{sec:K-1,sec:energy-1,sec:scalar-field,sec:estimates-linear-perturbations}). Whenever this happens, it will be clearly demarcated, as it greatly changes the meaning of the quantities!}
\begin{equation}\label{eq:outline-lin-wave}
    \partial_{\u u}\partial_{\u v}\lin{\u\phi} +\frac{\u\lambda}{ \u r}\partial_{\u u}\lin{\u \phi} + \frac{\u\nu}{\u r}\partial_{\u v}\lin{\u\phi} = \lin F,\quad \lin F\doteq -\left(\frac{\lin{\u\lambda}}{\u r}+\frac{\u\lambda\lin {\u r}}{\u r^2}\right)\partial_{\u u}\u\phi-\left(\frac{\lin{\u \nu}}{\u r}+\frac{\u\nu\lin{\u r}}{\u r^2}\right)\partial_{\u v}\u\phi.
\end{equation}
Let us assume, for this \cref{sec:outline-interlude} only, that the geometric component of $\lin{\u{\mathcal S}}$ is given entirely by the linearized Reissner--Nordstr\"om contribution from \cref{sec:intro-linear-estimates}. That is, let us assume for now that $\lin{\u r}{}_\dagger$, etc., vanish identically. 

It turns out that the linearization procedure here is already nontrivial at the level of the initial data. Recall from \cref{sec:outline-role-linearized} that $\lin{\u\phi}$ was defined by first putting the entire family $\mathcal S[\Psi(z)]$ into $L_i$-normalized teleological gauge and then linearizing. However, since the gauge itself depends on the solution via the procedure of \cref{sec:intro-gauge}, $\u{\lin\phi}$ ``sees'' the linearization of the gauge, even on $\u C{}_\ing$. By unraveling all of the definitions and using \eqref{eq:outline-lin-r-diamond} and \eqref{eq:outline-lin-nu-diamond}, we obtain
\begin{equation}\label{eq:outline-lin-phi-data}
 \big|\lin{\u\phi}|_{\u C{}_\ing} \big| \les |\lin{\mathscr P}_i| D'^{-1}|_{\u C{}_\ing}+\mathrm{better} ,\quad  \big|\partial_{\u u} \lin{\u\phi}|_{\u C{}_\ing} \big| \les |\lin{\mathscr P}_i|+\mathrm{better}.
\end{equation}
See already \cref{lem:lin-phi-data} for the precise calculation, which also shows that these bounds can be saturated. Therefore, when linearizing about extremal Reissner--Nordstr\"om, where $D'^{-1}\sim D^{-1/2}\sim \u u - \u v+1$ towards the horizon, there exist $\lin\Psi$ such that $\big|\lin{\u\phi}|_{\u C{}_\ing} \big|\sim \u u$ and $\big|\partial_{\u u} \lin{\u\phi}|_{\u C{}_\ing} \big|\sim 1$.\footnote{Note that $\partial_{\u u}$ is the \emph{degenerate} derivative in these coordinates: the nondegenerate derivative $Y\lin{\u\phi}|_{\u C{}_\ing}$ blows up like $\u u^2$!} In particular, such a perturbation would have 
\begin{equation*}
    \u{\mathcal E}{}_0^{(M,\pm 1,L_i)}[\lin{\u\phi}](1)\sim \int_1^{L_i} D^{-1}(r_\diamond(\u u, 1))\, d\u u\sim L_i.
\end{equation*}
We conclude that \emph{when linearizing around extremal Reissner--Nordstr\"om, $\lin{\u\phi}$ might not have bounded degenerate energy!} This is the origin of the growing dyadic estimate \eqref{eq:intro-dyadic-growing}. The initial outgoing flux of $\lin{\u\phi}$ does not contain any surprises and we do not discuss it further here. 

In the full range of parameters $|\varrho|\le 1+\alpha L_i^{-2}$, \eqref{eq:outline-lin-phi-data} yields the initial guess
\begin{equation}\label{eq:outline-initial-energy-guess}
     \u{\mathcal E}{}_p^{(M,\varrho,L_i)}[\lin{\u\phi}](1,\u u)\les 1+|\lin{\mathscr P}_i| D_1'^{-p-1}(\u u),
\end{equation}
where $D_1'(\u u)$ denotes $D'(r_\diamond)$ evaluated at $(\u u,1)$. Here we have introduced an additional $\u u$-cutoff in the energy:
\begin{equation*}
    \underline{\mathcal E}{}_p^{(M,\varrho,L_i)}[\lin{\u\phi}](\tau,\u u)  \doteq\int_{\u C{}_\tau\cap \mathcal R_{\u u{}_f}\cap\{\u u'\le \u u\}} h_p(r_\diamond)(\partial_{\u u}(\u r\lin{\u\phi}))^2\,d\u u'+\cdots,
\end{equation*}
compare with \eqref{eq:E-bar-p-outline}. 

Let us now assume that $\u\phi$ vanishes identically (so that we are really perturbing the vacuum) and hence $\lin F$ vanishes identically. Then we can input the initial energy bound \eqref{eq:outline-initial-energy-guess} into the scalar field estimates of \cref{sec:intro-unifying} and obtain, by a variant of the standard pigeonhole argument, the \emph{mixed growth/decay estimate}
\begin{equation}\label{eq:outline-lin-scalar-field-1}
    \mathcal E{}_p^{(M,\varrho,L_i)}[\lin{\u\phi}](\tau) + \underline{\mathcal E}{}_p^{(M,\varrho,L_i)}[\lin{\u\phi}](\tau,\u u)\les \tau^{-p_\mathrm{max}+p} \big(1+|\lin{\mathscr P}_i| D_1'^{-p_\mathrm{max}-1}(\u u)\big)
\end{equation}
for $p\in \{0\}\cup[\delta,p_\mathrm{max}]$ and $\tau\le L_i$, for any $p_\mathrm{max}\le 3-\delta$. Note that we have an equitable trade-off between $\tau$-decay and $D'^{-1}$-growth towards the horizon determined by $p_\mathrm{max}$! This is because the horizon growth is determined by the initial data assumption \eqref{eq:outline-initial-energy-guess} for $p=p_\mathrm{max}$: the overall constants in a \cite{dafermos2010new}-style dyadic pigeonholing argument depend on the starting point at $p=p_\mathrm{max}$.

Note that $D_1'^{-1}$ behaves the worst in the extremal case, where it is $\sim \u u$. In the far region, $\tau = \u u$, so that \eqref{eq:outline-lin-scalar-field-1} is exactly consistent with the claimed growth estimate \eqref{eq:intro-dyadic-growing} at $\mathcal I^+$. 

This is essentially already the full story when the data $\Psi$ we perturb is vacuum. In general, however, the inhomogeneity $\lin F$ in \eqref{eq:outline-lin-wave} will present a significant obstacle: we have to estimate the error terms $\mathbb E_p^\mathrm{inhom}[\lin{\u\phi},\lin F]$ from \cref{sec:intro-energy-estimates} to be consistent with \eqref{eq:outline-lin-scalar-field-1}. Already by inspecting the form of $\lin F$ and substituting in \eqref{eq:outline-lin-r-diamond}  and \eqref{eq:outline-lin-nu-diamond}, we see that $\lin F$ does not necessarily decay towards $\mathcal H^+$ in the extremal case, which means that we will have to prove growing estimates for $\mathbb E_p^\mathrm{inhom}[\lin{\u\phi},\lin F]$ as well. Finally, as remarked in \cref{sec:intro-energy-estimates}, the energy estimates proved in \cref{sec:energy-1} only close if $p_\mathrm{max}\ge 1+\delta$. To prove \cref{thm:intro-foliation}, it actually suffices to take $p_\mathrm{max}=1+\delta$, but for the refined estimates required to prove \cref{thm:intro-scaling,thm:intro-instability}, we will take $p_\mathrm{max}=3-3\delta$ (some small loss does seem necessary). 

In the course of the proof, we do eventually show that the linearized dagger error terms and presence of $\lin F$ do not change the initial predictions \eqref{eq:outline-initial-energy-guess} and \eqref{eq:outline-lin-scalar-field-1}, although this is far from obvious at this stage. 

\subsubsection{The main continuity argument}\label{sec:intro-linear-estimates-II}

To prove \cref{ingredient:Wi-existence,ingredient:parameter-ratio}, we employ a continuity argument in $\ve_0$. Namely, we construct simultaneously the implicit function $\mathscr W_i$ as well as prove estimates for $\lin{\mathscr P_i}$.

\begin{defn} \label{def:outline-E}
    For $\lin A\ge 1$ and $i\ge 1$, let $\mathfrak E(i,\lin A)\subset (0,\ve_0]$ be the set of $\ve>0$ such that there exists a $C^1_b$ function $\mathscr W_i:\cyl(\ve,1+\alpha L_i^{-2})\to\Bbb R$ satisfying the conclusion of \cref{ingredient:Wi-existence} with $\ve_0$ replaced by $\ve$ and such that if $\Psi\in \check{\mathscr W}_i(\cyl(\ve,1+\alpha L_i^{-2}))$ and $\lin\Psi\in\mathfrak Z$ with $\|\lin\Psi\|_\mathfrak{Z} = 1$, then the estimates {\mathtoolsset{showonlyrefs=false} \allowdisplaybreaks
\begin{align}
\label{eq:lin-r-bootstrap-outline} |\lin{\u r}(L_i,L_i)|&\le 10,\\
   \label{eq:lin-M-bootstrap-outline}   |\lin {\mathscr M}_i|&\le 10,\\
  \label{eq:lin-P-bootstrap-outline}      |\lin{\mathscr P}_i|&\le 10,\\
   \label{eq:lin-boot-1-outline}   |\lin{\u{r}}{}_\dagger|& \le \lin A\big(\tau^{-2+3\delta}+|\lin{\mathscr P}_i|\tau^{-2+3\delta}D_1'^{-2+3\delta/2}\big)\mathbf 1_{\{\u u\ge \u v\}}+\lin A r_\diamond^\delta\mathbf 1_{\{\u v\ge \u u\}},\\
|\u{\lin\lambda}|&  \le \lin A\big(D^{1-\delta/2}+\tau^{-2+3\delta}+|\lin{\mathscr P}_i|\big[D'^{-2+3\delta/2}D^{1-3\delta/4}+\tau^{-2+3\delta} D_1'^{-2+3\delta/2} \big]\big)+\lin A r_\diamond^{-1+\delta}\mathbf 1_{\{\u v\ge \u u\}},\label{eq:lin-boot-2-outline} \\
  \label{eq:lin-boot-3-outline}  |\u{\lin \nu}| & \le \lin A\big( D^{1-\delta/2}+|\lin{\mathscr P}_i|\big[D'^{-2+\delta}D^{1-\delta/2}+\tau^{-1+3\delta}DD_1'^{-2+3\delta/2}\big]\big)+\lin A r_\diamond^{-1}\mathbf 1_{\{\u v\ge \u u\}}
\end{align}}
    hold on $\mathcal R_{L_i}$.
\end{defn}

We then prove the most fundamental estimate:

\begin{ingredient}\label{ingredient:lin-A}
    For $\lin A$ sufficiently large, $\ve_0$ sufficiently small depending on $M_0$ and $\delta$, $\ve\in \mathfrak E(i,\lin A)$, and $\Psi,\lin\Psi$ as in \cref{def:outline-E}, the estimates \eqref{eq:lin-r-bootstrap-outline}--\eqref{eq:lin-boot-3-outline} hold with an additional overall factor of $\frac 12$ on the right-hand side. Moreover, 
    \begin{equation}\label{eq:lin-varpi-dagger-decay-outline}
       |\lin{\u\varpi}{}_\dagger| \les \lin A\ve \tau^{-1+\delta/2}
    \end{equation}
    on $\mathcal R_{L_i}\cap\{\u v\ge \u u\}$. 
\end{ingredient}

\begin{proof}[Outline of the proof] The bootstrap assumptions \eqref{eq:lin-r-bootstrap-outline}--\eqref{eq:lin-P-bootstrap-outline} facilitate the expansions \eqref{eq:outline-lin-r-diamond} and \eqref{eq:outline-lin-nu-diamond}, together with some control on the linearized background in the far region. Once $\lin r_\diamond$ is controlled, an initial estimate for $\lin{\u r}$ (which will later be improved) can be given using \eqref{eq:lin-boot-1-outline}.

    The initial energy for $\lin{\u\phi}$ can be estimated using $\lin{\u r}$, $\lin{\u\lambda}$, and $\lin{\u\nu}$ by \cref{lem:lin-phi-data} as mentioned earlier. We obtain exactly \eqref{eq:outline-initial-energy-guess} with a factor of $\lin A^2$ on the right-hand side. We can also estimate $\lin F$, the inhomogeneity in the linearized wave equation \eqref{eq:outline-lin-wave}. Note that $\lin F$ also depends on $\u\phi$, the scalar field of the solution we are perturbing around. 

Now we apply the inhomogeneous energy estimates from \cref{sec:intro-energy-estimates} to $\lin{\u\phi}$. Using the aforementioned estimate for $\lin F$ and the stability estimates for $\u\phi$ obtained in the proof of \cref{thm:intro-uniform-stability}, we can absorb the error terms $\mathbb E_p^\mathrm{inhom}[\lin{\u\phi},\lin F]$ into the good bulks and fluxes for $p\le p_\mathrm{max}=3-3\delta$. These error estimates are carried out in \cref{sec:linear-errors}. Then, we can prove the energy decay/growth estimates \eqref{eq:outline-lin-scalar-field-1} with an additional factor of $\lin A^2$ on the right-hand side via a variant of the standard pigeonhole argument; see already \cref{sec:linear-scalar-field}.  

With energy estimates for $\lin{\u\phi}$ at hand, we proceed to estimate the linearized geometric quantities, essentially by linearizing the procedure from \cref{sec:outline-geometry}. We improve the constants in \eqref{eq:lin-r-bootstrap-outline}--\eqref{eq:lin-boot-3-outline} using the bilinearity of the EMSF system in $\phi$: in the linearized equations, $\lin{\u\phi}$ always comes multiplied by $\u\phi$, which gains a factor of $\ve$ and improves the bootstrap constant. See already \cref{sec:linear-geometric} for details. The precise geometric estimates obtained in this step will be crucial for proving \cref{thm:intro-scaling,thm:intro-instability}.
\end{proof}

\begin{rk}
    Note that the quantities involved in the bootstrap set $\mathfrak E$ are slightly different from the quantities in the bootstrap set $\mathcal K$. In the logic of the proof of \cref{thm:intro-uniform-stability}, the geometric estimates have to be established before the scalar field estimates, because of the geometric error terms in the energy estimates. Therefore, it made sense to bootstrap energies for $\u\phi$. Now, however, the energy estimates for $\lin{\u\phi}$ have already been proved, and we only need to estimate the error terms, which only involve $\u\phi$ (which has already been estimated), $\lin{\u r}$, $\lin{\u\nu}$, and $\lin{\u\lambda}$. 
\end{rk}

Given \cref{ingredient:lin-A}, we can wrap up the proof of \cref{thm:intro-foliation}. First, we prove estimates for $\lin{\mathscr P}_i$:

\begin{proof}[Sketch of the proof of \cref{ingredient:parameter-ratio}] 
We use \eqref{eq:outline-lin-u-rho} to estimate $\lin{\mathscr P}_i=\lin{\u\rho}(L_i,\infty)$. By definition and \eqref{eq:lin-varpi-dagger-decay-outline},
\[|\lin\varpi_\circ-\lin{\u\varpi}(L_i,\infty)|= |\lin{\u\varpi}{}_\dagger(1,1)-\lin{\u\varpi}{}_\dagger(L_i,\infty)|\les \lin A\ve.\]
This proves \eqref{eq:intro-Pi-bound} and \eqref{eq:intro-Pi-surj}. In the same way, using $\lin{\mathscr M}_i=\lin{\u\varpi}(L_i,\infty)$, we obtain 
\begin{equation*}
 |\lin{\u\varpi}(L_i,\infty)-\lin{\u\varpi}(L_{i-1},\infty)| = |\lin{\u\varpi}{}_\dagger(L_{i-1},\infty)|\les \lin A\ve L_i^{-1+\delta/2},
\end{equation*}
which can be easily used to prove \eqref{eq:intro-Pi-Cauchy}.
\end{proof}

Finally, we construct the implicit functions $\mathscr W_i$:

\begin{proof}[Outline of the proof of \cref{ingredient:Wi-existence}] We show via a continuity argument that $\mathfrak E(\lin A,i)=(0,\ve_0]$, provided $\lin A$ and $\ve_0$ are chosen so that \cref{ingredient:lin-A} applies. The estimates for $\mathscr W_i$ and the convergence to $\mathscr W_\infty$ (and properties thereof) then follow (not \emph{so} straightforwardly) from the estimates for $\mathscr P_i$ in \cref{ingredient:parameter-ratio} and implicit differentiation.

First, we show that $\mathfrak E(\lin A,i)\ne\emptyset$. For $|\sigma|\le 1+\alpha L_i^{-2}$, we define $\mathscr W_i(0,\sigma)=\sigma$, which is fixed by the relation \eqref{eq:intro-Wi-1}. Let $\mathfrak L_i\doteq \{x_0\}\times[-1-\alpha L_i^{-2},1+\alpha L_i^{-2}]\subset\mathfrak M$. For any $\Psi\in\mathfrak L_i$ and $\lin\Psi\in\mathfrak Z$ with $\|\lin\Psi\|_\mathfrak{Z}= 1$, the estimates \eqref{eq:lin-r-bootstrap-outline}--\eqref{eq:lin-boot-3-outline} hold with room in the constants by \cref{ingredient:lin-A}. By the semiglobal theory of \cref{sec:semiglobal-1}, $\mathscr P_i$ is $C^1_b$ on a neighborhood of $\mathfrak L_i$ and we can use the implicit function theorem to extend $\mathscr W_i:\cyl(\ve',1+\alpha L_i^{-2})\to\Bbb R$ for some $0<\ve'\le \ve_0$ satisfying \eqref{eq:intro-Wi-1}. By perhaps shrinking $\ve'$ and using ($r$-weighted) continuous dependence of $\lin{\u{\mathcal S}}[\Psi,\lin\Psi]$ in $\Psi$, \eqref{eq:lin-r-bootstrap-outline}--\eqref{eq:lin-boot-3-outline} hold and hence $\ve'\in\mathfrak E(\lin A,i)$. (To prove this quantitative version of continuous dependence, we actually use second-order perturbations as mentioned in \cref{rk:overview-second-order}.) 

We now show that given $\ve\in\mathfrak E(\lin A,i)$ with $\ve\le\ve_0$, we have $\ve+\eta\in\mathfrak E(\lin A,i)$ for some $\eta>0$ which may depend on $i$ but not $\ve$. So, given $\mathscr W_i$ defined on $\cyl(\ve,1+\alpha L_i^{-2})$, we use \cref{ingredient:parameter-ratio} to invert $\mathscr P_i$ locally, in a ball of $\mathfrak M$ whose radius depends only on $i$ through the general semiglobal theory, but not on $\ve$ (this determines $\eta$). Then the extension of $\mathscr W_i$ is constructed by patching together these local implicit functions, which completes the proof.
\end{proof}

This concludes the overview of the proof of \cref{thm:intro-foliation}.

\subsection{Overview of the proofs of \texorpdfstring{\cref{thm:intro-scaling,thm:intro-instability}}{Theorems III and IV}}\label{sec:overview-III-IV}

\subsubsection{Proof of the scaling laws} 

The scaling law for the final parameter ratio $\mathscr P_\infty$, \eqref{eq:intro-P-scaling}, simply follows from the transversality criterion \eqref{eq:intro-transversality} for $C^1$ foliations. Then the scaling laws for the final horizon area $\mathscr A$ and final horizon temperature $\mathscr T$, \eqref{eq:intro-A-scaling} and \eqref{eq:intro-T-scaling}, respectively, follow from the explicit formulas of these functions in terms of the final black hole parameters. In fact, in the detailed statement of the theorem in \cref{thm:scaling} below, we give precise asymptotic expansions of these quantities as $p\to p_*$. For details, see already \cref{sec:are-scaling}.

The proof of scaling for the event horizon location $\mathscr U$, \eqref{eq:intro-U-scaling}, is much more involved. For $\Psi\in\mathfrak M_\mathrm{black}$, the event horizon location is given by 
\begin{equation}\label{eq:intro-mathscr-U}
\mathscr U(\Psi)=1-\int_1^\infty \u \nu(\u u,1)\,d\u u,    
\end{equation}
where $\u \nu$ is written in the \emph{eschatological gauge} ($\u u{}_f=\infty$ in \cref{sec:intro-gauge}). As a part of the proof of \cref{thm:intro-foliation}, we prove the expansion
\begin{equation}\label{eq:intro-lin-nu-dagger}
    \lin{\u\nu} = \lin\nu_\diamond + O(\ve_0 |\lin\nu_\diamond|)  + \cdots
\end{equation}
on $\u C{}_\ing$, where $\cdots$ denotes terms which are better behaved towards the horizon than $\lin\nu_\diamond$; see already \eqref{lem:lin-nu-dagger}. Obtaining \eqref{eq:intro-lin-nu-dagger} with the error term in this form requires considerable care and taking advantage of many cancellations at the horizon in the linearized estimates of \cref{sec:linear-geometric} to optimize horizon weights. 

As mentioned in \cref{sec:intro-linear-estimates}, $\lin\nu_\diamond$ is in general not integrable in $\u u$ at extremality. This is consistent with \eqref{eq:intro-mathscr-U} since we are anyway claiming that $\mathscr U$ is not $C^1$, but only $C^{1/2}$. Along the curve $\Psi_{p}\in \mathfrak M_\mathrm{black}\cap\mathfrak M_\mathrm{nbhd}$, we write
\begin{equation}
    \mathscr U(\Psi_{p_2})-\mathscr U(\Psi_{p_1}) = \int_1^\infty
\big(\nu_\diamond(\u u,1,p_2)-\nu_\diamond(\u u,1,p_1)\big)\,d\u u+\int_1^\infty
\big(\u\nu_\dagger(\u u,1,p_2)-\u\nu_\dagger(\u u,1,p_1)\big)\,d\u u.\label{eq:mathscr-U-difference}\end{equation}
The first integral can be explicitly computed in terms of the final parameters of $\mathcal S[\Psi_{p_1}]$ and $\mathcal S[\Psi_{p_2}]$ by a simple calculation on Reissner--Nordstr\"om. For the second integral, we use the fundamental theorem of calculus, now with the notation $\lin{\cdot}=\partial_p$:
\begin{equation*}
    \int_1^\infty
\big|\u\nu_\dagger(\u u,1,p_2)-\u\nu_\dagger(\u u,1,p_1)\big|\,d\u u\les  \int_1^\infty \int_{p_1}^{p_2} |\lin{\u\nu}_\dagger (\u u,1,p)|\,dp d\u u\les \ve_0 \int_1^\infty \int_{p_1}^{p_2} |\lin\nu_\diamond (\u u,1,p)|\,dp d\u u,
\end{equation*}
where we have omitted the other error terms from \eqref{eq:intro-lin-nu-dagger}. A tricky calculation on Reissner--Nordstr\"om (\cref{lem:one-sided}), using transversality of $\Psi_p$, shows that $\lin\nu_\diamond$ has a sign \emph{up to terms better in $|p_1-p_2|$}, and hence the absolute value in the double integral can be removed up to adding a new harmless error term. Then one integrates in $p$ and absorbs this term into the first integral in \eqref{eq:mathscr-U-difference} using smallness of $\ve_0$. 

In fact, this argument gives a precise asymptotic expansion of $\mathscr U(\Psi_p)$ as $p\to p_*$, which is presented in \cref{thm:scaling}. For the detailed proof, see already \cref{sec:horizon-location-proof}. 

\subsubsection{Foliation of the threshold by the asymptotic Aretakis charge}\label{sec:outline-Aretakis-foliation}

We outline here the proof of \hyperref[thm:intro-instab-1]{part 1.}~of \cref{thm:intro-instability}.

Given a dynamical extremal black hole $\Psi\in\mathfrak M_\mathrm{stab}^{+1,-1}$, \cite[Theorem II]{AKU24} implies that $Y\psi|_{\mathcal H^+}$ attains a finite limit as $v\to\infty$, called the asymptotic Aretakis charge $H_0[\phi]$. We may therefore define a function \index{H3@$ \mathscr H$, Aretakis charge map}
\begin{align*}
        \mathscr H:\mathfrak M_\mathrm{stab}^{+1,-1}&\to \Bbb R\\
        \Psi &\mapsto H_0[\phi]. 
    \end{align*}
    We claim that $\mathscr H$ is a $C^1_b$ submersion, whence its level sets form a $C^1_b$ foliation of $\mathfrak M_\mathrm{stab}^{+1,-1}$. 

    The difficulty is showing that $\mathscr H\in C^1_b(\mathfrak M_\mathrm{stab}^{+1,-1})$. Once this is established, we can compute its linearization explicitly on extremal Reissner--Nordstr\"om to verify the submersion property, where we can use the conservation law discovered by Aretakis! 

    As in \cite{AKU24}, we use the method of characteristics to solve the wave equation for $Y\psi$ along $\mathcal H^+$. This gives the following representation formula for $\mathscr H$:
    \begin{equation}\label{eq:outline-H}
        \mathscr H(\Psi) = \exp\left(-\int_1^{\infty}\u\kappa\u\varkappa(\infty,\u v')\,d\u v'\right)Y\psi (u_{\mathcal H^+},1) + \int_1^\infty\exp\left(-\int_{\u v'}^{\infty}\u\kappa\u\varkappa(\infty,\u v'')\,d\u v''\right)\u\kappa\u\varkappa\u\phi(\infty,\u v')\,d \u v',
    \end{equation}
    where we again work in the eschatological gauge. By cutting the integrals off at $\u v=L_i$, we obtain a sequence of approximating maps $\mathscr H_i$; we aim to show that these maps are Cauchy in $C^1_b(\mathfrak M_\mathrm{stab}^{+1,-1})$. 
    
    Note that the quantities $\u \kappa$, $\u\varkappa$, and $\u\phi$ are evaluated at $\u u=\infty$, i.e., at the event horizon $\mathcal H^+$.\footnote{This is to be understood in a precise limiting sense.} 
    From the discussion in \cref{sec:intro-linear-estimates,sec:outline-interlude}, it is clear that for general linear perturbations, $\lin{\u\kappa}$, $\lin{\u\varkappa}$, and $\lin{\u\phi}$ do not actually extend regularly to $\mathcal H^+$ in the asymptotically extremal case. Indeed, already on the initial data, we generically expect $|\lin{\u\phi}|$ to blow up towards $\mathcal H^+$ like $\u u$. Moreover, the formulas of $\mathscr H$ and $\mathscr H_i$ depend on $u_{\mathcal H^+}=\mathscr U(\Psi)$, which is provably not differentiable, which follows from \cref{thm:intro-scaling}. 

    However, we are only asking for regularity of $\mathscr H_i$ and $\mathscr H$ \emph{tangent to the threshold} $\mathfrak M_\mathrm{stab}^{+1,-1}$. A vector $\lin\Psi\in\mathfrak Z$ tangent to $\mathfrak M_\mathrm{stab}^{+1,-1}$ satisfies
    \begin{equation*}
        \lin{\mathscr P}_\infty=0.
    \end{equation*}
    Using this condition in \eqref{eq:outline-lin-r-diamond}, \eqref{eq:outline-lin-nu-diamond}, \eqref{eq:outline-lin-scalar-field-1}, \eqref{eq:lin-boot-1-outline}, \eqref{eq:lin-boot-2-outline}, and \eqref{eq:lin-boot-3-outline}, shows that tangential linearizations behave significantly better than the generic transverse linearization. In fact, we show that the event horizon location function $\mathscr U$ is $C^1_b$ when restricted to $\mathfrak M_\mathrm{stab}^{+1,-1}$. In fact, the restrictions of the quantities $\u\kappa$, $\u\varkappa$, and $\u\phi$ to $\mathcal H^+$ can be linearized and estimated in tangential directions. 

    Given a vector $\lin\Psi\in T\mathfrak M_\mathrm{stab}^{+1,-1}$, we show that $\lin{\mathscr H}_i$ can be defined and that these are uniformly Cauchy in $i$, which implies the $C^1_b$-regularity of the limit $\mathscr H$ by standard arguments.
    
    For details, see already \cref{sec:Aretakis-generic}.

\subsubsection{Transient instabilities}

We outline here the proof of \hyperref[thm:intro-instab-2]{part 2.}~of \cref{thm:intro-instability}.

Using the method of characteristics again as for \eqref{eq:outline-H}, we have in general
\begin{equation}
        \u{Y\psi}(\infty,\u v) = \exp\left(-\int_1^{\u v}\u\kappa\u\varkappa(\infty,\u v')\,d\u v'\right)Y\psi (u_{\mathcal H^+},1) + \int_1^{\u v}\exp\left(-\int_{\u v'}^{\u v}\u\kappa\u\varkappa(\infty,\u v'')\,d\u v''\right)\u\kappa\u\varkappa\u\phi(\infty,\u v')\,d \u v'.
    \end{equation}
    From this, we expect the exponential redshift effect to not matter until the integrating factor starts to decay, which happens at the redshift time $\u v = \beta$. 

    We define the \emph{transient charge map}
    \begin{align*}
        \mathscr H_\flat:\bigcup_{\sigma\in(-1,1)}\mathfrak M_\mathrm{stab}^\sigma&\to \Bbb R\\
        \Psi &\mapsto e^{4\pi}\u{Y\psi}(\infty,\beta(\Psi)). 
    \end{align*}
    The normalizing factor $e^{4\pi}$ has to do with the fact that the redshift integrating factor on extremal Reissner--Nordstr\"om is identically one, but equals $e^{-4\pi}$ on subextremal Reissner--Nordstr\"om at the redshift time. 

    \hyperref[thm:intro-instab-2]{Part 2.}~of \cref{thm:intro-instability} is now proved in two steps: First, we show that $\mathscr H_\flat$ is a $C^{3\delta/4}_b$-extension of $\mathscr H$ off of the threshold. Therefore, the transient charge of $\Psi_p$ is related to the Aretakis charge $h$ of $\Psi_{p_*}$. Then, by using an detailed expansion for the redshift factor in the subextremal case obtained from the proof of \cref{thm:intro-uniform-stability}, we prove nondecay and growth up to the redshift time, depending explicitly on $h$.

    For details, see already \cref{sec:Aretakis-generic}.

\section{Asymptotically flat solutions of the Einstein--Maxwell-scalar field system}\label{sec:semiglobal-1}

In this section, we study general asymptotically flat solutions of the spherically symmetric EMSF system. In \cref{sec:char-IVP}, we recall the characteristic initial value problem for this system and the fundamental local well-posedness theory. In \cref{sec:perturbation-theory}, we discuss families of solutions and the equations of first- and second-order perturbations. In \cref{sec:AF}, we define asymptotically flat solutions, asymptotically flat characteristic data, asymptotically flat seed data, the moduli space $\mathfrak M$, and define various ``norms.'' In \cref{sec:local-in-u}, we show that asymptotically flat characteristic data give rise to ``semiglobal'' solutions, i.e., the maximal development contains a full double null rectangle going to null infinity $\mathcal I^+$. In \cref{sec:global-structure}, we give a proof of Dafermos' dichotomy for the EMSF model and use it to prove \cref{ingredient-noncollapse} from \cref{sec:intro-proof-uniform}. In \cref{sec:perturbations-1}, we study first- and second-order perturbations of asymptotically flat solutions in initial data gauge. In \cref{sec:teleological-gauge}, we define the teleologically normalized coordinates $\u u$ and $\u v{}_{\u u{}_f}$. In \cref{sec:diffeo-ests}, we prove various estimates for the diffeomorphisms between the teleological gauges and the initial data gauge. In \cref{sec:PT-Bondi-gauge}, we study first- and second-order perturbations of asymptotically flat solutions in the teleological gauges. Finally, in \cref{sec:cts-differentiability}, we show that the  parameter ratio at null infinity, $\mathscr P_{\u u{}_f}$, is $C^1$ for any finite $\u u{}_f$.

\subsection{The characteristic initial value problem and the local structure of solutions}\label{sec:char-IVP}

\subsubsection{Characteristic data and existence in thin slabs}

Given $u_0,u_1,v_0,v_1\in\Bbb R$ with $u_0<u_1$ and $v_0<v_1$, let \index{C@$\mathcal C(u_0,u_1,v_0,v_1)$, bifurcate null hypersurface} \index{Rauf0@$\mathcal R(u_0,u_1,v_0,v_1)\doteq  [u_0,u_1]\times [v_0,v_1]$, double null rectangle} \index{C2@$\u C{}_\ing$, initial ingoing cone} \index{C1@$C_\out$, initial outgoing cone}
\begin{align*}
  \mathcal C(u_0,u_1,v_0,v_1)  &\doteq (\{u_0\}\times[v_0,v_1])\cup([u_0,u_1]\times\{v_0\}), \\
   \mathcal R(u_0,u_1,v_0,v_1) &\doteq [u_0,u_1]\times [v_0,v_1].
\end{align*} We will omit the decoration $(u_0,u_1,v_0,v_1)$ from $\mathcal C$ and $\mathcal R$ when the implied meaning is clear. We often allow for $u_1$ or $v_1$ to be $+\infty$, but will make it clear when this choice is disallowed. It is convenient to introduce the shorthand $C_\out \doteq [u_0,u_1]\times\{v_0\}$ and $\u C{}_\ing\doteq \{u_0\}\times[v_0,v_1]$, so that $\mathcal C=C_\out\cup\u C{}_\ing$. A continuous function $f:\mathcal C\to \Bbb R$ is said to be $C^k$ if $f_\out\doteq f|_{C_\out}$ and $f_\ing\doteq f|_{\u C{}_\ing}$ are $C^k$ functions. 

\begin{defn} Let $k\in\Bbb N$ and $\mathcal C$ be a (possibly infinite) bifurcate characteristic cone as defined above.
  A $C^k$ \emph{(bifurcate) characteristic initial data set} for the EMSF system consists of $C^k$ functions $\Omega^2_\circ:\mathcal C\to\Bbb R_{>0}$ and $\phi_\circ:\mathcal C\to\Bbb R$, a $C^{k+1}$ function\footnote{The reason for the tilde here is to not conflict with the notation $r_\circ$ used in the definition of seed data, which is more fundamental in the logic of the paper.} $\tilde r_\circ :\mathcal C\to\Bbb R_{>0}$, together with a real number $Q$. The functions $\tilde r_\circ$, $\Omega^2_\circ$, and $\phi_\circ$ are assumed to satisfy \eqref{eq:Ray-u} on $\u C{}_\ing$ and \eqref{eq:Ray-v} on $C_\out$. Characteristic initial data sets are denoted by the symbol $\Psi_\#$. \index{Psi1@$\Psi_\#$, characteristic initial data set} \index{ph1@$\phi_\circ$, initial data for $\phi$}
\end{defn}

When considering initial data of finite regularity in $C^k$ spaces, the function space in which solutions live is given by the following:

\begin{defn} A function $f:\mathcal D\to\Bbb R$, where $\mathcal D\subset\Bbb R^2_{u,v}$ is open, is said to be of class $C_\star^k$ if each of the partial derivatives $\partial_u^{k_1}\partial_v^{k_2}f$, with $0\le k_1+k_2\le k+1$, $k_1\le k$, and $k_2\le k$ exist, are continuous, and satisfy the usual symmetry properties expected of functions of class $C^{k+1}$.
\end{defn}\index{C@$C^k_\star$, differentiability class with special mixed partial derivatives}

For instance, if $f\in C_\star^2$, then $\partial_u^3f$ need not exist, but $\partial_u\partial_v\partial_uf$ and $\partial_u^2\partial_v f$ exist, are continuous, and are equal. The set of functions in $C_\star^k$ with bounded derivatives has a natural Banach space structure.

We have the following fundamental local well-posedness result in \emph{thin characteristic slabs}.

\begin{prop}\label{prop:slab-existence}
    For any $L>0$ and $B>0$, there exists a constant $\ve_\mathrm{slab}>0$ with the following property. Let $( \tilde r_\circ,\Omega^2_\circ, \phi_\circ,Q)$ be a characteristic initial data set of the Einstein--Maxwell-scalar field system on $\mathcal C(u_0,u_1,v_0,v_1)$ with $0<u_1-u_0\le L$ and $0<v_0<v_1\le L$ satisfying
    \begin{equation*}
        \|{\log \tilde r_\circ}\|_{C^1(\mathcal C)}+\|{\log\Omega^2_\circ}\|_{C^1(\mathcal C)}+\|\phi_\circ\|_{C^1(\mathcal C)}+|Q|\le B.
    \end{equation*}
    Then there exists a unique solution $(r,\Omega^2,\phi,Q)$ to the spherically symmetric Einstein--Maxwell-scalar field system on the ``double slab''
    \begin{equation*}
       \mathcal R(u_0,u_0+\min\{\ve_\mathrm{slab},u_1-u_0\},v_0,v_1)\cup\mathcal R(u_0,u_1,v_0,v_0+\min\{\ve_\mathrm{slab},v_1-v_0\})
    \end{equation*}
    which extends the initial data, with $r\in C_\star^{k+1}$ and $\Omega^2,\phi\in C_\star^k$. Moreover, higher order norms of $r,\Omega^2$, and $\phi$ are bounded in terms of appropriate higher order initial data norms. 
\end{prop}

Solutions arising from characteristic data $\Psi_\#$ will generally be denoted by $\mathcal S[\Psi_\#]$.\index{S2@$\mathcal S[\Psi_\#]$, solution determined by the characteristic data $\Psi_\#$}

\begin{proof}[Sketch of proof] We outline the main steps of the proof.

\textsc{Solving the wave equations in small rectangles}: We define the vector-valued function $\bm\psi\doteq(r,\Omega^2,\phi)$ and reformulate \eqref{eq:r-wave}, \eqref{eq:Omega-wave}, and \eqref{eq:phi-wave-1} as 
\begin{equation}\label{eq:F}
    \partial_u\partial_v\bm\psi = F(\bm\psi,\partial\bm\psi),
\end{equation}
which can be solved by a standard iteration scheme (setting $\partial_u\partial_v\bm\psi_n=F(\bm\psi_{n-1},\partial\bm\psi_{n-1})$) in small characteristic rectangles. See for instance \cite[Appendix A]{KU24}, where the proof can be easily modified to show that $\{\bm\psi_n\}$ is Cauchy in $C_\star^k$ in the current context. 

\textsc{Propagation of constraints}: Given the solution $\bm\psi$, we infer Raychaudhuri's equations \eqref{eq:Ray-u} and \eqref{eq:Ray-v} by observing that \eqref{eq:r-wave}, \eqref{eq:Omega-wave}, and \eqref{eq:phi-wave-1} imply the following pair of conservation laws:
\begin{align*}
     \partial_v\left(r\nu(u,v)-r\nu(u_0,v)-\int_{u_0}^u\big(\nu^2+r\nu\partial_u{\log\Omega^2}-r^2(\partial_u\phi)^2\big)(u',v)\,du'\right)&=0,\\
      \partial_u\left(r\lambda(u,v)-r\lambda(u,v_0)-\int_{v_0}^v\big(\lambda^2+r\lambda\partial_v{\log\Omega^2}-r^2(\partial_v\phi)^2\big)(u,v')\,dv'\right)&=0.
\end{align*}
Since \eqref{eq:Ray-u} and \eqref{eq:Ray-v} hold on initial data by assumption, this shows that $r\in C_\star^{k+1}$ and that \eqref{eq:Ray-u} and \eqref{eq:Ray-v} hold everywhere. 

\textsc{Extending to thin slabs}: The nonlinearity in \eqref{eq:F} satisfies the classical null condition for derivatives: \eqref{eq:F} can be viewed as a linear ODE in $u$ for the $v$-derivative of $\bm\psi$, and vice-versa. This allows Gr\"onwall's inequality to be used in the ``long'' direction of the slab to extend the existence ``time.'' See \cite{Luk-local-existence} for an elaboration of this principle in a much more complicated setting. Alternatively, one can use the ``generalized extension principle'' for this model (see \cite{Kommemi13}) and argue as in \cite[Proposition 3.17]{KU24}.
\end{proof}

\begin{rk}
    The higher regularity of $r$ in \cref{prop:slab-existence} is due to Raychaudhuri's equations \eqref{eq:Ray-u} and \eqref{eq:Ray-v}, which allow us to control one more derivative of $r$ than $\Omega^2$ and $\phi$.
\end{rk}

\subsubsection{The maximal globally hyperbolic development}

We also have a natural notion of maximal globally hyperbolic development \cite{CBG69,Zorn-slayed} in spherical symmetry, which can be directly realized as a subset of the domain of dependence of $\mathcal C$ viewed as a subset of $(1+1)$-dimensional Minkowski space. \index{Qmax@$\mathcal Q_\mathrm{max}$, quotient space of the maximal development}

\begin{prop}
 Let $\Psi_\#$ be a characteristic initial data set of the Einstein--Maxwell-scalar field system on $\mathcal C(u_0,u_1,v_0,v_1)$, where $u_1$ and $v_1$ are allowed to take the value $+\infty$. Then there exists a set $\mathcal Q_\mathrm{max}\subset \mathcal R(u_0,u_1,v_0,v_1)$ with the following properties:
 \begin{enumerate}
 \item $\mathcal Q_\mathrm{max}$ is globally hyperbolic as a subset of $\Bbb R^{1+1}_{u,v}$ with Cauchy surface $\mathcal C$.
\item The data $\Psi_\#$ extends uniquely to a solution on $\mathcal Q_\mathrm{max}$.
\item $\mathcal Q_\mathrm{max}$ is maximal with respect to properties 1.~and 2.
\end{enumerate}
\end{prop}

We refer to $(\mathcal Q_{\max},r,\Omega^2,\phi,Q)$ as the \emph{maximal development} of $\Psi_\#$, denoted $\mathcal S_\mathrm{max}[\Psi_\#]$.\index{S6@$\mathcal S_\mathrm{max}$, maximal globally hyperbolic development} Since $\mathcal Q_\mathrm{max}$ is a subset of $\Bbb R^2$, we may entertain questions about the structure of its (topological) boundary. We first restrict attention to the following class of spacetimes:

\begin{defn}
    A spherically symmetric spacetime $(\mathcal Q,r,\Omega^2)$ with distinguished ``ingoing'' null coordinate $u$ possesses \emph{no antitrapped spheres of symmetry} if $\nu<0$ on $\mathcal Q$.
\end{defn}

By Raychaudhuri's equation \eqref{eq:Ray-u}, $\partial_u(\Omega^{-2}\partial_ur)\le 0$ for any solution of the EMSF system, and hence it suffices to verify this condition on the initial data surface. 

\begin{lem}
    Let $\mathcal S_\mathrm{max}$ be the maximal development of characteristic data $\Psi_\#$ on a bifurcate characteristic hypersurface $\mathcal C=C_\out\cup \u C{}_\ing$. If $\nu<0$ on $C_\out$, then $\nu<0$ on $\mathcal Q_\mathrm{max}$.
\end{lem}

Using the other Raychaudhuri equation, \eqref{eq:Ray-v}, we have the following fundamental monotonicity:
\begin{lem}
    If $\lambda(u,v)=0$ for some $(u,v)\in \mathcal Q_\mathrm{max}$, then $\lambda(u,v')\le 0$ for any $(u,v')\in \mathcal Q_\mathrm{max}$ with $v'\ge v$.
\end{lem}

\subsubsection{Gauge-normalized seed data}\label{sec:seed-data}

To define the moduli space of initial data $\mathfrak M$ for our main theorems, it is convenient to parametrize the space of bifurcate characteristic initial data as a linear space in such a way that certain gauge conditions are automatically satisfied on $\mathcal C$ in the maximal development. 

\begin{defn}
    Let $\mathcal C$ be a spherically symmetric bifurcate null hypersurface and $k\in\Bbb N$. A $C^k$ \emph{seed data set} is a quadruple \index{psi@$\Psi$, seed data set, element of $\mathfrak M$} \index{rho@$\rho_\circ$, $\rho$ at the initial bifurcation sphere} \index{pi1@$\varpi_\circ$, $\varpi$ at the initial bifurcation sphere} \index{r@$r_\circ$, $r$ at the initial bifurcation sphere}
    \begin{equation*}
        \Psi\doteq (\phi_\circ,r_\circ,\varpi_\circ,\rho_\circ),
    \end{equation*}
    where $\phi_\circ\in C^k(\mathcal C)$, $r_\circ\in\Bbb R_{>0}$, and  $\varpi_\circ,\rho_\circ\in\Bbb R$, subject to the condition
    \begin{equation*}
        \mu_\circ \doteq \frac{2\varpi_\circ}{r_\circ} - \frac{\varpi_\circ^2\rho_\circ^2}{r_\circ^2} <1.
    \end{equation*}
\end{defn}

\begin{prop}\label{prop:seed-data-generation}
   Let $\Psi=(\phi_\circ,r_\circ,\varpi_\circ,\rho_\circ)$ be a seed data set on $\mathcal C(u_0,u_1,v_0,v_1)$ with $u_1-u_0<r_\circ$. Then there exists a unique characteristic initial data set $\Psi_\#=(\tilde r_\circ, \Omega^2_\circ,\phi_\circ,Q)$ on $\mathcal C(u_0,u_1,v_0,v_1)$ such that the maximal development $\mathcal S_\mathrm{max}[\Psi_\#]=(\mathcal Q_\mathrm{max},r,\Omega^2,\phi,Q)$ of $\Psi$ has the following properties: 
   \begin{enumerate}
       \item $Q=\varpi_\circ\rho_\circ$, 
       \item $r(u_0,v_0)=r_\circ$,
       \item $\varpi(u_0,v_0)=\varpi_\circ$,
       \item $\nu=-1$ on $\u C{}_\ing$, and
       \item $\lambda= 1$ on $C_\out$.
   \end{enumerate}
\end{prop}
We refer to the characteristic data $\Psi_\#$ obtained from $\Psi$ in this manner as \emph{gauge-normalized} characteristic data determined by $\Psi$. It is also convenient to write $\mathcal S_\mathrm{max}[\Psi]$ for $\mathcal S_\mathrm{max}[\Psi_\#]$. \index{S1@$\mathcal S[\Psi]$, solution determined by the seed data $\Psi$}
\begin{proof} For $(u,v)\in[u_0,u_1]\times[v_0,v_1]$, set
\begin{equation*}
  \tilde r_{\circ,\ing}(u)\doteq r_\circ-u+u_0,\quad \tilde r_{\circ,\out}(v)\doteq r_\circ+v-v_0,
\end{equation*}
and then
\begin{equation*}
  \Omega^2_{\circ,\ing}(u)\doteq  \frac{4}{1-\mu_\circ} \exp\left(-\int_{u_0}^u\tilde r_{\circ,\ing} (\partial_u\phi_{\circ,\ing})^2\,du'\right),\quad   \mathring\Omega^2_\out(v)\doteq  \frac{4}{1-\mu_\circ} \exp\left(\int_{v_0}^v \tilde r_{\circ,\out} (\partial_v\phi_{\circ,\out})^2\,dv'\right).
\end{equation*}
Assembling these functions into a characteristic data set $( r_\circ,\Omega^2_\circ,\phi_\circ,Q)$, we can immediately verify that the constraints are satisfied and that conditions 1.--5.~are satisfied on the maximal development by definition.
\end{proof}

\subsection{Perturbation theory} \label{sec:perturbation-theory}

\subsubsection{``Smooth'' dependence on initial data and local Cauchy stability}

\begin{defn}
    A function $f:\mathcal D\times U\to\Bbb R$, where $\mathcal D\subset\Bbb R_{u,v}^2$ and $U\subset\Bbb R_z^d$, is said to be of class $C_\star^k$ if the partial derivatives 
$ \partial_u^{k_1}\partial_v^{k_2}\partial_z^{k_3}f$, $\partial_v\partial_u^{k}\partial_z^{k_3}f$, and $\partial_u\partial_v^k\partial_z^{k_3}f$ exist and are continuous, where $k_3$ is a multiindex, $0\le k_1+k_2\le k$, and $0\le |k_3|\le k$, and if a multiderivative $\partial^\alpha$ in $(u,v,z)$ is obtained from $\partial_u^{k_1}\partial_v^{k_2}\partial_z^{k_3}$, $\partial_v\partial_u^{k}\partial_z^{k_3}$, or $\partial_u\partial_v^k\partial_z^{k_3}$ by permutation of the partial derivative factors, then $\partial^\alpha f$ exists and is equal to the corresponding $\partial_u^{k_1}\partial_v^{k_2}\partial_z^{k_3}f$, $\partial_v\partial_u^{k}\partial_z^{k_3}f$, or $\partial_u\partial_v^k\partial_z^{k_3}f$.
\end{defn}\index{C@$C^k_\star$, differentiability class with special mixed partial derivatives} \index{z@$z$, parameter for families of initial data}

Again, the space of $C_\star^k$ functions in $(u,v,z)$ with bounded derivatives admits a natural Banach space structure. In practice, we will take $d\le 2$ and $k\le 2$. The following result is standard and states that $C_\star^k$ families of characteristic data give rise to $C_\star^k$ families of solutions.

\begin{prop}\label{prop:smooth-dependence} For any $L_1>0$, $L_2>0$, and $B>0$, there exist constants $\delta>0$ and $C>0$ with the following property.  Let $\mathcal S$ be a solution to the EMSF system on the finite double null rectangle $\mathcal R=\mathcal R(u_0,u_1,v_0,v_1)$ arising from $C^k$ characteristic data $\Psi_\#$ on $\mathcal C$, with $u_1-u_0\le L_1$, $v_1-v_0\le L_2$, and $\|\mathcal S\|_{C^{k+1}\times C^k\times C^k(\mathcal R)}\le B$. Let $z\mapsto \Psi_{\#}(z)$ be a $C_\star^k$ $d$-parameter family of characteristic data with $\Psi_{\#}(0)=\Psi_{\#}$ and satisfying $\|\Psi_{\#}(\cdot)\|_{C_\star^k}\le B$. Then $\mathcal S[\Psi_{\#}(z)]$ exists on $\mathcal R$ for $|z|<\delta$. Moreover, $r$, $\Omega^2$, and $\phi$, viewed as functions of $(u,v,z)$, are $C_\star^{k+1}$-, $C_\star^k$-, and $C_\star^k$-regular, respectively, and it holds that
\begin{equation}\label{eq:local-Cauchy-stab}
    \|(r,\Omega^2,\phi)(\cdot,z_1)-(r,\Omega^2,\phi)(\cdot,z_2)\|_{C^{k+1}\times C^k\times C^k(\mathcal R)}\le C|z_1-z_2|
\end{equation}
for any $|z_1|,|z_2|<\delta$.
\end{prop}

We shall refer to the existence of $\mathcal S[\Psi_{\#}(z)]$ over the fixed rectangle $\mathcal R$ for $|z|$ small and the estimate \eqref{eq:local-Cauchy-stab} as \emph{(local) Cauchy stability} of $\mathcal S[\Psi_\#]$. Although the result is standard, we give a slightly nonstandard proof via a direct iteration scheme in the space $C_\star^k$. We also prove \eqref{eq:local-Cauchy-stab} using the mean value theorem, an argument we will use repeatedly in this paper. 

\begin{proof}[Sketch of proof] Recall the notation $\bm\psi$ and $F$ from the sketch of the proof of \cref{prop:slab-existence}. We restrict attention to $k=1$, $d=1$ for simplicity. 

Denote the solution corresponding to $\Psi_{\#}(0)$ by $\bm\psi_0$ and solve iteratively
\begin{equation*}
    \partial_u\partial_v\bm\psi_n = F(\bm\psi_{n-1},\partial\bm\psi_{n-1}),
\end{equation*} for $n\ge 1$,
where $\bm\psi_n=\bm\psi_n(u,v,z)$, with initial data $(\bm\psi_n)_\circ = \Psi_{\#}(z)$. The solution is given explicitly by
\begin{equation}\label{eq:psi-n}
    \bm\psi_n(u,v,z) = \int_{u_0}^u\int_{v_0}^v F_{n-1}(u',v',z)\,du'dv' + \Psi_{\#,\ing}(u,z)+\Psi_{\#,\out}(v,z) - \Psi_{\#,\out}(v_0,z),
\end{equation}
where $F_k\doteq F(\bm\psi_k,\partial\bm\psi_k)$. By induction, $\bm\psi_n\in C_\star^1$.

We show that $\{\bm\psi_n\}$ is bounded in $C_\star^1$. Using the representation formula, it is easy to prove inductively the bounds 
\begin{equation*}
    \|\bm\psi_n-\bm\psi_0\|_{C^1_{u,v}(\mathcal R)} \le C_1 |z|e^{M(u+v)},\quad 
    \|\partial_z\bm\psi_n\|_{C^1_{u,v}(\mathcal R)}\le C_1 e^{M(u+v)}
\end{equation*}
for constants $C_1$ and $M$ sufficiently large and $|z|<\delta$ for $\delta$ sufficiently small. It follows that there exists a constant $C_2$ such that
\begin{equation}\label{eq:psi-n-1}
    \|\bm\psi_n\|_{C_\star^1(\mathcal R\times(-\delta,\delta))}\le C_2
\end{equation}
for every $n\ge 1$. We now show that $\{\bm\psi_n\}$ is Cauchy in $C_\star^1$. First, the bound \eqref{eq:psi-n-1} implies the existence of a constant $C_3$ such that
\begin{align*}
    \big(|\hat{\bm\psi}_n|+ |\partial_z\hat{\bm\psi}_n|\big)(u,v,z) &\le C_3\int_{u_0}^u\int_{v_0}^vZ_{n-1}\,dv'du',\\
     \big( |\partial_u\hat{\bm\psi}_n|+|\partial_u\partial_z\hat{\bm\psi}_n|\big)(u,v,z) &\le C_3\int_{v_0}^vZ_{n-1}\,dv',\\
     \big( |\partial_v\hat{\bm\psi}_n|+|\partial_v\partial_z\hat{\bm\psi}_n|\big)(u,v,z)&\le C_3\int_{u_0}^uZ_{n-1}\,du',
\end{align*}
where $\hat{\bm\psi}_n\doteq \bm\psi_n-\bm\psi_{n-1}$ for $n\ge 2$ and 
\begin{equation*}
    Z_n(u,v,z)\doteq |\hat{\bm\psi}_n|+ |\partial_z\hat{\bm\psi}_n|+|\partial_u\hat{\bm\psi}_n|+|\partial_u\partial_z\hat{\bm\psi}_n|+|\partial_v\hat{\bm\psi}_n|+|\partial_v\partial_z\hat{\bm\psi}_n|.
\end{equation*}
Iterating these inequalities, carrying out the repeated integrals, and using that $Z_2$ is bounded, we infer the existence of a constant $C_4$ such that
\begin{equation*}
    \sup_{\mathcal R\times(-\delta,\delta)}|Z_n|\le \frac{C_4^n}{n!}.
\end{equation*}
By a telescoping series argument, this implies that $\{\bm\psi_n\}$ is Cauchy in $C_\star^1$, which completes the proof of existence. To prove the estimate \eqref{eq:local-Cauchy-stab}, let $n\to\infty$ in \eqref{eq:psi-n-1} and use the mean value theorem in $z$. 
\end{proof}

\subsubsection{First-order perturbations}

\index{*@$\lin\cdot$, linearized quantity} Given a dynamical quantity such as $r$, $\gamma$, etc., depending on $z\in\Bbb R^d$ as in \cref{prop:smooth-dependence}, we denote $\lin r\doteq \partial_z r$, $\lin \gamma\doteq \partial_z\gamma$, etc. We call the collection of such quantities a \emph{first-order} or \emph{linear perturbation} of the solution. Note that for instance $\lin\lambda = \partial_v\lin r$, $\lin\nu=\partial_u\lin r$, $\lin{\partial_v\phi} = \partial_v\lin\phi$, and $\lin\psi = \lin r\phi+r\lin\phi$ since the $(u,v)$ and $z$ derivatives commute by construction. Since $Q$ is constant on $\mathcal Q$, $\lin Q$ is also constant on $\mathcal Q$ and can be read off from the initial data family. 

Note that each first-order quantity such as $\lin r$ or $\lin\phi$ is actually a $d$-tuple of functions on $\mathcal Q$. However, as we will see below, the evolution equations for each component are decoupled, i.e., the equations determining $\partial_{z_1}\phi$ are independent of $\partial_{z_2}r$, etc., treating the first-order perturbations as scalar quantities is only a minor abuse of notation. 

By linearizing the null constraints \eqref{eq:Ray-u} and \eqref{eq:Ray-v} on $\mathcal C$, we infer that, in the context of \cref{prop:smooth-dependence}, it holds that\footnote{By $\lin{\Omega}^2$ we shall mean $\partial_z(\Omega^2)$, not $(\partial_z\Omega)^2$.}  \begin{equation}\label{eq:lin-const-1}
        \partial_u\left(\frac{\lin\nu}{\Omega^2}-\frac{\nu}{\Omega^4}\lin\Omega^2\right) = -\frac{\lin r}{\Omega^2}(\partial_u\phi)^2 + \frac{r}{\Omega^4}\lin\Omega^2(\partial_u\phi)^2 - \frac{2r}{\Omega^2}\partial_u\phi\partial_u\lin\phi
    \end{equation}
    on $\u C{}_\ing$ and 
    \begin{equation}\label{eq:lin-const-2}
        \partial_v\left(\frac{\lin\lambda}{\Omega^2}-\frac{\lambda}{\Omega^4}\lin\Omega^2\right) = -\frac{\lin r}{\Omega^2}(\partial_v\phi)^2 + \frac{r}{\Omega^4}\lin\Omega^2(\partial_v\phi)^2 - \frac{2r}{\Omega^2}\partial_v\phi\partial_v\lin\phi
    \end{equation}
     on $C_\out$. These constraints are propagated by the linearized wave system and we record the following observation, which is not strictly necessary in the logic of this paper\footnote{Our first-order perturbations will always arise as genuine linearizations of solutions to the nonlinear system, so we do not strictly need an independent existence theory for the linearized equations as in \cite[Appendix A]{DHR19}.} but explains why the notation for $\lin{\mathcal S}$ used below is unambiguous. \index{Psi1@$\lin\Psi_\#$, linearized characteristic data set} 
     
\begin{lem}\label{lem:first-order-constraint}
    Let $\Psi_\#$ be characteristic data with $\mathcal S[\Psi_\#]$ defined over a (possibly infinite) double null rectangle $\mathcal R$ and let $\lin\Psi_\#$ be defined over $\mathcal C$ and satisfy the linearized Raychaudhuri equations \eqref{eq:lin-const-1} and \eqref{eq:lin-const-2}. Let
    \begin{equation}\label{eq:first-order-constraint-1}
        \lin\Psi_\#\mapsto (\lin r,\lin\Omega^2,\lin \phi,\lin Q) 
    \end{equation} denote the solution of the differentiated versions of \eqref{eq:r-wave}, \eqref{eq:Omega-wave}, and \eqref{eq:phi-wave-1} which attains the data $\lin\Psi_\#$ on $\mathcal C$. Then $(\lin r,\lin\Omega^2,\lin\phi)$ satisfy the once-differentiated constraint equations \eqref{eq:lin-const-1} and \eqref{eq:lin-const-2} everywhere in $\mathcal R$. Moreover, if $\lin\Psi_\#$ arises as $\partial_z\Psi_{\#}(z)|_{z=0}$ for a one-parameter family $z\mapsto \Psi_{\#}(z)$ as in \cref{prop:smooth-dependence}, then the first-order perturbations given by \eqref{eq:first-order-constraint-1} are indeed equal to the $z$-derivatives of $(r,\Omega^2,\phi,Q)$ at $z=0$. 
\end{lem}
\begin{proof}
    We explain why \eqref{eq:lin-const-1} is propagated and leave the rest of the lemma to the reader as an exercise. A very tedious calculation shows that if $(\lin r,\lin\Omega^2,\lin\phi)$ satisfy the differentiated versions of \eqref{eq:r-wave}, \eqref{eq:Omega-wave}, and \eqref{eq:phi-wave-1}, then 
    \begin{multline*}
        \partial_v\bigg(\lin r\nu(u,v)+r\lin\nu(u,v) -\lin r\nu(u_0,v)- r\lin \nu(u_0,v)\\  - \int_{u_0}^u\left(2\nu\lin\nu + \lin r\nu\partial_u{\log\Omega^2}+r\lin\nu\partial_u{\log\Omega}^2+r\nu\Omega^{-2}\partial_u\lin\Omega^2-2r\lin r(\partial_u\phi)^2-2r^2\partial_u\phi \partial_u\lin\phi\right)\,du' \bigg) = 0,
    \end{multline*}
    which, as in the proof of \cref{prop:slab-existence}, shows that \eqref{eq:lin-const-1} holds on $\mathcal R$. 
\end{proof}

This lemma justifies the following notation and convention: Given a characteristic data set $\lin\Psi_\#$ on $\mathcal C$ which satisfies the constraints \eqref{eq:Ray-u} and \eqref{eq:Ray-v} and exists on $\mathcal R$, and a $\lin\Psi_\#$ as in the lemma satisfying the linearized constraints \eqref{eq:lin-const-1} and \eqref{eq:lin-const-2}, we define \index{S3@$\lin{\mathcal S}$, linearized solution}
\[\lin{\mathcal S}[\Psi_\#,\lin\Psi_\#]\doteq(\mathcal R,\lin r,\lin\Omega^2,\lin\phi,\lin Q)\]
as given by the lemma. In other words, \cref{lem:first-order-constraint} allows us to ``forget'' about the one-parameter family $\Psi_{\#,z}$ used to generate $\lin\Psi_\#$ in the first place. 

The first order quantities satisfy all of the evolution equations obtained by applying $\partial_z$ to \eqref{eq:r-wave}--\eqref{eq:kappa-u}. Moreover, all of the definitions such as \eqref{eq:Hawking-mass} and \eqref{eq:mu-defn} can also be differentiated and then give relations between first order quantities. We shall require the equations
\begin{equation}\label{eq:first-order-wave}
    \partial_u\partial_v\lin\phi + \frac{\lambda}{r}\partial_u\lin\phi + \frac{\nu}{r} \partial_v\lin\phi =\left(\frac{\lambda \lin r}{r^2}- \frac{\lin\lambda}{r}\right)\partial_u\phi +\left(\frac{\nu \lin r}{r^2}- \frac{\lin\nu}{r}\right)\partial_v\phi,
\end{equation}
\begin{equation}\label{eq:lin-psi}
    \partial_u\partial_v\lin\psi =  \nu\varkappa\phi\lin\kappa + \kappa\varkappa\phi\lin\nu + \kappa\nu\phi\lin\varkappa +  \kappa\nu\varkappa \lin\phi,
\end{equation}
\begin{equation*}
      \lin\mu = \frac{2\lin\varpi}{r}-\frac{2Q\lin Q}{r^2}- 2\varkappa\lin r,\quad  \lin\varkappa = \frac{2\lin\varpi}{r^2} - \frac{4Q\lin Q}{r^3} + \left(\frac{6Q^2}{r^4}-\frac{4\varpi}{r^3}\right)\lin r,
\end{equation*}
\begin{equation}\label{eq:lin-lambda}
    \partial_v \lin r = \lin \lambda = \kappa\varkappa\lin r + (1-\mu)\lin \kappa - \frac{2\kappa}{r}\lin\varpi+\frac{2\kappa Q}{r^2}\lin Q,
\end{equation}
\begin{equation}\label{eq:lin-nu}
    \partial_u \lin r = \lin \nu =\gamma\varkappa\lin r+ (1-\mu ) \lin \gamma - \frac{2\gamma}{r} \lin \varpi + \frac{2\gamma Q}{r^2} \lin Q ,
\end{equation}
\begin{equation}\label{eq:dvlnu}
    \partial_v\lin\nu = \partial_u\lin\lambda = \lin\kappa \nu\varkappa+ \kappa\lin\nu\varkappa + \kappa\nu\lin\varkappa,
\end{equation}
\begin{equation}\label{eq:dvlnu-ratio}
\partial_v  \left( \frac{\lin \nu}{\nu} \right) = \lin \kappa \varkappa + \kappa \lin r \left( - \frac{4\varpi}{r^3} + \frac{6Q^2}{r^4} \right) + \frac{2\kappa \lin \varpi}{r^2} - \frac{4\kappa Q \lin Q}{r^3} ,
\end{equation}
\begin{equation}\label{eq:dulkappa}
    \partial_u\lin\kappa = \left(\frac{\kappa\lin r}{\nu}+ \frac{r\lin\kappa}{\nu} - \frac{r\kappa\lin \nu}{\nu^2}\right)(\partial_u\phi)^2 +\frac{2r\kappa}{\nu}\partial_u\phi\partial_u\lin\phi,
\end{equation}
\begin{equation}\label{eq:dulkappa-log}
    \partial_u\left(\frac{\lin\kappa}{\kappa}\right) = \left(\frac{\lin r}{\nu}- \frac{r\lin \nu}{\nu^2}\right)(\partial_u\phi)^2 +\frac{2r}{\nu}\partial_u\phi\partial_u\lin\phi,
\end{equation}
\begin{equation}\label{eq:dvlgamma}
    \partial_v\lin\gamma = \left(\frac{\gamma\lin r}{\lambda}+ \frac{r\lin\gamma}{\lambda} -\frac{r\gamma\lin \lambda}{\lambda^2}\right)(\partial_v\phi)^2 +\frac{2r\gamma}{\lambda}\partial_v\phi\partial_v\lin\phi,
\end{equation}
\begin{equation}\label{eq:dvlgamma-log}
    \partial_v\left(\frac{\lin\gamma}{\gamma} \right)= \left(\frac{\lin r}{\lambda}-\frac{r\lin \lambda}{\lambda^2}\right)(\partial_v\phi)^2 +\frac{2r}{\lambda}\partial_v\phi\partial_v\lin\phi,
\end{equation}
\begin{equation}\label{eq:dulvarpi}
    \partial_u\lin\varpi = \frac{r\lin r}{\gamma}(\partial_u\phi)^2 - \frac{r^2\lin\gamma}{2\gamma^2}(\partial_u\phi)^2 + \frac{r^2}{\gamma}\partial_u\phi\partial_u\lin\phi,
    \end{equation}
    \begin{equation}\label{eq:dulvarpi-near-region}
    \partial_u\lin\varpi =  -\lin\mu\frac{r^2}{2\nu}(\partial_u\phi)^2 + (1-\mu)\frac{r\lin r}{\nu}(\partial_u\phi)^2-(1-\mu)\frac{r^2\lin\nu}{2\nu^2}(\partial_u\phi)^2 + (1-\mu)\frac{r^2}{\nu}\partial_u\phi\partial_u\lin\phi,
\end{equation}
\begin{equation}
    \partial_v\lin\varpi = \frac{r\lin r}{\kappa}(\partial_v\phi)^2 - \frac{r^2\lin\kappa}{2\kappa^2}(\partial_v\phi)^2 + \frac{r^2}{\kappa}\partial_v\phi\partial_v\lin\phi.\label{eq:first-order-varpi-v}
\end{equation}

\subsubsection{Second-order perturbations} 

We also require second-order perturbations, which consist of $d\times d$-matrix valued quantities on $\mathcal Q$. The equations satisfied by second-order perturbations necessarily involve the first-order perturbations from the previous section but now mixed derivatives will involve products of different first-order components. 

For clarity, we restrict now to the case $d=2$, so that $z=(z_1,z_2)$ and $\lin\Psi_\#$ splits as $\lin \Psi_\# =\big(
        \lin \Psi_\#^1,
        \lin \Psi_\#^2 \big),$  where $\lin\Psi_\#^i$ is associated to $z_i$. Note also that we have a splitting    $\lin{\mathcal S}[\Psi_\#,\lin\Psi_\#]=\big(
        \lin{\mathcal S}[\Psi_\#,\lin\Psi_\#^1],
        \lin{\mathcal S}[\Psi_\#,\lin\Psi_\#^2] \big),$   since the equations governing $\partial_{z_1}$ and $\partial_{z_2}$ quantities decouple. 

In the proof of the main theorems, we only require the mixed partial derivative $\partial_{z_1}\partial_{z_2}$. Accordingly, we denote $\qd r\doteq \partial_{z_1}\partial_{z_2} r$, $\qd\gamma\doteq \partial_{z_1}\partial_{z_2}\gamma$, etc. Again, $\qd Q$ can be immediately read off from data. All other quantities satisfy appropriate twice-differentiated versions of the equations \eqref{eq:r-wave}--\eqref{eq:kappa-u}. \index{*@$\qd\cdot$, second-order quantity}

To capture the algebraic structure of these equations, we introduce the symmetric inner product 
  \begin{equation*}
        \lin p\cdot \lin q \doteq \tfrac 12(\partial_{z_1}p\partial_{z_2}q+\partial_{z_2}p\partial_{z_1}q) 
    \end{equation*}
    for any linearized quantities $\lin p=(    \partial_{z_1}p ,  \partial_{z_2}p)$ and $\lin q= (\partial_{z_1}q,     \partial_{z_2}q )$. 
    For example, the second-order versions of $\partial_up = p^2$ and $\partial_v q = pq$ become
    \begin{equation*}
        \partial_u\qd p = 2p\qd p + 2\lin p\cdot \lin p\quad\text{and}\quad  \partial_v\qd q = q\qd p+p\qd q + 2\lin p\cdot \lin q
    \end{equation*}
 respectively.

 By differentiating the null constraints \eqref{eq:Ray-u} and \eqref{eq:Ray-v} on $\mathcal C$ twice, we infer that, in the context of \cref{prop:smooth-dependence}, it holds that
\begin{multline}\label{eq:qd-Ray-u}
   \partial_u\left(\frac{\qd\nu}{\Omega^2}-\frac{\nu}{\Omega^4}\qd\Omega^2+\frac{2\nu}{\Omega^6}\lin\Omega^2\cdot\lin\Omega^2-\frac{2}{\Omega^4}\lin\nu\cdot\lin\Omega^2\right) = -\frac{\qd r}{\Omega^2}(\partial_u\phi)^2 + \frac{r}{\Omega^4}\qd\Omega^2(\partial_u\phi)^2 - \frac{2r}{\Omega^2}\partial_u\phi\partial_u\qd\phi  \\ + 2\frac{\lin r\cdot\lin\Omega^2}{\Omega^4}(\partial_u\phi)^2 - \frac{4}{\Omega^2}\lin r\cdot\partial_u\lin\phi\partial_u\phi - \frac{2r}{\Omega^6}\lin\Omega^2\cdot\lin\Omega^2(\partial_u\phi)^2 + \frac{4r}{\Omega^4} \lin\Omega^2\cdot\partial_u\lin\phi\partial_u\phi - \frac{2r}{\Omega^2}\partial_u\lin\phi\cdot\partial_u\lin\phi
\end{multline}
on $\u C{}_\ing$ and
\begin{multline}\label{eq:qd-Ray-v}
   \partial_v\left(\frac{\qd\lambda}{\Omega^2}-\frac{\lambda}{\Omega^4}\qd\Omega^2+\frac{2\lambda}{\Omega^6}\lin\Omega^2\cdot\lin\Omega^2-\frac{2}{\Omega^4}\lin\lambda\cdot\lin\Omega^2\right) = -\frac{\qd r}{\Omega^2}(\partial_v\phi)^2 + \frac{r}{\Omega^4}\qd\Omega^2(\partial_v\phi)^2 - \frac{2r}{\Omega^2}\partial_v\phi\partial_v\qd\phi  \\ + 2\frac{\lin r\cdot\lin\Omega^2}{\Omega^4}(\partial_v\phi)^2 - \frac{4}{\Omega^2}\lin r\cdot\partial_v\lin\phi\partial_v\phi - \frac{2r}{\Omega^6}\lin\Omega^2\cdot\lin\Omega^2(\partial_v\phi)^2 + \frac{4r}{\Omega^4} \lin\Omega^2\cdot\partial_v\lin\phi\partial_v\phi - \frac{2r}{\Omega^2}\partial_v\lin\phi\cdot\partial_v\lin\phi
\end{multline}
on $C_\out$. Again, these constraints are propagated by the twice-differentiated Einstein equations. We also have the following analogue of \cref{lem:first-order-constraint}: \index{Psi2@$\qd\Psi_\#$, second-order perturbed characteristic data set}

\begin{lem}\label{lem:second-order-constraint}
    Let $\Psi_\#$ be characteristic data with $\mathcal S[\Psi_\#]$ defined over a (possibly infinite) double null rectangle $\mathcal R$, let $\lin\Psi_\#=(\lin\Psi_\#^1,\lin\Psi_\#^2)$ be defined over $\mathcal C$ and satisfy the linearized Raychaudhuri equations \eqref{eq:lin-const-1} and \eqref{eq:lin-const-2}, and let $\qd\Psi_\#$ be defined over $\mathcal C$ and satisfy the twice-differentiated Raychaudhuri equations \eqref{eq:qd-Ray-u} and \eqref{eq:qd-Ray-v}. Let
    \begin{equation}\label{eq:second-order-constraint-1}
        \qd\Psi_\#\mapsto (\qd r,\qd \Omega^2,\qd \phi,\qd Q) 
    \end{equation} denote the solution of the twice-differentiated versions of \eqref{eq:r-wave}, \eqref{eq:Omega-wave}, and \eqref{eq:phi-wave-1} which attains the data $\qd\Psi_\#$ on $\mathcal C$. Then $(\lin r,\lin\Omega^2,\lin\phi)$ satisfy the once-differentiated constraint equations \eqref{eq:lin-const-1} and \eqref{eq:lin-const-2} everywhere in $\mathcal R$. Moreover, if $\qd\Psi_\#$ arises as $\partial_{z_1}\partial_{z_2}\Psi_{\#}(z)|_{z=0}$ for a two-parameter family $z\mapsto \Psi_{\#}(z)$ as in \cref{prop:smooth-dependence}, then the second-order perturbations given by \eqref{eq:second-order-constraint-1} are indeed equal to the mixed $(z_1,z_2)$-derivatives of $(r,\Omega^2,\phi,Q)$ at $z=0$. 
\end{lem}

We define \index{S4@$\qd{\mathcal S}$, second-order perturbed solution}
\begin{equation*}
    \qd{\mathcal S}[\Psi_\#,\lin\Psi_\#,\qd\Psi_\#]\doteq (\mathcal R,\qd r,\qd\Omega^2,\qd\phi,\qd Q)
\end{equation*}
as given by the lemma. The second-order perturbations satisfy the following equations:
\begin{equation*}
\begin{split}
\partial_u\partial_v \qd\phi + \frac{\lambda}{r} \partial_u \qd\phi + \frac{\nu}{r} \partial_v \qd\phi & = \lin r\cdot \left( \frac{\lin \lambda}{r^2} \partial_u \phi + \frac{\lin \nu}{r^2} \partial_v \phi \right) -  \left( \frac{2\lambda}{r^3} \partial_u \phi + \frac{2\nu}{r^3} \partial_v \phi \right)\lin r\cdot\lin r - \frac{2\lin \lambda}{r} \cdot\partial_u \lin \phi - \frac{2\lin \nu}{r} \cdot\partial_v \lin \phi\\& \quad + \left( \frac{\lambda}{r^2} \partial_u \lin{\phi} + \frac{\nu}{r^2} \partial_v \lin \phi \right)\cdot\lin r - \frac{\qd\lambda}{r} \partial_u \phi - \frac{\qd\nu}{r} \partial_v \phi +\left( \frac{\lambda}{r^2} \partial_u \phi + \frac{\nu}{r^2} \partial_v \phi \right) \qd r ,
\end{split}
\end{equation*}
\begin{equation*}
\qd \mu = \frac{2\qd \varpi}{r} -2  \varkappa\qd r - \lin r\cdot \left( 2 \lin \varkappa + \frac{2\lin \varpi}{r^2} \right), 
\end{equation*}
\begin{equation*}
\qd \varkappa = \frac{2\qd \varpi}{r^2} -\left( \frac{4\varpi}{r^3} - \frac{6Q^2}{r^4} \right)  \qd r-\frac{4Q}{r^3}\qd Q - \frac{8}{r^3}\lin\varpi\cdot\lin r + \left(\frac{12\varpi}{r^4} - \frac{24Q^2}{r^5}\right)\lin r\cdot\lin r -\frac{4}{r^2}\lin Q\cdot\lin Q + \frac{24Q}{r^4}\lin Q\cdot\lin r,
\end{equation*}
\begin{equation*}
    \partial_v \qd r = \qd\lambda =   (1-\mu)\qd\kappa - \kappa\qd\mu -2\lin\kappa\cdot\lin\mu,
\end{equation*}
\begin{equation*}
\partial_v \qd \nu=\partial_v\qd\lambda =  \kappa \varkappa \qd \nu +  \varkappa \nu \qd \kappa  + \kappa \nu  \qd \varkappa + 2  \varkappa \lin \kappa\cdot \lin \nu + 2 \kappa \lin \varkappa\cdot \lin \nu + 2 \nu\lin \kappa\cdot \lin \varkappa,
\end{equation*}
\begin{equation*}
    \begin{split}
        \partial_u\qd\kappa & = \left(\frac{\kappa\qd r}{\nu}+\frac{2\lin r\cdot\lin\kappa}{\nu}+\frac{r\qd\kappa}{\nu}-\frac{2\kappa\lin r\cdot\lin\nu}{\nu^2}-\frac{2r\lin\kappa\cdot\lin\nu}{\nu^2}-\frac{r\kappa\qd\nu}{\nu^2}+\frac{2r\kappa\lin\nu\cdot\lin\nu}{\nu^3}\right)(\partial_u\phi)^2\\
        &\quad + 2\left(\frac{\kappa\lin r}{\nu}+\frac{r\lin\kappa}{\nu}-\frac{r\kappa\lin\nu}{\nu^2}\right)\cdot\partial_u\phi\partial_u\lin\phi + \frac{r\kappa}{\nu}\big(\partial_u\lin\phi\cdot\partial_u\lin\phi+2\partial_u\phi\partial_u\qd\phi\big),
    \end{split}
\end{equation*}
\begin{equation*}
    \begin{split}
        \partial_v\qd\gamma & = \left(\frac{\gamma\qd r}{\lambda}+\frac{2\lin r\cdot\lin\gamma}{\lambda}+\frac{r\qd\gamma}{\lambda}-\frac{2\gamma\lin r\cdot\lin\lambda}{\lambda^2}-\frac{2r\lin\gamma\cdot\lin\lambda}{\lambda^2}-\frac{r\gamma\qd\lambda}{\lambda^2}+\frac{2r\gamma\lin\lambda\cdot\lin\lambda}{\lambda^3}\right)(\partial_v\phi)^2\\
        &\quad + 2\left(\frac{\gamma\lin r}{\lambda}+\frac{r\lin\gamma}{\lambda}-\frac{r\gamma\lin\lambda}{\lambda^2}\right)\cdot\partial_v\phi\partial_v\lin\phi + \frac{r\gamma}{\lambda}\big(\partial_v\lin\phi\cdot\partial_v\lin\phi+2\partial_v\phi\partial_v\qd\phi\big),
    \end{split}
\end{equation*}
\begin{equation*}
    \begin{split}
        \partial_u\qd\varpi & = \left(\frac{\lin r\cdot\lin r+r\qd r}{\gamma}+\frac{2r\lin r\cdot\lin \gamma}{\gamma^2}-\frac{r^2\qd\gamma}{2\gamma^2}+\frac{r^2\lin\gamma\cdot\lin\gamma}{\gamma^3}\right)(\partial_u\phi)^2 \\ 
&\quad + 2\left(\frac{r\lin r}{\gamma}-\frac{r^2\lin\gamma}{2\gamma^2}\right)\partial_u\phi\partial_u\lin\phi+\frac{r^2}{2\gamma}\big(\partial_u\lin\phi\cdot\partial_u\lin\phi+2\partial_u\phi\partial_u\qd\phi\big).
    \end{split}
\end{equation*}

\subsection{Asymptotic flatness}\label{sec:AF}

\subsubsection{Definitions and ``norms''}\label{sec:AF-defns}

Let $u_0$ and $u_1$ be finite retarded times with $u_0<u_1$. We consider in the following the bifurcate null hypersurface $\mathcal C=\mathcal C(u_0,u_1,1,v_1)$ with $1<v_1\le\infty$ and its domain of influence $\mathcal R=\mathcal R(u_0,u_1,1,v_1)$. Note that we have fixed $v_0=1$, but only later will we fix $u_0=1$. 

\begin{defn}\label{def:AF-data} A $C^2$ characteristic data set $\Psi_\#$ on $\mathcal C$ is said to be \emph{asymptotically flat} if $v_1=\infty$, $\nu<0$ and $\lambda>0$ on $C_\out$, and if the quantity\footnote{Note that $\Psi_\#\mapsto\|\Psi_\#\|_{\mathcal C,1}$ is not a norm (and $\Psi_\#$ is anyway not an element of a vector space), but it is convenient to use the norm notation for the measure of the size of $\Psi_\#$. A similar comment applies to the quantity measuring the size of a solution on $\mathcal R$. However, the seed data $\Psi$ \emph{is} an element of a linear space, and $\Psi\mapsto\|\Psi\|_{\mathcal C}$, defined below, \emph{is} a norm.}
\begin{multline*}
       \|\Psi_\#\|_{\mathcal C} \doteq |Q|+\sup_{\mathcal C}\big(|{\log (r/v)}|+|\varpi|+|\psi|\big)+\sup_{\underline C{}_\ing}\big(|{\log(-\nu)}|+|\partial_u\nu|+|\partial_u^2\nu|+|\partial_u\phi|+|\partial_u^2\phi|\big) \\+ \sup_{C_\out}\big(|{\log\lambda}|+|r^2\partial_v\lambda|+|r^2\partial_v^2\lambda|+|{\log\kappa}|+|r^2\partial_v\kappa|+|r^2\partial_v^2\kappa|+|r^2\partial_v\psi|+|r^2\partial_v^2\psi|\big)
\end{multline*}
is finite. \index{Psi3@$\lVert\Psi_\#\rVert_\mathcal C$, $r$-weighted characteristic data norm}
\end{defn}

\begin{rk}
    Strictly speaking, $\Psi_\#=(r_\circ,\Omega^2_\circ,\phi_\circ,Q)$ on its own does not contain all of the information necessary to compute $\|\Psi_\#\|_{\mathcal C}$ immediately. One first has to solve the Einstein equations along $\mathcal C$ to recover $\varpi|_\mathcal{C}$, and then all of the quantities can be estimated. To see this in action, we refer the reader already to \cref{sec:AF-seed-data}.
\end{rk}

Next, we define what it means for a \emph{semiglobal} solution (i.e., one that exists on the full rectangle $\mathcal R$ with $v_1=\infty$) to be asymptotically flat. 

\begin{defn} A solution $\mathcal S=(\mathcal R,r,\Omega^2,\phi,Q)$ on $\mathcal R$ is said to be \emph{asymptotically flat} if $v_1=\infty$, $\nu<0$ and $\lambda>0$ on $\mathcal R$, and if the quantity
\begin{multline*}
   \|\mathcal S\|_{\mathcal R} \doteq |Q|+  \sup_{\mathcal R}\big(|{\log (r/v)}|+|\varpi|+|{\log(-\nu)}|+|\partial_u\nu|+|\partial_u^2\nu|+|{\log\lambda}|+|r^2\partial_v\lambda|+|r^2\partial_v^2\lambda|+|{\log(-\gamma)}|+|\partial_u\gamma|\\+|\partial_u^2\gamma|+|{\log\kappa}|+|r^2\partial_v\kappa|+|r^2\partial_v^2\kappa|+|\psi|+|r\partial_u\phi|+|r\partial_u^2\phi|+|r^2\partial_v\psi|+|r^2\partial_v^2\psi|\big)
\end{multline*}
    is finite. \index{S7@$\lVert\mathcal S\rVert_{\mathcal R}$, $r$-weighted solution norm}
\end{defn}

In the context of this definition, we only use $\|\mathcal S\|_\mathcal{R}$ with $v_1=\infty$, but it will be useful to consider it with $v_1$ finite for certain a priori estimates. 

\begin{rk}\label{rk:phi-from-norm} By differentiating the relation $\psi=r\phi$, we obtain 
    \begin{equation*}
      \sup_{\mathcal R}\big(  |r^2\partial_v\phi|+|r^2\partial_v^2\phi|+|\partial_u\psi|+|\partial_u^2\psi|\big)\les \|\mathcal S\|_{\mathcal R}
    \end{equation*} for any solution where the right-hand side is finite. Here the implicit constant is purely numerical, i.e., does not depend on $u_0$, $u_1$, $v_1$, or $\mathcal S$. Moreover, using the evolution equations, we obtain also
\begin{multline*}
    \sup_{\mathcal R}\big(|\partial_u\varpi|+|\partial_u^2\varpi|+|r^2\partial_v\varpi|+|r^2\partial_v^2\varpi|+|r^2\partial_u\partial_v\varpi|+|\partial_u\lambda|+|r^2\partial_u^2\lambda|+|r^3\partial_u\partial_v\lambda|+|r^2\partial_v\nu|\\+|r^3\partial_v^2\nu|+|r^2\partial_u\partial_v\nu|+|r^3\partial_v\gamma|+|r^3\partial_v^2\gamma|+|r^3\partial_u\partial_v\gamma|+|r\partial_u\kappa|+|r\partial_u^2\kappa|+|r^2\partial_u\partial_v\kappa|\\+|r^2\varkappa|+|r^2\partial_u\varkappa|+|r^2\partial_u^2\varkappa|+|r^3\partial_v\varkappa|+|r^4\partial_v^2\varkappa|+|r^{3}\partial_u\partial_v\varkappa|\big)\les \|\mathcal S\|_{\mathcal R}.
\end{multline*}
\end{rk}

\begin{rk}
    If the solution $\mathcal S$ arises from characteristic data $\Psi_\#$, then $ \|\mathcal S\|_{\mathcal R}\ge \|\Psi_\#\|_{\mathcal C}$.  
\end{rk}

We define also the weighted distance between two solutions $\mathcal S_i=(\mathcal R,r_i,\Omega^2_i,\phi_i,Q_i)$ ($i=1,2$) on the same rectangle $\mathcal R$ by
\begin{equation*}
    \|\mathcal S_1-\mathcal S_2\|_{\mathcal R}\doteq |Q_1-Q_2|+\sup_\mathcal R\big(|{\log(r_1/r_2)}|+ |\varpi_1-\varpi_2|+\cdots\big),
\end{equation*}
where $\cdots$ runs over all terms in the definition of $\|\mathcal S_i\|_\mathcal{R}$.

\subsubsection{The moduli space of seed data \texorpdfstring{$\mathfrak M$}{M}}\label{sec:AF-seed-data}

Let $\mathcal C$ be a semiglobal bifurcate null hypersurface $(v_1=\infty)$. We define the Banach space \index{F@$\mathfrak F$, space of scalar field initial data}  \index{phi2@$\lVert\phi_\circ\rVert_\mathfrak{F}$, initial scalar field norm}
\begin{equation*}
    \mathfrak F(\mathcal C)\doteq \{\phi_\circ\in C^2(\mathcal C):\|\phi_\circ\|_{\mathfrak F(\mathcal C)}<\infty\},
\end{equation*}
where the norm is defined by\index{F@$\lVert\cdot\rVert_\mathfrak{F(\mathcal C)}$, weighted $C^2$ norm on $\mathfrak{F}$} \begin{equation}\label{eq:norm-defn}
    \|\phi_\circ\|_\mathfrak{F(\mathcal C)}\doteq \sup_{C_\out}\big(|v\phi_{\circ,\out}|+|v^2\partial_v(v\phi_{\circ,\out})|+|v^2\partial_v^2(v\phi_{\circ,\out})|\big) + \sup_{\u C{}_\ing}\big(|\phi_{\circ,\ing}|+|\partial_u\phi_{\circ,\ing}|+|\partial_u^2\phi_{\circ,\ing}|\big).
\end{equation}
Next, we define the Banach space \index{X@$\mathfrak X$, ``horizontal'' Banach space} \index{x0@$x$, element of $\mathfrak X$} \index{x2@$\lVert x\rVert_\mathfrak{X}$, norm on $\mathfrak X$}
\begin{equation*}
    \mathfrak X(\mathcal C)\doteq \mathfrak F(\mathcal C)\times\Bbb R^2,
\end{equation*} 
with norm
\begin{equation*}
 \|x\|_{\mathfrak X(\mathcal C)}\doteq \|\phi_\circ\|_{\mathfrak F(\mathcal C)}+|r_\circ|+|\varpi_\circ|,
\end{equation*} where $x=(\phi_\circ,r_\circ,\varpi_\circ)$.
Finally, we define the Banach space \index{Z@$\mathfrak Z$, Banach space on which $\mathfrak M$ is modeled}
\begin{equation*}
    \mathfrak Z(\mathcal C)\doteq \mathfrak X(\mathcal C)\times\Bbb R,
\end{equation*}
with norm
\begin{equation*}
    \|\Psi\|_{\mathfrak Z(\mathcal C)}\doteq \|x\|_{\mathfrak X(\mathcal C)}+|\rho_\circ|,
\end{equation*}
where $\Psi=(x,\rho_\circ)=(\phi_\circ,r_\circ,\varpi_\circ,\rho_\circ)$. The space $\mathfrak X(\mathcal C)$ will not be used until \cref{sec:statements}.

 Elements of $\mathfrak Z(\mathcal C)$ give rise to gauge-normalized seed data in the sense of \cref{sec:seed-data}  when $r_\circ>u_1-u_0$ and $1-\mu_\circ>0$. By \cref{prop:seed-data-generation}, to each $\Psi\in \mathfrak Z(\mathcal C)$ satisfying these conditions, we can associate a genuine characteristic data set $\Psi_\#$ on $\mathcal C$. \index{M@$\mathfrak M$, the moduli space}
\begin{defn}\label{defn:mod-space} The \emph{moduli space of seed data}\index{M@$\mathfrak M$, the moduli space} on $\mathcal C$ is given by
    \begin{equation*}
        \mathfrak M(\mathcal C)\doteq \{\Psi\in \mathfrak Z(\mathcal C):r_\circ>u_1-u_0, \mu_\circ<1, \text{and }\|\Psi_\#\|_\mathcal{C}<\infty\}.
    \end{equation*}
 We endow $\mathfrak M(\mathcal C)$ with the subspace topology in $\mathfrak Z(\mathcal C)$ and define additionally 
    \begin{equation*}
\|\Psi\|_\mathcal{C}\doteq \|\Psi\|_\mathfrak{Z(\mathcal C)} + \frac{1}{1-\mu_\circ} + \frac{1}{r_\circ - (u_1-u_0)}.
\end{equation*}
\end{defn}

The following lemma gives a simple criterion for membership in $\mathfrak M(\mathcal C)$.

\begin{lem}\label{lem:AF-data-generation} For any $C\ge 1$ and $L>0$, there exist constants $\ve>0$ and $B\ge 1$ with the following property. Let $\Psi=(\phi_\circ, r_\circ,\varpi_\circ,\rho_\circ)\in \mathfrak Z$ on a semiglobal $\mathcal C$ with $u_1-u_0\le L$. If $\|(0,r_\circ,\varpi_\circ,\rho_\circ)\|_\mathcal{C}\le C$ and $\|\phi_\circ\|_\mathfrak{F}\le\ve$, then $\Psi\in\mathfrak M(\mathcal C)$ and it holds that
\begin{equation*}
    \|\Psi_\#\|_{\mathcal C}\le B.
\end{equation*}
\end{lem}
\begin{proof} In this proof, we allow the implicit constant in $\les$ to depend on $C$ and $L$. By \cref{prop:seed-data-generation}, we obtain a characteristic data set $\Psi_\#$ on all of $\mathcal C$. In order to estimate $\|\Psi_\#\|_\mathcal{C}$, we need to show that $\inf_{C_\out}(1-\mu)>0$ on $C_\out$ and that $\varpi$ is bounded on $\mathcal C$. Since $\lambda=1$ on $C_\out$ and $\nu=-1$ on $\u C{}_\ing$, we have 
    \begin{align*}
        \partial_v\varpi &= -r(\partial_v\phi)^2\varpi + \tfrac 12(r^2+Q^2)(\partial_v\phi)^2 \quad\text{on }C_\out,\\
        \partial_u\varpi &= r(\partial_u\phi)^2\varpi - \tfrac 12(r^2+Q^2)(\partial_u\phi)^2\quad\text{on }\u C{}_\ing.
    \end{align*}
 Using an integrating factor, we solve this for
    \begin{align}\label{eq:varpi-data-out}
        \varpi(u_0,v) &= \exp\left(-\int_{v_0}^vr(\partial_v\phi)^2\,dv'\right)\varpi_\circ + \int_{v_0}^v \exp\left(-\int_{v'}^vr(\partial_v\phi)^2\,dv''\right)\tfrac 12(r^2+Q^2)(\partial_v\phi)^2\,dv',\\
        \label{eq:varpi-data-in}\varpi(u,v_0)&=\exp\left(\int_{u_0}^ur(\partial_u\phi)^2\,du'\right)\varpi_\circ - \int_{u_0}^u \exp\left(\int_{u'}^ur(\partial_u\phi)^2\,du''\right)\tfrac 12(r^2+Q^2)(\partial_u\phi)^2\,du'
    \end{align} on $\mathcal C$. Therefore, for $\ve$ sufficiently small, $|\varpi-\varpi_\circ|\les \ve^2$. In particular, $1-\mu$ is bounded below by a positive constant on $C_\out$ since $1-\mu_\circ>0$. Moreover, this allows us to bound $\kappa=(1-\mu)^{-1}$ on $C_\out$. By taking derivatives, we also bound $r^2\partial_v\kappa$ and $r^2\partial_v^2\kappa$. Finally, $\Psi_\#$ has $\nu<0$ on $C_\out$ since $\partial_v|\gamma|>0$, so $\gamma<0$ on $C_\out$ and hence $\nu = (1-\mu)\gamma<0$.
\end{proof}

From the proof of the lemma, we obtain also the following fundamental result.

\begin{prop} \label{prop:MopeninZ}
For any semiglobal bifurcate null hypersurface $\mathcal C$,    $\mathfrak M(\mathcal C)$ is open in $\mathfrak Z(\mathcal C)$.
\end{prop}
\begin{proof}
    Let $\Psi\in\mathfrak M(\mathcal C)$. We show that if $\Psi'\in \mathfrak Z(\mathcal C)$ with $\eta\doteq\|\Psi'-\Psi\|_{\mathfrak Z(\mathcal C)}$ sufficiently small, then $\Psi'\in\mathfrak M(\mathcal C)$. It suffices to show that $1-\mu'$ is bounded below by a positive constant on $C_\out$, where $\mu'$ is computed relative to $\Psi_\#'$. By inspection of \eqref{eq:varpi-data-out} and the simple observation that $|r-r'|\le |r_\circ-r_\circ'|$, we have $|\mu-\mu'|\le C(\|\Psi\|_\mathcal{C})\eta$ on $C_\out$ for $\eta$ sufficiently small, which proves the claim.
\end{proof}

\begin{rk}
    The norm $\|\phi_\circ\|_{\mathfrak F(\mathcal C)}$ actually allows for the estimate $|\partial_v^2\psi|\les r^{-3}\|\phi_\circ\|_{\mathfrak F(\mathcal C)}$ on $C_\out$, which is a full power of $r$ stronger than what is assumed in the norm $\|\Psi_\#\|_\mathcal{C}$. In general, we have made no effort to track  the optimal $r$-weights in \cref{sec:semiglobal-1}. The power $v^3$ in  $\|\phi_\circ\|_{\mathfrak F(\mathcal C)}$ is actually only ever used once in this paper, namely in \cref{lem:lin-phi-data} below.
\end{rk}

\subsection{Local existence of semiglobal solutions}\label{sec:local-in-u}

\subsubsection{The fundamental soft estimate}

The following proposition captures the fact that the spherically symmetric Einstein--Maxwell-scalar field system is ``subcritical'' away from $r=0$ and can be completely controlled when $1-\mu = 1-\frac{2m}{r}$ is bounded below. If a black hole forms, then $1-\mu$ degenerates (as we will show), but the following a priori estimate is very useful in our bootstrap argument. These calculations are well known in the literature, see for instance \cite[Proposition 5.1]{Price-law}.

\begin{prop}\label{prop:fundamental-AF} For any $A\ge 1$, $c>0$, and $L>0$ there exists a constant $B(A,c,L)\ge 1$ with the following property. Let $\mathcal S$ be a solution of the spherically symmetric Einstein--Maxwell-scalar field system defined on the double null rectangle $\mathcal R=\mathcal R(u_0,u_1,v_0,v_1)$, with $u_1-u_0\le L$ and satisfying $\|\Psi_\#\|_\mathcal{C}\le A$ and 
\begin{equation}\label{eq:1-mu-assn}
    \inf_{\mathcal R}(1-\mu) \ge c.
\end{equation}
Then it holds that
\begin{equation*}
    \|\mathcal S\|_\mathcal{R}\le B. 
\end{equation*}
\end{prop}
\begin{proof} In this proof, we allow the implicit constant in $\les$ to depend on $A$, $c$, and $L$.  From the definitions \eqref{eq:greek-letters} and the assumptions $\Omega^2>0$, $\nu<0$, and $1-\mu>0$, we see that $\lambda>0$ on $\mathcal R$. It follows from \eqref{eq:m-u} that $\partial_um\le 0$ and $\partial_vm\ge 0$ and hence
    \begin{equation}\label{eq:mass-monotonicity}
        m(u_1,v_0)\le m(u,v)\le m(u_0,v_1).
    \end{equation}
     Therefore, $r\gtrsim 1$ and $1-\mu\les 1$. Using \eqref{eq:m-u}, \eqref{eq:1-mu-assn}, and \eqref{eq:mass-monotonicity}, we estimate
\begin{equation}\label{eq:u-energy-soft}
    \sup_{v_0\le v \le v_1}\int_{u_0}^{u_1} \frac{r^2}{-\nu}(\partial_u\phi)^2\,du\les   \sup_{v_0\le v \le v_1}\int_{u_0}^{u_1} \frac{r^2}{-2\gamma}(\partial_u\phi)^2\,du\le |m(u_0,v_1)-m(u_1,v_0)|\les 1.
\end{equation}
Using the lower bound on $r$, we then have
\begin{equation}\label{eq:kappa-soft}
    |{\log\kappa(u,v)}|\le  |{\log\kappa(u_0,v)}|+\int_{[u_0,u]\times\{v\}}\frac{r}{-\nu}(\partial_u\phi)^2\,du'\les 1
\end{equation}
on $\mathcal R$. Using \eqref{eq:m-u} and \eqref{eq:mass-monotonicity} again, it follows that $
    \lambda = \kappa(1-\mu) \sim 1$ and 
\begin{equation}\label{eq:v-energy-soft}
      \sup_{u_0\le u\le u_1}\int_{v_0}^{v_1} r^2(\partial_v\phi)^2\,dv\les   \sup_{u_0\le u\le u_1}\int_{v_0}^{v_1} \frac{r^2}{2\kappa}(\partial_v\phi)^2\,dv \les 1.
\end{equation}

Integrating \eqref{eq:nu-v}, we have
    \begin{equation}\label{eq:log-nu}
 |      { \log(-\nu(u,v))} |\le |{ \log(-\nu(u,1))}|+\int_{\{u\}\times[v_0,v]} \kappa|\varkappa|\,dv.
    \end{equation}
    Since $|\varkappa|\les r^{-2}$ and $\lambda\sim 1$, this gives $|{\log(-\nu)}|\les 1$ on $\mathcal R.$ Arguing as in \eqref{eq:kappa-soft}, using now \eqref{eq:v-energy-soft}, we have
    \begin{equation*}
        |{\log(-\gamma(u,v))}| \le |{\log(-\gamma(u,v_0))}|+ \int_{\{u\}\times[v_0,v]} \frac{r}{\lambda}(\partial_v\phi)^2\,dv'\les 1.
    \end{equation*}

We now estimate $\phi$ and its first derivatives pointwise. Using the fundamental theorem of calculus, we estimate \begin{equation*}
    |\phi(u,v)|\le |\phi(u_0,v)|+\int_{u_0}^u|\partial_u\phi|\,du'\les r(u_0,v)^{-1} + \left(\int_{u_0}^u \frac{r^2}{-\nu}(\partial_u\phi)^2\,du'\right)^{1/2}\left(\int_{u_0}^u\frac{-\nu}{r^2}\,du'\right)^{1/2} \les r(u,v)^{-1}
\end{equation*}
which shows $|\psi|\les 1$ on $\mathcal R$.
 Plugging this back into \eqref{eq:wave-equation-psi} and integrating in $u$ and $v$ respectively, we obtain $|\partial_v\psi|\les r^{-2}$ and $|\partial_u\psi|\les 1$. By differentiating $\phi = r^{-1}\psi$, we obtain $|r^2\partial_v\phi|\les 1$ and $|r\partial_u\phi|\les 1$.

    We now proceed to bound the higher order quantities in $\|\mathcal S\|_{\mathcal R}$. To estimate $\partial_u\gamma$, we differentiate \eqref{eq:kappa-u} in $u$ and use \eqref{eq:phi-wave-1} and \eqref{eq:nu-v} to estimate
    \begin{equation}\label{eq:partial-u-gamma-1}
        |\partial_v\partial_u\gamma| = \left|\frac{\nu \gamma}{\lambda}(\partial_v\phi)^2+\frac{r\partial_u \gamma}{\lambda}(\partial_v\phi)^2-\frac{r\gamma}{\lambda^2}\kappa\nu\varkappa(\partial_v\phi)^2-\frac{r\gamma}{\lambda}2\partial_v\phi\left(\frac{\nu\partial_v\phi}{r}+\frac{\lambda\partial_u\phi}{r}\right)\right|\les r^{-3}\big(1+|\partial_u\gamma|\big).
    \end{equation}
    We then obtain $|\partial_u\gamma|\les 1$ by Gr\"onwall's inequality. Similarly, we estimate 
    \begin{equation*}
        |\partial_v\partial_u\nu|\les r^{-2}\big(1+ |\partial_u\nu|\big),\quad
 |\partial_u\partial_v\lambda|\les r^{-2}(1+|\partial_v\lambda|),\quad
             |\partial_u\partial_v\kappa|\les r^{-2}(1+|\partial_v\kappa|),
    \end{equation*}
  from which we infer $   |\partial_u\nu|+|r^2\partial_v\lambda|+|r^2\partial_v\kappa|\les 1$.

    We may also differentiate \eqref{eq:wave-equation-psi} in $v$ to obtain $|\partial_u\partial_v^2\psi|\les r^{-4}$, whence $|\partial_v^2\psi|\les r^{-2}$ by integration in $u$. We differentiate  \eqref{eq:wave-equation-psi} in $u$ to obtain $|\partial_v\partial_u^2\psi|\les r^{-3}$, whence $|\partial_u^2\psi|\les 1$ by integration in $u$. By differentiating $\phi= r^{-1}\psi$ again, we estimate $|r\partial_u^2\phi|+|r^2\partial_v^2\phi|\les 1$.

    We can now proceed to differentiate the evolution equations for $\partial_u\gamma$, $\partial_u\nu$, $\partial_v\lambda$, and $\partial_v\kappa$ once more, obtaining for instance 
    \begin{equation}\label{eq:partial-u-gamma-2}
           |\partial_v\partial_u^2\gamma| \les r^{-3}\big(1+|\partial_u^2\gamma|\big)
    \end{equation}
    after a tedious calculation. Altogether we obtain $   |\partial_u^2\nu|+|r^2\partial_v^2\lambda|+|\partial_u^2\gamma|+|r^2\partial_v^2\kappa|\les 1$, which completes the proof of the proposition. \end{proof}

Sending $v_1\to\infty$, we have:

\begin{cor}\label{cor:fundamental-AF}
    Let $A\ge 1$, $c>0$, and $L>0$. Let $\mathcal S$ be a solution of the spherically symmetric Einstein--Maxwell-scalar field system defined on the semiglobal double null rectangle $\mathcal R(u_0,u_1,v_0,\infty)$, with $u_1-u_0\le L$ and satisfying $\|\Psi_\#\|_\mathcal{C}\le A$ and 
$    \inf_{\mathcal R}(1-\mu) \ge c.$ Then $\mathcal S$ is asymptotically flat on $\mathcal R$ and it holds that $
    \|\mathcal S\|_\mathcal{R}\le B(A,c,L). $
\end{cor}

 Using \cref{prop:fundamental-AF}, we obtain the following standard breakdown criterion (stated in the contrapositive) for the spherically symmetric EMSF system \cite{TheBVPaper,dafermos2005naked}:

\begin{cor}[Weak extension principle]\label{cor:weak-extension}
    Let $\mathcal S$ be a solution of the spherically symmetric Einstein--Maxwell-scalar field system defined over an open set $\mathcal Q\subset \Bbb R^2_{u,v}$. Suppose that the following hold:
    \begin{enumerate}
        \item $\mathcal R'\doteq \mathcal R(u_0,u_1,v_0,v_1)\setminus\{(u_1,v_1)\}\subset\mathcal Q$.
        \item $\nu<0$ on $\mathcal Q$.
        \item $\lambda\ge 0$ on $\mathcal R'$.
    \end{enumerate}
    Then $\mathcal S$ extends to a neighborhood of $(u_1,v_1)\in\overline{\mathcal Q}$.
\end{cor}

\subsubsection{Local existence in an outgoing slab}

We now show that any asymptotically flat characteristic data gives rise to a solution which exists in a thin characteristic slab going to null infinity $\mathcal I^+$.  

\begin{prop}[Local existence in an outgoing slab]\label{prop:slab-1} 
    For any $A\ge 1$, there exist constants $B_\loc(A)\ge 1$ and $\eta_\loc(A)>0$ such that if $\Psi_\#$ is an asymptotically flat characteristic initial data set on $\mathcal C=\mathcal C(u_0,u_1,1,\infty)$ satisfying $\|\Psi_\#\|_{\mathcal C}\le A$ and $u_1-u_0\le \eta_\mathrm{loc}$, then the solution $\mathcal S[\Psi_\#]$ with initial data $\Psi_\#$ exists on $\mathcal R= \mathcal R(u_0,u_1,1,\infty)$ and satisfies
    \begin{equation}
        \|\mathcal S[\Psi_\#]\|_{\mathcal R}\le B_\loc.\label{eq:semiglobal-S-1}
    \end{equation}
\end{prop}

\begin{proof}
By assumption, there exists a constant $c>0$ such that
\begin{equation}\label{eq:1-mu-data-assn}
    \inf_{\{u_0\}\times [1,\infty) }(1-\mu) = c.
\end{equation}
We argue by continuity in the domains $\mathcal R(v_*)=\mathcal R\cap\{v\le v_*\}$, where $v_*\ge 1$. Assuming the solution exists on $\mathcal R(v_*)$, we make there the bootstrap assumption
\begin{equation}
    1-\mu \ge \tfrac 13 c.\label{eq:1-mu-bootstrap}
\end{equation}
By \cref{prop:fundamental-AF}, it holds that $\|\mathcal S\|_{\mathcal R(v_*)}\le B'\doteq B(A,\tfrac 13 c,\eta)$ and therefore $|\partial_u(1-\mu)|\le C(B')$. Integrating this identity from $u=u_0$, using \eqref{eq:1-mu-data-assn}, and choosing $\eta$ small depending on $c$ and $B'$, we improve the constant on the right-hand side of \eqref{eq:1-mu-bootstrap} to $\frac 23$. Therefore, by a standard continuity argument and \cref{prop:slab-existence}, we can extend the solution to $\mathcal R(v_*+\ve)$ for some $\ve>0$, while still satisfying \eqref{eq:1-mu-bootstrap}. Therefore, we may take $v_*=\infty$ and conclude the proof with $B_\loc = B'$. 
\end{proof}

    \subsection{Global structure of asymptotically flat solutions}\label{sec:global-structure}

With a basic semiglobal existence theory at hand, we can now consider the \emph{maximal} region on which a development of asymptotically flat data remains asymptotically flat in the sense of \cref{sec:AF-defns}. This leads naturally to the notion of the \emph{Bondi lifetime} of the solution and \emph{null infinity} $\mathcal I^+$, which we define in \cref{sec:lifetime}. We then recall and prove Dafermos' theorem \cite{dafermos-trapped-surface}, \cref{thm:dichotomy}, that the maximal development either contains a black hole or remains regular but fails to collapse in \cref{sec:dichotomy}.\footnote{The proof we give here applies to any spherically symmetric Einstein-matter system for which the weak extension principle and dominant energy condition hold, as in \cite{dafermos-trapped-surface}.} For more details, especially about how $\mathcal I^+$ fits into the study of the entire boundary of $\mathcal S_\mathrm{max}$, we refer to Kommemi \cite{Kommemi13}. In 
\cref{sec:Bondi-mass}, we define the \emph{Bondi mass} function of $\mathcal I^+$ and prove some basic properties. Finally, in \cref{sec:superextremal}, we use ideas from the proof of Dafermos' theorem to show that \emph{superextremal solutions are dispersive} in the spherically symmetric EMSF model.

\subsubsection{The Bondi lifetime and null infinity \texorpdfstring{$\mathcal I^+$}{I+}}\label{sec:lifetime}

 Let $\mathcal S_\mathrm{max}=(\mathcal Q_\mathrm{max},r,\Omega^2,\phi,Q)$ be the maximal development of asymptotically flat characteristic data on the semiglobal bifurcate null hypersurface $\mathcal C=\mathcal C(1,u_1,1,\infty)$.\footnote{We now fix $u_0=1$.} Then 
    \begin{equation*}
        u_\Box\doteq \sup\{u_*\in[1,u_1]:\mathcal R(1,u_*,1,\infty)\subset\mathcal Q_\mathrm{max}\text{ and $\mathcal S_\mathrm{max}$ is asymptotically flat on $\mathcal R(1,u_*,1,\infty)$}\}
    \end{equation*}
    defines a number in $(1,u_1]$ by \cref{prop:slab-1}. We immediately note:
    \begin{lem}\label{lem:u-box-basic}
    For any $\mathcal S_\mathrm{max}$ arising from asymptotically flat data,  $\mathcal R=\mathcal R(u_0,u_\Box,1,\infty)\subset\mathcal Q_\mathrm{max}$, we have $\lambda\ge 0$ on $\mathcal R$.
\end{lem}
\begin{proof}  By definition of $u_\Box$, $\lambda>0$ on $[u_0,u_\Box)\times[1,\infty)$. Using the weak extension principle \cref{cor:weak-extension}, we see that $\mathcal R=\mathcal R(u_0,u_\Box,1,\infty)\subset\mathcal Q_\mathrm{max}$ and $\lambda\ge 0$ on $\mathcal R$. 
\end{proof}

\begin{defn}\label{def:lifetime}
The \emph{Bondi lifetime} of $\mathcal S_\mathrm{max}$ is defined to be \index{L@$\mathcal L_\mathrm{Bondi}$, Bondi lifetime}
\begin{equation}\label{eq:lifetime}
    \mathcal L_\mathrm{Bondi}\doteq -\lim_{v\to\infty}\int_{1}^{u_\Box}\gamma(u',v)\,du'.
\end{equation}
\end{defn}

\begin{lem} The limit in \eqref{eq:lifetime} either exists and defines a positive real number or diverges to $+\infty$. Moreover, $\mathcal L_\mathrm{Bondi}$ is invariant under gauge transformations of the $(u,v)$ coordinates as defined in \cref{sec:double-null-gauge}.
\end{lem}
\begin{proof} For any $v<\infty$, $\gamma$ takes values in $\Bbb R_{<0}$ on $[1,u_\Box)\times\{v\}$ by definition of $u_\Box$. Therefore, $\int_{1}^{u_\Box}\gamma(u',v)\,du'\in \Bbb R_{<0}\cup\{-\infty\}$ is well-defined. Now recall the monotonicity $\partial_v\gamma\le 0$, which gives the desired property of the limit as $v\to\infty$. Note further that each of the integrals $\int_{1}^{u_\Box}\gamma(u',v)\,du'$ are gauge invariant by the change of variables formula and \eqref{eq:COV-3}. 
\end{proof}

\begin{defn} \emph{Future null infinity}\index{I@$\mathcal I^+$, future null infinity} of $\mathcal S_\mathrm{max}$ is the subset $\mathcal I^+\subset \overline{\Bbb R}{}^2$ given by
    \begin{equation*}
        \mathcal I^+\doteq \begin{cases}
          [1,u_\Box]\times\{v=\infty\}  & \text{if }\mathcal L_\mathrm{Bondi}<\infty \\
           [1,u_\Box)\times\{v=\infty\} & \text{if }\mathcal L_\mathrm{Bondi}=\infty \\
        \end{cases},
    \end{equation*}
    which is to be thought of as a subset of the boundary of $\mathcal Q_\mathrm{max}$ in $ \overline{\Bbb R}{}^2$.
\end{defn}

\subsubsection{The fundamental dichotomy}\label{sec:dichotomy}

We now prove the fundamental dichotomy. This result is not specific to the spherically symmetric EMSF system as it relies only on \cref{prop:fundamental-AF} (which is true for many other matter models in spherical symmetry), but it is very important that $r(u_1,1)>0$ in the definition of the moduli space $\mathfrak M$. 

\begin{thm}[Dafermos] \label{thm:dichotomy}
Let $\mathcal C=\mathcal C(1,u_1,1,\infty)$ be a semiglobal bifurcate null hypersurface with moduli space of asymptotically flat seed data $\mathfrak M=\mathfrak M(\mathcal C)$. Let \index{M@$\mathfrak M_\mathrm{black}$, data leading to the formation of a black hole} \index{M@$\mathfrak M_\mathrm{non}$, noncollapsing data}
  \begin{align*}
       \mathfrak M_\mathrm{black}&\doteq \{\Psi\in \mathfrak M:r(u_\Box,\cdot)\emph{ is bounded}\},
       \\
       \mathfrak M_\mathrm{non}&\doteq \{\Psi\in \mathfrak M:r(u_\Box,\cdot)\emph{ is unbounded}\}.
  \end{align*}
  Then $\mathfrak M=\mathfrak M_\mathrm{black}\sqcup\mathfrak M_\mathrm{non}$ and
  \begin{align*}
       \mathfrak M_\mathrm{black}= \{\Psi\in \mathfrak M:\mathcal L_\mathrm{Bondi}=\infty\},\\
       \mathfrak M_\mathrm{non}= \{\Psi\in \mathfrak M:\mathcal L_\mathrm{Bondi}<\infty\}.
  \end{align*}
  Moreover, if $\Psi\in \mathfrak M_\mathrm{non}$, then $u_\Box = u_1$ and $\mathcal S_\mathrm{max}[\Psi]$ is asymptotically flat on $\mathcal R(u_0,u_1,1,\infty)$.
\end{thm}

\begin{rk} When $\Psi\in \mathfrak M_\mathrm{black}$, we write $u_{\mathcal H^+}\doteq u_\Box$ and
   $  \mathcal Q_\mathrm{max}\times S^2\cap\{u\ge u_{\mathcal H^+}\}= \mathcal Q_\mathrm{max}\times S^2\setminus J^-(\mathcal I^+\times S^2)$, where $J^-$ is taken with respect to the $(3+1)$-dimensional metric $g=g_\mathcal{Q}+r^2g_{S^2}$, is a \emph{black hole} region. The cone $\{u=u_{\mathcal H^+}\}$, the boundary of the black hole, is called the \emph{event horizon} $\mathcal H^+$. We define \index{H6@$\mathcal H^+$, event horizon} \index{uhorizon@$u_{\mathcal H^+}$, location of the event horizon} \index{r@$r_{\mathcal H^+}$, final horizon area radius} \index{pi2@$\varpi_{\mathcal H^+}$, final horizon renormalized mass}
   \begin{equation}\label{eq:rH+-defn}
       r_{\mathcal H^+}\doteq \sup_{\mathcal H^+} r= \lim_{v\to\infty} r(u_{\mathcal H^+},v),\quad \varpi_{\mathcal H^+}\doteq \sup_{\mathcal H^+} \varpi= \lim_{v\to\infty} \varpi(u_{\mathcal H^+},v).
   \end{equation}
   These limits exist by monotonicity: $\partial_vr\ge 0$ and $\partial_v\varpi\ge 0$ on $\mathcal H^+$.
         \end{rk}
\begin{rk}
The theorem implies that $\mathcal I^+$ has the alternative (more standard) characterization as
\begin{equation}\label{eq:I+-alt}
    \mathcal I^+ =\big\{(u,\infty):u\in [u_0,u_1],\lim_{v\to\infty} r(u,v)=\infty\big\}.
\end{equation}
\end{rk}
\begin{rk}
        With slightly more work, one can upgrade the statement $\mathfrak M_\mathrm{black}=\{\mathcal L_\mathrm{Bondi}=\infty\}$ to the statement that $\mathcal I^+$ is \emph{complete} in the sense of Christodoulou \cite{christodoulou1999global}. See \cite{dafermos-trapped-surface} for details. 
\end{rk}

\begin{proof}[Proof of \cref{thm:dichotomy}] Suppose first that $\Psi\in\mathfrak M_\mathrm{black}$. Since $0\le 1-\mu\les 1$ on $\mathcal R=\mathcal R(u_0,u_\Box,1,\infty)$, we have
\begin{equation*}
    \mathcal L_\mathrm{Bondi} =\lim_{v\to\infty}\int_{u_0}^{u_\Box}|\gamma(u',v)|\,du' \gtrsim \limsup_{v\to\infty} \int_{u_0}^{u_\Box}|\nu(u',v)|\,du' = \limsup_{v\to\infty}\big(r(u_0,v)-r(u_\Box,v)\big) = \infty.
\end{equation*}

Next, suppose that $\Psi\in\mathfrak M_\mathrm{non}$. We argue that $\inf_{\mathcal R}(1-\mu)>0$. If not, there exists a sequence $(u_i,v_i)\in\mathcal R$ such that $(1-\mu)(u_i,v_i)\to 0$ as $i\to\infty$. By definition of $\mathfrak M_\mathrm{non}$, $r$ is unbounded along every outgoing cone in $\mathcal R$ and therefore $1-\mu>0$ on $\mathcal R$. Since $m$ is bounded on $\mathcal R$, this implies $r(u_i,v_i)$ is bounded and hence $r(u_\Box,v_i)\le r(u_i,v_i)$ is also bounded. Therefore, $v_i$ must be bounded by definition of $\mathfrak M_\mathrm{non}$. After passing to a subsequence, we may assume $(u_i,v_i)$ converges, which gives a contradiction. By \cref{cor:fundamental-AF}, $\mathcal S_\mathrm{max}$ is now asymptotically flat on $\mathcal R$. Moreover, $u_\Box = u_1$, for otherwise we could extend $\mathcal S$ to an asymptotically flat solution on $\mathcal R(u_0,u_\Box+\eta,1,\infty)$ for some $\eta>0$ by \cref{prop:slab-1}. Finally, $\mathcal L_\mathrm{Bondi}<\infty$,  since $\gamma\sim -1$  by asymptotic flatness. 

We have shown that $\mathfrak M_\mathrm{black}\subset \{\mathcal L_\mathrm{Bondi}=\infty\}$ and $\mathfrak M_\mathrm{non}\subset \{\mathcal L_\mathrm{Bondi}<\infty\}$ and since $\mathfrak M$ is the disjoint union of $ \{\mathcal L_\mathrm{Bondi}=\infty\}$ and $  \{\mathcal L_\mathrm{Bondi}<\infty\}$, we conclude the proof of the theorem.
\end{proof}

The proof shows that if $\Psi\in \mathfrak M_\mathrm{black}$, then 
  \begin{equation}\label{eq:1-mu-exterior}
      \inf_{\mathcal R(1,u_{\mathcal H^+},1,\infty)}(1-\mu)= 0.
  \end{equation}
  Following \cite[Proposition 5.2]{Price-law}, we upgrade this to the following much stronger statement, which uses the structure of the EMSF system in a crucial way.

\begin{prop}\label{prop:1-mu-horizon}
For any $\Psi\in \mathfrak M_\mathrm{black}$,
  \begin{equation}\label{eq:1-mu-horizon}
      \lim_{v\to\infty}(1-\mu)(u_{\mathcal H^+},v)= 0.
  \end{equation}
\end{prop}
\begin{proof}
     We prove \eqref{eq:1-mu-horizon} by contradiction. Suppose that $(1-\mu)(u_{\mathcal H^+},v)\ge 2c$ for all $v\ge 1$ for some $c>0$. Therefore, $1-\frac{2\varpi_{\mathcal H^+}}{r_{\mathcal H^+}}+\frac{Q^2}{r_{\mathcal H^+}^2}\ge 2c$ and there exists an $\eta>0$ such that $1-\frac{2\varpi'}{r'}+\frac{Q^2}{r'^2}\ge c$ for all $|r'-r_{\mathcal H^+}|\le \eta$ and $|\varpi'-\varpi_{\mathcal H^+}|\le \eta$. We now define the set
    \begin{equation*}
        \mathcal N\doteq \{(u,v)\in\mathcal R(u_0,u_{\mathcal H^+},1,\infty):|r(u,v)-r_{\mathcal H^+}|\le\eta,|\varpi(u,v)-\varpi_{\mathcal H^+}|\le\eta\}.
    \end{equation*}
    By construction, $\mathcal H^+\cap\{v\ge v_0\}\subset\mathcal N$ for some $v_0\ge 1$ and
    \begin{equation}\label{eq:1-mu-N-lower}
        \inf_\mathcal N(1-\mu)\ge c.
    \end{equation}
    Note that $\mathcal N$ has the following causality property: if $(u,v)\in\mathcal N$, then $(u,v')\in\mathcal N$ for every $v'\in [v_0,v]$ and $(u',v)\in\mathcal N$ for every $u'\in[u,u_{\mathcal H^+}]$. 

    For $v\ge v_0$, put $\bf u(v) = \inf\{u\in \mathcal R(u_0,u_{\mathcal H^+},1,\infty) :(u,v)\in\mathcal N\}$, which is nondecreasing by the causality properties of $\mathcal N$. By definition and monotonicities, for every $v\ge v_0$, one of the following holds:
    \begin{equation}\label{eq:N-cont-1}
        r(\bf u(v),v) = r_{\mathcal H^+}+\eta\quad\text{or}\quad   \varpi(\bf u(v),v) = \varpi_{\mathcal H^+}+\eta.
    \end{equation} Since $r\to\infty$ along any outgoing cone with $u<u_{\mathcal H^+}$, we see $\bf u(v)\to u_{\mathcal H^+}$ as $v\to\infty$. We now show that 
    \begin{align}
        \label{eq:N-cont-2} \lim_{v\to\infty}r(\bf u(v),v) &= r_{\mathcal H^+},\\
        \label{eq:N-cont-3}\lim_{v\to\infty}\varpi(\bf u(v),v) &= \varpi_{\mathcal H^+},
    \end{align}
    which will contradict \eqref{eq:N-cont-1}. 

    \textsc{Proof of \eqref{eq:N-cont-2}}: By \eqref{eq:1-mu-N-lower} and the boundedness of $r$ and $\varpi$, we have $|\kappa\varkappa|\les \lambda$ in $\mathcal N$. It follows that along any outgoing cone in $\mathcal N$, the integral of $|\kappa\varkappa|$ is uniformly bounded. Therefore, by \eqref{eq:log-nu}, 
    \begin{equation}\label{eq:N-r-diff}
        |r(\bf u(v),v)-r(u_{\mathcal H^+},v)|\le \int_{\bf u(v)}^{u_{\mathcal H^+}} |\nu(u',v)|\,du'\les \int_{\bf u(v)}^{u_{\mathcal H^+}} |\nu(u',v_0)|\,du'\les u_{\mathcal H^+}-\bf u(v).
    \end{equation}
    Sending $v\to\infty$ proves \eqref{eq:N-cont-2}. 

    \textsc{Proof of \eqref{eq:N-cont-3}}: First, we estimate $Y\phi$ in $\mathcal N$, where $Y\doteq \nu^{-1}\partial_u$. By \eqref{eq:phi-wave-1} and \eqref{eq:nu-v}, 
\begin{equation*}
    \partial_v(rY\phi) = -\partial_v\phi - \kappa\varkappa rY\phi. 
\end{equation*}
Using an integrating factor, this is solved as
\begin{equation*}
    rY\phi(u,v) = \exp\left(-\int_{v_0}^v\varkappa\kappa(u,v')\,dv'\right)\left[rY\phi(u,v_0)-\int_{v_0}^v\exp\left(-\int_{v_0}^{v'}\varkappa\kappa(u,v'')\,dv''\right)\partial_v\phi(u,v')\,dv'\right]
\end{equation*}
Using monotonicity of the mass as in \cref{prop:fundamental-AF}, we have
\begin{equation*}
    \int_{v_0}^v|\partial_v\phi(u,v')|\,dv'\le \left( \int_{v_0}^v\kappa^{-1}(\partial_v\phi)^2(u,v')\,dv'\right)^{1/2}\left( \int_{v_0}^v\kappa(u,v')\,dv'\right)^{1/2}\les 1
\end{equation*}
for $(u,v)\in\mathcal N$. It follows that $|Y\phi|\les 1$ in $\mathcal N$. Using now \eqref{eq:varpi-u} in the form $\partial_u\varpi = \frac 12 r^2(1-\mu)\nu (Y\phi)^2$, together with \eqref{eq:N-r-diff}, we find
\begin{equation*}
    |\varpi(\bf u(v),v)-\varpi(u_{\mathcal H^+},v)|\les \int_{\bf u(v)}^{u_{\mathcal H^+}}|\nu| (Y\phi)^2(u',v)\,du' \les \int_{\bf u(v)}^{u_{\mathcal H^+}} |\nu(u',v_0)|\,du'\les u_{\mathcal H^+}-\bf u(v),
\end{equation*}
    which gives \eqref{eq:N-cont-3}. 

    \textsc{Completing the proof}: This argument shows that there exists a sequence $ v_i\to\infty$ such that $(1-\mu)(u_{\mathcal H^+},v_i)\to 0$ as $i\to\infty$. Therefore, 
    \begin{equation}\label{eq:final-parameters}
        1- \frac{2\varpi_{\mathcal H^+}}{r_{\mathcal H^+}}+\frac{Q^2}{r_{\mathcal H^+}^2} = 0
    \end{equation}
    and so again by monotonicities, \eqref{eq:1-mu-horizon} holds along all sequences tending to $\infty$.
\end{proof}

\begin{cor}\label{cor:final-parameters}
    For any $\Psi\in \mathfrak M_\mathrm{black}$, $|Q|\le \varpi_{\mathcal H^+}$ and 
    \begin{equation}\label{eq:final-radius}
        r_{\mathcal H^+} = \varpi_{\mathcal H^+} + \sqrt{\varpi_{\mathcal H^+}^2-Q^2}.
    \end{equation}
\end{cor}
\begin{proof}
    The first claim simply follows from the algebraic properties of \eqref{eq:final-parameters}. To prove \eqref{eq:final-radius} suppose instead that $|Q|<\varpi_{\mathcal H^+}$ and $ r_{\mathcal H^+} = \varpi_{\mathcal H^+} - \sqrt{\varpi_{\mathcal H^+}^2-Q^2}$ (the other option consistent with \eqref{eq:final-parameters}). Since $1-\mu\ge 0$ in $\mathcal R(1,u_{\mathcal H^+},1,\infty)$, $\varpi \le \frac r2(1+\frac{Q^2}{r^2})$. If $r\le r_{\mathcal H^+}+\eta$, then 
    \begin{equation*}
        \varpi \le \frac{r_{\mathcal H^+}+\eta}{2}\left(1+\frac{Q^2}{r_{\mathcal H^+}^2}+O(\eta)\right)\le \varpi_{\mathcal H^+}+O(\eta), 
    \end{equation*}
    where we used that $r_{\mathcal H^+}\le \varpi_{\mathcal H^+}$ and $r_{\mathcal H^+}\le |Q|$. Since
    \begin{equation*}
        \lim_{v\to\infty}\varkappa (u_{\mathcal H^+},v) = - 2r_{\mathcal H^+}^{-2}\sqrt{\varpi_{\mathcal H^+}^2-Q^2}<0,
    \end{equation*}
    it follows that for $\eta$ sufficiently small, $\varkappa\le -c$ for some $c>0$ on the set $\{|r- r_{\mathcal H^+}|\le\eta\}\cap \mathcal R(1,u_{\mathcal H^+},1,\infty)$. By \eqref{eq:nu-v}, $\partial_v\nu \ge 0$ and hence $|\nu|\les 1$ on this set. Arguing as in the proof of \cref{prop:1-mu-horizon}, we now obtain a contradiction by integrating $\nu$ between $r=r_{\mathcal H^+}+\eta$ and $\mathcal H^+$ along ingoing cones with $v\to\infty$. 
\end{proof}

\subsubsection{The Bondi mass function \texorpdfstring{$M_{\mathcal I^+}$}{M I+}}\label{sec:Bondi-mass}

Since $ \overline{\Bbb R}{}^2$ comes with a natural topology, it makes sense to extend certain functions on $\mathcal Q_\mathrm{max}$ to $\mathcal I^+$. For instance:
\begin{lem}\label{lem:mass-monotonicity}
For any $(u,\infty)\in\mathcal I^+$, the limit \index{M@$M_{\mathcal I^+}$, Bondi mass function}
\begin{equation}\label{eq:Bondi-mass}
M_{\mathcal I^+}(u)\doteq \lim_{v\to\infty}m(u,v)
\end{equation}
exists and defines a nonincreasing $C^1$ function on $\mathcal I^+$. If $(u_1,\infty)\in \mathcal I^+$ and $\|\mathcal S\|_{\mathcal R(1,u_1,1,\infty)}\le B$, then 
\begin{equation}\label{eq:mass-loss-bound}
    \sup_{[1,u_1]}|\partial_u M_{\mathcal I^+}|\le C,
\end{equation}
where the constant $C$ depends only on $B$ and $u_1$.
\end{lem}

The function $M_{\mathcal I^+}$ is called the \emph{Bondi mass function} of $\mathcal S_\mathrm{max}$. 

\begin{proof}
Let $(u_1,\infty)\in \mathcal I^+ $. Then, we have by \eqref{eq:m-u} and asymptotic flatness that $|\partial_v m |\les v^{-2}$ and $|\partial_v\partial_um|\les v^{-2}$ uniformly on $\mathcal R(u_0,u_1,1,\infty)$. Hence, $M_{\mathcal I^+}$ is well-defined and $C^1$ on $[u_0,u_1]$, and \eqref{eq:mass-loss-bound} is automatic. Finally, since $\partial_u m\le 0$ by $\gamma,\nu\le 0$, this property is inherited by $M_{\mathcal I^+}$. 
\end{proof}

 Note that by \eqref{eq:I+-alt}, it also holds that
\begin{equation*}
M_{\mathcal I^+}(u)=\lim_{v\to\infty}\varpi(u,v).
\end{equation*}
If $\Psi\in\mathfrak M_\mathrm{black}$, then we define the \emph{final Bondi mass} \index{M@$M_\infty$, final Bondi mass}
\begin{equation*}
M_\infty \doteq \lim_{u\to  u_{\mathcal H^+}}M_{\mathcal I^+}(u),
\end{equation*}
which is well-defined by \cref{lem:mass-monotonicity}. It is convenient to extend the domain of $M_{\mathcal I^+}$ to include $u_{\mathcal H^+}$ and to write $M_{\mathcal I^+}(u_{\mathcal H^+})=M_\infty$. 

\begin{lem}[Penrose inequality]\label{lem:Penrose}
    For any $\Psi\in \mathfrak M_\mathrm{black}$, $r\le 2M_\infty$ on $\mathcal H^+$. 
\end{lem}
\begin{proof}
    By \eqref{eq:final-radius}, $r\le r_{\mathcal H^+}\le 2\varpi_{\mathcal H^+}$. Let $v\ge 1$ and $u<u_{\mathcal H^+}$. Since $\varpi$ is a $C^1$ function, $\varpi(u_{\mathcal H^+},v)=\varpi(u,v)+O_v(|u-u_{\mathcal H^+}|)$. By monotonicity, $\varpi(u_{\mathcal H^+},v)\le M_{\mathcal I^+}(u)+O_v(|u-u_{\mathcal H^+}|)$. Sending $u\nearrow u_{\mathcal H^+}$ and then $v\to\infty$ yields $\varpi_{\mathcal H^+}\le M_\infty$ and completes the proof. 
\end{proof}

\begin{rk}
    In the context of our main theorems in this paper, it will in fact be true that $\varpi_{\mathcal H^+}=M_\infty$ (and both masses are the final Reissner--Nordstr\"om mass). In fact, in the ``weakly subextremal'' case $r_{\mathcal H^+}\ne |Q|$, this equality holds under no additional assumptions by \cite{Price-law}. A positive resolution of \cref{quest:large-data-1} would in particular imply that $\varpi_{\mathcal H^+}=M_\infty$ for \emph{all} solutions in $\mathfrak M_\mathrm{black}$.
\end{rk}

\begin{rk}
    We have given here a very short proof of the horizon Penrose inequality (in fact, a stronger inequality) using \cref{cor:final-parameters}, which relies on special structure of the EMSF system. For completeness, we also give a proof using only the much weaker \eqref{eq:1-mu-exterior}.
\begin{proof}[Alternative proof]
    By \eqref{eq:1-mu-exterior}, there exists a sequence $(u_i,v_i)\in \mathcal R(1,u_{\mathcal H^+},1,\infty)$ such that $(1-\mu)(u_i,v_i)\to 0$ as $i\to\infty$. After passing to a subsequence, we may assume that $u_i\to u_\infty\in [1,u_{\mathcal H^+}]$ and either $v_i\to v_\infty<\infty$ or $v_i\to\infty$ as $i\to\infty$. Since $1-\mu$ is bounded below by a positive number on $\mathcal R(1,u_*,1,\infty)$ for every $u_*<u_{\mathcal H^+}$, $u_\infty=u_{\mathcal H^+}$. Using monotonicities of $r$ and $m$, it holds that
    \begin{equation}\label{eq:r-error}
r(u_{\mathcal H^+},v_i)\le 2M_{\mathcal I^+}(u_i)+r(u_i,v_i)(1-\mu)(u_i,v_i). 
\end{equation}
As in the proof of \cref{thm:dichotomy}, $r(u_i,v_i)$ is bounded. 

    If $v_i\to\infty$, then by passing to the limit in \eqref{eq:r-error} and using monotonicity of $r$ again, it holds that $\sup_{\mathcal H^+}r\le 2M_\infty$. If $v_i\to v_\infty$, then $r(u_{\mathcal H^+},v_\infty)\le 2M_\infty$. Since $(1-\mu)(u_{\mathcal H^+},v_\infty)=0$, $r$ is constant on $\mathcal H^+$ past $v_\infty$ and again the desired conclusion holds.
\end{proof}

\end{rk}

\subsubsection{The parameter ratio at null infinity and superextremality}\label{sec:superextremal}

A well-known sufficient condition for lying in the class $\mathfrak M_\mathrm{black}$ (which is obvious in the present setting but is valid in a much more general context) is that $\mathcal S_\mathrm{max}$ possesses a trapped surface \cite{Hawking72BH}. In this work, we identify a sufficient condition for lying in the class $\mathfrak M_\mathrm{non}$, which crucially uses the structure of the spherically symmetric EMSF system.

Consider a solution $\mathcal S$ for which $M_{\mathcal I^+}$ is nonvanishing along $\mathcal I^+$. In the black hole case, $M_\infty>0$ by the Penrose inequality, so it follows that \index{P@$P_{\mathcal I^+}$, asymptotic parameter ratio function} \index{P@$P_\infty$, final parameter ratio}
\begin{equation*}
     P_{\mathcal I^+}(u)\doteq\frac{Q}{M_{\mathcal I^+}(u)}
\end{equation*}
is defined for every $u\in [1,u_\Box]$, where as before we set $ P_\infty\doteq  P_{\mathcal I^+}(u_{\mathcal{H}^+})$. This is called the \emph{final parameter ratio} of the black hole. By \cref{lem:mass-monotonicity},
    \begin{equation}\label{eq:P-monotonicity}
        | P_{\mathcal I^+}(u'')| \ge | P_{\mathcal I^+} (u')|
    \end{equation}
    for every $u',u''\in[1,u_\Box]$ with $u'\le u''$. 

\begin{thm}[Superextremal solutions fail to collapse]\label{thm:noncollapse}
    Let $\mathcal S_\mathrm{max}$ be the maximal development of asymptotically flat seed data $\Psi\in \mathfrak M$. If \begin{equation}
|P_{\mathcal I^+}(u_*)|>1\label{eq:superext-1}
    \end{equation} for some $u_*\in [1,u_\Box]$, then $\Psi\in\mathfrak M_\mathrm{non}$. 
\end{thm}
\begin{proof} Consider the region $\mathcal R=\mathcal R(u_*,u_\Box,1,\infty)\subset \mathcal Q_\mathrm{max}$. Since $\lambda\ge 0$ and $\nu<0$ on $\mathcal R$, $\varpi\le M_{\mathcal I^+}(u_*)$ on $\mathcal R$ by \eqref{eq:varpi-u}. Since $M_{\mathcal I^+}(u_*)<|Q|$, the rational function $x\mapsto 1-\frac{2\varpi(u,v)}{x}+\frac{Q^2}{x^2}$ is bounded below by the constant $c\doteq 1-P_{\mathcal I^+}(u_*)^{-2}>0$, for every $(u,v)\in\mathcal R$ and $x>0$. It follows that $1-\mu\ge c$ on $\mathcal R$ and hence the solution is asymptotically flat over $\mathcal R$ by \cref{cor:fundamental-AF}. Therefore, $u_\Box = u_1$ and $\Psi\in \mathfrak M_\mathrm{non}$. \end{proof}

\begin{rk}
The constancy of $Q$ is crucial for this statement: it suffices to measure $P_{\mathcal I^+}$ at one point on $\mathcal I^+$ to infer that $1-\mu$ is uniformly positive. The statement is false for spherically symmetric matter models with dynamical charge.
\end{rk}

We can already apply this criterion to rule out black hole formation from a large class of initial data:

\begin{prop}\label{prop:a-priori-noncollapse} For any $B\ge 1$ and $\theta>0$, there exists a constant $\ve(B,\theta)>0$ with the following property. If $\Psi=(\phi_\circ,r_\circ,\varpi_\circ,\rho_\circ)\in\mathfrak M$ satisfies $\|(0,r_\circ,\varpi_\circ,\rho_\circ)\|_\mathcal{C}\le B$, $\|\phi_\circ\|_\mathfrak{F}\le \ve$, and $|\rho_\circ|>1+\theta$, then $\Psi\in\mathfrak M_\mathrm{non}$.
\end{prop}
\begin{proof}
 By \cref{lem:AF-data-generation} and its proof, $|\varpi-\varpi_\circ|\les_B \ve^2$ on $C_\out$. Therefore, $| P_{\mathcal I^+}(u_0)-\rho_\circ|\les_B \ve^2$ and hence $|P_{\mathcal I^+}(u_0)|\ge 1+\frac\theta2>1$ for $\ve$ sufficiently small and we may apply \cref{thm:noncollapse}.
\end{proof}

\subsection{Perturbations of asymptotically flat solutions I: initial data gauge}\label{sec:perturbations-1}

We now turn to the issue of estimating first- and second-order perturbations associated to an asymptotically flat solution of the spherically symmetric EMCSF system. In \cref{sec:norms-1}, we define norms for asymptotically flat perturbations and initial data. In \cref{sec:perturbing-AF-data}, we show how to generate initial data for the first- and second-order perturbation systems from seed data. In \cref{sec:first-order-perturbations-estimates,sec:second-order-perturbations-estimates}, we prove estimates for asymptotically flat first- and second-order perturbations, respectively. In \cref{sec:semiglobal-cauchy}, we use the linearized estimates to prove a semiglobal Cauchy stability result. Finally, in \cref{sec:threshold-a-priori}, we give a basic a priori characterization of the black hole threshold $\partial\mathfrak M_\mathrm{black}$.

\subsubsection{Norms}\label{sec:norms-1}
Let $\mathcal C=\mathcal C(1,u_1,1,\infty)$ be a semiglobal bifurcate null hypersurface with associated semiglobal double null rectangle $\mathcal R=\mathcal R(1,u_1,1,\infty)$. Let $\lin\Psi_\#$ be linearized characteristic data on $\mathcal C$ and $\lin{\mathcal S}$ and $\qd{\mathcal S}$ be first- and second-order perturbations of a solution on $\mathcal R$. We define the initial data norms
\begin{multline*}
       \|\lin\Psi_\#\|_{\mathcal C} \doteq |\lin Q|+\sup_{\mathcal C}\big(|\lin r|+|\lin\varpi|+|\lin\psi|\big) \\+ \sup_{C_\out}\big(|r\lin\lambda|+|r\partial_v\lin\lambda|+|r\partial_v^2\lin\lambda|+|r\lin\kappa|+|r\partial_v\lin\kappa|+|r\partial_v^2\lin\kappa|+|r^2\partial_v\lin\psi|+|r^2\partial_v^2\lin\psi|\big)\\+\sup_{\underline C{}_\ing}\big(|\lin\nu|+|\partial_u\lin\nu|+|\partial_u^2\lin\nu|+|\lin\gamma|+|\partial_u\lin\gamma|+|\partial_u^2\lin\gamma|+|\partial_u\lin\phi|+|\partial_u^2\lin\phi|\big),
\end{multline*}
\begin{multline*}
       \|\qd\Psi_\#\|_{\mathcal C} \doteq |\qd Q|+\sup_{\mathcal C}\big(|\qd r|+|\qd\varpi|+|\qd\psi|\big) \\+ \sup_{C_\out}\big(|r\qd\lambda|+|r\partial_v\qd\lambda|+|r\partial_v^2\qd\lambda|+|r\qd\kappa|+|r\partial_v\qd\kappa|+|r\partial_v^2\qd\kappa|+|r^2\partial_v\qd\psi|+|r^2\partial_v^2\qd\psi|\big)\\+\sup_{\underline C{}_\ing}\big(|\qd\nu|+|\partial_u\qd\nu|+|\partial_u^2\qd\nu|+|\qd\gamma|+|\partial_u\qd\gamma|+|\partial_u^2\qd\gamma|+|\partial_u\qd\phi|+|\partial_u^2\qd\phi|\big),
\end{multline*}
and say that $\lin\Psi_\#$ (respectively, $\qd\Psi_\#$) is \emph{asymptotically flat} if $\|\lin\Psi_\#\|_\mathcal{C}$ (respectively, $\|\qd\Psi_\#\|_\mathcal{C}$) is finite. For the first- and second-order variations we define
\begin{multline*}
    \|\lin{\mathcal S}\|_{\mathcal R}\doteq |\lin Q|+\sup_\mathcal{R}\big(|\lin r|+|\lin\varpi|+|\lin\nu|+|\partial_u\lin\nu|+|\partial_u^2\lin\nu|+|r\lin\lambda|+|r^2\partial_v\lin\lambda|+|r^2\partial_v^2\lin\lambda|+|r\lin\kappa| \\ \qquad\qquad+|r^2\partial_v\lin\kappa|+|r^2\partial_v^2\lin\kappa|
+|\lin\gamma|+|\partial_u\lin\gamma|+|\partial_u^2\lin\gamma|+|r\lin\phi|+|r^2\partial_v\lin\psi|+|r^2\partial_v^2\lin\psi|+|\partial_u\lin\phi|+|\partial_u^2\lin\phi|\big),
\end{multline*}\vspace{-5mm}
\begin{multline*}
  \|\qd{\mathcal S}\|_{\mathcal R} \doteq |\qd Q|    +\sup_\mathcal{R}\big(|\qd r|+|\qd\varpi|+|\qd\nu|+|\partial_u\qd\nu|+|\partial_u^2\qd\nu|+|r\qd\lambda|+|r^2\partial_v\qd\lambda|+|r^2\partial_v^2\qd\lambda|+|r\qd\kappa| \\ \qquad\qquad+|r^2\partial_v\qd\kappa|+|r^2\partial_v^2\qd\kappa|
+|\qd\gamma|+|\partial_u\qd\gamma|+|\partial_u^2\qd\gamma|+|r\qd\phi|+|r^2\partial_v\qd\psi|+|r^2\partial_v^2\qd\psi|+|\partial_u\qd\phi|+|\partial_u^2\qd\phi|\big)
\end{multline*}
and say that $\lin{\mathcal S}$ (respectively, $\qd{\mathcal S}$) is \emph{asymptotically flat} if $\|\lin{\mathcal S}\|_\mathcal{R}$ (respectively, $\|\qd{\mathcal S}\|_{\mathcal R}$) is finite.

\subsubsection{Perturbations of asymptotically flat seed data}\label{sec:perturbing-AF-data}

The way we generate first-order deformations of characteristic data $\lin\Psi_\#$ in practice is by considering families of seed data, which live in a linear space. The most important construction will be the following linear one-parameter family: Let $\Psi\in\mathfrak M$ (for a semiglobal $\mathcal C$) and let $\lin\Psi\in \mathfrak Z$. Then $\Psi(z)\doteq \Psi+z\lin\Psi\in \mathfrak M$ for $|z|$ small by \cref{prop:MopeninZ} and it is obvious that $\lin\Psi_\#\doteq\partial_z\Psi_{\#}(z)|_{z=0}$ is indeed a linearized characteristic data set as required by \cref{lem:first-order-constraint}. The following lemma shows that the linear map $\lin\Psi\mapsto \lin\Psi_\#$ is bounded. 

\begin{lem}\label{lem:AF-data-lin}
  For any $u_1\ge 1$ and $B\ge1$, there exists a constant $\lin C_\#(u_1,B)\ge 1$ with the following property. Let $\Psi$ be a seed data set on $\mathcal C$ satisfying $\|\Psi\|_\mathcal{C}\le B$. Let $\lin\Psi\in \mathfrak Z$ and define $\lin\Psi_\#$ as above. Then $\lin\Psi_\#$ is asymptotically flat and 
  \begin{equation}\label{eq:data-linearization-bounded}
    \|\lin\Psi_\#\|_{\mathcal C}\le \lin C_\#\|\lin\Psi\|_{\mathfrak Z}.
\end{equation}
\end{lem} 
\begin{proof} Let $\Psi=(\phi_\circ,r_\circ,\varpi_\circ,\rho_\circ)$ and $\lin\Psi=(\lin\phi_\circ,\lin r_\circ,\lin \varpi_\circ,\lin\rho_\circ)$. Note that $Q=\rho_\circ\varpi_\circ$, hence by linearizing we have
\begin{equation}\label{eq:linearization-of-Q}
    \lin Q = \rho_\circ\lin\varpi_\circ + \varpi_\circ\lin\rho_\circ,
\end{equation}
which implies $|\lin Q|\les_B |\lin\varpi_\circ|+|\lin\rho_\circ|$. Since $\lambda=1$ on $C_\out$ for every $z$,  $\lin r=\lin r_\circ$ and $\lin\lambda$ vanishes identically on $C_\out$. Linearizing \eqref{eq:varpi-data-out}, we obtain $|\lin\varpi|\les_B \|\lin\Psi\|_{\mathfrak Z}$ on $C_\out$. Linearizing the gauge condition $\kappa = (1-\mu)^{-1}$, we obtain $|\lin\kappa|\les_B r^{-1}|\lin\varpi|+r^{-2}|\lin Q|\les_B r^{-1}\|\lin\Psi\|_{\mathfrak Z}$. We leave the rest of the estimates to the reader.
\end{proof}

For second-order perturbations, we again consider a linear construction using seed data. Let $\Psi\in\mathfrak M$ and $\lin\Psi_1,\lin\Psi_2\in\mathfrak Z$. Then $\Psi(z_1,z_2)\doteq \Psi+z_1\lin\Psi_1+z_2\lin\Psi_2\in\mathfrak M$ for $|z_1|+|z_2|$ small by \cref{prop:MopeninZ} and it is obvious that $\qd\Psi_\#\doteq \partial_{z_1}\partial_{z_2}\Psi_\#(z_1,z_2)|_{z_1=z_2=0}$ satisfies the hypotheses of \cref{lem:second-order-constraint}. Note of course that now $\lin\Psi_\#$ has two components, coming from the $z_1$ and $z_2$ derivatives separately, and the estimate \eqref{eq:data-linearization-bounded} holds for each component. The following lemma shows that the \emph{bilinear} map $(\lin\Psi_1,\lin\Psi_2)\mapsto \qd\Psi_\#$ is bounded. 

\begin{lem}\label{lem:AF-data-second-lin}
  For any $u_1\ge 1$ and $B\ge1$, there exists a constant $\qd C_\#(u_1,B)\ge 1$ with the following property. Let $\Psi$ be a seed data set on $\mathcal C$ satisfying $\|\Psi\|_\mathcal{C}\le B$. Let $\lin\Psi_1,\lin\Psi_2\in \mathfrak Z$ and define $\qd\Psi_\#$ as above. Then $\qd\Psi_\#$ is asymptotically flat and 
  \begin{equation}\label{eq:data-second-linearization-bounded}
    \|\qd\Psi_\#\|_{\mathcal C}\le \qd C_\#\|\lin\Psi_1\|_{\mathfrak Z}\|\lin\Psi_2\|_{\mathfrak Z}.
\end{equation}
\end{lem} 
\begin{proof}
   Let $\qd\Psi_\#=(\qd r,\qd\Omega^2,\qd\phi,\qd Q)$. By construction, $\qd r$ and $\qd\phi$ vanish identically and $\qd Q = 2 \lin\rho_\circ\cdot\lin\varpi_\circ.$ Differentiating \eqref{eq:varpi-data-out} and using $\qd\varpi_\circ=0$, we find that 
   \begin{equation*}
       |\qd\varpi(1,v)| = \int_1^v\exp\left(-\int_{v'}^vr(\partial_v\phi)^2\,dv''\right)|\lin\rho_\circ\cdot\lin\varpi_\circ|(\partial_v\phi)^2\,dv'\les_B \|\lin\Psi_1\|_{\mathfrak Z}\|\lin\Psi_2\|_{\mathfrak Z}.
   \end{equation*}
Continuing in this way, we can estimate  all of the terms in $\|\qd\Psi_\#\|_\mathcal{C}$ and we leave the rest of the estimates to the reader. 
\end{proof}

\subsubsection{Estimates for first-order perturbations}\label{sec:first-order-perturbations-estimates}

By \cref{lem:first-order-constraint}, $\lin\Psi_\#\mapsto \lin{\mathcal S}[\Psi_\#,\lin\Psi_\#]$ is linear. The following proposition shows that this map is bounded in an appropriate sense.

\begin{prop}\label{prop:first-order-ID} For any $u_1\ge 1$ and $B\ge1$, there exists a constant $\lin C(u_1,B)\ge1$ with the following property.   Let $\mathcal S$ be an asymptotically flat solution on $\mathcal R=\mathcal R(1,u_1,1,\infty)$ satisfying the bound $\|\mathcal S\|_{\mathcal R}\le B$. Let $\lin{\mathcal S}$ be the first-order perturbation of $\mathcal S$ obtained from perturbing the data $\Psi_\#$ by the asymptotically flat first-order variation $\lin{\Psi}_\#$ according to \cref{lem:first-order-constraint}. Then $\lin{\mathcal S}$ is asymptotically flat and it holds that
\begin{equation*}
    \|\lin{\mathcal S}[\Psi_\#,\lin\Psi_\#]\|_{\mathcal R}\le \lin C\|\lin{\Psi}_\#\|_{\mathcal C}.
\end{equation*}
\end{prop}

\begin{proof} Note that $\lin Q$ is read off from the data, recall \eqref{eq:linearization-of-Q}. For the remaining quantities in $\|\lin{\mathcal S}\|_\mathcal{R}$, we continue via a continuity argument in the domains $\mathcal R(u_*,v_*)\doteq [1,u_*]\times[1,v_*]\subset\mathcal R$.
    For $K_1,K_2,M\ge 1$ to be chosen later, we make the bootstrap assumptions
    \begin{gather*}
       |\lin\varpi|,\quad |r\lin\kappa|\le K_1e^{Mu}\|\lin\Psi_\#\|_{\mathcal C}, \\
        |\lin\nu|,\quad |r\lin\phi|\le  K_2e^{Mu}\|\lin\Psi_\#\|_{\mathcal C}
    \end{gather*}
on $\mathcal R(u_*,v_*)$. We show that for appropriate choices of $K_1,K_2$, and $M$, these estimates hold with constant $\frac 12$ on the right-hand side in place of $1$. We omit the standard continuity argument that then lets us take $u_*=u_1$ and $v_*=\infty$. Until we specify otherwise, the notation $\les$ does not allow for dependence on $K_1$, $K_2$, and $M$, but it does allow for dependence on the asymptotic flatness parameter $B$ and the time $u_1$. 

First, we obtain $|\lin r|\les (K_2M^{-1}e^{Mu}+1)\|\lin\Psi_\#\|_{\mathcal C}\les e^{Mu}\|\lin\Psi_\#\|_{\mathcal C}$ on $\mathcal R(u_*,v_*)$ for $M$ sufficiently large by integrating the bootstrap assumption for $\lin\nu$. We then have
\begin{equation}\label{eq:lin-psi-r-weight-2}
    |\partial_u\partial_v\lin\psi|\les r^{-3}|\lin\kappa|+r^{-3}|\lin\nu|+r^{-4}|\lin r|+r^{-3}|\lin\varpi|+r^{-4}|\lin Q|+r^{-2}|\lin\phi|\les (K_1+K_2)r^{-3}e^{Mu}\|\lin\Psi_\#\|_{\mathcal C},
\end{equation} by \eqref{eq:lin-psi}, and so integrating in $u$ gives $|\partial_v\lin\psi|\les r^{-3}e^{Mu}\|\lin\Psi_\#\|_{\mathcal C}$ on $\mathcal R(u_*,v_*)$ for $M$ sufficiently large and hence $|\lin\psi|\les e^{Mu}\|\lin\Psi_\#\|_{\mathcal C}$ on $\mathcal R(u_*,v_*)$.

Next, we estimate
\[|\partial_u\lin\lambda|\les r^{-3}|\lin r|+r^{-2}|\lin\kappa|+r^{-2}|\lin\nu|+r^{-2}|\lin\varpi|+r^{-3}|\lin Q|\les (K_1+K_2)r^{-2}e^{Mu}\|\lin\Psi_\#\|_{\mathcal C}, \]
\eqref{eq:dvlnu}
so by integrating in $u$, we obtain $|\lin\lambda|\les r^{-1}e^{Mu}\|\lin\Psi_\#\|_{\mathcal C}$ on $\mathcal R(u_*,v_*)$. We can improve the bootstrap assumption for $\lin \nu$ by integrating this estimate in $v$, since $\partial_u\lin\lambda=\partial_v\lin\nu$.
 It also follows that \[|\partial_v\lin\phi|\les r^{-2}|\lin\psi|+r^{-1}|\partial_v\lin\psi|+r^{-3}|\lin r|+r^{-2}|\lin\lambda| \les r^{-2}e^{Mu}\|\lin\Psi_\#\|_{\mathcal C}\] on $\mathcal R(u_*,v_*)$ and we improve the bootstrap assumption on $\lin\phi$ for $K_2$ sufficiently large by integrating in $v$. Integrating \eqref{eq:lin-psi-r-weight-2} in $v$, we obtain 
 \[ |\partial_u\lin\phi|\les r^{-2}|\lin\psi|+r^{-1}|\partial_u\lin\psi|+r^{-2}|\lin r|+ r^{-2}|\lin\nu|\les (K_1+K_2)r^{-1}e^{Mu}\|\lin\Psi_\#\|_{\mathcal C}\] on $\mathcal R(u_*,v_*)$ and hence can integrate
 \[|\partial_u\lin\varpi|\les  r^{-1}|\lin r|+|\lin\gamma|+r|\partial_u\lin\phi|\les (K_1+K_2)e^{Mu}\|\lin\Psi_\#\|_\mathcal{C}\]
  to improve the bootstrap assumption for $\lin\varpi$. 
  
  We close the bootstrap argument by using Gr\"onwall's inequality on 
  \[|\partial_u\lin\kappa|\les r^{-2}|\lin r|+r^{-1}|\lin\kappa|+r^{-1}|\lin\nu|+|\partial_u\lin\phi|\les r^{-1}|\lin\kappa| + e^{Mu}r^{-1}\|\lin\Psi_\#\|_{\mathcal C}\]
 to obtain $|\lin\kappa|\les r^{-1} e^{Mu}\|\lin\Psi_\#\|_{\mathcal C}$. We now fix $K_1$, $K_2$, and $M$ so that the desired estimates hold on $\mathcal R$ after performing a standard continuity argument. We now allow the $\les$ notation to depend on these constants. 

To finish the low-order estimates, we integrate
\begin{equation}\label{eq:lin-gamma-weights-1}
    |\partial_v\lin\gamma|\les r^{-4}|\lin r|+r^{-3}|\lin\gamma|+r^{-3}|\lin\lambda|+r^{-1}|\partial_v\lin\phi|\les r^{-3}\|\lin\Psi_\#\|_\mathcal{C}
\end{equation}
to obtain $|\lin\gamma|\les \|\lin\Psi_\#\|_\mathcal{C}$ on $\mathcal R$. 

The higher-order estimates proceed similarly as in \cref{prop:fundamental-AF}, but several of the $r$-weights are in fact better and we omit the detailed proof. For later reference, we do note however the estimate
\begin{equation}\label{eq:lin-gamma-weights-2}
    |\partial_v\partial_u\lin\gamma| + |\partial_v\partial_u^2\lin\gamma|\les r^{-3}\|\lin\Psi_\#\|_\mathcal{C}
\end{equation}
that is obtained along the way. 
\end{proof}

When $\Psi_\#$ and $\lin\Psi_\#$ arise from the procedure of \cref{sec:perturbing-AF-data}, we write
\[\lin{\mathcal S}[\Psi,\lin\Psi]\doteq \lin{\mathcal S}[\Psi_\#,\lin\Psi_\#].\]
By combining \cref{lem:AF-data-lin} and \cref{prop:first-order-ID}, we obtain
\begin{equation*}
    \|\lin{\mathcal S}[\Psi,\lin\Psi]\|_\mathcal{R}\le \lin C\lin C_\#\|\lin\Psi\|_\mathfrak{Z},
\end{equation*}
which we will use many times below.

\subsubsection{Estimates for second-order perturbations}\label{sec:second-order-perturbations-estimates}

By \cref{lem:second-order-constraint}, $(\lin\Psi_\#^1,\lin\Psi_\#^2)\mapsto \qd{\mathcal S}[\Psi_\#,\lin\Psi_\#,\qd\Psi_\#]$ is bilinear and $\qd\Psi_\#\mapsto \qd{\mathcal S}[\Psi_\#,\lin\Psi_\#,\qd\Psi_\#]$ is linear. The following proposition shows that this map is bounded in an appropriate sense.

\begin{prop}\label{prop:second-order-ID} For any $u_1\ge 1$ and $B\ge1$, there exists a constant $\qd C(u_1,B)\ge1$ with the following property.   Let $\mathcal S$ be an asymptotically flat solution on $\mathcal R$ satisfying the bound $\|\mathcal S\|_{\mathcal R}\le B$. Let $\qd{\mathcal S}$ be the second-order perturbation of $\mathcal S$ obtained from perturbing $\Psi_\#$ by the pair asymptotically flat first-order variations $(\lin{\Psi}_{\#}^1,\lin{\Psi}_{\#}^2)$ and the asymptotically flat second-order variation $\qd\Psi_\#$. Then $\qd{\mathcal S}$ is asymptotically flat and it holds that
\begin{equation*}
    \|\qd{\mathcal S}[\Psi_\#,\lin\Psi_{\#},\qd\Psi_{\#}]\|_{\mathcal R}\le \qd C\big(\|\lin{\Psi}_{\#}^1\|_{\mathcal C}\|\lin{\Psi}_{\#}^2\|_{\mathcal C}+\|\qd\Psi_\#\|_\mathcal{C}\big).
\end{equation*}
\end{prop}
We omit the proof, as it is similar to the $\lin{\mathcal S}$ case. From the proof, we later require the estimate
\begin{equation}\label{eq:qd-gamma-r-weight}
    |\partial_v\qd\gamma|+ |\partial_v\partial_u\qd\gamma| + |\partial_v\partial_u^2\qd\gamma| \les r^{-3}\big(\|\lin{\Psi}_{\#}^1\|_{\mathcal C}\|\lin{\Psi}_{\#}^2\|_{\mathcal C}+\|\qd\Psi_\#\|_\mathcal{C}\big),
\end{equation} which is analogous to \eqref{eq:lin-gamma-weights-1} and \eqref{eq:lin-gamma-weights-2}.

When $\Psi_\#$, $\lin\Psi_\#$, and $\qd\Psi_\#$ arise from the procedure of \cref{sec:perturbing-AF-data}, we write
\[\qd{\mathcal S}[\Psi,\lin\Psi_1,\lin\Psi_2]\doteq \lin{\mathcal S}[\Psi_\#,\lin\Psi_\#,\qd\Psi_\#].\] 
By combining \cref{lem:AF-data-second-lin} and \cref{prop:second-order-ID}, we obtain
\begin{equation*}
    \|\qd{\mathcal S}[\Psi,\lin\Psi_1,\lin\Psi_2]\|_\mathcal{R}\le \qd C(\lin C_\#^2+\qd C_\#)\|\lin\Psi_1\|_\mathfrak{Z}\|\lin\Psi_2\|_\mathfrak{Z},
\end{equation*}
which we will use many times below.

\subsubsection{Semiglobal Cauchy stability and the topology of \texorpdfstring{$\mathfrak M_\mathrm{black}$}{Mblack} and \texorpdfstring{$\mathfrak M_\mathrm{non}$}{Mnon}}\label{sec:semiglobal-cauchy}

Using \cref{lem:AF-data-lin} and \cref{prop:first-order-ID}, we can now easily prove a semiglobal Cauchy stability statement, which has an important corollary.  

\begin{prop}\label{prop:semiglobal-Cauchy-ID}
    For any $u_1\ge 1$ and $B\ge 1$, there exist constants $\eta>0$ and $C\ge 1$ with the following property. Let $\Psi\in\mathfrak M$ for which the solution $\mathcal S[\Psi]$ is asymptotically flat on $\mathcal R$ and satisfies $\|\mathcal S[\Psi]\|_{\mathcal R}\le B$. If $\Psi'\in \mathfrak M$ satisfies $\|\Psi'-\Psi\|_{\mathfrak Z}\le \eta$, then $\mathcal S[\Psi']$ is asymptotically flat on $\mathcal R$, with $\|\mathcal S[\Psi']\|_{\mathcal R}\le 2B$, and it holds that
    \begin{equation*}
        \|\mathcal S[\Psi]-\mathcal S[\Psi']\|_{\mathcal R}\le C\|\Psi-\Psi'\|_{\mathfrak Z}.
    \end{equation*}
\end{prop}
\begin{proof}
    Let $\Psi(z)\doteq (1-z)\Psi+z\Psi'$ for $z\in[0,1]$. Then $\lin\Psi\doteq\partial_z\Psi(z) = \Psi-\Psi'$ and $\|\lin\Psi\|_\mathfrak{Z}\le \eta$. For any $v_*\ge 1$, we define the set
    \begin{equation*}
        \mathcal Z(v_*)\doteq \{z\in[0,1]:\text{$\mathcal S[\Psi_z]$ exists on $\mathcal R(v_*)$ and $\|\mathcal S[\Psi_z]\|_{\mathcal R(v_*)}\le 2B$}\},
    \end{equation*}
    where $\mathcal R(v_*)\doteq [1,u_1]\times[1,v_*]$. Given any quantity $q$ in the definition of $\|\mathcal S[\Psi(z)]\|_{\mathcal R(v_*)}$, we have that $\partial_z q =\lin q$, where $\lin q$ is computed from $\lin{\mathcal S}[\Psi(z),\lin\Psi]$.

    Now let $z_*\in \mathcal Z(v_*)$. By \cref{lem:AF-data-lin} and \cref{prop:first-order-ID}, $|\lin q|\les \|\lin\Psi\|_\mathfrak{Z} \les \eta$ on $\mathcal R(v_*)$ for $z\in [0,z_*]$, so by integrating in $z$ we can improve the bootstrap assumption to $\|\mathcal S[\Psi_z]\|_{\mathcal R(v_*)}\le \frac 32B$ for $\eta$ sufficiently small. Now a standard continuity argument using \cref{prop:smooth-dependence} shows that $\mathcal Z(v_*)=[0,1]$. Since $v_*$ was arbitrary, this proves the proposition. 
\end{proof}

\begin{cor}\label{cor:M-topology} For any semiglobal bifurcate null hypersurface $\mathcal C$, $\mathfrak M_\mathrm{non}$ is open in $\mathfrak M$ and $\mathfrak M_\mathrm{black}$ is closed in $\mathfrak M$.
\end{cor}
\begin{proof}
    By \cref{thm:dichotomy}, $\mathfrak M_\mathrm{non}$ consists precisely of those solutions which are asymptotically flat on the full domain of dependence of $\mathcal C$. Therefore, by \cref{prop:semiglobal-Cauchy-ID}, $\mathfrak M_\mathrm{non}$ is open in $\mathfrak M$. Since $\mathfrak M_\mathrm{black}= \mathfrak M\setminus \mathfrak M_\mathrm{non}$, it is closed in $\mathfrak M$. 
\end{proof}

\subsubsection{Weakly extremal black holes and an a priori characterization of \texorpdfstring{$\partial\mathfrak M_\mathrm{black}$}{dMblack}}\label{sec:threshold-a-priori}

In this section, we give a general initial characterization of the \emph{black hole threshold} in $\mathfrak M$,
\begin{equation*}
    \partial\mathfrak M_\mathrm{black}\doteq \mathfrak M_\mathrm{black}\setminus \Int(\mathfrak M_\mathrm{black}),
\end{equation*}
where $\Int(A)$ denotes the interior of a set $A\subset\mathfrak M$. 

Recall the final horizon values $r_{\mathcal H^+}$ and $\varpi_{\mathcal H^+}$ of a black hole solution in $\mathfrak M$. We define the set of \emph{weakly extremal black holes} as \index{M@$\mathfrak M_\mathrm{weak\,ext}$, weakly extremal black hole data}
\begin{equation*}
    \mathfrak M_\mathrm{weak\,ext}\doteq \{\Psi\in\mathfrak M_\mathrm{black}:r_{\mathcal H^+}=\varpi_{\mathcal H^+}=|Q|\}.
\end{equation*}
Using monotonicities of the EMSF system, we have the following general result:
\begin{prop}\label{prop:weak-ext-properties}
    If $\Psi\in\mathfrak M_\mathrm{weak\,ext}$, then $\mathcal Q_\mathrm{max}=\mathcal R(1,u_1,1,\infty)$, $\lambda\ge 0$ on $\mathcal Q_\mathrm{max}$, and the Hawking mass $m$ is bounded up to the Cauchy horizon $\mathcal{CH}^+$. Moreover, if $\lambda(u_*,v_*)=0$ for some $(u_*,v_*)\in\mathcal Q_\mathrm{max}$, then $u_*=u_{\mathcal H^+}$ and $v\mapsto \phi(u_{\mathcal H^+},v)$ is constant for $v\ge v_*$. 
\end{prop}
\begin{proof}
The claims about $\lambda$, $\mathcal Q_\mathrm{max}$, and $\phi$ follow from the same argument used to prove \cite[Theorem 1, Part 5.]{AKU24} (which uses an unpublished argument of Kommemi that is not restricted to the case of small data, see also \cite[Appendix A]{LukOhI}). Boundedness of the Hawking mass follows by combining the fact that $r\sim 1$ towards $\mathcal{CH}^+$ with $1-\frac{2m}{r}\ge 0$. 
\end{proof}

A special role is played by the following set of \emph{marginal} black hole spacetimes: \index{M@$\mathfrak M_\mathrm{marg\,black}$, marginal black hole data}
\begin{equation*}
    \mathfrak M_\mathrm{marg\,black}\doteq\{\Psi\in\mathfrak M_\mathrm{black}:u_{\mathcal H^+}=u_1\}.
\end{equation*}

\begin{thm}
    For any semiglobal bifurcate null hypersurface $\mathcal C$ with moduli space $\mathfrak M$, it holds that
    \begin{equation*}
            \partial\mathfrak M_\mathrm{black}\subset \mathfrak M_\mathrm{marg\,black}\cup \mathfrak M_\mathrm{weak\,ext}. 
    \end{equation*}
\end{thm}
For the statement of this theorem, we again stress the importance of the condition $r(u_1,1)>0$ in the definition of $\mathfrak M$. 
\begin{proof}
    By \cref{cor:M-topology}, $\partial \mathfrak M_\mathrm{black}\subset\mathfrak M_\mathrm{black}$. Let $\Psi\in \partial \mathfrak M_\mathrm{black}$. Assume $u_{\mathcal H^+}<u_1$, else $\Psi\in\mathfrak M_\mathrm{marg\,black}$ by definition. If $r_{\mathcal H^+}=|Q|$, then $\varpi_{\mathcal H^+}=|Q|$ by \cref{cor:final-parameters}, and hence $\Psi\in\mathfrak M_\mathrm{weak\,ext}$. 

    It remains to rule out the case $r_{\mathcal H^+}\ne |Q|$. By \cite[Theorem 1.1]{Price-law}, the hypotheses of \cite[Proposition 7.1]{dafermos2005interior} are fulfilled and the black hole interior contains a strictly trapped surface ($\lambda<0$ somewhere). However, since $\mathcal S_\mathrm{max}[\Psi]$ can be approximated by solutions which have $\lambda>0$ everywhere, this gives a contradiction.
\end{proof}

\begin{rk}\label{rk:meeting}
For any $\mathcal C=\mathcal C(1,u_1,1,\infty)$, $\partial\mathfrak M_\mathrm{black}\cap \mathfrak M_\mathrm{marg\,black}\ne\emptyset$ and $\partial\mathfrak M_\mathrm{black}\cap \mathfrak M_\mathrm{weak\,ext}\ne\emptyset$. Indeed, for any $M>0$, \[\Psi_p^{(1)}\doteq (0,2M+u_1-1-p,M,0) \] has the property that $\Psi^{(1)}_p\in \mathfrak M_\mathrm{black}$ for $p\ge 0$, $\Psi^{(1)}_p\in \mathfrak M_\mathrm{non}$ for $p<0$, and $u_{\mathcal H^+}=u_1$ at $p=0$. Likewise, for any $M>0$, \[\Psi_p^{(2)}\doteq (0,M+u_1-1,M,1-p) \] has the property that $\Psi^{(2)}_p\in \mathfrak M_\mathrm{black}$ for $p\ge 0$, $\Psi^{(2)}_p\in \mathfrak M_\mathrm{non}$ for $p<0$, and $r_{\mathcal H^+}=M$ at $p=0$. Note that in fact $\Psi^{(2)}_0\in \mathfrak M_\mathrm{marg\,black}\cap \mathfrak M_\mathrm{weak\,ext}$!
\end{rk}

\subsection{The teleologically normalized gauges}\label{sec:teleological-gauge}

In this section, we define a method for consistently assigning a teleological gauge to an asymptotically flat solution that is helpful for proving black hole stability. First, in \cref{sec:Bondi}, we define \emph{Bondi time}, which normalizes the retarded time coordinate at null infinity $\mathcal I^+$ in a canonical manner. In \cref{sec:admissible}, we show that solutions which exist for at least some fixed Bondi time form an open subset of the moduli space, which is crucial for later continuity arguments. In \cref{sec:advanced}, we define a family of advanced time coordinates that are normalized along specially chosen curves. Finally, in \cref{sec:eschatological-definition}, we study some properties of these teleological gauges in the case a black hole actually forms. 

\subsubsection{The Bondi time function \texorpdfstring{$\mathfrak u$}{u}}\label{sec:Bondi}

\begin{prop}\label{prop:Bondi-time}
  Let $\mathcal C$ be a semiglobal bifurcate null hypersurface. For any $\Psi\in \mathfrak M(\mathcal C)$, 
    \begin{equation}\label{eq:def-mathfrak-u}
    \mathfrak u(u)\doteq 1-\lim_{v\to\infty}\int_1^u\gamma(u',v)\,du'
\end{equation}
defines a $C^3$ diffeomorphism $\mathfrak u:[1,u_1]\to[1,\mathcal L_\mathrm{Bondi}+1]${\index{u1@$\mathfrak u$, Bondi time function}} if $\Psi\in\mathfrak M_\mathrm{non}$ and $\mathfrak u:[1,u_{\mathcal H^+})\to[1,\infty)$ if $\Psi\in\mathfrak M_\mathrm{black}$.
\end{prop}
\begin{proof} Let $u_*<u_\Box$. Then $\mathcal S_\mathrm{max}[\Psi]$ is asymptotically flat on $\mathcal R(u_*)= \mathcal R(1,u_*,1,\infty)$ and $|\partial_v\partial_u^{\le 2}\gamma|\les_{u_*} r^{-3}$ on $\mathcal R(u_*)$ by 
\eqref{eq:kappa-u}, \eqref{eq:partial-u-gamma-1}, and \eqref{eq:partial-u-gamma-2}. Moreover, $\gamma\sim_{u_*}-1$, so by the dominated convergence theorem, $\mathfrak u:[1,u_*]\to \Bbb R$ is a $C^3$ diffeomorphism onto its image. Now letting $u_*\nearrow u_\Box$ and using the definition of $\mathcal L_\mathrm{Bondi}$, the proof of the proposition is complete. 
\end{proof}

\begin{defn}
    The function $\mathfrak u$ is called the \emph{Bondi time function} of $\mathcal S_\mathrm{max}[\Psi]$. We write the associated retarded time coordinate as $\u u\doteq\mathfrak u(u)${\index{u2@$\u u$, Bondi normalized retarded time}\index{u1@$\u{\mathfrak u}\doteq \mathfrak u^{-1}$}} and the inverse diffeomorphism as $\u{\mathfrak u}\doteq \mathfrak u^{-1}$.
\end{defn}

\subsubsection{\texorpdfstring{$\u u{}_f$}{u bar f}-admissible solutions}\label{sec:admissible}

For later use, we define now an important notion of solutions that have Bondi lifetime bounded from below. 

\begin{defn} 
    We say that a solution $\mathcal S$ on the semiglobal rectangle $\mathcal R$ is $\u u{}_f$-\emph{admissible} if $\mathcal L_\mathrm{Bondi}>\u u{}_f-1$. The set of $\u u{}_f$-admissible solutions\index{M@$\mathfrak M_{\u u{}_f}(\mathcal C)$, $\u u{}_f$-admissible solutions} arising from seed data in $\mathfrak M(\mathcal C)$ is denoted by
    \begin{equation*}
        \mathfrak M_{\u u{}_f}(\mathcal C)\doteq \{\Psi\in \mathfrak M(\mathcal C):\mathcal L_\mathrm{Bondi}>\u u{}_f-1\}.
    \end{equation*}
    Note that $\u{\mathfrak u}(\u u{}_f)< u_1$ and $\mathcal S[\Psi]$ is asymptotically flat over $\mathcal R(1,\u{\mathfrak u}(\u u{}_f),1,\infty)$ if $\Psi$ is $\u u{}_f$-admissible.
\end{defn}

Again by \cref{prop:semiglobal-Cauchy-ID}, these sets are open in $\mathfrak M$, which we quantify in the following lemma. For $\eta>0$ and $\Psi\in\mathfrak Z$, we denote by $B_\eta^\mathfrak{Z}(\Psi)$ the open $\eta$-ball in the $\|\cdot\|_\mathfrak{Z}$-norm centered at $\Psi$. 
\begin{lem}\label{lem:Mi-open-quantitative}
    For any $B\ge 1$, $\u u{}_f\ge1$, and $\varsigma>0$, there exists an $\eta>0$ such that if $\Psi\in \mathfrak M_{\u u{}_f}$, $u_1>\u{\mathfrak u}(\u u{}_f)+\varsigma$, and $\|\mathcal S[\Psi]\|_{\mathcal R(1,\u{\mathfrak u}(\u u{}_f),1,\infty)}\le B$, then $B^\mathfrak{Z}_{\eta}(\Psi)\subset\mathfrak M_{\u u{}_f}$. 
\end{lem}
\begin{proof}
    By \cref{prop:slab-1}, there exists a $\tilde\varsigma=\tilde\varsigma(B)\in (0,\varsigma)$ such that $\mathcal S[\Psi]$ extends to the rectangle $\mathcal R(\varsigma)=\mathcal R(1,\u{\mathfrak u}(\u u{}_f)+\tilde\varsigma,1,\infty)$ and satisfies $\|\mathcal S[\Psi]\|_{\mathcal R(\varsigma)}\le \tilde B$, where $\tilde B=\tilde B(B)$. By \cref{prop:semiglobal-Cauchy-ID}, we obtain an $\eta_0=\eta_0(\u{\mathfrak u}(\u u{}_f)+\tilde\varsigma,\tilde B)>0$ such that any $\Psi'\in B_{\eta_0}^\mathfrak{Z}(\Psi)$ has a solution on $\mathcal R(\varsigma)$ and satisfies there $\|\mathcal S[\Psi']\|_{\mathcal R(\varsigma)}\le 2\tilde B$. Since $\gamma\sim_{\tilde B} -1$ on $\mathcal R_\varsigma$ for $\mathcal S[\Psi]$, $\mathfrak u(\tilde u_f)-\u u{}_f>\xi$, where $\tilde u_f\doteq \u{\mathfrak u}(\u u{}_f)+\tilde\varsigma$ and $\xi=\xi(\tilde B,\tilde\varsigma)>0$.
    
    Let $\Psi'\in B_{\eta}^\mathfrak{Z}(\Psi)$, where $0<\eta\le \eta_0$, and define $\Psi(z)\doteq (1-z)\Psi +z\Psi'$ for $z\in [0,1]$. We will show below in \cref{lem:regularity-of-u} that if we consider $\mathfrak u$ as a joint function of $u$ and $z$, then $|\partial_z\mathfrak u|\les_{\tilde B} \|\Psi-\Psi'\|_\mathfrak{Z}$. Therefore, integrating this estimate, we obtain $\mathfrak u(\tilde u_f,1)-\u u{}_f>\frac 12\xi$ for $\eta$ sufficiently small, which shows that $\mathcal L_\mathrm{Bondi}> \u u{}_f-1$ for $\Psi'$. This completes the proof.
\end{proof}

\subsubsection{The advanced time coordinate \texorpdfstring{$\mathfrak v_{\u u{}_f}$}{v u bar f}}\label{sec:advanced}

Let $f:\Bbb R\to\Bbb R$ be a $C^\infty$ function such that $f(x)=x$ for $x\le -2$, $f(x)= 0$ for $x\ge 0$, and $f'\ge 0$ and $f''\le 0$ everywhere. For any $\u u{}_f\in\Bbb R$, define $\mathfrak g_{\u u{}_f}:\Bbb R\to\Bbb R$ by 
\begin{equation*}
    \mathfrak g_{\u u{}_f}(x) \doteq f(x-\u u{}_f-1)+\u u{}_f.
\end{equation*}
Note that
\begin{align}
 \label{eq:g-est-1} |\partial_x\mathfrak g_{\u u{}_f}|   &\les \mathbf 1_{\{x\le \u u{}_f+1\}},\\
 \label{eq:g-est-2} |\partial_{\u u{}_f}\mathfrak g_{\u u{}_f}|  &\les \mathbf 1_{\{x\ge \u u{}_f-1\}}.
\end{align}
\ul{This family of functions is fixed for the remainder of the paper.} The fact that $\partial_x\mathfrak g_{\u u{}_f}$ is only supported for $x\le \u u{}_f+1$ will be used crucially in essentially all of the estimates in \cref{sec:diffeo-ests} below.

We now use these functions to define a family of curves in (what will be) the teleological double null coordinate system.
\begin{lem}
    Let $\mathcal S$ be an asymptotically flat solution arising from seed data $\Psi\in \mathfrak M_{\u u{}_f}$ for some $\u u{}_f\ge 2$. Then the solution $\mathcal G^v$ to the ordinary differential equation
    \begin{equation}\label{eq:def-mathcal-G}
    \partial_s\mathcal G^v_{\u u{}_f}(s) = \frac{1}{\kappa(\u{\mathfrak u}(\mathfrak g_{\u u{}_f}(s)),\mathcal G^v_{\u u{}_f}(s))}
\end{equation}
 with initial condition $\mathcal G^v_{\u u{}_f}(1)=1$ exists and defines a $C^3$ diffeomorphism $\mathcal G^v_{\u u{}_f}:[1,\infty)\to [1,\infty)$.  
\end{lem}
\begin{proof}
    This is clear from the fact that $\kappa$ is bounded from above and below on $\mathcal R(1,\u{\mathfrak u}(\u u{}_f),1,\infty)$ by definition of asymptotic flatness.
\end{proof}
\begin{rk}
    We require $\u u{}_f\ge 2$ so that $\u{\mathfrak u}(\mathfrak g_{\u u{}_f}(1))=\u{\mathfrak u}(1)=1$.
\end{rk}

The following is now automatic:
\begin{prop}
    For any $\Psi\in \mathfrak M_{\u u{}_f}$ with $\u u{}_f\ge 2$, the formula
    \begin{equation}\label{eq:def-mathfrak-v}
    \mathfrak v(v,\u u{}_f)\doteq \mathfrak v_{\u u{}_f}(v) \doteq 1+\int_1^v \kappa(\u{\mathfrak u}(\mathfrak g_{\u u{}_f}(\hat{\mathcal G}{}^v_{\u u{}_f}(v'))),v')\,dv',
\end{equation} where $\hat{\mathcal G}^v_{\u u{}f}\doteq (\mathcal G^v_{\u u{}_f})^{-1}$,
defines a $C^3$ diffeomorphism $\mathfrak v(\cdot,u_f):[1,\infty)\to[1,\infty)$.
\end{prop}

\begin{defn}
    The function $\mathfrak v_{\u u{}_f}$ is called the \emph{$\u u{}_f$-normalized advanced time function}{\index{v2@$\mathfrak v_{\u u{}_f}$, $\u u{}_f$-normalized advanced time function}\index{v3@$\u{\mathfrak v}{}_{\u u{}_f}\doteq \mathfrak v{}_{\u u{}_f}^{-1}$}} of $\mathcal S_\mathrm{max}$. We write the associated advanced time coordinate as $\u v\doteq\mathfrak v_{\u u{}_f}(v)${\index{v4@$\u v{}_{\u u{}_f}$ (often $\u v$), $\u u{}_f$-normalized advanced time}} and the inverse diffeomorphism as $\u{\mathfrak v}{}_{\u u{}_f}\doteq \mathfrak v^{-1}_{\u u{}_f}$. The pair $(\mathfrak u,\mathfrak v_{\u u{}_f})${\index{u4@$(\mathfrak u,\mathfrak v_{\u u{}_f})$, $\u u{}_f$-normalized teleological gauge}} is called the \emph{$\u u{}_f$-normalized teleological gauge} of $\mathcal S_\mathrm{max}$ and defines a $C^3$ diffeomorphism 
    \begin{equation*}
        (\mathfrak u,\mathfrak v_{\u u{}_f}): \mathcal R(1,\u{\mathfrak u}(\u u{}_f),1,\infty)\to \mathcal R(1,\u u{}_f,1,\infty).
    \end{equation*}
    We will often omit the $\u u{}_f$ subscript on $\u v{}_{\u u{}_f}$, etc., when it has been fixed and the meaning is clear. 
\end{defn}

\subsubsection{Putting the solution in teleological gauge}\label{sec:putting}

Let $\Psi\in \mathfrak M_{\u u{}_f}$ with $\u u{}_f\ge 2$. Using the change-of-variables formulas \eqref{eq:COV-1}--\eqref{eq:COV-3}, we now define\index{_@$\u \cdot$, quantity in $\u u{}_f$-normalized teleological gauge}
\begin{align*}
    \u r(\u u,\u v,\u u{}_f)&\doteq r(\u{\mathfrak u}(\u u),\u{\mathfrak v}{}_{\u u{}_f}(\u v)),\\
     \u \varpi(\u u,\u v,\u u{}_f)&\doteq \varpi(\u{\mathfrak u}(\u u),\u{\mathfrak v}{}_{\u u{}_f}(\u v)),\\
     \u \phi(\u u,\u v,\u u{}_f)&\doteq \phi(\u{\mathfrak u}(\u u),\u{\mathfrak v}{}_{\u u{}_f}(\u v)),\\
     \u \nu(\u u,\u v,\u u{}_f)&\doteq \partial_{\u u}\u{\mathfrak u}(\u u) \nu(\u{\mathfrak u}(\u u),\u{\mathfrak v}{}_{\u u{}_f}(\u v)),\\
     \u \lambda(\u u,\u v,\u u{}_f)&\doteq \partial_{\u v}\u{\mathfrak v}_{\u u{}_f}(\u v) \lambda(\u{\mathfrak u}(\u u),\u{\mathfrak v}{}_{\u u{}_f}(\u v)),\\
     \u \gamma(\u u,\u v,\u u{}_f)&\doteq \partial_{\u u}\u{\mathfrak u}(\u u)\gamma(\u{\mathfrak u}(\u u),\u{\mathfrak v}{}_{\u u{}_f}(\u v)),\\
     \u \kappa(\u u,\u v,\u u{}_f)&\doteq\partial_{\u v}\u{\mathfrak v}{}_{\u u{}_f}(\u v) \kappa(\u{\mathfrak u}(\u u),\u{\mathfrak v}{}_{\u u{}_f}(\u v)),
\end{align*}
which gives a solution to the spherically symmetric EMSF system on $\mathcal R(1,\u u{}_f,1,\infty)$. We denote this solution by $\u{\mathcal S}{}_{\u u{}_f}[\Psi]$\index{S5@$\u{\mathcal S}{}_{\u u{}_f}[\Psi]$, $\mathcal S[\Psi]$ in $\u u{}_f$-normalized teleological gauge} and say that $\mathcal S[\Psi]$ has been written \emph{in $\u u{}_f$-normalized teleological gauge}. We also often write $\u r{}_{\u u{}_f}$ for $r(\cdot,\cdot,\u u{}_f)$, etc. 

\begin{prop}\label{prop:Bondi-conditions} 
    Let $\Psi\in \mathfrak M_{\u u{}_f}$ with $\u u{}_f\ge 2$. Put $\mathcal S[\Psi]$ into $\u u{}_f$-normalized teleological gauge. Then null infinity $\mathcal I^+$ of $\u{\mathcal S}{}_{\u u{}_f}[\Psi]$ is Bondi normalized in the sense that
    \begin{equation}\label{eq:gamma-BC}
        \lim_{\u v\to\infty} \u \nu{}_{\u u{}_f}(\u u,\u v)=   \lim_{\u v\to\infty} \u \gamma{}_{\u u{}_f}(\u u,\u v)=-1
    \end{equation}
    for every $\u u\in [1,\u u{}_f]$. The $\u v$-gauge is normalized by the condition that 
    \begin{equation}\label{eq:kappa-BC}
        \u \kappa{}_{\u u{}_f}\big|_{\mathcal G_{\u u{}_f}} = 1,
    \end{equation}
    where $\mathcal G_{\u u{}_f}$\index{G@$\mathcal G_{\u u{}_f}$, curve where $\u \kappa_{\u u{}_f} \equiv 1$} is the curve in the $(\u u,\u v)$-plane defined by $s\mapsto (\mathfrak g_{\u u{}_f}(s),s)$.
 \end{prop}
\begin{proof} \textsc{Proof of \eqref{eq:gamma-BC}}: We implicitly differentiate $\mathfrak u(\u{\mathfrak u}(\u u))=\u u$ and use the definition \eqref{eq:def-mathfrak-u} to derive 
\begin{equation}\label{eq:du-mathfrak-u-inverse}
    \partial_{\u u}\u{\mathfrak u}(\u u) = -\lim_{v\to\infty} \frac{1}{\gamma(\u{\mathfrak u}(\u u),v)}.
\end{equation}
Therefore, 
\begin{equation*}
      \lim_{\u v\to\infty} \u \gamma{}_{\u u{}_f}(\u u,\u v)= -\frac{\displaystyle\lim_{\u v\to\infty}\gamma(\u{\mathfrak u}(\u u),\u{\mathfrak v}{}_{\u u{}_f}(\u v))}{\displaystyle \lim_{v\to\infty}\gamma(\u{\mathfrak u}(\u u),v)}=-1
\end{equation*}
as $\u{\mathfrak v}{}_{\u u{}_f}(\u v)\to \infty$ as $v\to\infty$. Since $\u \mu{}_{\u u{}_f}\to 0$ as $\u v\to \infty$ in any asymptotically flat solution, we also obtain \eqref{eq:gamma-BC} for $\u \nu{}_{\u u{}_f}$ by the definition $\u \nu{}_{\u u{}_f}=(1-\u \mu{}_{\u u{}_f})\u \gamma{}_{\u u{}_f}$.

\textsc{Proof of \eqref{eq:kappa-BC}}: We differentiate \eqref{eq:def-mathfrak-v} and observe that
\begin{equation}\label{eq:dv-mathfrak-v}
    (\partial_v\mathfrak v_{\u u{}_f})(\mathcal G^v_{\u u{}_f}(s))= \kappa(\mathcal G^{u}_{\u u{}_f}(s),\mathcal G^v_{\u u{}_f}(s)), 
\end{equation}
where $ \mathcal G^{u}_{\u u{}_f}\doteq \u{\mathfrak u}\circ\mathfrak g_{\u u{}_f}$. We claim that $\mathfrak v_{\u u{}_f}(\mathcal G^v_{\u u{}_f}(s))=s$ for all $s\ge 1$. Indeed, using \eqref{eq:dv-mathfrak-v} and the definition \eqref{eq:def-mathcal-G} we simply compute
\begin{equation*}
    \partial_s\big(\mathfrak v_{\u u{}_f}(\mathcal G^v_{\u u{}_f}(s))\big) =  (\partial_v\mathfrak v_{\u u{}_f})(\mathcal G^v(s))\cdot\partial_s\mathcal G^v_{\u u{}_f} = 1.
\end{equation*} Therefore, $\mathcal G^v_{\u u{}_f}= \u{\mathfrak v}{}_{\u u{}_f}$. Now by implicitly differentiating $\mathfrak v_{\u u{}_f}(\u{\mathfrak v}{}_{\u u{}_f}(\u v))=\u v$ we obtain
\begin{equation}\label{eq:dv-mathfrak-v-inverse}
    \partial_{\u v}\u{\mathfrak v}{}_{\u u{}_f}(\u v) = \frac{1}{\kappa(\u{\mathfrak u}(\mathfrak g_{\u u{}_f}(\hat{\mathcal G}^v_{\u u{}_f}(\u{\mathfrak v}{}_{\u u{}_f}(\u v)))),\u{\mathfrak v}{}_{\u u{}_f}(\u v))} = \frac{1}{\kappa(\u{\mathfrak u}(\mathfrak g_{\u u{}_f}(\u v))),\u{\mathfrak v}{}_{\u u{}_f}(\u v))}
\end{equation}
and hence finally
\begin{equation*}
    \u \kappa{}_{\u u{}_f}(\mathfrak g_{\u u{}_f}(s),s) =1.\qedhere
\end{equation*}
\end{proof}
\subsubsection{Eschatological gauge in the black hole case}\label{sec:eschatological-definition}

When $\Psi\in \mathfrak M_\mathrm{black}$, then $\Psi\in \mathfrak M_{\u u{}_f}$ for every $\u u{}_f\ge 2$ by \cref{thm:dichotomy}, in fact,
\begin{equation*}
    \mathfrak M_\mathrm{black}=\bigcap_{\u u{}_f\ge 2} \mathfrak M_{\u u{}_f}.
\end{equation*} 
This means that we can put $\mathcal S[\Psi]$ into $\u u{}_f$-normalized teleological gauge for any $\u u{}_f\ge 2$. It is then natural to ask what happens if we let $\u u{}_f\to\infty$, but we first make an additional definition:
\begin{equation*}
    \mathfrak M_\mathrm{black}^{\kappa\sim 1}\doteq \{\Psi\in\mathfrak M_\mathrm{black}:\text{there exists $C>0$ such that $C^{-1}\le \kappa\le C$ on $\mathcal R(1,u_{\mathcal H^+},1,\infty)$}\}.
\end{equation*}
All of the black hole spacetimes constructed in this paper will lie in this set. 

\begin{rk}
    A positive resolution of the orbital stability portion of \cref{quest:large-data-1} would in particular imply that $ \mathfrak M_\mathrm{black}^{\kappa\sim 1} =  \mathfrak M_\mathrm{black}$ for the spherically symmetric EMSF model.
\end{rk}

\begin{lem}\label{lem:eschatology}
    If $\Psi\in \mathfrak M_\mathrm{black}^{\kappa\sim 1}$, the diffeomorphisms $\mathfrak v_{\u u{}_f}$ converge uniformly on compact sets to a $C^3$ diffeomorphism $\mathfrak v_\infty:[1,\infty)\to [1,\infty)$ as $\u u{}_f\to\infty$.
\end{lem}
\begin{defn}
    For $\Psi\in \mathfrak M_\mathrm{black}^{\kappa\sim 1}$, the pair $(\mathfrak u,\mathfrak v_\infty)$, which defines a $C^3$ diffeomorphism
    \begin{equation*}
        (\mathfrak u,\mathfrak v_\infty):[1,u_{\mathcal H^+})\times[1,\infty) \to [1,\infty)\times[1,\infty),
    \end{equation*}
    is called the \emph{eschatological gauge}\index{u5@$(\mathfrak u,\mathfrak v_\infty)$, eschatological gauge}.
\end{defn}
\begin{proof}[Proof of \cref{lem:eschatology}] Let $2\le \u u{}_f\le \u u{}_f'$. Note that for $1\le s\le \u u{}_f-1$, $\mathcal G^v_{\u u{}_f}(s)= \mathcal G^v_{\u u{}_f'}(s)$. Indeed, this follows immediately from the definition \eqref{eq:def-mathcal-G} and the fact that $  \mathfrak g_{\u u{}_f}(s)=  \mathfrak g_{\u u{}_f'}(s)$ for $1\le s\le \u u{}_f-1$. As shown in the proof of \cref{prop:Bondi-conditions}, $\mathfrak v_{\u u{}_f}^{-1}=\u{\mathfrak v}{}_{\u u{}_f}= \mathcal G^v_{\u u{}_f}$, so these inverse maps are eventually stationary on any compact set. By definition of $\mathfrak M_\mathrm{black}^{\kappa\sim 1}$, $\kappa\sim 1$ on $\mathcal R(1,u_{\mathcal H^+},1,\infty)$, so that $\mathfrak v_{\u u{}_f}^{-1}(\u u{}_f-1)=\mathfrak v_{\u u{}_f'}^{-1}(\u u{}_f-1)\sim \u u{}_f-1$. It follows $\mathfrak v_{\u u{}_f}$ is also eventually stationary on any compact set and hence converges to a $C^3$ diffeomorphism $[1,\infty)\to[1,\infty)$ as $\u u{}_f\to\infty$. \end{proof} 

In the eschatological gauge, the curve $\mathcal G_{\u u{}_f}$ degenerates to the diagonal $\Gamma=\{\u u=\u v\}$\index{g@$\Gamma=\{\u u=\u v\}$}, and \eqref{eq:kappa-BC} becomes
\begin{equation*}
    \u\kappa{}_{\infty}\big|_\Gamma = 1.
\end{equation*}

\subsection{Estimates for the diffeomorphisms}\label{sec:diffeo-ests}

In this section, we provide some fundamental estimates for the diffeomorphisms $\mathfrak u$ and $\mathfrak v_{\u u{}_f}$, and their derivatives, for families of asymptotically flat solutions.

Given families of seed data sets $z\mapsto \Psi(z)\in \mathfrak M_{\u u{}_f}$ or $(z_1,z_2)\mapsto \Psi(z_1,z_2)\in \mathfrak M_{\u u{}_f}$,  we can straightforwardly add dependence on these parameters to the definitions of $\mathfrak u$ and $\mathfrak v_{\u u{}_f}$. Indeed, we add dependence on $z$ or $(z_1,z_2)$ to the solutions as in \cref{sec:perturbation-theory} and then define
\begin{equation}\label{eq:diffeo-fam-1}
    \mathfrak u(u,z)\doteq 1-\lim_{v\to\infty}\int_1^u \gamma(u',v,z)\,du',
\end{equation}
\begin{equation}\label{eq:diffeo-fam-2}
    \partial_s\mathcal G^v_{\u u{}_f}(s,z)= \frac{1}{\kappa(\u{\mathfrak u}(\mathfrak g_{\u u{}_f}(s),z),\mathcal G^v_{\u u{}_f}(s,z),z)},
\end{equation}
\begin{equation}\label{eq:diffeo-fam-3}
     \mathfrak v(v,\u u{}_f,z) \doteq 1+\int_1^v \kappa(\u{\mathfrak u}(\mathfrak g_{\u u{}_f}(\hat{\mathcal G}^v_{\u u{}_f}(v',z)),z),v',z)\,dv',
\end{equation}
and similarly for the two-variable case, where $\u{\mathfrak u}$ denotes the inverse taken in the $u$ variable and $\hat{\mathcal G}^v_{\u u{}_f}$ denotes the inverse taken in the $s$ variable.

We make the following assumption in the following sections to quantify our estimates: 

\begin{ass}\label{ass:family} Let $\mathcal R$ be a semiglobal rectangle. Let $B\ge 1$ and $\u u{}_f\ge 2$. Let $\Psi\in \mathfrak M_{\u u{}_f}$, $\Psi_1,\Psi_2\in\mathfrak Z$, and define the two-parameter family $\Psi(z_1,z_2)= \Psi+z_1\Psi_1+z_2\Psi_2$. Let $U$ be a neighborhood of $(0,0)$ in $\Bbb R^2$ so that $\Psi(z_1,z_2)\in \mathfrak M_{\u u{}_f}$ and $\|\mathcal S[\Psi(z_1,z_2)]\|_{\mathcal R}\le B$ for all $(z_1,z_2)\in U$. Implicit constants in the $\les$ notation are only allowed to depend on $B$ and $\u u{}_f$.
\end{ass}

\subsubsection{Estimates for \texorpdfstring{$\mathfrak u$}{mathfrak u} and \texorpdfstring{$\u{\mathfrak u}$}{underline mathfrak u}}

Using \eqref{eq:partial-u-gamma-1},\eqref{eq:partial-u-gamma-2}, \eqref{eq:lin-gamma-weights-1}, \eqref{eq:lin-gamma-weights-2}, and \eqref{eq:qd-gamma-r-weight}, we easily observe:
\begin{lem}
    Under \cref{ass:family}, $\gamma$ extends to null infinity $\mathcal I^+$ in a $C_\star^2$-fashion. That is, the function
    \begin{equation*}
        \gamma_{\mathcal I^+}(u,z_1,z_2)\doteq \lim_{v\to\infty}\gamma(u,v,z_1,z_2)
    \end{equation*} is $C_\star^2$-regular and it holds that 
    \begin{equation*}
        \gamma_{\mathcal I^+}\sim -1,\qquad |\partial_u^{\le 2}\gamma_{\mathcal I^+}|\les 1,\qquad |\partial_u^{\le 2}\lin \gamma_{\mathcal I^+}|\les\|\lin\Psi_1\|_\mathfrak{Z}+\|\lin\Psi_2\|_\mathfrak{Z},\qquad |\partial_u^{\le 2}\qd \gamma_{\mathcal I^+}|\les  \|\lin\Psi_1\|_\mathfrak{Z}\|\lin\Psi_2\|_\mathfrak{Z} 
    \end{equation*}
    on $[1,u_1]\times U$. 
\end{lem} With this notation, we can write \begin{equation}\label{eq:formula-for-u-diffeo}\partial_u\mathfrak u(u,z_1,z_2)=-\gamma_{\mathcal I^+}(u,z_1,z_2)\quad\text{and}\quad \partial_{\u u}\u{\mathfrak u}(\u u,z_1,z_2)= - \frac{1}{\gamma_{\mathcal I^+}(\u{\mathfrak u}(\u u,z_1,z_2),z_1,z_2)}.\end{equation}
This immediately gives:
\begin{lem}\label{lem:regularity-of-u}
    Under \cref{ass:family}, $\mathfrak u\in C_\star^{3}$ and it holds that
\begin{equation*}
\partial_u\mathfrak u \sim 1,\qquad 
    |\partial_u^{\le 3}\mathfrak u|\les 1,\qquad
    |\partial_u^{\le 3}\lin{\mathfrak u}|\les\|\lin\Psi_1\|_\mathfrak{Z}+\|\lin\Psi_2\|_\mathfrak{Z} ,\qquad
    |\partial_u^{\le 3}\qd{\mathfrak u}|\les  \|\lin\Psi_1\|_\mathfrak{Z}\|\lin\Psi_2\|_\mathfrak{Z} 
\end{equation*}
    on $[1,u_1]\times U$.
\end{lem}
By implicit differentiation of the relation $\mathfrak u(\u{\mathfrak u}(\u u,z_1,z_2),z_1,z_2)=\u u$, we obtain:
\begin{lem}\label{lem:mathfrak-u-inverse}
     Under \cref{ass:family}, $\u{\mathfrak u}\in C_\star^{1}$ and it holds that 
   \begin{equation*}
    |\partial_{\u u}^{\le 1}\u{\mathfrak u}|\les 1,\qquad
    |\partial_{\u u}^{\le 1}\lin{\u{\mathfrak u}}|\les\|\lin\Psi_1\|_\mathfrak{Z}+\|\lin\Psi_2\|_\mathfrak{Z} ,\qquad
    |\partial_{\u u}^{\le 1}\qd{\u{\mathfrak u}}|\les  \|\lin\Psi_1\|_\mathfrak{Z}\|\lin\Psi_2\|_\mathfrak{Z} 
\end{equation*}
    on $[1,\u u{}_f]\times U$.
\end{lem}

Finally, we require a lemma quantitatively linking the quantities $\gamma$, $\gamma_{\mathcal I^+}$, and $\nu$ asymptotically.
    \begin{lem}\label{lem:gamma-nu-asymptotics}  Under \cref{ass:family}, it holds that
        \begin{align*}
           \partial_u^{\le 2}\gamma & =\partial_u^{\le 2}\gamma_{\mathcal I^+}+O(v^{-2}),\\
           \partial_u^{\le 2}\nu & = \partial_u^{\le 2}\gamma+O(v^{-1}),\\
           \partial_u^{\le 2}\lin\gamma & =\partial_u^{\le 2}\lin\gamma_{\mathcal I^+}+O\big(v^{-2}(\|\lin\Psi_1\|_\mathfrak{Z}+\|\lin\Psi_2\|_\mathfrak{Z})\big),\\
           \partial_u^{\le 2}\lin\nu & = \partial_u^{\le 2}\lin\gamma+O\big(v^{-1}(\|\lin\Psi_1\|_\mathfrak{Z}+\|\lin\Psi_2\|_\mathfrak{Z})\big),\\
           \partial_u^{\le 2}\qd\gamma & =\partial_u^{\le 2}\qd\gamma_{\mathcal I^+}+O\big(v^{-2}\|\lin\Psi_1\|_\mathfrak{Z}\|\lin\Psi_2\|_\mathfrak{Z}\big),\\
           \partial_u^{\le 2}\qd\nu & = \partial_u^{\le 2}\qd\gamma+O\big(v^{-1}\|\lin\Psi_1\|_\mathfrak{Z}\|\lin\Psi_2\|_\mathfrak{Z}\big)
        \end{align*}
        on $\mathcal R\times U$.
    \end{lem}
    \begin{proof}
        The asymptotics linking $\gamma$ and $\gamma_{\mathcal I^+}$ simply follow by integrating \eqref{eq:partial-u-gamma-1},\eqref{eq:partial-u-gamma-2}, \eqref{eq:lin-gamma-weights-1}, \eqref{eq:lin-gamma-weights-2}, and \eqref{eq:qd-gamma-r-weight}. To relate $\nu$ and $\gamma$, we write $\nu= (1-\mu)\gamma$ (which is the definition of $\gamma$), and differentiate this relation, and use asymptotic flatness and \cref{prop:first-order-ID,prop:second-order-ID}.
    \end{proof}

\subsubsection{Estimates for \texorpdfstring{$\mathcal G_{\u u{}_f}$}{G uf}}

\begin{lem}\label{lem:G-est-1}
      Under \cref{ass:family}, it holds that 
\begin{gather*}
    \mathcal G^v_{\u u{}_f}\sim s,\qquad \partial_s\mathcal G^v_{\u u{}_f}\sim 1,\qquad |\partial_s^2\mathcal G^v_{\u u{}_f}|\les s^{-2},\\
    |\lin{\mathcal G}^v_{\u u{}_f}|\les (\log s)\big(\|\lin\Psi_1\|_\mathfrak{Z}+\|\lin\Psi_2\|_\mathfrak{Z}\big) ,\qquad  |\partial_s\lin{\mathcal G}^v_{\u u{}_f}|\les s^{-1}\big(\|\lin\Psi_1\|_\mathfrak{Z}+\|\lin\Psi_2\|_\mathfrak{Z}\big) ,\\
    |\qd{\mathcal G}^v_{\u u{}_f}|\les (\log s)\|\lin\Psi_1\|_\mathfrak{Z}\|\lin\Psi_2\|_\mathfrak{Z}
\end{gather*}
      on $ [1,\infty)\times U$.
\end{lem}
\begin{proof} Since $\kappa\sim 1$, the first two estimates in the first line follow immediately from \eqref{eq:diffeo-fam-2}. To estimate $\partial_s^2\mathcal G^v_{\u u{}_f}$, we commute \eqref{eq:diffeo-fam-2} with $\partial_s$ and estimate, using asymptotic flatness and \eqref{eq:g-est-1},
\begin{equation*}
    |\partial_s^2\mathcal G^v_{\u u{}_f}|\les |\partial_u\kappa|\mathbf 1_{\{s\les 1\}} + |\partial_v\kappa| |\partial_s\mathcal G^v_{\u u{}_f}|\les s^{-2}.
\end{equation*}
To prove the second line, we commute \eqref{eq:diffeo-fam-2} with $\partial_z$ and use \eqref{lem:mathfrak-u-inverse} to obtain
\begin{equation*}
    |\partial_s\lin{\mathcal G}^v_{\u u{}_f}|\les |\partial_u\kappa||\lin{\u{\mathfrak u}}|+ |\partial_v\kappa||\lin{\mathcal G}^v_{\u u{}_f}|+|\lin\kappa|\les s^{-1}\big(\|\lin\Psi_1\|_\mathfrak{Z}+\|\lin\Psi_2\|_\mathfrak{Z}\big)+s^{-2}|\lin{\mathcal G}^v_{\u u{}_f}|.
\end{equation*} The desired estimate then follows from Gr\"onwall's lemma. The proof of the third line is similar and we omit the details. 
\end{proof}

\begin{lem}\label{lem:G-est-2}
    Under \cref{ass:family}, it holds that 
    \begin{equation*}
        |\partial_{\u u{}_f}\mathcal G^v_{\u u{}_f}|\les 1+\log s,\qquad
        |\partial_{\u u{}_f}\hat{\mathcal G}^v_{\u u{}_f}|\les (1+\log v)^{-1}
    \end{equation*}
    on $[1,\infty)\times U$. 
\end{lem}
\begin{proof}
    We differentiate \eqref{eq:diffeo-fam-2} in $\u u{}_f$ and use the assumptions $|\partial_u\kappa|\les r^{-1}\les v^{-1}$ and $|\partial_v\kappa|\les r^{-2}\les v^{-2}$ to obtain 
    \[|\partial_s(\partial_{\u u{}_f}\mathcal G^v_{\u u{}_f}(s))|\les s^{-2}|\partial_u\mathcal G^v_{\u u{}_f}(s)| + s^{-1}.\]
    An application of Gr\"onwall's inequality yields the first inequality. The second follows from implicit differentiation of the definition of $\hat{\mathcal G}^v_{\u u{}_f}$.  
\end{proof}

\subsubsection{Estimates for \texorpdfstring{$\mathfrak v$}{mathfrak v} and \texorpdfstring{$\u{\mathfrak v}$}{underline mathfrak v}}

\begin{lem}\label{lem:mathfrak-v-est}
    Under \cref{ass:family}, $\mathfrak v{}_{\u u{}_f}\in C_\star^{3}$ and it holds that
\begin{gather*}
    \mathfrak v_{\u u{}_f}\sim v,\qquad \partial_v \mathfrak v_{\u u{}_f}\sim 1,\qquad |\partial_v^2\mathfrak v_{\u u{}_f}|+|\partial_v^3\mathfrak v_{\u u{}_f}|\les v^{-2},\\
    |\lin{\mathfrak v}_{\u u{}_f}|\les(\log v)\big(\|\lin\Psi_1\|_\mathfrak{Z}+\|\lin\Psi_2\|_\mathfrak{Z}\big) ,\qquad |\partial_v\lin{\mathfrak v}_{\u u{}_f}|+|\partial_v^2\lin{\mathfrak v}_{\u u{}_f}|\les v^{-1}\big(\|\lin\Psi_1\|_\mathfrak{Z}+\|\lin\Psi_2\|_\mathfrak{Z} \big),\\
    |\qd{\mathfrak v}_{\u u{}_f}|\les(\log v)\|\lin\Psi_1\|_\mathfrak{Z}\|\lin\Psi_2\|_\mathfrak{Z} ,\qquad |\partial_v\qd{\mathfrak v}_{\u u{}_f}|\les v^{-1}\|\lin\Psi_1\|_\mathfrak{Z}\|\lin\Psi_2\|_\mathfrak{Z}
\end{gather*}
    on $[1,\infty)\times U$.
\end{lem}
\begin{proof}
By differentiating \eqref{eq:diffeo-fam-3}, we find
\begin{equation*}
    \partial_v\mathfrak v{}_{\u u{}_f}(v,z_1,z_2) = \kappa(\u{\mathfrak u}(\mathfrak g{}_{\u u{}_f}(\hat{\mathcal G}^v_{\u u{}_f}(v,z_1,z_2)),z_1,z_2),v,z_1,z_2),
\end{equation*}
which proves the first two inequalities in the first line since $\kappa\sim 1$. Differentiating in $v$, we obtain
\begin{equation*}
    |\partial_v^2\mathfrak v{}_{\u u{}_f}|\les |\partial_u\kappa| \mathbf 1_{\{v\les 1\}} + |\partial_v\kappa| \les v^{-2}.
\end{equation*}
The other estimates are proved similarly and we omit the details. 
\end{proof}
Again, by implicit differentiation, we have:
\begin{lem}\label{lem:mathfrak-v-inverse-est}
     Under \cref{ass:family}, $\u{\mathfrak v}{}_{\u u{}_f}\in C_\star^{1}$ and it holds that 
   \begin{equation*}
        |\lin{\u{\mathfrak v}}{}_{\u u{}_f}|\les(\log \u v)\big(\|\lin\Psi_1\|_\mathfrak{Z}+\|\lin\Psi_2\|_\mathfrak{Z}\big),\qquad   |\qd{\u{\mathfrak v}}{}_{\u u{}_f}|\les(\log \u v)\|\lin\Psi_1\|_\mathfrak{Z}\|\lin\Psi_2\|_\mathfrak{Z}
   \end{equation*}
    on $[1,\infty)\times U$.
\end{lem}

\begin{lem}\label{lem:v-update}
    Under \cref{ass:family}, it holds that 
\begin{gather*}
    |\partial_{\u u{}_f}\mathfrak v{}_{\u u{}_f}|\les \log v,\qquad |\partial_{\u u{}_f}\partial_v\mathfrak v{}_{\u u{}_f}|\les v^{-1},\\
       |\partial_{\u u{}_f}\u{\mathfrak v}{}_{\u u{}_f}|\les \log \u v,\qquad |\partial_{\u u{}_f}\partial_{\u v}\u{\mathfrak v}{}_{\u u{}_f}|\les \u v^{-1}
\end{gather*}
    on $[1,\infty)\times U$. 
\end{lem}
\begin{proof}
  We differentiate \eqref{eq:diffeo-fam-3} in $\u u{}_f$, use the assumptions $|\partial_u\kappa|\les v^{-1}$ and $|\partial_v\kappa|\les v^{-2}$, the estimates \eqref{eq:g-est-1} and \eqref{eq:g-est-2}, and \cref{lem:G-est-1,lem:G-est-2} to 
  obtain 
  \[|\partial_{\u u{}_f}\mathfrak v_{\u u{}_f}|\les \int_1^v v'^{-1}\,dv'= \log v.\]
The second estimate follows from implicit differentiation as usual.
\end{proof}

    \subsection{Perturbations of asymptotically flat solutions II: teleological gauge}\label{sec:PT-Bondi-gauge}

    We now face the issue of perturbing an asymptotically flat solution after it has been placed into teleological gauge. We do this by considering families of solutions as in \cref{sec:diffeo-ests}, putting them all into $\u u{}_f$-normalized teleological gauge as in \cref{sec:putting}, and then applying $\partial_{z_1}$ and $\partial_{z_2}$ with $\u u$ and $\u v$ fixed. 

\subsubsection{First-order perturbations}
    
\begin{prop}\label{prop:boundary-conditions} Let $\u u{}_f\ge 2$, $\Psi=(\phi_\circ,r_\circ,\varpi_\circ,\rho_\circ)\in \mathfrak M_{\u u{}_f}$, and $\lin\Psi=(\lin\phi_\circ,\lin r_\circ,\lin\varpi_\circ,\lin\rho_\circ)\in \mathfrak Z$. Consider the one-parameter family of solutions $\mathcal S[\Psi(z)]$, where $\Psi(z)\doteq \Psi+z\lin\Psi$. Then $\Psi(z)\in \mathfrak M_{\u u{}_f}$ for $z$ small and we may consider the one-parameter family of $\u u{}_f$ teleologically normalized solutions $\u{\mathcal S}{}_{\u u{}_f}[\Psi(z)]$ on the domain $\mathcal R(1,\u u{}_f,1,\infty)$. Defining the linearization operation $\lin\cdot$ to be application of $\partial_z$ to quantities in $\u{\mathcal S}{}_{\u u{}_f}[\Psi(z)]$ with $\u u$ and $\u v$ fixed, it holds that
    \begin{align}
     \label{eq:lin-r-BC}   \lin{\u r}{}_{\u u{}_f}(1,1,z)&= \lin  r_\circ,\\
       \label{eq:lin-varpi-BC}      \lin{\u \varpi}{}_{\u u{}_f}(1,1,z)&=\lin{\varpi}_\circ,\\
      \label{eq:lin-kappa-BC}  \lin{\u\kappa}{}_{\u u{}_f} \big|_{\mathcal G_{\u u{}_f}}&=0,\\
             \label{eq:lin-gamma-BC}   \lim_{\u v\to\infty}\lin{\u\gamma}{}_{\u u{}_f}& = 0
    \end{align}
    for all $z$ for which $\u{\mathcal S}{}_{\u u{}_f}[\Psi(z)]$ is defined.
\end{prop}
\begin{proof} By definition, the diffeomorphisms $\mathfrak u(\cdot,z)\times \mathfrak v_{\u u{}_f}(\cdot,z)$ fix the point $(1,1)$. It follows that
\[\u r{}_{\u u{}_f}(1,1,z)= r(1,1,z)=r_\circ + z\lin r_\circ,\quad\text{and}\quad \u \varpi{}_{\u u{}_f}(1,1,z)= \varpi(1,1,z)=\varpi_\circ + z\lin \varpi_\circ,\]
from which \eqref{eq:lin-r-BC} and \eqref{eq:lin-varpi-BC} follow immediately. Likewise, \eqref{eq:lin-kappa-BC} follows immediately from differentiating \eqref{eq:kappa-BC}, noting that the curve $\mathcal G_{\u u{}_f}$ is \emph{independent of $z$} by construction of the gauge. 

    \textsc{Proof of \eqref{eq:lin-gamma-BC}}: We begin with the definition
    \[\u{\gamma}{}_{\u u{}_f}(\u u,\u v,z)=-\frac{\gamma(\u{\mathfrak u}(\u u ,z),\u{\mathfrak v}{}_{\u u{}_f}(\u v,z),z)}{\gamma_{\mathcal I^+}(\u{\mathfrak u}(\u u,z),z)}\]
    and differentiate in $z$ to obtain
    \begin{align*}
      \partial_z\u{\gamma}{}_{\u u{}_f} &= \frac{\gamma}{\gamma_{\mathcal I^+}^2} \big(\partial_u\gamma_{\mathcal I^+}\lin{\u{\mathfrak u}}+\lin\gamma_{\mathcal I^+}\big) -\frac{1}{\gamma_{\mathcal I^+}}\big(\partial_u\gamma \,\lin{\u{\mathfrak u}}+\lin\gamma\big) + \partial_v\gamma \,\lin{\u{\mathfrak v}}{}_{\u u{}_f}\\ & = \left(\frac{\gamma}{\gamma_{\mathcal I^+}^2} \partial_u\gamma_{\mathcal I^+}-\frac{1}{\gamma_{\mathcal I^+}}\partial_u\gamma\right)\lin{\u{\mathfrak u}} + \left(\frac{\gamma}{\gamma_{\mathcal I^+}^2}\lin\gamma_{\mathcal I^+}-\frac{1}{\gamma_{\mathcal I^+}}\lin\gamma\right) + \partial_v\gamma \,\lin{\u{\mathfrak v}}{}_{\u u{}_f},   
    \end{align*}
    where all implicit arguments in $u$ are evaluated at $\u{\mathfrak u}(\u u,z)$ and implicit arguments in $v$ are evaluated at $\u{\mathfrak v}{}_{\u u{}_f}(\u v,z)$. For $z$ small, the family $\Psi(z)$ satisfies \cref{ass:family} by \cref{lem:Mi-open-quantitative} and its proof and hence we may use \cref{lem:mathfrak-u-inverse,lem:gamma-nu-asymptotics,lem:mathfrak-v-inverse-est} to estimate $|\partial_z\u{\gamma}{}_{\u u{}_f}|\les \u v^{-1}$, which immediately implies \eqref{eq:lin-gamma-BC}.
\end{proof}

\begin{rk}
    Since the Einstein equations are invariant to changes of the double null gauge, the family of $\u u{}_f$-normalized quantities satisfy the same set of equations as the initial-data normalized quantities. It follows that they also satisfy the same linearized (and second-order) equations and then \cref{prop:boundary-conditions} will be used to define boundary conditions for the transport equations. We will wait until \cref{sec:estimates-linear-perturbations} to actually use the linearized equations in teleological gauge, however, as all of the required ``local'' results for our soft arguments can be derived by differentiating the definitions of $\u u{}_f$-normalized quantities and using the initial data gauge estimates as we did in the previous proposition. 
\end{rk}

We collect now various useful linearized estimates.

\begin{prop}\label{prop:teleological-1}
    Under \cref{ass:family}, it holds that
{\allowdisplaybreaks\begin{align}
    |\lin{\u r}{}_{\u u{}_f}|&\les ({\log \u v}) \|\lin\Psi\|_\mathfrak{Z},\\
        |\lin{\u \nu}{}_{\u u{}_f}|&\les \|\lin\Psi\|_\mathfrak{Z},\\
         \label{eq:lin-varpi-1}   |\lin{\u\varpi}{}_{\u u{}_f}| &\les \|\lin\Psi\|_\mathfrak{Z},\\
      \label{eq:lin-varpi-2}  |\partial_{\u v}\lin{\u\varpi}{}_{\u u{}_f}| &\les \u v^{-2}(\log \u v)\|\lin\Psi\|_\mathfrak{Z}\\
       |\lin{\u \psi}{}_{\u u{}_f}|&\les \u v^{-1}\|\lin\Psi\|_\mathfrak{Z},\\
        |\partial_{\u u}\lin{\u\psi}{}_{\u u{}_f}| &\les \u v^{-1}\|\lin\Psi\|_\mathfrak{Z},\\
         |\partial_{\u v}\lin{\u\psi}{}_{\u u{}_f}| &\les \u v^{-2}({\log \u v})\|\lin\Psi\|_\mathfrak{Z}
\end{align}}
    on $\mathcal R(1,\u u{}_f,1,\infty)\times U$.
\end{prop}
\begin{proof}
 We explain how to prove \eqref{eq:lin-varpi-1} and \eqref{eq:lin-varpi-2} as the others are similar. We differentiate the definition $\u\varpi{}_{\u u{}_f}(\u u,\u v,z)= \varpi(\u{\mathfrak u}(\u u ,z),\u{\mathfrak v}{}_{\u u{}_f}(\u v,z),z)$ in $z$ and use asymptotic flatness to obtain 
    \begin{equation}\label{eq:lin-varpi-formula}|\lin{\u\varpi}{}_{\u u{}_f}|\les |\partial_u\varpi||\lin{\u{\mathfrak u}}|+|\partial_v\varpi||\lin{\u{\mathfrak v}}{}_{\u u{}_f}|+|\lin\varpi|\les \|\lin\Psi\|_\mathfrak{Z}.\end{equation}
    Differentiating first in $\u v$ and then in $z$, and using \eqref{eq:dv-mathfrak-v-inverse} gives
    \[|\partial_{\u v}\lin{\u\varpi}{}_{\u u{}_f}|\les |\partial_v\varpi|\big(|\partial_u\kappa||\lin{\u{\mathfrak u}}|+|\partial_v\kappa||\lin{\u{\mathfrak v}}{}_{\u u{}_f}|\big)+|\partial_u\partial_v\varpi||\lin{\u{\mathfrak u}}|+ |\partial_v^2\varpi||\lin{\u{\mathfrak v}}{}_{\u u{}_f}|+|\partial_v\lin\varpi|.\]
The only term we have not yet estimated is $\partial_v\lin\varpi$, but we can readily read off that $|\partial_v\lin\varpi|\les v^{-2}\|\lin\Psi\|_\mathfrak{Z}$ from \eqref{eq:first-order-varpi-v}. This completes the proof.
\end{proof}

The following lemma will be used to control the initial energy of $\lin{\u\phi}{}_{\u u{}_f}$ in \cref{sec:estimates-linear-perturbations} later. 
\begin{lem}\label{lem:lin-phi-data} Under \cref{ass:family}, it holds that
\begin{align}
 |\u{r}{}_{\u u{}_f} \lin{\u\phi}{}_{\u u{}_f}(1,\u v) |   & \les \|\phi_\circ\|_\mathfrak{F}\u{r}{}_{\u u{}_f}^{-1}|\lin{\u r}{}_{\u u{}_f}(1,\u v)-\lin r_\circ|+\|\lin\phi_\circ\|_\mathfrak{F},\label{eq:linphidata1} \\
   |\u{r}{}_{\u u{}_f}^2\partial_{\u v} \lin{\u\phi}{}_{\u u{}_f}(1,\u v) |  & \les \|\phi_\circ\|_\mathfrak{F}|\u{\lambda}_{\u u{}_f}(1,\u v)||\lin{\u r}{}_{\u u{}_f}(1,\u v)-\lin r_\circ|+\|\phi_\circ\|_\mathfrak{F}|\lin{\u \lambda}{}_{\u u{}_f}(1,\u v)| + \|\lin\phi_\circ\|_\mathfrak{F}|\u\lambda{}_{\u u{}_f}(1,\u v)|, \label{eq:linphidata2}\\
  |\u r_{\u u{}_f}^2\partial_{\u v}(\u r{}_{\u u{}_f} \lin{\u\phi}{}_{\u u{}_f})(1,\u v) | & \les    \| \phi_{\circ}\|_{\mathfrak F} |\u{\lambda}{}_{\u u_f}(1,\u v)| |\lin{\u r}{}_{\u u{}_f}(1,\u v)-\lin r_\circ|+ \| \phi_{\circ}\|_{\mathfrak F}  \u{r}{}_{\u u{}_f} |\lin{\u \lambda}_{\u u{}_f}(1,\u v)|  + \| \lin \phi_\circ\|_{\mathfrak F} |\u{\lambda}{}_{\u u_f}(1,\u v)|, \label{eq:linphidata3}\\  | \partial_{\u u}\lin{\u\phi}{}_{\u u{}_f}(\u u,1) |   & \les \|\phi_\circ\|_\mathfrak{F}|\u{\nu}_{\u u{}_f}(\u u,1)||\lin{\u r}{}_{\u u{}_f}(\u u,1)-\lin r_\circ|+\|\phi_\circ\|_\mathfrak{F}|\lin{\u \nu}{}_{\u u{}_f}(\u u,1)| + \|\lin\phi_\circ\|_\mathfrak{F}|\u\nu{}_{\u u{}_f}(\u u,1)|  \label{eq:linphidata4}
\end{align}   at $z=0$, where the implicit constants depend only on an upper bound for $r_\circ$. 
\end{lem}
\begin{proof}
    Recall the definitions \[\u{\nu}{}_{\u u{}_f}(\u u,\u v,z)= \partial_{\u u}\u{\mathfrak u}(\u u,z)\nu(\u{\mathfrak u}(\u u,z),\u{\mathfrak v}{}_{\u u{}_f}(\u v,z),z)\quad\text{and}\quad\u{\lambda}{}_{\u u{}_f}(\u u,\u v,z)= \partial_{\u v}\u{\mathfrak v}{}_{\u u{}_f}(\u v,z)\lambda(\u{\mathfrak u}(\u u,z),\u{\mathfrak v}{}_{\u u{}_f}(\u v,z),z).\]
    Now set $\u v=1$ in the first equation, $\u u =1$ in the second, and recall the normalization conditions $\u{\mathfrak u}(1,z)=1$, $\u{\mathfrak v}{}_{\u u{}_f}(1,z)=1$, $\nu(u,1,z)=-1$, and $\lambda(1,v,z)=1$ to see that 
    \begin{equation}\label{eq:diffeo-form}
        \partial_{\u u}\u{\mathfrak u}(\u u,z) = -\u{\nu}{}_{\u u{}_f}(\u u,1,z)\quad\text{and}\quad \partial_{\u v}\u{\mathfrak v}{}_{\u u{}_f}(\u v,z)=\u{\lambda}{}_{\u u{}_f}(1,\u v,z).
    \end{equation}
    By integration in $\u u$ and $\u v$, respectively, we then obtain
    \begin{equation}\label{eq:inverse-gauge-formula}
        \u{\mathfrak u}(\u u,z) =1 - \u r{}_{\u u{}_f}(\u u,1,z)+ r_\circ +z\lin r_\circ  \quad\text{and}\quad \u{\mathfrak v}{}_{\u u{}_f}(\u v,z) =1 + \u r{}_{\u u{}_f}(1,\u v,z)- r_\circ -z\lin r_\circ.
    \end{equation}

By definition,  $\u{\phi}{}_{\u u{}_f}(\u u,\u v,z)=\phi(\u{\mathfrak u}(\u u,z),\u{\mathfrak v}{}_{\u u{}_f}(\u v,z),z)$, 
from which we compute 
    \begin{align*} \lin{\u\phi}{}_{\u u{}_f}(1,\u v)  
& =  (\partial_z \phi_\circ) (1,\u{\mathfrak v}{}_{\u u{}_f}(\u v,z),z) + (\partial_v \phi_\circ) ( 1,\u{\mathfrak v}{}_{\u u{}_f}(\u v,z),z)( \partial_z \u{\mathfrak v}{}_{\u u{}_f} ) (\u v,z)\\
 & =  (\partial_z \phi_\circ) (1,\u{\mathfrak v}{}_{\u u{}_f}(\u v,z),z) + (\partial_v \phi_\circ) ( 1,\u{\mathfrak v}{}_{\u u{}_f}(\u v,z),z)(\lin{\u r}{}_{\u u{}_f}(1,\u v,z)-\lin r_\circ)
    \end{align*} 
    which upon multiplying by $\u r{}_{\u u{}_f}(1,\u v,z)$ and using \eqref{eq:norm-defn} gives \eqref{eq:linphidata1}. Taking one derivative gives
\begin{align*}
         \partial_{\u v} \lin{\u\phi}{}_{\u u{}_f}(1,\u v,z)  
&=  (\partial_z\partial_v  \phi_\circ) (1,\u{\mathfrak v}{}_{\u u{}_f}(\u v,z),z) \partial_{\u v} \u{\mathfrak v}{}_{\u u{}_f}(\u v,z)\\
 &\quad+ (\partial_v^2 \phi_\circ) ( 1,\u{\mathfrak v}{}_{\u u{}_f}(\u v,z),z)( \partial_z \u{\mathfrak v}{}_{\u u{}_f} ) (\u v,z) \partial_{\u v} \u{\mathfrak v}{}_{\u u{}_f}(\u v,z)\\
 &\quad+ (\partial_v \phi_\circ)( 1,\u{\mathfrak v}{}_{\u u{}_f}(\u v,z),z)( \partial_z \partial_{\u v}\u{\mathfrak v}{}_{\u u{}_f} ) (\u v,z)\\ 
 &=   (\partial_z\partial_v  \phi_\circ) (1,\u{\mathfrak v}{}_{\u u{}_f}(\u v,z),z) \u{\lambda}{}_{\u u{}_f}(1,\u v,z)\\
 &\quad+ (\partial_v^2 \phi_\circ) ( 1,\u{\mathfrak v}{}_{\u u{}_f}(\u v,z),z)
(\lin{\u r}{}_{\u u{}_f}(1,\u v,z)-\lin r_\circ)
 \u{\lambda}{}_{\u u{}_f}(1,\u v,z)\\
 &\quad+ (\partial_v \phi_\circ)( 1,\u{\mathfrak v}{}_{\u u{}_f}(\u v,z),z) \lin{\u \lambda}_{\u u{}_f}(1,\u v,z).
\end{align*}
We note that  \[|v^2 \partial_v \phi_\circ|\leq |v \phi_\circ| + |v \partial_v (v \phi_\circ)|\les \| \phi_\circ \|_\mathfrak F\quad \text{and}\quad |v^3 \partial_v^2  \phi_\circ|\leq |v^2 \partial_v(v\phi_\circ)| + 2|v^2 \partial_v \phi_\circ|\les \| \phi_\circ \|_\mathfrak F,\] so multiplying $  \partial_{\u v} \lin{\u\phi}{}_{\u u{}_f}(1,\u v,z)  $   by $\u r{}_{\u u{}_f}^2(1,\u v,z)$ and using \eqref{eq:norm-defn} gives \eqref{eq:linphidata2}. 
For \eqref{eq:linphidata3}, we compute  
\begin{align*}
    \partial_{\u v}(\u r{}_{\u u{}_f} \lin{\u\phi}{}_{\u u{}_f})(1,\u v,z)  
   & =  \partial_{\u v}\u r{}_{\u u{}_f}(1,\u v,z) \lin{\u\phi}{}_{\u u{}_f}(1,\u v) + \u r{}_{\u u{}_f}(1,\u v,z) \partial_{\u v}\lin{\u\phi}{}_{\u u{}_f}(1,\u v,z)
    \\  & =   \u{\lambda}{}_{\u u{}_f}(1,\u v,z) \Big[ (\partial_z \phi_\circ) (1,\u{\mathfrak v}{}_{\u u{}_f}(\u v,z),z) + (\partial_v \phi_\circ) ( 1,\u{\mathfrak v}{}_{\u u{}_f}(\u v,z),z)(\lin{\u r}{}_{\u u{}_f}(1,\u v,z)-\lin r_\circ)\Big] \\
    &\quad + \u r{}_{\u u{}_f}(1,\u v,z)  \Big[  (\partial_z\partial_v  \phi_\circ) (1,\u{\mathfrak v}{}_{\u u{}_f}(\u v,z),z) \u{\lambda}{}_{\u u{}_f}(1,\u v,z)\\
    &\quad+ (\partial_v^2 \phi_\circ) ( 1,\u{\mathfrak v}{}_{\u u{}_f}(\u v,z),z)
(\lin{\u r}{}_{\u u{}_f}(1,\u v,z)-\lin r_\circ)
 \u{\lambda}{}_{\u u{}_f}(1,\u v,z)\\
 &+ (\partial_v \phi_\circ)( 1,\u{\mathfrak v}{}_{\u u{}_f}(\u v,z),z) \lin{\u \lambda}_{\u u{}_f}(1,\u v,z)\Big]
 \\ & =   \u{\lambda}{}_{\u u{}_f}(1,\u v,z) ( \partial_v (r \partial_z \phi_\circ) )(1,\u{\mathfrak v}{}_{\u u{}_f}(\u v,z),z)  \\
    & \quad+ \u{\lambda}{}_{\u u{}_f}(1,\u v,z) (\lin{\u r}{}_{\u u{}_f}(1,\u v)-\lin r_\circ) ( \partial_v \phi_\circ + r \partial_v^2 \phi_\circ) ( 1,\u{\mathfrak v}{}_{\u u{}_f}(\u v,z),z)\\
 &\quad +   (r \partial_v \phi_\circ)( 1,\u{\mathfrak v}{}_{\u u{}_f}(\u v,z),z) \lin{\u \lambda}_{\u u{}_f}(1,\u v,z).
\end{align*}
We note that \[ \partial_v (r \partial_z \phi_\circ) = \partial_v ( v \partial_z\phi_\circ) + (r_\circ + z \lin r_\circ-1 ) \partial_v \partial_z\phi_\circ \] 
so that  \[ |\u r{}_{\u u{}_f}^2(1,\u v)  \partial_v (r \partial_z \phi_\circ) |\lesssim \| \lin \phi_\circ \|_{\mathfrak F}.\] Similarly, using again that  $v^2 \partial_v \phi_\circ$ and $v^3 \partial_v^2 \phi_\circ$ are bounded by  $\| \phi_\circ \|_\mathfrak F$, we control the other two terms and obtain \eqref{eq:linphidata3}. 

For \eqref{eq:linphidata4}---analogously to $ \partial_{\u v} \lin{\u\phi}{}_{\u u{}_f}(1,\u v)$ above---we have
\begin{align*}
    \partial_{\u u}\lin{\u\phi}{}_{\u u{}_f}(\u u,1)  
&=  (\partial_z\partial_u  \phi_\circ) (\u{\mathfrak u}(\u u,z),1,z) \partial_{\u u} \u{\mathfrak u}(\u u,z)\\
    &\quad+ (\partial_u^2 \phi_\circ) ( \u{\mathfrak u}(\u u,z),1,z)( \partial_z \u{\mathfrak u} ) (\u u,z) \partial_{\u u} \u{\mathfrak u}(\u u,z)\\
    &\quad+ (\partial_u \phi_\circ)( \u{\mathfrak u}(\u u,z),1,z)( \partial_z \partial_{\u u}\u{\mathfrak u} ) (\u u,z)\\ 
&=   (\partial_z\partial_u  \phi_\circ) (\u{\mathfrak u}(\u u,z),1,z) (-\u{\nu}{}_{\u u{}_f}(\u u,1,z))\\
    &\quad+ (\partial_u^2 \phi_\circ) ( \u{\mathfrak u}(\u u,z),1,z)
(\lin r_\circ-\lin{\u r}{}_{\u u{}_f}(\u u,1,z)) (-\u{\nu}{}_{\u u{}_f}(\u u,1,z))\\
 &\quad+ (\partial_u \phi_\circ)( \u{\mathfrak u}(\u u,z),1,z) (-\lin{\u \nu}{}_{\u u{}_f}(\u u,1,z)),
\end{align*}
which completes the proof upon using the definition \eqref{eq:norm-defn} again.
\end{proof}

\subsubsection{Second-order perturbations}

In the following proposition, the first three estimates will be used later  in \cref{sec:proof-of-main-thm} as a part of a continuity argument for first-order perturbations, and the fourth estimate will be used to show that the Bondi mass is $C^1$ in \cref{sec:cts-differentiability}.
\begin{prop}\label{prop:second-order-teleology} Define the $\qd\cdot$ operation on teleologically normalized quantities analogously to $\lin\cdot$ in \cref{prop:boundary-conditions}. Under \cref{ass:family}, it holds that
    \begin{align}
   \label{eq:qd-r-est}  |\qd{\u r}{}_{\u u{}_f}|& \les (\log \u v)\|\lin\Psi_1\|_\mathfrak{Z}\|\lin\Psi_2\|_\mathfrak{Z} ,\\
      \label{eq:qd-nu-est}  |\qd{\u\nu}{}_{\u u{}_f}|& \les  \u v^{-1}\|\lin\Psi_1\|_\mathfrak{Z}\|\lin\Psi_2\|_\mathfrak{Z}   ,\\
     \label{eq:qd-lambda-est}      |\qd{\u\lambda}{}_{\u u{}_f}|& \les  \u v^{-1}\|\lin\Psi_1\|_\mathfrak{Z}\|\lin\Psi_2\|_\mathfrak{Z},\\
   \label{eq:varpi-2}  |\qd{\u \varpi}{}_{\u u{}_f}|& \les \|\lin\Psi_1\|_\mathfrak{Z}\|\lin\Psi_2\|_\mathfrak{Z}
    \end{align}
    on $\mathcal R(1,\u u{}_f,1,\infty)\times U$.
\end{prop}
\begin{proof}
    \textsc{Proof of \eqref{eq:qd-r-est}}:  We begin with the definition $\u r{}_{\u u{}_f}(\u u,\u v, z_1,z_2) = r(\u{\mathfrak u}(\u u,z_1,z_2),\u{\mathfrak v}{}_{\u u{}_f}(\u v, z_1,z_2),z_1,z_2)$
     and differentiate once in $z_i$ to obtain $\lin{\u r}{}_{\u u{}_f}= \nu \,\lin{\u{\mathfrak u}}+\lambda \,\lin{\u{\mathfrak v}}{}_{\u u{}_f} +\lin r$,  where all implicit arguments in $u$ are evaluated at $\u{\mathfrak u}(\u u,z_1,z_2)$ and implicit arguments in $v$ are evaluated at $\u{\mathfrak v}{}_{\u u{}_f}(\u v,z_1,z_2)$. We will adopt this notational convention for the rest of the proof. 
    We differentiate again to obtain
    \[ \qd{\u r}{}_{\u u{}_f}= \big(\partial_u\nu\,\lin{\u{\mathfrak u}}+\partial_v\nu\,\lin{\u{\mathfrak v}}{}_{\u u{}_f}+\lin\nu\big)\cdot\lin{\u{\mathfrak u}}+\nu\, \qd{\u{\mathfrak u}}+ \big(\partial_u\lambda\,\lin{\u{\mathfrak u}}+\partial_v\lambda\,\lin{\u{\mathfrak v}}{}_{\u u{}_f}+\lin\lambda\big)\cdot\lin{\u{\mathfrak v}}{}_{\u u{}_f}+\lambda\, \qd{\u{\mathfrak v}}{}_{\u u{}_f} + \lin\nu\cdot \lin{\u{\mathfrak u}} +\lin\lambda\cdot\lin{\u{\mathfrak v}}{}_{\u u{}_f}+ \qd r .\]
    Now \eqref{eq:qd-r-est} follows from \cref{lem:mathfrak-u-inverse,lem:mathfrak-v-inverse-est} and asymptotic flatness.

    \textsc{Proof of \eqref{eq:qd-nu-est}}: This estimate is much more involved and requires tracking various cancellations due to \cref{lem:gamma-nu-asymptotics}. We begin with the definition
    \[\u \nu{}_{\u u{}_f}(\u u,\u v, z_1,z_2) = -\frac{\nu(\u{\mathfrak u}(\u u,z_1,z_2),\u{\mathfrak v}{}_{\u u{}_f}(\u v, z_1,z_2),z_1,z_2)}{\gamma_{\mathcal I^+}(\u{\mathfrak u}(\u u,z_1,z_2), z_1,z_2)}\]
    and differentiate one to obtain
    \begin{align*}
        \lin{\u\nu}{}_{\u u{}_f}& = \frac{\nu}{\gamma_{\mathcal I^+}^2}\big(\partial_u\gamma_{\mathcal I^+}\lin{\u{\mathfrak u}}+ \lin\gamma_{\mathcal I^+}\big) - \frac{1}{\gamma_{\mathcal I^+}}\big(\partial_u\nu\,\lin{\u{\mathfrak u}}+ \partial_v\nu\,\lin{\u{\mathfrak v}}{}_{\u u{}_f}+\lin\nu\big)\\&=\left[\frac{\nu}{\gamma_{\mathcal I^+}^2} \partial_u\gamma_{\mathcal I^+}-\frac{1}{\gamma_{\mathcal I^+}}\partial_u\nu\right]\lin{\u{\mathfrak u}} + \left[\frac{\nu}{\gamma_{\mathcal I^+}^2}\lin\gamma_{\mathcal I^+}-\frac{1}{\gamma_{\mathcal I^+}}\lin\nu\right] -\frac{1}{\gamma_{\mathcal I^+}} \partial_v\nu \,\lin{\u{\mathfrak v}}{}_{\u u{}_f}. 
    \end{align*}
    Note that by invoking \cref{lem:gamma-nu-asymptotics}, we can see from this that $|\lin{\u\nu}{}_{\u u{}_f}|\les \u v^{-1}\big(\|\lin\Psi_1\|_\mathfrak{Z}+\|\lin\Psi_2\|_\mathfrak{Z} \big).$ 
    We differentiate again to obtain
\begin{align*}
         \qd{\u\nu}{}_{\u u{}_f}& = \left[\frac{\partial_u\gamma_{\mathcal I^+}}{\gamma_{\mathcal I^+}^2}\big(\partial_u\nu\,\lin{\u{\mathfrak u}}+\partial_v\nu\,\lin{\u{\mathfrak v}}{}_{\u u{}_f}+\lin\nu\big)-2\frac{\nu\partial_u\gamma_{\mathcal I^+}}{\gamma_{\mathcal I^+}^3}\big(\partial_u\gamma_{\mathcal I^+}\lin{\u{\mathfrak u}}+\lin\gamma_{\mathcal I^+}\big)+\frac{\nu}{\gamma_{\mathcal I^+}^2}\big(\partial_u^2\gamma_{\mathcal I^+}\lin{\u{\mathfrak u}}+\partial_u\lin\gamma_{\mathcal I^+}\big)\right.\\
         &\qquad +\left. \frac{\partial_u\nu}{\gamma_{\mathcal I^+}^2} \big(\partial_u\gamma_{\mathcal I^+}\lin{\u{\mathfrak u}}+\lin\gamma_{\mathcal I^+}\big) - \frac{1}{\gamma_{\mathcal I^+}}\big(\partial_u^2\nu\,\lin{\u{\mathfrak u}}+\partial_u\partial_v\nu \,\lin{\u{\mathfrak v}}{}_{\u u{}_f}+\partial_u\lin\nu\big)\right]\cdot \lin{\u{\mathfrak u}}\\
         &\quad  + \left[  \frac{1}{\gamma^2_{\mathcal I^+}}\big(\partial_u\nu \,\lin{\u{\mathfrak u}}+\partial_v\nu\,\lin{\u{\mathfrak v}}{}_{\u u{}_f}+\lin\nu\big)\cdot \lin\gamma_{\mathcal I^+} - 2\frac{\nu}{\gamma^3_{\mathcal I^+}}\big(\partial_u\gamma_{\mathcal I^+}\,\lin{\u{\mathfrak u}}+\lin\gamma_{\mathcal I^+}\big)\cdot\lin\gamma_{\mathcal I^+} +\frac{\nu}{\gamma^2_{\mathcal I^+}}\big(\partial_u\lin\gamma_{\mathcal I^+}\cdot\lin{\u{\mathfrak u}}+\qd\gamma_{\mathcal I^+}\big)\right.\\
         & \qquad +\left.\frac{1}{\gamma^2_{\mathcal I^+}}\big(\partial_u\gamma_{\mathcal I^+}\lin{\u{\mathfrak u}}+\lin\gamma_{\mathcal I^+}\big)\cdot\lin\nu -\frac{1}{\gamma_{\mathcal I^+}}\big(\partial_u\lin\nu\cdot\lin{\u{\mathfrak u}}+\partial_v\lin\nu\cdot\lin{\u{\mathfrak v}}{}_{\u u{}_f}+\qd\nu\big) \right] \\
         & \quad +\left[\frac{\partial_v\nu}{\gamma^2_{\mathcal I^+}} \big(\partial_u\gamma_{\mathcal I^+}\lin{\u{\mathfrak u}}+\lin\gamma_{\mathcal I^+}\big)\cdot \lin{\u{\mathfrak v}}{}_{\u u{}_f} - \frac{1}{\gamma_{\mathcal I^+}}\big(\partial_u\partial_v\nu\,\lin{\u{\mathfrak u}}+ \partial_v^2\nu \,\lin{\u{\mathfrak v}}{}_{\u u{}_f} +\partial_v\lin\nu\big)\cdot \lin{\u{\mathfrak v}}{}_{\u u{}_f} -\frac{\partial_v\nu}{\gamma_{\mathcal I^+}}\qd{\u{\mathfrak v}}{}_{\u u{}_f}  \right]\\
         & \doteq  [1]\cdot \lin{\u{\mathfrak u}} + [2] + [3].
    \end{align*}
    We regroup the terms in $[1]$ as follows:
    \begin{align*}
        [1]& = \left\{\left(\frac{\partial_u\nu}{\gamma_{\mathcal I^+}^2}-2\frac{\nu}{\gamma_{\mathcal I^+}^3}\partial_u\gamma_{\mathcal I^+}\right)\big(\partial_u\gamma_{\mathcal I^+}\lin{\u{\mathfrak u}}+\lin\gamma_{\mathcal I^+}\big)+ \frac{\partial_u\gamma_{\mathcal I^+}}{\gamma^2_{\mathcal I^+}}\big(\partial_u\nu \,\lin{\u{\mathfrak u}}+\lin\nu\big)\right\},\\
        &\quad  + \left\{\left(\frac{\nu}{\gamma^2_{\mathcal I^+}}\partial_u^2\gamma_{\mathcal I^+}-\frac{1}{\gamma_{\mathcal I^+}}\partial_u^2\nu\right) \lin{\u{\mathfrak u}}\right\}+ \left\{\frac{\nu}{\gamma^2_{\mathcal I^+}}\partial_u\lin\gamma_{\mathcal I^+}-\frac{1}{\gamma_{\mathcal I^+}}\partial_u\lin\nu\right\}+ \frac{\partial_u\gamma_{\mathcal I^+}}{\gamma^2_{\mathcal I^+}}\partial_v\nu \,\lin{\u{\mathfrak v}}{}_{\u u{}_f} - \frac{1}{\gamma_{\mathcal I^+}}\partial_u\partial_v\nu \,\lin{\u{\mathfrak v}}{}_{\u u{}_f}.
    \end{align*}
    The terms in curly braces cancel up to $O(\u v^{-1}\|\lin\Psi_1\|_\mathfrak{Z}\|\lin\Psi_2\|_\mathfrak{Z} )$ errors by \cref{lem:gamma-nu-asymptotics} and the final two terms also decay by asymptotic flatness. Similarly, we regroup the terms in $[2]$ as
\begin{align*}
    [2] & =\left\{\left(\frac{1}{\gamma_{\mathcal I^+}^2}\lin \nu-\frac{2\nu}{\gamma_{\mathcal I^+}^3}\lin\gamma_{\mathcal I^+}\right)\cdot\big(\partial_u\gamma_{\mathcal I^+}\lin{\u{\mathfrak u}}+\lin\gamma_{\mathcal I^+}\big)+\frac{\partial_u\nu}{\gamma_{\mathcal I^+}^2}\lin{\u{\mathfrak u}}\cdot\lin\gamma_{\mathcal I^+}+\frac{1}{\gamma_{\mathcal I^+}^2}\lin\nu\cdot\lin\gamma_{\mathcal I^+}\right\} ,\\
    & \quad +\left\{\frac{\nu}{\gamma^2_{\mathcal I^+}}\partial_u\lin\gamma_{\mathcal I^+}- \frac{1}{\gamma_{\mathcal I^+}}\partial_u\lin\nu\right\}\cdot\lin{\u{\mathfrak u}}+ \left\{\frac{\nu}{\gamma^2_{\mathcal I^+}}\qd\gamma_{\mathcal I^+}- \frac{1}{\gamma_{\mathcal I^+}}\qd\nu\right\}+ \frac{1}{\gamma_{\mathcal I^+}^2}\partial_v\nu\, \lin{\u{\mathfrak v}}{}_{\u u{}_f}\cdot \lin\gamma_{\mathcal I^+} -\frac{1}{\gamma_{\mathcal I^+}}\partial_v\lin\nu\cdot \lin{\u{\mathfrak v}}{}_{\u u{}_f}.
\end{align*}
Again, all of the terms in curly braces cancel out up to decaying errors and the final two terms decay by asymptotic flatness. Finally, every term in $[3]$ decays by asymptotic flatness, which completes the proof of \eqref{eq:qd-lambda-est}.
    
    \textsc{Proof of \eqref{eq:qd-lambda-est}}:
We begin with the definition
\[\u\lambda{}_{\u u{}_f}(\u u,\u v, z_1,z_2) = \frac{\lambda(\u{\mathfrak u}(\u u,z_1,z_2),\u{\mathfrak v}{}_{\u u{}_f}(\u v, z_1,z_2),z_1,z_2)}{\kappa(\u{\mathfrak u}(\mathfrak g_{\u u{}_f}(\u v),z_1,z_2),\u{\mathfrak v}{}_{\u u{}_f}(\u v, z_1,z_2),z_1,z_2)}.\] The calculation for this quantity is complicated by the fact that the $u$-arguments of the numerator and denominator are different, but in view of the uniform bounds of \cref{lem:mathfrak-u-inverse}, we may ignore this issue. 
We differentiate once to obtain 
\begin{align*}\lin{\u\lambda}{}_{\u u{}_f}&= -\frac{\lambda}{\kappa^2}\big(\partial_u\kappa \,\lin{\u{\mathfrak u}}+\partial_v\kappa\,\lin{\u{\mathfrak v}}{}_{\u u{}_f}+\lin\kappa\big)+ \frac{1}{\kappa}\big(\partial_u\lambda \,\lin{\u{\mathfrak u}}+\partial_v\lambda\,\lin{\u{\mathfrak v}}{}_{\u u{}_f}+\lin\lambda\big)\\&=-\frac{\u\lambda{}_{\u u{}_f}}{\kappa}\big(\partial_u\kappa \,\lin{\u{\mathfrak u}}+\partial_v\kappa\,\lin{\u{\mathfrak v}}{}_{\u u{}_f}+\lin\kappa\big)+ \frac{1}{\kappa}\big(\partial_u\lambda \,\lin{\u{\mathfrak u}}+\partial_v\lambda\,\lin{\u{\mathfrak v}}{}_{\u u{}_f}+\lin\lambda\big).\end{align*} Note that each of these terms decays at least like $O\big(\u v^{-1}(\|\lin\Psi_1\|_\mathfrak{Z}+\|\lin\Psi_2\|_\mathfrak{Z})\big)$. We differentiate a second time to obtain
\begin{align*}
  \qd{\u\lambda}{}_{\u u{}_f}& = -\frac{\lin{\u\lambda}{}_{\u u{}_f}}{\kappa}\big(\partial_u\kappa \,\lin{\u{\mathfrak u}}+\partial_v\kappa\,\lin{\u{\mathfrak v}}{}_{\u u{}_f}+\lin\kappa\big) + \frac{\u\lambda{}_{\u u{}_f}}{\kappa^2}\big(\partial_u\kappa \,\lin{\u{\mathfrak u}}+\partial_v\kappa\,\lin{\u{\mathfrak v}}{}_{\u u{}_f}+\lin\kappa\big)\cdot\big(\partial_u\kappa \,\lin{\u{\mathfrak u}}+\partial_v\kappa\,\lin{\u{\mathfrak v}}{}_{\u u{}_f}+\lin\kappa\big) \\
  & \quad -\frac{\lin{\u\lambda}{}_{\u u{}_f}}{\kappa}\big(
   \partial_u^2\kappa \,\lin{\u{\mathfrak u}}\cdot \lin{\u{\mathfrak u}}  + 2\partial_u\partial_v\kappa \,\lin{\u{\mathfrak v}}{}_{\u u{}_f}\cdot\lin{\u{\mathfrak u}} + 2 \partial_u\lin\kappa \cdot\lin{\u{\mathfrak u}}   +\partial_v^2\kappa\,\lin{\u{\mathfrak v}}{}_{\u u{}_f}\cdot \lin{\u{\mathfrak v}}{}_{\u u{}_f}+2\partial_v\lin\kappa\cdot\lin{\u{\mathfrak v}}{}_{\u u{}_f}
  +\qd\kappa \big)\\
  & \quad - \frac{1}{\kappa^2}\big(\partial_u\kappa \,\lin{\u{\mathfrak u}}+\partial_v\kappa\,\lin{\u{\mathfrak v}}{}_{\u u{}_f}+\lin\kappa\big)\cdot\big(\partial_u\lambda \,\lin{\u{\mathfrak u}}+\partial_v\lambda\,\lin{\u{\mathfrak v}}{}_{\u u{}_f}+\lin\lambda\big)\\
  & \quad +\frac{1}{\kappa}\big(
   \partial_u^2\lambda \,\lin{\u{\mathfrak u}}\cdot \lin{\u{\mathfrak u}}  + 2\partial_u\partial_v\lambda \,\lin{\u{\mathfrak v}}{}_{\u u{}_f}\cdot\lin{\u{\mathfrak u}} + 2 \partial_u\lin\lambda \cdot\lin{\u{\mathfrak u}}   +\partial_v^2\lambda\,\lin{\u{\mathfrak v}}{}_{\u u{}_f}\cdot \lin{\u{\mathfrak v}}{}_{\u u{}_f}+2\partial_v\lin\lambda\cdot\lin{\u{\mathfrak v}}{}_{\u u{}_f}
+\qd\lambda
  \big).
\end{align*}
One again checks that each of these terms decays at least like $O(\u v^{-1}\|\lin\Psi_1\|_\mathfrak{Z}\|\lin\Psi_2\|_\mathfrak{Z})$, which proves \eqref{eq:qd-lambda-est}.

\textsc{Proof of \eqref{eq:varpi-2}}: Computing similarly to $\u r{}_{\u u{}_f}$, we obtain 
\begin{multline*}
 \qd{\u \varpi}{}_{\u u{}_f}= \big(\partial_u^2\varpi\,\lin{\u{\mathfrak u}}+\partial_v\partial_u\varpi\,\lin{\u{\mathfrak v}}{}_{\u u{}_f}+\partial_u\lin\varpi\big)\cdot\lin{\u{\mathfrak u}}+\partial_u\varpi\, \qd{\u{\mathfrak u}} \\+ \big(\partial_u\partial_v\varpi\,\lin{\u{\mathfrak u}}+\partial_v^2\varpi\,\lin{\u{\mathfrak v}}{}_{\u u{}_f}+\partial_v\lin\varpi\big)\cdot\lin{\u{\mathfrak v}}{}_{\u u{}_f}+\partial_v\varpi\, \qd{\u{\mathfrak v}}{}_{\u u{}_f} + \partial_u\lin\varpi\cdot \lin{\u{\mathfrak u}} +\partial_v\lin\varpi\cdot\lin{\u{\mathfrak v}}{}_{\u u{}_f}+ \qd \varpi   
\end{multline*}
and quickly conclude \eqref{eq:varpi-2} using all of the estimates at our disposal.
\end{proof}

\subsubsection{Updating \texorpdfstring{$\u u{}_f$}{u bar f}}
In the proof of dyadic iteration, \cref{prop:dyadic-iteration} below, we will be required to show that the ``bootstrap conditions'' defining the decay classes $\mathcal K$ and $\tilde{\mathcal K}$ are continuous with respect to the ``bootstrap time'' $\u u{}_f$. To do this, we use the following:
\begin{prop}\label{prop:updating-uf} Under \cref{ass:family}, it holds that
\begin{equation*}
    |\partial_{\u u{}_f} \u r{}_{\u u{}_f}|  \les \log \u v,
\end{equation*}
\begin{equation*}
        |\partial_{\u u{}_f} \u \varpi{}_{\u u{}_f}|+   |\partial_{\u u{}_f} \u\nu{}_{\u u{}_f}|+   |\partial_{\u u{}_f}\u\phi{}_{\u u{}_f}|+  |\partial_{\u u{}_f}\partial_{\u v}\u\psi{}_{\u u{}_f}|+ |\partial_{\u u{}_f}\partial_{\u u}\u\phi{}_{\u u{}_f}| \les \u v^{-2}\log \u v,
\end{equation*}
    on $\mathcal R(1,\u u{}_f,1,\infty)\times U$.
\end{prop}
\begin{proof}
    This is a tedious, but straightforward, calculation using asymptotic flatness and \cref{lem:v-update}.
\end{proof}

\subsection{Continuous differentiability of the asymptotic mass and parameter ratio}\label{sec:cts-differentiability}

Recall from \cref{sec:global-structure} the Bondi mass function $M_{\mathcal I^+}$ and asymptotic parameter ratio function $P_{\mathcal I^+}$. For any $1\le \u u\le \u u{}_f$, we now define the functions \index{M@$\mathscr M_{\u u}$, $\mathscr M_\infty$ Bondi mass function at time $\u u$ and final Bondi mass function} \index{P@$\mathscr P_\infty$, (signed) final parameter ratio} \index{P@$\mathscr P_{\u u}$, (signed) parameter ratio at time $\u u$}
\begin{align*}
    \mathscr M_{\u u} : \mathfrak M_{\u u{}_f} &\to \Bbb R& \mathscr P_{\u u} : \mathfrak M_{\u u{}_f} &\to \Bbb R\\
    \Psi &\mapsto M{}_{\mathcal I^+}(\u{\mathfrak u}(\u u)), &     \Psi &\mapsto P{}_{\mathcal I^+}(\u{\mathfrak u}(\u u)).
\end{align*}
The goal of this section is to prove the following fundamental result.
\begin{thm}\label{thm:C1-finite-time}
     For any $\u u{}_f\ge2$ and $\u u\in [1,\u u{}_f]$, $\mathscr M_{\u u}$ and $\mathscr P_{\u u}$ are $C^1$ functions on $\mathfrak M_{\u u{}_f}$ in the Fr\'echet sense.
\end{thm}

In the following section we recall the notion of a $C^1$ function on a Banach space.

\subsubsection{Gateaux and Fr\'echet derivatives} 

We record here some useful definitions and results from nonlinear functional analysis. For a general reference, see for instance \cite{drabek2013methods}.

In this section, $(X,\|\cdot\|_X)$ and $(Y,\|\cdot\|_Y)$ denote real Banach spaces (in our cases of interest, either $Y=X$ or $Y=\Bbb R$), $U\subset X$ is an open set, and $f:U\to Y$ is a function. The space of bounded linear maps $L:X\to Y$, $\mathcal L(X,Y)$, is endowed with the norm topology (i.e., given the norm $\|L\|_{\mathcal L(X,Y)}\doteq \sup_{x\in X,\|x\|_X\le 1}\|Lx\|_Y$). We denote the dual by $X^*\doteq \mathcal L(X,\Bbb R)$.

\begin{defn}
    The function $f$ is \emph{Gateaux differentiable} at $x\in U$ in the direction $v\in X$ if the function $z\mapsto f(x+zv)$ ($z\in\Bbb R$) is differentiable at $z=0$ in the sense that the limit
    \begin{equation*}
        Df(x,v)\doteq   \lim_{z\to 0}\frac{f(x+zv)-f(x)}{z}
    \end{equation*} exists. The map $Df(x,\cdot)$ is called the \emph{Gateaux derivative} of $f$ at $x$.     We say that $Df(x,\cdot)$ is \emph{linear} if  $Df(x,v)$ exists for every $v\in X$ and is a linear map $X\to Y$.
\end{defn}

In general, $Df(x,\cdot)$ always respects scaling (i.e., $Df(x,av)=aDf(x,v)$ for $a\in\Bbb R$) but need not respect addition. Moreover, even if it is linear, it need not be bounded. 

\begin{defn}
    The function $f$ is \emph{Fr\'echet differentiable} at $x\in U$ if there exists a bounded linear operator $f'(x)\in\mathcal L(X,Y)$ such that 
    \begin{equation*}
        \lim_{\|h\|_X\to 0} \|h\|^{-1}_X\|f(x+h)-f(x)-f'(x)h\|_Y = 0.
    \end{equation*}
    The map $f'(x)$ is called the \emph{Fr\'echet derivative} of $f$ at $x$. 
\end{defn}

\begin{defn}
    The function $f$ is said to be \emph{continuously differentiable in $U$} (in symbols, $f\in C^1(U)$) if $f$ is Fr\'echet differentiable for every $x\in U$ and $f':U\to \mathcal L(X,Y)$ is continuous. We set
    \begin{equation*}
         \|f\|_{C^1(U)}  \doteq \sup_{U}\|f\|_Y+\sup_U\|f'\|_{\mathcal L(X,Y)},
    \end{equation*}
    provided the right-hand side is finite, in which case we write that $f\in C^1_b(U)$.
\end{defn}

The following proposition gives our fundamental criterion for Fr\'echet differentiability in terms of the Gateaux derivative, see \cite[Theorem 3.2.15]{drabek2013methods} for a proof.

\begin{prop}\label{prop:differentiability-criterion}
    If $f:U\to Y$ is Fr\'echet differentiable at $x\in U$, it is Gateaux differentiable at $x$ for all directions $v\in X$. Conversely, if $Df(x,v)$ exists for every $(x,v)\in U\times X$, $Df(x,\cdot)$ is linear for every $x\in U$, and $x\mapsto Df(x,\cdot)$ defines a continuous map $U\to \mathcal L(X,Y)$, then $f\in C^1(U)$ and $f'(x)v= Df(x,v)$ for every $(x,v)\in U\times X$.
\end{prop}

We use this to prove a useful criterion for interchanging limits and Fr\'echet derivatives. 

\begin{lem}\label{lem:sequences-derivatives}
    Let $U$ be convex and let $\{f_n\}$ be a sequence of $C^1$ functions on $U$. Suppose the sequence $\{f_n(x_0)\}$ converges for some $x_0\in U$ and that 
    \begin{equation}\label{eq:app-deriv-bounded}
       \sup_{n\ge 1}\|f_n\|_{C^1(U)}<\infty,
    \end{equation}
    \begin{equation}\label{eq:G-Cauchy}
        \lim_{m,n\to\infty}\sup_{x\in U}\sup_{\|v\|_X\le 1}\|Df_n(x,v)-Df_m(x,v)\|_Y=0.
    \end{equation}
    Then $f\doteq\lim_{n\to\infty}f_n$ converges uniformly on bounded sets, is $C^1$ in $U$, and
    \begin{equation*}
        f'(x)v = \lim_{n\to\infty} f'_n(x)v
    \end{equation*}
    for every $(x,v)\in U\times X$.
\end{lem}
\begin{proof}
    Without loss of generality, we may take $U$ to be bounded.  We first show that $\{f_n\}$ is Cauchy in $C^0_b(U)$. Indeed, by the mean value theorem, we have
    \begin{multline*}
        \|f_n(x)-f_m(x)\|_Y \le \|f_n(x)-f_m(x)-f_n(x_0)+f_m(x_0)\|_Y+\|f_n(x_0)+f_m(x_0)\|_Y\\
        \le \sup_{z\in[0,1]}\big\|Df_n((1-z)x_0+zx,x-x_0)-Df_m((1-z)x_0+zx,x-x_0)\big\|_Y+\|f_n(x_0)+f_m(x_0)\|_Y,
    \end{multline*}
    which by \eqref{eq:G-Cauchy} shows that $\sup_{U}\|f_n-f_m\|_Y\to 0$ as $n,m\to\infty$ since $\|x-x_0\|_X$ is bounded for $x\in U$. Let $f=\lim f_n$.

    By \eqref{eq:app-deriv-bounded} and \eqref{eq:G-Cauchy}, the functions $x\mapsto Df_n(x,\cdot)$ form a Cauchy sequence in the Banach space of bounded continuous functions $U\to \mathcal L(X,Y)$. Put $h(x,v)=\lim_{n\to\infty}Df_n(x,v)$, which is also linear in $v$ and continuous as a map $U\to \mathcal L(X,Y)$. We argue now that for each $(x,v)\in U\times X$, $f$ is Gateaux differentiable at $x$ in the direction of $v$ and $ Df(x,v)=h(x,v).$ The proof of the lemma is then complete by \cref{prop:differentiability-criterion}. 

    To prove the claim, fix $(x,v)\in U\times X$ and put
    \begin{equation*}
        H_n(z)\doteq \frac{f_n(x+zv)-f_n(x)}{z},\quad H(z)\doteq \frac{f(x+zv)-f(x)}{z}
    \end{equation*}
    for $z\in\Bbb R$ such that $x+zv\in U$. Clearly, $\lim_{z\to 0}H_n(z)=Df_n(x,v)$. Arguing as in the first part of the proof, the sequence $\{H_n\}$ is Cauchy in $C^0$ of a closed interval of $\Bbb R$ containing $z=0$ and hence converges uniformly. Since $f_n\to f$ uniformly, $H_n\to H$ uniformly for $z\ne 0$. It follows that
    \begin{equation*}
        \lim_{z\to 0}H(z) = \lim_{n\to\infty}H_n(0) = h(x,v),
    \end{equation*}
    as was to be shown. \end{proof}

\subsubsection{Proof of \texorpdfstring{\cref{thm:C1-finite-time}}{Theorem 3.81}}

\begin{proof}[Proof of \cref{thm:C1-finite-time}] It clearly suffices to prove the theorem for $\mathscr M_{\u u}$ since $\Psi\mapsto Q$ is $C^\infty$. 

Let $\u v{}_n\to \infty$ be an arbitrary sequence with $\u v{}_n\ge 1$ for every $n$. For $\Psi\in\mathfrak M_{\u u{}_f}$ and $\u u\in[1,\u u{}_f]$, set $f_n(\Psi)=\u \varpi{}_{\u u{}_f}(\u u,\u v{}_n)$ (computed in $\u{\mathcal S}{}_{\u u{}_f}[\Psi]$). Clearly $f_n(\Psi)\to \mathscr M_{\u u}(\Psi)$ as $n\to\infty$ pointwise on  $\mathfrak M_{\u u{}_f}$ by definition of $\mathscr M_{\u u}$.

\textsc{Proof that $f_n$ is $C^1$}: We use \cref{prop:differentiability-criterion}. In fact, we will use \eqref{eq:varpi-2} to show that $\Psi\mapsto Df_n(\Psi,\cdot)$ exists for every $n$ and is  \emph{locally Lipschitz}.

Fix $\Psi_0\in \mathfrak M_{\u u{}_f}$. By \cref{lem:Mi-open-quantitative} and its proof, we may assume that for some $\eta>0$, $B_\eta^\mathfrak{Z}(\Psi_0)\subset\mathfrak M_{\u u{}_f}$ and that \cref{ass:family} holds on this ball. Let $\Psi\in B^\mathfrak{Z}_\eta(\Psi_0)$, $\lin\Psi_1,\lin\Psi_2\in \mathfrak Z$ with $\|\lin\Psi_1\|_\mathfrak{Z},\|\lin\Psi_2\|_\mathfrak{Z}\le 1$, and define the two-parameter family $\Psi(z_1,z_2)=\Psi+z_1\lin\Psi_1+z_2\lin\Psi_2$. By the definition of Gateaux derivative, it holds that
\begin{equation*}
    Df_n(\Psi+z_2\lin\Psi_2,\lin\Psi_1) = \partial_{z_1} \u{\varpi}{}_{\u u{}_f}(\u u,\u v{}_n,0,z_2).
\end{equation*} 
We now differentiate in $z_2$, estimate $|\partial_{z_1}\partial_{z_2}\u{\varpi}{}_{\u u{}_f}|\les 1$ using \eqref{eq:varpi-2}, and then integrate in $z_2$ to obtain 
\begin{equation}\label{eq:gateaux-approx}
    |Df_n(\Psi,\lin\Psi_1)-Df_n(\Psi+z_2\lin\Psi_2,\lin\Psi_1)|\les |z_2|,
\end{equation} for $z_2$ small, 
where the implicit constant depends only on $\Psi_0$ and $\eta$. Letting $\lin\Psi_2$ vary over the unit ball of $\mathfrak Z$, is it clear that $\Psi'\mapsto Df_n(\Psi',\cdot)$ is continuous at $\Psi$. Since $\Psi$ was arbitrary, the map is continuous in $B^\mathfrak{Z}_\eta(\Psi_0)$ and the claim is proved. 

\textsc{Proof that $f_n\to\mathscr M_{\u u}$ in $C^1$}: We use \cref{lem:sequences-derivatives}. By \eqref{eq:lin-varpi-2} and the mean value theorem, we have that 
\begin{equation*}
    |Df_n(\Psi,\lin\Psi)-Df_m(\Psi,\lin\Psi)|\les \u v{}_n^{-1}\log \u v{}_n
\end{equation*}
for $\Psi\in B^\mathfrak{Z}_\eta(\Psi_0)$, $\|\lin\Psi\|_\mathfrak{Z}\le 1$, and $m\ge n$. This completes the proof of the theorem. \end{proof}
    
From the proof, we obtain:
\begin{cor}\label{cor:P'-formula} For any $\u u{}_f\ge 2$, $\Psi=(\phi_\circ,r_\circ,\varpi_\circ,\rho_\circ)\in \mathfrak M_{\u u{}_f}$, $\lin\Psi=(\lin\phi_\circ,\lin r_\circ,\lin\varpi_\circ,\lin\rho_\circ)\in\mathfrak Z$, and $1\le \u u\le \u u{}_f$, it holds that 
    \begin{align}
       \label{eq:M'-formula} \mathscr M_{\u u}'(\Psi)\lin\Psi & = \lim_{\u v\to\infty}\lin{\u\varpi}{}_{\u u{}_f}(\u u,\u v),\\
       \label{eq:P'-formula} \mathscr P_{\u u}'(\Psi)\lin\Psi&= \frac{\lin \rho_\circ \varpi_\circ + \rho_\circ \lin\varpi_\circ}{\mathscr M_{\u u}(\Psi)}-\frac{\mathscr P_{\u u}(\Psi)}{\mathscr M_{\u u}(\Psi)}\mathscr M_{\u u}'(\Psi)\lin\Psi,
    \end{align}
    where $\lin{\u{\varpi}}{}_{\u u{}_f}$ denotes the linearized $\varpi$ associated to $\lin{\u{\mathcal S}}{}_{\u u{}_f}[\Psi,\lin\Psi]$.
\end{cor}

\begin{rk}\label{rk:wrong-time}
 The formula \eqref{eq:M'-formula} might appear slightly suspicious since the function $M_{\u u}$ manifestly does not depend on $\u u{}_f$, but $\lin{\u{\varpi}}{}_{\u u{}_f}$ does through $\mathfrak v_{\u u{}_f}$. Note, however, that by \eqref{eq:lin-varpi-formula} the term in $\lin{\u{\varpi}}{}_{\u u{}_f}$ that involves $\mathfrak v_{\u u{}_f}$ vanishes anyway in the limit $\u v\to \infty$, so that \eqref{eq:M'-formula} is indeed valid for any $\u u{}_f\ge \u u$.
\end{rk}

By taking the limit $n\to\infty$ in \eqref{eq:gateaux-approx} and recalling the dependencies of the implicit constants involved, we obtain:

\begin{cor}\label{cor:P-Lipschitz} For any $B\ge 1$, $u_1\ge1$, and $\u u{}_f\ge 2$, there exists a constant $C\ge1$ with the following property. Let $\Psi\in \mathfrak M$ and suppose $\eta>0$ has the property that $B_\eta^\mathfrak{Z}(\Psi)\subset\mathfrak M_{\u u{}_f}$  and $\|\mathcal S[\Psi']\|_{\mathcal R(1,u_1,1,\infty)}\le B$ for every $\Psi'\in B_\eta^\mathfrak{Z}(\Psi)$. Then it holds that 
\begin{align*} 
|\mathscr M_{\u u}(\Psi_1)-\mathscr M_{\u u}(\Psi_2)|+   \|\mathscr M'_{\u u}(\Psi_1)-\mathscr M'_{\u u}(\Psi_2)\|_{\mathfrak Z^*}&\le  C\|\Psi_1-\Psi_2\|_\mathfrak{Z},\\
  |\mathscr P_{\u u}(\Psi_1)-\mathscr P_{\u u}(\Psi_2)|+   \|\mathscr P'_{\u u}(\Psi_1)-\mathscr P'_{\u u}(\Psi_2)\|_{\mathfrak Z^*}&\le  C\|\Psi_1-\Psi_2\|_\mathfrak{Z}
\end{align*}
for every $1\le \u u\le \u u{}_f$ and $\Psi_1,\Psi_2\in B_\eta^\mathfrak{Z}(\Psi)$.
\end{cor}

\section{The Reissner--Nordstr\"om family}\label{sec:RN}

In this section, we collect fundamental facts about the Reissner--Nordstr\"om family of solutions. We review the basic definitions and features of the metrics in \cref{sec:geometry-RN} and show how they arise in the context of seed data from \cref{sec:semiglobal-1}. In \cref{sec:RN-geo-estimates}, we prove uniform estimates for the Reissner--Nordstr\"om geometry that will be fundamental for the proofs of all of the main theorems. Finally, in \cref{sec:linearized-RN}, we present the linearization of Reissner--Nordstr\"om in the sense of \cref{sec:PT-Bondi-gauge}.

\subsection{The geometry of Reissner--Nordstr\"om}\label{sec:geometry-RN}

\subsubsection{The Reissner--Nordstr\"om metric}

 \begin{figure}
\centering{
\def\svgwidth{31pc}
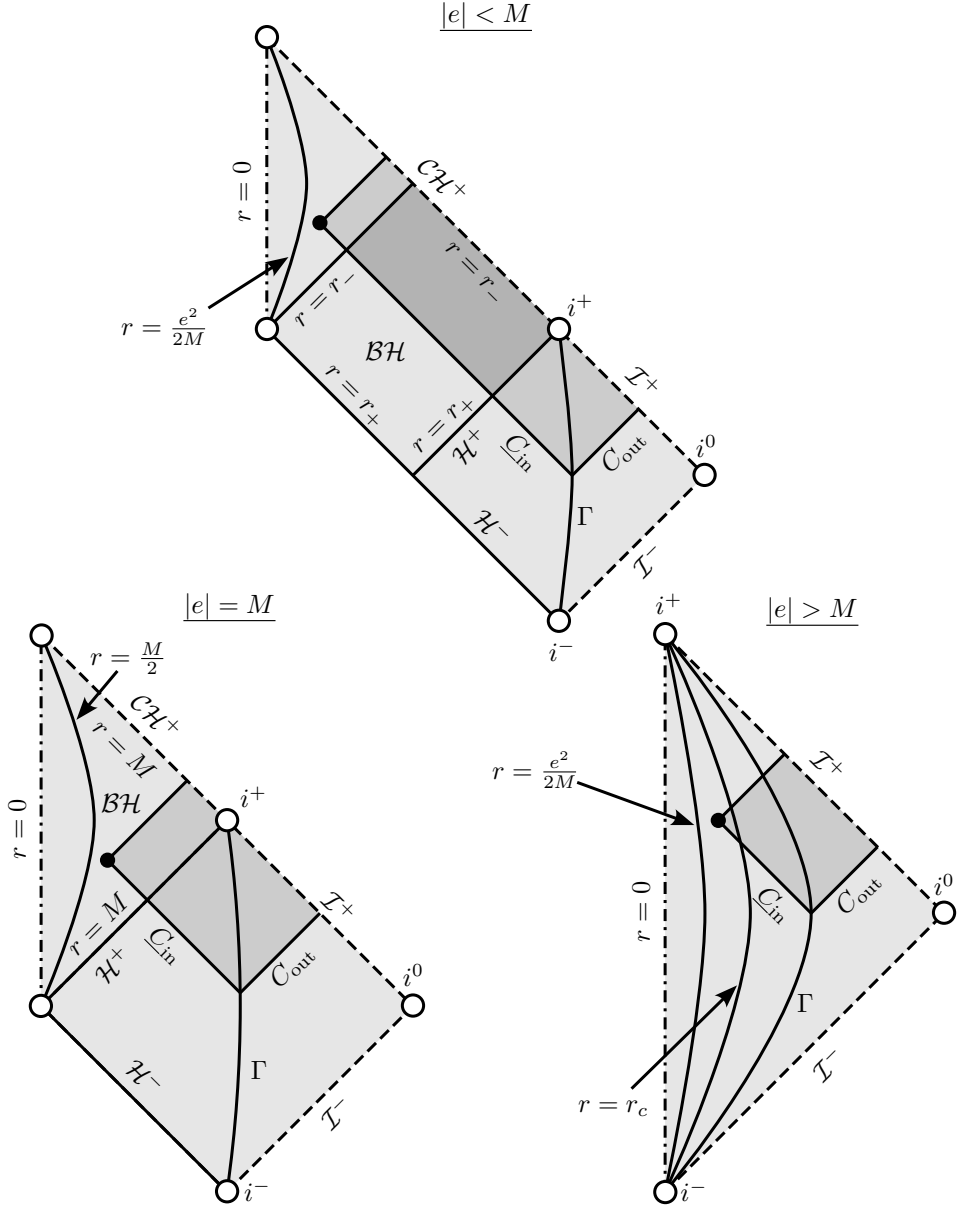}
\caption{Penrose diagrams depicting the relevant portions of the maximally extended Reissner--Nordstr\"om solutions relevant for this work. The shaded regions represent developments of Reissner--Nordstr\"om seed data on a bifurcate null hypersurface $C_\out\cup\u C{}_\ing$ (see already  \cref{sec:RN-seed-data} below) as $|\rho_\circ|$ varies through $1$. In each case, the timelike curve $\Gamma$ represents a constant $r=100M$ curve. As $|e|\nearrow M$, the ``trapped region'' in the subextremal black hole interior disappears, although this is obscured by the conformal compactification used to draw the diagrams. Note that the extremal black hole interior does not contain any trapped surfaces. In the superextremal case, we have drawn the curve $r=r_c$, which is defined in \cref{sec:RN-geo-estimates}. Most of our superextremal estimates will only be valid for $r\ge r_c$. We have drawn in each case the curve $r= \frac{e^2}{2M}$, where the Hawking mass $m$ vanishes. The region between $r=0$ and $r=\frac{e^2}{2M}$ always has negative Hawking mass and is therefore completely inaccessible in gravitational collapse. Note, however, that in the subextremal case the relevant region goes through the \emph{left} Cauchy horizon, which is not traditionally considered an important part of the maximal analytic extension, but does play a role in near-extremal dynamics. Note also that $\frac{e^2}{2M}<r_c$ if $|e|< \sqrt{4+\sqrt{8}}M$, which is always the case in this paper.}
\label{fig:RN}
\end{figure}

 Let $M>0$ and $\varrho\in\Bbb R$. The Reissner--Nordstr\"om metric written in Schwarzschild coordinates $(t,r,\vartheta,\varphi)$ is
\begin{equation*}
     g_{M,\varrho}= -D\,dt^2+D^{-1}\,dr^2+r^2g_{S^2},
\end{equation*}
 where
 \begin{equation*}
     D=D(r;M,\varrho)\doteq 1-\frac{2M}{r}+\frac{\varrho^2M^2}{r^2}.
 \end{equation*}
 This metric solves the coupled Einstein--Maxwell system with electromagnetic field strength tensor
 \begin{equation*}
  \mathbf   F = \frac{e}{r^2}dr\wedge dt,
 \end{equation*}
 where the charge is defined by
 \begin{equation*}
     e = \varrho M.
 \end{equation*}
 
 Recall the terminology \emph{subextremal} $|\varrho|<1$, \emph{extremal} $|\varrho|=1$, and \emph{superextremal} $|\varrho|>1$.
  When $|\varrho|\le 1$ and $g_{M,\varrho}$ analytically extends to describe a black hole spacetime, Schwarzschild coordinates cover the domain of outer communication of the black hole for $(t,r) \in \mathbb R \times (r_+,\infty)$, where $r_+ \doteq  M(1+\sqrt{1-\varrho^2})$. In the superextremal case $|\varrho|>1$, $(t,r)\in \Bbb R\times (0,\infty)$ cover the maximal analytic extension. When $|\varrho|\le 1$, the Cauchy horizon of the black hole is located at $r_-\doteq M(1-\sqrt{1-\varrho^2})$. Note that in this case
  \begin{equation*}
      D= \frac{(r-r_+)(r-r_-)}{r^2}.
  \end{equation*}

  In order to cover the event horizon $\mathcal H^+$ located at $r=r_+$ in the black hole case, we introduce the \emph{tortoise coordinate} \begin{equation}\label{eq:tortoise}
    r_*(r;M,\varrho)\doteq \int^r \frac{dr'}{D(r';M,\varrho)},
\end{equation} where we have left the integration constant ambiguous. Note that $r_*\to \infty$ as $r\to\infty$ and $r_*\to-\infty$ as $r\to r_+$. By defining the advanced time coordinate $v\doteq \frac 12(t+r_*)$, we bring $g_{M,e}$ into the \emph{ingoing Eddington--Finkelstein} form 
\begin{equation*}
    g_{M,\varrho}= - 4Ddv^2+4dvdr+r^2g_{S^2},
\end{equation*} which is regular across $\mathcal H^+$. The vector field $T\doteq \frac 12\partial_v$ is the time-translation Killing vector field (equal to $\partial_t$ in Schwarzschild coordinates in the domain of outer communication).
In these coordinates, the vector field $Y\doteq \partial_r$ is everywhere past-directed null and is transverse to $\mathcal H^+$. It is also translation-invariant in the sense that 
$  \mathcal L_TY=0.$

In order to put the Reissner--Nordstr\"om metric in double null gauge, we define the retarded time coordinate $u\doteq \frac 12(t-r_*)$, so that the metric takes the \emph{Eddington--Finkelstein double null} form
\begin{equation*}
    g_{M,e}=-4D\,dudv+r^2g_{S^2}.
\end{equation*} The area-radius $r$ is now an implicit function of $u$ and $v$ and depends also on $M$ and $\varrho$ in a nontrivial manner. These coordinates once again only cover the domain of outer communication if $|\varrho|\le 1$. The event horizon $\mathcal H^+$ formally corresponds to $u=+\infty$. For all values of $M$ and $\varrho$, null infinity $\mathcal I^+$ formally corresponds to $v=+\infty$. From the identity $r_*=v-u$, we infer that in these coordinates, $\partial_ur=-D$ and $\partial_vr=D$. Since clearly $\Omega^2=4D$ and $D=1-\mu$, this implies that 
\begin{equation*}
    \gamma=-1\quad\text{and}\quad \kappa = 1
\end{equation*}
in these coordinates. Since $\partial_ur=-D$, we have that $Y=(\partial_ur)^{-1}\partial_u$.

Finally, by simply changing the origin of the $(u,v)$ coordinates, and changing notation to $(\u u,\u v)$ for consistency below, we observe: 

\begin{lem}\label{lem:existence-anchoring}\index{rd@$r_\diamond$, background area radius}
    Let $M>0$ and $\varrho\in\Bbb R$ be Reissner--Nordstr\"om parameters. Then for any $R_0\in (r_+,\infty)$ if $|\varrho|\le 1$, or $R_0>0$ if $|\varrho|>1$, there exists a unique function $r_\diamond:\Bbb R^2\to (r_+,\infty)$ if $|\varrho|\le 1$, or $r_\diamond:H\to (0,\infty)$ if $|\varrho|>1$, where $H$ is a halfspace $\{\u v-\u u> c\}$, such that 
    \begin{equation}\label{eq:ERN-EF}
        g_\diamond\doteq -4D(r_\diamond(\u u,\u v))\,d\u ud\u v+ r^2_\diamond(\u u,\u v)g_{S^2}
    \end{equation}
    on $\Bbb R^2\times S^2$ or $H\times S^2$ is isometric to the appropriate region of Reissner--Nordstr\"om  with parameters $(M,\varrho)$ and such that 
    \begin{equation*}\varpi_\diamond=M,\quad Q_\diamond = e=\varrho M,\quad
      \partial_{\u u} r_\diamond=-D(r_\diamond),\quad \partial_{\u v}r_\diamond=D(r_\diamond),\quad\text{and}\quad    r_\diamond(1,1)=R_0.
    \end{equation*}
    The function $r_\diamond$ is constant along lines $\u u-\u v=\mathrm{const}$.
\end{lem}

As mentioned above, the Eddington--Finkelstein double null coordinates defined in this section do not cover the event horizon $\mathcal H^+$ when $|\varrho|\le 1$, but it can be formally attached as the null hypersurface $u=+\infty$. With this understanding, we may extend geometric quantities associated to the metric \eqref{eq:ERN-EF} to the horizon by setting
\begin{equation}\label{eq:extension-to-horizon}
     r_\diamond(\infty,\cdot)=r_+,\quad   \nu_\diamond(\infty,\cdot)=\lambda_\diamond(\infty,\cdot)=0,\quad \varkappa_\diamond(\infty,\cdot) = \frac{2}{r_+^2}\left(M-\frac{e^2}{r_+}\right).
\end{equation}

\subsubsection{Reissner--Nordstr\"om arising from seed data}\label{sec:RN-seed-data}

Let $\mathcal C=\mathcal C(1,u_1,1,\infty)$ be a semiglobal bifurcate null hypersurface with domain of dependence $\mathcal R$. If $r_\circ>0$, $\varpi_\circ>0$, and $\rho_\circ\in\Bbb R$ satisfy the relations $r_\circ > u_1 - 1$ and
\begin{equation}\label{eq:RN-condition-1}
    \frac{2\varpi_\circ}{r_\circ}-\frac{\varpi_\circ^2 \rho_\circ^2}{r_\circ^2}<1,
\end{equation}
then $\Psi=(0,r_\circ,\varpi_\circ,\rho_\circ)\in\mathfrak M(\mathcal C)$ and $\mathcal S_\mathrm{max}[\Psi]$ exists on $\mathcal R$ and is isometric to the appropriate portion of Reissner--Nordstr\"om with parameters $(M,\varrho)=(\varpi_\circ,\rho_\circ)$. 

If $|\rho_\circ|\le 1$, then \eqref{eq:RN-condition-1} is satisfied if $r_\circ > \varpi_\circ(1+\sqrt{1-\rho_\circ^2})=r_+$ and is satisfied for any $(r_\circ,\varpi_\circ)$ if $|\rho_\circ|>1$. In fact, when $|\rho_\circ|>1$, $\Psi\in\mathfrak M_\mathrm{non}$. Note that the condition \begin{equation}\label{eq:u-1-condition}r_\circ>u_1-1\end{equation} means that $r>0$ at the end of $\u C{}_\ing$.
It is not necessarily the case that $\Psi\in\mathfrak M_\mathrm{black}$ when $|\rho_\circ|\le 1$, however, which requires that additionally
\begin{equation}\label{eq:RN-condition-2}
    r_\circ + 1 - u_1 \le r_+.
\end{equation} Note that this is a condition on both $r_\circ$ and $u_1$! 
If this condition holds, the event horizon is located at 
\begin{equation*}
    u_{\mathcal H^+}= 1+ r_\circ - r_+,
\end{equation*}
which is $\le u_1$ by \eqref{eq:RN-condition-2}. For any $|\rho_\circ|\le 1$, $\lambda(u_{\mathcal H^+},v)=0$ for every $v\ge 1$. Recall the inner horizon radius $r_- = \varpi_\circ(1-\sqrt{1-\rho_\circ^2})$, which differs from $r_+$ when $|\rho_\circ|<1$. If $u\in(1+r_\circ-r_-,1+r_\circ-r_+)$, then $\lambda(u,v)<0$ for every $v\ge 1$. If \begin{equation}\label{eq:RN-condition-3}
     r_\circ + 1 - u_1 \le r_-,
\end{equation} then $\lambda(1+r_\circ-r_-,v)=0$ for every $v\ge 1$.

In the setting of the main theorems, we will always assume the conditions \eqref{eq:RN-condition-1} and \eqref{eq:RN-condition-2}, and \eqref{eq:RN-condition-3} will be satisfied for $|\rho_\circ|$ close to one. Fix for the moment $\varpi_\circ$, $r_\circ$, and $u_1$ such that \eqref{eq:RN-condition-1} and \eqref{eq:RN-condition-2} hold for all $|\rho_\circ|\le 1$. By varying $\rho_\circ$, we obtain a one-parameter family $\rho_\circ\mapsto\Psi(\rho_\circ)$ of Reissner--Nordstr\"om data. By \cref{thm:dichotomy}, $\mathcal L_\mathrm{Bondi}<\infty$ if $|\rho_\circ|>1$ and $\mathcal L_\mathrm{Bondi}=\infty$ if $|\rho_\circ|\le 1$. Moreover, $\mathcal L_\mathrm{Bondi}\to\infty$ as $|\rho_\circ|\searrow 1$. 

Consider now $\mathcal S_\mathrm{max}[\Psi(\rho_\circ)] = (\mathcal R,r,\Omega^2,\phi,Q)$. For any $\u u{}_f\in[2,\mathcal L_\mathrm{Bondi}+1]$, we may consider the teleologically normalized solution $\u{\mathcal S}{}_{\u u{}_f}[\Psi(\rho_\circ)]$. It then holds that
\begin{equation*}
    \u r{}_{\u u{}_f}= r_\diamond\quad\text{and}\quad \u{\Omega}{}^2_{\u u{}_f}=4D(r_\diamond;M,\varrho)\quad\text{on }\mathcal R(1,\u u{}_f,1,\infty),
\end{equation*}
where $r_\diamond$ is as in \cref{lem:existence-anchoring} with $(u_0,v_0)=(1,1)$, $R_0=r_\circ$, and $(M,\varrho)=(\varpi_\circ,\varrho_\circ)$. Note that $\u u{}_f$ determines only the domain of the teleologically normalized solution---the values of the functions $\u r{}_{\u u{}_f}$ and $\u{\Omega}{}^2_{\u u{}_f}$ are unaffected.

\subsection{Basic estimates for the   Reissner--Nordström geometry}\label{sec:RN-geo-estimates}

Let $M>0$ and $\varrho\in\Bbb R$. We introduce the notation \[\mathfrak t \doteq M\sqrt{|1-\varrho^2|},\qquad r_c\doteq M+\mathfrak t\] and consider the region of Reissner--Nordstr\"om where\index{t@$\mathfrak t$, non-extremality parameter}~\index{r@$r_c\doteq M+ \sqrt{\lvert M^2-e^2\rvert} = M + \mathfrak t$}
\begin{equation}\label{eq:exterior-condition}
r \geq r_c.
\end{equation}
We will further assume throughout the section the condition  
\begin{equation} \label{eq:sub-supextremal-condition}
   0\le|\varrho|\le \tfrac 54.
\end{equation}
 The conditions \eqref{eq:exterior-condition}  and \eqref{eq:sub-supextremal-condition} capture exactly the region of the Reissner--Nordström family for which the $h_p$-hierarchy of \cref{sec:hp-hierarchy} holds.

\subsubsection{Properties of \texorpdfstring{$D$}{D}}

\begin{lem}
\label{lem:estimates-on-RN-geometry}
Under the assumptions \eqref{eq:exterior-condition}  and \eqref{eq:sub-supextremal-condition}, we have   
\begin{align}
   D&\geq \frac{ (r-r_c + \mathfrak t  \mathbf{1}_{\{|\varrho|\ge 1\}})(r-r_c+\mathfrak t)}{ r^2},\label{eq:lower-bound-on1-mu}\\
     \label{eq:1-mu-hierarchy-computation}
D' &\geq \frac{M}{2r^3} (r-r_c + \mathfrak t) \geq\frac{M}{4  r^2} \sqrt{D},\\
      D &\geq \frac{2r}{M} D' (r-r_c) ,\label{eq:D'r-rc}
\\      \label{eq:upper-bound-varkappa}
D'&\leq 2  r^{-2} M \sqrt{D} + 4 r^{-3} M \mathfrak t \mathbf{1}_{\{|\varrho|\le 1\}},\\
    D'' & \leq 6r^{-2} \label{eq:upper-bound-on-deriative-of-varkappa},
\end{align}
where $'\doteq \frac{d}{dr}$.  
\end{lem}
Note in particular that $D'\ge 0$ in the full range of parameters considered when $r\ge r_c$. 
\begin{proof}
We begin by showing \eqref{eq:lower-bound-on1-mu}.  In the case $|\varrho|\geq 1$,  we estimate \begin{equation*} D  =   r^{-2}\big(  ( r- M)^2+\mathfrak t^2\big) \geq   r^{-2}  ( r- M)^2= r^{-2} ( r - r_c+\mathfrak t)^2,\end{equation*}
   while in the case of $|\varrho|\leq 1$ we estimate
    \begin{equation*}D =   r^{-2}( r - r_+)( r-r_-) =  r^{-2}( r - r_c)( r-r_c + 2\mathfrak t) \geq  r^{-2}( r - r_c)( r-r_c + \mathfrak t) . \end{equation*}
  
  For \eqref{eq:1-mu-hierarchy-computation}, we first write 
    \begin{align*}
       D'  & = 2  r^{-3}( M  r - e^2)  =2  r^{-3} (  M( r-r_c) + M^2 + M \mathfrak t  - e^2 ) \\ &\geq \frac{2M}{r^3} \left( r-r_c + \mathfrak t - \frac{\mathfrak t^2}{M}\mathbf{1}_{\{|\varrho|\geq 1\}}\right)  
        \geq  \frac{M}{2r^3}(r-r_c+\mathfrak t), 
    \end{align*}
    where we used in the last step that  $1\leq |\varrho|\leq \frac 54 $ implies $ 3 \mathfrak t -4 M^{-1}\mathfrak t^2 \geq 0$, so that 
    \[ 
    2  \mathfrak t - 2 \frac{\mathfrak t^2}{M}\mathbf{1}_{\{|\varrho|\geq 1\}}
   \geq \frac 12 \left(  \mathfrak t  + 3 \mathfrak t - 4 M^{-1}\mathfrak t^2 {M}\mathbf{1}_{\{|\varrho|\geq 1\}}\right) \geq \frac 12 \mathfrak t. \]
 The  second inequality in \eqref{eq:1-mu-hierarchy-computation} follows now from 
    \begin{equation*}
    D  = r^{-2}\big( ( r-M)^2 -M^2 + e^2 \big) \leq r^{-2} \big( ( r-r_c + \mathfrak t)^2 + \mathfrak t^2 \big) \leq 2 r^{-2} (r-r_c+\mathfrak t)^2.
    \end{equation*}
    To show \eqref{eq:D'r-rc} we observe that from  \eqref{eq:lower-bound-on1-mu}  and \eqref{eq:1-mu-hierarchy-computation} we obtain 
\[ D \geq  r^{-2} (r-r_c)(r-r_c + \mathfrak t) \geq \frac{2r}{M}D' (r-r_c) . \] 

Next, the upper bound \eqref{eq:upper-bound-varkappa} for $D'$ follows from
\begin{align*}
       D'  & =2  r^{-3} (  M( r-r_c) + M^2 + M \mathfrak t  - e^2 )\leq 2 r^{-3} M (r-r_c) + 4 r^{-3} M\mathfrak t \leq 2M r^{-2} \sqrt D + 4 r^{-3} M \mathfrak t,
    \end{align*}
    where we used $M^2-e^2\leq M \mathfrak t$ in the first step and \eqref{eq:lower-bound-on1-mu} in the second step. 
    Note that in the case $|\varrho|\geq 1$, this improves to 
    \[ D' \leq 2M r^{-3}   (r-r_c +\mathfrak t)\leq 2 M r^{-2} \sqrt{D}\]
    which shows \eqref{eq:upper-bound-varkappa}.

Finally, to show \eqref{eq:upper-bound-on-deriative-of-varkappa} we note that 
\begin{equation*}
   r^2  D'' = -\frac{4M}{ r} + \frac{6e^2}{ r^2}  
\end{equation*}
which, by inspection, attains its maximum on $\{ r\geq r_c\}$ at the boundary. For $r =\infty$ the inequality \eqref{eq:upper-bound-on-deriative-of-varkappa} is clearly true so it suffices to show it for $r=r_c$. Using $M\leq r_c  $ and $|\varrho|\leq \frac 54 $, we estimate
\begin{equation*}
     r^2  D'' \leq r_c^{-2} \left( - 4 M r_c + 6 e^2\right) \leq M^{-2} ( - 4 M^2 + 6 e^2) \leq 6. \qedhere
\end{equation*}
\end{proof}

\begin{lem}\label{lem:monotonicity-D'}
Under the assumptions \eqref{eq:exterior-condition}  and \eqref{eq:sub-supextremal-condition} we have   
\begin{align}
D(r_1)&\le D(r_2),\label{eq:D-monotonicity}\\
D'(r_1)& \le\left(\frac{r_2}{r_1}\right)^2 D'(r_2) \label{eq:almost-monotonicity}
\end{align}
for $r_c\le r_1\le r_2$. 
\end{lem}
\begin{proof}
This follows immediately from the observations that $D'\ge 0$ and $(r^2D')' = e^2r^{-2}\ge 0$ when $r\ge r_c$. 
\end{proof}

\subsubsection{The functions \texorpdfstring{$h_p$}{hp}}\label{sec:hp-defn}

For $M>0$ and $p\in\Bbb R$, we define \index{hp@$h_p$, horizon multiplier}
\begin{equation}
    h_p(r;M,\varrho)\doteq 20 D^{-p/2}+ |p| M   D'D^{-p/2-1/2}. \label{eq:hp-defn}
\end{equation}
This function will be essential to our uniform estimates for the wave equation on Reissner--Nordstr\"om. We now estimate the bulk term generated by the horizon multiplier $h_p\partial_u$ for $p\geq0$. 

First, using \eqref{eq:1-mu-hierarchy-computation}, we derive immediately:

\begin{lem}
Under the assumptions \eqref{eq:exterior-condition}  and \eqref{eq:sub-supextremal-condition} we have
\begin{align}
\label{eq:hp-upper}h_p&\le(80r^2M^{-1}+|p|M) D'D^{-p/2-1/2},\\
\label{eq:hp-hierarchy-property}D'h_p &\ge \frac{4|p|M}{80r^4+|p-1|M^2r^2}h_{p-1}
\end{align}
for $p\in\Bbb R$. 
\end{lem}

\begin{rk} Note that \eqref{eq:hp-upper} shows that the second term in \eqref{eq:hp-defn} controls $h_p$ from above and below when $p\ne 0$, but we include the first term for the later estimates to be consistent across $p=0$ and in order to have the clean statement \eqref{eq:hp-main-estimate} below. 
\end{rk}
 
\begin{prop} \label{prop:estimate-on-hp} 
Under the assumptions \eqref{eq:exterior-condition}  and \eqref{eq:sub-supextremal-condition} we have
    \begin{equation}
    -Dh_p' \ge \frac{pM^2}{160r^2+pM^2}D'h_p \label{eq:hp-main-estimate}
    \end{equation}
        for $p\ge 0$. 
  \end{prop}
  \begin{proof}
        We compute
\begin{align*}
  h_p'& = -10 p   D'D^{-p/2-1}+pM D'' D^{-p/2-1/2}- p M \left(\frac p2 + \frac 12\right)  (D')^2 D^{-p/2-3/2}\\
  & \leq -10p    D'D^{-p/2-1}+ 8p   D' D^{-p/2-1}- p M \left(\frac p2 + \frac 12\right)  (D')^2 D^{-p/2-3/2}\\ & \leq- \frac{pM}{2} (D')^2 D^{-p/2-3/2},
\end{align*}
where we used that $ pM D'' D^{-p/2-1/2} \leq  8p  D' D^{-p/2-1}$ which follows from \eqref{eq:upper-bound-on-deriative-of-varkappa} and \eqref{eq:1-mu-hierarchy-computation}. Multiplying by $-D$ and using \eqref{eq:hp-upper}, we obtain \eqref{eq:hp-main-estimate}.    \end{proof}

\subsubsection{The function \texorpdfstring{$w_\delta$}{w delta}}

In the Morawetz estimate, we will make use of the weight function \index{w@$w_\delta$, weight function for Morawetz estimate}
\begin{equation} \label{eq:def-w-delta}
w_\delta(r;M,\varrho) \doteq D^{\delta/2}(1+   D'^{-\delta} \mathbf{1}_{\{|\varrho|\leq 1\}}),\end{equation}
where $\delta>0$ is a constant. In the following lemma, the implicit constants are allowed to depend on $\delta$.
\begin{lem} 
\label{lem:morawetz-horizon-weight}
Under the assumptions \eqref{eq:exterior-condition}  and \eqref{eq:sub-supextremal-condition} we have
    \begin{equation}
    \label{eq:lower-bound-w}
    w_\delta' \gtrsim D' D^{\delta/2 -1} (1+ D'^{-\delta} \mathbf{1}_{\{ 5 D \leq  (MD')^2 \}}) 
    \end{equation}
    as well as 
    \begin{equation}\label{eq:upper-bouds-wd}
    w_\delta \les 1, \quad {w_\delta'}^{-1} D (D')^2 \les D^{2-\delta/2} D'. 
    \end{equation}
\end{lem}  
\begin{proof} We compute
    \begin{equation}\label{eq:w-prime}
    w_\delta' = \frac{\delta}{2} D^{\delta/2 -1} {D'} \left( 1+  D'^{-\delta} \mathbf{1}_{\{|\varrho|\leq 1\}} (1 - 2  D D'' D'^{-2} ) \right)\end{equation}
    and consider the cases $|\varrho|>1$ and $|\varrho|\leq 1$, separately. 
In the case $|\varrho|> 1$,  we infer from \eqref{eq:upper-bound-varkappa} that  \[D'^2 \leq 4 r^{-4} M^2 D \leq \frac{4}{M^2} D
\] so we have 
$\mathbf{1}_{\{5 D \leq  (MD')^2 \}} \mathbf{1}_{\{|\varrho|>1 \}} \equiv 0$. In particular, in view of \eqref{eq:w-prime}, the desired \eqref{eq:lower-bound-w} is clearly satisfied. 

In the case $|\varrho|\leq 1$, we first note that  $M^2 D'' \leq 2 $ which follows from observing that  $D''$ is maximized on $[r_+,\infty)$ for $r=e=M$. Using this, we estimate \[ 1-\frac{2 D D''}{ D'^2} \mathbf{1}_{\{ 5 D \leq  (MD')^2 \}}
\geq 1-\frac{4 D }{ M^2 D'^2} \mathbf{1}_{\{ 5 D \leq  (MD')^2 \}}
 \geq \frac 15.\] 
We can conclude \eqref{eq:lower-bound-w} once we show that $1 - 2  D D'' D'^{-2} \geq 0$. To show that it suffices to show that  \[  \tilde w  \doteq  r^6(D'^2 - 2 D D'')\geq 0  \]   which holds true since  \[\tilde w' = 24 (r-M) (M r - e^2) \geq 0, \text{ and } \tilde w(r_+) = 4 (M^2-e^2) (4 M^2 - e^2 + 2 M r_+) \geq 0.\]

Finally, the bounds \eqref{eq:upper-bouds-wd} are direct consequences of \eqref{eq:w-prime} and \cref{lem:estimates-on-RN-geometry}.
\end{proof}

\subsubsection{The \texorpdfstring{$\mathcal J_{a,b}$}{Jab} integrals}

It will be useful to derive some estimates on the integrals\index{J@$\mathcal J_{a,b}$, radial singular weight integrals}
\begin{equation}\label{eq:mathcal-J}
    \mathcal J_{a,b}(r;M,\varrho)\doteq \int^{100M}_{r} \frac{1}{D(x;M,\varrho)^a D'(x;M,\varrho)^b} \,dx,
\end{equation} where $a,b\in\Bbb R$ and $r>r_c$. To simplify notation, we will suppress the dependency of $\mathcal J_{a,b}$ on $M$ and $\varrho$ in the following. Our estimates will depend on lower and upper bounds for $M$, which we quantify as follows. Given $M_0>0$, we assume that \index{M@$M_0$, fixed mass scale}
\begin{equation}
    \label{eq:mass-condition} \tfrac 12 M_0\le M\le 2M_0.
\end{equation}
 
\begin{prop}   \label{prop:J-integrals} Let  $(M,\varrho)$ satisfy \eqref{eq:sub-supextremal-condition} and \eqref{eq:mass-condition}, and let $r\in (r_c,100M]$. 
    \begin{enumerate}
    \item If $0\leq a < 1$ and $2a+b> 1$, then 
    \[
     \mathcal J_{a, b}( r) \lesssim  \left(\frac{1}{1-a} + \frac{1}{2a+b-1} \right) D'( r)^{-2a-b+1}.
    \]
    \item If $a=1$ and $b>-1$, then
    \[  \mathcal J_{1, b}( r) \lesssim D'(r)^{-1-b} \left( \left|\log\left(\frac{D'(r)^2}{D( r)}\right)\right| + \frac{1}{1+b} \right).\]
             \item If $a>1$ and $b\geq -1$, then     \[ \mathcal J _{a, b} (r)\lesssim \frac{1}{a-1} D'( r)^{-b-1} D( r)^{1-a} . \]
    \end{enumerate} 
    In this proposition and its proof, the implicit constant in $\les$ depends only on $M_0$. 
\end{prop}

    \begin{proof} 
    Put $y\doteq x - r_c$.  By \cref{lem:estimates-on-RN-geometry} in the region $0\leq y\leq 100M -r_c$, we have $D(x) \gtrsim y (y+\mathfrak t)$ and 
    $ D'(x) \sim y+\mathfrak t$.
        Hence,
        \begin{equation*}
            \mathcal J_{a,b} \lesssim \int_{r - r_c}^{100M-r_c} y^{-a} (y+\mathfrak t)^{-a-b} dy. 
        \end{equation*}
{\scshape{Case 1:} $0\leq a < 1$ and $2a+b >1$.}   We estimate
        \begin{align*}
\mathcal J_{a,b} \lesssim \int_{ r - r_c}^{100M-r_c} y^{-a} (y+\mathfrak t)^{-a-b} & dy   = \int_{ r - r_c}^{100M-r_c} y^{-a} (y+\mathfrak t)^{-a-b}\mathbf{1}_{\{y \leq \mathfrak t\} }  dy \\
& \qquad +  \int_{ r - r_c}^{100M-r_c} y^{-a} (y+\mathfrak t)^{-a-b}\mathbf{1}_{\{y \geq \mathfrak t\} }  dy  \\
& \lesssim \mathfrak t^{-a-b} \int_{ r - r_c}^{100M-r_c} y^{-a} \mathbf{1}_{\{y \leq \mathfrak t\} }  dy +   \int_{ r - r_c}^{100M-r_c} y^{-2a-b} \mathbf{1}_{\{y \geq \mathfrak t\} }  dy\\
& \lesssim  \frac{\mathfrak t^{-a-b}}{1-a} \left[ \mathfrak t^{-a+1} - ( r - r_c)^{-a+1}  \right] \mathbf{1}_{ \{  r - r_c \leq \mathfrak t\}} + \frac{1}{2a+b-1} \mathfrak t^{-2a-b+1} \mathbf{1}_{ \{  r - r_c \leq \mathfrak t\}} \\ &\qquad + \frac{1}{2a+b-1} ( r - r_c)^{-2a-b+1} \mathbf{1}_{\{  r - r_c \geq \mathfrak t \} }\\
& \lesssim \left(\frac{1}{1-a} + \frac{1}{2a+b-1} \right) D'( r)^{-2a-b+1}.
        \end{align*}
        
{\scshape{Case 2:} $a=1$ and $b>-1$.} We argue similarly as in the previous case but the integral on the second line gives 
\begin{align*}
    \mathcal J_{1,b}( r)& \lesssim \mathfrak t^{-1-b} \int_{ r - r_c}^{100M-r_c} y^{-1} \mathbf{1}_{\{y\leq \mathfrak t\}} dy + \int_{ r-r_c}^{100M-r_c} y^{-2 -b} \mathbf{1}_{\{y\geq \mathfrak t\}} dy\\ &  \lesssim \left[\mathfrak t^{-1-b} \log\left( \frac{\mathfrak t}{ r- r_c}\right) +\frac{1}{1+b} \mathfrak t^{-1-b} \right] \mathbf{1}_{\{ r - r_c \leq \mathfrak t \}} + \frac{1}{1+b} ( r - r_c)^{-1-b}  \mathbf{1}_{\{ r - r_c \geq \mathfrak t \}} \\
    & \lesssim D'( r)^{-1-b} \left( \log\left(\frac{D'( r)^2}{D(r)}\right) + \frac{1}{1+b} \right)  \mathbf{1}_{\{ r - r_c \leq \mathfrak t \}}+\frac{1}{1+b} D'( r)^{-1-b} \mathbf{1}_{\{ r - r_c \geq \mathfrak t \}}\\
    & \lesssim D'( r)^{-1-b} \left( \left|\log\left(\frac{D'( r)^2}{D(r)}\right) \right|+ \frac{1}{1+b} \right),
\end{align*}
where we used \cref{lem:estimates-on-RN-geometry}. 
    
{\scshape{Case 3:} $a>1$ and $b\geq -1$.}  We use \cref{lem:monotonicity-D'} to estimate
\begin{equation*}
    \mathcal J_{a,b}( r) = \int_{ r}^{100M} \frac{1}{D(x)^a D'(x)^b}dx \lesssim \frac{1}{D'( r)^{b+1}} \int_{ r }^{100M} \frac{D'(x)}{D(x)^a} dx \lesssim \frac{1}{a-1} D'( r)^{-b-1} D( r)^{1-a}. \qedhere
\end{equation*}
    \end{proof}

\subsubsection{Some estimates in coordinates}

In the proof of the main theorem, we will need to analyze spacetimes which look superextremal on some timescale, and so we will need to ensure that the condition \eqref{eq:exterior-condition} holds on the semiglobal slab $\mathcal R$. We will now present a simple condition that ensures this. 

Let $M$ and $\varrho$ be Reissner--Nordstr\"om parameters and consider one of the Reissner--Nordstr\"om area radius functions $r_\diamond$ from \cref{lem:existence-anchoring}. We impose the condition \begin{equation}\label{eq:bif-sphere-condition}
 \left|\frac{r_\diamond(1,1)}{M}-100\right| \le 2,
\end{equation}
where we take $(u_0,v_0)=(1,1)$. 
Under this condition and  \eqref{eq:sub-supextremal-condition}, it holds that
\begin{equation}
\tfrac{9}{10}\le D(r_\diamond(1,1)) \le 1.\label{eq:D-bif-sphere}
\end{equation}

\begin{lem}\label{lem:rgeqrc} \index{uinfty@$u_\infty(M,\varrho)$, superextremal retarded time upper bound}
Let $(M,\varrho)$ satisfy \eqref{eq:sub-supextremal-condition}, $r_\diamond$ satisfy \eqref{eq:bif-sphere-condition}, and assume that \begin{equation}\label{eq:defn-u-infty}
u_1\le u_\infty(M,\varrho) \doteq \sup\{u\geq 1: (u-1)^{-2} \geq 64\pi^{-2}
 (\varrho^2 - 1) M^{-2}\}.
\end{equation} Then $r_\diamond> r_c$ on $\mathcal R(1,u_1,1,\infty)$. 
\end{lem}
\begin{proof}
Since $r_c = r_{+}$ when $|\varrho|\leq 1$, the statement is true for $|\varrho|\leq 1$ by definition. Therefore, suppose $|\varrho|> 1$. 
 Since $ \nu <0$ and $ \lambda >0$ on superextremal Reissner--Nordstr\"om, it suffices to show that $ r_\diamond(u_\infty,1) \geq r_c$ which follows once we show that 
 \begin{equation}\label{eq:u-infty-1-estimate}
   u_\infty - 1 <    \int^{r_\diamond(1,1)}_{r_c} \frac{dr'}{D( r')}  .
 \end{equation}
In order to prove \eqref{eq:u-infty-1-estimate}, we compute:
\begin{align*}
 \int_{r_c}^{r_\diamond(1,1)} \frac{dr'}{D(r')} & = \int_{r_c}^{r_\diamond(1,1)} \frac{{ r }^2}{( r - M)^2 + e^2-M^2} d r \geq  \int_{\sqrt{e^2-M^2}}^{r_\diamond(1,1) - M} \frac{M^2}{x^2 + e^2 - M^2 } dx \\
    & =\frac{M^2}{\sqrt{e^2-M^2}}  \int_{1}^{\frac{r_\diamond(1,1) - M}{\sqrt{e^2-M^2}}}\frac{dx}{x^2 + 1} = \frac{M^2}{ \sqrt{e^2-M^2}}\left(  \arctan\left[ \frac{r_\diamond(1,1)-M}{  \sqrt{e^2-M^2}}\right] - \frac{\pi}{4} \right) > \frac{M^2 \pi}{8\sqrt{e^2-M^2}},
\end{align*}
where we used that $\arctan\left[\frac{r_\diamond(1,1) - M}{\sqrt{e^2-M^2}}\right] > \frac{3}{8}\pi$ which follows from \eqref{eq:bif-sphere-condition}. This concludes the proof in view of the definition of $u_\infty$ in \eqref{eq:defn-u-infty}.
\end{proof}

The following lemma will be useful for defining the approximate stable manifolds in \cref{sec:proof-of-main-thm}.

  \begin{lem}\label{lem:def-of-alpha}
      There exists a constant $\alpha>0$, which depends only on $M_0$, such that if $\frac 12M_0\le M\le 2M_0$ and $|\varrho|\le 1+\alpha L_i^{-2}$ for some $i\ge 1$, where $L_i\doteq 2^i$\index{L@$L_i\doteq 2^i$, dyadic time scale}, then $u_\infty(M,\varrho)\ge L_i$.
  \end{lem}
  \begin{proof}
    Inspection of \eqref{eq:defn-u-infty}. 
  \end{proof}

The following estimate will be used later to apply the theory of \cref{sec:semiglobal-1} to spacetimes in the class $\mathcal K$ later in \cref{sec:K-2}. 

\begin{lem} Let  $(M,e)$ satisfy \eqref{eq:sub-supextremal-condition}, $r_\diamond$ satisfy \eqref{eq:bif-sphere-condition}, and $u_1\le u_\infty(M,\varrho)$. Then 
    \begin{equation}\label{eq:D-lower-bound}
        D(r_\diamond)\ge \tfrac{9}{10}e^{-2M^{-1}(u-1)}
    \end{equation}
    on $\mathcal R(1,u_1,1,\infty)$.
\end{lem}
\begin{proof} Using $\partial_u{\log D} = -D'$ and $0\le D'\le 2r_c^{-2}M\le 2M^{-1}$, we estimate 
\begin{equation*}
|{\log D(r_\diamond(u,1))}|\le |{\log D(r_\diamond(1,1))}| + \int_1^u  D'\,du' \le |{\log D(r_\diamond(1,1))}|+ 2M^{-1}(u-1).
\end{equation*}
Inverting and using \eqref{eq:D-monotonicity} and \eqref{eq:D-bif-sphere} gives the claim. \end{proof}

In \cref{sec:estimates-linear-perturbations}, we shall require the following estimates. 

\begin{lem}\label{lem:D_1'-est} Let  $(M,\varrho)$ satisfy \eqref{eq:sub-supextremal-condition}, $r_\diamond$ satisfy \eqref{eq:bif-sphere-condition}, and $u_1\ge u_\infty(M,e)$. Then 
    \begin{equation}
       \big (D'(r_\diamond(u,1))\big)^{-1}\le C( u-1) + CM^{-1}\label{eq:D_1'-est}
    \end{equation}
    for $1\le u \le u_1$, where the constant $C$ does not depend on $M$ or $u$.
\end{lem}
\begin{proof}
    We compute $\partial_u(D')^{-1}=(D')^{-2}D''D$ and estimate $(D')^{-2}D''D\le 96M^{-2}r^2_\diamond\le C$ using \cref{lem:estimates-on-RN-geometry}, so the claim follows by integration. 
\end{proof}

\begin{lem}\label{lem:D-quadratic-decay}  Let $(M,\varrho)$ satisfy \eqref{eq:sub-supextremal-condition} and \eqref{eq:mass-condition}, $r_\diamond$ satisfy \eqref{eq:bif-sphere-condition}, and $u_1\ge u_\infty(M,e)$. Then there exists a constant $C$ depending only on $M_0$ such that
    \begin{equation}\label{eq:D-decay}
        D(r_\diamond(u,1))\le Cu^{-2}
    \end{equation}
    for $1\le u\le u_1$.
\end{lem}
\begin{proof}
We use the definition of $r_*$, \eqref{eq:tortoise}, and the Eddington--Finkelstein $(u,v)$ coordinates, together with \cref{prop:J-integrals} to estimate 
\begin{equation*}
    u\les \mathcal J_{1,0}(r_\diamond(u,1))\les D'^{-1}\big(1+|{\log(D'^2/D)}|\big),
\end{equation*} where the right-hand side is evaluated at $r_\diamond(u,1)$. It follows that
\begin{equation*}
    u^2 D\les D'^{-2}D\big(1+|{\log(D'^2/D)}|^2\big)\les 1,
\end{equation*}
where we used \eqref{eq:1-mu-hierarchy-computation} and the boundedness of $x^{1/2}\log x$ near $x=0$.
\end{proof}

\begin{lem}\label{lem:F-monotonicity} 
   Under the hypotheses of \cref{lem:D-quadratic-decay}, there exists a constant $C$ with the following properties. Let $F(u)\doteq u^{-1}D'(r_\diamond (u,1))^{-1}$. If $1\le u\le u'\le u_1$, then \begin{equation}
        F(u')\le CF(u).\label{eq:F-monotonicity}
    \end{equation}
\end{lem}
\begin{proof} In this proof, it is implied that $D$, $D'$, and $D''$ are evaluated at $r_\diamond(u,1)$. Let $\eta>0$ be a constant to be determined and set $A_\eta\doteq \{u\ge 1:D\ge \eta^2 D'^2\}$. For $u\in A_\eta$, \eqref{eq:D-decay} implies 
\begin{equation*}
    F^{-1}= uD' \le \eta^{-1}uD^{1/2}\les \eta^{-1}.
\end{equation*}
    Next, we compute 
    \begin{equation*}
        F' = u^{-1} D'^{-2}DD''-u^{-2}D'^{-1}= u^{-2}D'^{-2}D^{1/2}(uD''D^{1/2}-D'D^{-1/2}).
    \end{equation*}
    For $u\in A_\eta^c$, it follows that
    \begin{equation*}
        F' \leq u^{-2}D'^{-2}D^{1/2}(uD''  D^{1/2} -\eta^{-1})<0
    \end{equation*} for $\eta$ sufficiently small, where we used \eqref{eq:D-decay} and that $|D''|\les 1$. With $\eta$ now fixed, using also \eqref{eq:D_1'-est}, we see that $F\sim 1$ in $A_\eta$. The claim now readily follows. 
\end{proof}

For a proof of the following lemma, see for instance \cite[Lemma 9.3]{AKU24} (or simply note that $D'\sim D^{1/2}$ and hence $\partial_uD^{-1/2}\sim D^{-1}$ in extremal Reissner--Nordstr\"om).
\begin{lem}\label{lem:r-M-extremal}
   Let $|\varrho| = 1$, $M$ satisfy \eqref{eq:mass-condition}, and $r_\diamond$ satisfy \eqref{eq:bif-sphere-condition}. Then it holds that 
    \begin{equation*}
        (r_\diamond -M)^{-1}\sim u-v+1
    \end{equation*}
    for $u\ge v$, where the implicit constant depends only on $M_0$. 
\end{lem}

\subsection{The linearized Reissner--Nordström family}\label{sec:linearized-RN}

Recall from \cref{sec:RN-seed-data} how Reissner--Nordstr\"om arises from seed data on $\mathcal C$ in the formalism of \cref{sec:semiglobal-1}. With this setup, we now study linearized perturbations of Reissner--Nordstr\"om in teleological gauge. Note, crucially, that the worst behavior of $\lin{\u r}$, $\lin{\u \nu}$, and $\lin{\u\lambda}$ for $r$ close to $r_c$ comes explicitly from the $\lin\rho_\circ$ component of $\lin\Psi$. This distinction will be fundamental to the main linearized estimates in \cref{sec:estimates-linear-perturbations}.

\begin{prop}\label{prop:linear-RN} Let $\Psi = (0, r_\circ, \varpi_\circ, \rho_\circ)$, where we assume \eqref{eq:sub-supextremal-condition}, \eqref{eq:mass-condition}, and \eqref{eq:bif-sphere-condition}, with $(M,\varrho)=(\varpi_\circ,\rho_\circ)$. Let $\lin\Psi=(\lin\phi_\circ,\lin r_\circ,\lin\varpi_\circ,\lin\rho_\circ)\in\mathfrak Z$ and consider the linearized solution $\lin{\u{\mathcal S}}{}_{\u u{}_f}[\Psi,\lin\Psi]$ constructed in \cref{sec:PT-Bondi-gauge}, where $1\le \u u{}_f\le u_\infty(M,\varrho)$. Then the linearized geometric quantities are independent of $\lin\phi_\circ$ and it holds that
\begin{gather}
\label{eq:DHR-1}\lin{\u\varpi} = \lin M,\qquad \lin Q = M\lin\varrho+\varrho \lin M,\qquad \lin{\u\gamma} = \lin{\u\kappa} = 0,\\
\label{eq:DHR-2}|\lin{\u r}| \les \big(D'^{-1}|\lin\varrho|+\|\lin\Psi_0\|_\mathfrak{Z}\big)\mathbf 1_{\{\u u \ge \u v\}}+(1+|{\log r_\diamond}|)\|\lin\Psi_0\|_{\mathfrak Z}\mathbf 1_{\{\u v\ge \u u\}},\\
\label{eq:DHR-3}|\lin{\u \nu}|+|\lin{\u \lambda}| \les D\big(D'^{-2}|\lin\varrho|+\|\lin\Psi_0\|_{\mathfrak Z}\big)\big(1+|{\log(D'^{-2}D)}|\big)\mathbf 1_{\{\u u\ge \u v\}}+r^{-1}_\diamond\|\lin\Psi_0\|_{\mathfrak Z}\mathbf 1_{\{\u v\ge\u u\}}
\end{gather}
on $\mathcal R$, where $\lin\Psi_0\doteq(0,\lin r_\circ,\lin\varpi_\circ,\lin\rho_\circ)$, $(\lin M,\lin\varrho)\doteq (\lin\varpi_\circ,\lin\rho_\circ)$, $D$ and $D'$ are evaluated at $(r_\diamond,M,\varrho)$, and we have omitted the subscript $\u u{}_f$ as the linearized quantities are independent of it. Each linearized geometric quantity is constant along lines $\u u - \u v=\mathrm{const}$. Moreover, $\lin{\u\phi}$ satisfies the free wave equation on the Reissner--Nordstr\"om background.
\end{prop}

 Because the EMSF system is quadratic in $\phi$, the linearized scalar field decouples when perturbing exact Reissner--Nordstr\"om. Therefore, $\lin{\u r}$ actually equals the linearization of $r_\diamond$, \emph{within the Reissner--Nordstr\"om family}, if one parametrizes the Reissner--Nordstro\"om family by $M$, $\varrho$, and $r_\diamond(1,1)$. With this in mind, we define the notation 
  \begin{equation}\label{eq:lin-diamond-defn}
  \lin r_\diamond\doteq \lin{\u r},\qquad \lin \lambda_\diamond \doteq \lin{\u\lambda},\qquad \lin\nu_\diamond\doteq \lin{\u\nu},
  \end{equation}
  where $\lin{\u r}$, $\lin{\u\lambda}$, and $\lin{\u\nu}$ are as in \cref{prop:linear-RN}.

\begin{proof}[Proof of \cref{prop:linear-RN}] Since $\u\phi$ vanishes identically for $\u{\mathcal S}[\Psi]$, the linearized Einstein equations give immediately
\begin{equation*}
\partial_{\u u}\lin{\u\varpi} = \partial_{\u v}\lin{\u\varpi} = \partial_{\u v}\lin{\u\gamma} = \partial_{\u u}\lin{\u\kappa} = 0 
\end{equation*}
on $\mathcal R$, from which \eqref{eq:DHR-1} easily follows after using the boundary conditions from \cref{prop:boundary-conditions}. Since $\phi$ enters into the Einstein equations only via these equations, $\lin{\u r}$, $\lin{\u\nu}$, and $\lin{\u\lambda}$ are clearly independent of $\lin{\u\phi}$ and we now use the notation \eqref{eq:lin-diamond-defn}. Also, $\lin{\u\phi}$ solves the free wave equation by \eqref{eq:first-order-wave}.

By \eqref{eq:lin-lambda} and \eqref{eq:DHR-1}, we compute
\begin{equation}\label{eq:lin-lambda-diamond} 
   \lin\lambda_\diamond= \partial_{\u v}\lin r_\diamond = D'\lin r_\diamond - \frac{2}{r_\diamond}\lin{\u\varpi} + \frac{2QM}{r_\diamond^2}\lin\varrho + \frac{2Q\varrho}{r_\diamond^2}\lin M = D'\left(\lin r_\diamond - \frac{r_\diamond}{M}\lin M \right) + \frac{2QM}{r_\diamond^2}\lin\varrho.
\end{equation}
Similarly, 
\begin{equation} \label{eq:lin-nu-diamond}
   \lin\nu_\diamond = -D'\left(\lin r_\diamond - \frac{r_\diamond}{M}\lin M \right) - \frac{2QM}{r_\diamond^2}\lin\varrho.
\end{equation}
Note that therefore $\lin\lambda_\diamond = -\lin \nu_\diamond$. By adding these identities, we obtain $(\partial_{\u u}+\partial_{\u v})\lin r_\diamond=0$ (which was one of the claims) and hence $\lin r_\diamond(u,u)=\lin r_\circ$ for $u\ge 1$. Inserting this back into \eqref{eq:lin-lambda-diamond} and \eqref{eq:lin-nu-diamond}, we see that $|\lin{\u\lambda}|+|\lin{\u\nu}|\les \|\lin\Psi\|_{\mathfrak Z}$ on $\{\u u=\u v\}$. 

We now estimate $\lin r_\diamond$, $\lin\lambda_\diamond$, and $\lin\nu_\diamond$ for $v\ge u$. Note that
\begin{equation}\label{eq:equation_for-linr}
    \partial_{\u v} \left( \frac{\lin r_\diamond}{D} \right)    = - \frac{2}{D r_\diamond}\lin M+ \frac{2Q}{D r^2_\diamond}  {\lin{Q}}.
\end{equation}
Hence, upon integration and using that $\frac{9}{10} \leq D \leq 1$  as well as $\u\lambda =D$, we obtain 
\begin{equation*}
    |\lin r_\diamond(\u u,\u v) -\lin r_\circ | \lesssim  \int_{r_\circ}^{r_\diamond(\u u,\u v)} \left( \frac{2}{D^2 r' }|\lin M| + \frac{2|Q|}{D^2 r'^2} |\lin Q| \right)  dr'  \lesssim \log\left(\frac{r_\diamond(\u u,\u v)}{r_\circ}\right) |\lin M| + | \lin Q|. 
\end{equation*}
Inserting this back into \eqref{eq:lin-lambda-diamond}, we obtain
$|\lin \lambda_\diamond|\lesssim D'|\lin r_\diamond|+r^{-1}_\diamond\|\lin\Psi\|_{\mathfrak Z}\lesssim r^{-1}_\diamond \| \lin\Psi\|_{\mathfrak Z} $
for $v\ge u$. Since $\lin\nu_\diamond=- \lin\lambda_\diamond$, we have the same estimate for $\lin\nu_\diamond$. 

For the region $u\ge v$ we use the dual of \eqref{eq:equation_for-linr}, which together with \eqref{eq:DHR-1} gives 
\begin{equation}\label{eq:du-r-diamond}
    \partial_{\u u} \left( \frac{\lin r_\diamond}{D}\right) = \frac{D'}{D}\frac{r_\diamond}{M}\lin M - \frac{2QM}{Dr_\diamond^2}\lin\varrho.
\end{equation}
Integrating from $\{\u u = \u v\}$, we obtain
\begin{equation*}
    \lin r_\diamond(\u u, \u v) = D\frac{\lin r_\circ}{D(r_\circ)} + D\int_{r_\diamond}^{r_\circ} \frac{D'}{D^2}\frac{r'}{M}\lin M\,dr' -D \int_{r_\diamond}^{r_\circ} \frac{2QM}{D^2r'^2}\lin\varrho\,dr',
\end{equation*} where $r_\diamond$ and $D(r_\diamond)$ (not under the integral) on the right-hand side are evaluated at $(\u u,\u v)$. For the middle integral, we use \eqref{eq:D'r-rc} to compute
\begin{equation*}
    \int_{r_\diamond}^{r_\circ} \frac{D'}{D^2}\frac{r'}{M}\lin M\,dr'=  \frac{r_c}{M}\lin M\int_{r_\diamond}^{r_\circ} \frac{D'}{D^2}\,dr'+  \int_{r_\diamond}^{r_\circ} \frac{D'}{D^2}\frac{(r'-r_c)}{M}\lin M\,dr'= \frac{r_c}{M}\lin M\left(\frac{1}{D(r_\diamond)}-\frac{1}{D(r_\circ)}\right) + O\big(\mathcal J_{1,0}(r_\diamond)\lin M\big).
\end{equation*}
We conclude that
\begin{equation}\label{eq:r-1-diamond-proof}
    \lin r_\diamond = \frac{r_c}{M}\lin M + O\big(D\mathcal J_{1,0}(r_\diamond)\|\lin\Psi_0\|_\mathfrak{Z}\big) + O\big(D\mathcal J_{2,0}(r_\diamond)|\lin\varrho|\big),
\end{equation}
which implies \eqref{eq:DHR-2} after using \cref{prop:J-integrals}. 

To prove \eqref{eq:DHR-3}, we split $\lin\nu_\diamond=\lin\nu_{\diamond,1}+\lin\nu_{\diamond,2}$, where $\lin\nu_{\diamond,i}$ is taken from $\lin{\u{\mathcal S}}{}_{\u u{}_f}[\Psi,\lin\Psi_0^i]$ with $\lin\Psi_0^1=(0,\lin r_\circ,\lin M,0)$ and $\lin\Psi_0^2=(0,0,0,\lin\varrho)$. To estimate $\lin \nu_{\diamond,1}$, insert \eqref{eq:r-1-diamond-proof} back into \eqref{eq:lin-nu-diamond}, which yields
\begin{equation}\label{eq:lin-nu-diamond-1}
    \lin\nu_{\diamond,1} = -D'\left(\frac{r_c-r_\diamond}{M}\lin M + O\big(D\mathcal J_{1,0}(r_\diamond)\|\lin\Psi_0\|_\mathfrak{Z}\big) \right) =O\big( D\big(1+|{\log(D'^{-2}D)}|\big)\|\lin\Psi_0^1\|_\mathfrak{Z}\big).
\end{equation}
In order to estimate $\lin\nu_{\diamond,2}$, we use \eqref{eq:dvlnu-ratio} to write
\begin{equation}\label{eq:dv-lin-nu-diamond}
    \partial_{\u v} \left(\frac{\lin\nu_{\diamond,2}}{-D}\right) = D'' \lin r_{\diamond,2} - \frac{4 QM}{r^3_\diamond} \lin\varrho.
\end{equation}
Since $|\lin r_{\diamond,2}|\les D'^{-1}|\lin\varrho|$, integration yields $
    |\lin\nu_{\diamond,2}|\les D(1+ \mathcal J_{1,1}) |\lin\varrho|$, and so a final application of \cref{prop:J-integrals} completes the proof of \eqref{eq:DHR-3}.
\end{proof}

We now record several useful refined estimates. First, from \eqref{eq:r-1-diamond-proof}, we have:
\begin{lem}\label{lem:lin-r-expansion}
    Under the hypotheses of \cref{prop:linear-RN}, we have the expansion
\begin{equation}\label{eq:lin-r-expansion}
    \lin r_\diamond= \frac{r_c}{M}\lin M+O\big(D'^{-1}|\lin\varrho|\big) +O\big(D'^{-1} D\big(1+|{\log(D'^{-2}D)}|\big)\|\lin\Psi_0\|_\mathfrak{Z}\big)
\end{equation}
for $\u u\ge \u v$.
\end{lem}

\begin{lem}\label{lem:lin-varkappa-diamond-estimate}
  Under the hypotheses of \cref{prop:linear-RN}, we have 
\begin{equation}\label{eq:lin-varkappa-diamond-estimate}
   |\lin\varkappa_\diamond|\les D'\|\lin\Psi_0\|_\mathfrak{Z} + D'^{-1}|\lin\varrho|
\end{equation}
for $\u u\ge \u v$.
\end{lem}
\begin{proof} Using \eqref{eq:DHR-1} and \eqref{eq:lin-r-expansion}, we compute
\begin{align*}
    \lin\varkappa_\diamond &= \frac{2}{r_\diamond^2}\lin M - \frac{4Q}{r_\diamond^3}\lin Q + \left(\frac{6Q^2}{r_\diamond^4}-\frac{4M}{r_\diamond^3}\right)\lin r_\diamond \\
    & = - \frac{4QM}{r_\diamond^3}\lin\varrho + \left[\frac{2}{r_\diamond^2}-\frac{4Q^2}{M r_\diamond^3}+\frac{6Q^2r_c}{Mr_\diamond^4}-\frac{4r_c}{r_\diamond^3}\right]\lin M  + O(D'^{-1}|\lin\varrho|)+O\big(D'^{-1} D\big(1+|{\log(D'^{-2}D)}|\big)\|\lin\Psi_0\|_\mathfrak{Z}\big).
\end{align*}
Estimating $|{\log x}|\les_\delta x^{-\delta/2}$ for $\delta>0$ and $x$ small, we estimate the final term by 
\begin{equation*}
    \les_\delta D'^{-1}D\cdot D'^\delta D^{-\delta/2}\|\lin\Psi_0\|_\mathfrak{Z}\les D'\|\lin\Psi_0\|_\mathfrak{Z}.
\end{equation*}

The term in square brackets can be written as $2M^{-1}r_\diamond^{-2}q$, where $q(r_\diamond)\doteq Mr_\diamond^2-2(Q^2+Mr_c)r_\diamond+3Q^2$. We claim that $q(r_\diamond) = O(D')$, which would complete the proof of \eqref{eq:lin-varkappa-diamond-estimate}. To prove the claim, we observe that $Q^2=M^2+O(\mathfrak t^2)$ by definition and that $\mathfrak t \les D'$ by \eqref{eq:1-mu-hierarchy-computation}. Therefore, $q(M)=O(\mathfrak t)$ and so
\begin{equation*}
   | q(r_\diamond)| \le |q(r_\diamond)-q(M)|+|q(M)|\les (r_\diamond -M) +\mathfrak t\les (r_\diamond - r_c)+\mathfrak t\les D'.\qedhere
\end{equation*}
\end{proof}

The following estimate will be crucial for absorbing a dangerous error term in the proof of \cref{thm:intro-scaling}.
\begin{lem}\label{lem:one-sided} Under the hypotheses of \cref{prop:linear-RN}, there exists a constant $C\ge 1$ depending only on $M_0$ such that if $\lin\varrho\ge 0$, then the one-sided bound
\begin{equation}\label{eq:lin-nu-diamond-one-sided}
    \lin\nu_\diamond(\u u,1) \le CD\u u\lin\varrho+C\u u^{-3/2}\|\lin\Psi_0\|_\mathfrak{Z}
\end{equation}
holds for every $\u u\ge 1$.
\end{lem}
\begin{proof} First, observe that for $|\varrho|$ bounded away from 1, \eqref{eq:DHR-3} and \eqref{eq:D-decay} already prove a two-sided bound that beats \eqref{eq:lin-nu-diamond-one-sided}. Hence, we focus on the case when $|\varrho|$ is close to 1. 

We now split $\lin r_\diamond = \lin r_{\diamond,1}+\lin r_{\diamond,2}$ and $\lin\nu_\diamond = \lin\nu_{\diamond,1}+\lin\nu_{\diamond,2}$ as in the proof of \cref{prop:linear-RN}. According to \eqref{eq:lin-nu-diamond-1} and \eqref{eq:D-decay}, we have the good two-sided bound
\begin{equation*}
    |\lin\nu_{\diamond,1}|\les D\cdot D'^{1/2}D^{-1/4} \|\lin\Psi_0\|_\mathfrak{Z}\les \u u^{-3/2} \|\lin\Psi_0\|_\mathfrak{Z},
\end{equation*}
so we can further restrict attention to $\lin \nu_{\diamond,2}$. By scaling, we may assume $\lin\varrho =1$. Using \eqref{eq:du-r-diamond}, we obtain 
$\lin r_{\diamond,2} =-\theta+O(D)$, where $\theta\gtrsim  D\u u^3\ge 0$, and we used \eqref{eq:D-decay} to estimate $\int D^{-1}\,d\u u'\gtrsim \u u^3$. For extremal Reissner--Nordstr\"om, $D''(r_+)=2M^{-2}$, so for $|\varrho|$ sufficiently close to $1$, we have $D''\gtrsim 1 - O(D^{1/2})$ by Taylor's theorem, where we used \eqref{eq:lower-bound-on1-mu} to estimate $r-r_+\les D^{1/2}$. It follows from \eqref{eq:dv-lin-nu-diamond} that
\begin{equation*}
    \lin\nu_{\diamond,2}(\u u,\u v) = D (\u u,\u v)\cdot\left.\frac{\lin\nu_{\diamond,2}}{D}\right|_{\Gamma}+ D(\u u,\u v)\int^{\u u}_{\u v}\left(D''\lin r_{\diamond,2}- \frac{4QM}{r_\diamond^3}\right)\,d\u v'
\end{equation*}
The worst behaved term, involving $\theta$, enters as $-\theta + O(D^{1/2}\theta)$ and $D^{1/2}\theta\les 1$, so we in fact have
\begin{equation*}
    \lin \nu_{\diamond,2} \les D + D\int_{\u v}^{\u u}\,d\u v'\les D \u u
\end{equation*}
as desired.
\end{proof}

\section{Setup and statements of the main theorems}\label{sec:statements} 

With the preliminaries out of the way, we are now ready to state precisely the main theorems of the paper. In \cref{sec:definitions}, we give the final definitions required for the main theorems. In \cref{sec:uniform-stability}, we state precisely the refined dichotomy and uniform stability \cref{thm:intro-uniform-stability}. Then, in \cref{sec:threshold}, we state the precise versions of \cref{thm:intro-foliation,thm:intro-scaling,thm:intro-instability}.

\subsection{Definitions}\label{sec:definitions}

\subsubsection{Fixing the domain and scale}\label{sec:fixing}

\index{M@$M_0$, fixed mass scale}
Fix for the remainder of the paper a number $M_0>0$ and let \index{U@$U_*\doteq \frac{995}{10}M_0+1$, final $u$ for the setup of the main theorems}
\begin{equation}\label{eq:U*}U_*\doteq \frac{995}{10}M_0+1.\end{equation}
We now fix the semiglobal bifurcate null hypersurface $\mathcal C=\mathcal C(1,U_*,1,\infty)$ for the rest of the paper. On $\mathcal C$, we consider the moduli space of characteristic seed data $\mathfrak M$, which is an open subset of the Banach space $\mathfrak Z(\mathcal C)=\mathfrak X(\mathcal C)\times\Bbb R$. We refer the reader back to \cref{sec:AF} for the definitions of these spaces. 

Next, we fix a basepoint \index{x1@$x_0$, basepoint in $\mathfrak X$}
\begin{equation*}
    x_0\doteq (0,100M_0,M_0)\in \mathfrak X.
\end{equation*}
The distinguished Reissner--Nordstr\"om family we wish to consider, in the sense of \cref{sec:RN-seed-data}, is given by the family\index{R@$ \mathfrak R\doteq  \{(x_0,\rho_\circ):\rho_\circ\in\Bbb R\}\subset \mathfrak M$, RN family}
\begin{equation*}
   \mathfrak R\doteq  \{(x_0,\rho_\circ):\rho_\circ\in\Bbb R\}.
\end{equation*}
Note that the conditions \eqref{eq:RN-condition-1}, \eqref{eq:u-1-condition}, and \eqref{eq:RN-condition-2} are satisfied by the definitions of $U_*$ and $x_0$. In particular, $\mathfrak R\subset\mathfrak M$. If a black hole forms ($|\rho_\circ|\le 1$), the event horizon is located at 
\begin{equation}\label{eq:RN-horizon-loc}
    u_{\mathcal H^+}^0(\rho_\circ)\doteq 1+99M_0-M_0\sqrt{1-\rho_\circ^2} < U_*.
\end{equation}

For a point $x\in \mathfrak X$ and $\ve>0$, let $B^\mathfrak{X}_\ve(x)$ denote the open $\|\cdot\|_\mathfrak{X}$-ball of radius $\ve$ around $x$. In order to quantify neighborhoods of $\mathfrak R$, we introduce the \emph{cylinder}\index{cyl@$ \cyl(\ve,\ell)$, cylinder in moduli space of radius $\ve$ and length $\ell$} $\cyl(\ve,\ell)$, parametrized by a \emph{radius} $\ve>0$ and \emph{length} $\ell>0$:
\begin{equation*}
    \cyl(\ve,\ell)\doteq B^\mathfrak{X}_{\ve}(x_0)\times [-\ell,\ell]\subset\mathfrak Z.
\end{equation*}
We note the following basic result which is immediate from \cref{lem:AF-data-generation}:
\begin{lem}\label{lem:basic-cylinder}
    For any $\ell>0$, there exists an $\ve_\loc(M_0,\ell)>0$ , such that
    \begin{equation*}
        \cyl(\ve_\loc,\ell)\subset\mathfrak M.
    \end{equation*}
\end{lem}

\begin{rk}
    Note already that if $\ell<1$ and $\ve$ is sufficiently small (depending on $\ell$!), then all solutions in $\cyl(\ve,\ell)$ lie in $\mathfrak M_\mathrm{black}$ and converge to asymptotically \emph{subextremal} Reissner--Nordstr\"om black holes by \cite{Price-law,luk2019strong}. 
\end{rk}

Using \cref{prop:a-priori-noncollapse}, we obtain:
\begin{prop}\label{prop:difference-of-cylinders}
    There exists a sequence of positive constants $\{\ve_j(M_0)\}_{j\ge 1}$ such that
    \begin{equation*}
        \cyl(\ve_j,2^{j+1})\setminus \cyl(\ve_j,2^j)\subset\mathfrak M_\mathrm{non}.
    \end{equation*}
\end{prop}
\begin{rk}
    We need to possibly shrink $\ve_j$ as $j$ increases to ensure that $\cyl(\ve_j,2^{j+1})\setminus \cyl(\ve_j,2^j)$ is a subset of $\mathfrak M$ (\cref{lem:AF-data-generation}). Moreover, the $\ve$ required to stave off collapse can depend on $\rho_\circ$, which forces us to possibly make $\ve_j$ smaller again to apply \cref{prop:a-priori-noncollapse}.
\end{rk}

The set $\mathfrak M_\mathrm{nbhd}$\index{M@$\mathfrak M_\mathrm{nbhd}\subset\mathfrak M$, neighborhood of the Reissner--Nordstr\"om family} in \cref{thm:intro-uniform-stability,thm:intro-foliation,thm:intro-scaling,thm:intro-instability} is given by 
\begin{equation*}
    \mathfrak M_\mathrm{nbhd} \doteq \cyl(\ve_0,2) \cup \bigcup_{j\ge 1}\big(  {\cyl(\ve_j,2^{j+1})}\setminus \cyl(\ve_j,2^j)\big),
\end{equation*}
where $\ve_0>0$ will be chosen according to \cref{thm:uniform-stability-RN,thm:dichotomy-revisited,thm:main,thm:scaling,thm:instabilities} below, depending only on $M_0$ and an arbitrary parameter $\delta>0$. It is clear that $\mathfrak M_\mathrm{nbhd}$ is a neighborhood of the Reissner--Nordstr\"om family $\mathfrak R$ as promised. By \cref{prop:difference-of-cylinders}, $\mathfrak M_\mathrm{nbhd}\setminus\cyl(\ve_0,2)\subset\mathfrak M_\mathrm{non}$, so in the rest of the paper we only need to focus on the behavior of solutions in $\cyl(\ve_0,2)$. 

\subsubsection{The anchoring condition and definitions of energies}\label{sec:anchoring}

We now describe the background Reissner--Nordstr\"om anchoring procedure. Let $2\le \u u{}_f\le\infty$ and take $\Psi\in \mathfrak M_{\u u_f}$. Then $\mathcal L_\mathrm{Bondi}(\Psi)\ge 1$ and $\mathcal S[\Psi]$ may be put into the $\u u{}_f$-normalized teleological gauge of \cref{sec:teleological-gauge} on \index{Rauf1@$\mathcal R_{\u u{}_f}\doteq\mathcal R(1,\u u{}_f,1,\infty)$}
\begin{equation}\label{eq:Ruf-defn}
   \mathcal R_{\u u{}_f}\doteq\mathcal R(1,\u u{}_f,1,\infty).
\end{equation} Let $M>0$ and $\varrho\in\Bbb R$ be Reissner--Nordstr\"om parameters and suppose that $\u u{}_f\le u_\infty(M,\varrho)$, which was defined in \eqref{eq:defn-u-infty} above. 

Using \cref{lem:existence-anchoring}, we now assign a Reissner--Nordstr\"om background $(r_\diamond,\Omega^2_\diamond)$ with parameters $(M,\varrho)$ by setting 
\begin{equation}\label{eq:finite-anchoring}
r_\diamond (1, 1) = \u r{}_{\u u{}_f} (\u u{}_f,\u u_f)
\end{equation} when $\u u{}_f<\infty$, 
where $\u r{}_{\u u{}_f}$ is the area-radius function of $\u{\mathcal S}_{\u u{}_f}[\Psi]$. Note that $r_\diamond$ is constant along $\{\u u=\u v\}$, so $r_\diamond (\u u{}_f,\u u_f) = \u r (\u u{}_f,\u u_f)$ as well. When $\u u{}_f=\infty$, we assume that $\Psi\in\mathfrak M_\mathrm{black}^{\kappa\sim 1}$ so that the eschatological gauge of \cref{sec:eschatological-definition} exists, we set
\begin{equation}\label{eq:eschatological-anchoring}
r_\diamond(1,1) = \lim_{\u u\to\infty} \u r{}_\infty(\u u,\u u),
\end{equation}
where we also need to assume that this limit exists.  

\begin{defn}The $(r_\diamond,\Omega^2_\diamond)$ defined by this process is called the \emph{$(M,\varrho,\u u{}_f)$-anchored background solution} associated to $\Psi$.\end{defn}

We now adopt the following conventions:
\begin{itemize}
    \item \index{*@$\diamond$, background quantity} Diamond quantities such as $\lambda_\diamond$, $\varpi_\diamond=M$, or $\kappa_\diamond=1$ correspond to those of $(r_\diamond,\Omega^2_\diamond)$.
    \item \index{*@$\dagger$, difference between dynamical and background quantity} Differences are denoted with a dagger, such as $\u r{}_{\u u{}_f\dagger}\doteq \u r{}_{\u u{}_f}- r_\diamond$ or $\u\gamma{}_{\u u{}_f\dagger} \doteq \u\gamma{}_{\u u{}_f}+1$.
    \item When $\u u{}_f$ is fixed in a section, we will often omit the subscript on $\u u{}_f$-teleologically normalized quantities.
    \item When $\u u{}_f$ varies, it will be convenient to recall the dependence of $r_\diamond$ on $\u u{}_f$ and we write $(r_{\u u{}_f\diamond},\Omega^2_{\u u{}_f\diamond})$.
\end{itemize}

\begin{figure}
\centering{
\def\svgwidth{16pc}
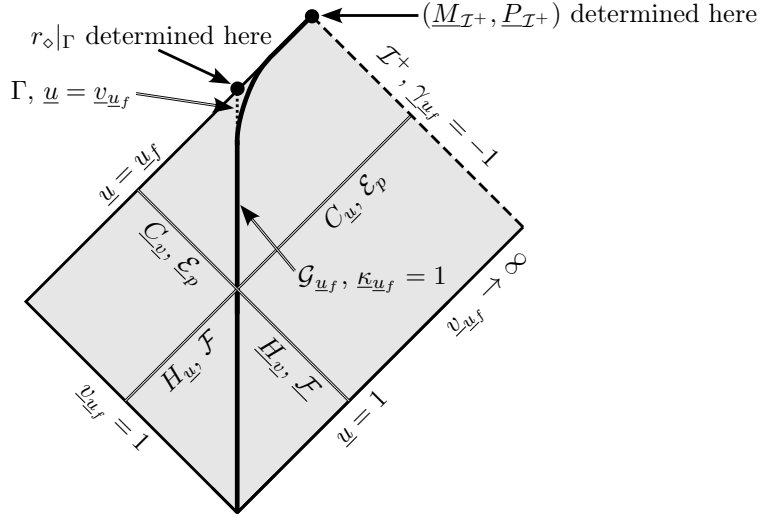}
\caption{A Penrose diagram showing the teleological normalizations, null hypersurfaces, energies, and anchoring conditions in our ``bootstrap'' domain $\mathcal R_{\u u{}_f}$. The dotted vertical line is the diagonal $\Gamma=\{\u u=\u v{}_{\u u{}_f}\}$. The function $\tau$ measures advanced time to the left of $\Gamma$ and retarded time to the right of $\Gamma$.}
\label{fig:bootstrap-setup}
\end{figure}

We may now define the fundamental weighted energy norms for the scalar field. Some of the norms will depend explicitly on the background solution $r_\diamond$ in a nontrivial manner. We fix once and for all an arbitrary parameter 
\begin{equation}\index{delta@$\delta$, fixed small parameter}
    0<\delta<\tfrac{1}{100}.\label{eq:defn-delta}
\end{equation}
We also define a time function on $\Bbb R^2_{\u u,\u v}$ by\index{tau@$\tau\doteq\min\{\u u, \u v\}$, time function}
\begin{equation*}
\tau(\u u, \u v)\doteq \min\{\u u , \u v\}
\end{equation*}
and the null hypersurfaces
\begin{align*}\index{C1@$C_{\u u}$, outgoing cone in the far region}\index{C2@$\u C{}_{\u v}$, ingoing cone in the near region}\index{H8@$H_{\u u}$, outgoing cone in the near region}\index{H9@$\u H{}_{\u v}$, ingoing cone in the far region}
   C_{\u u}&\doteq \{\u u\}\times [\u u,\infty) ,& \u C{}_{\u v}&\doteq [\u v,\infty)\times\{\u v\},\\
   H_{\u u}&\doteq \{\u u\}\times [1,\u u],&\underline H{}_{\u v}&\doteq [1,\u v]\times\{\u v\}.
\end{align*}
Note that we will use $\tau$ as both a parameter and a function, but this should not cause any confusion. Refer to \cref{fig:bootstrap-setup} for a Penrose diagram summarizing the gauge condition, anchoring condition, the hypersurfaces defined above, and the energies defined below.

\begin{defn}\label{def-of-energies} Let $2\le \u u{}_f\le \infty$ and let $\Psi\in \mathfrak M_{\u u{}_f}$ with $(M,\varrho,\u u{}_f)$-anchored background solution $(r_\diamond, \Omega^2_\diamond)$. Let $\zeta:\mathcal R_{\u u{}_f}\to \Bbb R$ be a $C^1$ function.\footnote{In the context of the proof of the main theorem, $\zeta$ will either be $\u \phi{}_{\u u{}_f}$ or $\lin{\u\phi}{}_{\u u{}_f}$.} For $\tau,\tau'\in[1,\u u{}_f]$, $p\in\{0\}\cup[\delta,3-\delta]$, and $(\u u,\u v)\in\mathcal R_{\u u{}_f}$, we define:
\begin{enumerate}
    \item The $h_p$-weighted energy flux: \index{Ep1@$\underline{\mathcal E}{}_p^{(M,\varrho,\u u{}_f)}$, $h_p$-weighted energy}
\begin{equation*}
    \underline{\mathcal E}{}_p^{(M,\varrho,\u u{}_f)}[\zeta](\tau)\doteq \int_{\u C{}_\tau\cap\mathcal R}\big(h_p(r_\diamond)((\partial_{\u u}\u\zeta)^2+(\partial_{\u u}(\u r{}_{\u u{}_f}\zeta))^2)+D(r_\diamond)\u\zeta^2\big)\,d\u u,
\end{equation*}
where the function $h_p$ was defined in \cref{sec:hp-defn}.

   \item  The $r^p$-weighted energy flux: \index{Ep2@$\mathcal E{}_p^{(M,\varrho,\u u{}_f)}$, $r^p$-weighted energy}
   \begin{equation*}
   \mathcal E_p^{(M,\varrho,\u u_f)}[\zeta](\tau)\doteq \int_{C_\tau} \big(\mathbf 1_{\{p>0\}}\u r{}_{\u u{}_f}^p(\partial_{\u v}(\u r{}_{\u u{}_f}\zeta))^2 +\u r{}_{\u u{}_f}^{2}(\partial_{\u v}\zeta)^2+  \zeta^2\big)\,d\u v.
   \end{equation*}        

        \item The energy flux along outgoing cones in the near region: \index{F1@$\mathcal F^{(M,\varrho,\u u{}_f)}$, outgoing energy in near region}
        \begin{equation*}
           \mathcal F^{(M,\varrho,\u u{}_f)}[\zeta](\u u,\tau')\doteq \int_{H_{\u u}\cap\{\tau\ge\tau'\}}\big((\partial_{\u v}\zeta)^2+D(r_\diamond)\zeta^2\big)\,d\u v.
        \end{equation*}
        \item The energy flux along ingoing cones in the far region:\index{F2@$\u{\mathcal F}^{(M,\varrho,\u u{}_f)}$, ingoing energy in far region}
        \begin{equation*}
            \underline{\mathcal F}^{(M,\varrho,\u u{}_f)}[\zeta](\u v,\tau')\doteq \int_{\underline H{}_{\u v}\cap\{\tau\ge\tau'\} \cap\mathcal R}\big(\u r{}_{\u u{}_f}^2(\partial_{\u u}\zeta)^2+\zeta^2\big)\,d\u u.
        \end{equation*}
    \end{enumerate}
\end{defn}

\subsubsection{Definitions of \texorpdfstring{$\mathcal K$}{K} and \texorpdfstring{$\tilde{\mathcal K}$}{K tilde}}

We now define the set of initial data whose evolutions decay to Reissner--Nordstr\"om on a certain timescale. We first restrict attention to parameters which satisfy \eqref{eq:mass-condition}, \eqref{eq:sub-supextremal-condition}, and the hypotheses of \cref{lem:rgeqrc}.

\begin{defn}
 The triple $(M,\varrho,\u u{}_f)$ is \emph{admissible} if $\frac 12 M_0\le M\le 2M_0$, $|\varrho|\le \frac 54$, and $2\le \u u{}_f\le u_\infty(M,\varrho)$.
\end{defn}

\begin{defn}\label{def:K}\index{K1@$\mathcal K(\ve,M,\varrho,\u u{}_f)$, strict decay class} \index{K2@$\tilde{\mathcal K}(\ve,M,\varrho,\u u{}_f)$, relaxed decay class} \index{A1@$A$, bootstrap constant in $\mathcal K$ and $\tilde{\mathcal K}$, see also $A_\mathrm{unif}$}
Let $\ve>0$, $A\ge 1$, and $(M,\varrho,\u u{}_f)$ be an admissible triple, where we allow $\u u{}_f=\infty$. We define
\[\mathcal K(\ve,M,\varrho,\u u{}_f)\subset\mathfrak M\]
to consist of those seed data $\Psi\in \mathfrak M$ for which the following holds:
\begin{enumerate}
    \item\label{K-prop-1} $\Psi$ is $\u u{}_f$-admissible. If $\u u{}_f=\infty$, we require additionally:
    \begin{enumerate}
        \item[1.i.]\label{K-prop-1i} $\kappa$ is bounded above and below by positive constants on $\mathcal R_\infty$, so that $\Psi\in \mathfrak M_\mathrm{black}^{\kappa\sim 1}$.
        \item[1.ii.]\label{K-prop-1ii} The limit in \eqref{eq:eschatological-anchoring} exists.
    \end{enumerate}
    \item\label{K-prop-2} The initial data satisfies
\begin{equation}\label{eq:new-initial-smallness}
    \Psi\in\cyl(\ve,2).
\end{equation}
 
 \item\label{K-prop-3} The Bondi mass satisfies \begin{equation}\label{eq:varpi-exact}
            \u M{}_{\mathcal I^+}(\u u{}_f)= M
        \end{equation}
        at the final time. When $\u u{}_f=\infty$, this is to be interpreted as $M_\infty=M$.
        \item\label{K-prop-4} The charge-to-mass ratio satisfies \begin{equation}\label{eq:P-exact}
          \u P{}_{\mathcal I^+}(\u u{}_f)  =\varrho
        \end{equation}
        at the final time. When $\u u{}_f=\infty$, this is to be interpreted as $P_\infty=\varrho$.
        \item\label{K-prop-5} Let $(r_\diamond,\Omega^2_\diamond)$ denote the $(M,\varrho,\u u{}_f)$-anchored background solution associated to $\Psi$ and define the diamond and dagger quantities as in \cref{sec:anchoring}. It then holds that
        \begin{align}
         \label{eq:K-nu-1}   |\u \nu{}_{\u u{}_f\dagger}|&\le A^2\ve^{3/2}D(r_\diamond)\tau^{-1+\delta} ,\\
             \label{eq:K-r-1}    |\u r{}_{\u u{}_f\dagger}| &\le  A^2\ve^{3/2}\big(\tau^{-2+\delta}+r_\diamond^\delta\u u{}_f^{-2+\delta}\big),\\
             \label{eq:K-varpi-1}    |\u\varpi{}_{\u u{}_f\dagger}| &\le A\ve^{3/2}\tau^{-3+\delta}
        \end{align}
     on $\mathcal R_{\u u{}_f}$.
        \item\label{K-prop-6} The energies satisfy
        \begin{align}
              \label{eq:K-energy-1}    \mathcal E_p^{(M,\varrho,\u u{}_f)}[\u \phi{}_{\u u{}_f}](\tau)&\le A\ve^2 \tau^{-3+\delta+p},\\
              \label{eq:K-energy-2}    \underline{\mathcal E}{}_p^{(M,\varrho,\u u{}_f)}[\u \phi{}_{\u u{}_f}](\tau)&\le A\ve^2\tau^{-3+\delta+p},\\
              \label{eq:K-energy-3}    \mathcal F^{(M,\varrho,\u u{}_f)}[\u \phi{}_{\u u{}_f}](\u u,\tau)& \le A\ve^2\tau^{-3+\delta},\\
              \label{eq:K-energy-4}    \underline{\mathcal F}^{(M,\varrho,\u u{}_f)}[\u \phi{}_{\u u{}_f}](\u v,\tau)& \le A\ve^2\tau^{-3+\delta}
        \end{align}
      for every $\tau\in[1,\u u{}_f]$, $p\in\{0\}\cup[\delta,3-\delta]$, and $(\u u,\u v)\in\mathcal R_{\u u{}_f}$.
\end{enumerate}
\end{defn}

\begin{rk}
    The notation for the set $\mathcal K(\ve,M,\varrho,\u u{}_f)$ (and $\tilde{\mathcal K}$ below) does not involve the constant $A$ to reduce visual clutter, as it will be fixed \index{A2@$A_\mathrm{unif}$, fixed uniform bootstrap constant in $\mathcal K$ and $\tilde{\mathcal K}$} to be a large number $A_\mathrm{unif}(M_0,\delta)$ in \cref{thm:uniform-stability-RN}.
\end{rk}

\begin{defn}\label{def:K-tilde}
Let $\ve>0$, $A\ge 1$, and $(M,\varrho,\u u{}_f)$ be an admissible triple, where we allow $\u u{}_f=\infty$. We define
\[\tilde{\mathcal K}(\ve,M,\varrho,\u u{}_f)\subset\mathfrak M\]
to consist of those seed data $\Psi\in \mathfrak M$ for which conditions \hyperref[K-prop-1]{1.}, \hyperref[K-prop-2]{2.}, \hyperref[K-prop-5]{5.}, and \hyperref[K-prop-6]{6.}~of \cref{def:K} hold, but conditions \hyperref[K-prop-3]{3.}~and \hyperref[K-prop-4]{4.}~are relaxed to:
\begin{enumerate}
    \item[$\tilde 3$.]\label{K-prop-3tilde} The Bondi mass satisfies  \begin{equation}\label{eq:varpi-modulation}
    |\u M{}_{\mathcal I^+}(\u u{}_f)-M|\le 10\ve^{3/2}\u u{}_f^{-3+\delta}
\end{equation}
        at the final time. 
        
    \item[$\tilde 4$.]\label{K-prop-4tilde} The charge-to-mass ratio satisfies \begin{equation}\label{eq:P-modulation}
    |\u P{}_{\mathcal I^+}(\u u{}_f)-\varrho|\le 10M_0^{-1}\ve^{3/2}\u u{}_f^{-3+\delta}
\end{equation}
        at the final time. 
\end{enumerate}
\end{defn}

Note that 
\begin{equation*}
    \tilde{\mathcal K}(\ve,M,\varrho,\u u{}_f)\supsetneq \mathcal K(\ve,M,\varrho,\u u{}_f).
\end{equation*} in general for finite $\u u{}_f$, but
\begin{equation*}
    \tilde{\mathcal K}(\ve,M,\varrho,\infty)= \mathcal K(\ve,M,\varrho,\infty).
\end{equation*}

\subsection{Uniform asymptotic stability of Reissner--Nordstr\"om}\label{sec:uniform-stability} 

\subsubsection{Uniform decay estimates}

The following is the first of the detailed main theorems of this paper and includes the uniform decay aspect of \cref{thm:intro-uniform-stability} from the introduction.

\begin{thm}[Uniform decay estimates]\label{thm:uniform-stability-RN}
    There exist constants $\ve_0>0$ and $A_\mathrm{unif}\ge 1$, depending only on $M_0$ and $\delta$, such that if $\ve\le \ve_0$, $(M,\varrho,\u u{}_f)$ is an admissible parameter triple, $A=A_\mathrm{unif}$ in \cref{def:K,def:K-tilde}, and $\Psi\in \mathcal K(\ve,M,\varrho,\u u{}_f)$ or $\Psi\in \tilde{\mathcal K}(\ve,M,\varrho,\u u{}_f)$, then the following holds:
 \begin{enumerate}
 \item \underline{Improved ``bootstrap assumptions''}: The estimates \eqref{eq:K-nu-1}--\eqref{eq:K-energy-4} in the definitions of $\mathcal K$ and $\tilde{\mathcal K}$ hold with ``improved'' constants $\frac 12 A_\mathrm{unif}$ and $\frac 12 A_\mathrm{unif}^2$ on the right-hand sides instead of $A_\mathrm{unif}$ and $ A_\mathrm{unif}^2$, respectively.
 
      \item \underline{Decay of the geometry}: It holds that 
      \begin{equation}\label{eq:uniform-decay-1}
          |\u\gamma{}_{\u u{}_f\dagger}|\les\ve^2r^{-1}_\diamond\tau^{-3+\delta},\qquad |\u\kappa{}_{\u u{}_f\dagger}|\les \ve^2\min\{\tau^{-3+\delta},r^{-1}_\diamond\tau^{-2+\delta}\},\qquad |\u\nu{}_{\u u{}_f\dagger}|\les \ve^{3/2}r^{-1}_\diamond\tau^{-2+\delta}
      \end{equation}
      on $\mathcal R_{\u u{}_f}\cap\{\u v\ge \u u\}$ and 
 {   \mathtoolsset{showonlyrefs=false}  \begin{gather}
     \label{eq:uniform-decay-2}     |\u r{}_{\u u{}_f\dagger}|\les \ve^{3/2}\big(\tau^{-2+\delta}+r_\diamond^\delta u_f^{-2+\delta}\big),\qquad |\u\lambda{}_{\u u{}_f\dagger}|\les \ve^{3/2}r^{-2}_\diamond\tau^{-2+\delta}+\ve^{3/2}\min\{\tau^{-3+\delta},r^{-1}_\diamond\tau^{-2+\delta}\},\\ \label{eq:uniform-decay-3}|\u\nu{}_{\u u{}_f\dagger}|\les\ve^{3/2}D\tau^{-1+\delta},\qquad  |\u\kappa{}_{\u u{}_f\dagger}|\les \ve^2\tau^{-1+\delta},\qquad |D'\u\kappa{}_{\u u{}_f\dagger}|\les \ve^2\tau^{-2+\delta},\qquad |D\u\kappa{}_{\u u{}_f\dagger}|\les \ve^2\tau^{-3+\delta},\\ |\u\varpi{}_{\u u{}_f\dagger}|\les \ve^{3/2}\tau^{-3+\delta}\label{eq:uniform-decay-4}
      \end{gather}}
    on $\mathcal R_{\u u{}_f}$, where here the notation $\les$ means the implicit constant depends only on $M_0$ and $\delta$. When $\u u{}_f=\infty$, these estimates extend to $\mathcal H^+$ using the convention \eqref{eq:extension-to-horizon}.

      \item \underline{Pointwise estimates for the scalar field}: It holds that
\begin{align*}
|\u\psi{}_{\u u{}_f}|&\les \ve\tau^{-1+\delta/2},\\
|\sqrt{r^3_\diamond D'}\u\phi{}_{\u u{}_f}|&\les \ve\tau^{-3/2+3\delta/4},\\
|\u r{}_{\u u{}_f}^2\partial_{\u v}\u\phi{}_{\u u{}_f}|+|\u r{}_{\u u{}_f}^2\partial_{\u v}\u\psi{}_{\u u{}_f}|+|\u rY\u \phi{}_{\u u{}_f}|+|Y\u \psi{}_{\u u{}_f}|&\les \ve
\end{align*}
      on $\mathcal R_{\u u{}_f}$, where $Y\doteq \u\nu{}_{\u u{}_f}^{-1}\partial_{\u u}$ and $\u\psi{}_{\u u{}_f}\doteq \u r{}_{\u u{}_f} \u\phi{}_{\u u{}_f}$.

    \item \ul{Orbital stability}: The mass parameter $M$ satisfies 
    \begin{equation}\label{eq:mass-orbital-stability}
        |M-\varpi_\circ|\les \ve^{3/2}. 
    \end{equation}
    With respect to the anchored Reissner--Nordstr\"om solution $r_\diamond$ with parameters $(M,\varrho)$, the $p=3-\delta$ energy of the scalar field is bounded by its initial value:
    \begin{equation}\label{eq:orb-stab-1}
        \sup_{1\le\tau\le \u u{}_f}\big(\mathcal E^{(M,\varrho,\u u{}_f)}_{3-\delta}[\u\phi{}_{\u u{}_f}](\tau)+\u{\mathcal E}{}^{(M,\varrho,\u u{}_f)}_{3-\delta}[\u\phi{}_{\u u{}_f}](\tau)\big) \les \mathcal E^{(M,\varrho,\u u{}_f)}_{3-\delta}[\u\phi{}_{\u u{}_f}](1)+\u{\mathcal E}{}^{(M,\varrho,\u u{}_f)}_{3-\delta}[\u\phi{}_{\u u{}_f}](1).
    \end{equation}
    Moreover, the scalar field is pointwise bounded in $C^1$ in $\mathcal R_{\u u{}_f}$ in terms of its initial values,
    \begin{multline}\label{eq:orb-stab-2}
        \sup_{\mathcal R_{\u u{}_f}}\big(|\u\psi{}_{\u u{}_f}|+|\u r{}_{\u u{}_f}^2\partial_{\u v}\u \psi{}_{\u u{}_f}|+|\u r{}_{\u u{}_f}^2\partial_{\u v}\u\phi{}_{\u u{}_f}|+|Y\u\psi{}_{\u u{}_f}|+|\u r{}_{\u u{}_f}Y\u\phi{}_{\u u{}_f}|\big) \\ 
        \les  \sup_{\mathcal R_{\u u{}_f}\cap \mathcal C}\big(|\u\psi{}_{\u u{}_f}|+|\u r{}_{\u u{}_f}^2\partial_{\u v}\u \psi{}_{\u u{}_f}|+|\u r{}_{\u u{}_f}^2\partial_{\u v}\u\phi{}_{\u u{}_f}|+|Y\u\psi{}_{\u u{}_f}|+|\u r{}_{\u u{}_f}Y\u\phi{}_{\u u{}_f}|\big).
    \end{multline}
    The right-hand side of \eqref{eq:orb-stab-1} is $\les \ve^2$ and the right-hand side of \eqref{eq:orb-stab-2} is $\les\ve$.

 \item \underline{Improved estimates when $\u u{}_f=\infty$}: If $\u u{}_f=\infty$, then all of the estimates in \eqref{eq:uniform-decay-1}--\eqref{eq:uniform-decay-4} hold with $\ve^{3/2}$ replaced by $\ve^2$.

  \item \underline{Stability of the location of the event horizon}: Recall the reference Reissner--Nordstr\"om black hole family $\mathfrak R$ from \cref{sec:fixing}, with event horizon located at $u^0_{\mathcal H^+}$ in the initial data gauge. If $\u u{}_f=\infty$, the event horizon location $u_{\mathcal H^+}$ satisfies
    \begin{equation}\label{eq:event-horizon-estimate}
      |u_{\mathcal H^+}-u^0_{\mathcal H^+}(M,\varrho)|\les \ve.
  \end{equation}
\end{enumerate}
\end{thm}

The proof of this theorem is given in \cref{sec:proof-uniform-stab}.

\begin{rk} If $\u u{}_f=\infty$ and $|\varrho| =1$, then $\Psi\in \mathfrak M_\mathrm{weak\,ext}$ and  \cref{prop:weak-ext-properties} applies.
\end{rk}

\begin{figure}
\centering{
\def\svgwidth{13pc}
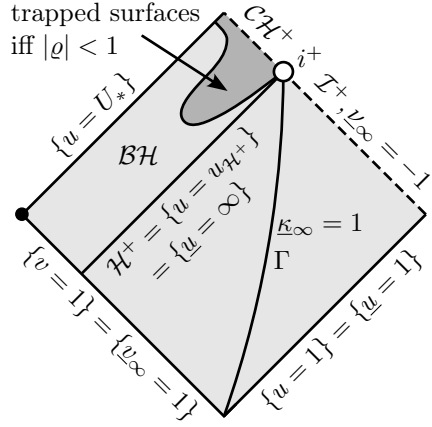}
\caption{A Penrose diagram depicting the maximal development of seed data in the black hole case $\u u{}_f=\infty$. The existence of the Cauchy horizon when $\mathscr P_\infty\ne 0$ and the statement about trapped surfaces follow from \cite{Dafermos-thesis,Price-law,dafermos2005interior} and \cref{prop:weak-ext-properties}.}
\label{fig:stability}
\end{figure}

\subsubsection{Revisiting the fundamental dichotomy}

\cref{thm:dichotomy} states that solutions in the spherically symmetric EMSF model (arising from data in our moduli space $\mathfrak M$) either fail to collapse or form black holes. The following theorem refines this by showing that in a neighborhood of $\mathfrak R$, solutions either fail to collapse because they become superextremal along $\mathcal I^+$, or collapse to form an asymptotically Reissner--Nordstr\"om black hole.  

\begin{thm}\label{thm:dichotomy-revisited}    
        There exists a constant $\ve_0>0$, depending only on $M_0$ and $\delta$, such that the following refined version of the fundamental dichotomy \cref{thm:dichotomy} holds:
    \begin{enumerate}
        \item  $\Psi\in \cyl(\ve_0,2)\cap\mathfrak M_\mathrm{non}$ if and only if $|\u{P}_{\mathcal I^+}(L_i)|>1$ for some $i\ge 0$.
        \item $\Psi\in \cyl(\ve_0,2)\cap \mathfrak M_\mathrm{black}$ if and only if  $\Psi\in\mathcal K(\ve,M_i,\varrho_i,L_i)$ for every $i\ge 1$, where $\ve\le \ve_0$, $\{(M_i,\varrho_i)\}_{i\ge 1}$ are Reissner--Nordstr\"om parameters with $|\varrho_i|\le 1$, and $A=A_\mathrm{unif}$ in \cref{def:K}. Moreover, the sequence $(M_i,\varrho_i)$ has a limit $(M_\infty,\varrho_\infty)$ and it holds that $\Psi\in \mathcal K(\ve,M_\infty,\varrho_\infty,\infty)$. 
    \end{enumerate}
\end{thm}

Together with \cref{thm:uniform-stability-RN}, this completes the detailed version of \cref{thm:intro-uniform-stability}. The proof of this theorem is given in \cref{sec:revisiting}.

\subsection{The isologous foliation and the extremal threshold}\label{sec:threshold}

We now state the precise version of \cref{thm:intro-foliation}. 

Recall that given a subset $U\subset\mathfrak Z$ and a function $f:U \to \mathbb R$, we define its \emph{augmentation} $ \check f: U \to \mathfrak Z$ by $\check f(x,\rho_\circ) = (x,f(x,\rho_\circ))$. Given $U\subset \mathfrak Z$ which is not open, we say that $f\in C^1(U)$ if there exists an open neighborhood $\mathcal U$ of $U$ such that $f$ extends to a $C^1$ function on $\mathcal U$. We say that $f\in C^1_b(U)$ if there exists a $C^1_b$ extension. The sets $\cyl(\ve,\ell)$ are neither open nor closed, and if $f\in C^1_b(\cyl(\ve,\ell))$, with $F:\mathcal O\to\Bbb R$ a $C^1_b$ extension ($\mathcal O\supset \cyl(\ve,\ell)$ open), then $\|f\|_{C^1(\cyl(\ve,\ell))}\doteq \sup_{\cyl(\ve,\ell)} \big(|F|+|F'|\big)$. This norm is independent of $F$.

\begin{thm}\label{thm:main}
    Let $M_0>0$ and $0<\delta<\frac{1}{100}$. Then there exist a constant $\ve_0>0$ depending on $M_0$ and $\delta$, and a $C^1_b$ function
    \begin{equation*}
        \mathscr W_\infty : \cyl(\ve_0,1)\to \Bbb R
    \end{equation*} depending on $M_0$, with the following properties:
    \begin{enumerate}
        \item \ul{Existence of stable manifolds}:\index{M@$\mathfrak M_\mathrm{stab}^\sigma$, stable manifold with final parameter ratio $\sigma$, isologous hypersurface} For $\sigma \in [-1,1]$, put 
        \begin{equation*}
        \mathfrak M_\mathrm{stab}^\sigma \doteq \{(x,\mathscr W_\infty(x,\sigma)):x\in B^\mathfrak{X}_{\ve_0}(x_0)\}
        \end{equation*} i.e., $\mathfrak M^\sigma_\stab$ is the graph in $\mathfrak Z=\mathfrak X\times\Bbb R$ of the map $x\mapsto \mathscr W_\infty(x,\sigma)$ over $B^\mathfrak{X}_{\ve_0}(x_0)$.  Then $\mathfrak M_\mathrm{stab}^\sigma\subset\mathfrak M_\mathrm{black}$ and if $\Psi\in \mathfrak M_\mathrm{stab}^\sigma$, then $\Psi\in \mathcal K(\ve, M ,\sigma,\infty)$, where $\ve\le \ve_0$. In other words, elements of $\mathfrak M_\stab^\sigma$ form asymptotically Reissner--Nordstr\"om black holes with charge-to-mass ratio $\sigma$. In particular, the conclusion of \cref{thm:uniform-stability-RN} applies to such a $\Psi$.

        \item \ul{Foliation of the black hole region}: The subset of black hole solutions in $\cyl(\ve_0,2)$ is exactly the image of the augmented map $\check{\mathscr W}_\infty$ and the noncollapsing set is the complement of its image:
        \begin{align*}
            \mathfrak M_\mathrm{black}\cap \cyl(\ve_0,2)& = \check {\mathscr W}_\infty(\cyl(\ve_0,1)),\\
            \mathfrak M_\mathrm{non}\cap \cyl(\ve_0,2)& = \cyl(\ve_0,2)\setminus \check {\mathscr W}_\infty(\cyl(\ve_0,1)).
        \end{align*}
        Moreover, $\check {\mathscr W}_\infty:\cyl(\ve_0,1)\to  \mathfrak M_\mathrm{black}\cap \cyl(\ve_0,2)$ is a $C^1_b$ diffeomorphism. 
        \item \ul{The threshold property}: The black hole threshold in $\cyl(\ve_0,2)$ equals the extremal threshold: 
        \begin{equation*}
            \partial \mathfrak M_\mathrm{black}\cap \cyl(\ve_0,2) = \partial \mathfrak M_\mathrm{non}\cap \cyl(\ve_0,2) = \mathfrak M_\stab^{+1}\sqcup \mathfrak M_\stab^{-1},
        \end{equation*}
        where $\partial$ denotes the topological boundary in the topology of $\mathfrak M$. 
    \end{enumerate}
\end{thm}

For short, we write $\mathfrak M_\stab^{+1,-1}\doteq \mathfrak M_\stab^{+1}\sqcup \mathfrak M_\stab^{-1}$. Let \begin{align*}
        \mathscr P_\infty:\mathfrak M_\mathrm{black}&\to [-1,1]\\
        \Psi &\mapsto P_\mathcal{I^+}(u_{\mathcal H^+})
    \end{align*}
    denote the final parameter ratio map. Then the isologous property of $\mathfrak M_\mathrm{stab}^\sigma$ is captured by
    \begin{equation*}
        \mathscr P_\infty(x,\mathscr W_\infty(x,\sigma))=\sigma
    \end{equation*}
    for every $x\in B^\mathfrak{X}_{\ve_0}(x_0)$. Moreover, we have the following important corollary of this construction:

\begin{cor}\label{cor:transversality} Under the hypotheses of \cref{thm:main}, the following holds:

\begin{enumerate}
    \item \ul{Regularity of $\mathscr P_\infty$ and $\mathscr M_\infty$}: $\mathfrak M_\mathrm{black}\cap\cyl(\ve_0,2)$ is a  connected $C^1_b$ manifold with boundary in $\cyl(\ve_0,2)$ of full dimension. Moreover, it holds that
    \begin{equation}\label{eq:P-M-regularity}
        \mathscr P_\infty,\mathscr M_\infty\in C^1\big(\mathfrak M_\mathrm{black}\cap\cyl(\ve_0,2)\big).
    \end{equation}
    \item \ul{Transversality criterion}: Let $p\mapsto \Psi_p$ be a $C^1$ curve in $\cyl(\ve_0,2)$ such that $\Psi_p\in\mathfrak M_\mathrm{black}$ for $p\le p_*$ and $\Psi_{p_*}\in\mathfrak M^{+1,-1}_\mathrm{stab}$. Then $\Psi_p$ is transverse to $\mathfrak M^{+1,-1}_\mathrm{stab}$ at $p_*$ if
    \begin{equation}\label{eq:transversality-criterion}
        \left.\frac{d^-}{dp^-}\right|_{p=p_*}\mathscr P_\infty(\Psi_p)\ne 0,
    \end{equation}
    where $\frac{d^-}{dp^-}$ denotes the left-hand derivative. 
\end{enumerate}
\end{cor}

The proofs of \cref{thm:main} and \cref{cor:transversality} are given in \cref{sec:thm-2-proof}.

\subsection{Universal scaling laws near the threshold}

We now state the precise version of \cref{thm:intro-scaling}. Given $\Psi\in\mathfrak M_\mathrm{black}\cap\cyl(\ve_0,2)$, $\mathcal S_\mathrm{max}[\Psi]$ contains a black hole which asymptotes to a Reissner--Nordstr\"om solution with parameters $(M,\sigma)$, $|\sigma|\le 1$. Therefore, we may define the \emph{final event horizon area} 
\begin{align*}
        \mathscr A:\mathfrak M_\mathrm{black}\cap\cyl(\ve_0,2)&\to \mathbb (0,\infty)\\
        \Psi &\mapsto 4\pi r_+^2(M,\sigma)=4\pi M^2\big(1+\sqrt{1-\sigma^2}\big)^2
    \end{align*}
and the \emph{final event horizon temperature}
    \begin{align*}
        \mathscr T:\mathfrak M_\mathrm{black}\cap\cyl(\ve_0,2)&\to [0,\infty)\\
      \displaystyle  \Psi &\mapsto \frac{\bm\kappa(M,\sigma)}{2\pi}=\frac{D'(r_+;M,\sigma)}{4\pi} = \frac{\sqrt{1-\sigma^2}}{2\pi M(1+\sqrt{1-\sigma^2})^2},
    \end{align*}
    where $\bm\kappa$ is the surface gravity of the Reissner--Nordstr\"om event horizon. Finally, we define the \emph{event horizon location function}
    \begin{align*}
        \mathscr U:\mathfrak M_\mathrm{black}\cap\cyl(\ve_0,2)&\to [1,U_*]\\
        \Psi &\mapsto u_{\mathcal H^+},
    \end{align*}
    where $u_{\mathcal H^+}$ is the (initial data gauge) $u$-coordinate of $\mathcal H^+$.

\begin{thm}\label{thm:scaling} For $\ve_0>0$ sufficiently small depending on $M_0$, the following holds:
\begin{enumerate}
    \item \ul{H\"older regularity of $\mathscr A$, $\mathscr T$, and $\mathscr U$}: It holds that
\begin{equation}\label{eq:Holder}
     \mathscr A,\mathscr T,\mathscr U\in C^{1/2}_b\big(\mathfrak M_\mathrm{black}\cap\cyl(\ve_0,2)\big)
\end{equation}
and 
     \begin{equation}\label{eq:C1-sub}
      \mathscr A,\mathscr T,\mathscr U\in C^1_b\big(\{|\mathscr P_\infty|\le \ell\}\cap\cyl(\ve_0,2)\big)
    \end{equation} for every $\ell <1$.
    \item \ul{Improved regularity tangent to the foliation}: For every $\sigma\in[-1,1]$, 
      \begin{equation}\label{eq:C1-tangent}
      \mathscr A,\mathscr T,\mathscr U\in C^1_b(\mathfrak M_\mathrm{stab}^\sigma),
    \end{equation} and the bound on the $C^1$ norm is uniform in $\sigma$.

    \item \ul{Sharp scaling laws}: Let $p\mapsto \Psi_p$ be a $C^1$ curve in $\cyl(\ve_0,2)$ such that $\Psi_p\in\mathfrak M_\mathrm{black}$ for $p\le p_*$ and $\Psi_{p_*}\in\mathfrak M^{+1,-1}_\mathrm{stab}$.
    Then it holds that
    \begin{align}\label{eq:A-as}
        \mathscr A(\Psi_p)-\mathscr A(\Psi_{p_*}) &=+ \sqrt{2c} 8\pi \mathscr M(\Psi_{p_*})^2|p-p_*|^{1/2}+o(|p-p_*|^{1/2}),\\
   \label{eq:T-as}
        \mathscr T(\Psi_p) &= \frac{\sqrt{2c}}{2\pi \mathscr M_\infty(\Psi_{p_*})}|p-p_*|^{1/2}+o(|p-p_*|^{1/2}),\\\label{eq:U-as}
        \mathscr U(\Psi_p)-\mathscr U(\Psi_{p_*}) &=  - \sqrt{2c}(1+O(\ve_0))\mathscr M_\infty(\Psi_{p_*})|p-p_*|^{1/2}+o(|p-p_*|^{1/2})
    \end{align} as $p\nearrow p_*$, 
    where $c\doteq  \big|{\left.\frac{d^-}{dp^-}\right|_{p=p_*}\mathscr P_\infty(\Psi_p)}\big|$. 
\end{enumerate}
\end{thm}

The proof of this theorem is given in \cref{sec:scaling-proofs}.

\begin{rk}
   The statement in \cref{thm:intro-scaling}  follows now from \cref{cor:transversality} since $c> 0$ if $p\mapsto \Psi_p$ crosses the threshold transversally. Since the threshold is $C^1$, transversality is indeed a generic condition.  
\end{rk}
\begin{rk}
    The expansions \eqref{eq:A-as}--\eqref{eq:U-as} imply that
    \begin{equation*}
      \mathscr A,\mathscr T,\mathscr U\notin C^{\alpha}\big(\mathfrak M_\mathrm{black}\cap\cyl(\ve_0,2)\big).
    \end{equation*}
    for any $\alpha > \frac 12$. 
\end{rk}

\subsection{Horizon instabilities}

Finally, we state the precise version of \cref{thm:intro-instability}. For any $\Psi\in\mathfrak M_\mathrm{stab}^{+1,-1}$, we showed in \cite{AKU24} that the \emph{asymptotic Aretakis charge} \index{H1@$H_0[\phi]$, asymptotic Aretakis charge}
\begin{equation*}
    H_0[\phi]\doteq \lim_{v\to\infty}Y\psi|_{\mathcal H^+}(v)
\end{equation*}
exists and is finite. In the setting of this paper, it is useful to think of the Aretakis charge as the map \index{H3@$ \mathscr H$, Aretakis charge map}
\begin{align*}
        \mathscr H:\mathfrak M_\mathrm{stab}^{+1,-1}&\to \Bbb R\\
        \Psi &\mapsto H_0[\phi]. 
    \end{align*}
    We also define the sets\index{H5@$\mathfrak H_h\doteq \{\Psi\in \mathfrak M_\mathrm{ext}:H_0[\phi]=h\}$, level set in extremal manifold}
    \begin{equation*}
        \mathfrak H_h\doteq \{\Psi\in \mathfrak M_\mathrm{stab}^{+1,-1}:\mathscr H(\Psi)=h\}
    \end{equation*}
    for $h\in \Bbb R$.  

In order to state the transient instability for $\Psi \in \mathfrak M_\mathrm{sub}\cap \cyl(\ve_0,2)$ close to the extremal threshold, we define the inverse temperature \index{beta@$\beta$, inverse temperature, redshift time, transient time scale}
\begin{equation*}
    \beta(\Psi)\doteq \frac{1}{\mathscr T(\Psi)}
\end{equation*}
for $\Psi \in \mathfrak M_{\mathrm{sub}} \cap\cyl(\ve_0,2)$ which is the \emph{transient time scale} on which the solution $\mathcal S[\Psi]$ features the transient horizon instability. (In the following we will often omit the explicit dependence of $\beta$ on $\Psi$.)  

For $\Psi \in \mathfrak M_{\mathrm{sub}} \cap\cyl(\ve_0,2)$ we define the \emph{transient Aretakis charge} $H_0^\flat[\phi]$ as \index{H2@$H_0^{\flat}[\phi]$, transient Aretakis charge} 
\begin{equation*}
    H_0^\flat[\phi] \doteq  e^{4\pi} Y\psi|_{\mathcal H^+}(\beta).
\end{equation*}
 We will prove in \cref{prop:H-is-continuous} that the dynamical transient Aretakis charge map \index{H4@$\mathscr H_\flat$, transient Aretakis charge map}
 \begin{align*}
    \mathscr H_{\flat}: \mathfrak M_{\mathrm{sub}} \cap\cyl(\ve_0,2) &\to \mathbb R\\
    \Psi   &\mapsto H_0^\flat[\phi],
\end{align*}
extends $\mathscr H$ to $\mathfrak M_\mathrm{sub}\cap \cyl(\ve_0,2)$ as a $C^{3\delta/4}_b$ function.

\begin{thm}\label{thm:instabilities} For $\ve_0>0$ sufficiently small depending on $M_0$ and $\delta$, the following holds:
    \begin{enumerate}
        \item \label{thm:instability-1}\ul{The Aretakis instability}:\footnote{This part of the theorem already follows directly from the proof of Theorem 2 in \cite{AKU24}.} If $\Psi\in \mathcal K(\ve,M,\pm 1,\infty)\cap\mathfrak H_h$ with $\ve\le\ve_0$, we have the expansions
        \begin{align}
            Y\psi|_{\mathcal H^+}&= h + O(\ve^3v^{-1+\delta}),\\
            Y^2\psi|_{\mathcal H^+}&= -2M^{-2}hv + O(\ve v^{-1+\delta}),\\
            \Ric(Y,Y)|_{\mathcal H^+}&= 2M^{-2}h^2 + O(\ve^2v^{-1+\delta/2}) ,\\
          \nabla_Y\Ric(Y,Y)|_{\mathcal H^+}&= - 8M^{-4}h^2 v+ O(\ve^2 v^{-1+\delta})
        \end{align}
    
        \item \label{thm:instability-2} \ul{Regularity of the Aretakis charge map}: The map $\mathscr H:\mathfrak M_\mathrm{stab}^{+1,-1}\to\Bbb R$ is a $C^1$ submersion. Therefore, the level sets $\{\mathfrak H_h\}_{h\in\Bbb R}$ form a $C^1$ foliation of the extremal threshold $\mathfrak M_\mathrm{stab}^{+1,-1}$. In particular, $\mathfrak H_0$ is a $C^1$ hypersurface in $\mathfrak M_\mathrm{stab}^{+1,-1}$, and the Aretakis instability is activated on the open and dense set of asymptotically extremal solutions $\mathfrak M_\mathrm{stab}^{+1,-1}\setminus\mathfrak H_0$. 
        \item \label{thm:instability-3} \ul{Transient horizon instability}: If $\Psi\in \mathcal K(\ve,M,\sigma,\infty)$ with $\ve\le\ve_0$, $|\sigma|<1$, and $ h_\flat = \mathscr H_\flat [\Psi]$, we have the expansions  
        {\mathtoolsset{showonlyrefs=false}
        \begin{align} \label{eq:transient-sec5-1}
            Y\psi|_{\mathcal H^+}(v)&= e^{-4 \pi v /\beta  }  h_\flat + O(\ve v^{-1+\delta}),\\   
            Y^2\psi|_{\mathcal H^+}(v)&= -\frac{\sigma^2 M^2}{2 \pi r_+^4}   \beta e^{ - 4\pi v / \beta  }  \left( 1- e^{ -4\pi(v -1)/\beta } \right) h_\flat + O(\ve   v^{\delta} \min \{1, \beta  v^{-1}\}) ,\\
            \Ric(Y,Y)|_{\mathcal H^+}(v)&=2r_+^{-2} e^{-8 \pi v /\beta  }  h_\flat^2 + O( \ve^2 v^{-1+\delta/2}),\\
              \nabla_Y {\Ric}(Y,Y)|_{\mathcal H^+}( v)&= -\frac{2\sigma^2 M^2}{\pi r_+^6} \beta e^{-8 \pi v /\beta  } \left( 1- e^{ -4\pi( v -1)/\beta } \right) h_\flat^2 + O(\ve^2   v^{\delta} \min \{1, \beta v^{-1}\}) \label{eq:transient-sec5-4}
        \end{align}}
        for $v\geq 1$.

    Moreover, let $p\mapsto \Psi_p$ be a $C^1$ curve in $\cyl(\ve_0,2)$ such that $\Psi_p\in\mathfrak M_\mathrm{sub}$ for $p< p_*$ and $\Psi_{p_*}\in\mathfrak H_h\subset \mathfrak M^{+1,-1}_\mathrm{stab}$. Assume that the curve intersects $\mathfrak M_{\mathrm{stab}}^{+1,-1}$ transversally, i.e.\ $c\doteq  \big|{\left.\frac{d^-}{dp^-}\right|_{p=p_*}\mathscr P_\infty(\Psi_p)}\big| \neq 0$. 
         Then,  we have
         \begin{align}  \label{eq:transient-curve-1}
           Y\psi|_{\mathcal H^+} (v)& = e^{-4 \pi v /\beta  }  h  +O(\ve_0 c^{-3\delta/4} v^{-3\delta/2}),\\ \label{eq:transient-curve-2}
             | Y^2\psi|_{\mathcal H^+} (v)| & \gtrsim  |h|v  + O(\ve_0 c^{-3\delta/4} v^{1-3\delta/2})  \text{ for } v \leq \beta,    \\  \label{eq:transient-curve-3}
               \Ric(Y,Y)|_{\mathcal H^+}(v)& =2r_+^{-2} e^{-8 \pi v /\beta  }  h^2+O( \ve_0^2 c^{-3\delta/4} v^{-3\delta/2}),\\ \label{eq:transient-curve-4}
                | \nabla_Y\Ric(Y,Y)|_{\mathcal H^+}(v)| & \gtrsim h^2v+O( \ve_0^2 c^{-3\delta/4} v^{1-3\delta/2})\text{ for } v \leq \beta
         \end{align}
for $p<p_*$ sufficiently close to $p_*$. The meaning of ``sufficiently close'' may depend on the interpolating family, but the implicit constants do not.

    \end{enumerate}
\end{thm}
The proof of this theorem is given in \cref{sec:Aretakis-generic}.

\section{Geometric estimates for solutions in \texorpdfstring{$\tilde{\mathcal K}$}{K tilde}}\label{sec:K-1}

In this section, we prove fundamental decay estimates for the geometry of solutions arising from data in the class $\tilde{\mathcal K}(\ve,M,\varrho,\u u{}_f)$. We begin by stating our assumptions and conventions in \cref{sec:K-conventions}. In \cref{sec:preliminaries}, we give some preliminary results and estimate $\u\kappa{}_{\u u{}_f\dagger}$, $\u\gamma{}_{\u u{}_f\dagger}$, $(1-\u\mu){}_{\u u{}_f\dagger}$, $\u\lambda{}_{\u u{}_f\dagger}$, and $\u\nu{}_{\u u{}_f\dagger}$ (only in the far region). Then, in \cref{sec:improving-geometry}, we estimate $\u r{}_{\u u{}_f\dagger}$, $\u\nu{}_{\u u{}_f\dagger}$ (in the near region), and $\u\varpi{}_{\u u{}_f}-\u M{}_{\mathcal I^+}(\u u{}_f)$.

\subsection{Assumptions and conventions for \texorpdfstring{\cref{sec:K-1,sec:energy-1,sec:scalar-field}}{Sections~\getrefnumber{sec:K-1} to \getrefnumber{sec:scalar-field}}}\label{sec:K-conventions}

\begin{ass}\label{ass:K-tilde}
    Let $0<\ve\le\ve_\loc$, where $\ve_\loc$ is given by \cref{lem:basic-cylinder} for $\ell = 2$, $(M,\varrho,\u u{}_f)$ be an admissible parameter triple, let $A\ge 1$ in \cref{def:K-tilde}, and let $\Psi\in\tilde{\mathcal K}(\ve,M,\varrho,\u u{}_f)$. In every statement, we allow $\ve$ to be chosen small depending only on $M_0$, $\delta$, and $A$.
\end{ass}

In these sections, we will work with developments of such a $\Psi$ in the $\u u{}_f$-teleological gauge $(\u u, \u v)$ on $\mathcal R_{\u u_f}\doteq \mathcal R(1,\u u{}_f,1,\infty)$. We set \index{Rauf2@$  \mathcal R_{\u u_f}^{\le}\doteq \mathcal R_{\u u_f}\cap\{\u v\le \u u\}$}
 \index{Rauf3@$\mathcal R_{\u u_f}^{\ge}\doteq \mathcal R_{\u u_f}\cap\{\u v\ge \u u\}$}

\begin{equation*}
    \mathcal R_{\u u_f}^{\le}\doteq \mathcal R_{\u u_f}\cap\{\u v\le \u u\},\qquad  \mathcal R_{\u u_f}^{\ge}\doteq \mathcal R_{\u u_f}\cap\{\u v\ge \u u\}.
\end{equation*}
For ease of reading, we will not always repeat all of these hypotheses in the lemmas and propositions below and will \ul{omit the underlines and anchoring time} on the coordinates $(\u u,\u v)$ and the teleologically normalized quantities such as $\u\phi{}_{\u u{}_f}$ and $\u \lambda{}_{\u u{}_f}$. To emphasize: in these sections, unless otherwise stated, \ul{all quantities are written in the $u_f$-teleological gauge}. The background Reissner--Nordstr\"om solution is still written as $(r_\diamond,\Omega^2_\diamond)$. We will also omit the superscript $(M,\varrho, u_f)$ on the energy quantities $\mathcal E_p$, $\u{\mathcal E}{}_p$, $\mathcal F$, and $\u{\mathcal F}$.

 In order to keep track of the constant $A$ in \cref{def:K-tilde}, the notations $a\les b$, $a\gtrsim b$, and $a\sim b$ mean that the estimate does not (additionally) depend on $\ve$, $M$, $e$, $u_f$, $u$, $v$, $\Psi$, or $A$.

\subsection{Preliminary estimates}\label{sec:preliminaries}

   Immediately from the definitions and standard monotonicity properties of the equations, we have:
   \begin{lem}\label{lem:basic-estimates} Under \cref{ass:K-tilde}, it holds that
    \begin{equation*}
           \tfrac{9}{10} \le\frac{\varpi}{M}\le \tfrac{11}{10},\quad \tfrac{9}{10}\le\frac{r}{r_\diamond}\le \tfrac{11}{10},\quad
       \tfrac{9}{10} \le\frac{\nu}{-D}\le \tfrac{11}{10},\quad
        \lambda > 0, \quad |Q_\dagger|\les \ve^{3/2}u{}_f^{-3+\delta}     
    \end{equation*}
    in $\mathcal R_{u_f}$, 
    \begin{equation*}
        1-\mu\ge\tfrac 34,\quad
    \lambda\ge \tfrac 12,\quad
    -2\le\nu\le -\tfrac 12 
    \end{equation*}
    in $\mathcal R_{u_f}^{\ge}$, and
    \begin{equation*}
        r\sim 1 \quad\text{on}\quad \mathcal R_{u_f}\cap\{v\le u_f+1\}.
    \end{equation*}
\end{lem}

We will often use these estimates without comment. We can now bound the asymptotically flat norm of the solution in the initial data gauge, which will be useful for continuity arguments later. 
\begin{lem} \label{lem:K-AF-est}
    For any $u_f\ge 2$, there  exists a constant $B=B(u_f)$ such that under \cref{ass:K-tilde},
    \begin{equation}\label{eq:K-AF-est}
        \|\mathcal S[\Psi]\|_{\mathcal R_{u_f}}\le B.
    \end{equation}
\end{lem}
\begin{proof}
   Since $\gamma<0$, $\partial_v\gamma\le 0$, and $\gamma(u,v)\to -1$ as $v\to\infty$ by \cref{prop:Bondi-conditions}, we have
   \begin{equation*}
       1-\mu = \frac{-\nu}{-\gamma}\ge -\nu \ge \tfrac {9}{10} D
   \end{equation*}
    on $\mathcal R_{u_f}$. By \eqref{eq:D-lower-bound}, the condition \eqref{eq:1-mu-assn} is fulfilled for $c$ depending only on $u_f$ and the estimate for $ \|\mathcal S[\Psi]\|_{\mathcal R_{u_f}}$ follows from \cref{prop:fundamental-AF}. 
\end{proof}

 We now use the assumption on the energy decay of $\phi$ to derive bounds for $\kappa_\dagger$ and $\gamma_\dagger$. 

\begin{lem}\label{lem:kappa-tilde-estimate} Under \cref{ass:K-tilde}, it holds that
    \begin{align}
   \label{eq:check-kappa-1}   |\kappa_\dagger|  &\les A\ve^2\tau^{-1+\delta},\\
   \label{eq:check-kappa-2}  D' |\kappa_\dagger|   &\les A\ve^2\tau^{-2+\delta},\\
 \label{eq:check-kappa-3}    D |\kappa_\dagger|   &\les A\ve^2\tau^{-3+\delta},
    \end{align}
    in $\mathcal R^{\le}_{u_f}$ and 
    \begin{equation}
      \label{eq:check-kappa-4}      |\kappa_\dagger|\les A\ve^2\min\{\tau^{-3+\delta},r^{-1}\tau^{-2+\delta}\}\\
            \end{equation}    in $\mathcal R^{\ge}_{u_f}$.
\end{lem}
\begin{proof} By \eqref{eq:kappa-u} and the bootstrap assumptions,
    \begin{equation}\label{eq:kappa-aux-1}
        |\partial_u\kappa_\dagger| = \frac{r|1+\kappa_\dagger|}{|\nu|}(\partial_u\phi)^2\les r(1+|\kappa_\dagger|)D^{-1}(\partial_u\phi)^2.
    \end{equation}
    
\textsc{Estimate on $\Gamma$:} By construction of the gauge, $\kappa_\dagger=0$ on $\mathcal G_{u_f}$, which coincides with $\Gamma=\{u=v\}$ for $v\le u_f-1$. For $\{u_f-1\le v\le u_f\}$, we integrate \eqref{eq:kappa-aux-1} to the future from $\mathcal G_{u_f}$, use Gr\"onwall's inequality, and note that $\tau(\mathcal G_{u_f}(v),v)\sim \tau(v,v)$ to obtain
\begin{equation*}
    |\kappa_\dagger(v,v)|\les \u{\mathcal F}[\phi](v,\tau(\mathcal G_{u_f}(v),v))\exp\left(\u{\mathcal F}[\phi](v,\tau(\mathcal G_{u_f}(v),v))\right) \les A\ve^2 \tau^{-3+\delta}(v,v).
\end{equation*}
It follows that
\begin{equation}\label{eq:kappa-dagger-Gamma}
    \big|\kappa_\dagger|_\Gamma\big|\les A\ve^2\tau^{-3+\delta}.
\end{equation}

    \textsc{Estimates for $u\ge v$}: 
    Integrating \eqref{eq:kappa-aux-1} to the future from $\Gamma$, and using \eqref{eq:kappa-dagger-Gamma} and Gr\"onwall's inequality yields
    \begin{equation*}
        |\kappa_\dagger|\les A\ve^2\tau^{-3+\delta}+ \underline{\mathcal E}{}_2[\phi](\tau)\exp\big(\underline{\mathcal E}{}_2[\phi](\tau)\big)\les A\ve^2\tau^{-1+\delta},
    \end{equation*}
    which gives \eqref{eq:check-kappa-1}. In particular,  $\kappa\sim 1$, which we will frequently use without comment in the sequel. Next, we multiply \eqref{eq:kappa-aux-1} by $D'$ and use the almost-monotonicity \eqref{eq:almost-monotonicity} to estimate
    \begin{align*}
        D'|\kappa_\dagger|(u,v)& \les A\ve^2\tau^{-3+\delta}+ D'( r_\diamond(u,v))\int_{\underline C{}_v\cap\{u'\le u\}}D^{-1}(\partial_u\phi)^2\,du'\\ &\les A\ve^2\tau^{-3+\delta}+\int_{\underline C{}_v\cap\{u'\le u\}}D'D^{-1}(\partial_u\phi)^2\,du'\\ &\les A\ve^2\tau^{-3+\delta}+\underline{\mathcal E}{}_1[\phi](\tau)\les A\ve^2\tau^{-2+\delta},
    \end{align*}
    which gives \eqref{eq:check-kappa-2}. Finally, \eqref{eq:check-kappa-3} is obtained by multiplying \eqref{eq:kappa-aux-1} by $D$ and using \eqref{eq:D-monotonicity} to estimate
    \begin{equation*}
        D|\kappa_\dagger|(u,v)\les A\ve^2\tau^{-3+\delta}+ \int_{\underline C{}_v\cap\{u'\le u\}}
        (\partial_u\phi)^2\,du'\les A\ve^2\tau^{-3+\delta}+\underline{\mathcal E}{}_0[\phi](\tau)\les A\ve^2\tau^{-3+\delta}.
    \end{equation*}

      \textsc{Estimate for $v\ge u$}: Arguing as before, we estimate
      \begin{equation*}
            |\kappa_\dagger(v,v)|\les A\ve^2\tau^{-3+\delta}+ \u{\mathcal F}[\phi](v,\tau)\exp\left(\u{\mathcal F}[\phi](v,\tau)\right) \les A\ve^2 \tau^{-3+\delta},
      \end{equation*}
which gives half of \eqref{eq:check-kappa-4}. To prove the other half, we estimate
      \begin{equation}
          |\partial_u(r\kappa_\dagger)| \les |\kappa_\dagger| + r^2(\partial_u\phi)^2
      \end{equation}
      and integrate again from $\Gamma$, using now the previous estimate for $\kappa_\dagger$.
\end{proof}

\begin{lem}\label{lem:gamma-tilde-estimate}  Under \cref{ass:K-tilde}, it holds that
\begin{equation}      \label{eq:gamma-check-1} 
    |\gamma_\dagger|\les A\ve^2 r^{-1}\tau^{-3+\delta}
\end{equation}
    in $\mathcal R^{\ge}_{u_f}$.
\end{lem}
\begin{proof}
    This is proved using \eqref{eq:kappa-u} as in \cite[Lemma 5.3]{AKU24}, noting that $\gamma_\dagger(u,v)\to 0$ as $v\to\infty$ for every $u\in[1,u_f]$ by the Bondi gauge condition (\cref{prop:Bondi-conditions}). 
\end{proof}

Following \cite[Section 5.2.4]{AKU24}, we now derive expansions of $\mu_\dagger$, $\nu_\dagger$, and $\lambda_\dagger$ in terms of $D$ and $D'$, with rapidly decaying error. 

\begin{prop}\label{prop:Taylor}      Under \cref{ass:K-tilde}, it holds that
\begin{align}
   \label{eq:Taylor-1-mu} 1-\mu &= D+D'r_\dagger +O\big(A\ve^{3/2} r^{-1}\tau^{-3+\delta}\big),\\
   \label{eq:Taylor-lambda}  \lambda & = D + \kappa D' r_\dagger + O(A\ve^{3/2}\min\{\tau^{-3+\delta},r^{-1}\tau^{-2+\delta}\})
\end{align}
on $\mathcal R_{u_f}$ and 
 \begin{equation}
       \label{eq:Taylor-nu}   \nu =-D+ \gamma D'r_\dagger+ O(A\ve^{3/2}r^{-1}\tau^{-3+\delta})
      \end{equation}
      on $\mathcal R_{u_f}^{\ge}$.
\end{prop}
\begin{proof}
    For any $\eta>0$, we Taylor expand \begin{equation}\label{eq:r-taylor}\frac{1}{r^{\eta}}=\frac{1}{r_\diamond^{\eta}}-  \frac{\eta}{r_\diamond^{\eta+1}} r_\dagger+O_\eta\left(\frac{| r_\dagger|^2}{r_\diamond^{\eta+2}}\right)\end{equation} and immediately compute
    \begin{align*}
             (1-\mu) - D=-\frac{2\varpi}{r}+\frac{2M}{r_\diamond}+\frac{Q^2}{r^2}-\frac{e^2}{r_\diamond^2} & = -\frac{2\varpi_\dagger}{r}+\frac{2Mr_\dagger}{r_\diamond ^2}+O\left(\frac{|r_\dagger|^2}{r_\diamond^3}\right)- \frac{2e^2r_\dagger}{r_\diamond^3}+\frac{2eQ_\dagger+Q_\dagger^2}{r_\diamond^2}+O\left(\frac{|r_\dagger|^2}{r_\diamond^4}\right)\\ 
             &= D'r_\dagger + O(r_\diamond^{-1}|\varpi_\dagger|)+O(r_\diamond^{-3}|r_\dagger|^2)+O(r_\diamond ^{-2}|Q_\dagger|)
    \end{align*}
    for $\ve$ sufficiently small, which gives \eqref{eq:Taylor-1-mu} after using \cref{lem:basic-estimates}, \eqref{eq:K-r-1}, and \eqref{eq:K-varpi-1}.  For $\lambda$, we use \eqref{eq:check-kappa-3},  \eqref{eq:check-kappa-4}, and \eqref{eq:Taylor-1-mu} to compute
    \begin{equation*}
        \lambda-D = \kappa(1-\mu)-D = D\kappa_\dagger+\kappa\big(D' r_\dagger +O\big(A\ve^{3/2} r^{-1}\tau^{-3+\delta}\big)\big) = \kappa D' r_\dagger + O(A\ve^{3/2}\min\{\tau^{-3+\delta},r^{-1}\tau^{-2+\delta}\}).
    \end{equation*}
The argument for $\nu$ is similar, using instead \eqref{eq:gamma-check-1}.
\end{proof}

\subsection{Improving the estimates for \texorpdfstring{$\nu$}{nu}, \texorpdfstring{$r$}{r}, and \texorpdfstring{$\varpi$}{varpi}}\label{sec:improving-geometry}

\begin{lem}\label{lem:r-tilde}
    Under \cref{ass:K-tilde}, it holds that
    \begin{equation}\label{eq:r-tilde-est}
         |r_\dagger| \les A\ve^{3/2} \big(\tau^{-2+\delta}+\big(1+|{\log r_\diamond}|\big)u_f^{-2+\delta}\big)
    \end{equation}
    on $\mathcal R_{u_f}$.
\end{lem}
\begin{proof} Let $1\le u_*< u_f$. 

    \textsc{Estimate along $\Gamma$}:  
    Using \eqref{eq:check-kappa-3}, \eqref{eq:gamma-check-1}, \eqref{eq:Taylor-lambda}, and \eqref{eq:Taylor-nu}, we estimate
    \begin{equation}\label{eq:r-deriv-along-Gamma}
    \partial_v\big(r_\dagger(v,v)\big) = \lambda_\dagger+\nu_\dagger = (\kappa+\gamma)D'r_\dagger + O(A\ve^{3/2}\tau^{-3+\delta}) = O(A\ve^{3/2}\tau^{-3+\delta})
    \end{equation}
    along $\Gamma$ for $\ve$ sufficiently small. Integrating backwards, we obtain
    \begin{equation}\label{eq:r-dagger-Gamma}
    \big|r_\dagger|_{\Gamma\cap\{u\le u_*\}}\big| \les |r_\dagger(u_*,u_*)| + A\ve^{3/2}\tau^{-2+\delta}.
    \end{equation}
    
    \textsc{Estimate for $u\ge v$}: Since $\lambda -D = \partial_vr_\dagger$, we may view \eqref{eq:Taylor-lambda} as an ODE for $r_\dagger$ in $v$. For $(u,v),(u,v_2)\in \mathcal R_{u_f}^{\le }$, we use an integrating factor to write 
    \begin{equation}\label{eq:r-tilde-1}
         r_\dagger(u,v) = \exp\left(-\int_v^{v_2}\kappa D'\,dv'\right) r_\dagger(u,v_2) + \int_v^{v_2}\exp\left(-\int_v^{v'}\kappa D'\,dv''\right)O(A\ve^{3/2}\min\{\tau^{-3+\delta},r^{-1}\tau^{-2+\delta}\}) \,dv'.
    \end{equation}
    Fixing $v_2 = u$ and using \eqref{eq:1-mu-hierarchy-computation} (which implies that $\kappa D'\ge 0$) and \eqref{eq:r-dagger-Gamma}, we then conclude
    \begin{equation}\label{eq:r-tilde-3}
        |r_\dagger|\les |r_\dagger(u_*,u_*)|+ A\ve^{3/2}\tau^{-2+\delta}
    \end{equation}
    on $\mathcal R_{u_f}^{\le}\cap\{u\le u_*\}$. 

    \textsc{Estimate for $v\ge u$}: We first recycle \eqref{eq:r-tilde-1} and use the observation that $D'\les r_\diamond^{-2}$ is integrable in $v$ to estimate
    \begin{equation}\label{eq:r-tilde-2}
        \big|r_\dagger|_{C_{u_*}}\big| \les |r_\dagger(u_*,u_*)|+ A\ve^{3/2}\big(1+|{\log r_\diamond}|\big)u_*^{-2+\delta}.
    \end{equation}
    Since $\nu+D=\partial_ur_\dagger$, we may view \eqref{eq:Taylor-nu} as an ODE for $ r_\dagger$ in $u$. For $(u,v),(u_2,v)\in \mathcal R_{u_f}^{\ge}$, we use an integrating factor to write 
    \begin{equation*}
        r_\dagger(u,v) =  \exp\left(-\int_u^{u_2}\gamma D'\,du'\right)r_\dagger(u_2,v) + \int_u^{u_2}\exp\left(-\int_u^{u'}\gamma D'\,du''\right)O(A\ve^{3/2} r^{-1}\tau^{-3+\delta}) \,du'.
    \end{equation*}
    The integrating factors are bounded using again $D'\les r_\diamond^{-2}$. Fixing $(u_2,v)\in \Gamma\cup C_{u_*}$ and using \eqref{eq:r-dagger-Gamma} and \eqref{eq:r-tilde-2}, we obtain
\begin{equation}\label{eq:r-tilde-4}
    |r_\dagger| \les  |r_\dagger(u_*,u_*)|+A\ve^{3/2}\big(1+|{\log r_\diamond}|\big)u_*^{-2+\delta} + A\ve^{3/2}\tau^{-2+\delta}
\end{equation}
    in $\mathcal R_{u_f}^{\ge}\cap\{u\le u_*\}$. 
    
  We now use the anchoring properties \eqref{eq:finite-anchoring} and \eqref{eq:eschatological-anchoring}, which imply that $r_\dagger(u_*,u_*)\to 0$ as $u_*\nearrow u_f$, to conclude \eqref{eq:r-tilde-est} from \eqref{eq:r-tilde-3} and \eqref{eq:r-tilde-4}.
  \end{proof}

Using \eqref{eq:r-taylor}, \eqref{eq:r-tilde-est}, and the definition \eqref{eq:varkappa}, we have the following expansion for $\varkappa$:
\begin{lem}
    Under \cref{ass:K-tilde}, it holds that
\begin{align}
 \label{eq:varkappa-expansion-detailed}   \varkappa &=D'+\left(\frac{6Q^2}{r_\diamond^4}-\frac{4M}{r_\diamond^3}\right)r_\dagger +O(A\ve^{3/2}r^{-2}\tau^{-3+\delta})\\
    &= D' + O\big( A \ve^{3/2} r^{-2} \tau^{-3+\delta} + A \ve^{3/2} r^{-3} \tau^{-2+\delta}\big)\label{eq:varkappa-expansion}
\end{align}
on $\mathcal R_{u_f}$.
\end{lem}

This is now used to prove:

\begin{lem}\label{lem:nu-tilde}  Under \cref{ass:K-tilde}, it holds that
    \begin{equation}\label{eq:nu-est-main}
        \left|1+\frac{\nu}{D}\right|\les \mathbf 1_{\{u\ge v\}} A\ve^{3/2}\tau^{-1+\delta} + \mathbf 1_{\{v\ge u\}} A\ve^{3/2} r^{-1}\tau^{-2+\delta}
    \end{equation}
    on $\mathcal R_{u_f}$.
\end{lem}
\begin{proof}
    \textsc{Estimate for $v\ge u$}: We use the expansion \eqref{eq:Taylor-nu} and \cref{lem:r-tilde} to estimate
    \begin{equation}
        |\nu_\dagger|\les  r^{-2}|r_\dagger| + A\ve^{3/2} r^{-1}\tau^{-3+\delta}\les A\ve^{3/2} r^{-1}\tau^{-2+\delta},
    \end{equation}
    which implies \eqref{eq:nu-est-main} for $v\ge u$ after dividing by $ -\nu_\diamond= D\sim 1$.

    \textsc{Estimate for $u\ge v$}: We use the equation \eqref{eq:nu-v}, the identity $\partial_v\nu_\diamond = -D'D$, \eqref{eq:check-kappa-2}, and the expansion \eqref{eq:varkappa-expansion} to estimate
    \begin{equation}\label{eq:nu-help-1}
        \left|\partial_v{\log\left(\frac{\nu}{\nu_\diamond}\right)}\right| = |\kappa \varkappa - D'| \les A\ve^{3/2}\tau^{-2+\delta} + D'|\tilde\kappa| \les A\ve^{3/2}\tau^{-2+\delta}.
    \end{equation}
    Since we have already shown that \eqref{eq:nu-est-main} holds along $\Gamma$, we can integrate \eqref{eq:nu-help-1} down to conclude that \eqref{eq:nu-est-main} holds in all of $\mathcal R_{u_f}$.
\end{proof}

Finally, we estimate the deviation of $\varpi$ from $M_{\mathcal I^+}(u_f)$.

\begin{lem}\label{lem:varpi-tilde}
Under \cref{ass:K-tilde}, it holds that
    \begin{equation}\label{eq:varpi-difference}
        |\varpi - M_{\mathcal I^+}(u_f)|\les A\ve^2\tau^{-3+\delta}
    \end{equation}
    on $\mathcal R_{u_f}$.
\end{lem}
\begin{proof} Let $1\le u_*<u_f$.

    \textsc{Estimate for $u=u_*$}: By \eqref{eq:varpi-u} and \cref{lem:kappa-tilde-estimate}, we have $|\partial_v\varpi|\les r^2(\partial_v\phi)^2$ on $\mathcal R_{u_f}$. Since $M_{\mathcal I^+}(u_*)=\lim_{v\to\infty}\varpi(u_*,v)$, we integrate this inequality and use \eqref{eq:K-energy-1} to obtain
    \begin{equation*}
     \sup_{v\ge u_*}   |\varpi(u_*,v)-M_{\mathcal I^+}(u_*)|\les A\ve^2 u_*^{-3+\delta}.
    \end{equation*}
    We can continue integrating down, using now \eqref{eq:K-energy-3}, to obtain 
     \begin{equation}\label{eq:varpi-tilde-1}
      |\varpi(u_*,v)-M_{\mathcal I^+}(u_*)|\les A\ve^2 u_*^{-3+\delta} + A\ve^2\tau^{-3+\delta}(u_*,v)\les  A\ve^2\tau^{-3+\delta}(u_*,v).
    \end{equation}

    \textsc{Estimate away from $\{u=u_*\}$}: We now use \eqref{eq:varpi-u} to estimate $\varpi$ to the past of $\{u=u_f\}$. By \eqref{eq:Taylor-1-mu}, we have
    \begin{equation*}
        |\partial_u\varpi|\les r^2(\partial_u\phi)^2 + A\ve^2\tau^{-2+\delta}r^2D'D^{-1}(\partial_u\phi)^2 + A\ve^2\tau^{-3+\delta} r^2D^{-1}(\partial_u\phi)^2.
    \end{equation*}
    When $v\le u\le u_*$, we have by definition of the $h_p$-energy that
    \begin{equation}\label{eq:varpi-tilde-2}
        |\varpi(u,v)-\varpi(u_*,v)|\les \u{\mathcal E}{}_0[\phi](\tau) + A\ve^2\tau^{-2+\delta}  \u{\mathcal E}{}_1[\phi](\tau)+A\ve^2\tau^{-3+\delta}  \u{\mathcal E}{}_2[\phi](\tau)\les A\ve^2\tau^{-3+\delta}
    \end{equation}
    for $\ve$ sufficiently small. When $r\ge R$, then $|\partial_u\varpi|$ is controlled by the integrand of  $\u{\mathcal F}$, so we may use \eqref{eq:K-energy-4}. Combining this with \eqref{eq:varpi-tilde-1} and \eqref{eq:varpi-tilde-2} and sending $u_*\nearrow u_f$ yields \eqref{eq:varpi-difference} as desired.
\end{proof}

\section{A priori energy estimates on solutions in \texorpdfstring{$\tilde{\mathcal K}$}{K tilde}}\label{sec:energy-1}

In this section, we derive the fundamental hierarchies of energy estimates for the scalar field which were used to improve the estimates for the geometry in \cref{sec:K-1}. These estimates apply uniformly for all admissible parameters. We begin in \cref{sec:wave-setup} by fixing the setup for our energy estimates. In \cref{sec:Hardy}, we record some basic Hardy inequalities. In \cref{sec:Kodama}, we prove a basic degenerate energy boundedness statement for $\phi$ which is analogous to the usual $T$-energy estimate on Reissner--Nordstr\"om. In \cref{sec:ILED}, we prove Morawetz estimates for $\phi$, i.e., weighted spacetime $L^2$ bounds for $\phi$ and its derivatives. In \cref{sec:hp-hierarchy}, we prove $h_p$-weighted energy estimates for $r$ close to $r_c$. Finally, in  \cref{sec:rp-hierarchy} we prove $r^p$-weighted energy estimates in the far region.

\subsection{Setup}\label{sec:wave-setup}

 In this section, we adopt the notational conventions outlined in \cref{sec:K-conventions}. Let $\Psi\in\tilde{\mathcal K}(\ve,M,\varrho,\ve)$ be as in \cref{ass:K-tilde}; we assume throughout that this assumption holds. We may therefore assume that all of the results of \cref{sec:K-1} hold (we will continue to use some of the more basic results without comment). 

Let $\zeta\in C_\star^1(\mathcal R_{u{}_f})$, $F\in C^0(\mathcal R_{u{}_f})$, and suppose that $\zeta$\index{zeta@$\zeta$, solution to inhomogeneous wave equation} satisfies the inhomogeneous wave equation  \index{F@$F$, inhomogeneity in wave equation for $\zeta$} 
\begin{equation}\label{eq:inhomog-wave}
    \partial_u\partial_v \zeta  + \frac{\lambda}{r}\partial_u \zeta + \frac{\nu}{r}\partial_v \zeta = F
\end{equation} on $\mathcal R_{u{}_f}$. We write $ \xi \doteq  r \zeta$\index{xi@$\xi\doteq r\zeta$, rescaled solution to inhomogeneous wave equation} which satisfies 
\begin{equation}\label{eq:inhomog-wave-xi}
    \partial_u \partial_v \xi  =\frac{\kappa \nu \varkappa}{r}  \xi + r F .
\end{equation}

The energy estimates in this section will take place on regions $\mathcal D\subset\mathcal R_{u_f}$ defined as follows: Let $1\le u_1\le u_2\le u_f$, $(u_1,v_1),(u_2,v_2)\in\Gamma\cap\mathcal R_{u_f}$,  $u_2'>u_2$, and $v_2'>v_2$. We then define \index{D1@$\mathcal D$, butterfly region} \index{D2@$ \mathcal D^{\le}\doteq \mathcal D\cap\mathcal R^{\le}_{\u u{}_f}$} \index{D3@$\mathcal D^{\ge}\doteq\mathcal D\cap\mathcal R^{\ge}_{\u u{}_f}$}
\begin{equation*}
    \mathcal D\doteq [u_1,u_2]\times[v_1,v_2']\cup [u_1,u_2']\times[v_1,v_2]
\end{equation*}
and 
\begin{equation*}
    \mathcal D^{\le}\doteq \mathcal D\cap\mathcal R^{\le}_{u_f},\quad  \mathcal D^{\ge}\doteq\mathcal D\cap\mathcal R^{\ge}_{u_f}.
\end{equation*}
The null segments constituting $\partial \mathcal D$ are numbered I--VI as depicted in \cref{fig:butterfly} below. We define $\tau_1$ to be the value of $\tau$ on $\mathrm{I}\cup\mathrm{II}$ and $\tau_2$ to be the value of $\tau$ on $\mathrm{IV}\cup\mathrm{V}$.

 \begin{figure}
\centering{
\def\svgwidth{12pc}
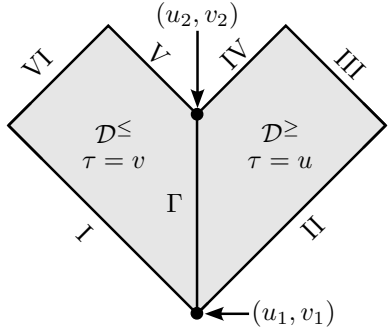}
\caption{A Penrose diagram depicting the region $\mathcal D$ and the hypersurfaces I--VI used in the energy estimates in this section.}
\label{fig:butterfly}
\end{figure}

\subsection{Hardy inequalities}\label{sec:Hardy}

\begin{lem}[$D$-weighted Hardy inequalities]\label{lem:hp-Hardy} Under \cref{ass:K-tilde}, for any $q\ge \delta$ and $f\in C^1(\mathcal D)$, it holds that
\begin{align}
    \label{eq:hp-Hardy-1}
        \int_{\underline C{}_v\cap\mathcal D}D'D^{q/2}f^2\,du&\les \int_{\underline C{}_v\cap\mathcal D}D^{q/2-1/2}(\partial_uf)^2\,du + \int_{\underline C{}_v\cap\mathcal D} Df^2\,du,\\
     \label{eq:hp-Hardy-2}   \iint_{\mathcal D^\le} D'D^{q/2}f^2\,dudv &\les \iint_{\mathcal D^\le}D^{q/2-1/2} (\partial_uf)^2\,dudv + \int_{\Gamma\cap\mathcal D}f^2\,du.
\end{align}
\end{lem}
\begin{proof} 
Let $[u_1,u_2]\times\{v\}\subset\mathcal D^{\le}$. Since $\partial_uD^{q/2} = -\frac q2D'D^{q/2}$, integrating by parts gives
    \begin{equation*}
        -\int_{u_1}^{u_2} \frac q2D'D^{q/2}f^2\,du = (D^{q/2}f^2)(u_2)-(D^{q/2}f^2)(u_1)-2\int_{u_1}^{u_2}D^{q/2} f\partial_uf\,du,
    \end{equation*}
    and hence 
    \begin{align}
       \nonumber q\int_{u_1}^{u_2}D'D^{q/2}f^2\,du + (D^{q/2}f^2)(u_2)&\les q^{-1}\int_{u_1}^{u_2}D'^{-1}D^{q/2}(\partial_uf)^2\,du + (D^{q/2}f^2)(u_1),\\
        &\les q^{-1}\int_{u_1}^{u_2}D^{q/2-1/2}(\partial_uf)^2\,du + (D^{q/2}f^2)(u_1)\label{eq:hp-hardy-proof}
    \end{align}
    by Young's inequality and \eqref{lem:estimates-on-RN-geometry}. To prove \eqref{eq:hp-Hardy-1}, one can handle the boundary term by applying \eqref{eq:hp-hardy-proof} to a cut-off version of $f$, as in \cite[Lemma 6.15]{AKU24}; we omit the details here. We obtain \eqref{eq:hp-Hardy-2} by setting $u_1=v$ in \eqref{eq:hp-hardy-proof} and integrating in $v$. 
\end{proof}
\begin{rk}
 Recall the function $h_p=h_p(r_\diamond)$ which was defined in \cref{sec:hp-defn}. Then \eqref{eq:hp-Hardy-1} implies 
    \begin{equation*}
         \int_{\underline C{}_v\cap\mathcal D}h_{p-2}f^2\,du\les \int_{\underline C{}_v\cap\mathcal D}h_p(\partial_uf)^2\,du + \int_{\underline C{}_v\cap\mathcal D} Df^2\,du
    \end{equation*}
    for $p\le 1-\delta$. 
\end{rk}

\begin{lem}\label{lem:zeta-xi-1}
Under \cref{ass:K-tilde}, for any $0\le p\le 3-\delta$ and $\zeta\in C^1(\mathcal D)$, it holds that
\begin{align}
     \label{eq:zeta-xi-1}   \int_{\underline C{}_v\cap\mathcal D}h_p(\partial_u\zeta)^2\,du&\les \int_{\underline C{}_v\cap\mathcal D}h_p(\partial_u\xi)^2\,du + \int_{\underline C{}_v\cap\mathcal D} D\zeta^2\,du,\\
      \label{eq:zeta-xi-2} \iint_{\mathcal D^\le}h_{p-1}(\partial_u\zeta)^2\,dudv &\les \iint_{\mathcal D^\le} h_{p-1}(\partial_u\xi)^2\,dudv + \int_{\Gamma\cap\mathcal D}\zeta^2\,du,
\end{align}
where, recall, $\xi\doteq r\zeta$. 
\end{lem}
\begin{proof}
    By \cref{lem:basic-estimates}, $(\partial_u\zeta)^2\les (\partial_u\xi)^2+D^2\xi^2$ in $\mathcal D^{\le}$. We then apply \eqref{eq:hp-Hardy-1} with $f=\xi$ and $q= -p+3\ge \delta$ to obtain
    \begin{multline*}
        \int_{\u C{}_v\cap\mathcal D} h_pD^2 \xi^2\,du \les  \int_{\u C{}_v\cap\mathcal D} h_p(\partial_u\xi)^2\,du+\int_{\u C{}_v\cap\mathcal D} D'D^{-p/2+3/2} \xi^2\,du \\\les  \int_{\u C{}_v\cap\mathcal D}\big(h_p+D^{-p/2+1}\big)(\partial_u\xi)^2\,du +  \int_{\u C{}_v\cap\mathcal D}D\xi^2\,du,
    \end{multline*}
    which immediately implies \eqref{eq:zeta-xi-1}. To prove \eqref{eq:zeta-xi-2}, use instead \eqref{eq:hp-Hardy-2}.
\end{proof}

\begin{lem}[$r$-weighted Hardy inequalities] Under \cref{ass:K-tilde}, for any $q\le 1- \delta$ and $f\in C^1(\mathcal D)$, it holds that
\begin{align}
    \label{eq:rp-Hardy-1}
        \int_{ C_u\cap\mathcal D}r^qf^2\,dv&\les \int_{ C_u\cap\mathcal D}r^q(\partial_v(rf))^2\,dv + \int_{C_u\cap\mathcal D} f^2\,dv,\\
     \label{eq:rp-Hardy-2}   \iint_{\mathcal D^\ge} r^qf^2\,dudv &\les \iint_{\mathcal D^\ge}r^q (\partial_v(rf))^2\,dudv + \int_{\Gamma\cap\mathcal D}f^2\,dv.
\end{align}
\end{lem}
\begin{proof}
    The proof is similar to that of \cref{lem:hp-Hardy}, but we write $r^qf^2=r^{q-2}(rf)^2$ and use $\partial_vr^{q-2}$ in place of $\partial_u D^{q/2}$.
\end{proof}

\begin{lem}
    Under \cref{ass:K-tilde}, for any $0\le p\le 1-\delta$ and $\zeta\in C^1(\mathcal D)$, it holds that
\begin{align}
     \label{eq:zeta-xi-3}   \int_{C_u\cap\mathcal D}r^{p+2}(\partial_v\zeta)^2\,dv&\les \int_{C_u\cap\mathcal D}r^{p+1}(\partial_v\xi)^2\,dv + \int_{C_u\cap\mathcal D} \zeta^2\,dv,\\
      \label{eq:zeta-xi-4} \iint_{\mathcal D^\ge}r^{p-1}(\partial_v\zeta)^2\,dudv &\les \iint_{\mathcal D^\ge} r^{p-1}(\partial_v\xi)^2\,dudv + \int_{\Gamma\cap\mathcal D}\zeta^2\,dv.
\end{align}
\end{lem}
\begin{proof} By \cref{lem:basic-estimates}, 
$(\partial_v\zeta)^2\les r^{-2}(\partial_v\xi)^2+r^{-2}\zeta^2$
    in $\mathcal D^\ge$. We then apply \eqref{eq:rp-Hardy-1} with $f=\zeta$ and $q=p$ to obtain
    \begin{equation*}
        \int_{C_u\cap\mathcal D}r^{p+2}(\partial_v\zeta)^2\,dv\les  \int_{C_u\cap\mathcal D} r^p(\partial_u\xi)^2\,dv +  \int_{C_u\cap\mathcal D} r^{p}\zeta^2\,dv'\les  \int_{C_u\cap\mathcal D} r^p(\partial_v\xi)^2\,dv' +  \int_{C_u\cap\mathcal D}\zeta^2\,dv.
    \end{equation*}
    To prove \eqref{eq:zeta-xi-4}, use instead \eqref{eq:rp-Hardy-2}.
\end{proof}

\subsection{The degenerate energy estimate}\label{sec:Kodama}

\begin{prop}\label{prop:T-energy} Under \cref{ass:K-tilde}, for any $\zeta\in C_\star^1(\mathcal R_{u_f})$ satisfying \eqref{eq:inhomog-wave} and domain $\mathcal D\subset\mathcal R_{u_f}$ as depicted in \cref{fig:butterfly}, it holds that
    \begin{multline}\label{eq:T-est}
\int_\mathrm{III}\big(r^2(\partial_u\zeta)^2+\zeta^2\big)\,du+ \int_\mathrm{IV}\big(r^2(\partial_v\zeta)^2+\zeta^2\big)\,dv+\int_\mathrm{V}\big((\partial_u\zeta)^2+D \zeta^2\big)\,du  + \int_\mathrm{VI}\big((\partial_v\zeta)^2+D\zeta^2\big)\,dv\\  \les  \int_\mathrm{I}\big((\partial_u\zeta)^2+D\zeta^2\big)\,du + \int_\mathrm{II}\big(r^2(\partial_v\zeta)^2+\zeta^2\big)\,dv +\iint_{\mathcal{R}} r^2\big( | \partial_u\zeta |+| \partial_v \zeta |\big) | F | \, dudv \\ 
+\ve \iint_{\mathcal D^\ge}\big((\partial_u\zeta)^2 +(\partial_v\zeta)^2 \big)\,dudv+\ve\iint_{\mathcal D^\le} D'D^{\delta/2}(\partial_v\zeta)^2\, dudv\\+\ve\iint_{\mathcal D^\le}\big(\tau^{-2+2\delta}h_{-3+\delta}+\tau^{-4+2\delta}h_{-1+\delta}+\tau^{-6+2\delta}h_{1+\delta}\big)(\partial_u\zeta)^2\,dudv  .
    \end{multline}
    \end{prop}

    \begin{proof} We compute the expression\footnote{This corresponds to using the background time-translation Killing vector field $\frac 12\partial_u+\frac 12\partial_v$ as a multiplier.} $\partial_v(r^2(\partial_u\zeta)^2)+\partial_u(r^2(\partial_v\zeta)^2)$ using \eqref{eq:inhomog-wave} and add the null Lagrangian identity $-\partial_v\partial_u (\sigma (r_\diamond-r_c) \zeta^2 ) + \partial_u \partial_v (\sigma (r_\diamond - r_c)\zeta^2)=0$, where $\sigma$ is a constant, to obtain
\begin{multline}\label{eq:T-est-1}
 \partial_v \big(r^2 (\partial_u \zeta)^2+ \sigma D \zeta^2 - 2 \sigma (r_\diamond - r_c) \zeta \partial_u \zeta \big) + \partial_u \big(r^2 (\partial_v \zeta)^2 + \sigma D \zeta^2  + 2 \sigma (r_\diamond - r_c) \zeta \partial_v \zeta \big)\\
 = - 2 r(\nu + \lambda) \partial_u \zeta \partial_v \zeta +  2r^2 F(\partial_u+\partial_v)\zeta . 
\end{multline}

The terms being differentiated on the left-hand side are denoted by $I_u$ and $I_v$, respectively. We estimate the $I_u$ flux term on the left-hand side as \begin{equation*}
  I_u=   r^2 (\partial_u \zeta)^2 + \sigma D \zeta^2 - 2 \sigma(r_\diamond - r_c)  \zeta \partial_u \zeta \gtrsim  (r^2 - \sqrt{\sigma}\, r_\diamond^{2} ) (\partial_u \zeta)^2  + \sigma D \zeta^2 -  \sigma^{3/2} r_\diamond^{-2} ( r_\diamond-r_c)^2 \zeta^2 \gtrsim_\sigma r^2 (\partial_u \zeta)^2 +  D \zeta^2, 
\end{equation*}
for $\sigma$ and $\ve$ sufficiently small, where we used \eqref{eq:lower-bound-on1-mu}. Analogously, we also obtain an upper bound, giving  altogether  \begin{equation*}
    I_u \sim r^2(\partial_u \zeta)^2 + D \zeta^2.
\end{equation*}
Similarly, we have
\begin{equation*}
    I_v   \sim  r^2 (\partial_v \zeta)^2 +  D \zeta^2.
\end{equation*}
Therefore, integrating \eqref{eq:T-est-1} over $\mathcal D$ yields
    \begin{multline}\label{eq:T-est-2}
\int_\mathrm{III}\big(r^2(\partial_u\zeta)^2+\zeta^2\big)\,du+ \int_\mathrm{IV}\big(r^2(\partial_v\zeta)^2+\zeta^2\big)\,dv+\int_\mathrm{V}\big((\partial_u\zeta)^2+D \zeta^2\big)\,du  + \int_\mathrm{VI}\big((\partial_v\zeta)^2+D\zeta^2\big)\,dv\\  \les  \int_\mathrm{I}\big((\partial_u\zeta)^2+D\zeta^2\big)\,du + \int_\mathrm{II}\big(r^2(\partial_v\zeta)^2+\zeta^2\big)\,dv +\iint_{\mathcal D} \big(r|\nu+\lambda||\partial_u\zeta||\partial_v\zeta|+r^2(|\partial_u\zeta|+|\partial_v\zeta|)  | F |\big) \, dudv.
\end{multline}

The first bulk term on the right is estimated by using \eqref{eq:Taylor-lambda}, \eqref{eq:r-tilde-est}, and \eqref{eq:nu-est-main} to derive 
\begin{align}
\nonumber    |\lambda+\nu| &=\big|D+\kappa D'\tilde r +O(\ve\min\{ r^{-1}\tau^{-2+\delta},\tau^{-3+\delta}\})   -D\big(1+O(\mathbf 1_{\{u\ge v\}} A\ve^{3/2}\tau^{-1+\delta} + \mathbf 1_{\{v\ge u\}} A\ve^{3/2}\tau^{-2+\delta})\big)\big|\\&\les \mathbf 1_{\{u\ge v\}}\ve\big(D\tau^{-1+\delta}+ D'\tau^{-2+\delta}+\tau^{-3+\delta}\big)+\mathbf 1_{\{v\ge u\}}\ve r^{-1}\tau^{-2+\delta}.\label{eq:lambda+nu}
\end{align}
In $\mathcal D^{\le}$, we use Young's inequality and \cref{lem:estimates-on-RN-geometry} to write 
\begin{multline}
    |\lambda+\nu||\partial_u\zeta||\partial_v\zeta|\les \ve^{-1}|\lambda+\nu|^2 D'^{-1}D^{-\delta/2}(\partial_u\zeta)^2 + \ve D'D^{\delta/2}(\partial_v\zeta)^2\\ + \ve\big(\tau^{-2+2\delta}h_{-3+\delta}+\tau^{-4+2\delta}h_{-1+\delta} +\tau^{-6+2\delta}h_{1+\delta}\big)(\partial_u\zeta)^2 + \ve D'D^{\delta/2}(\partial_v\zeta)^2.
\end{multline}
On the other hand, in $\mathcal D^{\ge}$, we use \eqref{eq:lambda+nu} to estimate $r|\lambda+\nu||\partial_u\zeta||\partial_v\zeta|\les \ve((\partial_u\zeta)^2+(\partial_v\zeta)^2)$. Putting these together and using Young's inequality on the final term in \eqref{eq:T-est-2} yields \eqref{eq:T-est} as desired. 
    \end{proof}

\subsection{Integrated local energy decay}\label{sec:ILED}

\begin{prop}[The Morawetz estimate]\label{prop:Morawetz}  Under \cref{ass:K-tilde}, for any $\zeta\in C_\star^1(\mathcal R_{u_f})$ satisfying \eqref{eq:inhomog-wave} and domain $\mathcal D\subset\mathcal R_{u_f}$ as depicted in \cref{fig:butterfly}, it holds that
 \begin{multline}\label{eq:Mor-final}
 \int_\mathrm{III}\big(r^2(\partial_u\zeta)^2+\zeta^2\big)\,du+ \int_\mathrm{IV}\big(r^2(\partial_v\zeta)^2+\zeta^2\big)\,dv+\int_\mathrm{V}\big((\partial_u\zeta)^2+D \zeta^2\big)\,du  + \int_\mathrm{VI}\big((\partial_v\zeta)^2+D\zeta^2\big)\,dv\\
    +\iint_{\mathcal D^\le}   \big(D' D^{\delta/2}  \left( \left(1+ D'^{-\delta} \mathbf{1}_{\{ 5 D\leq (M D')^2\} }\right)  \left( ( \partial_u \xi )^2 + ( \partial_v \xi )^2\right) + ( \partial_u \zeta )^2   + ( \partial_v \zeta )^2 \right) + D' D^{1+\delta/2} \zeta^2 \big) \, dudv \\  +\iint_{\mathcal D^\ge}  \big( r^{1-\delta} \big( ( \partial_u \zeta )^2 + ( \partial_v \zeta )^2\big)+r^{-1-\delta}  \big( ( \partial_u \xi )^2 + ( \partial_v \xi )^2\big) + r^{-1-\delta} \zeta^2 \big) \, dudv  
 +\int_{\Gamma\cap\mathcal D} \big(  (\partial_u \zeta )^2  +  (\partial_v \zeta )^2 + \zeta^2 \big) \, du  \\\les   \int_\mathrm{I}\big((\partial_u\zeta)^2+D\zeta^2\big)\,du + \int_\mathrm{II}\big(r^2(\partial_v\zeta)^2+\zeta^2\big)\,dv +\iint_{\mathcal D} r^2 \big( |\partial_u\zeta| + |\partial_v \zeta| + r^{-1} D |\zeta| \big) | F | \, dudv
 \\
+\ve\iint_{\mathcal D^\le}\big(\tau^{-2+2\delta}h_{-3+\delta}+\tau^{-4+2\delta}h_{-1+\delta}+\tau^{-6+2\delta}h_{1+\delta}\big)(\partial_u\zeta)^2\,dudv .
\end{multline}
\end{prop}

\begin{rk}
    The general proof strategy for the Morawetz estimate here (except for the improvement of the weights at the horizon) is the same as in \cite{AKU24} and we refer the reader there for comments on the proof. The improvement of the horizon weights in the exact extremal case is due to the first author of the present paper, Aretakis, and Gajic  \cite{AAG20}. The version presented here, valid in the full range of admissible parameters, is novel. 
\end{rk}

It is convenient to define
\begin{equation*}
    \mathbb M(\zeta,F,\mathcal D)\doteq \text{RHS of \eqref{eq:Mor-final}}.
\end{equation*}

\subsubsection{The initial estimate}

We recall the following version of the divergence theorem in the plane to fix signs. Let $j^u,j^v$, and $Q$ be regular functions on $\mathcal D$ satisfying $\partial_uj^u+\partial_vj^v=Q$. Then
    \begin{equation}\label{eq:divergence}
        \iint_{\mathcal D^{\ge}} Q\,dudv= \int_\mathrm{IV}j^u\,dv-\int_\mathrm{II}j^u\,dv+\int_\mathrm{III}j^v\,du+\int_{\Gamma\cap\mathcal D} (j^u-j^v)\,du.
    \end{equation} 
    
\begin{lem}\label{lem:Mor-version-1}
    Under the hypotheses of \cref{prop:Morawetz}, it holds that
 \begin{equation}\label{eq:Mor-version-1}
    \iint_{\mathcal D}  \frac{D}{r^2} \big( ( \partial_u \zeta )^2 +( \partial_v \zeta )^2\big)\, dudv  +\int_{\Gamma\cap\mathcal D} \big(  (\partial_v \zeta )^2  +  (\partial_u \zeta )^2 \big) \, du  \les  \textnormal{RHS of } \eqref{eq:T-est}.
\end{equation}
\end{lem}
    
\begin{proof} \textsc{The bulk term}:
    We use the wave equation \eqref{eq:inhomog-wave} obtain\footnote{This corresponds to using the standard vector field $X=-r_\diamond^{-3}\partial_u+r_\diamond^{-3}\partial_v$ as a multiplier.} 
    \begin{equation}\label{eq:Mor-proof-1}
        -\partial_v \left( \frac{r^2}{r_\diamond^3}   (\partial_u \zeta)^2 \right) + \partial_u \left(\frac{r^2}{r_\diamond^3} (\partial_v \zeta)^2 \right) =  3 \frac{r^2}{r_\diamond^4} D \big(   (\partial_u \zeta)^2 + (\partial_v \zeta)^2\big) - 2 \frac{r}{r_\diamond^3} (\lambda-\nu)  \partial_u \zeta  \partial_v \zeta  - 2 \frac{r^2}{r_\diamond^3} (  \partial_u \zeta - \partial_v \zeta) F.
    \end{equation}
Arguing as in \eqref{eq:lambda+nu}, we estimate
\begin{equation}\label{eq:Mor-proof-2}
    |(\lambda-\nu)-2D|\les \mathbf 1_{\{u\ge v\}}\ve\big( D\tau^{-1+\delta}+D'\tau^{-2+\delta}+\tau^{-3+\delta}\big)+\mathbf 1_{\{v\ge u\}}\ve r^{-1}\tau^{-2+\delta}.
\end{equation}
By Young's inequality and \cref{lem:basic-estimates}, we have
\begin{equation}\label{eq:Mor-proof-3}
    3 \frac{r^2}{r_\diamond^4} D \big(   (\partial_u \zeta)^2 + (\partial_v \zeta)^2\big) -4\frac{r}{r_\diamond^3} D\partial_u\zeta\partial_v\zeta \ge \frac{r}{r_\diamond^3}\left(3\frac{r}{r_\diamond}-2\right)D\big(   (\partial_u \zeta)^2 + (\partial_v \zeta)^2\big) \gtrsim \frac{D}{r^2} \big(   (\partial_u \zeta)^2 + (\partial_v \zeta)^2\big) .
\end{equation}
Integrating \eqref{eq:Mor-proof-1} over $\mathcal R$, using \eqref{eq:Mor-proof-2} and \eqref{eq:Mor-proof-3}, and applying \cref{prop:T-energy} yields the first half of \eqref{eq:Mor-version-1}.

\textsc{The term along $\Gamma$}: Integrate \eqref{eq:Mor-proof-1} over $\mathcal D^\ge$ and apply \eqref{eq:divergence}. We then can use the half of the Morawetz estimate already proved to estimate 
\begin{equation*}
    \int_{\Gamma\cap\mathcal D}\big(  (\partial_v \zeta )^2  +  (\partial_u \zeta )^2 \big) \, du \sim \int_{\Gamma\cap\mathcal D}\left(  \frac{r^2}{r_\diamond^3}(\partial_v\zeta)^2+ \frac{r^2}{r_\diamond^3}(\partial_u\zeta)^2\right)  du \les \textnormal{RHS of } \eqref{eq:T-est}.
\end{equation*}
This completes the proof of \eqref{eq:Mor-version-1}. \end{proof}

\subsubsection{Improving the weights}

\begin{lem}
    Under the hypotheses of \cref{prop:Morawetz}, it holds that
    \begin{equation}\label{eq:Mor-sharp-r}
         \iint_{\mathcal D^\ge}   \big(r^{1-\delta} \big(( \partial_u \zeta )^2 +( \partial_v \zeta )^2\big)+r^{-1-\delta}\zeta^2\big)\, dudv  +\int_{\Gamma\cap\mathcal D} \zeta^2 \, du  \les  \textnormal{RHS of }\eqref{eq:T-est}  +\iint_{\mathcal D^\ge} r|\zeta||F|\,dudv.
    \end{equation}
\end{lem}
\begin{proof} For any function $f=f(r)$, a tedious calculation using \eqref{eq:inhomog-wave} yields
\begin{multline*}
      - r^2f'\big(\lambda(\partial_u\phi)^2-\nu(\partial_v\phi)^2\big)-2fr(\lambda-\nu-2)\partial_u\phi\partial_v\phi+2\big(rf''\lambda+2(rf'-f)\kappa\varkappa\big)(-\nu)\phi^2\\=\partial_u\big(r^2f(\partial_v\phi)^2-2rf'\lambda\phi^2+\partial_v(rf\phi^2)\big)-\partial_v\big(r^2f(\partial_u\phi)^2+2rf'\nu\phi^2-\partial_u(rf\phi^2)\big) + 2 r^2 f ( \partial_u \phi - \partial_v \phi ) F - 4rf\phi F.
\end{multline*}
We integrate this over $\mathcal D^\ge$ with the choice $f(r)=-1+\rho^{-1}\chi(r)r^{-\delta}$, where $\rho \ge 1$ is a large constant to be determined and $\chi$ is a cutoff function such that $\chi(r)=0$ for $r\le \rho$, $\chi(r)=1$ for $r\ge 2\rho$, and such that $|\chi|\les \rho^{-1}$ and $|\chi''|\les\rho^{-2}$.

By \eqref{eq:divergence}, we have
\begin{multline*}
    \iint_{\mathcal D^\ge}(N_1+N_2+Z\zeta^2)\,dudv + \iint_{\mathcal D^\ge}\big(2r^2f(\partial_v\phi-\partial_u\phi)F+4rf\phi F\big)\,dudv \\ = \int_{\Gamma\cap\mathcal D}(j^u-j^v)\,du +\int_\mathrm{IV}j^u\,dv - \int_\mathrm{II}j^u\,dv + \int_\mathrm{III}j^v\,du,
\end{multline*}
where the bulk terms are given by 
    \begin{align*}
    N_1&\doteq \rho^{-1}r^{1-\delta}(\delta\chi -\chi' r)\big(\lambda(\partial_u\zeta)^2-\nu (\partial_v\zeta )^2\big),\\
          N_2&\doteq 2(1-\rho^{-1}\chi r^{-\delta})r(\lambda-\nu-2)\partial_u \zeta  \partial_v\zeta,\\
            Z &\doteq  \rho^{-1}\lambda(-\nu)\left(\chi'' r^{1-\delta}-2\delta \chi'r^{-\delta}+\delta(\delta+1)\chi r^{-1-\delta}\right)-2\left(1+\rho^{-1}\chi' r^{1-\delta}-(\delta+1)\rho^{-1}\chi r^{-\delta}\right)2\kappa\nu\varkappa,
            \end{align*}
            and the fluxes are given by 
  \begin{equation*}
        j^u\doteq r^2f(\partial_v\zeta)^2+rf'\lambda\zeta^2+\lambda f\zeta^2+2rf\zeta  \partial_v\zeta ,\quad  j^v\doteq -r^2f(\partial_u\zeta )^2-rf'\nu\zeta^2+\nu f\zeta^2+2rf\zeta  \partial_u\zeta.
    \end{equation*}

\textsc{Estimate for the zeroth order bulk}: Arguing exactly as in the proof of \cite[Lemma 6.13]{AKU24}, we have that $Z\gtrsim \rho^{-1}r^{-1-\delta}$ for $\rho$ sufficiently large, whence
\begin{equation}\label{eq:Z-bulk}
    \iint_{\mathcal D^\ge}Z\zeta^2\,dudv \gtrsim \rho^{-1}\iint_{\mathcal D^\ge}r^{-1-\delta}\zeta^2\,dudv.
\end{equation}

\textsc{Estimates for the first order bulks}:  Let $N_0\doteq r^{-2}\big((\partial_u\zeta)^2+(\partial_v\zeta)^2\big)$. It is immediate that $N_1\gtrsim \delta\rho^{-1}r^{1-\delta}\big((\partial_u\zeta)^2+(\partial_v\zeta)^2\big)$ for $r\ge 2\rho$. We use \eqref{eq:Mor-proof-2} to estimate
\begin{equation*}
    |\lambda-\nu-2|\les |\lambda-\nu-2D|+2|D-1|\les r^{-1}
\end{equation*}
and hence obtain  $|N_2|\les |\partial_u\zeta||\partial_v\zeta|$. For $r\le \rho^2$, we have $|\partial_u\zeta|\partial_v\zeta|\les R^4N_0$ and for $r\ge \rho^2$, we have $|\partial_u\phi||\partial_v\phi|\les \delta^{-1}\rho^{-1+\delta}N_1$. Therefore, for any $b>0$ we may choose $\rho$ sufficiently large that 
\begin{equation*}
    |N_2|\les R^4N_0+b\mathbf 1_{\{r\ge 2\rho\}}N_1.
\end{equation*}
Choosing $b$ sufficiently small (and $\rho$ sufficiently large), it holds that
\begin{equation*}
    \iint_{\mathcal D\cap\{r\ge 2\rho\}} (N_1+N_2)\,dudv \gtrsim \rho^{-1}\iint_{\mathcal D\cap\{r\ge 2\rho\}}\rho^{-1}r^{1-\delta}\big((\partial_u\zeta)^2+(\partial_v\zeta)^2\big) \,dudv - R^4\iint_{\mathcal D\cap\{r\ge 2\rho\}} N_0\,dudv.
\end{equation*}

\textsc{Estimates for the null fluxes}: These terms are all estimated using \cref{prop:T-energy}.

\textsc{Estimate for the flux along $\Gamma$}: On $\Gamma$, we have by definition
\begin{equation*}
    j^v|_\Gamma = r^2(\partial_u\zeta)^2-\nu\zeta^2-2r\zeta\partial_u\zeta,\quad j^u|_\Gamma = -r^2(\partial_v\zeta)^2-\lambda\zeta^2-2r\zeta\partial_v\zeta.
\end{equation*}
Therefore,
\begin{equation*}
    -\int_{\Gamma\cap\mathcal D}( j^u-j^v)\,du\gtrsim \int_{\Gamma\cap\mathcal D}\zeta^2\,ds - \int_{\Gamma\cap\mathcal D}\big((\partial_u\zeta)^2+(\partial_v\zeta)^2\big)\,ds.
\end{equation*}

Putting all of these estimates together, and using that the integral of $N_0$ is controlled by the previous estimate \eqref{eq:Mor-version-1}, we arrive at \eqref{eq:Mor-sharp-r}.\end{proof}

\begin{lem}    Under the hypotheses of \cref{prop:Morawetz}, it holds that
    \begin{multline}\label{eq:Mor-sharp-horizon}
         \iint_{\mathcal D^\le}D' D^{\delta/2} ( 1+ D'^{-\delta} \mathbf{1}_{\{ 5 D \leq  M^2 D'^2\}} ) \big((\partial_u\xi)^2+(\partial_v\xi)^2 +(\partial_u\zeta)^2+(\partial_v\zeta)^2 \big) + D' D^{1+\delta/2} \xi^2 \big)\,dudv\\ \les \textnormal{RHS of }\eqref{eq:Mor-sharp-r} +\iint_{\mathcal D^\le} D  |\zeta||F|\,dudv .
    \end{multline}
\end{lem}
 \begin{proof}
 From the wave equation \eqref{eq:inhomog-wave-xi}, we derive the identity
\begin{equation*}
      \partial_v (w_\delta (\partial_u \xi)^2 ) -   \partial_u ( w_\delta (\partial_v \xi)^2)= D w_\delta' \big((\partial_u\xi)^2+(\partial_v\xi)^2\big)+2w_\delta (\partial_u\xi-\partial_v\xi)\left(  \frac{\kappa \nu \varkappa}{r} \xi +  rF\right) ,
\end{equation*}
where we recall the function $w_\delta$ from \eqref{eq:def-w-delta}.
Applying \cref{lem:morawetz-horizon-weight}, integrating over $\mathcal D^\le$, using Young's inequality, \eqref{eq:varkappa-expansion}, \eqref{eq:T-est}, \eqref{eq:Mor-sharp-r}, we obtain
\begin{multline}\label{eq:Mor-proof-5}
   \iint_{\mathcal D^\le} D' D^{\delta/2} ( 1+ D'^{-\delta} \mathbf{1}_{\{ 5 D \leq  M^2 D'^2\}} ) \big((\partial_u\xi)^2+(\partial_v\xi)^2\big)\,dudv\les   \iint_{\mathcal D^\le} D w_\delta' \big((\partial_u\xi)^2+(\partial_v\xi)^2\big)\,dudv \\ 
   \les \text{RHS of \eqref{eq:Mor-sharp-r}} + \iint_{\mathcal D^\le} |\partial_u\xi-\partial_v\xi||F|\,dudv  + \iint_{\mathcal D^\le} \big({D'} D^{2-\delta/2}+\ve \tau^{-4+2\delta}D'^{-1}D^{2-\delta/2}\big)\xi^2\,dudv.
\end{multline}
Using $\xi^2 \lesssim \zeta^2$, \eqref{eq:hp-Hardy-2}, and \eqref{eq:Mor-version-1}, we estimate 
\begin{equation*}
    \iint_{\mathcal D^\le}D'D^{2-\delta/2}\xi^2\,dudv 
    \les \text{RHS of \eqref{eq:Mor-sharp-r}} + \iint_{\mathcal D^\le}D^{3/2-\delta/2}(\partial_u\zeta)^2\,dudv\les \text{RHS of \eqref{eq:Mor-sharp-r}}.
\end{equation*}
Similarly, the last term in \eqref{eq:Mor-proof-5} is estimated as
\begin{align*}
    \iint_{\mathcal D^\le}\ve \tau^{-4+2\delta}(D')^{-1}D^{2-\delta/2}\xi^2\,dudv 
  &   \les \text{RHS of \eqref{eq:Mor-sharp-r}} + \ve\iint_{\mathcal D^\le} \tau^{-4+2\delta} D^{1/2-\delta/2} (\partial_u\zeta)^2\,dudv\\ & \les \text{RHS of \eqref{eq:Mor-sharp-r}}. 
\end{align*}

Finally, the term involving $\xi^2$ in \eqref{eq:Mor-sharp-horizon} is now estimated by \eqref{eq:hp-Hardy-2}. This then also gives the desired control for $(\partial_u \zeta)^2$ and $(\partial_v \zeta)^2 $ which concludes the proof. 
\end{proof}

\begin{proof}[Proof of \cref{prop:Morawetz}]
    Add \eqref{eq:T-est}, \eqref{eq:Mor-sharp-r}, and \eqref{eq:Mor-sharp-horizon}. For $\ve$ sufficiently small, the third line of \eqref{eq:T-est} can be absorbed into the Morawetz bulk, and we arrive at \eqref{eq:Mor-final}.
\end{proof}

\subsection{The \texorpdfstring{$h_p$}{hp}-hierarchy}
\label{sec:hp-hierarchy}

\begin{prop}[The $h_p$-hierarchy]
\label{prop:hp} 
Under \cref{ass:K-tilde}, for any $p\in [\delta,3-\delta]$, $\zeta\in C_\star^1(\mathcal R_{u_f})$ satisfying \eqref{eq:inhomog-wave}, and domain $\mathcal D\subset\mathcal R_{u_f}$ as depicted in \cref{fig:butterfly}, it holds that
\begin{multline}\label{eq:hp}
\int_\mathrm{V}h_p(\partial_u\xi)^2\,du + \iint_{\mathcal D^\le} D'h_p(\partial_u\xi)^2\,dudv \\ \les \mathbb M(\zeta,F,\mathcal D)+\int_{I} h_p (\partial_u \xi)^2 du + \iint_{\mathcal D^\le}h_p|\partial_u\xi||F|\,dudv + \iint_{\mathcal D^\le}\ve\tau^{-4+2\delta}h_p(\partial_u\xi)^2\,dudv.
\end{multline}
\end{prop}
\begin{rk}\label{rk:hp-strong-bulk}
    Using \eqref{eq:hp-hierarchy-property}, the bulk term on the left-hand side of \eqref{eq:hp} can be used to estimate 
    \begin{equation*}
        \iint_{\mathcal D^\le} D'h_p(\partial_u\xi)^2\,dudv \gtrsim \iint_{\mathcal D^\le} h_{p-1}(\partial_u\xi)^2\,dudv, 
    \end{equation*}
    which shows that we get a \emph{hierarchy} of $h_p$-weighted inequalities, but the stronger bulk term will be important in \cref{sec:estimates-linear-perturbations} later. 
\end{rk}
    \begin{proof}[Proof of \cref{prop:hp}] We evaluate the expression
    \begin{equation*}
        \iint_{\mathcal D^\le}\partial_v\big(h_p(\partial_u\xi)^2\big)\,dudv
    \end{equation*}
    first by integrating by parts and then by using the wave equation \eqref{eq:inhomog-wave-xi}. This leads to the identity
    \begin{multline*}
        \int_\mathrm{V}h_p(\partial_u\xi)^2\,du - \int_\mathrm{I}h_p(\partial_u\xi)^2\,du - \int_{\Gamma\cap\mathcal D} h_p(\partial_u\xi)^2\,du \\ 
        = -\iint_{\mathcal D^\le}(-\partial_vh_p)(\partial_u\xi)^2\,dudv + \iint_{\mathcal D^\le}2\kappa\nu\varkappa h_p\zeta\partial_u\xi\,dudv \iint_{\mathcal D^\le}2r h_p\partial_u\xi F\,dudv.
    \end{multline*}
    Since $-\partial_vh_p=-Dh_p'$, \eqref{eq:varkappa-expansion}, \eqref{eq:Mor-final}, and \cref{prop:estimate-on-hp}  imply
    \begin{multline}
         \int_\mathrm{V}h_p(\partial_u\xi)^2\,du + \iint_{\mathcal D^\le} D'h_p(\partial_u\xi)^2\,dudv \\ \les \mathbb M(\zeta,F,\mathcal D)+\int_{I} h_p (\partial_u \xi)^2 du + \iint_{\mathcal D^\le}h_p|\partial_u\xi||F|\,dudv + \iint_{\mathcal D^\le}h_p(D'+\ve\tau^{-2+\delta})D|\xi||\partial_u\xi|\,dudv.\label{eq:hp-proof-1}
    \end{multline}

We now handle the first term arising from the final integral in \eqref{eq:hp-proof-1}. For $b>0$, we estimate
    \begin{equation}\label{eq:hp-proof-2}
        h_pD'D|\xi||\partial_u\xi|\les bD'h_p(\partial_u\xi)^2 +b^{-1}h_pD^2\xi^2\les bD'h_p(\partial_u\xi)^2 + b^{-1}D'D^{-p/2+3/2}\xi^2,
    \end{equation}
    where we also used $D^2 h_p\lesssim D' D^{-p/2 + 3/2}$ which follows from \eqref{eq:hp-upper}.
Using \eqref{eq:hp-proof-2}  in \eqref{eq:hp-proof-1}, we can absorb the first term of \eqref{eq:hp-proof-2} into the bulk by choosing $b$ sufficiently small and estimate the second term using \eqref{eq:hp-Hardy-2} (with $q= 3-p\geq\delta$), to obtain
\begin{equation*}
    \iint_{\mathcal D^\le}D'D^{-p/2+3/2}\xi^2\,dudv \les \mathbb M(\zeta,F,\mathcal R) + \iint_{\mathcal D^\le}D^{-p/2+1}(\partial_u\xi)^2\,dudv.
\end{equation*}
Note that $D^{-p/2+1}\le b D^{-p/2}+b^{-p/2}D$ for any $b>0$, so the contribution from the first term can be absorbed into the bulk by choosing $b$ sufficiently small, and the contribution from the second term is estimated again using \eqref{eq:Mor-final}. 

The final term in \eqref{eq:hp-proof-1} is estimated by
\begin{equation*}
    \iint_{\mathcal D^\le} \ve\tau^{-2+\delta}Dh_p|\xi||\partial_u\xi|\,dudv \les \iint_{\mathcal D^\le}\ve \tau^{-4+2\delta}h_p(\partial_u\xi)^2\,dudv+\iint_{\mathcal D^\le}\ve D^2h_p\xi^2\,dudv.
\end{equation*}
The first term here gives rise to the final term in \eqref{eq:hp} and the second term is estimated as in \eqref{eq:hp-proof-2}. This completes the proof of \eqref{eq:hp}.
    \end{proof}

\subsection{The \texorpdfstring{$r^p$}{rp}-hierarchy}\label{sec:rp-hierarchy}

\begin{prop}\label{prop:rp} Under \cref{ass:K-tilde}, for any $p\in [\delta,3-\delta]$, $\zeta\in C_\star^1(\mathcal R_{u_f})$ satisfying \eqref{eq:inhomog-wave}, and domain $\mathcal D\subset\mathcal R_{u_f}$ as depicted in \cref{fig:butterfly}, it holds that
\begin{equation}
    \label{eq:rp}
\int_\mathrm{IV}r^p(\partial_v\xi)^2\,dv+ 
\iint_{\mathcal D^\ge}r^{p-1}(\partial_v\xi)^2\,dudv 
\les \mathbb M(\zeta,F,\mathcal D)+ \int_\mathrm{II}r^p(\partial_v\xi)^2\,dv + \iint_{\mathcal D^\ge}r^{p+1}|\partial_v\xi||F|\,dudv.
\end{equation}
\end{prop}
\begin{proof}
The proof proceeds exactly as in \cite[Proposition 6.19]{AKU24}. We evaluate the expression 
  \begin{equation*}
        \iint_{\mathcal D^\ge} \partial_u\big(r^p(\partial_v\xi)^2\big)\,dudv
    \end{equation*}
    first by integrating by parts and then by using the wave equation \eqref{eq:inhomog-wave-xi}. This leads to the identity
    \begin{multline*}
        \int_\mathrm{IV}r^p(\partial_v\xi)^2\,dv-\int_\mathrm{II}r^p(\partial_v\xi)^2\,dv+\int_{\Gamma\cap\mathcal D} r^p(\partial_v\xi)^2\,du \\ = -\iint_{\mathcal D^\ge}pr^{p-1}(-\nu)(\partial_v\xi)^2\,dudv +\iint_{\mathcal D^\ge}2\kappa\nu\varkappa r^{p-1}\xi\partial_v\xi\,dudv +\iint_{\mathcal D^\ge}2r^{p+1}\partial_v\xi F\,dudv.
    \end{multline*}
    Using the estimate $|\varkappa|\les r^{-2}$, the Hardy inequality \eqref{eq:rp-Hardy-2}, and the Morawetz estimate \eqref{eq:Mor-final}, the middle term on the right-hand side can be absorbed into the bulk, up to a term $\les\mathbb M(\zeta,F,\mathcal D)$. See the proof of \cite[Lemma 6.23]{AKU24} for details.
\end{proof}

\section{Boundedness and decay of the scalar field}\label{sec:scalar-field}

In this section, we show that the scalar field $\phi$ on a $\tilde{\mathcal K}$ solution is bounded and decays. In \cref{sec:improved-energy}, we improve the constant in the ``bootstrap assumptions'' \eqref{eq:K-energy-1}--\eqref{eq:K-energy-4}. Then, in \cref{sec:pointwise}, we prove pointwise estimates for $\phi$ and its derivatives. 

\subsection{Improving the energy estimates for \texorpdfstring{$\phi$}{phi}}\label{sec:improved-energy}

\begin{prop}\label{prop:improved-energy}
    Under \cref{ass:K-tilde}, it holds that 
{\mathtoolsset{showonlyrefs=false}\begin{align}
\label{eq:K-imp-1}\u{\mathcal E}{}_p[\phi](\tau)  & \les \ve^2\tau^{-3+\delta+p} ,\\
\label{eq:K-imp-2} \mathcal E_p[\phi](\tau) & \les\ve^2\tau^{-3+\delta+p} ,\\
\label{eq:K-imp-3}\u{\mathcal F}[\phi](v,\tau)  & \les\ve^2\tau^{-3+\delta},\\
\label{eq:K-imp-4}\mathcal F[\phi](u,\tau)  & \les \ve^2\tau^{-3+\delta}
\end{align}}
 for every $p\in\{0\}\cup[\delta,3-\delta]$, $1\le \tau\le u_f$, and $(u,v)\in\mathcal R_{u_f}$.
\end{prop}

The proof begins with the following \emph{hierarchy property} of the energies, which makes full use of all of the energy estimates proved in \cref{sec:energy-1}.

\begin{lem}\label{lem:decay-hierarchy-phi}   Under \cref{ass:K-tilde}, it holds that
    \begin{equation}\label{eq:master-boundedness}
        \sup_{\tau_1\le\tau\le\tau_2}\big( \mathcal E_p[\phi](\tau)+\underline{\mathcal E}{}_p[\phi](\tau)\big)\les \mathcal E_p[\phi](\tau_1)+\underline{\mathcal E}{}_p[\phi](\tau_1)+ A\ve^3\tau_1^{-4+3\delta+p}
    \end{equation}
    for every $1\le \tau_1\le \tau_2\le u_f$ and $p\in\{0\}\cup[\delta,3-\delta]$, and
    \begin{equation}\label{eq:master-integral}
        \int_{\tau_1}^{\tau_2}\big( \mathcal E_{p-1}[\phi](\tau)+\underline{\mathcal E}{}_{p-1}[\phi](\tau)\big)\,d\tau \les \mathcal E_p[\phi](\tau_1)+\underline{\mathcal E}{}_p[\phi](\tau_1)+ A\ve^3\tau_1^{-4+3\delta+p}
    \end{equation}
    for every $1\le \tau_1\le \tau_2\le u_f$ and $p\in[\delta,3-\delta]$.
\end{lem}

\begin{proof} Using \eqref{eq:K-energy-2}, we estimate
\begin{multline*}
       \iint_{\mathcal R_{\le R}}\big(\tau^{-2+2\delta}h_{-3+\delta}+\tau^{-4+2\delta}h_{-1+\delta}+\tau^{-6+2\delta}h_{1+\delta}\big)(\partial_u\phi)^2\,dudv \\
       \le \int_{\tau_1}^{\tau_2}\big(\tau^{-2+2\delta}\u{\mathcal E}{}_0[\phi](\tau)+\tau^{-4+2\delta}\u{\mathcal E}{}_0[\phi](\tau)+\tau^{-6+2\delta}\u{\mathcal E}{}_{1+\delta}[\phi](\tau)\big)\,d\tau\les A\ve^2\tau^{-4+3\delta}_1
\end{multline*}
and
\begin{equation*}
     \iint_{\mathcal R_{\le R}}\tau^{-4+2\delta}h_p(\partial_u\psi)^2\,dudv\le \int_{\tau_1}^{\tau_2}\tau^{-4+2\delta}\u{\mathcal E}{}_p[\phi](\tau)\,d\tau \les A\ve^2\tau_1^{-4+3\delta+p}.
\end{equation*}
Therefore, we arrive at \eqref{eq:master-boundedness} and \eqref{eq:master-integral} by using \cref{rk:hp-strong-bulk} and \cref{prop:Morawetz,prop:hp,prop:rp} (and estimating the $\partial_u\phi$ terms in $\u{\mathcal E}{}_p[\phi]$ using \cref{lem:zeta-xi-1}).
\end{proof}

We note the following interpolation lemma:
\begin{lem}\label{lem:interpolation} Under \cref{ass:K-tilde}, for any  $\delta\le p_1<p<p_2\le 3-\delta$, $1\le\tau\le u_f$, and $\zeta\in C^1(\mathcal R_{u_f})$, it holds that
\begin{align*}
      \mathcal E_p[\zeta](\tau)&\les \big(\mathcal E_{p_1}[\zeta](\tau))^\frac{p_2-p}{p_2-p_1}\big(\mathcal E_{p_2}[\zeta](\tau))^\frac{p-p_1}{p_2-p_1},\\
     \underline{\mathcal E}{}_p[\zeta](\tau) &\les \big(\underline{\mathcal E}{}_{p_1}[\zeta](\tau))^\frac{p_2-p}{p_2-p_1}\big(\underline{\mathcal E}{}_{p_2}[\zeta](\tau))^\frac{p-p_1}{p_2-p_1}.
  \end{align*}
\end{lem}
\begin{proof}
    This is a consequence of the following general inequality for a nonnegative function $w$ on a measure space $(X,\mu)$, which is  immediately obtained from H\"older's inequality:
    \begin{equation*}
        \int_X w^p\,d\mu\le\left(\int_X w^{p_1}\,d\mu\right)^\frac{p_2-p}{p_2-p_1}\left(\int_X w^{p_2}\,d\mu\right)^\frac{p-p_1}{p_2-p_1}.\qedhere
    \end{equation*}
\end{proof}

Next, we estimate the initial energy in terms of the $\mathfrak F$-norm of $\phi_\circ$. 

\begin{lem}\label{lem:initial-energy-est}  Under \cref{ass:K-tilde}, for every $p\in\{0\}\cup[\delta,3-\delta]$ it holds that
    \begin{equation}
        \mathcal E_p[\phi](1)+\underline{\mathcal E}{}_p[\phi](1)\les \ve^2. \label{eq:data-estimate}
    \end{equation}
\end{lem}
\begin{proof} 
   Because $\lambda\sim 1$ on $C_\out$  and $\nu\sim -D$ on $\underline C{}_\ing$, it follows from the definition of $\|\phi_\circ\|_\mathfrak{F}$ that
\begin{align}
    \label{eq:data-aux-1}
|r\phi|+r^2|\partial_v\phi|+r^2|\partial_v\psi|&\les\|\phi_\circ\|_\mathfrak{F}\le\ve \quad\text{on }C_\out,\\
    \label{eq:data-aux-2}
        |\phi|+D^{-1}|\partial_u\phi|+D^{-1}|\partial_u\psi|&\les \|\phi_\circ\|_\mathfrak{F}\le \ve\quad\text{on }\u C{}_\ing.
\end{align}
The bound \eqref{eq:data-estimate} for $\mathcal E_p(1)$ is now immediate. The bound for $\underline{\mathcal E}{}_p(1)$ follows once we observe that
\[ \int_{\u C{}_0}h_{p}D^2\,du\les\int_{\u C{}_0}h_{3-\delta}D^2\,du\les \int_{r_c}^{100M_0} D'D^{-1+\delta/2}\,dr'\les D^{\delta/2}\big|^{100M_0}_{r_c}\les 1.\qedhere\]
\end{proof}

\begin{proof}[Proof of \cref{prop:improved-energy}] In view of \cref{lem:decay-hierarchy-phi,lem:interpolation,lem:initial-energy-est}, the estimates \eqref{eq:K-imp-1}--\eqref{eq:K-imp-4} follow from a standard application of the pigeonhole principle. See \cite[Section 7.1]{AKU24} for a detailed and exactly analogous proof, or \cref{sec:linear-scalar-field} below for a detailed proof in a more complicated setting.
\end{proof}

For use in \cref{sec:estimates-linear-perturbations}, we require:
\begin{lem}\label{lem:phi-estimates}
Under \cref{ass:K-tilde}, for any $p\in [\delta,3-\delta]$ and domain $\mathcal D\subset\mathcal R_{u_f}$ as depicted in \cref{fig:butterfly}, it holds that
\begin{align}   \label{eq:Mor-decay-final-far}
\iint_{\mathcal D^{\ge}}\big(r^{1-\delta}\big((\partial_u\phi)^2+(\partial_v\phi)^2\big)+ r^{-1-\delta}\phi^2\big)\,dudv& \les \ve^2\tau_1^{-3+\delta},\\
    \label{eq:Mor-decay-final-near}
     \iint_{\mathcal D^\le}D' D^{\delta/2}( 1+ D'^{-\delta} \mathbf{1}_{\{ 5 D \leq  M^2 D'^2\}} ) \big((\partial_u\phi)^2+(\partial_v\phi)^2\big)\,dudv&\les \ve^2\tau_1^{-3+\delta},\\
     \iint_{\mathcal D^\ge}r^{1+\delta}(\partial_v\phi)^2\,dudv &\les\ve^2\tau^{-3+2\delta}_1,\label{eq:delta-decay-final-far}\\
     \label{eq:hp-decay-final}
        \iint_{\mathcal D^\le}D'^2 D^{-p/2-1/2} (\partial_u\phi)^2\,dudv&\les \ve^2\tau_1^{-3+\delta+p}.
\end{align}
\end{lem}
\begin{proof}
    This follows from  \cref{prop:Morawetz,prop:hp,prop:rp} after estimating the error terms as in the proof of \cref{lem:decay-hierarchy-phi}. 
\end{proof}

\subsection{Pointwise estimates for the scalar field}\label{sec:pointwise}

\begin{prop}\label{prop:pw-estimates}
  Under \cref{ass:K-tilde}, we have the pointwise decay estimates
\begin{align}
 \label{est:phi_bound}    \big|\sqrt{r^3D'}\phi\big|&\les  \ve \tau^{-3/2+3\delta/4},\\
\label{est:psi_bound}    |\psi|&\les \ve\tau^{-1+\delta/2}
\end{align}
and the pointwise boundedness estimates
\begin{align}
  \label{est:vpsi_bound}   |r^2\partial_v\psi|&\les \ve,\\
 \label{est:vphi_bound}   |r^2\partial_v\phi|&\les \ve,\\
\label{est:upsi_bound}    |Y\psi|&\les \ve,\\
  \label{est:uphi_bound}  |rY\phi|&\les \ve
\end{align}
on $\mathcal R_{u_f}$, where 
\begin{equation*}
    Y\doteq \nu^{-1}\partial_u.
\end{equation*}
\end{prop}
\begin{proof} \textsc{Proof of \eqref{est:phi_bound} for $u\ge v$}: Let $\chi_1=\chi_1(y)$, where $y=v-u$, be a smooth cutoff satisfying $\chi_1(y)=1$ for $y\le -1$ and $\chi_1(y)=0$ for $y\ge 0$. For $(u,v)\in\mathcal R_{u_f}$ with $y\le -1$, the segment $[v,u]\times\{v\}$ is contained in $\mathcal R_{u_f}$. Therefore, using the support property of $\chi'$ and the estimates \eqref{eq:upper-bound-on-deriative-of-varkappa} and \eqref{eq:hp-Hardy-1}, we find
\begin{align*}
    |D'\phi^2(u,v)|&=\left|\int_{u^R(v)}^u\big(\chi_1'\nu D'\phi^2-\chi_1D''D\phi^2+2\chi_1D'\phi\partial_u\phi\big)\,du'\right|\\
    &\les \u{\mathcal E}{}_0[\phi](\tau)+ \left(\int_{v}^uD'D^{\delta/2+1/2}\phi^2\,du'\right)^{1/2}\left(\int_{v}^uD'D^{-\delta/2-1/2}(\partial_u\phi)^2\,du'\right)^{1/2}\\
    &\les \u{\mathcal E}{}_0[\phi](\tau)+\big(\u{\mathcal E}{}_0[\phi](\tau)\big)^{1/2}\big(\u{\mathcal E}{}_\delta[\phi](\tau)\big)^{1/2}\les \ve^2 \tau^{-3+3\delta/2}.
\end{align*}
We can estimate $\phi$ for $y$ close to $0$ by the $p=0$ energy, see the proof of \cite[Proposition 7.4]{AKU24} for details. 

\textsc{Proof of \eqref{est:psi_bound} for $u\ge v$}: For $y\le -1$, we have again
\begin{align*}
    \psi^2(u,v)  &= \left|\int_{v}^u \big(\chi'_1\nu\phi^2+\chi_1\psi\partial_u\psi\big)\,du'\right| \les \u{\mathcal E}{}_0[\phi](\tau) + \left(\int_{v}^u D\psi^2\,du'\right)^{1/2}\left(\int_{v}^uD^{-1}(\partial_u\psi)^2\,du'\right)^{1/2} \\ 
    & \les \u{\mathcal E}{}_0[\phi](\tau)+\big(\u{\mathcal E}{}_0[\phi](\tau)\big)^{1/2}\big(\u{\mathcal E}{}_2[\phi](\tau)\big)^{1/2}\les \ve^2 \tau^{-2+\delta}.
\end{align*}

\textsc{Proof of \eqref{est:phi_bound} and \eqref{est:psi_bound} for $v\ge u$}: As in the proof of \cite[Proposition 7.4]{AKU24}, we use the fundamental theorem of calculus and Cauchy--Schwarz to estimate
\begin{equation*}
    r^{-\beta}\psi^2\les \mathcal E_0[\phi](\tau)+ \big(\mathcal E_0[\phi](\tau)\big)^{1/2}\big(\mathcal E{}_{2-2\beta}[\phi](\tau)\big)^{1/2}\les \ve^2\tau^{-2-\beta+\delta}
\end{equation*}
on $\mathcal R_{u_f}^\ge$ for $\beta\in\{0,1\}$.

\textsc{Proof of \eqref{est:vpsi_bound}}: Using \eqref{eq:wave-equation-psi}, we compute
\begin{equation*}
    \partial_u(r^2\partial_v\psi) = \frac{2\nu}{r}(r^2\partial_v\psi) + r\kappa\nu\varkappa\psi
\end{equation*}
which can be solved for
\begin{equation}\label{eq:bounded-aux-1}
   ( r^2\partial_v\psi)(u,v) = \frac{r^2(u,v)}{r^2(0,v)}  ( r^2\partial_v\psi)(0,v)+r^2(u,v)\int_0^u \left(\frac{\kappa\nu\varkappa}{r}\psi\right)(u',v) \,du'
\end{equation} using an integrating factor (note that $\exp(\int_{u_1}^{u_2}\frac{2\nu}{r}(u',v)\,du')=r^2(u_2,v)/r^2(u_1,v)$). Since $r(u_2,v)\le r(u_1,v)$ for $u_1\le u_2$ and 
\begin{equation*}
    r^2(u,v)\int_0^u r^{-3}(u',v)\,(-\nu)du'\les 1,
\end{equation*}
we readily obtain \eqref{est:vpsi_bound} from \eqref{eq:data-aux-1}, \eqref{est:psi_bound},  \eqref{eq:bounded-aux-1}, and the geometric estimates. 

\textsc{Proof of \eqref{est:vphi_bound}}: This follows immediately from the identity
\begin{equation*}
    r^2\partial_v\phi=r\partial_v\psi-\lambda\phi
\end{equation*}
and the previously proved estimates. 

\textsc{Proof of \eqref{est:upsi_bound}}: Using  \eqref{eq:nu-v} and \eqref{eq:wave-equation-psi}, we compute
\begin{equation}\label{eq:Ypsi-negative-1}
    \partial_v(Y\psi)= -\kappa\varkappa Y\psi + \kappa\varkappa\phi,
\end{equation}
which can be solved for
\begin{equation}\label{eq:Ypsi-negative-2}
    Y\psi(u,v)=\exp\left(-\int_0^v \kappa\varkappa\,dv'\right)Y\psi(u,0)+\int_0^v\exp\left(-\int_{v'}^v  \kappa\varkappa\,dv''\right)  \kappa\varkappa\phi\,dv'
\end{equation}
using an integrating factor, where the integrands are evaluated at constant $u$. Since $\kappa D'\ge 0$ on $\mathcal D_{u_f}$ and \eqref{eq:varkappa-expansion} holds, we have
\begin{equation*}
    \exp\left(-\int_{v_1}^{v_2}\kappa\varkappa\,dv'\right)\le \exp\left(-\int_{v_1}^{v_2}\kappa\tilde\varkappa\,dv'\right)\les 1+\int_{v_1}^{v_2}|\tilde\varkappa|\,dv'\les 1+A\ve^{3/2}v_1^{-1+\delta}\les 1
\end{equation*}
for $\ve$ sufficiently small. Next, we estimate 
\begin{equation*}
     \int_0^v |\varkappa|| \phi|\,dv'\les \int_0^v \big(D'+r^{-2}\tau^{-2+\delta}\big)|\phi|\,dv'\les \ve 
\end{equation*}  
by \eqref{est:phi_bound} and \eqref{est:psi_bound}. Therefore, \eqref{eq:Ypsi-negative-2} and \eqref{eq:data-aux-2} yield \eqref{est:upsi_bound} as desired. 

\textsc{Proof of \eqref{est:uphi_bound}}: This follows immediately from the identity 
\begin{equation*}
    r\frac{\partial_u\phi}{-\nu}= \frac{\partial_u\psi}{-\nu}+\phi
\end{equation*} and the previously proved estimates. \end{proof}

\section{Uniform asymptotic stability of the Reissner--Nordstr\"om family}\label{sec:K-2}

In this section, we complete the proof of \cref{thm:uniform-stability-RN,thm:dichotomy-revisited}. In \cref{sec:proof-uniform-stab}, we combine the results from \cref{sec:K-1,sec:energy-1,sec:scalar-field} to prove \cref{thm:uniform-stability-RN}. In \cref{sec:continuity-arguments}, we use \cref{thm:uniform-stability-RN} and the theory from \cref{sec:semiglobal-1} to establish dyadic iteration and modulation statements (recall \cref{ingredient:iteration} from \cref{sec:intro-proof-overview}). In \cref{sec:revisiting}, we prove \cref{thm:dichotomy-revisited}. 

Starting in this section, we suspend the convention from \cref{sec:K-conventions} about dropping underlines and anchoring times, as a crucial part of the arguments here will involve updating $\u u{}_f$. 

\subsection{The proof of \texorpdfstring{\cref{thm:uniform-stability-RN}}{Theorem 5.11}}\label{sec:proof-uniform-stab}

We now complete the proof of the uniform decay estimates for the Reissner--Nordstr\"om family. 

\begin{proof}[Proof of \cref{thm:uniform-stability-RN}]
\textsc{Choosing the uniform constants}: By \cref{lem:r-tilde,lem:nu-tilde}, we have the estimates
\begin{align*}
  \left|1+\frac{\u\nu{}_{\u u{}_f}}{D}\right|  & \les A\ve^{3/2}\tau^{-1+\delta},\\
    |\u r{}_{\u u{}_f\dagger}|  & \les A\ve^{3/2}\big(\tau^{-2+\delta}+r_\diamond^\delta u_f^{-2+\delta}\big),
\end{align*}
in $\mathcal R_{\u u{}_f}$, where the implicit constant does not depend on $A$. Therefore, by choosing $A$ sufficiently large, these estimates imply 
\begin{align*}
  \left|1+\frac{\u\nu{}_{\u u{}_f}}{D}\right|  & \le \tfrac 12 A^2\ve^{3/2}\tau^{-1+\delta},\\
  |\u r{}_\dagger|  & \le \tfrac 12 A^2\ve^{3/2}\big(\tau^{-2+\delta}+r_\diamond^\delta u_f^{-2+\delta}\big)
\end{align*}
as desired. For $\varpi$, we estimate using \cref{lem:varpi-tilde} and the modulation condition \eqref{eq:varpi-modulation}
\begin{equation}\label{eq:varpi-mod-1}
    |\u\varpi{}_{\u u{}_f\dagger}| \le |\u\varpi{}_{\u u{}_f\dagger}-\u M{}_{\mathcal I^+}(\u u{}_f)|+|\u M{}_{\mathcal I^+}(\u u{}_f)-M|\les A\ve^2\tau^{-3+\delta}+\ve^{3/2}u_f^{-3+\delta}\le (A\ve^{1/2}+1)\ve^{3/2}\tau^{-3+\delta}.
\end{equation}
Therefore, by choosing $A$ sufficiently large and $\ve_0\ge \ve$ sufficiently small, we have
\begin{equation*}
   |\u\varpi{}_{\u u{}_f\dagger}|\le \tfrac 12 A\ve^{3/2}\tau^{-3+\delta} 
\end{equation*}
in $\mathcal R_{\u u{}_f}$ as desired. Finally, by choosing $A$ larger than twice the maximum of the implicit constants on the right-hand sides of the energy estimates \eqref{eq:K-imp-1}--\eqref{eq:K-imp-4}, we improve \eqref{eq:K-energy-1}--\eqref{eq:K-energy-4} to the constant $\frac 12 A$. Now simply let $A_\mathrm{unif}$ be sufficiently large and $\ve_0$ be sufficiently small that all of these conditions are satisfied, and these clearly only depend on $M_0$ and $\delta$. This proves part 1.~of the theorem.

\textsc{Proofs of 2. and 3.}: These follow immediately from the estimates proved in \cref{sec:K-1} and \cref{prop:pw-estimates}.

\textsc{Proof of orbital stability}: The inequality \eqref{eq:mass-orbital-stability} follows immediately from evaluating the estimate for $\u \varpi{}_{\u u{}_f\dagger}$ at $(1,1)$. To prove orbital stability of the energy, \eqref{eq:orb-stab-1}, one can perform an auxiliary bootstrap argument to replace $\ve$ in all of the estimates by $\breve \ve\doteq \mathcal E^{(M,\varrho,\u u{}_f)}_{3-\delta}[\u\phi{}_{\u u{}_f}](1)+\u{\mathcal E}{}^{(M,\varrho,\u u{}_f)}_{3-\delta}[\u\phi{}_{\u u{}_f}](1)$. This closes because the only place where $\ve$ actually enters directly into the estimates is via the initial data estimate \eqref{eq:data-estimate}, and so replacing it with $\breve\ve\les \ve$ works just as well. We refer to \cite[Section 
8.3.5]{AKU24} for details. Pointwise orbital stability, \eqref{eq:orb-stab-2}, follows immediately from the fact that \cref{prop:pw-estimates} was proved using the method of characteristics and that the nonlinear errors are controlled by the weighted $C^1$ norm of $\phi_\circ$.
    
\textsc{Improving the power of $\ve$ when $u_f=\infty$}: Revisiting \eqref{eq:varpi-mod-1}, we see that $|\u\varpi{}_{\u u{}_f\dagger}|\les A\ve^2\tau^{-3+\delta}$ when $\u u_f=\infty$. Combining this with the fact that now $Q_\dagger=0$, we can immediately improve the power of $\ve$ in \cref{prop:Taylor} to $\ve^2$. Note that the estimates for $\u\kappa{}_\dagger$ and $\u\gamma{}_\dagger$ in \cref{lem:kappa-tilde-estimate,lem:gamma-tilde-estimate} already have $\ve^2$. Revisiting the proofs of \cref{lem:r-tilde,lem:nu-tilde}, we see that all powers of $\ve$ are now improved to $\ve^2$, which proves the claim.

\textsc{Estimate for the location of $\mathcal H^+$}: Since $\nu(\cdot,1)=-1$ in the initial data gauge, we have
\begin{equation}\label{eq:horizon-location-formula}
    u_{\mathcal H^+}-1 = -\int_1^{u_{\mathcal H^+}}\nu(u',1)\,du' = r_\circ - r(u_{\mathcal H^+},1).
\end{equation}
By \eqref{eq:new-initial-smallness} and \eqref{eq:mass-orbital-stability}, $r_\circ= 100M_0+O(\ve)$ and $M=M_0+O(\ve)$, and by the geometric estimates, \[r(u_{\mathcal H^+},1)= r_\diamond(\infty,1)+O(\ve^2)=r_+(M,\varrho)+O(\ve^2)=M\big(1+\sqrt{1-\varrho^2}\big)+O(\ve^2).\] 
Combining these estimates shows
\[u_{\mathcal H^+}= 1+ r_\circ -r(u_{\mathcal H^+},1)= 1+99M_0-M_0\sqrt{1-\varrho^2}+O(\ve),\]
as desired. This completes the proof of \cref{thm:uniform-stability-RN}. \end{proof}

\begin{cor}[Uniform Bondi mass loss formula]\label{cor:Bondi-mass-loss}
    Under the hypotheses of \cref{thm:uniform-stability-RN}, it holds that
    \begin{align}
 \label{eq:Bondi-mass-loss}   |\u M{}_{\mathcal I^+}(\u u{}_1)-\u M{}_{\mathcal I^+}(\u u{}_2)| &\les\ve^2 \u u{}_1^{-3+\delta},\\
      |\u{P}{}_{\mathcal I^+}(\u u{}_1)-\u{P}{}_{\mathcal I^+}(\u u{}_2)| &\les\ve^2 \u u{}_1^{-3+\delta},\label{eq:P-change}
    \end{align}
    for every $1\le \u u{}_1\le \u u{}_2\le \u u{}_f$, where again the implicit constants depend only on $M_0$ and $\delta$. 
\end{cor}
\begin{proof}
    This follows immediately from the proof of \cref{lem:varpi-tilde} and the definition of $\u M{}_{\mathcal I^+}$.
\end{proof}

\subsection{The fundamental continuity arguments}\label{sec:continuity-arguments}

\begin{center}
    \emph{For the remainder of this paper, we put $A=A_\mathrm{unif}$ from \cref{thm:uniform-stability-RN} in \cref{def:K,def:K-tilde}.}
\end{center}

\subsubsection{Dyadic iteration}

\begin{prop}\label{prop:dyadic-iteration}
  If $\ve_0>0$ is chosen sufficiently small, depending only on $M_0$ and $\delta$, the following holds. Suppose $\ve\le\ve_0$, $(M,\varrho,\u u{}_f)$ is an admissible parameter triple, and $2\le \u u{}_f\le \frac 12 u_\infty(M,\varrho)$. Then it holds that
   \begin{equation*}
   \mathcal K(\ve,M,\varrho,\u u{}_f)\subset \tilde{\mathcal K}(\ve,M,\varrho,2\u u{}_f).
   \end{equation*}
\end{prop}
\begin{proof}
    Let $\Psi\in \mathcal K(\ve,M,\varrho,\u u{}_f)$. The proposition only has content when $\u u{}_f<\infty$, so assume that this is the case. We define the set
    \begin{equation*}
        \mathcal Z\doteq \{\u u{}_*\in[\u u{}_f,2\u u{}_f]:\Psi\in \tilde{\mathcal K}(\ve,M,\varrho,\u u{}_*)\}
    \end{equation*}
    and aim to show that $\mathcal Z=[\u u{}_f,2\u u{}_f]$ by a continuity argument. By the assumption on $\u u{}_f$, $(M,\varrho,\u u{}_*)$ is an admissible parameter triple for every $\u u{}_*\in [\u u{}_f,2\u u{}_f]$. We immediately have that $\mathcal Z$ is nonempty because $\u u{}_f\in\mathcal Z$. We show that there exists an $\eta_0>0$, depending perhaps on $2\u u{}_f$ and $\ve$, so that if $\u u{}_*\in \mathcal Z$, then $\u u{}_*+\eta\in \mathcal Z$ for every $0<\eta\le \eta_0$. Once this has been established, it follows that $\mathcal Z=[\u u{}_f,2 \u u{}_f]$ and hence the proposition is proved. We will extensively use the local theory of \cref{sec:semiglobal-1} and hence \ul{we allow the implicit constant in $\les$ to depend additionally on $2\u u{}_f$} in this proof.  
    
   Let $\u u{}_*\in\mathcal Z$. First, note that by \cref{lem:K-AF-est}, $\|\mathcal S\|_{\mathcal R(1,\u{\mathfrak u}(\u u{}_*),1,\infty)}\les 1$. In order to increment the solution forward in time, we need to show that the number $\u{\mathfrak u}(\u u{}_*)$ is quantitatively bounded away from $U_*$ which defines the bifurcate null hypersurface $\mathcal C$  (recall \eqref{eq:U*}). By the same logic that led to \eqref{eq:horizon-location-formula}, we have
   \begin{equation*}
       \u{\mathfrak u}(\u u{}_*)-1= r_\circ-\u r{}_{\u u{}_*}(\u u{}_*,1) = 100M_0 - r_\diamond (\u u{}_*,1) +O(\ve) \le 99M_0+O(\ve),
   \end{equation*}
   where we used that $r_\diamond(\u u{}_*,1)\ge r_c\ge M= M_0+O(\ve)$. Therefore, $\u{\mathfrak u}(\u u{}_*)\le \theta U_*$ for $\ve$ sufficiently small, where $\theta\in (0,1)$ is a universal constant. We prove the various conditions in \cref{def:K-tilde}.

\textsc{Proof of \hyperref[K-prop-1]{condition 1.}}:   Since $\u{\mathfrak u}(\u u{}_*)$ stays uniformly away from $U_*$, by \cref{prop:semiglobal-Cauchy-ID} and \cref{lem:Mi-open-quantitative} there exists an $\eta_0>0$, depending only on $2\u u{}_f$, such that $\Psi\in \mathfrak M_{\u u{}_*+\eta_0}$ and $\|\mathcal S\|_{\mathcal R(1,\u{\mathfrak u}(\u u{}_*+\eta_0),1,\infty)}\les 1$.

\textsc{Proof of \hyperref[K-prop-2]{condition 2.}}: This is inherited from the assumption $\Psi\in \tilde{\mathcal K}(\ve,M,\varrho,\u u{}_f)$.

\textsc{Proof of \hyperref[K-prop-3tilde]{condition $\tilde 3$.}}: Let $C$ be the implicit constant in \eqref{eq:Bondi-mass-loss}. Using \eqref{eq:Bondi-mass-loss} and \eqref{eq:varpi-exact}, and by taking $\ve_0$ small in terms of $C$, we estimate
\begin{equation*}
    |\u M{}_{\mathcal I^+}(\u u{}_*)-M|= |\u M{}_{\mathcal I^+}(\u u{}_*)-\u M{}_{\mathcal I^+}(\u u{}_f)|\le (C\ve^{1/2})\left(\frac{\u u{}_*}{\u u{}_f}\right)^{3-\delta}\cdot \ve^{3/2}\u u{}_*^{-3+\delta} \le \ve^{3/2}\u u{}_*^{-3+\delta},
\end{equation*}
which is \eqref{eq:varpi-modulation} (with a much better constant). By \eqref{eq:mass-loss-bound},
\begin{equation*}
       |\u M{}_{\mathcal I^+}(\u u{}_*+\eta)-M|\le \ve^{3/2}\u u{}_*^{-3+\delta} + C(B)\eta \le 2\ve^{3/2} (\u u{}_*+\eta)^{-3+\delta}
\end{equation*}
for $\eta\le\eta_0$ sufficiently small depending also on $\ve$ now. 

   \textsc{Proof of \hyperref[K-prop-4tilde]{condition $\tilde 4$.}}: We compute
   \begin{equation*}
    |\u P{}_{\mathcal I^+}(\u u{}_*+\eta)-\varrho |= \left|\frac{\varrho}{\u M{}_{\mathcal I^+}(\u u{}_*+\eta)}\big(M-\u M{}_{\mathcal I^+}(\u u{}_*+\eta)\big)\right|\le 5\ve^{3/2}(\u u{}_*+\eta)^{-3+\delta},
   \end{equation*}
   \textsc{Proof of \hyperref[K-prop-5]{condition 5.}}: We need to show that the estimates \eqref{eq:K-nu-1}--\eqref{eq:K-varpi-1} hold on $\mathcal R_{\u u{}_*+\eta}$, for $\mathcal S$ written in the $(\u u{}_*+\eta)$-normalized teleological gauge, for any $0<\eta\le\eta_0$ sufficiently small. By \cref{thm:uniform-stability-RN}, we may assume that these estimates hold on $\mathcal R_{\u u{}_*}$, when $\mathcal S$ is written in the $\u u{}_*$-normalized teleological gauge, with constant $\frac 12$ on the right-hand side in place of $1$. Note that the anchored background solution $(r_\diamond,\Omega_\diamond^2)$ depends on the gauge normalization time. As such, we write $r_{\diamond,\u u{}_*+\eta}$, etc., for clarity. 
   
We argue that
\begin{align}
    \label{eq:dyadic-proof-1} \sup_{\mathcal R_{\u u{}_*}}|\u \nu{}_{\u u{}_*+\eta}+D(r_{\diamond, \u u{}_*+\eta})| &\le \tfrac 34 A^2_\mathrm{unif}\ve^{3/2}D(r_{\diamond, \u u{}_*+\eta})\tau^{-1+\delta},\\
  \label{eq:dyadic-proof-2}    \sup_{\mathcal R_{\u u{}_*}}|\u r{}_{\u u{}_*+\eta}-r_{\diamond, \u u{}_*+\eta}| &\le \tfrac 34 A^2_\mathrm{unif}\ve^{3/2}\big(\tau^{-2+\delta}+r^\delta_{\diamond, \u u{}_*+\eta}\u u{}_f^{-2+\delta}\big),\\
    \label{eq:dyadic-proof-3}  \sup_{\mathcal R_{\u u{}_*}}|\u \varpi{}_{\u u{}_*+\eta}-M| &\le \tfrac 34 A_\mathrm{unif}\ve^{3/2}\tau^{-3+\delta}
\end{align}
  for $\eta$ sufficiently small depending only on $2\u u{}_f$. We prove only \eqref{eq:dyadic-proof-1}, as the proofs of \eqref{eq:dyadic-proof-2}  and \eqref{eq:dyadic-proof-3} are similar (in fact simpler).

  Before proving \eqref{eq:dyadic-proof-1}, however, we need to relate $r_{\diamond, \u u{}_*}$ and $r_{\diamond, \u u{}_*+\eta}$. We have 
    \begin{equation*}
       |\u r{}_{\u u{}_*+\eta}(\u u{}_*+\eta,\u u{}_*+\eta) - \u r{}_{\u u{}_*}(\u u{}_*,\u u{}_*)|\le   |\u r{}_{\u u{}_*+\eta}(\u u{}_*+\eta,\u u{}_*+\eta) - \u r{}_{\u u{}_*+\eta}(\u u{}_*,\u u{}_*)|+   |\u r{}_{\u u{}_*+\eta}(\u u{}_*,\u u{}_*) - \u r{}_{\u u{}_*}(\u u{}_*,\u u{}_*)|.
    \end{equation*}
    We can estimate the first term by $\les \eta$ using the mean value theorem and the bound $|\u \nu{}_{\u u{}_*+\eta}|+|\u \lambda{}_{\u u{}_*+\eta}|\les 1$ and the second term by $\les\eta$ using the mean value theorem and \cref{prop:updating-uf}. It follows from the anchoring condition \eqref{eq:finite-anchoring} for the background solutions that
    \begin{equation*}
        |r_{\diamond,\u u{}_*+\eta}(1,1)-r_{\diamond,\u u{}_*}(1,1)|\les \eta.
    \end{equation*}
    We can now use the linear theory of Reissner--Nordstr\"om, \cref{prop:linear-RN}, and the mean value theorem, to estimate 
\begin{equation*}
    |r_{\diamond,\u u{}_*+\eta}-r_{\diamond,\u u{}_*}|\les \eta\big(1+|{\log \u v}|\big)
\end{equation*}
    on $\mathcal R_{\u u{}_*+\eta}$. (Note that the bad weight $D'^{-1}(r_\diamond)$ in \eqref{eq:DHR-2} is $\les 1$ by our conventions for implicit constants in this proof.) Since $D'(r_\diamond)\les r_\diamond^{-2}$, we can use again \cref{prop:updating-uf} to estimate
    \begin{multline*}
     |\u \nu{}_{\u u{}_*+\eta}+D(r_{\diamond, \u u{}_*+\eta})|\les |\u \nu{}_{\u u{}_*+\eta}-\u \nu{}_{\u u{}_*}| + |\u \nu{}_{\u u{}_*}+D(r_{\diamond, \u u{}_*})| +  |D(r_{\diamond, \u u{}_*+\eta})-D(r_{\diamond, \u u{}_*})|   \\ \les \eta \u v^{-2}\log\u v + \tfrac 12 A_\mathrm{unif}\ve^{3/2}\big(D(r_{\diamond, \u u{}_*+\eta})+|D(r_{\diamond, \u u{}_*+\eta})-D(r_{\diamond, \u u{}_*})|\big)\tau^{-1+\delta} + \eta \u v^{-2}\big(1+|{\log \u v}|\big)\\
     \les \tfrac 34 A^2_\mathrm{unif}\ve^{3/2}D(r_{\diamond, \u u{}_*+\eta})\tau^{-1+\delta}
    \end{multline*}
    on $\mathcal R_{\u u{}_*}$, for $\eta$ sufficiently small depending on $\ve$, $A_\mathrm{unif}$, and $2 \u u{}_f\ge \u u{}_*$. Note that we also used the positivity of $D$, quantified in \eqref{eq:D-lower-bound}, to absorb the terms involving $\eta$. This completes the proof of \eqref{eq:dyadic-proof-1}. 

    We now show that \eqref{eq:K-nu-1}--\eqref{eq:K-varpi-1} hold on $\mathcal R_{\u u{}_*+\eta}$ in the $(\u u{}_*+\eta)$-normalized teleological gauge, if $\eta$ is sufficiently small. Again, we only prove \eqref{eq:K-nu-1}. It is easy to see that $|\partial_{\u u}\u \nu{}_{\u u{}_*+\eta}|\les 1$ on $\mathcal R_{\u u{}_*+\eta}$ using the estimates of \cref{sec:semiglobal-1}, so we can integrate this estimate in $\u u$ and use \eqref{eq:dyadic-proof-1} on $\{\u u=\u u{}_*\}$ to prove
    \begin{equation*}
        \sup_{\mathcal R_{\u u{}_*+\eta}\setminus \mathcal R_{\u u{}_*}}|\u \nu{}_{\u u{}_*+\eta}+D(r_{\diamond, \u u{}_*+\eta})|\le A^2_\mathrm{unif}\ve^{3/2}D(r_{\diamond, \u u{}_*+\eta})\tau^{-1+\delta}
    \end{equation*} for $\eta$ sufficiently small,
    as desired.

    \textsc{Proof of \hyperref[K-prop-6]{condition 6.}}: Finally, we move to the energy estimates \eqref{eq:K-energy-1}--\eqref{eq:K-energy-4}. Again, we only prove one of them, namely \eqref{eq:K-energy-1}. As before, this is proved in two steps: first, we show that (for any $p\in \{0\}\cup[\delta,3-\delta]$)
    \begin{equation}\label{eq:updating-energy-1}
        \mathcal E_p^{(M,\varrho,\u u{}_*+\eta)}[\u\phi{}_{\u u{}_*+\eta}](\tau)\le \tfrac 34 A_\mathrm{unif}\tau^{-3+\delta+p}
    \end{equation}
    for $\tau\in[1, \u u{}_*]$ and then
    \begin{equation}\label{eq:updating-energy-2}
        \mathcal E_p^{(M,\varrho,\u u{}_*+\eta)}[\u\phi{}_{\u u{}_*+\eta}](\tau)\le  A_\mathrm{unif}\tau^{-3+\delta+p}
    \end{equation}
    for $\tau\in[1, \u u{}_*+\eta]$.

   To prove \eqref{eq:updating-energy-1}, we use \cref{prop:updating-uf} to compute, for any $\tau \in[1,\u u{}_*]$,
    \begin{equation*}
     |\partial_\eta \mathcal E_p^{(M,\varrho,\u u{}_*+\eta)}[\u\phi{}_{\u u{}_*+\eta}](\tau)|\les\int_\tau^\infty \big(\u r{}_{\u u{}_*+\eta}^{p-1}|\partial_\eta \u r{}_{\u u{}_*+\eta}|(\partial_{\u v}\u\psi{}_{\u u{}_*+\eta})^2+\u r{}_{\u u{}_*+\eta}^{p}|\partial_{\u v}\u\psi{}_{\u u{}_*+\eta}||\partial_\eta \partial_{\u v}\u\psi{}_{\u u{}_*+\eta}|+\cdots\big)\,d\u v \les 1.
   \end{equation*}
   By integrating in $\eta$, we obtain \eqref{eq:updating-energy-1} for $\eta$ sufficiently small. Then to prove \eqref{eq:updating-energy-2}, we estimate
   \begin{multline*}
       \left|\left(\int_{\{\u u{}_*+\eta\}\times[\u u{}_*,\infty)}-\int_{\{\u u{}_*\}\times[\u u{}_*,\infty)}\right)\big(\u r{}_{\u u{}_*+\eta}^{p}(\partial_{\u v}\u\psi{}_{\u u{}_*+\eta})^2+\cdots\big)\,d\u v\right| \\\les \iint_{[\u u{}_*,\u u{}_*+\eta]\times[\u u{}_*,\infty)}\big|\partial_{\u u}\big(\u r{}_{\u u{}_*+\eta}^{p}(\partial_{\u v}\u\psi{}_{\u u{}_*+\eta})^2+\cdots\big)\big| \,d\u u d\u v  \les \iint_{[\u u{}_*,\u u{}_*+\eta]\times[\u u{}_*,\infty)} (\u v^{p-5}+\u v^{-2})\,d\u u d\u v\les  \eta.
   \end{multline*}
   For $\eta$ sufficiently small, this proves \eqref{eq:updating-energy-2} and hence concludes the proof of the proposition.  
\end{proof}

\subsubsection{Modulation}

\begin{prop}\label{prop:modulation} If $\ve_0>0$ is chosen sufficiently small, depending only on $M_0$ and $\delta$, the following holds. Suppose $\ve\le\ve_0$, $(M,\varrho,\u u{}_f)$ is an admissible parameter triple, and $\Psi\in\tilde{\mathcal K}(\ve,M,\varrho,\u u{}_f)$. Then 
 \begin{equation*}
     \Psi\in \mathcal K(\ve,\u M{}_{\mathcal I^+}(\u u{}_f),\u P{}_{\mathcal I^+}(\u u{}_f),\u u{}_f).
 \end{equation*}
\end{prop}
\begin{proof}
    For $z\in [0,1]$, let $M_z\doteq (1-z)M+z\u M{}_{\mathcal I^+}(\u u{}_f)$ and $\varrho_z\doteq (1-z)\varrho +z\u P{}_{\mathcal I^+}(\u u{}_f)$. We define the set
    \begin{equation*}
        \mathcal Z\doteq \{z\in [0,1]:\Psi\in\tilde{\mathcal K}(\ve,M_z,\varrho_z,\u u{}_f)\}
    \end{equation*}
    and aim to show that $\mathcal Z=[0,1]$ by a continuity argument. Clearly, $\mathcal Z$ is nonempty since $0\in \mathcal Z$. As in the proof of \cref{prop:dyadic-iteration}, we show that there exists an $\eta_0>0$, depending perhaps on $\u u{}_f$, such that if $z_*\in \mathcal Z$, then $z_*+\eta\in\mathcal Z$ for every $0<\eta\le \min\{\eta_0,1-z_*\}$. 

  Let $z_*\in \mathcal Z$.  Since conditions \hyperref[K-prop-1]{1.} and \hyperref[K-prop-2]{2.} in \cref{def:K-tilde} are independent of $(M,\varrho)$, there is nothing to check here--they are satisfied for any $\eta\in (0,1-z_*]$. To verify conditions \hyperref[K-prop-3tilde]{$\tilde 3$.} and \hyperref[K-prop-4tilde]{$\tilde 4$.}, we compute
    \begin{align*}
        |\u M{}_{\mathcal I^+}(\u u{}_f)-M_{z_*+\eta}|&=(1-z_*-\eta)|\u M{}_{\mathcal I^+}(\u u{}_f)-M|\le (1-z_*-\eta)10\ve^{3/2}\u u{}_f^{-3+\delta},\\
        |\u P{}_{\mathcal I^+}(\u u{}_f)-\varrho_{z_*+\eta}|&=(1-z_*-\eta)|\u P{}_{\mathcal I^+}(\u u{}_f)-\varrho|\le (1-z_*-\eta)10M_0^{-1}\ve^{3/2}\u u{}_f^{-3+\delta},
    \end{align*}
    so again these are satisfied for any $\eta\in (0,1-z_*]$.  

    Conditions \hyperref[K-prop-5]{5.} and \hyperref[K-prop-6]{6.} depend on $z$ through the dependence of $r_\diamond$ on $(M,\varrho)$ since the gauge is fixed. Using \cref{prop:linear-RN} and \cref{thm:uniform-stability-RN} as in the proof of \cref{prop:dyadic-iteration}, we can show that these are satisfied for $z_*+\eta$, for some $\eta>0$. We omit the details. 

    Now that we have shown that $\mathcal Z=[0,1]$, the proof is complete because the conditions \hyperref[K-prop-3]{3.} and \hyperref[K-prop-4]{4.} in the definition of $\mathcal K$ are satisfied at $z=1$ by construction. 
\end{proof}

\subsection{Revisiting the fundamental dichotomy}\label{sec:revisiting}

\subsubsection{Getting started}

We begin with two preparatory lemmas that will serve as the base case for an induction argument in the proof of \cref{thm:dichotomy-revisited}. 

\begin{lem}\label{lem:first-step-K} There exist $\theta_0>0$ and $\ve_0>0$, depending only on $M_0$ and $\delta$, such that
    \begin{equation*}
        \cyl(\ve,1+\theta_0)\subset\bigcup_{(M,\varrho,2)\,\mathrm{admissible}} \mathcal K(\ve,M,\varrho,2),
    \end{equation*}
   for every $\ve\le \ve_0$.
\end{lem}
\begin{rk}
The proof of this lemma is not, as one might imagine, a bootstrap argument in time like the proof of \cref{prop:dyadic-iteration}. This is because our teleological gauges are only defined for $\u u{}_f\ge 2$ because of the curves $\mathcal G_{\u u{}_f}$ used to define the $\u v$-coordinate. Instead, we essentially perform a bootstrap argument \emph{in $\ve$}, a technique that will later be used again when constructing stable manifolds in the proof of \cref{thm:main}. 
\end{rk}
\begin{proof}[Proof of \cref{lem:first-step-K}] Let $\Psi=(\phi_\circ,r_\circ,\varpi_\circ,\rho_\circ)\in   \cyl(\ve,1+\theta_0)$, where $\ve\le \ve_0$ and $\theta_0\doteq \frac 18$. For $z\in [0,1]$, let $\Psi_z\doteq (z\phi_\circ,r_\circ,\varpi_\circ,\rho_\circ)$, which defines a continuous path in $\mathfrak M$ for $\ve_0$ and $\theta_0$ chosen sufficiently small. We define the set
\begin{equation*} 
\mathcal Z\doteq \{z\in [0,1]:\Psi_z\in\tilde{\mathcal K}(\ve,\varpi_\circ,\rho_\circ,2)\}.
\end{equation*} 

By construction, $\Psi_0$ generates a member of the Reissner--Nordstr\"om family and hence $0\in\mathcal Z$. To show that $\mathcal Z=[0,1]$, we may now argue as in the proofs of \cref{prop:dyadic-iteration,prop:modulation} using the theory of \cref{sec:semiglobal-1} (in particular \cref{prop:teleological-1}) and \cref{thm:uniform-stability-RN} with $\u u{}_f=2$. These arguments show that $\Psi=\Psi_1\in\tilde{\mathcal K}(\ve,\varpi_\circ,\rho_\circ,2)$ and after modulating we see that $\Psi\in\mathcal K(\ve,\u M{}_{\mathcal I^+}(2),\u P{}_{\mathcal I^+}(2),2)$. By choosing $\ve_0$ possibly smaller, we can ensure that $|\u P{}_{\mathcal I^+}(2)|\le \frac 54$, which is required to be admissible. 
\end{proof}

\begin{lem}\label{lem:data-non-1} For $\ve_0>0$ sufficiently small, depending only on $M_0$, it holds that
    \begin{equation*}
   |\mathscr P_0|>1 \quad\text{on}\quad \cyl(\ve_0,2)\setminus  \cyl(\ve_0,1+\theta_0)
    \end{equation*}
    and hence 
    \begin{equation*}
         \cyl(\ve_0,2)\setminus  \cyl(\ve_0,1+\theta_0)\subset\mathfrak M_\mathrm{non}.
    \end{equation*}
\end{lem}
\begin{proof}
This follows immediately from \cref{prop:a-priori-noncollapse} (and its proof).  
\end{proof}

\subsubsection{The proof of \texorpdfstring{\cref{thm:dichotomy-revisited}}{Theorem 5.13}}\label{sec:proof-refined-dichotomy}

\begin{proof}[Proof of \cref{thm:dichotomy-revisited}]
    We will show that any $\Psi\in \cyl(\ve,2)$, where $\ve\le \ve_0$ is sufficiently small, either fails to collapse because $|\u{P}_{\mathcal I^+}(L_i)|>1$ for some dyadic time $L_i\doteq 2^i$, or $\Psi\in \mathcal K(\ve,M_i,\varrho_i,L_i)$ for all $i\ge 1$, for some sequence of non-superextremal ratios $|\varrho_i|\le 1$. In the latter case, we will show convergence to a Reissner--Nordstr\"om black hole with parameters $(\lim M_i,\lim \varrho_i)$. 

Let $\ve_0>0$ be sufficiently small that \cref{thm:uniform-stability-RN}, \cref{prop:dyadic-iteration}, \cref{prop:modulation}, \cref{lem:first-step-K}, and \cref{lem:data-non-1} apply for any $\ve\le \ve_0$. Let $\Psi\in \cyl(\ve,2)$. 

First, we note that either $\Psi\in \cyl(\ve,2)\setminus \cyl(\ve,1+\theta_0)$, in which case $|\u P_{\mathcal I^+}(L_0)|>1$ and the solution fails to collapse by \cref{lem:data-non-1}, or $\Psi\in \mathcal K(\ve,M_1,\varrho_1,L_1)$ by \cref{lem:first-step-K}, where $(M_1,\varrho_1)=(\u M{}_{\mathcal I^+}(L_1),\u P{}_{\mathcal I^+}(L_1))$. 
If $|\varrho_1|>1$ in the latter case, then $\Psi\in\mathfrak M_\mathrm{non}$.

Therefore, we may argue by induction: we assume that $\Psi\in \mathcal K(\ve,M_i,\varrho_i,L_i)$ for some index $i\ge 1$ and admissible triple $(M_i,\varrho_i,L_i)$ satisfying $|\varrho_i|\le 1$. Then $u_\infty(M_i,\varrho_i)=\infty$ and $\Psi\in \tilde{\mathcal K}(\ve,M_i,\varrho_i,L_{i+1})$ by \cref{prop:dyadic-iteration}. By \cref{prop:modulation}, $\Psi\in \mathcal K(\ve,\u M{}_{\mathcal I^+}(L_{i+1}),\u P{}_{\mathcal I^+}(L_{i+1}),L_{i+1})$. If $|\u P{}_{\mathcal I^+}(L_{i+1})|>1$, then $\Psi\in\mathfrak M_\mathrm{non}$ and we stop. Otherwise, set $(M_{i+1},\varrho_{i+1}) = (\u M{}_{\mathcal I^+}(L_{i+1}),\u P{}_{\mathcal I^+}(L_{i+1}))$ and repeat this process. 

We now analyze the case when this process does not terminate at a finite $i$. Let $\{(M_i,\varrho_i)\}_{i\ge 1}$ be the sequence of parameters at dyadic times. By the mass loss formula, \cref{cor:Bondi-mass-loss}, $\lim_{i\to\infty}(M_i,\varrho_i)\doteq (M_\infty,\varrho_\infty)$ exists and $|\varrho_\infty|\le 1$. We now claim that $\Psi\in \mathcal K(\ve,M_\infty,\varrho_\infty,\infty)$.

\textsc{Proof of \hyperref[K-prop-1]{condition 1.}}: Clearly $\Psi\in \mathfrak M_{\u u{}_f}$ for every $\u u{}_f\ge 2$, so $\Psi\in \mathfrak M_\mathrm{black}$ by \cref{thm:dichotomy}. We use \eqref{eq:diffeo-form} and the $i$-independent estimates $\u\lambda{}_{L_i}\sim \u\kappa{}_{L_i}\sim 1$ on $\mathcal R_{L_i}$ to infer that $\Psi\in\mathfrak M_\mathrm{black}^{\kappa\sim 1}$ and hence the eschatological gauge construction of \cref{sec:eschatological-definition} is defined for $\mathcal S[\Psi]$. While this proves \hyperref[K-prop-1i]{condition 1.i.}, the proof of \hyperref[K-prop-1ii]{condition 1.ii.} is more involved. 

We have to compare $r_{\diamond,L_i}$ and $r_{\diamond,L_{i+1}}$ as $i\to\infty$. By construction of the teleological $\u v$-coordinate, note that the diffeomorphisms $\u{\mathfrak u}\times \u{\mathfrak v}{}_{L_i}$ and $\u{\mathfrak u}\times \u{\mathfrak v}{}_{L_j}$, for $j\ge i$, coincide on the set $[1,L_i-1]^2$ (recall the proof of \cref{lem:eschatology}). By \eqref{eq:r-deriv-along-Gamma}, $   \big| \partial_{\u v}\u r{}_{L_i}|_\Gamma\big|\les \ve^{3/2}\tau^{-3+\delta}$
for every $i$. Therefore, we may estimate
\begin{align*}
        |r_{\diamond,L_i}(1,1)-r_{\diamond,L_i+1}(1,1)|&= |\u r{}_{L_i}(L_i,L_i)- \u r{}_{L_{i+1}}(L_{i+1},L_{i+1})|\\
        &\le |\u r{}_{L_i}(L_i,L_i)- \u r{}_{L_{i}}(L_{i}-1,L_{i}-1)| +\underbrace{|\u r{}_{L_{i}}(L_{i}-1,L_{i}-1)- \u r{}_{L_{i+1}}(L_{i}-1,L_{i}-1)|}_{=0}\\&\qquad +|\u r{}_{L_{i+1}}(L_i-1,L_i-1)- \u r{}_{L_{i+1}}(L_{i+1},L_{i+1})|\\
        &\les \ve^{3/2}L_i^{-2+\delta}.
\end{align*}

It follows that, $r_{\diamond,L_i}(1,1)$ has a limit as $i\to\infty$, which defines the candidate for $r_{\diamond,\infty}$ by \cref{lem:existence-anchoring}. We now show that \eqref{eq:eschatological-anchoring} is satisfied for this choice, i.e., that
\begin{equation*}
    \lim_{\u u\to\infty }\u r{}_\infty(\u u, \u u)= r_{\diamond,\infty}(1,1)= \lim_{i\to\infty}r_{\diamond,L_i}(1,1).
\end{equation*}
Indeed, if $\u u\le L_i-1$, then 
\begin{align*}
    |\u r{}_\infty(\u u,\u u)-r_{\diamond,\infty}(1,1)| &=  |\u r{}_{L_i}(\u u,\u u)-r_{\diamond,\infty}(1,1)|\\ &\le |\u r{}_{L_i}(\u u,\u u)-r_{\diamond,L_i}(1,1)|+|\u r{}_{\diamond,L_i}(1,1)-r_{\diamond,\infty}(1,1)| \\&\les \ve^{3/2}\u u^{-2+\delta}+\ve^{3/2}L_i^{-2+\delta}\les \ve^{3/2}\u u^{-2+\delta},
\end{align*}
which proves the claim as we may now let $\u u\to\infty$.

\textsc{Proof of \hyperref[K-prop-2]{condition 2.}}: This is a hypothesis of the theorem.

\textsc{Proof of \hyperref[K-prop-3]{conditions 3.} and \hyperref[K-prop-4]{4.}}: These hold by definition. 

\textsc{Proof of \hyperref[K-prop-5]{condition 5.}}: These arguments in the proof of \hyperref[K-prop-1ii]{condition 1.ii.}~show that $\u \nu{}_{\dagger,L_i}$, $\u r{}_{\dagger,L_i}$, and $\u\varpi{}_{\dagger,L_i}$ converge on compact subsets of $\mathcal R_\infty$ as $i\to\infty$, and hence \hyperref[K-prop-5]{condition 5.} is satisfied. 

\textsc{Proof of \hyperref[K-prop-6]{condition 6.}}: This follows again from the arguments in the proof of \hyperref[K-prop-1ii]{condition 1.ii.} together with Fatou's lemma. 

Therefore, $\Psi$ satisfies all of the conditions of \cref{def:K} and $\Psi\in \mathcal K(\ve,M_\infty,\varrho_\infty,\infty)$. This completes the proof of the theorem. \end{proof}

\section{Constructing the stable manifolds and the proof of the main theorem} \label{sec:proof-of-main-thm}

In this section, we prove the main theorem of the paper, \cref{thm:main}, modulo important estimates that are deferred until \cref{sec:estimates-linear-perturbations}. It could be helpful at this point to review the proof outline in \cref{sec:overview-II}. We begin in \cref{sec:dyadic-inversion} by constructing the functions $\mathscr W_i$ which are implicit functions for $\mathscr P_i$. In \cref{sec:eschatological-foliation}, we show that the $\mathscr W_i$'s converge to a $C^1$ function $\mathscr W_\infty$ whose graphs are stable manifolds for the family of Reissner--Nordstr\"om black holes. Finally, in \cref{sec:main-thm-proof}, we prove that solutions outside of the image of $\check{\mathscr W}_\infty$ fail to collapse, which completes the proof of \cref{thm:main}.

\subsection{Implicit functions for the parameter ratio at dyadic times}\label{sec:dyadic-inversion}

\subsubsection{Existence of \texorpdfstring{$\mathscr W_i$}{Wi}}

Recall the constant $\alpha>0$ from \cref{lem:def-of-alpha} which will now play an important role. We also recall the decomposition of the space $\mathfrak Z$ as $\mathfrak X\times\Bbb R$ from \cref{sec:AF-seed-data}. In the following, we will denote the variable of the $\Bbb R$-factor by $\sigma$ when the function in question is $\mathscr W_i$ or $\mathscr W_\infty$. 

\begin{prop}\label{prop:Wi} If $\ve_0>0$ is sufficiently small, depending only on $M_0$ and $\delta$, the following holds. For every $i \ge1$, there exists a $C^1$ function\footnote{Recall the definition of $C^1$ functions on the non-open set $\cyl(\ve,\ell)$ which was given in \cref{sec:threshold}.}
  \begin{equation*}
      \mathscr W_i:\cyl(\ve_0,1+\alpha L_i^{-2})\to \Bbb R,
  \end{equation*}
  satisfying the estimates\index{W@$\mathscr W_i$, implicit function for $\mathscr P_i$}
    \begin{align}
   \label{eq:Wi-est-1}    \|\mathscr W_i\|_{C^1(\cyl(\ve_0,1+\alpha L_i^{-2}))} &\les 1,\\
  \label{eq:Wi-est-2}  \sup_{\cyl(\ve_0,1+\alpha L_i^{-2})}|\partial_\sigma\mathscr W_i-1|  &\les \ve_0,
    \end{align}
 such that if $(x,\sigma)\in \cyl(\ve,1+\alpha L_i^{-2})$, then 
 \begin{equation}\label{eq:Wi-defn}
     (x,\mathscr W_i(x,\sigma)) \in \mathcal K(\ve,M,\sigma,L_i),
 \end{equation} where $(M,\sigma,L_i)$ is an admissible triple.
\end{prop}

After unraveling definitions, the condition \eqref{eq:Wi-defn} implies 
\begin{equation}\label{eq:Pi-inversion}
    \check{\mathscr P}_i\circ \check{\mathscr W}_i = \id_{\cyl(\ve_0,1+\alpha L_i^{-2})},
\end{equation}
where we recall the augmentation $\check{\mathscr P}_i$ is defined by $\check{\mathscr P}_i(x,\varrho_\circ)=(x,\mathscr P_i(x,\rho_\circ))$ and similarly for $\check{\mathscr W}_i$.  \cref{prop:Wi} will be proved below in \cref{sec:proof-Wi} via a continuity argument. The proof relies on nontrivial estimates for the linearized solutions in the class $\mathcal K$ which will be proved in \cref{sec:estimates-linear-perturbations}. 

\begin{rk}
    The condition \eqref{eq:Wi-est-2} implies that the augmented map $\check{\mathscr W}_i$ is a diffeomorphism onto its image for $\ve_0$ sufficiently small (and we will always assume this is the case).
\end{rk}

\begin{rk}
   We will see in the proof that the condition \eqref{eq:Pi-inversion} means that $\mathscr W_i$ is uniquely determined by the dyadic scaling $L_i=2^i$, but we could equally well work with a scaling $\tilde L_i = \mathfrak d^i$ for any real $\mathfrak d>1$. This would produce a different set of $\tilde{\mathscr W}_i$'s, but result in the same $\mathscr W_\infty$ below.
\end{rk}

\begin{rk}\label{rk:W-normalization}
    The condition \eqref{eq:Wi-defn} uniquely determines $\mathscr W_i$ for $x=x_0$. Indeed, since $(x_0,\rho_\circ)$ evolves into Reissner--Nordstr\"om with charge-to-mass ratio $\rho_\circ$, it holds that $  \mathscr W_i(x_0,\sigma) = \sigma $  for $|\sigma|\le 1+\alpha L_i^{-2}$. 
\end{rk}

\subsubsection{The linearized background \texorpdfstring{$\lin r_\diamond$}{lin r diamond}}\label{sec:lin-r-diamond}

A critical part of the proof of \cref{prop:Wi} is to define the \emph{linearized anchored background solution} $\lin r_\diamond$. Let $i\ge 1$ and $\Psi\in\mathcal K(\ve,M,\varrho,L_i)$. Given $\lin\Psi\in \mathfrak Z$, we may define as usual the one-parameter family $z\mapsto \Psi_z\doteq \Psi+z\lin\Psi$. For $z$ small, $\Psi_z\in\mathfrak M_{L_i}$ as well, and set $(M_z,\varrho_z)=(\mathscr M_i(\Psi_z),\mathscr P_i(\Psi_z))$. We may now apply the anchoring procedure of \cref{sec:anchoring} to obtain a one-parameter family $z\mapsto r_\diamond(\u u,\u v,z)$ of $(M_z,\varrho_z,L_i)$-anchored background solutions on $\mathcal R_{L_i}$. We now define the linearized background by setting
\begin{equation}\label{eq:lin-r-diamond-defn}
    \lin r_\diamond \doteq \partial_z r_\diamond.
\end{equation}

At $z=0$, this is equivalent to the following procedure: Consider the $L_i$-teleologically normalized linearized solution $\lin{\u{\mathcal S}}{}_{L_i}[\Psi,\lin\Psi]$ defined in \cref{sec:PT-Bondi-gauge}. Now define
\begin{equation*}
 \Psi_\diamond\doteq (0,\u r{}_{L_i}(L_i,L_i),M,\varrho)  ,\quad  \lin\Psi_\diamond\doteq (0,\lin{\u r}{}_{L_i}(L_i,L_i),\lin{\mathscr M}_i,\lin{\mathscr P}_i),
\end{equation*}
and consider the linearized solution $\lin{\u{\mathcal S}}{}_{L_i}[\Psi_\diamond,\lin\Psi_\diamond]$. After unraveling the definitions, we see that this linearized solution agrees with \eqref{eq:lin-r-diamond-defn}. Note that $\lin \Psi_\diamond$ depends linearly on $\lin\Psi$.

As in \cref{sec:anchoring}, we define difference quantities with a dagger, such as $\lin{\u r}{}_{\u u{}_f\dagger}\doteq \lin{\u r}{}_{\u u{}_f}- \lin r_\diamond$ or $\lin{\u\gamma}{}_{\u u{}_f\dagger} \doteq \lin{\u\gamma}{}_{\u u{}_f}$. By \eqref{eq:DHR-1}, we have the crucial relation
\begin{equation}\label{eq:lin-P_i-bootstrap-formula}
   \lin Q=\varrho\lin{\mathscr M}_i+M\lin{\mathscr P}_i.
\end{equation} From \cref{prop:linear-RN} and \cref{lem:lin-r-expansion} we obtain:

\begin{lem}\label{lem:lin-r-diamond}
    Let $(M,\varrho,L_i)$ be admissible and define $\lin{\u{\mathcal S}}{}_{L_i}[\Psi_\diamond,\lin\Psi_\diamond]$ by the above procedure. Then it holds that
    \begin{gather}
     \label{eq:lin-r-diamond-10}  |\lin r_\diamond|\les \big(\|\lin\Psi_\diamond\|_\mathfrak{Z} + D'^{-1}|\lin{\mathscr P}_i|\big)\mathbf 1_{\{\u u\ge\u v\}} +r_\diamond^\delta\|\lin\Psi_\diamond\|_\mathfrak{Z}\mathbf 1_{\{\u v\ge\u u\}},\\
    \label{eq:lin-nu-diamond-10}   |\lin\nu_\diamond| = |\lin\lambda_\diamond|\les \big(D'^{\delta}D^{1-\delta/2} \|\lin\Psi_\diamond\|_\mathfrak{Z}+D'^{-2+\delta}D^{1-\delta/2}|\lin{\mathscr P}_i|\big)\mathbf 1_{\{\u u\ge\u v\}}+r_\diamond^{-1}\|\lin\Psi_\diamond\|_\mathfrak{Z}\mathbf 1_{\{\u v\ge\u u\}}
    \end{gather}
   in $\mathcal R_{L_i}$ and 
   \begin{equation}\label{eq:lin-r-expansion-10}
       \lin r_\diamond = \frac{r_c}{M}\lin{\mathscr M}_i + O(D'^{-1}|\lin{\mathscr P}_i|)+O(D'\|\lin\Psi_\diamond\|_\mathfrak{Z})
   \end{equation}
   in $\mathcal R_{L_i}^{\le}$. 
\end{lem}

  \subsubsection{The bootstrap set \texorpdfstring{$\mathfrak E$}{E}} 

  Let $\ve_0>0$ be small enough so that \cref{thm:uniform-stability-RN,thm:dichotomy-revisited} and their proofs apply. 
  
\begin{defn}\label{def:mathfrak-E-new} For $\lin A\ge 1$ and $i\ge 1$, we define the set $\mathfrak E(i, \lin A)$ to be the set of $\ve\in (0,\ve_0]$ for which the following holds:
\begin{enumerate} 
    \item\label{E-prop-1} For every $\ve\in \mathfrak E(i,\lin A)$, there exists a $C^1$ function $\mathscr W_i:\cyl(\ve,1+\alpha L_i^{-2})\to \Bbb R$ such that if $(x,\sigma)\in \cyl(\ve,1+\alpha L_i^{-2})$, then $\check{\mathscr W}_i(x,\sigma)\in\mathcal K(\ve,M,\sigma,L_i)$ for an admissible parameter triple $(M,\sigma,L_i)$.

    \item\label{E-prop-2}  Let $\ve\in \mathfrak E(i, \lin A)$ and take $\Psi\in\check{\mathscr W}_i(\cyl(\ve,1+\alpha L_i^{-2}))$. For any $\lin \Psi\in\mathfrak Z$, we can then define the linearized solution $\lin{\u{\mathcal S}}{}_{L_i}[\Psi,\lin\Psi]$ in $L_i$-normalized teleological gauge. If $\|\lin\Psi\|_\mathfrak{Z}\le 1$, it holds that 
\begin{align}
\label{eq:lin-r-bootstrap} |\lin{\u r}{}_{L_i}(L_i,L_i)|&\le 10,\\
   \label{eq:lin-M-bootstrap}   |\lin {\mathscr M}_i|&\le 10,\\
  \label{eq:lin-P-bootstrap}      |\lin{\mathscr P}_i|&\le 10.
\end{align}
    \item\label{E-prop-3}  For $\lin{\u{\mathcal S}}{}_{L_i}[\Psi,\lin\Psi]$ as above, we assume that the estimates
    \begin{align}
 \label{eq:lin-boot-1-new}   |\u{\lin r}{}_{L_i}-\lin r_\diamond|& \le \lin A\big(\tau^{-2+3\delta}+|\lin{\mathscr P}_i|\tau^{-2+3\delta}D_1'^{-2+3\delta/2}\big),\\
|\u{\lin\lambda}{}_{L_i}|&  \le \lin A\big(D^{1-\delta/2}+\tau^{-2+3\delta}+|\lin{\mathscr P}_i|\big[D'^{-2+3\delta/2}D^{1-3\delta/4}+\tau^{-2+3\delta} D_1'^{-2+3\delta/2} \big]\big),\label{eq:lin-boot-2-new} \\
  \label{eq:lin-boot-3-new}  |\u{\lin \nu}{}_{L_i}| & \le \lin A\big( D^{1-\delta/2}+|\lin{\mathscr P}_i|\big[D'^{-2+\delta}D^{1-\delta/2}+\tau^{-1+3\delta}DD_1'^{-2+3\delta/2}\big]\big)
 \end{align}
 hold on $\mathcal R_{L_i}^\le$ and
{\mathtoolsset{showonlyrefs=false}
\begin{align}
 \label{eq:lin-boot-4-new}   |\u{\lin r}_{L_i}|&\le \lin A r_\diamond^\delta, \\
 \label{eq:lin-boot-5-new}   |\u{\lin\lambda}_{L_i}|
    & \leq  \lin A  r^{-1+\delta}_\diamond ,\\
  \label{eq:lin-boot-6-new}  |\u{\lin\nu}_{L_i}|&\le \lin Ar^{-1}_\diamond 
\end{align}
}
   hold  on $\mathcal R_{L_i}^\ge$. In these estimates, $r_\diamond$ is anchored according to the procedure of \cref{sec:anchoring} and \index{D1@$D_1'(u)\doteq D'(r_\diamond(u,1))$}
   \begin{equation}\label{eq:D1'-defn}
       D_1'(\u u)\doteq D'(r_\diamond(\u u,1);M,\varrho).
   \end{equation}
\end{enumerate}
\end{defn}

\begin{rk} For any $\lin A\ge 1$ sufficiently large depending on $M_0$ and $\delta$, $\mathfrak E(i, \lin A)$ is nonempty for every $i\ge 1$. While one could prove this here immediately using the theory of \cref{sec:semiglobal-1}, it is more convenient to show it in the proof of \cref{prop:Wi} below.
\end{rk}

\begin{rk}
    Strictly speaking, the wording of \cref{def:mathfrak-E-new} allows for the function $\mathscr W_i$ to depend on $\ve$, but its uniqueness will follow from the construction below. 
\end{rk}

\subsubsection{The main linearized estimates}

The main analytic input for the proof of \cref{prop:Wi} consists of the following two propositions:

\begin{prop}\label{prop:lin-boot-improvement}
   For $\lin A\ge1$ sufficiently large, depending only on $M_0$ and $\delta$, and $\ve_0>0$ sufficiently small, depending also on $\lin A$, the following holds. If $\ve\in (0,\ve_0]\cap \mathfrak E(\lin A,i)$ for any $i\ge 1$, then \eqref{eq:lin-r-bootstrap}--\eqref{eq:lin-P-bootstrap} hold with $10$ replaced by $5$ and \eqref{eq:lin-boot-1-new}--\eqref{eq:lin-boot-6-new} hold with $\lin A$ replaced by $\frac 12 \lin A$. 
\end{prop}

\begin{prop}\label{prop:Pi-est}
    Under the assumptions of \cref{prop:lin-boot-improvement}, the following holds. The functions $\mathscr P_i$ and $\mathscr M_i$, which are $C^1$ on an open set containing $\check{\mathscr W}_i(\cyl(\ve,1+\alpha L_i^{-2}))$ by the theory of \cref{sec:semiglobal-1}, satisfy
 \begin{align}\label{eq:Pi-est-1}
  \sup_{\check{\mathscr W}_i(\cyl(\ve_0,1+\alpha L_i^{-2}))}\|\mathscr P_i'\|_{\mathfrak Z^*}&\les \lin A,\\
   \label{eq:Pi-est-2} \sup_{\check{\mathscr W}_i(\cyl(\ve_0,1+\alpha L_i^{-2}))}\|\mathscr P_i'-\mathscr P_j'\|_{\mathfrak Z^*}&\les \lin A\ve^2 L_j^{-1+\delta/2}\quad\text{for every }1\le j\le i,\\
  \label{eq:Pi-est-3}   \sup_{\check{\mathscr W}_i(\cyl(\ve_0,1+\alpha L_i^{-2}))}|\partial_{\rho_\circ}\mathscr P_i-1|&\les \lin A\ve,\\
    \label{eq:Mi-est-1} \sup_{\check{\mathscr W}_i(\cyl(\ve_0,1+\alpha L_i^{-2}))}\|\mathscr M_i'-\mathscr M_j'\|_{\mathfrak Z^*}&\les \lin A\ve^2 L_j^{-1+\delta/2}\quad\text{for every }1\le j\le i.
 \end{align}
\end{prop}

The proofs of \cref{prop:Pi-est,prop:lin-boot-improvement} are the content of \cref{sec:estimates-linear-perturbations} below. The final proofs are given in \cref{sec:completing-linear}.

\subsubsection{The implicit function theorem for \texorpdfstring{$C^{1,1}$}{C 1, 1} functions on Banach spaces}\label{sec:implicit-function}

We state here for reference the inverse function theorem for $C^{1,1}$ functions on Banach spaces, with an important quantitative refinement. For a proof, see for instance \cite[Theorem A.43]{lee2021geometric} (one simply tracks the dependence of constants in any standard proof of the inverse function theorem).

\begin{thm}\label{thm:inverse-FT} Let $r_0$, $C$, and $c_0$ be positive constants. Then there exist positive constants $r_0'$, $\rho_0$, and $C'$, where $C'$ is independent of $r_0$, such that the following holds. Let $(X,\|\cdot\|_X)$ be a Banach space, $U\subset X$ be open, $f:U\to X$ be $C^{1,1}$, and $B_{r_0}(x_0)\subset U$ be an open ball. Suppose that $\|f\|_{C^{1,1}(B_{r_0}(x_0))}\le C$ and that $f'(x_0)$ is invertible with $\|f'(x_0)v\|_X\ge c_0\|v\|_X$ for every $v\in X$. Then there exists an open set $B_{r_{0}'}(x_0)\subset\mathcal O\subset B_{r_0}(x_0)$ such that $f:\mathcal O\to B_{\rho_0}(y_0)$ is a $C^{1,1}$ diffeomorphism, where $y_0\doteq f(x_0)$. Moreover, if $w$ denotes the local inverse of $f$ on $B_{\rho_0}(y_0)$, then $\|w\|_{C^{1,1}(B_{\rho_0}(y_0))}\le C'$.
\end{thm}

\begin{rk}
The hypotheses that $f\in C^{1,1}$ (with an explicit bound on its $C^{1,1}$ norm) is only required to quantify the size of the ball on which the inverse $w$ exists. It can be replaced with other assumptions on the modulus of continuity of $f'$, but this is the most convenient formulation for our purposes. 
\end{rk}

We will apply this in the form of the implicit function theorem, whose setup and proof we now briefly recall. Let $U\subset Z\doteq X\times\Bbb R_z$ be open and $f:U\to \Bbb R$ be $C^{1,1}$. The augmented map $ \check f: U \to  Z,(x,z)\mapsto (x,f(x,z))$ is clearly a $C^{1,1}$ mapping as well. 
The derivative of $\check f$ is given by
\begin{equation*}
\check f' = \begin{pmatrix}
\id_X & 0 \\
\partial_xf & \partial_zf
\end{pmatrix}
\end{equation*}
and hence $\check f'$ is invertible if $\partial_zf\ne 0$. Let $W$ be the local inverse of $\check f$ given by the inverse function theorem. Then $W$ is the augmentation of a $C^{1,1}$ function $w:B\to\Bbb R$, for a ball $B\subset Z$.\footnote{\emph{Proof.} Split $W$ into its components along $X$ and $\Bbb R$ as $W=(W_1,W_2)$. Then $(x,z)=\check f(W(x,z))=(W_1(x,z),f(W_2(x,z)))$. By inspection, $W_1(x,z)=x$, so that $W=\check W_2$.} We call this function $w$ an \emph{implicit function} since it satisfies the relation $f(w(x,z))=z$ for every $(x,z)\in B$. The function $w$ inherits the estimate of its $C^{1,1}$ norm from that of $W$ given by the inverse function theorem. 

\begin{rk}\label{rk:local-uniqueness} Since $\check f$ is a diffeomorphism on the domain $\mathcal O$ given by \cref{thm:inverse-FT}, $W=\check w$ (and hence $w$) will clearly be uniquely defined on $B=\check f(\mathcal O)$. \end{rk}
  
\subsubsection{The proof of \texorpdfstring{\cref{prop:Wi}}{Proposition 10.1}}\label{sec:proof-Wi}

We begin the proof by fixing some crucial $i$-dependent constants using the theory of \cref{sec:semiglobal-1}. In order to fix the scale in the lemma, we let $\tilde\ve_0>0$ denote the smallness parameter for which all of the arguments of \cref{sec:K-2} apply.

\begin{lem}\label{lem:i-dependent-constants} For every $i\ge 1$, there exist positive constants $B(i)$, $\qd B(i)$, $\eta_0(i)$,  and $C(i)$, which depend also on $M_0$ and $\delta$, with the following properties:
 \begin{enumerate}
     \item $\|\mathcal S[\Psi]\|_{\mathcal R(1,\u{\mathfrak u}(L_i),1,\infty)}\le B(i)$ for every $\Psi\in \tilde{\mathcal K} (\tilde\ve_0,M,\varrho,L_i)$ with admissible $(M,\varrho,L_i)$.
     
     \item If $\Psi\in\mathcal K(\frac 12\tilde\ve_0,M,\varrho,L_i)$ for admissible $(M,\varrho,L_i)$, then $B^\mathfrak{Z}_{\eta_0(i)}(\Psi)\subset \tilde{\mathcal K}(\tilde\ve_0,M,\varrho,L_i)$.
     
         \item Let $\Psi$ be as in point 2. Then $\mathscr M_i,\mathscr P_i\in C^{1,1}(B^\mathfrak{Z}_{\eta_0(i)}(\Psi))$ and $\|(\mathscr M_i,\mathscr P_i)\|_{C^{1,1}(B^\mathfrak{Z}_{\eta_0(i)}(\Psi))}\le C(i)$.
     
     \item Let $\Psi$ be as in point  2., $\Psi'\in  B^\mathfrak{Z}_{\eta_0(i)}(\Psi)$, and $\lin \Psi_1,\lin\Psi_2\in \mathfrak Z$ with $\|\lin \Psi_1\|_\mathfrak{Z},\|\lin\Psi_2\|_\mathfrak{Z}\le 1$. Then for the $L_i$-teleologically normalized second-order perturbation  $\qd{\u{\mathcal S}}{}_{L_i}[\Psi',\lin\Psi_1,\lin\Psi_2]$, it holds that {\mathtoolsset{showonlyrefs=false}
 \begin{align}
   \label{eq:2nd-1}     |\qd{\u r}{}_{L_i}|& \le \qd B(i){\log\u v} ,\\
       \label{eq:2nd-2}      |\qd{\u \lambda}{}_{L_i}|& \le \qd B(i)\u v^{-1} ,\\
       \label{eq:2nd-3}      |\qd{\u \nu}{}_{L_i}|& \le \qd B(i) \u v^{-1}.
    \end{align}}
 \end{enumerate}
\end{lem}
\begin{proof} Point 1.~follows from \cref{lem:K-AF-est}. Point 2.~follows from a continuity argument using \cref{prop:teleological-1} as in \cref{sec:K-2}. Point 3.~follows from \cref{thm:C1-finite-time} and \cref{cor:P-Lipschitz}. Finally, point 4.~follows from \cref{prop:second-order-teleology}.
\end{proof}

 We also require the following lemma which relates the estimates for $\mathscr P_i$ in \cref{prop:Pi-est} to $\mathscr W_i$.

\begin{lem}   Under the assumptions of \cref{prop:lin-boot-improvement}, the following estimates hold:
       \begin{align}
 \label{eq:Wi-est-0}   \sup_{\cyl(\ve,1+\alpha L_i^{-2})}|\mathscr W_i|&\le 2   ,\\
   \label{eq:Wi-est-1'}   \sup_{\cyl(\ve,1+\alpha L_i^{-2})} \|\mathscr W_i'\|_{\mathfrak Z^*} &\les \lin A,\\
  \label{eq:Wi-est-2'}  \sup_{\cyl(\ve,1+\alpha L_i^{-2})}|\partial_\sigma\mathscr W_i-1|  &\les \lin A \ve.
    \end{align}
\end{lem}
\begin{proof} Implicit differentiation of \eqref{eq:Pi-inversion} gives
    \begin{equation}\label{eq:Wi'-formula}
        \mathscr W_i'(\Psi) = \big(\partial_{\rho_0}\mathscr P_i(\Psi)\big)^{-1}({-}\partial_x\mathscr P_i(\Psi),1).
    \end{equation}
Now \eqref{eq:Wi-est-1'} and \eqref{eq:Wi-est-2'} follow from this and \eqref{eq:Pi-est-1} and \eqref{eq:Pi-est-3}. Finally, \eqref{eq:Wi-est-0} follows from the normalization condition $\mathscr W_i(x_0,\sigma)=\sigma$ (recall \cref{rk:W-normalization}) and integrating the bound \eqref{eq:Wi-est-1'} for $\ve_0$ small depending on $\lin A$.
\end{proof}

\begin{proof}[Proof of \texorpdfstring{\cref{prop:Wi}}{Proposition 10.1}]
We aim to show that for $\lin A$ sufficiently large that  and $\ve_0\le\tilde\ve_0$ sufficiently small,  $\mathfrak E(i,\lin A)$ contains the interval $(0,\ve_0]$ for every $i\ge 1$. 
    
    \textsc{Proof that $\mathfrak E(i,\lin A)$ is nonempty, \hyperref[E-prop-1]{condition 1.}}: Let $\mathfrak L_i\doteq\{x_0\}\times [-1-\alpha L_i^{-2},1+\alpha L_i^{-2}]$. If $\Psi\in\mathfrak L_i$, then $\Psi\in \mathcal K(0,M_0,\rho_\circ,L_i)$ and hence by \cref{lem:i-dependent-constants}, $\mathscr P_i$ is $C^{1,1}$ on $\tilde{\mathfrak L}_i\doteq\bigcup_{\Psi\in\mathfrak L_i} B^\mathfrak{Z}_{\eta_0(i)}(\Psi)$ with finite $C^{1,1}$ norm. Moreover, as was observed in \cref{rk:W-normalization}, $\partial_{\rho_\circ}\mathscr P_i=1$ on $\mathfrak L_i$. Therefore, by the inverse function theorem, there exists a number $\eta_1(i)>0$ such that for every $\Upsilon\in\mathfrak L_i$ there exists an open set $\mathcal O_\Upsilon\subset\tilde{\mathfrak L}_i$ such that $\check{\mathscr P}_i:\mathcal O_\Upsilon\to B^\mathfrak{Z}_{\eta_1(i)}(\Upsilon)$ is a diffeomorphism. 
    
    Let $\mathscr W_\Upsilon:B^\mathfrak{Z}_{\eta_1(i)}(\Upsilon)\to\Bbb R$ be the implicit function associated to $\mathscr P_i$ on ${\mathcal O_\Upsilon}$. We claim that these functions glue together on $\bigcup_{\Upsilon\in\mathfrak L_i}B_{\eta_1(i)}^\mathfrak{Z}(\Upsilon)$ to give an implicit function for $\mathscr P_i$ on $\bigcup_{\Upsilon\in \mathfrak L_i}\mathcal O_\Upsilon$. By a standard argument, it suffices to check that if $\Upsilon_1,\Upsilon_2\in \mathfrak L_i$, then $\mathscr W_{\Upsilon_1}=\mathscr W_{\Upsilon_2}$ on $B_{\eta_1(i)}^\mathfrak{Z}(\Upsilon_1)\cap B_{\eta_1(i)}^\mathfrak{Z}(\Upsilon_2)$. Suppose that these two balls intersect (otherwise there is nothing to prove) and note that, as was observed above, they agree on $\mathfrak L_i\cap B_{\eta_1(i)}^\mathfrak{Z}(\Upsilon_1)\cap B_{\eta_1(i)}^\mathfrak{Z}(\Upsilon_2)$. Now let $\Upsilon'\in B_{\eta_1(i)}^\mathfrak{Z}(\Upsilon_1)\cap B_{\eta_1(i)}^\mathfrak{Z}(\Upsilon_2)$, which is convex, and let $z\mapsto \Upsilon(z)$, $z\in [0,1]$, be a line segment with $\Upsilon(0)\in  \mathfrak L_i\cap B_{\eta_1(i)}^\mathfrak{Z}(\Upsilon_1)\cap B_{\eta_1(i)}^\mathfrak{Z}(\Upsilon_2)$ and $\Upsilon(1)=\Upsilon'$. By a simple continuity argument, using \cref{rk:local-uniqueness}, we see that $\mathscr W_{\Upsilon_1}(\Upsilon(z))=\mathscr W_{\Upsilon_2}(\Upsilon(z))$ for every $z\in [0,1]$. This proves the claim. We now define $\mathscr W_i$ on $\bigcup_{\Upsilon\in\mathfrak L_i}B_{\eta_1(i)}^\mathfrak{Z}(\Upsilon)$ to be this glued function.
    
By construction, $\mathscr W_i$ is an implicit function for $\mathscr P_i$. Hence, if $(x,\sigma)\in \bigcup_{\Upsilon\in\mathfrak L_i}B_{\eta_1(i)}^\mathfrak{Z}(\Upsilon)$ and $\|x\|_\mathfrak{X}\le \ve\le \tilde\ve_0$, then $\Psi=\check{\mathscr W}_i(x,\sigma)\in \mathcal K(\ve,M,\sigma,L_i)$, after using part 2.~of \cref{lem:i-dependent-constants}  and applying \cref{prop:modulation}. Since $\cyl(\frac 12 \eta_1(i),1+\alpha L_i^{-2})\subset  \bigcup_{\Upsilon\in\mathfrak L_i}B_{\eta_1(i)}^\mathfrak{Z}(\Upsilon)$ by definition, this proves that \hyperref[E-prop-1]{condition 1.}~of \cref{def:mathfrak-E-new} is satisfied for $\ve\in (0,\frac 12\eta_1(i)]$ after possibly shrinking $\eta_1(i)$ to ensure that $\frac 12\eta_1(i)\le \tilde\ve_0$. 

\textsc{Proof that $\mathfrak E(i,\lin A)$ is nonempty, \hyperref[E-prop-2]{condition 2.}}: Let $\Psi\in\mathfrak L_i$ and let $\lin\Psi\in\mathfrak Z$ with $\|\lin\Psi\|_\mathfrak{Z}\le 1$. By \cref{prop:linear-RN} and its proof, $\lin{\u r}{}_{L_i}=\lin r_\diamond$ (which was defined in \cref{sec:lin-r-diamond}) and hence 
\[\lin{\u r}{}_{L_i} = \lin r_\diamond(L_i,L_i)=\lin r_\diamond(1,1)=\lin r_\circ,\]
which shows that \eqref{eq:lin-r-bootstrap} holds with constant $1$. By the same argument, $\lin{\mathscr M}_i=\lin\varpi_\circ$ and $\lin{\mathscr P}_i=\lin\rho_\circ$, so that \eqref{eq:lin-M-bootstrap} and \eqref{eq:lin-P-bootstrap} hold with constants $1$ as well. By parts 3.~and 4.~of \cref{lem:i-dependent-constants}, we see that \eqref{eq:lin-r-bootstrap}--\eqref{eq:lin-P-bootstrap} then hold for $\Psi\in\tilde{\mathfrak L}_i$ after possibly shrinking $\eta_0(i)$. (Note that making $\eta_0(i)$ smaller also possibly shrinks $\eta_1(i)$, but it remains positive.) Since $\check{\mathscr W}_i$ maps $\cyl(\frac 12\eta_1(i),1+\alpha L_i^{-2})$ into $\tilde{\mathfrak L}_i$, we have verified \hyperref[E-prop-2]{condition 2.}~for $\ve\in(0,\frac 12\eta_1(i)]$.

\textsc{Proof that $\mathfrak E(i,\lin A)$ is nonempty, \hyperref[E-prop-3]{condition 3.}}: Let $\tilde C$ denote the implicit constant in the inequalities \eqref{eq:lin-r-diamond-10} and \eqref{eq:lin-nu-diamond-10}. As above, for $\Psi\in\mathfrak L_i$, $\lin{\u r}{}_{L_i}=\lin r_\diamond$, $\lin{\u \nu}{}_{L_i}=\lin\nu_\diamond$, and $\lin{\u\lambda}{}_{L_i}=\lin\lambda_\diamond$. By choosing $\lin A\ge2\tilde C$, the estimates \eqref{eq:lin-boot-1-new}--\eqref{eq:lin-boot-6-new} hold with constant $\frac 12 \lin A$. We now argue that they hold with constant $\lin A$ on $\tilde{\mathfrak L}_i$. 

We consider the variation of $\lin{\mathcal S}{}_{L_i}[\Psi,\lin\Psi]$ along $\lin \Psi'\in\mathfrak Z$ and use part 4.~of \cref{lem:i-dependent-constants} to estimate it. Note that by the discussion in \cref{sec:lin-r-diamond}, $\qd r_\diamond$, the variation of $\lin r_\diamond$ along $\lin\Psi'$, is given by the $r$-component of $\qd{\u{\mathcal S}}{}_{L_i}[\Psi_\diamond,\lin\Psi_\diamond,\lin\Psi']$, which can also be estimated by \eqref{eq:2nd-3}. Therefore, by integrating \eqref{eq:2nd-1}--\eqref{eq:2nd-3} in the $z_2$ variable and choosing $\eta_0(i)$ perhaps smaller depending on $\tilde C$, $\qd B(i)$, and $i$ (note that $D'^{-1}$ and $D$ are bounded below by positive constants depending on $i$ in the near region), we see that \eqref{eq:lin-boot-1-new}--\eqref{eq:lin-boot-6-new} hold on $\tilde{\mathfrak L}_i$.
   
   This proves that \hyperref[E-prop-3]{condition 3.} is satisfied and hence  $(0,\frac 12\eta_1(i)]\subset \mathfrak E(\lin A,\ve_0)$ for $\lin A\ge 2\tilde C$.

\textsc{Proof that $\mathfrak E(i,\lin A)\supset (0,\ve_0]$, \hyperref[E-prop-1]{condition 1.}}: We now show that for $\lin A$ sufficiently large, $\ve_0$ sufficiently small, and by possibly shrinking $\eta_1(i)$, it holds that if $\ve\in \mathfrak E(i,\lin A)$ and $\ve\le \ve_0$, then $\ve+\frac 12 \eta_1(i)\in \mathfrak E(i,\lin A)$. We assume that $\ve_0\le \frac 12\tilde \ve_0$.  

    Arguing as in the proof of nonemptiness, $\mathscr P_i$ is $C^{1,1}$ on the neighborhood
    \[\mathfrak B\doteq \bigcup_{\Psi\in \check{\mathscr W}_i(\cyl(\ve,1+\alpha L_i^{-2}))}B^\mathfrak{Z}_{\eta_0(i)}(\Psi)\]
    and has bounded $C^{1,1}$ norm. Using now \eqref{eq:Pi-est-3}, we obtain that $\partial_{\rho_\circ}\mathscr P_i\ge \frac 12$ on $\mathfrak B$ for $\ve_0$ sufficiently small depending on $\lin A$. Using the inverse function theorem, we now obtain a new\footnote{We get a ``new'' radius since the lower bound is only $\frac 12$ and not $1$ away from $x=x_0$.} $\eta_1(i)>0$ such that for every $\Upsilon\in \cyl(\ve,1+\alpha L_i^{-2})$, there exists an open set $\mathcal O_\Upsilon\subset\mathfrak B$ such that $\check{\mathscr P}_i:\mathcal O_\Upsilon\to B^\mathfrak{Z}_{\eta_1(i)}(\Upsilon)$ is a diffeomorphism. Using the convexity of intersections of balls again, we see that these local implicit functions extend $\mathscr W_i$ to $\cyl(\ve+\frac 12\eta_1(i),1+\alpha L_i^{-2})$. This verifies \hyperref[E-prop-1]{condition 1.}~for $\ve+\frac 12\eta_1(i)$.

\textsc{Proof that $\mathfrak E(i,\lin A)\supset (0,\ve_0]$, \hyperref[E-prop-2]{conditions 2.}~and \hyperref[E-prop-2]{3.}}: Take $\ve_0$ and $\lin A$ such that the conclusion of \cref{prop:lin-boot-improvement} holds. Then, by integrating \eqref{eq:2nd-1}--\eqref{eq:2nd-3} in $z_2$ again, we obtain \eqref{eq:lin-r-bootstrap}--\eqref{eq:lin-boot-6-new} on $\mathfrak B$ and hence on $\check{\mathscr W}_i(\cyl(\ve+\frac 12\eta_1(i),1+\alpha L_i^{-2}))$. This concludes the proof that $\ve+\frac 12\eta_1(i)\in \mathfrak E(i,\lin A)$ and hence the proof of the proposition. 
\end{proof}

  \subsection{Construction of the eschatological foliation}\label{sec:eschatological-foliation}
  
  At this point, we fix $\lin A$ (depending only on $M_0$ and $\delta$) sufficiently large that \cref{prop:Pi-est,prop:lin-boot-improvement} hold and $\ve_0$ sufficiently small that 
  \begin{equation*}
   \tfrac 12 \le \partial_{\rho_\circ}\mathscr P_i\le 2
  \end{equation*} on $\check{\mathscr W}_i(\cyl(\ve_0,1+\alpha L_i^{-2}))$ for every $i\ge 1$. We now allow the implicit constant in $\les$ to depend on $\lin A$. 

\begin{prop}\label{prop:eschatological-foliation}
       As $i\to\infty$, the functions  $\mathscr W_i$ given by \cref{prop:Wi}, when restricted to $\cyl(\ve_0,1)$, converge uniformly to a $C^1$ function\footnote{We emphasize that the convergence is only proved to be in $C^0$, but that the limiting function is indeed $C^1$.}\index{W@$\mathscr W_\infty$, implicit function for $\mathscr P_\infty$ whose graph defines the stable manifolds} 
     \begin{equation*}
         \mathscr W_\infty:\cyl(\ve_0,1)\to \Bbb R
     \end{equation*}
      satisfying the estimates
    \begin{align}
   \label{eq:W-infinity-est-1}    \|\mathscr W_\infty\|_{C^1(\cyl(\ve_0,1))} &\les 1,\\
  \label{eq:W-infinity-est-2}  \sup_{\cyl(\ve_0,1)}|\partial_\sigma\mathscr W_\infty-1|  &\les \ve_0.
    \end{align}
    Moreover, $\check{\mathscr W}_\infty:\cyl(\ve_0,1)\to \mathfrak M_\mathrm{black}$ and if $(x,\sigma)\in \cyl(\ve,1)$ for $\ve\le\ve_0$, then 
 \begin{equation}\label{eq:Winfty-defn}
     (x,\mathscr W_\infty(x,\sigma)) \in \mathcal K(\ve,M,\sigma,\infty).
 \end{equation} 
\end{prop}

By the characterization of $\mathcal K(\ve,M,\sigma,\infty)$ given by \cref{thm:uniform-stability-RN}, the image of $B_{\ve_0}^\mathfrak{X}(x_0)$ under $x\mapsto \mathscr W_\infty(x,\sigma)$ is indeed a stable manifold for Reissner--Nordstr\"om black holes with charge-to-mass ratio $\sigma$. 

\subsubsection{Convergence of \texorpdfstring{$\mathscr W_i$}{Wi}}

First, we argue that $\{\mathscr W_i\}_{i\ge 1}$ forms a Cauchy sequence in $C^0$.

\begin{lem}\label{lem:W-i-convergence} For every $1\le i\le j$, it holds that 
\begin{equation*}
    \sup_{\cyl(\ve_0,1+\alpha L_j^{-2})}|\mathscr W_i-\mathscr W_j|\les \ve^2_0L_i^{-3+\delta}. 
\end{equation*} 
\end{lem}
\begin{proof}
     Since $i\le j$, the compositions $\check{\mathscr W}_i\circ\check{\mathscr P}_j\circ\check{\mathscr W}_j$ and $\check{\mathscr W}_i\circ\check{\mathscr P}_i\circ\check{\mathscr W}_j$ are defined on $\cyl(\ve_0,1+\alpha L_j^{-2})$. The first composition is easy to understand since $\check{\mathscr P}_j\circ\check{\mathscr W}_j$ is the identity on $\cyl(\ve_0,1+\alpha L_j^{-2})$ and clearly $\cyl(\ve_0,1+\alpha L_j^{-2})\subset \cyl(\ve_0,1+\alpha L_i^{-2})=\dom(\mathscr W_i)$. The second composition is slightly more subtle: the image of $\cyl(\ve_0,1+\alpha L_j^{-2})$ under $\check{\mathscr W}_j$ consists of $\Psi$'s with $|\mathscr P_j(\Psi)|\le 1+\alpha L_j^{-2}$. By \eqref{eq:P-monotonicity}, $|\mathscr P_i(\Psi)|\le 1+\alpha L_j^{-2}\le 1+\alpha L_i^{-2}$. 
     It follows that $\check {\mathscr P}_i\circ \check{\mathscr W}_j:\cyl(\ve_0,1+\alpha L_j^{-2})\to \cyl(\ve_0,1+\alpha L_i^{-2})=\dom (\mathscr W_i)$. 
     
     For $\Upsilon\in \cyl(\ve_0,1+\alpha L_j^{-2})$, we may then estimate
    \begin{align*}
    |\mathscr W_i(\Upsilon)-\mathscr W_j(\Upsilon)|  &=\| \check{\mathscr W}_i(\Upsilon)-\check{\mathscr W}_j(\Upsilon) \|_\mathfrak{Z}\\ &=\|\check{\mathscr W}_i(\check{\mathscr P}_j(\check{\mathscr W}_j(\Upsilon)))-\check{\mathscr W}_i(\check{\mathscr P}_i(\check{\mathscr W}_j(\Upsilon)))\|_\mathfrak{Z}\\
        &\les \|\check{\mathscr P}_j(\check{\mathscr W}_j(\Upsilon))-\check{\mathscr P}_i(\check{\mathscr W}_j(\Upsilon))\|_\mathfrak{Z}\\
        &\les \ve_0^2 L_i^{-3+\delta},
    \end{align*}
    where in the second to last step we used \eqref{eq:Wi-est-1} and the mean value theorem, and in the last step we used the decay estimate \eqref{eq:P-change}.
\end{proof}

Let
\begin{equation*}
    \mathscr W_\infty\doteq \lim_{i\to\infty}\mathscr W_i\quad\text{on}\quad \cyl(\ve_0,1),
\end{equation*} which exists by \cref{lem:W-i-convergence}. 
It is elementary that $\mathscr W_\infty$ is a continuous function. Next, we show that the functions $\mathscr W_i$ are monotone in $i$. 

\begin{lem}\label{lem:Wi-monotonicity}
    If $\Upsilon\in \cyl(\ve_0,1+\alpha L_{j}^{-2})$ for $j\in \Bbb N\cup\{\infty\}$ and $i\le j$, then 
    \begin{equation}\label{eq:W-monotonicity}
        |\mathscr W_{j}(\Upsilon)|\le |\mathscr W_i(\Upsilon)|.
    \end{equation}
\end{lem}
\begin{proof} We prove \eqref{eq:W-monotonicity} for $j=i+1$, the general case then follows by iteration. 

    Let $\Upsilon=(x,\sigma)\in\cyl(\ve_0,1+\alpha L_{i+1}^{-2}) $ and suppose that $|\mathscr W_{i+1}(\Upsilon)|> |\mathscr W_i(\Upsilon)|$. Note that $\mathscr W_i(\Upsilon)$ and $\mathscr W_{i+1}(\Upsilon)$ have the same sign, which we may assume to be positive without loss of generality. Since $\mathscr W_{i+1}(x,0)=0$ and $\sigma\mapsto \mathscr W_{i+1}(x,\sigma)$ is strictly monotone by \eqref{eq:Wi-est-2}, by the intermediate value theorem there exists a $\sigma'\in(0,\sigma)$ such that $\mathscr W_{i+1}(\Upsilon')=\mathscr W_i(\Upsilon)$, where $\Upsilon'\doteq(x,\sigma')$. We then have
    \begin{equation*}
        \sigma' = \mathscr P_{i+1}(\check{\mathscr W}_{i+1}(\Upsilon'))\ge \mathscr P_{i}(\check{\mathscr W}_{i+1}(\Upsilon')) = \mathscr P_i(\check{\mathscr W}_i(\Upsilon)) =\sigma,
    \end{equation*}
    where we used \eqref{eq:P-monotonicity}, which gives a contradiction. Therefore, \eqref{eq:W-monotonicity} holds.
\end{proof}

The estimate \eqref{eq:W-monotonicity} implies 
\begin{equation}\label{eq:domain-monotonicity}
   \check{\mathscr W}_j(\cyl(\ve_0,1+\alpha L_j^{-2})) \subset\check{\mathscr W}_i(\cyl(\ve_0,1+\alpha L_j^{-2})) 
\end{equation} for $i\le j$, which will be quite useful. Finally, we characterize the image of $\check{\mathscr W}_\infty$. 

\begin{lem}\label{lem:domain-of-W-infty} $
    \check{\mathscr W}_\infty:\cyl(\ve_0,1)\to \mathfrak M_\mathrm{black}$ and if $(x,\sigma)\in \cyl(\ve,1)$ for $\ve\le \ve_0$, then \eqref{eq:Winfty-defn} holds.
\end{lem}
\begin{proof} Let $\Upsilon\in \cyl(\ve,1)$, where $\ve\le \ve_0$. For $i\in\Bbb N\cup\{\infty\}$, set $\Psi_i=\check{\mathscr W}_i(\Upsilon)$. Note that $\|\Psi_\infty-\Psi_i\|_\mathfrak{Z}\les \ve^2L_i^{-3+\delta}$ by \cref{lem:W-i-convergence}, so $\mathcal L_\mathrm{Bondi}(\Psi_\infty)=\infty$ by \cref{lem:Mi-open-quantitative}. Hence, $\Psi_\infty\in\mathfrak M_\mathrm{black}$ by the fundamental dichotomy \cref{thm:dichotomy}. In fact, by the refined dichotomy \cref{thm:dichotomy-revisited}, $\Psi_\infty\in \mathcal K(\ve,M_\infty,\varrho_\infty,\infty)$ for some $|\varrho_\infty|\le 1$. We now argue that $\varrho_\infty\doteq\mathscr P_\infty(\Psi_\infty)=\sigma$. 

We may assume that $\sigma>0$ without loss of generality. By \cref{lem:Wi-monotonicity}, $0<\mathscr W_\infty(\Upsilon)\le \mathscr W_i(\Upsilon)$, and therefore $
      \{x\}\times[\mathscr W_\infty(\Upsilon),\mathscr W_i(\Upsilon)]\subset \check{\mathscr W}_i(\cyl(\ve,1)) $. We may now use \eqref{eq:Pi-est-3} to estimate
    \begin{equation*}
        |\mathscr P_i(\check{\mathscr W}_\infty(\Upsilon))-\mathscr P_i(\check{\mathscr W}_i(\Upsilon))|\les|\mathscr W_\infty(\Upsilon)-\mathscr W_i(\Upsilon)|\les \ve^2L_i^{-3+\delta}.
    \end{equation*}
    Since $\mathscr P_i(\check{\mathscr W}_i(\Upsilon))=\sigma$, this shows that 
    \begin{equation*}
        \mathscr P_\infty(\check{\mathscr W}_\infty(\Upsilon))\doteq \lim_{i\to \infty} \mathscr P_i(\check{\mathscr W}_\infty(\Upsilon)) = \sigma
    \end{equation*}
as desired. 
\end{proof}

\subsubsection{Regularity of \texorpdfstring{$\mathscr W_\infty$}{W infinity}}

First, we identify the candidate for the derivative of $\mathscr P_\infty$. Immediately from \eqref{eq:Pi-est-2} and the fact that $\mathfrak Z^*$ is a Banach space, we have:

\begin{lem}
    For any $\Psi\in \check {\mathscr W}_\infty(\cyl(\ve_0,1))$, the limit
    \begin{equation*}
        \delta \mathscr P_\infty(\Psi)\doteq \lim_{i\to\infty} \mathscr P'_i(\Psi),
    \end{equation*}
    where $\mathscr P'_i(\Psi)$ denotes the Fr\'echet derivative of $\mathscr P_i$ at $\Psi$, exists in the operator norm topology on $\mathfrak Z^*$. In fact, it holds that
    \begin{equation}\label{eq:Pi'-convergence}
       \sup_{\Psi\in\cyl(\ve_0,1)} \|\delta\mathscr P_\infty(\Psi)-\mathscr P_i'(\Psi)\|_{\mathfrak Z^*}\les \ve_0^2L_i^{-1+\delta/2}.
    \end{equation}
\end{lem}

Recall the splitting $\mathfrak Z=\mathfrak X\times\Bbb R$. Given $\lin x\in \mathfrak X$, it is convenient to define  $\delta_x\mathscr P_\infty(\Psi)\in \mathfrak X^*$ by
\begin{equation*}
    \delta_x\mathscr P_\infty(\Psi)\lin x\doteq \delta\mathscr P_\infty(\Psi)(\lin x,0),
\end{equation*}
where we think of $(\lin x,0)\in \mathfrak Z$. Also, we define
\begin{equation*}
    \delta_{\rho_0}\mathscr P_\infty(\Psi)\doteq \delta\mathscr P_\infty(\Psi)(0,1),
\end{equation*}
where $0\in\mathfrak X$. Finally, we set 
\begin{align*}
  \mathscr Q_i(\Psi)   &\doteq \big(\partial_{\rho_0}\mathscr P_i(\Psi)\big)^{-1}\big({-} \partial_x\mathscr P_i(\Psi),1\big),\\
      \mathscr Q_\infty(\Psi)&\doteq \big(\delta_{\rho_0}\mathscr P_\infty(\Psi)\big)^{-1}\big({-} \delta_x\mathscr P_\infty(\Psi),1\big),
\end{align*}
which are of course all elements of $\mathfrak Z^*$. Note that $\mathscr Q_i$, for $i\in\Bbb N\cup\{\infty\}$, is defined on $\cyl(\ve_0,1+\alpha L_i^{-2})$.  

\begin{lem}\label{lem:Ai-properties} The functions $\mathscr Q_i$ have the following properties:
\begin{enumerate}
    \item For every $i\in \Bbb N$, $\Psi\mapsto \mathscr Q_i(\Psi)$ is a continuous map $\check{\mathscr W}_i(\cyl(\ve_0,1+\alpha L_i^{-2}))\to \mathfrak Z^*$.

   \item For every $i\in\Bbb N$, the Fr\'echet derivative of $\mathscr W_i$ on $\cyl(\ve_0,1+\alpha L_i^{-2})$ equals $\mathscr Q_i\circ \mathscr W_i$.

   \item For every $i\le j$,
    \begin{equation}\label{eq:Aj-convergence}
        \sup_{\Psi\in \check{\mathscr W}_\infty(\cyl(\ve_0,1+\alpha L_j^{-2}))}\|\mathscr Q_j(\Psi)-\mathscr Q_i(\Psi)\|_{\mathfrak Z^*}\les \ve^2_0 L_i^{-1+\delta/2}.
    \end{equation}

    \item  For every $\Upsilon\in\cyl(\ve_0,1)$, $\mathscr Q_j(\check{\mathscr W}_j(\Upsilon))\to\mathscr Q_\infty(\check{\mathscr W}_\infty(\Upsilon))$ in $\mathfrak Z^*$ as $j\to\infty$.

    \item The function $\mathscr Q_\infty\circ\check{\mathscr W}_\infty:\cyl(\ve_0,1)\to \mathfrak Z^*$ is continuous.
\end{enumerate}
\end{lem}
\begin{proof} Part 1.~follows from the fact that $\mathscr P_i$ is $C^1$ on a neighborhood of $\cyl(\ve_0,1+\alpha L_i^{-2})$. Part 2.~follows from implicit differentiation of the relation \eqref{eq:Pi-inversion}. Part 3.~follows from \eqref{eq:Pi'-convergence}.
  
    \textsc{Proof of part 4.}: Let $i\le j$ and take $\Upsilon\in\cyl(\ve_0,1)$. We estimate 
\begin{multline*}
       \|\mathscr Q_\infty(\check{\mathscr W}_\infty(\Upsilon))-\mathscr Q_j(\check{\mathscr W}_j(\Upsilon))\|_{\mathfrak Z^*}\le \|\mathscr Q_\infty(\check{\mathscr W}_\infty(\Upsilon))-\mathscr Q_i(\check{\mathscr W}_\infty(\Upsilon))\|_{\mathfrak Z^*} \\+\|\mathscr Q_i(\check{\mathscr W}_\infty(\Upsilon))-\mathscr Q_i(\check{\mathscr W}_j(\Upsilon))\|_{\mathfrak Z^*}+\|\mathscr Q_i(\check{\mathscr W}_j(\Upsilon))-\mathscr Q_j(\check{\mathscr W}_j(\Upsilon))\|_{\mathfrak Z^*},
    \end{multline*}
    where we note that the compositions are well-defined by \eqref{eq:domain-monotonicity}. For any $\eta>0$, the first and third terms are $\le \eta$ for $i$ sufficiently large by \eqref{eq:Aj-convergence}. For $i$ now fixed, we can take $j$ sufficiently large so that the middle term is $\le \eta$ by definition of $\mathscr W_\infty$ and part 1.~of the lemma. 

    \textsc{Proof of part 5.}: Let $\Upsilon_0,\Upsilon\in \cyl(\ve_0,1)$. We estimate
    \begin{multline*}
        \|\mathscr Q_\infty(\check{\mathscr W}_\infty(\Upsilon))-\mathscr Q_\infty(\check{\mathscr W}_\infty(\Upsilon_0))\|_{\mathfrak Z^*} \le \|\mathscr Q_\infty(\check{\mathscr W}_\infty(\Upsilon))-\mathscr Q_i(\check{\mathscr W}_\infty(\Upsilon))\|_{\mathfrak Z^*}\\+\|\mathscr Q_i(\check{\mathscr W}_\infty(\Upsilon))-\mathscr Q_i(\check{\mathscr W}_\infty(\Upsilon_0))\|_{\mathfrak Z^*}+\|\mathscr Q_i(\check{\mathscr W}_\infty(\Upsilon_0))-\mathscr Q_\infty(\check{\mathscr W}_\infty(\Upsilon_0))\|_{\mathfrak Z^*}.
    \end{multline*}
    Again, the first and third terms can be made arbitrarily small for $i$ large using \eqref{eq:Aj-convergence} and then the middle term can be made arbitrarily small for $\|\Upsilon-\Upsilon_0\|_\mathfrak{Z}$ small by using the continuity of $\mathscr W_\infty$ and part 1.~of the lemma. This shows that $\mathscr Q_\infty\circ \check{\mathscr W}_\infty$ is continuous at $\Upsilon_0$.
\end{proof}

We can finally show that $\mathscr W_\infty$ is $C^1$ in the sense that it can be extended to a neighborhood of $\cyl(\ve_0,1)$ in a $C^1$ fashion.

      \begin{lem}\label{lem:EW}
For $i\in\Bbb N\cup\{\infty\}$, define $E\mathscr W_i:\mathring\cyl(\ve_0,3)\to\Bbb R$ by
 \begin{equation*}
          E\mathscr W_i(x,\sigma)\doteq\begin{cases}
          3\mathscr W_i(x,-\sigma+2)-2\mathscr W_i(x,-2\sigma+3) & \text{if }1<\sigma<3\\
            \mathscr W_i(x,\sigma)  & \text{if }-1\le\sigma\le 1\\
            3\mathscr W_i(x,-\sigma-2)-2\mathscr W_i(x,-2\sigma-3) & \text{if }-3<\sigma<-1
          \end{cases}, 
      \end{equation*} where $\mathring\cyl(\ve_0,3)$ denotes the interior of $\cyl(\ve_0,3)$.
      Then the following holds:
\begin{enumerate}
    \item For every $i\in\Bbb N$, $E\mathscr W_i$ is a $C^1$ extension of $\mathscr W_i|_{\cyl(\ve_0,1)}$ to $\mathring\cyl(\ve_0,3)$. 
    \item $E\mathscr W_i$ converges uniformly to $E\mathscr W_\infty$ on $\mathring\cyl(\ve_0,3)$ as $i\to\infty$.
    \item $E\mathscr W_\infty$ is $C^1$ on $\mathring\cyl(\ve_0,3)$ and
    \begin{equation}\label{eq:formula-for-derivative}
              (E\mathscr W_\infty)'(\Upsilon) = \mathscr Q_\infty(\check{\mathscr W}_\infty(\Upsilon))\quad\text{for}\quad \Upsilon\in\cyl(\ve_0,1).
          \end{equation}
\end{enumerate}
      \end{lem}
\begin{proof} Part 1.~follows from the fact that $\mathscr W_i$ is $C^1$ on a neighborhood of $\cyl(\ve_0,1)$. Part 2.~follows immediately from \cref{lem:W-i-convergence}.

    \textsc{Proof of part 3.}: Let $\Upsilon_0\in \mathring\cyl(\ve_0,3)$ and take $\lin\Upsilon\in \mathfrak Z$ with $\|\lin\Upsilon\|_\mathfrak{Z}\le 1$. For $z\in\Bbb R$, set $\Upsilon(z)\doteq \Upsilon_0 + z\lin\Upsilon$. Since $E\mathscr W_i$ is $C^1$, we have
    \begin{equation}\label{eq:FTC-final-proof}
        E\mathscr W_i(\Upsilon_z)-E\mathscr W_i(\Upsilon_0) = \int_0^z (E\mathscr W_i)'(\Upsilon'(z'))\lin\Upsilon \,dz'
    \end{equation}
    for all $z$ such that $\Upsilon(z)\in \mathring\cyl(\ve_0,3)$. Writing $\lin\Upsilon=(\lin x,\lin\sigma)$, we compute
    \begin{equation*}
       (E\mathscr W_i)'(\Upsilon)\lin\Upsilon= \begin{cases}
          3\mathscr Q_i(\check{\mathscr W}_i(x,-\sigma+2))(\lin x,-\lin\sigma)-2\mathscr Q_i(\check{\mathscr W}_i(x,-2\sigma+3))(\lin x,-2\lin\sigma)  & \text{if }1<\sigma<3\\
            \mathscr Q_i(\check{\mathscr W}_i(x,\sigma))\lin\Upsilon  & \text{if }-1\le\sigma\le 1\\
            3\mathscr Q_i(\check{\mathscr W}_i(x,-\sigma-2))(\lin x,-\lin\sigma)-2\mathscr Q_i(\check{\mathscr W}_i(x,-2\sigma-3))(\lin x,-2\lin\sigma) & \text{if }-3<\sigma<-1
          \end{cases}
    \end{equation*} for $\Upsilon\in \mathring\cyl(\ve_0,3)$.
    For every $\Upsilon\in\mathring\cyl(\ve_0,3)$, we use part 4.~of \cref{lem:Ai-properties} to argue that
    \begin{multline*}
        \lim_{i\to\infty}(E\mathscr W_i)'(\Upsilon)\lin\Upsilon= \delta(E\mathscr W_\infty)(\Upsilon)\lin\Upsilon\\\doteq  \begin{cases}
          3\mathscr Q_\infty(\check{\mathscr W}_\infty(x,-\sigma+2))(\lin x,-\lin\sigma)-2\mathscr Q_\infty(\check{\mathscr W}_\infty(x,-2\sigma+3))(\lin x,-2\lin\sigma)  & \text{if }1<\sigma<3\\
            \mathscr Q_\infty(\check{\mathscr W}_\infty(x,\sigma))\lin\Upsilon  & \text{if }-1\le\sigma\le 1\\
            3\mathscr Q_\infty(\check{\mathscr W}_\infty(x,-\sigma-2))(\lin x,-\lin\sigma)-2\mathscr Q_\infty(\check{\mathscr W}_\infty(x,-2\sigma-3))(\lin x,-2\lin\sigma) & \text{if }-3<\sigma<-1
          \end{cases}.
    \end{multline*}
    Moreover, $\Upsilon\mapsto \delta(E\mathscr W_\infty)(\Upsilon) $ is continuous as a map $\mathring\cyl(\ve_0,3)\to\mathfrak Z^*$ by part 5.~of \cref{lem:Ai-properties}.

    By \eqref{eq:Wi-est-1}, $|(E\mathscr W_i)'(\Upsilon(z))\lin\Upsilon|\les 1$, and we apply the dominated convergence theorem to \eqref{eq:FTC-final-proof} to obtain
    \begin{equation}
        E\mathscr W_\infty(\Upsilon_z)-E\mathscr W_\infty(\Upsilon_0)=\int_0^z \delta(E\mathscr W_\infty)(\Upsilon'(z'))\lin\Upsilon \,dz'.
    \end{equation}
    It follows that the Gateaux derivative of $E\mathscr W_\infty$ is given by 
    \begin{equation}
      D(E\mathscr W_\infty)(\Upsilon_0,\lin\Upsilon)= \left. \frac{d}{dz}\right|_{z=0}  E\mathscr W_\infty(\Upsilon_z)=\delta(E\mathscr W_\infty)(\Upsilon_0)\lin\Upsilon.
    \end{equation}
    The claim now follows from \cref{prop:differentiability-criterion}.
\end{proof}

\begin{proof}[Proof of \cref{prop:eschatological-foliation}]
 Combine   \cref{lem:domain-of-W-infty} and part 3. of \cref{lem:EW}. 
\end{proof}

\subsection{Completing the proof of \texorpdfstring{\cref{thm:main}}{Theorem~\getrefnumber{thm:main}}}\label{sec:main-thm-proof}

\subsubsection{Failure to collapse beyond the extremal manifold}\label{sec:no-gaps}

\begin{prop}\label{prop:no-gaps} For $\ve_0>0$ sufficiently small depending only on $M_0$ and $\delta$, the following holds. If $\Psi=(x,\rho_\circ)\in \cyl(\ve_0,2)$ and $|\rho_\circ|> |\mathscr W_\infty(x,1)|$, then $\Psi\in \mathfrak M_\mathrm{non}$. In other words,
\begin{equation}\label{eq:classification-proof}
    \cyl(\ve_0,2)\setminus \check{\mathscr W}_\infty(\cyl(\ve_0,1)) = \mathfrak M_\mathrm{non}\cap \cyl(\ve_0,2).
\end{equation}
\end{prop}
\begin{proof} The proof rests on the following \emph{no gaps property} for the images of the $\mathscr W_i$'s. For every $i\in\Bbb N$ and $x\in B_{\ve_0}^\mathfrak{X}(x_0)$, it holds that
\begin{align}\label{eq:no-gaps-1}
    \mathscr W_i(x,1)&< \mathscr W_{i+1}(x,1+\alpha L_{i+1}^{-2}),\\
  \label{eq:no-gaps-2}  \mathscr W_i(x,-1)&> \mathscr W_{i+1}(x,-1-\alpha L_{i+1}^{-2}).
\end{align}
The content of these inequalities is that they go in the opposite direction of \eqref{eq:W-monotonicity}.

    \textsc{Proof of the no gaps property}: We only prove \eqref{eq:no-gaps-1}, as the proof of \eqref{eq:no-gaps-2} is similar. Let $\Psi\doteq \check{\mathscr W}_i(x,1)$ and $\Psi'\doteq \check{\mathscr W}_{i+1}(x,1+\alpha L_{i+1}^{-2})$. Then $\mathscr P_i(\Psi)=1$ and $\mathscr P_{i+1}(\Psi')=1+\alpha L_{i+1}^{-2}$. We can use dyadic iteration, \cref{prop:dyadic-iteration}, to extend $\mathcal S[\Psi]$ to Bondi time $L_{i+1}$ and measure $\mathscr P_{i+1}(\Psi)$. By \eqref{eq:P-change},
    \begin{equation*}
        \mathscr P_{i+1}(\Psi) = 1 + O(\ve^2_0L_i^{-3+\delta}),
    \end{equation*}
    so that $\mathscr P_{i+1}(\Psi)<\mathscr P_{i+1}(\Psi')$ for $\ve_0$ sufficiently small. Since $\rho_\circ\mapsto\mathscr P_{i+1}(x,\rho_\circ)$ is strictly monotone increasing, we have \eqref{eq:no-gaps-1}.

    \textsc{Completing the proof of the proposition}: The no gaps property implies that
\begin{align}
   \label{eq:no-gaps-3}  (\mathscr W_\infty(x,1),\mathscr W_1(x,1+\alpha L_1)]&= \bigcup_{i\ge 1}(\mathscr W_i(x,1),\mathscr W_i(x,1+\alpha L_i^{-2})],\\
   \label{eq:no-gaps-4}    [\mathscr W_1(x,-1-\alpha L_1),\mathscr W_\infty(x,-1))&= \bigcup_{i\ge 1}[\mathscr W_i(x,-1-\alpha L_i^{-2}),\mathscr W_i(x,1)).
\end{align}
Indeed, by \eqref{eq:W-monotonicity}, \eqref{eq:Wi-est-2}, and \eqref{eq:no-gaps-1},
\begin{equation*}
     \mathscr W_i(x,1)< \mathscr W_{i+1}(x,1+\alpha L_{i+1}^{-2})\le \mathscr W_{i}(x,1+\alpha L_{i+1}^{-2})<  \mathscr W_{i}(x,1+\alpha L_{i}^{-2}),
\end{equation*}
which implies that 
\begin{equation*}
    (\mathscr W_i(x,1),\mathscr W_i(x,1+\alpha L_i^{-2})]\cap (\mathscr W_{i+1}(x,1),\mathscr W_{i+1}(x,1+\alpha L_{i+1}^{-2})]\ne\emptyset. 
\end{equation*}
Using again \eqref{eq:W-monotonicity} and the definition of $\mathscr W_\infty$, this proves \eqref{eq:no-gaps-3}. The equality \eqref{eq:no-gaps-4} is proved similarly. 

By \eqref{eq:Wi-defn} and the fundamental dichotomy \cref{thm:dichotomy}, if $x\in B^\mathfrak{X}_{\ve_0}(x_0)$ and $\rho_\circ$ lies in either of the intervals \eqref{eq:no-gaps-3} or \eqref{eq:no-gaps-4}, then $(x,\rho_\circ)\in\mathfrak M_\mathrm{non}$.

We now argue that if $\ve_0$ is sufficiently small, $x\in B^\mathfrak{X}_{\ve_0}(x_0)$, and $\rho_\circ$ lies in either $[-1-\frac 12 \alpha L_1,\mathscr W_\infty(x,-1))$ or $(\mathscr W_\infty(x,1),1+\frac 12\alpha L_1]$, then $\rho_\circ$ lies in either \eqref{eq:no-gaps-3} or \eqref{eq:no-gaps-4}. Indeed, by \cref{rk:W-normalization} and \eqref{eq:Wi-est-1}, 
\[\mathscr W_1(x,\pm (1+\alpha L_1)) = \pm (1+\alpha L_1)+O(\ve_0)\]
on $B_{\ve_0}^\mathfrak{X}(x_0)$, which implies that these hypersurfaces are disjoint from the hyperplanes $\{\rho_\circ= \pm (1+\frac 12\alpha L_1)\}$ for $\ve_0$ sufficiently small. 

We have thus shown that for $\ve_0$ sufficiently small, and $\Psi\in \cyl(\ve_0,1+\frac 12\alpha L_1)$, either $\Psi\in \check{\mathscr W}_\infty(\cyl(\ve_0,1))$, or $\Psi\in \mathfrak M_\mathrm{non}$. By applying \cref{prop:a-priori-noncollapse} and perhaps shrinking $\ve_0$
once more, we also have that $\cyl(\ve_0,2)\setminus\cyl(\ve_0,1+\frac 12\alpha L_1)\subset\mathfrak M_\mathrm{non}$, which completes the classification \eqref{eq:classification-proof}. \end{proof}

\subsubsection{Putting everything together}\label{sec:thm-2-proof}

\begin{proof}[Proof of \cref{thm:main}]
    This is now immediate from \cref{prop:eschatological-foliation,prop:no-gaps}. 
\end{proof}

\begin{proof}[Proof of \cref{cor:transversality}] \textsc{Regularity of $\mathscr P_\infty$ and $\mathscr M_\infty$}: We work with the composed functions $\tilde{\mathscr P}_\infty\doteq \mathscr P_\infty\circ\check{\mathscr W}_\infty$ and $\tilde{\mathscr M}_\infty\doteq\mathscr M_\infty\circ\check{\mathscr W}_\infty$. Since $\check{\mathscr W}_\infty:\cyl(\ve_0,1)\to\mathfrak M_\mathrm{black}\cap\cyl(\ve_0,2)$ is a $C^1_b$ diffeomorphism, \eqref{eq:P-M-regularity} is equivalent to 
    \begin{equation*}
        \tilde{\mathscr P}_\infty,\tilde{\mathscr M}_\infty\in C^1_b\big({\cyl(\ve_0,1)}\big).
    \end{equation*}
     For $\tilde{\mathscr P}_\infty$, this is immediate since $\tilde{\mathscr P}_\infty(x,\sigma)=\sigma$. To prove it for $\tilde{\mathscr M}_\infty$, we use the approximating maps $\tilde{\mathscr M}_i\doteq \mathscr M_i\circ\check{\mathscr W}_\infty$, which uniformly converge to $\tilde{\mathscr M}_\infty$ on $\cyl(\ve_0,1)$ by our stability estimates. For $\Upsilon\in\cyl(\ve_0,1)$ and $\lin\Upsilon\in \mathfrak Z$, $\tilde{\mathscr M}_i(\Upsilon)\lin\Upsilon=\mathscr M_i'(\Psi)\lin\Psi$, where $\Psi=\check{\mathscr W}_\infty(\Upsilon)$ and $\lin\Psi=\check{\mathscr W}_\infty'(\Upsilon)\lin\Upsilon$. We can now use \cref{lem:sequences-derivatives} and \eqref{eq:Mi-est-1} to conclude that $\tilde{\mathscr M}_\infty$ is $C^1_b$.

    \textsc{Transversality criterion}: This follows from standard differential geometry, using the fact that $\mathfrak M_\mathrm{stab}^{+1,-1}$ is a regular level set of the $C^1$ function $\mathscr P_\infty$. 
\end{proof}

\section{Estimates for linear perturbations of \texorpdfstring{$\mathcal K$}{K} spacetimes}\label{sec:estimates-linear-perturbations}

The goal of this section is to prove \cref{prop:lin-boot-improvement,prop:Pi-est}. In \cref{sec:11-conventions}, we establish the assumptions and conventions for the section. In \cref{sec:linear-errors}, we estimate the inhomogeneous error terms in the energy estimates from \cref{sec:energy-1} for $\lin{\u\phi}{}_{L_i}$. In \cref{sec:linear-scalar-field}, we carry out a pigeonhole argument to close the energy estimates for $\lin{\u\phi}{}_{L_i}$. In \cref{sec:linear-geometric}, we estimate the linearized geometric quantities. Finally, in \cref{sec:completing-linear}, we prove \cref{prop:lin-boot-improvement,prop:Pi-est}.

\subsection{Assumptions and conventions for \texorpdfstring{\cref{sec:estimates-linear-perturbations}}{Section 11}}  \label{sec:11-conventions}

In this section, we reinstate the notational conventions \cref{sec:K-1,sec:energy-1,sec:scalar-field}. In particular, we again \ul{omit the underlines and normalization times} on the coordinates $(\u u,\u v)$, the teleologically normalized quantities such as $\u\phi{}_{L_i}$ and $\u \lambda{}_{L_i}$, and also the teleologically normalized linearized quantities such as $\lin{\u r}{}_{L_i}$ and $\lin{\u \gamma}{}_{L_i}$.

\begin{ass}\label{assumption-1}
    Let $0<\ve\le\ve_0$, where $\ve_0$ is such that the arguments of \cref{sec:K-2} apply. Let $i\ge 1$, $(M,\varrho,L_i)$ be an admissible parameter triple, $\lin A\ge 1$, $\Psi\in \mathcal K(\ve,M,\varrho,L_i)$, and  $\lin\Psi\in \mathfrak Z$ with $\|\lin\Psi\|_\mathfrak Z\le 1$. We assume that the $L_i$-normalized linearized solution $\lin{\u{\mathcal{S}}}{}_{L_i}[\Psi,\lin\Psi]$ satisfies the estimates \eqref{eq:lin-r-bootstrap}--\eqref{eq:lin-boot-6-new}.
    In every statement, we allow $\ve$ to be chosen small depending only on $M_0$, $\delta$, and $\lin A$. 
\end{ass}

 In order to keep track of the constant $A$ in \cref{def:K-tilde}, the notations $a\les b$, $a\gtrsim b$, and $a\sim b$ mean that the implicit constant does not (additionally) depend on $\ve$, $M$, $e$, $u_f$, $u$, $v$, $\Psi$, or $\lin A$.

We will use the following modified energies:
\begin{equation*}
    \u{\mathcal E}{}_p[\zeta](\tau,u')\doteq \int_{\u C(\tau)\cap\{u\le u'\}}\big(h_p(r_\diamond)((\partial_u \zeta)^2+(\partial_u(r\zeta))^2)+D(r_\diamond)\zeta^2\big)\,du,
\end{equation*}
where we cut off the ingoing cone at $u'\le L_i$. This is because of the singular nature of the energy estimates towards $\mathcal H^+=\{u=\infty\}$. 

We recall from \cref{sec:wave-setup} the domains $\mathcal D$ as depicted in \cref{fig:butterfly}. In particular, recall that the $u$-coordinate of the hypersurface $\mathrm{VI}$ is $u_2$. We recall
\begin{equation*}
    D_1'(u)\doteq D'(r_\diamond(u,1))
\end{equation*} from \eqref{eq:D1'-defn}.
For estimates involving a quantity on $\mathcal D$ on the left-hand side and a quantity involving $D_1'$ on the right-hand side, our convention is that $D_1'$ is always evaluated at $u_2$. 

\subsection{Inhomogeneous error terms in the energy estimates} \label{sec:linear-errors}

Using our bootstrap assumption on the geometry, we will prove energy estimates for the linearized scalar field $\lin \phi$, which satisfies
\begin{equation}\label{eq:lin-sf-eqn}
    \partial_u\partial_v\lin\phi = -\frac\lambda r\partial_u\lin\phi - \frac\nu r\partial_v\lin\phi + \lin F,
\end{equation}
on $\mathcal R_{L_i}$, where \index{F@$\lin F$, inhomogeneity in wave equation for $\lin\phi$}
\begin{equation}\label{eq:F1-def}
    \lin F\doteq -\left(\frac{\lin\lambda}{r}+\frac{\lambda\lin r}{r^2}\right)\partial_u\phi-\left(\frac{\lin\nu}{r}+\frac{\nu\lin r}{r^2}\right)\partial_v\phi.
\end{equation}
We apply the energy estimates of \cref{sec:energy-1} with $\zeta=\lin\phi$ and $F=\lin F$, so that \eqref{eq:lin-sf-eqn} is in the form \eqref{eq:inhomog-wave} with inhomogeneity given by \eqref{eq:F1-def}. We have to estimate the inhomogeneous errors (terms on the right-hand side involving $\lin F$) in \cref{prop:Morawetz,prop:hp,prop:rp}.

\subsubsection{Structure of the error terms}

We use the bootstrap assumptions \eqref{eq:lin-boot-1-new}--\eqref{eq:lin-boot-6-new} to estimate $\lin F$. First, we estimate $\lin r$ using the bootstrap assumption for $\lin r_\dagger$. 

\begin{lem}\label{lem:weighted-lin-r-est}
    Under \cref{assumption-1}, it holds that 
    \begin{align}
        \label{eq:lin-r-lin-nu-comparison-1}
        |D\lin r|&\les \textnormal{RHS of } \eqref{eq:lin-boot-3-new},\\
        |\lambda\lin r|&\les \textnormal{RHS of } \eqref{eq:lin-boot-2-new}
    \end{align}  in $\mathcal R^\le_{L_i}$.
\end{lem}
\begin{proof}
   Using \cref{lem:lin-r-diamond}, we estimate
    \begin{equation}
     |\lin r| \le |\lin r_\diamond| + |\lin r_\dagger|   \les \lin A\big(1+|\lin{\mathscr P}_i|\big[D'^{-1}+\tau^{-2+3\delta}D_1'^{-2+3\delta/2}\big]\big).  \label{eq:lin-r-alternate}
    \end{equation}
    The claim is then clear by inspection. 
\end{proof}

\begin{lem}
    \label{lem:lin-F-near-far-region} Under \cref{assumption-1}, it holds that 
\begin{multline*}
    \lin F^2\les\lin A^2\big(D^{2-\delta}+\tau^{-4+6\delta}+\lin{\mathscr P}_i^2\big[D^{2-3\delta/2}D_1'^{-4+3\delta}+\tau^{-4+6\delta}D_1'^{-4+3\delta}\big]\big) (\partial_u\phi)^2 \\ + \big(D^{2-\delta}+\lin{\mathscr P}_i^2\big[D'^{-4+2\delta}D^{2-\delta}+\tau^{-2+6\delta}D^2D_1'^{-4+3\delta}\big]\big)(\partial_v\phi)^2
\end{multline*}
 in $\mathcal R^{\le}_{L_i}$ and
\begin{equation*}
    \lin F^2 \les  \lin A^2 r^{-4+2\delta} \left( (\partial_u\phi)^2 + (\partial_v\phi)^2 \right).
\end{equation*}
 in $\mathcal R^{\ge}_{L_i}$.
\end{lem}
\begin{proof} The estimate in $\mathcal R_{L_i}^\ge$ is immediate from inspection of \eqref{eq:F1-def} and the bootstrap assumptions \eqref{eq:lin-boot-4-new}--\eqref{eq:lin-boot-6-new}. For the estimate in $\mathcal R_{L_i}^\le$, we use \cref{lem:weighted-lin-r-est} to estimate
\begin{equation*}
    |\lin F|\les (\textnormal{RHS of } \eqref{eq:lin-boot-2-new})|\partial_u\phi| + (\textnormal{RHS of } \eqref{eq:lin-boot-3-new})|\partial_v\phi|,
\end{equation*}
and squaring immediately yields the claim. 
\end{proof}

The inhomogeneous error terms occurring on the right-hand sides of  \eqref{eq:Mor-final}, \eqref{eq:hp}, and \eqref{eq:rp} are \index{Err1@$\lin{\mathbb E}_\mathrm{M}^\le(\mathcal D)$, near region Morawetz error for $\lin\phi$} \index{Err2@$ \lin{\u{\mathbb E}}{}_p(\mathcal D)$, $h_p$ error for $\lin\phi$} \index{Err3@$\lin{\mathbb E}_\mathrm{M}^\ge(\mathcal D)$, far region Morawetz error for $\lin\phi$} \index{Err4@$\lin{\mathbb E}{}_p(\mathcal D)$, $r^p$ error for $\lin\phi$}
\begin{align*}
    \lin{\mathbb E}_\mathrm{M}^\le(\mathcal D)& \doteq \iint_{\mathcal D^\le}\big(|\partial_u\lin\phi|+|\partial_v\lin\phi| + D |\lin \phi| \big)|\lin F|\,dudv,\\
        \lin{\u{\mathbb E}}{}_p(\mathcal D)&\doteq \iint_{\mathcal D^\le} h_p|\partial_u(r\lin\phi)||\lin F| \,dudv,\\
    \lin{\mathbb E}_\mathrm{M}^\ge(\mathcal D)&\doteq \iint_{\mathcal D^\ge}r^2\big(|\partial_v\lin \phi|+ |\partial_u\lin \phi| + r^{-1} |\lin \phi|\big)|\lin F|\,dudv,\\
    \lin{\mathbb E}{}_p(\mathcal D)&\doteq \iint_{\mathcal D^\ge} r^{p+1}|\partial_v(r\lin\phi)||\lin F| \,dudv. 
\end{align*}

\subsubsection{Morawetz error terms}

 \begin{lem}\label{lem:lin-energy-error-1} Under \cref{assumption-1}, for any domain $\mathcal D\subset\mathcal R_{L_i}$ as in \cref{fig:butterfly} and $b>0$ it holds that 
\begin{equation}
    \lin{\mathbb E}_\mathrm{M}^\le(\mathcal D)\lesssim   b\iint_{\mathcal D^\le}D'D^{\delta/2}\big((\partial_u\lin\phi)^2+(\partial_v\lin\phi)^2 + D^2 |\lin \phi|^2 \big)\,dudv + 
b^{-1} \lin A^2 \ve^2 \tau_1^{-3+\delta}\big(1+\lin{\mathscr P}_i^2D_1'^{-4+3\delta}\big).\label{eq:lin-Mora-1}
\end{equation}  
 \end{lem}
 \begin{proof} 
Using Young's inequality and \cref{lem:lin-F-near-far-region}, we estimate
\begin{align} \nonumber \lin{\mathbb E}_\mathrm{M}^\le(\mathcal D) 
 & \les b\iint_{\mathcal D^\le}D'D^{\delta/2}\big((\partial_u\lin\phi)^2+(\partial_v\lin\phi)^2 + D^2 |\lin \phi|^2 \big)\,dudv + 
b^{-1}\iint_{\mathcal D^\le}D'^{-1}D^{-\delta/2}\lin F^2\,dudv\\
& \les  b\iint_{\mathcal D^\le}D'D^{\delta/2}\big((\partial_u\lin\phi)^2+(\partial_v\lin\phi)^2 + D^2 |\lin \phi|^2 \big)\,dudv +  b^{-1}\lin A^2 \sum_{i=1}^7 I_i,\label{eq:lin-Mor-error-1}\end{align}
where 
\begingroup
\allowdisplaybreaks
\begin{align*}
   I_1 &\doteq \iint_{\mathcal D^{\le}}D'^{-1}D^{2-3\delta/2}(\partial_u\phi)^2\,dudv,\\
   I_2 &\doteq \iint_{\mathcal D^{\le}}\tau^{-4+6\delta}D'^{-1}D^{-\delta/2}(\partial_u\phi)^2\,dudv,\\
   I_3 &\doteq \iint_{\mathcal D^{\le}}\lin{\mathscr P}_i^2 D'^{-1}D^{2-2\delta}D_1'^{-4+3\delta}(\partial_u\phi)^2\,dudv,\\
   I_4 &\doteq \iint_{\mathcal D^{\le}}\lin{\mathscr P}_i^2\tau^{-4+6\delta}D'^{-1}D^{-\delta/2}D_1'^{-4+3\delta}(\partial_u\phi)^2\,dudv,\\
   I_5 &\doteq \iint_{\mathcal D^{\le}}D'^{-1}D^{2-3\delta/2}(\partial_v\phi)^2\,dudv,\\
   I_6 &\doteq \iint_{\mathcal D^{\le}}\lin{\mathscr P}_i^2 D'^{-1}D^{2-2\delta}D_1'^{-4+3\delta}(\partial_v\phi)^2\,dudv,\\
   I_7 &\doteq \iint_{\mathcal D^{\le}}\lin{\mathscr P}_i^2\tau^{-2+6\delta}D'^{-1}D^{2-\delta/2}D_1'^{-4+3\delta}(\partial_v\phi)^2\,dudv.
\end{align*}

\textsc{Estimate for $I_1$ and $I_5$}: Using \eqref{eq:Mor-decay-final-near}, we obtain 
\begin{equation*}
    I_1+I_5\les \iint_{D^\le}D'D^{1-3\delta/2}\big((\partial_u\phi)^2+(\partial_v\phi)^2\big)\,dudv\les\ve^2\tau_1^{-3+\delta}.
\end{equation*}

\textsc{Estimate for $I_2$}: Using \eqref{eq:K-imp-1}, we obtain
\begin{equation*}
    I_2 \les \iint_{\mathcal D^{\le}}\tau^{-4+6\delta}D^{-1/2-\delta/2}(\partial_u\phi)^2\,dudv\les\int_{\tau_1}^{\tau_2}\tau^{-4+6\delta}\u{\mathcal E}{}_{1+\delta}[\phi](\tau)\,d\tau \les \ve^2\tau_1^{-5+8\delta}.
\end{equation*}

\textsc{Estimate for $I_3$ and $I_6$}: Using \eqref{eq:Mor-decay-final-near}, we obtain
\begin{equation*}
    I_3+I_6\les \lin{\mathscr P}_i^2D_1'^{-4+3\delta}\iint_{\mathcal D^{\le}} D'D^{1-2\delta}\big((\partial_u\phi)^2+(\partial_v\phi)^2\big)\,dudv\les \lin{\mathscr P}_i^2\ve^2\tau_1^{-3+\delta}D_1'^{-4+3\delta}.
\end{equation*}

\textsc{Estimate for $I_4$}: Using \eqref{eq:K-imp-1}, we obtain
\begin{align*}
   I_4&\les \lin{\mathscr P}_i^2D_1'^{-4+3\delta} \iint_{\mathcal D^{\le}}\tau^{-4+6\delta}D^{-1/2-\delta/2}(\partial_u\phi)^2\,dudv\\&\les\lin{\mathscr P}_i^2D_1'^{-4+3\delta}\int_{\tau_1}^{\tau_2} \tau^{-4+6\delta}\u{\mathcal E}{}_{1+\delta}[\phi](\tau)\,d\tau \les \lin{\mathscr P}_i^2\ve^2\tau_1^{-5+8\delta}D_1'^{-4+3\delta}.
\end{align*}

\textsc{Estimate for $I_7$}: Using \eqref{eq:Mor-decay-final-near}, we obtain
\begin{equation*}
  I_7\les \lin{\mathscr P}_i^2D_1'^{-4+3\delta}  \iint_{\mathcal D^{\le}}\tau^{-2+6\delta}D'D^{1-\delta/2}(\partial_v\phi)^2\,dudv
  \les \lin{\mathscr P}_i^2\ve^2\tau_1^{-3+\delta}D_1'^{-4+3\delta}.
\end{equation*}
\endgroup Using these estimates in \eqref{eq:lin-Mor-error-1} completes the proof.
\end{proof}

\begin{lem}\label{lem:lin-energy-error-2} Under \cref{assumption-1}, for any domain $\mathcal D\subset\mathcal R_{L_i}$ as in \cref{fig:butterfly} and $b>0$ it holds that 
\begin{equation}\label{eq:lin-Mora-2}
      \lin{\mathbb E}_\mathrm{M}^\ge(\mathcal D)\les  b\iint_{\mathcal D^\ge}r^{1-\delta}\big(|\partial_u\lin\phi|^2+|\partial_v\lin\phi|^2 + r^{-2} |\lin \phi|^2 \big)\,dudv+b^{-1} \lin A^2 \ve^2 \tau_1^{-3+\delta}\big(1+\lin{\mathscr P}_i^2D_1'^{-4+3\delta}\big).
\end{equation}
\end{lem}
\begin{proof}
    This follows from Young's inequality, \cref{lem:lin-F-near-far-region}, and
\[
        \iint_{\mathcal D^\ge}r^{3+\delta}\lin F^2\,dudv\les \lin A^2  \iint_{\mathcal D^\ge}r^{-1+3\delta}\big((\partial_u\phi)^2 + (\partial_v\phi)^2\big)\,dudv\les  \lin A^2 \ve^2 \tau_1^{-3+\delta}\big(1+\lin{\mathscr P}_i^2D_1'^{-4+3\delta}\big)
\]
which follows from \eqref{eq:Mor-decay-final-far}.
\end{proof}

\subsubsection{The \texorpdfstring{$h_p$}{hp} error terms}

We perform the error estimates for $p\in [1, 3-3\delta]$. 

\begin{lem}\label{lem:lin-energy-error-3} Under \cref{assumption-1}, for any $p\in[1, 3-3\delta]$ and $b>0$, it holds that 
\begin{equation}\label{eq:hp-error}
       \lin{\u{\mathbb E}}{}_p(\mathcal D)  
        \les
         b\sup_{\tau_1\le\tau\le\tau_2}\u{\mathcal E}{}_p[\lin\phi](\tau,u_2)+b^{-1}\lin A^2\ve^2 \tau_1^{-3+2\delta+p}\big(1+\lin{\mathscr P}_i^2 D_1'^{-4+3\delta}\big).
    \end{equation}
\end{lem}
\begin{proof} Using Young's inequality, we have 
\begin{equation}
     \lin{\u{\mathbb E}}{}_p(\mathcal D)  \les  b  \iint_{\mathcal D^\le} \tau^{-1-\delta}h_p(\partial_u(r\lin\phi))^2\,dudv + b^{-1} \lin A^2\sum_{i=1}^7I_i\les  b  \sup_{\tau_1\leq \tau \leq \tau_2} \u{\mathcal E}{}_p[\lin \phi] (\tau,u_2)+ b^{-1} \lin A^2\sum_{i=1}^7I_i,
     \label{eq:lin-hp-error-1}
\end{equation}
where 
\begingroup\begin{align*}
    I_1&\doteq \iint_{\mathcal D^\le} \tau^{1+\delta}D'D^{-p/2+3/2-\delta} (\partial_u\phi)^2\,dudv,\\
    I_2&\doteq \iint_{\mathcal D^\le} \tau^{-3+7\delta}D'D^{-p/2-1/2} (\partial_u\phi)^2\,dudv,\\
    I_3&\doteq \iint_{\mathcal D^\le} \lin{\mathscr P}_i^2\tau^{1+\delta}D'D^{-p/2+3/2-3\delta/2} D_1'^{-4+3\delta}(\partial_u\phi)^2\,dudv,\\
    I_4&\doteq \iint_{\mathcal D^\le} \lin{\mathscr P}_i^2\tau^{-3+7\delta}D'D^{-p/2-1/2} D_1'^{-4+3\delta}(\partial_u\phi)^2\,dudv,\\
    I_5&\doteq \iint_{\mathcal D^\le} \tau^{1+\delta}D'D^{-p/2+3/2-\delta} (\partial_v\phi)^2\,dudv,\\
    I_6&\doteq \iint_{\mathcal D^\le} \lin{\mathscr P}_i^2\tau^{1+\delta}D'^{-3+2\delta}D^{-p/2+3/2-\delta} (\partial_v\phi)^2\,dudv,\\
    I_7&\doteq \iint_{\mathcal D^\le} \lin{\mathscr P}_i^2\tau^{-1+7\delta}D'D^{-p/2+3/2} D_1'^{-4+3\delta}(\partial_v\phi)^2\,dudv. 
\end{align*}\endgroup

 \textsc{Estimate for $I_1$ and $I_5$}: Recall the dyadic time steps $L_j \doteq 2^j$, set $\mathcal D_j^\le\doteq \mathcal D^\le\cap\{L_j\le\tau\le L_{j+1}\}$, and let $j_0\doteq \lfloor \log_2\tau_1\rfloor$. Using \eqref{eq:Mor-decay-final-near}, we obtain for $p=3-3\delta$
 \begin{equation*}
     I_1+I_5 = \iint_{\mathcal D^\le}\tau^{1+\delta}D'D^{\delta/2}\big((\partial_u\phi)^2+(\partial_v\phi)^2\big)\,dudv \les \sum_{j\ge j_0} L_j^{1+\delta} \cdot \ve^2L_j^{-3+\delta} \les \ve^2L_{j_0}^{-2+2\delta}\les \ve^2\tau_1^{-2+2\delta}.
 \end{equation*}

 \textsc{Estimate for $I_2$}: Using \eqref{eq:K-imp-1}, we obtain
 \begin{equation*}
     I_2\les \int_{\tau_1}^{\tau_2}\tau^{-3+7\delta}\u{\mathcal E}{}_p[\phi](\tau)\,d\tau \les\ve^2 \tau_1^{-5+8\delta+p}.
 \end{equation*}

 \textsc{Estimate for $I_3$}: Using \eqref{eq:K-imp-1}, we obtain 
 \begin{equation*}
     I_3\les \lin{\mathscr P}_i^2D_1'^{-4+3\delta}\int_{\tau_1}^{\tau_2}\tau^{1+\delta}\u{\mathcal E}{}_0[\phi](\tau)\,d\tau\les \lin{\mathscr P}_i^2\ve^2\tau^{-2+2\delta}D_1'^{-4+3\delta}.
 \end{equation*}

 \textsc{Estimate for $I_4$}: Arguing exactly as for $I_2$ after pulling out of the integral the term $D_1'^{-4+3\delta}$, we obtain
 \begin{equation*}
     I_4\les \lin{\mathscr P}_i^2 \ve^2\tau_1^{-5+8\delta+p}D_1'^{-4+3\delta}.
 \end{equation*}

  \textsc{Estimate for $I_6$}: We again estimate $I_6$ only for $p=3-3\delta$.  We split $I_6=I_6^A+I_6^B$, where in $I_6^A$ we restrict the domain of integration to $\{5D\le (MD')^2\}$ and in $I_6^B$ to $\{5D\ge (MD')^2\}$. By dyadically decomposing and using \eqref{eq:Mor-decay-final-near}, we obtain first
  \begin{equation*}
      I_6^B\les \lin{\mathscr P}_i^2D_1'^{-4+3\delta}\sum_{j\ge j_0}L_j^{1+\delta}\iint_{\mathcal D_j^\le\cap \{5D\le (MD')^2\}} D'^{1-\delta}D^{\delta/2}(\partial_v\phi)^2\,dudv\les \lin{\mathscr P}_i^2\ve^2 \tau_1^{-2+2\delta}D_1'^{-4+3\delta}.
  \end{equation*}
  Next, from the basic monotonicity of $D$, we have $D^{-1} \leq D^{-1}(u,L_i)$ on $\mathcal D^{\le}_i$. We can then estimate
    \begin{align*}
     I_6^B& \lesssim \lin{\mathscr P}_i^2   \sum_{i\ge i_0} L_i^{1+\delta} \iint_{\mathcal D_i^\le \cap\{ 5D\geq  (MD')^2 \} }  D^{-3/2+3\delta/2} (u,L_{i})    (\partial_v\phi)^2 \,dudv \\
  &  \lesssim \lin{\mathscr P}_i^2 \sum_{i\ge i_0}  L_i^{1+\delta}\int_{\u C{}_{L_i} \cap  \mathcal D^{\le} \cap \{ 5D\geq  (MD')^2 \}  } D^{-3/2+3\delta/2}  \,du\cdot \sup_{u\le u_2} \int_{\{u \} \times [L_i , L_{i+1} ]\cap \mathcal D^{\le}}  (\partial_v\phi)^2  dv 
   \\ & \lesssim\lin{\mathscr P}_i^2 \ve ^2     \tau_1^{-2+2\delta}D_1'^{-4+3\delta},
    \end{align*}
    where we used that 
    \[\int_{\u C{}_{L_i} \cap  \mathcal D^{\le} \cap\{ 5D \geq  (MD')^2 \}  } D^{-3/2+3\delta/2}  \,du\les D_1'^{-1}(u_2)\int_{\u C{}_{L_i} \cap \mathcal D^{\le} \cap \{ 5D\geq  (MD')^2 \}  } D'D^{-3/2+3\delta/2}  \,du\les D_1'(u_2)^{-4+3\delta}.\]

 \textsc{Estimate for $I_7$}: Using \eqref{eq:Mor-decay-final-near}, we obtain for $p=3-3\delta$
 \begin{equation*}
     I_7\les \lin{\mathscr P}_i^2 D_1'^{-4+3\delta}\iint_{\mathcal D^\le}D'D^{3\delta/2}(\partial_v\phi)^2\les \lin{\mathscr P}_i^2 \ve^2\tau_1^{-3+\delta}D_1'^{-4+3\delta}.
 \end{equation*}
Using these estimates in \eqref{eq:lin-hp-error-1} completes the proof.
\end{proof}

\subsubsection{The \texorpdfstring{$r^p$}{rp} error terms}

\begin{lem}[$r^p$ error terms]\label{lem:lin-energy-error-4} 
    Under \cref{assumption-1}, for any $p\in[1,3-3\delta]$ and $b>0$, it holds that 
    \begin{equation}\label{eq:lin-rp-error}
    \lin{\mathbb E}{}_p(\mathcal D)  \lesssim  
     b \sup_{\tau_1 \leq \tau \leq \tau_2} \mathcal E_p[\lin\phi](\tau) +b^{-1} \lin A^2 \ve^2 \tau_1^{-2+2\delta}.
\end{equation}
\end{lem}
\begin{proof} Using Young's inequality and \cref{lem:lin-F-near-far-region} again, we have
\begin{align*}
    \lin{\mathbb E}{}_p(\mathcal D) &\les b\iint_{\mathcal D^\ge}\tau^{-1-\delta}r^p(\partial_v(r\lin\phi))^2\,dudv+    b^{-1}\iint_{\mathcal D^\ge}\tau^{1+\delta}r^{p+2}\lin F^2\,dudv\\
    &\les b\sup_{\tau_1 \leq \tau \leq \tau_2} \mathcal E_p[\lin\phi](\tau) + b^{-1}\lin A^2\iint_{\mathcal D^\ge}\tau^{1+\delta}r^{1-\delta} \big( (\partial_u\phi)^2 + (\partial_v\phi)^2 \big)\,dudv,
\end{align*} where we used $p=3-3\delta$ in the inhomogeneous error term.
We decompose $\mathcal D^\ge$ into $\mathcal D_i^\ge\doteq \mathcal D\cap\{L_i\le\tau\le L_{i+1}\}$ and use \eqref{eq:Mor-decay-final-far} to estimate
\begin{equation*}
    \iint_{\mathcal D^\ge}\tau^{1+\delta}r^{1-\delta} \big( (\partial_u\phi)^2 + (\partial_v\phi)^2 \big)\,dudv \les \sum_{i\ge i_0}L_i^{1+\delta}\iint_{\mathcal D^\ge_i}r^{1-\delta} \big( (\partial_u\phi)^2 + (\partial_v\phi)^2 \big)\,dudv\les \ve^2 \tau_1^{-2+2\delta},
\end{equation*}
which completes the proof.
\end{proof}

\subsection{Estimates for the scalar field}\label{sec:linear-scalar-field}

\begin{prop}\label{prop:decay-linphi}
   Under \cref{assumption-1}, it holds that
\begin{align}
\label{eq:lin-phi-energy-1}\u{\mathcal E}{}_p[\lin\phi](\tau,u)  & \les \lin A^2\big(1+\lin{\mathscr P}_i^2D_1'(u)^{-4+3\delta}\big)\tau^{-3+3\delta+p},\\
\label{eq:lin-phi-energy-2} \mathcal E_p[\lin\phi](\tau) & \les \lin A^2\big(1+\lin{\mathscr P}_i^2D_1'(\tau)^{-4+3\delta}\big)\tau^{-3+3\delta+p} ,\\
\label{eq:lin-phi-energy-3} \u{\mathcal F}[\lin\phi](v,\tau)  & \les \lin A^2\big(1+\lin{\mathscr P}_i^2D_1'(\tau)^{-4+3\delta}\big)\tau^{-3+3\delta},\\
\label{eq:lin-phi-energy-4} \mathcal F[\lin\phi](u,\tau)  & \les \lin A^2\big(1+\lin{\mathscr P}_i^2D_1'(u)^{-4+3\delta}\big)\tau^{-3+3\delta}
\end{align}
 for every $p\in\{0\}\cup[\delta,3-3\delta]$, $1\le \tau\le L_i$, and $(u,v)\in\mathcal R_{L_i}$.
\end{prop}

We begin the proof by estimating the initial energy of $\lin \phi$. 
\begin{lem}\label{lem:lin-phi-initial-energy}  Under \cref{assumption-1}, it holds that
   \begin{equation}\label{eq:lin-phi-initial-energy}
       \mathcal E_p[\lin\phi](1) + \u{\mathcal E}{}_p[\lin\phi](1,u)\les \lin A^2\big(1+\lin{\mathscr P}_i^2 D_1'(u)^{-4+3\delta}\big)
   \end{equation}
    for every $p\in\{0\}\cup[\delta,3-3\delta]$ and $1\le u\le L_i$.
\end{lem}
\begin{proof} We argue as in \cref{lem:initial-energy-est}, using now \cref{lem:lin-phi-data} to estimate $\lin\phi$ and its derivatives on $\mathcal C$. Using the bootstrap assumptions for the linearized geometry, evaluated on $\mathcal C$, we obtain
\begin{equation*}
    \lin\phi^2\les \lin A^2 r^{-2+2\delta},\quad (\partial_v\lin\phi)^2\les \lin A^2 r^{-4+2\delta} ,\quad (\partial_v(r\lin\phi))^2\les \lin A^2 r^{-4+2\delta}
\end{equation*}
on $C_\out$ and 
\begin{equation*}
    \lin \phi^2\les \lin A^2\big(1+\lin{\mathscr P}_i^2D'^{-4+3\delta}\big),\quad (\partial_u\lin\phi)^2\les \lin A^2\big(D^{2-\delta}+\lin{\mathscr P}_i^2\big[D'^{-4+2\delta}D^{2-\delta}+D'^{-4+3\delta}D^2\big]\big)
\end{equation*}
on $\u C{}_\ing$. The bound \eqref{eq:lin-phi-initial-energy} for $\mathcal E_p[\lin \phi](1)$ is immediate from the first set of estimates. For $\u{\mathcal E}{}_p[\lin\phi](1,u)$, we use the second set to estimate 
\begin{equation*}
    \int_{\u C{}_\ing\cap\{u'\le u\}}D\big(1+\lin{\mathscr P}_i^2D'^{-4+3\delta}\big) \,du'\les \big(1+\lin{\mathscr P}_i^2D_1'^{-4+3\delta}(u)\big)\int_{\u C{}_\ing\cap\{u'\le u\}}D \,du'\les 1+\lin{\mathscr P}_i^2D_1'^{-4+3\delta}(u),
\end{equation*}\vspace{-5mm}
\begin{multline*}
    \int_{\u C{}_\ing\cap\{u'\le u\}} h_{3-3\delta}\big(D^{2-\delta}+\lin{\mathscr P}_i^2\big[D'^{-4+2\delta}D^{2-\delta}+D'^{-4+3\delta}D^2\big]\big)\,du' \\ \les \big(1+\lin{\mathscr P}_i^2D_1'^{-4+3\delta}(u)\big)\int_{\u C{}_\ing\cap\{u'\le u\}}D'D^{\delta/2}\,du'\les 1+\lin{\mathscr P}_i^2D_1'^{-4+3\delta}(u),
\end{multline*}
which yields the other half of \eqref{eq:lin-phi-initial-energy}. \end{proof}

\begin{proof}[Proof of \cref{prop:decay-linphi}] Fix $1\le u_2\le L_i$. We show that 
\begin{align}
\label{eq:lin-phi-energy-1-1}\u{\mathcal E}{}_p[\lin\phi](\tau,u_2)  & \les \lin A^2\big(1+\lin{\mathscr P}_i^2D_1'(u_2)^{-4+3\delta}\big)\tau^{-3+3\delta+p},\\
\label{eq:lin-phi-energy-2-1} \mathcal E_p[\lin\phi](\tau) & \les \lin A^2\big(1+\lin{\mathscr P}_i^2D_1'(u_2)^{-4+3\delta}\big)\tau^{-3+3\delta+p} ,\\
\label{eq:lin-phi-energy-3-1} \u{\mathcal F}[\lin\phi](v,\tau)  & \les \lin A^2\big(1+\lin{\mathscr P}_i^2D_1'(u_2)^{-4+3\delta}\big)\tau^{-3+3\delta},\\
\label{eq:lin-phi-energy-4-1} \mathcal F[\lin\phi](u,\tau)  & \les \lin A^2\big(1+\lin{\mathscr P}_i^2D_1'(u_2)^{-4+3\delta}\big)\tau^{-3+3\delta}
\end{align}
 for every $p\in\{0\}\cup[\delta,3-3\delta]$, $1\le \tau\le u_2$, and $(u,v)\in\mathcal R_{u_2}$. These estimates imply \eqref{eq:lin-phi-energy-1}--\eqref{eq:lin-phi-energy-4}. From now on in this proof, we omit the argument of $D_1'$, as it will always be evaluated at $u_2$.

 \textsc{Establishing the hierarchy}: We start with the Morawetz estimate, \cref{prop:Morawetz}. Using \cref{lem:lin-energy-error-1,lem:lin-energy-error-2} to estimate the inhomogeneous error terms, we find 
    \begin{multline*}
        \sup_{\tau_1\le\tau\le\tau_2}\big(\mathcal E_0[\lin\phi](\tau)+\underline{\mathcal E}{}_0[\lin\phi](\tau,u_2)\big)\les  \mathcal E_0[\lin\phi](\tau_1)+\underline{\mathcal E}{}_0[\lin\phi](\tau_1,u_2)+b^{-1}\lin A^2\ve^2\tau_1^{-3+\delta}\big(1+\lin{\mathscr P}_i^2D_1'^{-4+3\delta}\big)\\+\ve \int_{\tau_1}^{\tau_2}\big(\tau^{-2+2\delta}\underline{\mathcal E}{}_{0}[\lin\phi](\tau,u_2)+\tau^{-6+2\delta}\underline{\mathcal E}{}_{1+\delta}[\lin\phi](\tau,u_2)\big)\,d\tau
    \end{multline*}
    for $b$ sufficiently large, where we used the good Morawetz bulk to absorb the first error terms in \eqref{eq:lin-Mora-1} and \eqref{eq:lin-Mora-2}, and we used the trivial estimate $\tau^{-2+2\delta}h_{-3+\delta}+\tau^{-4+2\delta}h_{-1+\delta}\les \tau^{-2+\delta}$. Using the integrability of $\tau^{-2+2\delta}$ and then using $\ve$ to absorb the first term in the integral into the left-hand side, we simplify this to 
    \begin{multline*}
        \sup_{\tau_1\le\tau\le\tau_2}\big(\mathcal E_0[\lin\phi](\tau)+\underline{\mathcal E}{}_0[\lin\phi](\tau,u_2)\big)\les  \mathcal E_0[\lin\phi](\tau_1)+\underline{\mathcal E}{}_0[\lin\phi](\tau_1,u_2)+b^{-1}\lin A^2\ve^2\tau_1^{-3+\delta}\big(1+\lin{\mathscr P}_i^2D_1'^{-4+3\delta}\big)\\+\ve \tau_1^{-5+2\delta} \sup_{\tau_1\le\tau\le\tau_2} \underline{\mathcal E}{}_{1+\delta}[\lin\phi](\tau,u_2).
    \end{multline*}
    Note that the right-hand side of this estimate also bounds the $\mathbb M(\lin\phi,\lin F,\mathcal D)$ error terms in \cref{prop:hp,prop:rp}. Therefore, using the $h_p$ hierarchy property \eqref{eq:hp-hierarchy-property}, and the error estimates from \cref{lem:lin-energy-error-3,lem:lin-energy-error-4}, we obtain
     \begin{multline}
          \sup_{\tau_1\le\tau\le\tau_2}\big(\mathcal E_p[\lin\phi](\tau)+\underline{\mathcal E}{}_p[\lin\phi](\tau,u_2)\big)+\int_{\tau_1}^{\tau_2}\big(\mathcal E_{p-1}[\lin\phi](\tau)+\underline{\mathcal E}{}_{p-1}[\lin\phi](\tau,u_2)\big)\,d\tau\\\les \mathcal E_p[\lin\phi](\tau_1)+\underline{\mathcal E}{}_p[\lin\phi](\tau_1,u_2)+b^{-1}\lin A^2\ve^2\tau_1^{-3+2\delta+p}\big(1+\lin{\mathscr P}_i^2D_1'^{-4+3\delta}\big)+\ve \tau_1^{-5+2\delta} \sup_{\tau_1\le\tau\le\tau_2} \underline{\mathcal E}{}_{1+\delta}[\lin\phi](\tau,u_2),\label{eq:lin-hier-1}
     \end{multline} for $p\in[1,3-3\delta]$, where we took $b$ small to absorb the first error terms in \eqref{eq:hp-error} and \eqref{eq:lin-rp-error}. 

\textsc{Proofs of \eqref{eq:lin-phi-energy-1-1} and \eqref{eq:lin-phi-energy-2-1}}: First, we apply \eqref{eq:lin-hier-1} with $p=3-3\delta$ and $\tau_1=1$, use \cref{lem:lin-phi-initial-energy} to estimate the initial energy, and use $\underline{\mathcal E}{}_{1+\delta}[\lin\phi](\tau,u_2)\les \underline{\mathcal E}{}_{3-3\delta}[\lin\phi](\tau,u_2)$ to absorb the last term on the right-hand side, and take $\ve_0 \le b$, obtaining altogether 
\begin{equation}\label{eq:lin-phi-bounded}
    \mathcal E_{3-3\delta}[\lin\phi](\tau)+\underline{\mathcal E}{}_{3-3\delta}[\lin\phi](\tau,u_2)\les \lin A^2\big(1+\lin{\mathscr P}_i^2D_1'^{-4+3\delta}\big)
\end{equation} for every $\tau\le u_2$.
Using this, we improve \eqref{eq:lin-hier-1} to 
 \begin{multline}
          \sup_{\tau_1\le\tau\le\tau_2}\big(\mathcal E_p[\lin\phi](\tau)+\underline{\mathcal E}{}_p[\lin\phi](\tau,u_2)\big)+\int_{\tau_1}^{\tau_2}\big(\mathcal E_{p-1}[\lin\phi](\tau)+\underline{\mathcal E}{}_{p-1}[\lin\phi](\tau,u_2)\big)\,d\tau\\\les \mathcal E_p[\lin\phi](\tau_1)+\underline{\mathcal E}{}_p[\lin\phi](\tau_1,u_2)+\lin A^2\tau_1^{-3+2\delta+p}\big(1+\lin{\mathscr P}_i^2D_1'^{-4+3\delta}\big).\label{eq:lin-hier-2}
     \end{multline}

Set $J\doteq \lfloor {\log_2 u_2}\rfloor$. For each $j\in\{0,\dotsc,J-1\}$, we deduce from \eqref{eq:lin-phi-bounded} and \eqref{eq:lin-hier-2} for $p=3-3\delta$ on the interval $[L_j,L_{j+1}]$, together with the pigeonhole principle, the existence of a $\tau'_j\in[L_j,L_{j+1}]$ such that 
\begin{equation}
    \mathcal E_{2-\delta}[\lin\phi](\tau_j')+\underline{\mathcal E}{}_{2-\delta}[\lin\phi](\tau_j',u_2)\les (L_{j+1}-L_j)^{-1}\lin A^2\big(1+\lin{\mathscr P}_i^2D_1'^{-4+3\delta}\big)\les \lin A^2 L_j^{-1}\big(1+\lin{\mathscr P}_i^2D_1'^{-4+3\delta}\big).\label{eq:decay-aux-2}
\end{equation}
Using now \eqref{eq:lin-hier-2} for $p=2-\delta$ and the dyadicity of the sequence $L_j$, we upgrade \eqref{eq:decay-aux-2} to
\begin{equation}
     \mathcal E_{2-3\delta}[\lin\phi](\tau)+\underline{\mathcal E}{}_{2-3\delta}[\lin\phi](\tau,u_2)\les\lin A^2 \tau^{-1}\big(1+\lin{\mathscr P}_i^2D_1'^{-4+3\delta}\big)\label{eq:decay-aux-3}
\end{equation}
for every $\tau\in[1,u_2]$. Using \cref{lem:interpolation} to interpolate between \eqref{eq:lin-phi-bounded} and \eqref{eq:decay-aux-3}, we obtain
\begin{equation}
     \mathcal E_{2}[\lin\phi](\tau)+\underline{\mathcal E}{}_{2}[\lin\phi](\tau,u_2) \les \lin A^2 \tau^{-1+3\delta}\big(1+\lin{\mathscr P}_i^2D_1'^{-4+3\delta}\big)\label{eq:decay-aux-4}
\end{equation}
for every $\tau\in[1,u_2]$. We now apply \eqref{eq:decay-aux-4} and \eqref{eq:lin-hier-2} with $p=2$ to the intervals $[L_j,L_{j+1}]$ with $i\in\{0,\dotsc,J-1\}$ to find $\tau_j''\in[L_j,L_{j+1}]$ such that
\begin{align*}
     \mathcal E_{1}[\lin\phi](\tau_j'')+\underline{\mathcal E}{}_{1}[\lin\phi](\tau_j'',u_2)&\les (L_{j+1}-L_j)^{-1}\big( \mathcal E_p[\lin\phi](L_j)+\underline{\mathcal E}{}_p[\lin\phi](L_j,u_2)+\lin A^2L_j^{-1+2\delta}\big(1+\lin{\mathscr P}_i^2D_1'^{-4+3\delta}\big)\big)\\
     &\les \lin A^2L_j^{-2+3\delta}\big(1+\lin{\mathscr P}_i^2D_1'^{-4+3\delta}\big). 
\end{align*}
Using \eqref{eq:lin-hier-2}, this is again immediately upgraded to 
\begin{equation*}
      \mathcal E_{1}[\lin\phi](\tau)+\underline{\mathcal E}{}_{1}[\lin\phi](\tau,u_2)\les \lin A^2\tau^{-2+3\delta}\big(1+\lin{\mathscr P}_i^2D_1'^{-4+3\delta}\big)
\end{equation*}
for every $\tau\in[1,u_2]$. Repeating this argument once more, we conclude 
\begin{equation*}
      \mathcal E_{0}[\lin\phi](\tau)+\underline{\mathcal E}{}_{0}[\lin\phi](\tau,u_2)\les \lin A^2\tau^{-3+3\delta}\big(1+\lin{\mathscr P}_i^2D_1'^{-4+3\delta}\big)
\end{equation*}
for every $\tau\in[1,u_2]$. By interpolation, we finally have \eqref{eq:lin-phi-energy-1-1} and \eqref{eq:lin-phi-energy-2-1} for every $p\in\{0\}\cup[\delta,3-\delta]$ and $\tau\in[1,u_2]$. 

\textsc{Proofs of \eqref{eq:lin-phi-energy-3-1} and \eqref{eq:lin-phi-energy-4-1}}: Applying again \cref{prop:Morawetz} (specifically the estimate for the fluxes along III and VI), we obtain \eqref{eq:lin-phi-energy-3-1} and \eqref{eq:lin-phi-energy-4-1} by estimating the terms on the right-hand side of \eqref{eq:Mor-final} as we did for \eqref{eq:lin-hier-1}.
\end{proof}

\begin{lem}
    Under \cref{assumption-1}, for any domain $\mathcal D\subset\mathcal R_{L_i}$ as in \cref{fig:butterfly}, it holds that
    \begin{equation}\label{eq:lin-phi-energy-5}
        \iint_{\mathcal D^{\ge}} r^{1-\delta}\big((\partial_u\lin\phi)^2+(\partial_v\lin\phi)^2\big)\,dudv \les \lin A^2\tau^{-3+3\delta}\big(1+\lin{\mathscr P}_i^2D_1'^{-4+3\delta}\big).
    \end{equation}
\end{lem}
\begin{proof} This follows from \cref{prop:Morawetz} and the error estimates in the proof of \cref{prop:decay-linphi}.
\end{proof}

\begin{lem}\label{lem:lin-phi-pw}
     Under \cref{assumption-1}, it holds that
     \begin{equation}\label{eq:r-lin-phi-pointwise}
         |r\lin\phi|\les \lin A \tau^{-1+3\delta/2}\big(1+|\lin{\mathscr P}_i|D_1'^{-2+3\delta/2}\big)
     \end{equation}
     on $\mathcal R_{\u u{}_f}$.
\end{lem}
\begin{proof}
    To prove this, one repeats the proof of \eqref{est:phi_bound} in \cref{prop:pw-estimates}. In particular, one estimates $(r\lin \phi)^2$ by interpolating between the $p=0$ and $p=2$ energies, which gives the decay rate $\tau^{-2+3\delta}$. We leave the details to the reader. 
\end{proof}

\subsection{Estimates for the linearized geometric quantities}\label{sec:linear-geometric}

In this section, we use the energy estimates for $\lin\phi$ from the previous section to estimate $\lin\gamma$, $\lin\kappa$, $\lin\varpi$, $\lin r$, $\lin\lambda$, and $\lin\nu$. These estimates will allow us to close the bootstrap argument, i.e., prove \cref{prop:lin-boot-improvement,prop:Pi-est}.

\subsubsection{Estimates for \texorpdfstring{$\lin\gamma$}{lin gamma} and \texorpdfstring{$\lin\kappa$}{lin kappa}}

We begin by estimating $\lin\gamma$ (only in the far region). 

\begin{lem}\label{lem:lin-gamma-far} Under \cref{assumption-1}, it holds that
    \begin{align}
         \label{eq:lin-gamma-far}  |\lin\gamma|&\les  \lin A \ve r^{-1}\tau^{-3+2\delta}\big(1+|\lin{\mathscr P}_i|D_1'^{-2+3\delta/2}\big),\\
       \label{eq:lin-gamma-far-2}  |\lin\gamma|&\les  \lin A \ve r^{-1}\tau^{-1+\delta/2}
    \end{align}in $\mathcal R_{L_i}^\ge$, and along $\Gamma=\{u=v\}$ we have the integrated estimate
\begin{equation}\label{eq:lin-gamma-far-integrated}
    \int_u^{L_i} |\lin\gamma(u',u')| \,du'\les \lin A \ve \tau^{-3+5\delta/2}\big(1+|\lin{\mathscr P}_i|D_1'^{-2+3\delta/2}\big).
\end{equation}
\end{lem}

The proof requires a technical lemma, which will be used several more times in the sequel.
\begin{lem} \label{lem:D'-dyadic-sum}
    Under \cref{assumption-1}, it holds that 
    \begin{equation*}
      \sum_{j\ge j_0}L_j^{-2+3\delta/2-\alpha}D_1'(L_{j+1})^{-2+3\delta/2}  \les_\alpha u^{-2+3\delta/2-\alpha} D_1'(u)^{-2+3\delta/2}
    \end{equation*}
    for any $\alpha>0$, where $j_0=\lfloor{\log_2 u}\rfloor$.
\end{lem}
\begin{proof}
Recall the function $F\doteq u^{-1}D_1'$ from \cref{lem:F-monotonicity}. Using its almost-monotonicity, \eqref{eq:F-monotonicity}, we estimate
    \begin{multline*}
        \sum_{j\ge j_0} L_j^{-2+3\delta/2-\alpha}D_1'(L_{j+1})^{-2+3\delta/2} = \sum_{j\ge j_0}L_j^{-\alpha} F(L_{j+1})^{-2+3\delta/2}\\\les F(u)^{-2+3\delta/2}\sum_{j\ge j_0}L_j^{-\alpha} \les_\alpha u^{-3+3\delta/2-\alpha}D_1'(u)^{-2+3\delta/2},
    \end{multline*}
    which completes the proof.
\end{proof}

\begin{proof}[Proof of \cref{lem:lin-gamma-far}] Integrating \eqref{eq:dvlgamma-log} and using the boundary condition \eqref{eq:lin-gamma-BC} gives
    \begin{equation*}
        |r\lin\gamma|(u,v) \les\int_v^\infty \big(r(\partial_{v'}\phi)^2|\lin r|+r^{2}(\partial_{v'}\phi)^2|\lin\lambda|+r^{2}|\partial_{v'}\phi||\partial_{v'}\lin\phi|\big)\,dv'.
    \end{equation*} for $v\ge u$.
Then, using the bootstrap assumptions for $\lin r$ and $\lin \lambda$ and the energy decay \eqref{eq:lin-phi-energy-2}, we obtain
\begin{align*}
  |r\lin\gamma|(u,v) &\les  \lin A\int_v^\infty r^2 (\partial_{v'} \phi)^2 dv' + \left(\int_v^\infty r^2 (\partial_{v'} \phi)^2 dv '\int_v^\infty r^2 (\partial_{v'} \lin\phi)^2 dv'\right)^{1/2}\\
  &\les \lin A\mathcal E_0[\phi](\tau)+\big(\mathcal E_0[\phi](\tau)\big)^{1/2}\big(\mathcal E_0[\lin\phi](\tau)\big)^{1/2}\les \lin A \ve \tau^{-3+2\delta}+\lin A |\lin{\mathscr P}_i| \ve \tau^{-3+2\delta}D_1'^{-2+3\delta/2}.
\end{align*}
    Moreover, using \eqref{eq:delta-decay-final-far}, \eqref{eq:lin-phi-energy-5}, and \cref{lem:D'-dyadic-sum}, we have the integrated estimate
    \begin{align*}
        \int_{u}^{L_i}|\lin\gamma|(u',u')\,du'  &\les \iint_{\mathcal R^\ge_{L_i}\cap\{u'\ge u\}}\big((\partial_v\phi)^2|\lin r|+r(\partial_v\phi)^2|\lin\lambda|+r|\partial_v\phi||\partial_v\lin\phi|\big)\,du'dv \\ & \lesssim \lin A
        \iint_{\mathcal R^\ge_{L_i}\cap\{ u'\ge u\}}r^\delta(\partial_v\phi)^2 \,du'dv \\ &\hspace{-20mm}+ \sum_{j\ge \lfloor {\log_2 u}\rfloor} \left(\iint_{\mathcal R^\ge_{L_i}\cap\{L_j\le u'\le L_{j+1}\}} r^{1+\delta} (\partial_v  \phi)^2 \, du'dv \iint_{\mathcal R^\ge_{L_i}\cap\{L_j\le u'\le L_{j+1}\}} r^{1-\delta}(\partial_v \lin \phi)^2 \, du'dv\right)^{1/2} \\
         &\les  \lin A  \ve^2 \tau^{-3+\delta}+ \sum_{j\ge \lfloor \log_2u\rfloor}\ve L_j^{-3/2+\delta}\cdot \lin AL_j^{-3/2+3\delta/2}\big(1+|\lin{\mathscr P}_i|D_1'(L_{j+1})^{-2+3\delta/2}\big)\\
         &\les \lin A \ve \tau^{-3+5\delta/2}\big(1+|\lin{\mathscr P}_i|D_1'^{-2+3\delta/2}\big). 
    \end{align*}
    This proves \eqref{eq:lin-gamma-far}. Then \eqref{eq:lin-gamma-far-2} follows from $D_1'\les u$.
\end{proof}

Next, we estimate $\lin\kappa$ along $\Gamma$ and in the far region.
 
\begin{lem}\label{lem:lin-kappa-far}
    Under \cref{assumption-1}, it holds that
\begin{align}
  \label{eq:lin-kappa-far-1}   |\lin\kappa|  & \les \lin A\ve\tau^{-3+2\delta}\big(1+|\lin{\mathscr P}_i|D_1'^{-2+3\delta/2}\big),\\
   \label{eq:lin-kappa-far-2}  |\lin\kappa|  &\les \lin A\ve r^{-1+\delta}
\end{align}         in $\mathcal R_{L_i}^{\ge }$. Moreover, along $\Gamma$ it holds that
    \begin{equation}\label{eq:lin-kappa-integral}
        \int^{L_i}_u|\lin\kappa(u',u')|\,du'\les \lin A\ve\tau^{-3+2\delta}\big(1+|\lin{\mathscr P}_i|D_1'^{-2+3\delta/2}\big).
    \end{equation}
\end{lem}
\begin{proof} \textsc{Estimating $\lin\kappa$ along $\Gamma$}: By construction of the gauge, recall \cref{prop:boundary-conditions}, $\lin\kappa=0$ on $\mathcal G_{L_i}$, which coincides with $\Gamma=\{u=v\}$ for $v\le L_i-1$. For $\{L_i-1\le v\le L_i\}$, we integrate \eqref{eq:dulkappa-log} to the future from $\mathcal G_{u_f}$, and use Gr\"onwall's inequality, the bootstrap assumptions for $\lin r$ and $\lin \nu$, and \eqref{eq:lin-phi-energy-3} to estimate
\begin{align}
   \nonumber |\lin\kappa(v,v)|&\les \int_{v}^{\mathfrak g_{L_i}(v)}\big((\partial_u\phi)^2|\lin r|+(\partial_u\phi)^2 |\lin\nu|+|\partial_u\phi||\partial_u\lin\phi|\big)\,du \\&\les \lin A \ve \tau^{-3+2\delta}+\lin A|\lin{\mathscr P}_i|\ve \tau^{-3+2\delta} D_1'(v)^{-2+3\delta/2} .\label{eq:lin-kappa-Gamma}
\end{align}
Since the set where $\lin\kappa\ne 0$ along $\Gamma$ is of size 1, \eqref{eq:lin-kappa-integral} follows immediately. 

\textsc{Proof of \eqref{eq:lin-kappa-far-1}}: We use \eqref{eq:dulkappa-log}, the bootstrap assumptions for $\lin r$ and $\lin \lambda$, and \eqref{eq:lin-kappa-Gamma} to estimate \cref{lem:D'-dyadic-sum}
\begin{align*}
     | \lin \kappa(u,v) | &\les |\lin\kappa(v,v)|+ \int_u^v \big( (\partial_u \phi)^2(|\lin r| + r |\lin \nu| ) + r |\partial_u \phi| |\partial_u \lin \phi|\big) \,du'\\
     &\les \lin A \ve \tau^{-3+2\delta}\big(1+|\lin{\mathscr P}_i|D_1'^{-2+3\delta/2}\big)+\lin A \int_u^{v} r^\delta (\partial_u \phi)^2 \,du'+\int_u^v r |\partial_u\phi||\partial_u \lin\phi|\, du'\\
     &\les \lin A \ve \tau^{-3+2\delta}\big(1+|\lin{\mathscr P}_i|D_1'^{-2+3\delta/2}\big)+ \lin A \,  \u{\mathcal F}[\phi] (v,\tau)+\sum_{j\ge \lfloor {\log_2 u}\rfloor}\big( \u{\mathcal F}[\phi](v,L_j)  \u{\mathcal F}[\lin \phi](v,L_j)\big)^{1/2} \\
     &\les  \lin A \ve \tau^{-3+2\delta}\big(1+|\lin{\mathscr P}_i|D_1'^{-2+3\delta/2}\big).
\end{align*}
Since $D_1'(u)\les u$, this estimate also implies
\begin{equation*}
    |\lin\kappa|\les \lin A\ve \tau^{-1+\delta/2}.
\end{equation*}

    \textsc{Proof of \eqref{eq:lin-kappa-far-2}}: We use the bootstrap assumptions together with the previous estimate to obtain
    \begin{align}
       \nonumber |\partial_u(r^{1-\delta}\lin\kappa)|&\les r^{-\delta}|\lin\kappa| + r^{1-\delta}(|\lin r|+r|\lin\kappa|+r|\lin\nu|)(\partial_u\phi)^2+r^{2-\delta}|\partial_u\phi||\partial_u\lin\phi|\\&\les  \lin A\ve r^{-\delta}\tau^{-1+\delta/2} + \lin A r^{2-\delta} (\partial_u\phi)^2+r^{2-\delta}|\partial_u\phi||\partial_u\lin\phi|.\label{eq:lin-kappa-proof-1}
    \end{align}         
It holds that 
    \begin{equation*}
        \int_u^vr^{-\delta}\tau^{-1+\delta/2}\,du'\les 1
    \end{equation*}
         for every $v\ge u$. Indeed, in the region where $r\ge\tau$, the integrand is bounded by $\tau^{-1-\delta/2}=u^{-1-\delta/2}$, which is integrable, and in the region where $\tau\ge r$, the integrand is bounded by $r^{-1-\delta/2}$, which is also integrable since $|\nu|\sim 1$. Therefore, integrating \eqref{eq:lin-kappa-proof-1}, we obtain
         \begin{align*}
     | r^{1-\delta}\lin \kappa(u,v) | &\les |\lin\kappa(v,v)|+\lin A\ve\int_u^v r^{-\delta}\tau^{-1+\delta/2}\,du'+ \lin A\int_u^v r (\partial_u\phi)^2\,du'+\int_u^vr^{2-\delta}|\partial_u\phi||\partial_u\lin\phi|\,du'\\
     &\les \lin A\ve+ \lin A \,  \u{\mathcal F}[\phi] (v,\tau)+\sum_{j\ge \lfloor {\log_2 u}\rfloor}\big( \u{\mathcal F}[\phi](v,L_j)  \u{\mathcal F}[\lin \phi](v,L_j)\big)^{1/2}\les\lin A\ve,
\end{align*}
which completes the proof.\end{proof}

Finally, we estimate $\lin\kappa$ in the near region.

\begin{lem} \label{lem:lin-kappa-near}
   Under \cref{assumption-1}, it holds that
      \begin{align*}
        |\lin\kappa|&\les \lin A\ve\tau^{-1+3\delta}\big(1+|\lin{\mathscr P}_i| D_1'^{-2+3\delta/2}\big),\\
        |D'\lin\kappa|&\les \lin A\ve\tau^{-2+3\delta}\big(1+|\lin{\mathscr P}_i| D_1'^{-2+3\delta/2}\big) ,\\
        |D\lin\kappa|&\les \lin A\ve\tau^{-3+3\delta}\big(1+|\lin{\mathscr P}_i| D_1'^{-2+3\delta/2}\big) 
    \end{align*}
    in $\mathcal R^\le_{L_i}$. 
\end{lem}
\begin{proof} For any $(u,v)\in\mathcal R_{L_i}^\le$, we integrate \eqref{eq:dulkappa-log} and use the basic estimates on the geometry to obtain
  \begingroup
  \allowdisplaybreaks
    \begin{align*}
          |\lin\kappa|&\les |\lin\kappa(v,v)|+\int_{v}^u \big(D^{-1}(\partial_u\phi)^2|\lin r|+D^{-2}(\partial_u\phi)^2|\lin\nu|+D^{-1}|\partial_u\phi||\partial_u\lin\phi|\big)\,du',\\
           D' |\lin\kappa|&\les |\lin\kappa(v,v)|+\int_{v}^u\big(D'D^{-1}(\partial_u\phi)^2|\lin r|+D'D^{-2}(\partial_u\phi)^2|\lin\nu|+D'D^{-1}|\partial_u\phi||\partial_u\lin\phi|\big)\,du',\\
            D  |\lin\kappa|&\les|\lin\kappa(v,v)|+ \int_{v}^u\big((\partial_u\phi)^2|\lin r|+D^{-1}(\partial_u\phi)^2|\lin\nu|+|\partial_u\phi||\partial_u\lin\phi|\big)\,du'.
    \end{align*}
   The first term in each of these estimates is bounded by \eqref{eq:lin-kappa-Gamma}. We now estimate the integrals. For any $(a,b) \in \{ (0,0), (0,1), (1,0)\} $, \eqref{eq:lin-boot-3-new} and \eqref{eq:lin-r-lin-nu-comparison-1} give 
\begin{equation*}
    D'^bD^a  |\lin\kappa|\les |\lin\kappa(v,v)| + \lin AI_1 + \lin A|\lin{\mathscr P}_i|I_2+ \lin A|\lin{\mathscr P}_i|I_3+I_4,
\end{equation*}
where
\begin{align*}
  I_1  &\doteq \int_v^u D'^{b}D^{-1-3\delta/4+a}(\partial_u\phi)^2\,du',\\
  I_2  &\doteq \int_v^u D'^{-2+3\delta/2+b}D^{-1-3\delta/4+a}(\partial_u\phi)^2\,du',\\
  I_3 &\doteq \int_v^u \tau^{-1+3\delta}D'^bD^{-1+a}D_1'^{-2+3\delta/2}(\partial_u\phi)^2\,du',\\
  I_4  &\doteq \int_v^uD'^bD^{-1+a}|\partial_u\phi||\partial_u\lin\phi|\,du'.
\end{align*}
We estimate
\begin{gather*}
   I_1\les \u{\mathcal E}{}_{2+\delta-2a-b}[\phi](\tau)\les \ve^2\tau^{-1+2\delta -2a-b},\\
   I_2\les D_1'^{-2+3\delta/2} \int_v^u D'^{b}D^{-1-3\delta/4+a}(\partial_u\phi)^2\,du'\les \ve^2\tau^{-1+3\delta-2a-b} D_1'^{-2+3\delta/2},\\
   I_3 \les \tau^{-1+3\delta}D_1'^{-2+3\delta/2}\int_v^u D'^bD^{-1+a}(\partial_u\phi)^2\,du'\les \ve^2 \tau^{-2+4\delta-2a-b} D_1'^{-2+3\delta/2} ,\\
      I_4 \les \big( \u{\mathcal E}{}_{2-2a-b} [\phi](\tau) \u{\mathcal E}{}_{2-2a-b} [\lin \phi](\tau,u)\big)^{1/2} \les \lin A\ve \tau^{-1+2\delta-2a-b}\big(1+|\lin{\mathscr P}_i|D_1'^{-2+3\delta/2}\big).
\end{gather*}
\endgroup
This completes the proof.  \end{proof}

\subsubsection{Estimates for \texorpdfstring{$\lin\varpi_\dagger$}{lin varpi dagger}}

First, we provide the estimate for $\lin\varpi_\dagger$ in the far region.

\begin{lem}\label{lem:lin-varpi-dagger-far-region}
Under \cref{assumption-1}, it holds that
    \begin{align}\label{eq:lin-varpi-dagger-far}
        |\lin\varpi_\dagger|&\les \lin A\ve \tau^{-3+2\delta}\big(1+|\lin{\mathscr P}_i|D_1'^{-2+3\delta/2}\big),\\
        |\lin\varpi_\dagger|&\les \lin A\ve \tau^{-1+\delta/2}\label{eq:lin-varpi-second}
    \end{align}
   in $\mathcal R_{L_i}^\ge$.
\end{lem}
\begin{proof} Using the equations \eqref{eq:dulvarpi} and \eqref{eq:first-order-varpi-v}, the bootstrap assumptions, and \cref{lem:lin-gamma-far,lem:lin-kappa-far}, we obtain
\begin{align*}
    |\partial_u\lin\varpi_\dagger|&\les \lin A r^2(\partial_u\phi)^2+r^2|\partial_u\phi||\partial_u\lin\phi|,\\
    |\partial_v\lin\varpi_\dagger|&\les \lin A r^2(\partial_v\phi)^2+r^2|\partial_v\phi||\partial_v\lin\phi|
\end{align*}
in $\mathcal R_{L_i^\ge}$. Using the boundary condition $\lim_{v\to\infty}\lin\varpi_\dagger(L_i,v)=0$, \cref{prop:decay-linphi}, and \cref{lem:D'-dyadic-sum}, we obtain \eqref{eq:lin-varpi-dagger-far}. Then \eqref{eq:lin-varpi-second} follows from $D_1'\les u$.
\end{proof}

For estimates in the near region, we require the following lemma. 

\begin{lem}\label{lem:mu-cancellations}
   Under \cref{assumption-1}, it holds that
\begin{align}
\nonumber |\lin\mu|&\les |\lin\varpi_\dagger| + \lin A \tau^{-2+\delta} + \lin A \tau^{-2+3\delta} D'+D'^\delta D^{1-\delta/2}+ \lin A|\lin{\mathscr P}_i|\big(\tau^{-2+\delta}D'^{-1}\\&\qquad\qquad +\tau^{-2+3\delta}D'D_1'^{-2+3\delta/2}+\tau^{-4+4\delta}D_1'^{-2+3\delta/2}+D'^{-2+3\delta/2}D^{1-3\delta/4}\big), \label{eq:lin-mu-1}  \\
    |\lin\mu_\dagger+\varkappa\lin r_\dagger| & \les|\lin\varpi_\dagger| + \ve\tau^{-2+\delta}D' + \ve\tau^{-3+\delta} + |\lin{\mathscr P}_i|\ve\tau^{-2+\delta}D'^{-1} \label{eq:lin-mu-2} 
\end{align}
   in $\mathcal R_{L_i}^{\le}$.
\end{lem}

\begin{rk}
    The estimate \eqref{eq:lin-mu-1} is suboptimal in terms of $\ve$, but it suffices as written to close the bootstrap argument. The factors of $\ve$ in \eqref{eq:lin-mu-2} are crucial, however. 
\end{rk}

\begin{proof}[Proof of \cref{lem:mu-cancellations}]
    Linearizing the identity $\nu = -(1-\mu)$ around Reissner--Nordstr\"om, we see that $\lin \nu_\diamond = \lin\mu_\diamond$. It follows that $
        \lin\mu = \lin\nu_\diamond + \lin\mu_\dagger. $
    By \cref{lem:lin-r-diamond} and the bootstrap assumption for $\lin r_\dagger$, \eqref{eq:lin-mu-1} follows from \eqref{eq:lin-mu-2}. Using \eqref{eq:lin-P_i-bootstrap-formula}, we compute
\begin{multline*}
   \lin\mu_\dagger=  \frac{2}{r}\lin\varpi - \frac{2}{r_\diamond}\lin{\mathscr M}_i -\frac{2Q}{r^2}\lin Q +\frac{2Q}{r^2_\diamond}\lin Q -\varkappa \lin r+D'\lin r_\diamond \\ = \frac{2}{r}\lin\varpi_\dagger - \varkappa \lin r_\dagger -\left(\frac{2Q}{r^2}-\frac{2Q}{r_\diamond^2}\right)M\lin{\mathscr P}_i + \left[ \frac{2}{r} - \frac{2}{r_\diamond} -\frac{2Q^2}{Mr^2} +\frac{2Q^2}{Mr^2_\diamond}-\frac{r_c}{M}\varkappa_\dagger \right]\lin{\mathscr M}_i - \varkappa_\dagger \left(\lin r_\diamond-\frac{r_c}{M}\lin{\mathscr M}_i\right). 
\end{multline*}
Using \eqref{eq:r-taylor} and  \eqref{eq:varkappa-expansion-detailed}, we write the term in square brackets as
\begin{equation*}
    \left[\frac 2r-\cdots\right] = \left(- \frac{2}{r_\diamond^2} + \frac{4Q^2}{Mr_\diamond^3} + \frac{4r_c}{r_\diamond^3} -\frac{6Q^2r_c}{Mr_\diamond^4}\right) r_\dagger+ O(\ve\tau^{-3+\delta})= - \frac{2M_ir_\dagger}{Mr_\diamond^4}q +  O(\ve\tau^{-3+\delta}), 
\end{equation*}
where $q\doteq Mr_\diamond^2 - (2Q^2+2Mr_c)r_\diamond +3Q^2r_c$. Recall the quantity $\mathfrak t\doteq \sqrt{|M^2-Q^2|}$ from \cref{lem:estimates-on-RN-geometry}. Using $r_c= M+\mathfrak t$, $Q^2=M^2+O(\mathfrak t^2)$, $\mathfrak t = O(D')$, and $r_\diamond -M=(r_\diamond-r_c)+\mathfrak t = O(D^{1/2})+\mathfrak t = O(D')$, we observe that
\begin{equation*}
    q = Mr^2_\diamond-4M^2r_\diamond +3M^3 +O(D') = M(r_\diamond -M)(r_\diamond-3M)+O(D') = O(D'). 
\end{equation*}
Using now \eqref{eq:lin-r-expansion-10}, we conclude \eqref{eq:lin-mu-2}.
\end{proof}

We can now estimate $\lin\varpi_\dagger$ in the near region.

\begin{lem}\label{lem:lin-varpi-dagger-near-region}
Under \cref{assumption-1}, it holds that
\begin{equation}\label{eq:lin-varpi-dagger-near-region-estimate}
        |\lin\varpi_\dagger| \les \lin A\ve \tau^{-3+3\delta}\big(1+|\lin{\mathscr P}_i| D_1'^{-2+3\delta/2}\big)
    \end{equation}
   in $\mathcal R_{L_i}^\le$. 
\end{lem}
\begin{proof} We integrate $\partial_u\lin\varpi_\dagger$ from $\Gamma$, where we use \cref{lem:lin-varpi-dagger-far-region} to evaluate the boundary term. From \eqref{eq:dulvarpi-near-region} we readily obtain
\begin{equation}\label{eq:varpi-dagger-aux}
    |\lin\varpi_\dagger|\les  \lin A\ve \tau^{-3+2\delta}+\lin A|\lin{\mathscr P}_i|\ve D_1'^{-2+3\delta/2}\tau^{-3+2\delta}+I_1+I_2+I_3
\end{equation}
in $\mathcal R_{L_i}^\le$, where 
\begin{align*}
    I_1&\doteq \int^u_vD^{-1}(\partial_u\phi)^2|\lin\mu| \,du',\\
    I_2&\doteq \int^u_v |1-\mu|(\partial_u\phi)^2\big(D^{-1}|\lin r|+D^{-2}|\lin\nu|\big)\,du',\\
    I_3&\doteq \int^u_v |1-\mu|D^{-1}|\partial_u\phi||\partial_u\lin\phi|\,du'.
\end{align*}

\textsc{Estimate for $I_1$}: By \eqref{eq:lin-mu-1}, we have $I_1\les \lin A\sum_{j=1}^8J_1^j$, where {\allowdisplaybreaks
\begin{gather*}
    J_1^1\doteq \int_v^u D^{-1}(\partial_u\phi)^2|\lin\varpi_\dagger|\,du',\\
    J_1^2\doteq \int_v^u \tau^{-2+\delta}D^{-1}(\partial_u\phi)^2\,du'\les\tau^{-2+\delta}\u{\mathcal E}{}_2[\phi](\tau)\les \ve^2\tau^{-3+2\delta},\\
    J_1^3\doteq \int_v^u \tau^{-2+3\delta}D'D^{-1}(\partial_u\phi)^2\,du'\les\tau^{-2+3\delta}\u{\mathcal E}{}_1[\phi](\tau)\les\ve^2\tau^{-4+4\delta},\\
    J_1^4\doteq \int_v^u D'^\delta D^{-\delta/2}(\partial_u\phi)^2\,du'\les  \u{\mathcal E}{}_\delta[\phi](\tau)\les\ve^2\tau^{-3+2\delta},\\
    J_1^5\doteq \int_v^u |\lin{\mathscr P}_i|\tau^{-2+\delta}D'^{-1} (\partial_u\phi)^2\,du'\les  |\lin{\mathscr P}_i|\tau^{-2+\delta}D_1'^{-1}\u{\mathcal E}_2[\phi](\tau)\les |\lin{\mathscr P}_i|\ve^2\tau^{-3+2\delta}D_1'^{-1}, \\
    J_1^6\doteq \int_v^u|\lin{\mathscr P}_i|\tau^{-2+3\delta}D'D_1'^{-2+3\delta/2}(\partial_u\phi)^2\,du'\les |\lin{\mathscr P}_i|\tau^{-2+3\delta} D_1'^{-2+3\delta/2}\u{\mathcal E}{}_1[\phi](\tau)\les |\lin{\mathscr P}_i|\ve^2\tau^{-4+4\delta}D_1'^{-2+3\delta/2},\\
    J_1^7\doteq \int_v^u|\lin{\mathscr P}_i|\tau^{-4+4\delta}D_1'^{-2+3\delta/2}(\partial_u\phi)^2\,du' \les |\lin{\mathscr P}_i|\tau^{-4+4\delta} D_1'^{-2+3\delta/2}\u{\mathcal E}{}_2[\phi](\tau)\les |\lin{\mathscr P}_i|\ve^2\tau^{-5+5\delta}D_1'^{-2+3\delta/2},\\
    J_1^8\doteq \int_v^u |\lin{\mathscr P}_i| D'^{-2+3\delta/2}D^{-3\delta/4}(\partial_u\phi)^2\,du' \les |\lin{\mathscr P}_i|D_1'^{-2+3\delta/2}\u{\mathcal E}{}_{3\delta/2}[\phi](\tau)\les |\lin{\mathscr P}_i|\ve^2\tau^{-3+3\delta}D_1'^{-2+3\delta/2}.
\end{gather*}}
This gives
\begin{equation}\label{eq:varpi-dagger-I1}
    I_1\les \text{Gr\"onwall term}+ \lin A\ve^2\tau^{-3+3\delta} +\lin A |\lin{\mathscr P}_i|\ve^2\tau^{-3+3\delta}D_1'^{-2+3\delta/2}.
\end{equation}

\textsc{Estimate for $I_2$}: We use the geometric estimates and \eqref{eq:lin-r-lin-nu-comparison-1} to estimate
\begin{multline*}
    I_2\les \lin A\int_v^u\big(D+\tau^{-2+\delta}D'+\tau^{-3+\delta}\big)\\\times\big(D^{-1-\delta/2}+|\lin{\mathscr P}_i|D'^{-2+3\delta/2}D^{-1-3\delta/4}+|\lin{\mathscr P}_i|\tau^{-1+3\delta}D^{-1}D_1'^{-2+3\delta/2}\big)(\partial_u\phi)^2\,du'\les \lin A\sum_{i=1}^9 J_2^j,
\end{multline*}
where {\allowdisplaybreaks
\begin{gather*}
    J_2^1\doteq \int_v^u D^{-\delta/2}(\partial_u\phi)^2\,du' \les\u{\mathcal E}{}_{\delta}[\phi](\tau)\les \ve^2\tau^{-3+2\delta},\\
    J_2^2\doteq \int_v^u \tau^{-2+\delta}D'D^{-1-\delta/2}(\partial_u\phi)^2\,du' \les\tau^{-2+\delta}\u{\mathcal E}{}_{1+\delta}[\phi](\tau)\les \ve^2\tau^{-4+3\delta},\\
    J_2^3\doteq \int_v^u \tau^{-3+\delta}D^{-1-\delta/2}(\partial_u\phi)^2\,du' \les \tau^{-3+\delta}\u{\mathcal E}{}_{2+\delta}[\phi](\tau) \les\ve^2\tau^{-4+3\delta},\\
    J_2^4\doteq \int_v^u|\lin{\mathscr P}_i|D'^{-2+3\delta/2}D^{-3\delta/4} (\partial_u\phi)^2\,du' \les |\lin{\mathscr P}_i|D'^{-2+3\delta/2}\u{\mathcal E}{}_{3\delta/2}[\phi](\tau)\les |\lin{\mathscr P}_i|\ve^2\tau^{-3+3\delta}D_1'^{-2+3\delta/2} ,\\
    J_2^5\doteq \int_v^u|\lin{\mathscr P}_i|\tau^{-2+\delta}D'^{-1+3\delta/2}D^{-1-3\delta/4} (\partial_u\phi)^2\,du'\les |\lin{\mathscr P}_i|\tau^{-2+\delta}D'^{-2+3\delta/2}\u{\mathcal E}_{1+3\delta/2}[\phi](\tau)\\
    \hphantom{J_2^5\doteq \int_v^u|\lin{\mathscr P}_i|\tau^{-2+\delta}D'^{-1+3\delta/2}D^{-1-3\delta/4} (\partial_u\phi)^2\,du'\les} \les|\lin{\mathscr P}_i|\ve^2\tau^{-4+7\delta/2}D_1'^{-2+3\delta/2}, \\
J_2^6\doteq \int_v^u|\lin{\mathscr P}_i|\tau^{-3+\delta}D'^{-2+3\delta/2}D^{-1-3\delta/4} (\partial_u\phi)^2\,du' \les |\lin{\mathscr P}_i|\tau^{-3+\delta}D'^{-2+3\delta/2}\u{\mathcal E}{}_{2+3\delta/2}[\phi](\tau)\\ \hphantom{J_2^6\doteq \int_v^u|\lin{\mathscr P}_i|\tau^{-3+\delta}D'^{-2+3\delta/2}D^{-1-3\delta/4} (\partial_u\phi)^2\,du' \les}\les|\lin{\mathscr P}_i|\tau^{-4+5\delta/2}D_1'^{-2+3\delta/2}, \\
    J_2^7\doteq \int_v^u|\lin{\mathscr P}_i| \tau^{-1+3\delta}D_1'^{-2+3\delta/2}(\partial_u\phi)^2\,du'\les |\lin{\mathscr P}_i| \tau^{-1+3\delta}D_1'^{-2+3\delta/2}\u{\mathcal E}{}_0[\phi](\tau)\les |\lin{\mathscr P}_i|\ve^2\tau^{-4+4\delta}D_1'^{-2+3\delta/2} ,\\
    J_2^8\doteq \int_v^u|\lin{\mathscr P}_i| \tau^{-3+4\delta}D'D^{-1}D_1'^{-2+3\delta/2}(\partial_u\phi)^2\,du'\les|\lin{\mathscr P}_i| \tau^{-3+4\delta}D_1'^{-2+3\delta/2}\u{\mathcal E}{}_1[\phi](\tau) \les |\lin{\mathscr P}_i|\ve^2\tau^{-5+5\delta}D_1'^{-2+3\delta/2} ,\\
    J_2^9\doteq \int_v^u|\lin{\mathscr P}_i| \tau^{-4+4\delta}D^{-1}D_1'^{-2+3\delta/2}(\partial_u\phi)^2\,du'\les |\lin{\mathscr P}_i| \tau^{-4+4\delta}D_1'^{-2+3\delta/2}\u{\mathcal E}{}_2[\phi](\tau)\les |\lin{\mathscr P}_i| \ve^2\tau^{-5+5\delta}D_1'^{-2+3\delta/2}.
\end{gather*}}
It follows that
\begin{equation}\label{eq:varpi-dagger-I2}
    I_2\les \lin A\ve^2\tau^{-3+2\delta}+\lin A|\lin{\mathscr P}_i|\ve^2 \tau^{-3+3\delta}D_1'^{-2+3\delta/2}.
\end{equation}

\textsc{Estimate for $I_3$}: Using \eqref{eq:lin-phi-energy-1} and Cauchy--Schwarz, we estimate
\begin{align}
   \nonumber I_3 & \les \int_v^u \big(1+\ve\tau^{-2+\delta}D'D^{-1}+\ve\tau^{-3+\delta}D^{-1}\big)|\partial_u\phi||\partial_u\lin\phi|\,du' ,\\
 \nonumber   &\les \big(\u{\mathcal E}{}_0[\phi](\tau)\u{\mathcal E}{}_0[\lin\phi](\tau,u)\big)^{1/2}+\ve\tau^{-2+\delta}\big(\u{\mathcal E}{}_1[\phi](\tau)\u{\mathcal E}{}_1[\lin\phi](\tau,u)\big)^{1/2}+\ve\tau^{-3+\delta}\big(\u{\mathcal E}{}_2[\phi](\tau)\u{\mathcal E}{}_2[\lin\phi](\tau,u)\big)^{1/2}\\
    &\les \lin A\ve \tau^{-3+2\delta} + \lin A|\lin{\mathscr P}_i|\ve \tau^{-3+2\delta}D_1'^{-2+3\delta/2}.\label{eq:varpi-dagger-I3}
\end{align}

\textsc{Putting everything together}: Inserting \eqref{eq:varpi-dagger-I1}, \eqref{eq:varpi-dagger-I2}, and \eqref{eq:varpi-dagger-I3} into \eqref{eq:varpi-dagger-aux}, we can finally use Gr\"onwall's lemma to absorb the $\lin\varpi_\dagger$ term and conclude the proof of the lemma.
\end{proof}

\subsubsection{Estimates for \texorpdfstring{$\lin r_\dagger$}{lin r dagger}, \texorpdfstring{$\lin\lambda_\dagger$}{lin lambda dagger}, and \texorpdfstring{$\lin \nu_\dagger$}{lin nu dagger}}

First, we estimate $\lin r_\dagger$ along $\Gamma$. 

\begin{lem}\label{lem:lin-r-dagger-curve} 
  Under \cref{assumption-1}, it holds that
\begin{align}
 \label{eq:lin-r-dagger-curve-1}   |\lin r_\dagger|&\les \lin A \ve \tau^{-2+\delta}+\lin A|\lin{\mathscr P}_i|\ve\tau^{-3+5\delta/2}D_1'^{-2+3\delta/2} ,\\  \label{eq:lin-r-dagger-curve-2} 
    |\lin r_\dagger|&\les \lin A \ve \tau^{-1+\delta}
\end{align}
    along $\Gamma\cap \mathcal R_{L_i}$.
\end{lem}
\begin{proof}
    We parametrize $\Gamma$  by $s\in [1,u_f]$, where $u(s) = v(s) = s$. Then, along $\Gamma$, we have from  \eqref{eq:lin-lambda}, \eqref{eq:lin-nu},  \eqref{eq:lin-lambda-diamond}, and \eqref{eq:lin-nu-diamond} that
\begin{equation*}
    \frac{d}{ds}\lin r_\dagger = 2(\gamma+\kappa)\varkappa\lin r + (1-\mu)(\lin\gamma+\lin\kappa) - \frac{2(\gamma+\kappa)}{r}\lin\varpi + \frac{2(\gamma+\kappa)Q}{r^2}\lin Q.
\end{equation*}
From the geometric estimates and \eqref{eq:lin-varpi-second}, we obtain
\begin{equation}\label{eq:lin-r-dagger-derivative-curve}
    \left|\frac{d}{ds}\lin r_\dagger\right|\les \ve^2\tau^{-3+\delta}|\lin r_\dagger|+|\lin \gamma|+|\lin\kappa| + \ve^2\tau^{-3+\delta}(|\lin\varpi|+|\lin Q|) 
     \les \ve^2\tau^{-3+\delta}|\lin r_\dagger|+ \lin A \ve^2\tau^{-3+\delta}+|\lin \gamma|+|\lin\kappa|.
\end{equation}
Since $\lin r_\dagger(L_i,L_i)=0$, we can integrate this inequality, use Gr\"onwall's lemma, \eqref{eq:lin-gamma-far-integrated}, and \eqref{eq:lin-kappa-integral} to obtain \eqref{eq:lin-r-dagger-curve-1}. Then \eqref{eq:lin-r-dagger-curve-2} follows from $D_1'\les u$.  
\end{proof}

We now proceed to estimate $\lin r_\dagger$, $\lin\lambda_\dagger$, and $\lin\nu_\dagger$ in the far region.

\begin{lem}\label{lem:r-dagger-etc-far}
Under \cref{assumption-1}, it holds that
\begin{align}
   \label{eq:lin-r-dagger-far-region}   |\lin r_\dagger|&\les   \ve \lin A r^\delta,\\
   \label{eq:lin-lambda-dagger-far-region}   |\lin \lambda_\dagger |&\lesssim \ve \lin A r^{-1+\delta},\\
    \label{eq:lin-nu-dagger-far-region}   |\lin \nu_\dagger| &\lesssim \ve \lin A r^{-1}
\end{align}
    in $\mathcal R_{L_i}^\ge$. Moreover, it holds that
    \begin{equation}\label{eq:lin-nu-Gamma}
        \left|\frac{\lin\nu}{\nu}-\frac{\lin\nu_\diamond}{\nu_\diamond}\right|\les \lin A\ve \tau^{-2+\delta} +\lin A|\lin{\mathscr P}_i|\ve\tau^{-3+5\delta/2}D_1'^{-2+3\delta/2} 
    \end{equation}
    along $\mathcal R_{L_i}\cap\Gamma$.
\end{lem}

\begin{proof} \textsc{Proof of \eqref{eq:lin-r-dagger-far-region}:}
     By \eqref{eq:lin-lambda} and \eqref{eq:lin-lambda-diamond}, we have
     \begin{equation}\label{eq:lin-r-dagger-v}
         \partial_v\lin r_\dagger - 2\kappa\varkappa \lin r_\dagger = (2\kappa\varkappa-D') \lin r_\diamond +(1-\mu)\lin\kappa+\left(\frac{2}{r_\diamond}-\frac{2\kappa}{r}\right)\lin{\mathscr M}_i +\frac{2\kappa}{r}\lin\varpi_\dagger  +\left( \frac{2\kappa Q}{r^2}- \frac{2Q}{r_\diamond^2}\right)\lin Q.
     \end{equation}
     Hence, using the geometric estimates, \eqref{eq:lin-r-diamond-10}, \eqref{eq:lin-kappa-far-2}, and \eqref{eq:lin-varpi-second}, we obtain 
   \begin{equation} \label{eq:dv-lin-r-a-priori-far}
        |\partial_v\lin r_\dagger -2\kappa\varkappa\lin r_\dagger|\les  \ve r^{-2}|\lin r_\diamond|+|\lin\kappa|+\ve r^{-1}(|\lin{\mathscr M}_i|+|\lin Q|)+r^{-1}|\lin\varpi_\dagger|\les \lin A\ve r^{-1+\delta}.
   \end{equation}
Thus integrating this inequality in $v$ and using $|\kappa\varkappa|\les r^{-2}$, we obtain \eqref{eq:lin-r-dagger-far-region}.

\textsc{Proof of \eqref{eq:lin-lambda-dagger-far-region}:} Since $\lin\lambda_\dagger = \partial_v\lin r_\dagger$, \eqref{eq:dv-lin-r-a-priori-far} is immediately rearranged to yield
\begin{equation*}
    |\lin\lambda_\dagger|\les |\varkappa||\lin r_\dagger|+ \lin A\ve r^{-1+\delta}\les \lin A\ve r^{-1+\delta}
\end{equation*}
as desired, where we used \eqref{eq:lin-r-dagger-far-region}.

\textsc{Proofs of \eqref{eq:lin-nu-dagger-far-region} and \eqref{eq:lin-nu-Gamma}:} By \eqref{eq:lin-nu} and \eqref{eq:lin-nu-diamond}, we have
     \begin{equation}\label{eq:nu-1-diamond-1}
     \lin \nu_\dagger - 2\gamma\varkappa \lin r_\dagger = (2\gamma\varkappa+D') \lin r_\diamond +(1-\mu)\lin\gamma-\left(\frac{2}{r_\diamond}+\frac{2\gamma}{r}\right)\lin{\mathscr M}_i +\frac{2\gamma}{r}\lin\varpi_\dagger  +\left( \frac{2\gamma Q}{r^2}+ \frac{2Q}{r_\diamond^2}\right)\lin Q.
     \end{equation} 
     Using the geometric estimates, \eqref{eq:lin-r-diamond-10}, \eqref{eq:lin-gamma-far}, and \eqref{eq:lin-varpi-second} we obtain
     \begin{equation}\label{eq:lin-nu-dagger-far-a-priori}
         |\lin\nu_\dagger|\les \ve r^{-2}\tau^{-2+\delta} |\lin r_\diamond|+ r^{-2}|\lin r_\dagger|  + |\lin\gamma| +\ve r^{-2}\tau^{-2+\delta} (|\lin{\mathscr M}_i|+|\lin Q|) + r^{-1}|\lin\varpi_\dagger|.
     \end{equation}
    To prove \eqref{eq:lin-nu-dagger-far-region}, we simply insert \eqref{eq:lin-r-diamond-10}, \eqref{eq:lin-gamma-far-2}, \eqref{eq:lin-varpi-second}, and \eqref{eq:lin-r-dagger-far-region} into this estimate. On the other hand, we prove \eqref{eq:lin-nu-Gamma}, we use the identity
    \begin{equation}\label{eq:nu-difference-identity}
        \frac{\lin\nu}{\nu}-\frac{\lin\nu_\diamond}{\nu_\diamond} = \frac{\lin\nu_\dagger}{\nu} + \frac{\lin\nu_\diamond}{\nu}\left(1-\frac{\nu}{\nu_\diamond}\right)
    \end{equation}
    together with \eqref{eq:lin-r-diamond-10}, \eqref{eq:lin-gamma-far}, \eqref{eq:lin-varpi-dagger-far}, \eqref{eq:lin-r-dagger-curve-1}, and \eqref{eq:lin-nu-dagger-far-a-priori}. 
\end{proof}

Now we estimate $\lin r_\dagger$ and $\lin \lambda_\dagger$ in the near region.

\begin{lem} \label{lem:lin-r-dagger-lambda-dagger-near-region}
Under \cref{assumption-1}, it holds that
\begin{align}
  \label{eq:lin-r-dagger-near}  |\lin r_\dagger|&\les \lin A\ve\tau^{-2+3\delta}\big(1+|\lin{\mathscr P}_i|D_1'^{-2+3\delta/2}\big),\\
 |\lin\lambda_\dagger|&\les \lin A\ve \big(\tau^{-3+3\delta}+ \tau^{-2+3\delta}D'\big) +\lin A|\lin{\mathscr P}_i|\ve\big(\tau^{-2+\delta}D'^{-1}+\tau^{-2+3\delta}D' D_1'^{-2+3\delta/2}+\tau^{-3+3\delta} D_1'^{-2+3\delta/2} \big) \label{eq:lin-lambda-dagger-near}
\end{align}
    in $\mathcal R_{L_i}^\le$.
\end{lem}
\begin{proof} \textsc{Proof of \eqref{eq:lin-r-dagger-near}}: Using the identity $\partial_vr = \kappa(1-\mu)$, we derive the equation 
\begin{equation} \label{eq:lin-r-over-D}
   ( \partial_v-\varkappa)\left(\frac{\lin r_\dagger}{D}\right) = \frac{(1-\mu)\lin\kappa - \kappa_\dagger\lin \mu -(\lin\mu_\dagger+\varkappa\lin r_\dagger)}{D}. 
\end{equation}
Using \cref{lem:kappa-tilde-estimate,lem:lin-kappa-near,lem:mu-cancellations,lem:lin-varpi-dagger-near-region}, we have
\begin{align}
 \label{eq:1-mu-lin-kappa} |(1-\mu)\lin\kappa| &\les \lin A\ve\tau^{-3+3\delta}+\lin A|\lin{\mathscr P}_i|\ve\tau^{-3+3\delta}D_1'^{-2+3\delta/2},\\
\label{eq:kappa-dagger-lin-mu}|\kappa_\dagger\lin\mu|  & \les  \lin A\ve^2\tau^{-3+3\delta}+\lin A|\lin{\mathscr P}_i|\ve^2\tau^{-3+3\delta}D_1'^{-2+3\delta/2},\\
 |\lin\mu_\dagger+\varkappa\lin r_\dagger| & \les \lin A \ve\tau^{-2+\delta}D' + \lin A\ve\tau^{-3+3\delta}+ |\lin{\mathscr P}_i|\ve\tau^{-2+\delta}D'^{-1}+\lin A\ve\tau^{-3+3\delta}|\lin{\mathscr P}_i|D_1'^{-2+3\delta/2}. \label{eq:lin-mu-dagger-etc}
\end{align}
In obtaining \eqref{eq:kappa-dagger-lin-mu}, we have to take special care to handle the terms $D'^\delta D^{1-\delta/2}$ and $D'^{-2+3\delta/2}D^{1-3\delta/4}$ in \eqref{eq:lin-mu-1}. Indeed, we interpolate as follows:
\begin{align*}
 D'^\delta D^{1-\delta/2}   |\kappa_\dagger|& = D'^\delta |\kappa_\dagger|^{\delta/2}(D|\kappa_\dagger|)^{1-\delta/2}\les \ve^2D'^\delta \tau^{(-1+\delta)\delta/2+(-3+\delta)(1-\delta/2)}\les\ve^2\tau^{-3+2\delta},\\
 D'^{-2+3\delta/2} D^{1-3\delta/4}   |\kappa_\dagger|& =   D'^{-2+3\delta/2}|\kappa_\dagger|^{3\delta/4}(D|\kappa_\dagger|)^{1-3\delta/4}\\&\les \ve^2  D'^{-2+3\delta/2} \tau^{(-1+\delta)3\delta/4+(-3+\delta)(1-3\delta/4)}\les \ve^2 \tau^{-3+5\delta/2}D_1'^{-2+3\delta/2}.
\end{align*}
Altogether, this gives
\begin{equation*}
    \left|  ( \partial_v-\varkappa)\left(\frac{\lin r_\dagger}{D}\right)\right|\les \lin A \ve\tau^{-2+\delta}D'D^{-1} +\lin A\ve\tau^{-3+3\delta}D^{-1}+|\lin{\mathscr P}_i|\ve\tau^{-2+\delta}D'^{-1}D^{-1} + \lin A|\lin{\mathscr P}_i|\ve\tau^{-3+3\delta}D^{-1}D_1'^{-2+3\delta/2}.
\end{equation*}
    By the geometric estimates, $\varkappa$ has a good sign up to an integrable error, so we may integrate \eqref{eq:lin-r-over-D} from $\Gamma$, and use \eqref{eq:lin-r-dagger-curve-1} and Gr\"onwall's inequality to obtain 
\begin{equation*}
    |\lin r_\dagger(u,v)|\les D\big(\lin A\ve\tau^{-2+\delta}+\lin A|\lin{\mathscr P}_i|\ve\tau^{-3+5\delta/2}D_1'^{-2+3\delta/2}\big)+\lin A\ve I_1+\lin A\ve I_2+|\lin{\mathscr P}_i|\ve I_3 + \lin A|\lin{\mathscr P}_i|\ve I_4,
\end{equation*}
where
\begin{gather*}
    I_1\doteq D \int^u_v \tau^{-2+\delta}D'D^{-1} \,dv'\les\tau^{-2+\delta}D\int_{r_\diamond(u,v)}^{r_\diamond(v,v)} D^{-2}D' \,dr'\les \tau^{-2+\delta}D\cdot D^{-1} \les \tau^{-2+\delta},\\
    I_2\doteq D\int_v^u \tau^{-3+3\delta}D^{-1}\,dv'\les \tau^{-2+3\delta},\\
    I_3\doteq D\int_v^u\tau^{-2+\delta}D'^{-1}D^{-1}\,dv',\\
    I_4\doteq D\int_v^u\tau^{-3+3\delta}D^{-1}D_1'^{-2+3\delta/2}\,dv'\les \tau^{-2+3\delta}D_1'^{-2+3\delta/2}.
\end{gather*}
We estimate $I_3$ with H\"older's inequality as follows:
\begin{align*}
    I_3&\les DD_1'^{-2+3\delta/2}\int_v^u \tau^{-2+\delta}D'^{1-3\delta/2}D^{-1}\,dv'\\&\les DD_1'^{-2+3\delta/2}\left(\int_v^u\tau^{(-2+\delta)\frac{2}{3\delta}}\,dv'\right)^{3\delta/2} \left(\int_v^uD'D^{-2/(2-3\delta)}\,dv'\right)^{1-3\delta/2}\\&\les DD_1'^{-2+3\delta/2}\cdot \tau^{-2+5\delta/2}\cdot D^{-1} =  \tau^{-2+5\delta/2}D_1'^{-2+3\delta/2}.
\end{align*}
Combining these estimates, we conclude \eqref{eq:lin-r-dagger-near}.

\textsc{Proof of \eqref{eq:lin-r-dagger-near}}: Using $\lambda = \kappa(1-\mu)$ again, we find \[\lin\lambda_\dagger = (1-\mu)\lin\kappa - \kappa_\dagger\lin\mu-(\lin\mu_\dagger+\varkappa\lin r_\dagger)+\varkappa \lin r_\dagger.\]
Collecting \eqref{eq:lin-r-dagger-near} and \eqref{eq:1-mu-lin-kappa}--\eqref{eq:lin-mu-dagger-etc}, we arrive at \eqref{eq:lin-lambda-dagger-near}.
\end{proof}
    
We proceed with an important calculation of $\lin\varkappa$ in the near region, which will be used again later in \cref{sec:fine-properties}.

\begin{lem}\label{lem:lin-varkappa}
  Under \cref{assumption-1}, it holds that
    \begin{equation}  \label{eq:lin-varkappa-integral}      \int_{\{u\}\times[v_1,v_2]}|\varkappa\lin\kappa+\kappa\lin\varkappa-\lin\varkappa_\diamond|\,dv\les \lin A\ve\tau^{-1+3\delta}_1+\lin A|\lin{\mathscr P}_i|\ve\tau^{-1+3\delta}_1D_1'^{-2+3\delta/2}
    \end{equation}
    for every $\{u\}\times[v_1,v_2]\subset \mathcal R^{\le}_{L_i}$.
\end{lem}

\begin{proof} By \cref{lem:lin-varkappa-diamond-estimate}, $  |\lin\varkappa_\diamond|\les D' + |\lin{\mathscr P}_i| D'^{-1}$. Next, we simply expand to see that
\begin{align*}
    \lin\varkappa_\dagger & = \frac{2}{r^2}\lin\varpi - \frac{2}{r_\diamond^2}\lin{\mathscr M}_i - \frac{4Q}{r^3}\lin Q + \frac{4Q}{r_\diamond^3}\lin Q + \left(\frac{6Q^2}{r^4}-\frac{4\varpi}{r^3}\right)\lin r-\left(\frac{6Q^2}{r^4_\diamond}-\frac{4M}{r^3_\diamond}\right)\lin r_\diamond,\\
    & = O(|\lin\varpi_\dagger|) + O(|\lin r_\dagger|) + O(|r_\dagger|) + O(|\varpi_\dagger \lin r|+|r_\dagger\lin r|).
\end{align*}
Using \eqref{eq:lin-r-alternate}, \eqref{eq:lin-varpi-dagger-near-region-estimate}, and \eqref{eq:lin-r-dagger-near}, this yields
\begin{equation*}
    |\lin\varkappa_\dagger|\les \lin A\ve\tau^{-2+3\delta}  + \lin A|\lin{\mathscr P}_i|\ve\tau^{-2+3\delta}D_1'^{-2+3\delta/2}. 
\end{equation*} Moreover, by \cref{lem:kappa-tilde-estimate,lem:lin-kappa-near}, 
\begin{align*}
    |\varkappa\lin\kappa|&\les \lin A\ve\tau^{-2+3\delta}\big(1+|\lin{\mathscr P}_i|D_1'^{-2+3\delta/2}\big),\\
    |\kappa_\dagger\lin\varkappa|& \les \lin A\ve^2\tau^{-2+\delta} + \lin A|\lin{\mathscr P}_i|\tau^{-2+5\delta/2}D_1'^{-2+3\delta/2},
\end{align*}
where in the second estimate we used that
\begin{equation*}
    D'^{-1}|\kappa_\dagger|= D'^{-2+3\delta/2}|\kappa_\dagger|^{3\delta/2}|D'\kappa_\dagger|^{1-3\delta/2}\les\ve^2 \tau^{-2+5\delta/2}D_1'^{-2+3\delta/2}.
\end{equation*}
It follows that
\begin{equation}\label{eq:lin-varkappa-lin-kappa}
   | \varkappa\lin\kappa + \kappa\lin\varkappa - \lin\varkappa_\diamond| = |\varkappa\lin\kappa +\kappa_\dagger\lin\varkappa + \lin\varkappa_\dagger |\les \lin A\ve\tau^{-2+3\delta}+\lin A|\lin{\mathscr P}_i|\ve\tau^{-2+3\delta}D_1'^{-2+3\delta/2}.
\end{equation}
and therefore integrating yields \eqref{eq:lin-varkappa-integral}.
\end{proof}

Finally, we estimate $\lin \nu_\dagger$ in the near region.
\begin{lem}\label{lem:lin-nu-dagger}
Under \cref{assumption-1}, it holds that
\begin{equation}\label{eq:lin-nu-dagger}
    |\lin\nu_\dagger| \les \ve\tau^{-1+\delta}|\lin\nu_\diamond| +  \lin A\ve\tau^{-1+3\delta}D+\lin A|\lin{\mathscr P}_i|\ve\tau^{-1+3\delta} DD_1'^{-2+3\delta/2}
\end{equation}
    in $\mathcal R_{L_i}^\le$.
\end{lem}
\begin{proof}
    Using \eqref{eq:dvlnu-ratio}, we compute 
\begin{equation*}
    \partial_v\left(\frac{\lin\nu}{\nu}-\frac{\lin\nu_\diamond}{\nu_\diamond}\right) = \varkappa\lin\kappa+\kappa\lin\varkappa-\lin\varkappa_\diamond
\end{equation*}
and therefore
\begin{equation*}
    \lin\nu_\dagger(u,v) = \left(1-\frac{\nu}{\nu_\diamond}\right)\lin\nu_\diamond(u,v) + \nu(u,v)\left(\frac{\lin\nu}{\nu}-\frac{\lin\nu_\diamond}{\nu_\diamond}\right)(u,u) + \nu(u,v)\int_v^u\left(\varkappa\lin\kappa+\kappa\lin\varkappa-\lin\varkappa_\diamond\right)\,dv'.
\end{equation*}
We conclude by using \eqref{eq:lin-nu-Gamma} and \eqref{eq:lin-varkappa-integral}.\end{proof}

\subsection{Proofs of \texorpdfstring{\cref{prop:lin-boot-improvement,prop:Pi-est}}{Propositions 10.9 and 10.10}}\label{sec:completing-linear}

\begin{proof}[Proof of \cref{prop:lin-boot-improvement}] The idea for improving the constants in \hyperref[E-prop-2]{condition 2.}~is that we are given $\lin r_\circ$, $\lin \varpi_\circ$, and $\lin\rho_\circ$ as a part of $\lin\Psi$, and we can now use the linearized geometric estimates to relate these to $\lin r(L_i,L_i)$, $\lin{\mathscr M}_i$, and $\lin{\mathscr P}_i$.

\textsc{Improving \eqref{eq:lin-r-bootstrap}}: By construction and stationarity of the background, $\lin r(L_i,L_i)=\lin r_\diamond(L_i,L_i)=\lin r_\diamond(1,1)$ and $\lin r(1,1)=\lin r_\circ$. Therefore, 
\begin{equation*}
    |\lin r(L_i,L_i)|\le|\lin r_\circ|+ |\lin r_\dagger(L_i,L_i)|\le 1+C\lin A\ve L_i^{-1+\delta}\le 5,
\end{equation*} for $\ve_0$ sufficiently small, where we used \eqref{eq:lin-r-dagger-curve-2}.

\textsc{Improving \eqref{eq:lin-M-bootstrap}}: By construction, $\lin{\mathscr M}_i$ is constant and $\lin\varpi(1,1)=\lin\varpi_\circ$. Therefore,  we may  use \eqref{eq:lin-varpi-second} to estimate
\begin{equation*}
    |\lin{\mathscr M}_i|\le |\lin\varpi_\circ| + |\lin\varpi_\dagger(1,1)|\le 1+C\lin A\ve \le 5
\end{equation*}
for $\ve_0$ sufficiently small. 

\textsc{Improving \eqref{eq:lin-P-bootstrap}}: By \cref{cor:P'-formula} and \cref{cor:Bondi-mass-loss},
\begin{equation}\label{eq:Pi-expansion}
    \lin{\mathscr P}_i = \frac{\varpi_\circ\lin\rho_\circ+\rho_\circ\lin\varpi_\circ}{M}-\frac{\sigma}{M}\lin{\mathscr M}_i=(1+O(\ve_0^2))\lin\rho_\circ + \frac{\rho_\circ}{M}\lin\varpi_\dagger(1,1)+O(\ve_0^2)\lin{\mathscr M}_i,
\end{equation}
which is $\le 5$ in magnitude for $\ve_0$ sufficiently small. 

\textsc{Improving \eqref{eq:lin-boot-1-new}--\eqref{eq:lin-boot-3-new}}: To improve \eqref{eq:lin-boot-1-new}, we simply take $\ve_0$ small in \eqref{eq:lin-r-dagger-near}. To improve \eqref{eq:lin-boot-2-new}, write $|\lin{\u\lambda}{}_{L_i}|\le |\lin\lambda_\diamond|+|\lin{\u\lambda}{}_{L_i\dagger}|$, estimate the first term by \eqref{eq:lin-nu-diamond-10}, which does not involve $\lin A$ in view of \eqref{eq:lin-r-bootstrap}--\eqref{eq:lin-P-bootstrap}, and estimate the second term by \eqref{eq:lin-lambda-dagger-near}. After consolidating terms and taking $\lin A$ large and $\ve_0$ correspondingly small, we have improved \eqref{eq:lin-boot-2-new}. Improving \eqref{eq:lin-boot-3-new} works the same way, using now \eqref{eq:lin-nu-dagger}.

\textsc{Improving \eqref{eq:lin-boot-4-new}--\eqref{eq:lin-boot-6-new}}: Again, combine \cref{lem:lin-r-diamond,lem:r-dagger-etc-far}, take $\lin A$ large, and $\ve_0$ small. \end{proof}

\begin{proof}[Proof of \cref{prop:Pi-est}] Using \cref{thm:C1-finite-time}, we may compute the Fr\'echet derivatives using Gateaux derivatives, which have been estimated in this section. Indeed, \eqref{eq:Pi-est-1} and \eqref{eq:Pi-est-3} follow immediately from \eqref{eq:Pi-expansion} for $\ve_0$ sufficiently small. The proof of \eqref{eq:Pi-est-2} is slightly more involved. 

\textsc{Estimates for $\lin{\mathscr M}_i-\lin{\mathscr M}_j$ and $\lin{\mathscr P}_i-\lin{\mathscr P}_j$}: Let $j\le i$. By \cref{rk:wrong-time}, we may also compute $\lin{\mathscr M}_j$ with respect to the $L_i$-teleologically normalized solution. By \cref{cor:P'-formula} and the definitions,
\begin{equation*}
    \lin{\mathscr M}_i-\lin{\mathscr M}_j = \lim_{v\to\infty}\big(\lin{\mathscr M}_i-\lin\varpi(L_j,v)\big)=-\lim_{v\to\infty}\lin\varpi_\dagger(L_j,v),
\end{equation*}
which is $O(\lin A\ve L_j^{-1+\delta/2})$ by \cref{lem:lin-varpi-dagger-far-region}. By \cref{cor:P'-formula} again, 
\begin{equation*}
   \lin{\mathscr P}_i-\lin{\mathscr P}_j= (\varpi_\circ\lin\rho_\circ+\rho_\circ\lin\varpi_\circ)\left(\frac{1}{\mathscr M_i}-\frac{1}{\mathscr M_j}\right)-\frac{\mathscr P_i}{\mathscr M_i}\lin{\mathscr M}_i+\frac{\mathscr P_j}{\mathscr M_j}\lin{\mathscr M}_j,
\end{equation*}
so after using \cref{cor:Bondi-mass-loss}, we arrive at \eqref{eq:Pi-est-2}. \end{proof}

\section{Fine properties of the extremal threshold}\label{sec:fine-properties}

In this section, we prove \cref{thm:scaling,thm:instabilities}, which concern qualitative and quantitative behavior of solutions near the extremal threshold. In \cref{sec:revisiting-linearized}, we revisit \cref{sec:estimates-linear-perturbations} and make some slight refinements for use in our later analysis. In \cref{sec:scaling-proofs}, we prove \cref{thm:scaling}. The proof of \cref{thm:instabilities} is given in \cref{sec:Aretakis-generic}.

We will make extensive use of many of the estimates proved so far, and always take $\ve_0$ sufficiently small so that all of the previous arguments apply.

\subsection{Revisiting the linearized estimates}\label{sec:revisiting-linearized} 

\subsubsection{Taking \texorpdfstring{$i \to\infty$}{i to infty} in \texorpdfstring{\cref{sec:estimates-linear-perturbations}}{Section 11}}\label{sec:taking-limit}

Let $\Psi\in\mathcal K(\ve,M,\varrho,\infty)$ and $\lin\Psi\in\mathfrak Z$. In \cref{sec:estimates-linear-perturbations}, we obtained estimates for the components of $\lin{\u{\mathcal S}}{}_{L_i}[\Psi,\lin\Psi]$ for every finite $i\ge 1$, relative to the teleologically normalized linearized background $\lin{\u{\mathcal S}}{}_{\diamond,L_i}[\Psi_{\diamond,L_i},\lin\Psi_{\diamond,L_i}]$. As explained in \cref{sec:eschatological-definition}, the gauge changes are eventually stationary in $i$ on any compact subset of $\mathcal R_\infty$, and we can therefore make sense of $\lin{\u{\mathcal S}}{}_{\infty}[\Psi,\lin\Psi]$. 

To pass to the limit in the estimates, we need to make sense of the object $\lin{\u{\mathcal S}}{}_{\diamond,\infty}[\Psi_{\diamond,\infty},\lin\Psi_{\diamond,\infty}]$, which is formally the limit of $\lin{\u{\mathcal S}}{}_{\diamond,L_i}[\Psi_{\diamond,L_i},\lin\Psi_{\diamond,L_i}]$. By the proof of \cref{thm:dichotomy-revisited} in \cref{sec:proof-refined-dichotomy}, the limiting $\Psi_{\diamond,\infty}$ exists and $r_{\diamond,L_i}\to r_{\diamond,\infty}$ uniformly on compact sets. 

\begin{lem}
    Let $\Psi\in\mathcal K(\ve,M,\varrho,\infty)$. For any $\lin\Psi\in\mathfrak Z$, the limit $   \lin r_\star\doteq \lim_{\u u\to\infty}\lin{\u r}{}_\infty(\u u,\u u)$  exists. 
\end{lem}
\begin{proof}
This follows from \eqref{eq:lin-r-dagger-derivative-curve},  \eqref{eq:lin-r-dagger-curve-2}, the constancy of $\lin r_\diamond$ along $\Gamma$, and \eqref{eq:lin-gamma-far-integrated} and \eqref{eq:lin-kappa-integral}.
\end{proof}
We then set
\begin{equation*}
    \lin\Psi_{\diamond,\infty}\doteq(0,\lin r_\star,\lin{\mathscr M}_\infty,\lin{\mathscr P}_\infty),
\end{equation*} which we use to define $\lin r_{\diamond,\infty}$ as in \cref{sec:lin-r-diamond}. From \cref{prop:Pi-est} and its proof, it follows that $\lin\Psi_{\diamond,L_i}\to \lin\Psi_{\diamond,\infty}$ as $i\to\infty$. By chasing the definitions, it is then clear that $\lin r_{\diamond,L_i}\to \lin r_{\diamond,\infty}$ on compact subsets of $\mathcal R_\infty$ as $i\to\infty$. We conclude:

\begin{prop}
    For any $\Psi\in\mathcal K(\ve,M,\varrho,\infty)$ and $\lin\Psi\in\mathfrak Z$ with $\|\lin\Psi\|_\mathfrak{Z}\le 1$, the estimates of \cref{sec:estimates-linear-perturbations} hold for the eschatologically normalized solutions $\lin{\u{\mathcal S}}{}_{\infty}[\Psi,\lin\Psi]$ and $\lin{\u{\mathcal S}}{}_{\diamond,\infty}[\Psi_{\diamond,\infty},\lin\Psi_{\diamond,\infty}]$ on $\mathcal R_\infty$, with $\lin{\mathscr M}_\infty$ in place of $\lin{\mathscr M}_i$ and $\lin{\mathscr P}_\infty$ in place of $\lin{\mathscr P}_i$.
\end{prop}

We will assume the conclusion of this proposition without further comment in what follows. For example, when referring to a specific estimate such as \eqref{eq:lin-r-dagger-near} in this context, we mean 
\begin{equation*}
    |\lin{\u r}{}_{\infty}-\lin r_{\diamond,\infty}|\les \lin A\ve \tau^{-2+3\delta}\big(1+|\lin{\mathscr P}_\infty|D_1'^{-2+3\delta/2}\big)
\end{equation*}
in $\mathcal R_\infty^{\le}$.

\subsubsection{Conventions for \texorpdfstring{\cref{sec:fine-properties}}{Section 12}}

For the rest of the main body of the paper, we will only consider $\Psi\in \mathcal K(\ve,M,\varrho,\infty)$ and will always work with the eschatological gauge of \cref{sec:eschatological-definition}. Therefore, to reduce visual clutter, we write $\u r=\u r{}_\infty$, $\lin{\u\phi}=\lin{\u\phi}{}_\infty$, $r_\diamond = r_{\diamond,\infty}$, $\lin r_\diamond = \lin r_{\diamond,\infty}$, etc., from now on. The constant $\lin A$ is fixed sufficiently large for the proof of \cref{prop:Wi} to work.

\subsubsection{Linearizing tangent to \texorpdfstring{$\mathfrak M_\mathrm{stab}^{\sigma}$}{M stab sigma}}\label{sec:tangential-linearization}

In the proofs of \cref{thm:scaling,thm:instabilities}, it will be important to track the ``better'' behavior of linearized quantities tangent to the stable manifolds $\mathfrak M^\sigma_\mathrm{stab}$. Recall that $\mathfrak M^\sigma_\mathrm{stab}$ is the image of the $C^1$ map $\check{\mathscr W}_\infty|_\sigma\doteq \check{\mathscr W}_\infty(\cdot,\sigma):B^\mathfrak{X}_{\ve_0}(x_0)\to\mathfrak Z$. By standard differential geometry, \index{*@$\lint\cdot$, tangential linearization}
\begin{equation*}
    T_\Psi\mathfrak M_\mathrm{stab}^\sigma = \{\check{\mathscr W}_\infty|_\sigma'(\check{\mathscr P}_\infty(\Psi))\lin x:\lin x\in \mathfrak X\}
\end{equation*}
for any $\Psi\in\mathfrak M_\mathrm{stab}^\sigma$. We denote elements of $ T_\Psi\mathfrak M_\mathrm{stab}^\sigma$ by $\lint\Psi$ and the associated linearized solution by $\lint{\u{\mathcal S}}[\Psi,\lint\Psi]$. Since $\mathfrak M_\mathrm{stab}^\sigma$ is a level set of $\mathscr P_\infty$, we have the fundamental relation
\begin{equation*}
    \lint{\mathscr P}_\infty=0,
\end{equation*}
which greatly simplifies many of the estimates in \cref{sec:estimates-linear-perturbations}. In particular, since $D_1'^{-1}$ terms now do not appear, many tangentially linearized quantities will have regular limits as $\u u\to\infty$ in the extremal case.  
 
\subsection{Universal scaling laws}\label{sec:scaling-proofs}

\subsubsection{Final area and temperature}\label{sec:are-scaling}

Recall the final area and temperature functions
\begin{align*}
        \mathscr A:\mathfrak M_\mathrm{black}\cap\cyl(\ve_0,2)&\to \mathbb (0,\infty) & \mathscr T:\mathfrak M_\mathrm{black}\cap\cyl(\ve_0,2)&\to [0,\infty),\\
        \Psi &\mapsto 4\pi M^2\big(1+\sqrt{1-\sigma^2}\big)^2 & \displaystyle  \Psi &\mapsto  \frac{\sqrt{1-\sigma^2}}{2\pi M(1+\sqrt{1-\sigma^2})^2} .
    \end{align*}

\begin{proof}[Proof of \cref{thm:scaling} for $\mathscr A$ and $\mathscr T$] We consider the maps
\begin{align*}
      \tilde{\mathscr A}\doteq \mathscr A\circ\check{\mathscr W}_\infty:\cyl(\ve_0,1)&\to (0,\infty),\\
      \tilde{\mathscr T}\doteq \mathscr T\circ\check{\mathscr W}_\infty:\cyl(\ve_0,1)&\to [0,\infty),
\end{align*}
for which we prove the corresponding statements. Since $\check{\mathscr W}_\infty$ is a $C^1_b$ diffeomorphism, these will be equivalent to what we want for $\mathscr A$ and $\mathscr T$. 

\textsc{Proof of (H\"older) regularity}: For $(x,\sigma)\in\cyl(\ve_0,1)$, we compute 
\begin{align}
   \label{eq:tilde-A-formula} \tilde{\mathscr A}(x,\sigma) &= 4\pi\tilde{\mathscr M}_\infty(x,\sigma)^2\big(1+\sqrt{1-\sigma^2}\big)^2,\\
     \label{eq:tilde-T-formula}  \tilde{\mathscr T}(x,\sigma) &=\frac{\sqrt{1-\sigma^2}}{2\pi \tilde{\mathscr M}_\infty(x,\sigma)(1+\sqrt{1-\sigma^2})^2},
\end{align} where $\tilde{\mathscr M}_\infty\doteq \mathscr M_\infty\circ \check{\mathscr W}_\infty$. Since $\tilde{\mathscr M}_\infty$ is $C^1_b$ by \cref{cor:transversality}, we obtain
\begin{equation*}
    \tilde{\mathscr A},\tilde{\mathscr T}\in C^1(\{|\sigma|<1\})\cap C^{1/2}_b(\{|\sigma|\le 1\})
\end{equation*}
simply by inspection. When $\sigma$ is fixed, it is also clear that $x\mapsto \tilde{\mathscr A}(x,\sigma),\tilde{\mathscr T}(x,\sigma)$ are $C^1_b$.

\textsc{Proof of scaling laws}: Let $p\mapsto \Psi_p$ be a $C^1$ path so that $\Psi_p\in\mathfrak M_\mathrm{black}\cap\cyl(\ve_0,2)$ for $p\le p_*$. Let $\Upsilon_p=(x_p,\sigma_p)\doteq\check{\mathscr P}_\infty(\Psi_p)$, so that $p\mapsto \Upsilon_p$ is a $C^1$ path in $\cyl(\ve_0,1)$ for $p\le p_*$ with $\sigma=\pm 1$. By Taylor's theorem, $\sigma_p=\pm 1\mp c|p-p_*|+o(p)$ and $x_p=x_{p_*}+O_\mathfrak{X}(|p-p_*|)$ as $p\nearrow p_*$, where $c\doteq  \big|{\left.\frac{d^-}{dp^-}\right|_{p=p_*}\mathscr P_\infty(\Psi_p)}\big|$. Using regularity of $\tilde{\mathscr M}_\infty$ and expanding the square roots, we observe from \eqref{eq:tilde-A-formula} and \eqref{eq:tilde-T-formula} that
\begin{align*}
 \tilde{\mathscr A}(x_p,\sigma_p) &=  \tilde{\mathscr A}(x_{p_*},\pm 1) + \sqrt{2c}8\pi \tilde{\mathscr M}_\infty(x_{p_*},\pm 1)^2 |p-p_*|^{1/2} + o(|p-p_*|^{1/2}),\\
    \tilde{\mathscr T}(x_p,\sigma_p) &=  \underbrace{\tilde{\mathscr T}(x_{p_*},\pm 1)}_{=0}+\frac{\sqrt{2c}}{2\pi \tilde{\mathscr M}_\infty(x_{p_*},\pm 1)}|p-p_*|^{1/2}+o(|p-p_*|^{1/2}).
\end{align*}
as $p\nearrow p_*$. This gives exactly \eqref{eq:A-as} and \eqref{eq:T-as}. \end{proof}

\subsubsection{Location of the event horizon}\label{sec:horizon-location-proof}

Recall the notation $\mathscr U:\mathfrak M_\mathrm{black}\cap\cyl(\ve_0,2)\to (1,U_*)$ for the event horizon location in the initial data $u$-coordinate. The formula \eqref{eq:horizon-location-formula} for the horizon location can be interpreted in the eschatological gauge as
\begin{equation*}
     u_{\mathcal H^+}=1 -\int_1^{\infty}\u\nu(\u u,1)\,d\u u.
\end{equation*} As before, we set $\tilde{\mathscr U}\doteq \mathscr U\circ\check{\mathscr W}_\infty$. For $i\ge 1$, we define the functions
\begin{align*}
    \tilde{\mathscr U}_i:\cyl(\ve_0,1)&\to\Bbb R\\
    \Upsilon&\mapsto 1-\int_{1}^{L_i} \u\nu(\u u,1)\,d\u u,
\end{align*}
where $\u\nu$ is taken from $\u{\mathcal S}[\check{\mathscr W}_\infty(\Upsilon)]$. By the theory of \cref{sec:semiglobal-1}, each $\tilde{\mathscr U}_i$ is $C^1_b$ on $\cyl(\ve_0,1)$, with a (as we will see, necessarily) degenerating bound. Since $|\u\nu|\les D\les \u u^{-2}$ by the geometric estimates and \cref{lem:D-quadratic-decay}, the dominated convergence theorem implies immediately: 
\begin{lem}\label{lem:U-tilde-approx}
    $\tilde{\mathscr U}_i$ converges uniformly to $\tilde{\mathscr U}$ on $\cyl(\ve_0,1)$ as $i\to\infty$. 
\end{lem}

Using these approximants, we have:

\begin{lem}\label{lem:U-C1-sub}
    For every $\ell <1$, $\tilde{\mathscr U}\in  C^1_b(\cyl(\ve_0,\ell))$ and
    \begin{equation}\label{eq:U-subextremal-derivative}
        \sup_{\cyl(\ve_0,\ell)}\|\tilde{\mathscr U}'\|_{\mathfrak{Z}^*}\les (1-\ell)^{-1/2-\delta/2}.
    \end{equation}
\end{lem}
\begin{proof}
 Let $\Upsilon\in \cyl(\ve_0,\ell)$ and $\lin\Upsilon\in\mathfrak Z$ with $\|\lin\Upsilon\|_\mathfrak{Z}\le 1$. Then 
 \begin{equation}\label{eq:Ui-diff}
    \tilde{\mathscr U}_{i+1}'(\Upsilon)\lin\Upsilon -\tilde{\mathscr U}_i'(\Upsilon)\lin\Upsilon = - \int_{L_i}^{L_{i+1}}\lin{\u\nu}(\u u,1)\,d\u u,
 \end{equation}
 where $\lin{\u\nu}$ is taken from $\lin{\u{\mathcal S}}[\Psi,\lin\Psi]$ and $\lin\Psi\doteq \check{\mathscr W}_\infty'(\Upsilon)\lin\Upsilon$. Note that $\|\lin\Psi\|_\mathfrak{Z}\les 1$. By \eqref{eq:lin-boot-3-new}, \eqref{eq:D-decay}, and \eqref{eq:1-mu-hierarchy-computation} (in the form $D'\gtrsim \sqrt{1-\ell}$), we obtain
 \begin{equation*}
     |\lin{\u\nu}(\u u,1)| \les D'^{-2+\delta}D^{1-\delta/2} \les (1-\ell)^{-1/2-\delta/2}\u u^{-2+\delta}.
 \end{equation*}
Therefore, \eqref{eq:Ui-diff} is summable and the proof of the lemma is complete in view of  \cref{lem:sequences-derivatives,lem:U-tilde-approx}.
\end{proof}

\begin{lem}\label{lem:C1-tangent}
    For every $\sigma\in [-1,1]$, $\tilde{\mathscr U}|_\sigma\doteq \tilde{\mathscr U}(\cdot,\sigma)\in C^1_b(B^\mathfrak{X}_{\ve_0}(x_0))$, with a bound independent of $\sigma$. 
\end{lem}
\begin{proof}
    As explained in \cref{sec:tangential-linearization}, derivatives in $x$ of $\tilde{\mathscr U}_i|_\sigma\doteq\tilde{\mathscr U}_i(\cdot,\sigma)$ are computed using the tangential linearization. Indeed, let $(x,\sigma)\in\cyl(\ve_0,1)$ and $\lin x\in \mathfrak X$ with $\|\lin x\|_\mathfrak{X}\le 1$. Then 
    \begin{equation*}
        \tilde{\mathscr U}_{i+1}|_\sigma'(x)\lin x-\tilde{\mathscr U}_i|_\sigma'(x)\lin x = -\int_{L_i}^{L_{i+1}} \lint{\u\nu}(\u u,1)\,d\u u,
    \end{equation*}
   and the magnitude of this expression can be estimated using \eqref{eq:lin-boot-6-new} and \eqref{eq:D-decay} by
    \begin{equation*}
       \les \int_{L_i}^{L_{i+1}}  D^{1-\delta/2}\,d\u u\les   \int_{L_i}^{L_{i+1}} \u u^{-2+\delta}\,d\u u\les L_i^{-1+\delta},
    \end{equation*}
    which is summable and independent of $\sigma$. Therefore, the lemma follows from \cref{lem:sequences-derivatives,lem:U-tilde-approx}.
\end{proof}

\begin{prop}\label{prop:tilde-U-expansion} Let $[0,1]\ni p\mapsto \Upsilon_p=(y_p,\sigma_p)$ be a $C^1$ curve in $\cyl(\ve_0,1)$ with $\|{\frac{d}{dp}\Upsilon_p}\|_\mathfrak Z\le 1$ and $p\mapsto \sigma_p$ monotone.\footnote{We use here the notation $y_p\in  B^\mathfrak{X}_{\ve_0}(x_0)$ to avoid confusion with the basepoint $x_0$ at $p=0$.} Then it holds that 
\begin{multline}\label{eq:tilde-U-expansion}
   |\tilde{\mathscr U}(\Upsilon_p)-\tilde{\mathscr U}(\Upsilon_0)+ r_+(\tilde{\mathscr M}_\infty(\Upsilon_p),\sigma_p)- r_+(\tilde{\mathscr M}_\infty(\Upsilon_0),\sigma_0)|  \\\les \ve_0 |r_+(\tilde{\mathscr M}_\infty(\Upsilon_p),\sigma_p)- r_+(\tilde{\mathscr M}_\infty(\Upsilon_0),\sigma_0)|+\ve_0 p^{1/2} + p^{1/2+\delta/4}
\end{multline}
for $p\in [0,1]$. 
\end{prop}
\begin{proof} We compute
 \begin{align*}
        \tilde{\mathscr U}(\Upsilon_{p})-\tilde{\mathscr U}(\Upsilon_{0})&=\int_1^\infty \big(\u\nu(\u u,1,0)-\u\nu(\u u,1,p)\big)\,d\u u\\
        &=\int_1^\infty \big(\nu_\diamond(\u u,1,0)-\nu_\diamond(\u u,1,p)\big)\,d\u u + \int_1^\infty \big(\u\nu{}_\dagger(\u u,1,0)-\u\nu{}_\dagger(\u u,1,p)\big)\,d\u u\doteq I_1+I_2.
    \end{align*}
 Since
    \begin{equation*}
        \int_1^\infty \nu{}_\diamond(\u u,1,p)\,d\u u = r_+(\tilde{\mathscr M}_\infty(\Upsilon_p),\sigma_p)-r_\diamond(1,1,p),
    \end{equation*}
    we have
    \begin{align*}
      I_1 &= r_+(\tilde{\mathscr M}_\infty(\Upsilon_{0}),\sigma_{0})-r_+(\tilde{\mathscr M}_\infty(\Upsilon_{p}),\sigma_{p})-\int_{0}^{p}\lin r_\diamond(1,1,p')\,dp'\\
        &= r_+(\tilde{\mathscr M}_\infty(\Upsilon_{0}),\sigma_{0})-r_+(\tilde{\mathscr M}_\infty(\Upsilon_{p}),\sigma_{p}) + O(p)
    \end{align*}
 by definition of $\lin r_\diamond$ and \eqref{eq:lin-r-bootstrap}. 
 
 Now we use \eqref{eq:lin-nu-dagger} to estimate
    \begin{equation*}
        |I_2|\le \int_1^\infty\int_{0}^{p}|\lin{\u\nu}{}_\dagger(\u u,1,p')|\,dp' d\u u\les \int_1^\infty\int_{0}^{p}\big(\ve_0|\lin\nu_\diamond|+D+\sigma_p'D'^{-2+3\delta/2}D\big)(\u u,1,p')\,dp' d\u u\doteq \ve_0I_2^1+I_2^2+I_2^3,
    \end{equation*}
    where $\sigma_p'\doteq \frac{d\sigma_p}{dp}$, which we assume to be nonnegative without loss of generality. 
    
    \textsc{Estimate for $I_2^1$}: We use \cref{lem:one-sided}. Denote the positive  part of a function $f$ by $f_+$. Then \cref{lem:one-sided} implies $(\lin\nu_\diamond)_+\les D\u u \sigma_p'+ \u u^{-3/2}$. Since $|\lin\nu_\diamond|= 2(\lin\nu_\diamond)_+ - \lin\nu_\diamond$, we have
    \begin{equation*}
        I_2^1 \les -I_1+ \int_1^\infty \int_0^p \big(D\u u\sigma_p'+\u u^{-3/2}\big)\,dp'd\u u.
    \end{equation*}
    We simply integrate the second term in the integral to get $O(p)$. For the first, we use \eqref{eq:D-decay} and the change of variables formula to obtain 
    \begin{multline}
    \int_1^\infty \int_0^p D\u u\sigma_p'\,dp' d\u u\les \int_1^\infty \int_{\sigma_0}^{\sigma_p} D^{1/2} \,d\sigma d\u u\les \int_{\sigma_0}^{\sigma_p}D'^{-1}(r_+)\underbrace{\int^{100M_0}_{r_+} D'D^{-1/2} \, dr'}_{\les 1}d\sigma \\ \les\int_{\sigma_0}^{\sigma_p}(1-\sigma)^{-1/2}\,d\sigma \les (1-\sigma_{0})^{1/2}-(1-\sigma_{p})^{1/2}\les (\sigma_p-\sigma_0)^{1/2} \les p^{1/2}.\label{eq:need-to-revisit}
    \end{multline}
    
     \textsc{Estimate for $I_2^2$}: Using \eqref{eq:D-decay}, we have
\begin{equation*}
   I_2^2= \int_1^\infty\int_{0}^{p} D\,dp'd\u u\les \int_1^\infty\int_{0}^{p} \u u^{-2}\,dp'd\u u\les p.
\end{equation*}

      \textsc{Estimate for $I_2^3$}: we use \eqref{eq:1-mu-hierarchy-computation} as in the proof of \cref{lem:U-C1-sub} to bound $D'^{-2+3\delta/2}D\les D'^{\delta/2} D^{\delta/2}\les (1-\sigma_p)^{\delta/4} D^{\delta/2}$ and then use the change of variables formula to obtain
\begin{multline*}
    I_2^3 \les \int_{\sigma_{0}}^{\sigma_{p}}\int_1^\infty (1-\sigma)^{\delta/4} D^{\delta/2}\,d\u u d\sigma\les \int_{\sigma_{0}}^{\sigma_{p}} (1-\sigma)^{\delta/4} D'^{-1}(r_+)\underbrace{\int^{100M_0}_{r_+} D'D^{-1+\delta/2} \, dr'}_{\les 1}d\sigma \\ \les \int_{\sigma_{0}}^{\sigma_{p}} (1-\sigma)^{-1/2+\delta/4}\,d\sigma \les (1-\sigma_{0})^{1/2+\delta/4}-(1-\sigma_{p})^{1/2+\delta/4}\les (\sigma_p-\sigma_0)^{1/2+\delta/4} \les p^{1/2+\delta/4}.
\end{multline*}

Combining all of these estimates yields \eqref{eq:tilde-U-expansion} as desired.
\end{proof}

\begin{proof}[Proof of \cref{thm:scaling} for $\mathscr U$]
 The regularity statements \eqref{eq:Holder}, \eqref{eq:C1-sub}, and \eqref{eq:C1-tangent} follow from \cref{prop:tilde-U-expansion}, \cref{lem:U-C1-sub}, and \cref{lem:C1-tangent}, respectively. To prove the expansion \eqref{eq:U-as}, first expand
 \begin{equation}\label{eq:r+-expansion}
     r_+(\tilde{\mathscr M}_\infty(\Upsilon_p),\sigma_p)- r_+(\tilde{\mathscr M}_\infty(\Upsilon_0),\sigma_0) = \sqrt{2c}\mathscr M_\infty(\Psi_{p_*})|p-p_*|^{1/2}+o(|p-p_*|^{1/2}).
 \end{equation}
 Under the hypotheses of \cref{thm:scaling}, part 3., we revisit the proof of \eqref{eq:tilde-U-expansion}, specifically \eqref{eq:need-to-revisit}, and obtain 
\begin{equation*}
    \tilde{\mathscr U}(\Upsilon_p)-\tilde{\mathscr U}(\Upsilon_0) = (1+O(\ve_0))\big(r_+(\tilde{\mathscr M}_\infty(\Upsilon_0),\sigma_0)-  r_+(\tilde{\mathscr M}_\infty(\Upsilon_p),\sigma_p)\big) + \ve_0c|p-p_*|^{1/2} + o(|p-p_*|^{1/2})
\end{equation*}
as $p\nearrow p_*$. Together with \eqref{eq:r+-expansion}, this completes the proof of \eqref{eq:U-as}.
\end{proof}

\subsection{Genericity of the Aretakis instability}\label{sec:Aretakis-generic}

\subsubsection{The horizon redshift integrating factor}

Recall the gauge-invariant vector field 
\begin{equation*}
    Y\doteq \nu^{-1}\partial_u,
\end{equation*}
which equals $\partial_r$ in standard ingoing Eddington--Finkelstein coordinates $(v,r)$ in Reissner--Nordstr\"om. If $\Psi\in\mathfrak M_\mathrm{black}$, it is not hard to see that $\u{Y\psi}(\infty,\u v)\doteq\lim_{\u u\to\infty}\u{Y\psi}(\u u,\u v)$ and $\u{Y^2\psi}(\infty,\u v)\doteq\lim_{\u u\to\infty}\u{Y^2\psi}(\u u,\u v)$ exist for every $\u v$ and that 
\[Y\psi(u_{\mathcal H^+},v)=\u{Y\psi}(\infty,\mathfrak v(v)),\quad Y^2\psi(u_{\mathcal H^+},v)=\u{Y^2\psi}(\infty,\mathfrak v(v))\]
for every $v$. We compute $Y\psi(u_{\mathcal H^+},\cdot)$ using the equation \eqref{eq:Ypsi-negative-1}. As in \cref{sec:pointwise}, we obtain
\begin{equation}\label{eq:Ypsi-representation-formula-t}
    \u{Y\psi}(\infty,\u v{}_2) = R(\u v{}_1,\u v{}_2)Y\psi(u_{\mathcal H^+},\u v{}_1)+ \int_{\u v{}_1}^{\u v{}_2}R(\u v',\u v{}_2)\u\kappa\u\varkappa\u\phi(\infty,\u v')\,d\u v',
\end{equation} for every $1\le\u v{}_1\le\u v{}_2$,
where \index{R1@$R(\u v',\u v)$, horizon redshift integrating factor}
\begin{equation*}
    R(\u v',\u v)\doteq \exp\left(-\int_{\u v'}^{\u v}\u \kappa\u \varkappa(\infty,\u v'')\,d\u v''\right)
\end{equation*}
is the \emph{horizon redshift integrating factor}. By \eqref{eq:varkappa-expansion}, \cref{lem:kappa-tilde-estimate}, and \cref{prop:pw-estimates}, 
\begin{align}
\label{eq:varkappa-expansion-3}     \u\kappa \u\varkappa(\infty,\u v)&= D'(r_+)+O(\ve^2\u v^{-2+\delta}),\\
|\u\kappa \u\varkappa\u\phi(\infty,\u v)| & \les \ve D'(r_+)^{1/2}\u v^{-3/2+3\delta/4}+\ve^3 \tau^{-3+3\delta/2}. \label{eq:varkappa-phi-horizon}
\end{align}
The first expansion implies that $R(\u v',\u v)\les 1$ for any $\Psi\in \mathfrak M_\mathrm{black}\cap\cyl(\ve_0,2)$. We need some more delicate estimates for $R$, and separate out the extremal and subextremal cases.

\begin{lem}[The extremal case]\label{lem:extremal-redshift} Let $\Psi\in\mathfrak M_\mathrm{stab}^{+1,-1}$. Then it holds that
    \begin{equation}\label{eq:partial-v-redshift-extremal}
          |\partial_{\u v} R(\u v', \u v)|\les \ve^2  \u v^{-2+\delta}
    \end{equation}
    for all $1\le \u v'\le\u v$. Suppose further that $z\mapsto \Psi(z)$ is a $C^1$ curve in $\mathfrak M_\mathrm{stab}^{+1,-1}$ with $\Psi(0)=\Psi$ and $\|\lint\Psi\|_\mathfrak{Z}\le 1$.\footnote{Recall the convention $\lint\Psi=\lin\Psi$ from \cref{sec:tangential-linearization} which reminds us that $\lint \Psi$ is tangent to $\mathfrak M_\mathrm{stab}^{+1,-1}$.} Then $R$ is $C^1_\star$ in $(\u v',\u v,z)$ and it holds that
    \begin{align}
   \label{eq:lint-R}   |\lint R(\u v',\u v)|&\les \ve \u v'^{-1+3\delta} ,\\
     \label{eq:dv-lint-R}   |\partial_{\u v}\lint R( \u v',\u v)|&\les  \ve \u v^{-2+3\delta},
\end{align}
for all $1\le\u v'\le\u v$, where $\lint\cdot$ denotes $\partial_z$ with $(\u v',\u v)$ fixed. 
\end{lem}

For the regularity statement, we need a simple lemma.
\begin{lem}\label{lem:v-smoothness}
   Under the hypotheses of \cref{lem:extremal-redshift}, the eschatological function $\u{\mathfrak v}$ is $C^1$ jointly in $\u v$ and $\Psi$. Moreover, for every $\u v\ge 1$ it holds that
    \begin{equation}\label{eq:lint-u-v}
        |\lint{\u{\mathfrak v}}(\u v)|\les  \u v^{\delta}.
    \end{equation}
\end{lem}
\begin{proof}
    For bounded $v$, $\mathfrak v(v)$ is equal to any finite-time normalized $\mathfrak v_{\u u{}_f}(v)$ for large enough $\u u{}_f$ by the proof of \cref{lem:eschatology}. Regularity of $\u{\mathfrak v}$, the inverse map, is automatic. The estimate \eqref{eq:lint-u-v} is then obtained by linearizing \eqref{eq:inverse-gauge-formula} and using \eqref{eq:lin-boot-4-new}. 
\end{proof}

\begin{proof}[Proof of \cref{lem:extremal-redshift}] Since $\partial_{\u v} R(\u v',\u v)=-R(\u v',\u v)\u\kappa\u\varkappa(\infty,\u v)$, the estimate \eqref{eq:partial-v-redshift-extremal} follows immediately from \eqref{eq:varkappa-expansion-3} in the extremal case.

 \textsc{Regularity of $R$}: Fix $\Psi\in\mathfrak M_\mathrm{stab}^{+1,-1}$.  Since $(u_{\mathcal H^+},\u{\mathfrak v}(L_j))$ is a finite point in the maximal development of $\Psi$, $\kappa\varkappa$ is $C^1_\star$ as a function of $(u',v',\Psi')$ in a neighborhood of $(u_{\mathcal H^+},\u{\mathfrak v}(L_j),\Psi)$ by \cref{prop:smooth-dependence}. By \cref{lem:C1-tangent} and \cref{lem:v-smoothness}, $\Psi'\mapsto u_{\mathcal H^+}$ and $\Psi'\mapsto \u{\mathfrak v}(L_j)$ are also $C^1$, provided $\Psi'$ is restricted to $\mathfrak M_\mathrm{stab}^{+1,-1}$. It follows that $R$ is $C^1_\star$ in the variables $(\u v',\u v,\Psi)$, $\Psi\in\mathfrak M_\mathrm{stab}^{+1,-1}$.

\textsc{Proof of \eqref{eq:lint-R}}: Differentiating the definition of $R$, we find
\begin{equation}\label{eq:lint-R-formula}
    -R^{-1}\lint R(\u v',\u v) = \int_{\u v'}^{\u v}\lint{\u \kappa\u \varkappa}(\infty,\u v'')\,d\u v''.\end{equation}
Using \cref{lem:lin-varkappa} and the fact that $\lint\varkappa_\diamond=0$ on $\mathcal H^+$ in the extremal case (which follows from \cref{lem:lin-varkappa-diamond-estimate}), we conclude \eqref{eq:lint-R}. 

\textsc{Proof of \eqref{eq:dv-lint-R}}: We differentiate \eqref{eq:lint-R-formula} to obtain
\begin{equation*}
    |\partial_{\u v} \lint R(\u v',\u v)|\les |\partial_{\u v} R||\lint R|+ \lint{\u{\kappa\varkappa}}(\infty,\u v).
\end{equation*} We then conclude \eqref{eq:dv-lint-R} by using \eqref{eq:lin-varkappa-lin-kappa} and \eqref{eq:lint-R}. \end{proof}

For $\Psi\in\mathfrak M_\mathrm{sub}$, let $(M,\varrho)$ be the final Reissner--Nordstr\"om parameters with the associated lapse $D$. We define\footnote{In the thermodynamic analogy of black hole mechanics, where the surface gravity $\bm\kappa=\frac 12 D'(r_+)$ is proportional to the temperature, $\beta$ is the \emph{inverse temperature} (and is standard notation for this notion). Note that $\mathscr T=\beta^{-1}$.} \index{beta@$\beta$, inverse temperature, redshift time, transient time scale}
\begin{equation*}
 \beta(\Psi) \doteq 4 \pi  D'(r_+)^{-1},
\end{equation*} 
which we call the \emph{redshift time}.  In the following, we will just write $\beta$ and omit the explicit dependence on $\Psi$. It will also be useful to have a separate notation for the redshift integrating factor evaluated at the redshift time: \index{R2@$R_\flat(\u v')$, redshift integrating factor at the redshift time}
\begin{equation*}
    R_\flat(\u v')\doteq R(\u v',\beta).
\end{equation*}

\begin{lem}[The subextremal case]\label{lem:subextremal-redshift} Let $\Psi\in\mathfrak M_\mathrm{stab}^\sigma$ with $|\sigma|<1$. Then it holds that
\begin{align}
    \label{eq:redshift-bound-sub}
    R(\u v',\u v ) &= \exp\left( - \frac{4\pi}{\beta}( \u v -\u v') \right) \big( 1  + O(\ve^2  \u v'^{-1+\delta} ) \big),\\
    \label{eq:partial-v-redshift-subextremal}
    |\partial_{\u v} R(\u v',\u v)|&\les \beta^{-1} + \ve^2\u v^{-2+\delta}
\end{align}
 for all $1\le \u v'\le \u v$. Suppose further that $z\mapsto \Psi(z)$ is a $C^1$ curve in $\mathfrak M$ with $\Psi(0)=\Psi$ and $\|\lin\Psi\|_\mathfrak{Z}\le 1$. Then $R_\flat$ is $C^1_\star$ and it holds that
\begin{equation}\label{eq:lin-R-flat}
    |\lin R_\flat(\u v')|\les 1 + \beta \u v'|\lin{\mathscr P}_\infty| + \ve\beta^{2-3\delta/2}\u v'^{-2+3\delta}|\lin{\mathscr P}_\infty|
\end{equation}
 for all $1\le\u v'\le\beta$.
\end{lem}
\begin{proof} The expansion \eqref{eq:redshift-bound-sub} follows immediately from \eqref{eq:varkappa-expansion-3}, which also readily yields \eqref{eq:partial-v-redshift-subextremal} since $\partial_{\u v}R(\u v',\u v)=-R(\u v',\u v)\u\kappa\u\varkappa(\infty,\u v)$. 

\textsc{Proof of \eqref{eq:lin-R-flat}}: Regularity of $R_\flat$ along $\Psi(z)$ is proved by the same argument of \cref{lem:extremal-redshift}, using \cref{lem:U-C1-sub} instead of \cref{lem:C1-tangent}. To prove \eqref{eq:lin-R-flat}, we first compute
\begin{equation*}
    -R_\flat^{-1}\lin R_\flat(\u v') = \u\kappa\u\varkappa(\infty,\beta)\lin\beta + \int_{\u v'}^\beta(\lin{\u\kappa\u\varkappa})(\infty,\u v'') \,d\u v''.
\end{equation*}
Since $\varkappa_\diamond(\infty,\cdot)=D'(r_+)$, we have \begin{equation}\label{eq:lin-beta}
    \lin\beta = -\frac{\beta^2}{4\pi}\lin\varkappa_\diamond(\infty,\u v)
\end{equation}
for every $\u v$. Using now \eqref{eq:lin-varkappa-lin-kappa} and \eqref{eq:varkappa-expansion-3}, we find 
\begin{align*}
    -R_\flat^{-1}\lin R_\flat(\u v') &= \frac{4\pi}{\beta}\lin\beta + O(\ve^2\beta^{-2+\delta}) + \int_{\u v'}^\beta \left(-\frac{4\pi}{\beta^2}\lin\beta + O\big(\ve(1+|\lin{\mathscr P}_\infty|\beta^{2-3\delta/2})\u v''^{-2+3\delta}\big)\right)\,d\u v''\\
    &= \frac{4\pi \u v'}{\beta^2}\lin\beta+O\big(\ve(1+|\lin{\mathscr P}_\infty|\beta^{2-3\delta/2})\u v'^{-2+3\delta}\big)
\end{align*}
for all $1\le \u v'\le\beta$. Using \cref{lem:lin-varkappa-diamond-estimate} and \eqref{eq:lin-beta}, we estimate
\begin{equation}\label{eq:lin-beta-2}
    |\lin\beta|\les \beta^2(\beta^{-1}+\beta |\lin{\mathscr P}_\infty|),
\end{equation}
which finally gives \eqref{eq:lin-R-flat}.
\end{proof}

\subsubsection{The asymptotic Aretakis charge} 

 We recall the following computation from \cite[Section 9.1]{AKU24}:

\begin{lem} For any solution of the spherically symmetric EMSF system, it holds that
  {\mathtoolsset{showonlyrefs=false}  \begin{align}  \label{eq:RYY} \Ric(Y,Y) &= 2(Y\phi)^2,\\
  \label{eq:DYRYY}  \nabla_Y\Ric(Y,Y)&= 4 Y\phi Y^2\phi - 4r(Y\phi)^4,\\
        \partial_v(Y\psi)+(\kappa\varkappa) Y\psi &=   \kappa\varkappa\,\phi,\label{eq:dvYpsi}\\
     \partial_v(Y^2\psi)+(2 \kappa\varkappa) Y^2\psi &= -\frac{2\kappa Q^2}{r^4}Y\psi+E,\label{eq:dv2Ypsi}
\end{align}}
where
\begin{equation}
    E\doteq \frac{2\kappa Q^2}{r^3}\phi+(1-\mu)r\kappa (Y\phi)^3+\kappa\varkappa\left(r^2(Y\phi)^3-Y\phi\right).
    \label{eq:dv2Ypsi-error}
\end{equation}
\end{lem}

We have the following proposition, which was already proved in \cite[Section 9.1]{AKU24}. 

\begin{prop}\label{prop:dynamical-Aretakis}
    Let $\Psi\in \mathcal K(\ve,M,\pm 1,\infty)$ with $\ve\le\ve_0$. Then the following hold:  
   \begin{enumerate}[i)]
        \item \ul{Non-decay}: The ``asymptotic Aretakis charge'' \index{H1@$H_0[\phi]$, asymptotic Aretakis charge}
        \begin{equation}\label{eq:dynamical-charge}
            H_0[\phi]\doteq\lim_{\u v\to\infty} \u{Y\psi}(\infty,\u v)
        \end{equation}
        exists and it holds that 
\begin{align}
      \u{Y\psi}(\infty,\u v)&=H_0[\phi]+O(\ve^{3}\u v^{-1+\delta}),\label{eq:Aretakis-main}\\
      \u{\Ric}(Y,Y)(\infty,\u v)&=2M^{-2}\big(H_0[\phi]\big)^2+O( \ve^2 \u v^{-1+\delta/2})\label{eq:main-Ricci-1}
\end{align} for all $\u v\ge 1$.

        \item \ul{Conditional growth}: It holds that 
           \begin{align}
       \u{Y^2\psi}(\infty,\u v)&=-2M^{-2}H_0[\phi]\u v+O(\ve\u v^\delta) \label{eq:aretakis-main-2},\\
              \nabla_Y\u{\Ric}(Y,Y)(\infty,\u v)&=-8M^{-4}\big(H_0[\phi]\big)^2\u v + O(\ve^2\u v^\delta) \label{eq:main-Ricci-2}
        \end{align}
        for all $v\ge 1$.
    \end{enumerate}
\end{prop}
\begin{proof} \textsc{Non-decay}: By \eqref{eq:varkappa-phi-horizon} and \eqref{eq:Ypsi-representation-formula-t} with $\u v{}_2=\u v$ and $\u v{}_1=1$, 
\begin{equation*}
    |\partial_{\u v}\u{Y\psi}(\infty,\u v)|\les\ve |\partial_{\u v}R(1,\u v)|+|\u\kappa\u\varkappa\u\phi(\infty,\u v)| + \int_1^{\u v}|\partial_{\u v}R(\u v',\u v')||\u\kappa\u\varkappa\u\phi(\infty,\u v')|\,d\u v'\les \ve^3\u v^{-2+\delta},
\end{equation*}
which yields \eqref{eq:Aretakis-main} after integration. Now \eqref{eq:main-Ricci-1} follows from \eqref{eq:RYY}. 

\textsc{Conditional growth}: We solve \eqref{eq:dv2Ypsi} with an integrating factor to obtain
\begin{equation}\label{eq:Y2psi-formula}
    \u{Y^2\psi}(\infty,\u v{}_2) = R^2(\u v{}_1,\u v{}_2)\u{Y^2\psi}(\infty,\u v{}_1)+\int_{\u v{}_1}^{\u v{}_2}R^2(\u v',\u v{}_2)\left(-\frac{2\u\kappa Q^2}{\u r^4}\u{Y\psi} +\u E\right)(\infty,\u v')\,d\u v'
\end{equation}
for $1\le \u v{}_1\le \u v{}_2$. Using the geometric estimates, we have
\begin{equation}\label{eq:Y2-errors}
   \int_{\u v{}_1}^{\u v{}_2}\frac{2\u\kappa Q^2}{\u r^4}\u{Y\psi}(\infty,\u v')\,d\u v'=\frac{2}{M^2}H_0[\phi]\u v + O(\ve^3\u v^\delta) ,\quad |\u E(\infty,\u v')|\les \ve \u v^{-1+\delta/2},
\end{equation}
and \eqref{eq:aretakis-main-2} follows. Now \eqref{eq:main-Ricci-2} follows from \eqref{eq:DYRYY}. \end{proof}

\begin{defn} The \emph{Aretakis charge map} \index{H3@$ \mathscr H$, Aretakis charge map} is defined on the extremal stable manifold as
    \begin{align*}
        \mathscr H:\mathfrak M^{+1,-1}_\mathrm{stab}&\to \Bbb R\\
        \Psi &\mapsto H_0[\phi],
    \end{align*}
    where $H_0[\phi]$ is given by \eqref{eq:dynamical-charge}. Note that $|H_0[\phi]|\les\ve$.
\end{defn}

\begin{rk}
    The argument in \cite{AKU24} takes place in the initial data gauge $(u,v)$ ($(\hat u,\hat v)$ in the notation there) instead of the eschatological gauge. Using the following lemma, the expansions \eqref{eq:aretakis-main-2} and \eqref{eq:main-Ricci-2} are equivalent to those written in the initial data gauge in \cite{AKU24}.
\end{rk}

\begin{lem}
For any $\Psi\in \mathfrak M_\mathrm{black}\cap\cyl(\ve_0,2)$, the eschatological advanced time coordinate satisfies
    \begin{equation}\label{eq:mathfrak-v-expansion}
        \u{\mathfrak v}(\u v) = \u v + O({\log \u v})
    \end{equation}
    for all $\u v\ge 1$.
\end{lem}
\begin{proof}
    Since $\u\lambda\sim 1$ on $C_\out$, we already have $\u{\mathfrak v}(\u v)\sim v$ for all $\u v$ by \eqref{eq:diffeo-form}, which gives
    \begin{equation*}
        \u{\mathfrak v}(\u v) = 1 + \int_1^{\u v}\u\lambda(1,\u v')\,d\u v'.
    \end{equation*}
    By \eqref{eq:kappa-aux-1}, we have $\u\lambda(1,\u v)=(1-\u\mu)\u\kappa =(1+O(\u v^{-1}))(1+O(\u v^{-1}))=1+O(\u v^{-1})$, and \eqref{eq:mathfrak-v-expansion} readily follows.
\end{proof}

\subsubsection{Regularity of \texorpdfstring{$\mathscr H$}{H} and the proof of \texorpdfstring{\cref{thm:instabilities}}{Theorem 5.19}, Part 2.} \label{sec:regularity-of-H}

\begin{prop}\label{prop:H-C1}
  The Aretakis charge map $\mathscr H:\mathfrak M_\mathrm{stab}^{+1,-1}\to\Bbb R$ is $C^1_b$.
\end{prop}

\begin{proof}
     \textsc{The approximating maps $\mathscr H_j$}: For $j\ge 1$, we define
  \begin{align*}
        \mathscr H_j:\mathfrak M^{+1,-1}_\mathrm{stab}&\to \Bbb R\\
        \Psi &\mapsto Y\psi(u_{\mathcal H^+},\u{\mathfrak v}(L_j))
    \end{align*}
    and write $\tilde{\mathscr H}\doteq \mathscr H\circ\check{\mathscr W}_\infty$ and $\tilde{\mathscr H}_j\doteq \mathscr H_j\circ\check{\mathscr W}_\infty$. By \eqref{eq:Aretakis-main}, 
    \begin{equation}\label{eq:H-approx-conv}
        \sup_{B^\mathfrak{X}_{\ve_0}(x_0)}|\tilde{\mathscr H}-\tilde{\mathscr H}_j|\les \ve L_j^{-1+\delta},
    \end{equation}
    so that $\tilde{\mathscr H}_j\to \tilde{\mathscr H}$ uniformly. Arguing as in the proof of \cref{lem:extremal-redshift}, we see that $\tilde{\mathscr H}_j$ is $C^1$ on $B_{\ve_0}^\mathfrak{X}(x_0)$. From \eqref{eq:H-approx-conv}, it also follows that $\mathscr H$ and $\tilde{\mathscr H}$ are continuous.

  \textsc{The Cauchy property of $\tilde{\mathscr H}_j'$}: We show that the Fr\'echet derivatives $\tilde{\mathscr H}_j'$ are Cauchy in the space of bounded continuous maps $B^\mathfrak{X}_{\ve_0}(x_0)\to \mathfrak Z^*$. We compute $\tilde{\mathscr H}_j'$ using the Gateaux derivative as usual. Let $x\in B^\mathfrak{X}_{\ve_0}(x_0)$, $\Upsilon=(x,\pm 1)$, and $\lin x\in\mathfrak X$ with $\|\lin x\|_\mathfrak{X}\le 1$. Define $\lint\Psi=\check{\mathscr W}_\infty|_{\pm 1}'(x,\pm 1)\lin x$ as in \cref{sec:tangential-linearization}, which is tangent to $\mathfrak M^{+1,-1}_\mathrm{stab}$. By definition, it holds that
     \begin{equation*}
         \tilde{\mathscr H}_j'(\Upsilon)\lin x = \lint{\u{Y\psi}}(\infty, L_j).
     \end{equation*}
We will show that
\begin{equation}\label{eq:H'-diff}
    | \tilde{\mathscr H}_{j+1}'(\Upsilon)\lin x- \tilde{\mathscr H}_j'(\Upsilon)\lin x|\les \ve L_j^{-1+3\delta}
\end{equation}
 The proposition then follows from \cref{lem:sequences-derivatives}.

Using \eqref{eq:Ypsi-representation-formula-t}, we compute
\begin{equation}\label{eq:lint-Ypsi}
        \lint{\u{Y\psi}}(\infty,\u v)  = Y\psi(u_{\mathcal H^+},1) \lint R(1,\u v)+R(1,\u v)\lint{Y\psi}(u_{\mathcal H^+},1)-R(1,\u v)Y^2\psi(u_{\mathcal H^+},1)\lint{\mathscr U}
         +\int_1^{\u v}  \big(\u\kappa \u\varkappa\u\phi\lint R+ R (\lint{\u\kappa\u\varkappa\u\phi})\big)\,d\u v',
\end{equation}
where terms under the integral are evaluated at $(\infty,\u v')$ or $(\u v',\u v)$ where appropriate. By \eqref{eq:partial-v-redshift-extremal} and \eqref{eq:dv-lint-R}, 
\begin{equation}\label{eq:dyadic-redshift-estimate}
    |R(1,L_{j+1})-R(1,L_j)|\les \ve^2 L_j^{-1+\delta},\quad |\lint R(1,L_{j+1})-\lint R(1,L_{j})|\les \ve L_j^{-1+3\delta}.
\end{equation}
Then we compute
\begin{equation}\label{eq:derivative-of-integral-H}
   \left| \partial_{\u v}\int_1^{\u v}  \big(\u\kappa \u\varkappa\u\phi\lint R+ R (\lint{\u\kappa\u\varkappa\u\phi})\big)\,d\u v'\right|\le |(\lint{\u\kappa\u\varkappa\u\phi})(\infty, \u v)| + \int_1^{\u v}\big(|\u\kappa \u\varkappa\u\phi\partial_{\u v}\lint R|+ |\partial_{\u v}R(\lint{\u\kappa\u\varkappa\u\phi})|\big)\,d\u v'.
\end{equation}
Using \eqref{eq:r-lin-phi-pointwise} and \eqref{eq:lin-varkappa-lin-kappa}, we estimate $|(\lint{\u\kappa\u\varkappa\u\phi})(\infty, \u v)|\les \ve^2\tau^{-3+5\delta/2}$. (Note that we have implicitly used the regularity of $\mathscr U$ tangent to $\mathfrak M_\mathrm{stab}^{+1,-1}$ to infer that $\lim_{\u u\to\infty}\lint{\u\phi}(\u u,\u v)=\lint{\u\phi}(\infty,\u v)$ exists.) Therefore, using also \cref{lem:extremal-redshift} and integrating \eqref{eq:derivative-of-integral-H} from $\u v = L_j$ to $\u v=L_{j+1}$, we find
\begin{equation*}
     \left|\int_{L_j}^{L_{j+1}}  \big(\u\kappa \u\varkappa\u\phi\lint R+ R (\lint{\u\kappa\u\varkappa\u\phi})\big)\,d\u v'\right|\les \ve^2 L_j^{-2+5\delta/2}.
\end{equation*}
Together with \eqref{eq:dyadic-redshift-estimate}, this yields \eqref{eq:H'-diff} as desired. \end{proof}

\begin{proof}[Proof of \cref{thm:instabilities}, Part 2.] We show that for $\ve_0$ sufficiently small, $\mathscr H$ is a submersion on $\mathfrak M^{+1,-1}_\mathrm{stab}$. Therefore, its level sets $\mathfrak H_h$ give a $C^1$ foliation of $\mathfrak M_\mathrm{stab}^{+1,-1}$ by the implicit function theorem. Since $\tilde{\mathscr H}$ is a $C^1$ map to $\Bbb R$, it suffices to exhibit one $\lin x\in\mathfrak X$ such that $\tilde{\mathscr H}'(0,\pm 1)\lin x\ne 0$ and then take $\ve_0$ perhaps smaller.

Let $\lin\phi_\circ\in \mathfrak F$ be such that $\lin\phi_\circ = 1$ on $\u C{}_\ing$, and then set $\Upsilon=(0,\pm 1)\in\mathfrak Z$ and $\lin x=(\lin\phi_\circ,0,0)\in\mathfrak X$. By \cref{prop:linear-RN}, $\lint{\u\phi}$ solves the linear wave equation on extremal Reissner--Nordstr\"om and $\lint{\u\kappa}$ and $\lint{\u\varkappa}$ vanish identically. By the proofs of \cref{lem:C1-tangent,lem:extremal-redshift}, $\lint{\mathscr U}$ and $\lint R$ also vanish. Note finally that $\lint\psi(u,1)=100M_0-u+1$ and $u_{\mathcal H^+}=99M_0+1$. Hence, by \eqref{eq:lint-Ypsi}, $\tilde{\mathscr H}_j'(\Upsilon)\lin x = R(1, L_j)$, which is identically $1$ in extremal Reissner--Nordstr\"om. It follows that 
    \begin{equation*}
         \tilde{\mathscr H}'(\Upsilon)\lin x=\lim_{j\to\infty} \tilde{\mathscr H}_j'(\Upsilon)\lin x =1\ne 0,
    \end{equation*}
    which completes the proof.
\end{proof}

\subsubsection{Behavior of \texorpdfstring{$Y\psi$}{Y psi} along \texorpdfstring{$\mathcal H^+$}{H+} up to the redshift time}

Recall the set of asymptotically subextremal black holes $\mathfrak M_\mathrm{sub}\doteq \mathfrak M_\mathrm{black}\cap \{|\mathscr P_\infty|<1\}$. For $\Psi\in \mathfrak M_\mathrm{sub}$, we define the \emph{transient Aretakis charge} \index{H2@$H_0^{\flat}[\phi]$, transient Aretakis charge} \[ H_0^{\flat}[\phi] \doteq e^{4\pi}\u{Y\psi} (\infty, \beta).\]  
By \eqref{eq:Ypsi-representation-formula-t}, we have the representation formula
\begin{equation}\label{eq:Hb-formula}
    H_0^\flat[\phi] = e^{4\pi}R_\flat(1)Y\psi(u_{\mathcal H^+},1)+\int_1^{\beta}e^{4\pi}R_\flat(\u v')\u\kappa\u\varkappa\u\phi(\infty,\u v')\,d\u v'.
\end{equation}
Note that the geometric estimates and \cref{prop:pw-estimates} show that \[ |H_0^\flat[\phi]|\les \ve\] for $\Psi\in \mathfrak M_\mathrm{sub}$.
The normalization factor $e^{4\pi}$ is required for \cref{lem:H-boundary-limit} below. 

\begin{prop}\label{prop:transient-instability-1}
       Let $\Psi\in \mathcal K(\ve,M,\varrho,\infty)$ with $|\varrho|<1$ and  $\ve\le\ve_0$. Then the following hold: 

\begin{enumerate}[i)]
    \item \ul{Delayed decay:} It holds that 
        \begin{align} \label{eq:transient-instability-1}
          \u{ Y\psi}(\infty, \u v) &= e^{-4 \pi \u v /\beta  }   H_0^{\flat}[\phi]  + O(\ve  \u v^{-1 + \delta})\\
               \u{  \Ric}(Y,Y)(\infty,\u v)&=2r_+^{-2} e^{-8 \pi \u v /\beta  }  \big(H_0^\flat[\phi]\big)^2+O( \ve^2 \u v^{-1+\delta/2})\label{eq:transient-Ricci-1}
       \end{align}
       for $\u v\geq 1$. 

    \item \ul{Conditional growth up to the redshift time:}
It holds that
           \begin{align}
      \u{Y^2\psi}(\infty,\u v)&=- \frac{ Q^2}{2 \pi r_+^4} H_0^{\flat}[\phi]    \beta e^{ - 4\pi\u v / \beta  }  \left( 1- e^{ -4\pi(\u v -1)/\beta } \right)  + O(\ve  \u v^{\delta} \min \{1, \beta \u v^{-1}\}) \label{eq:aretakis-main-asy},\\
              \nabla_Y\u{\Ric}(Y,Y)(\infty, \u v)&= -\frac{2Q^2}{\pi r_+^6}\big(H_0^\flat[\phi]\big)^2\beta e^{-8 \pi \u v /\beta  } \left( 1- e^{ -4\pi(\u v -1)/\beta } \right)+ O(\ve^2  \u v^{\delta} \min \{1, \beta \u v^{-1}\}) \label{eq:main-Ricci-asy}
        \end{align}
        for all $\u v\ge 1$. In particular, for $1\le\u v\le\beta$, we have 
        \begin{equation}\label{eq:transient-aretakisY^2psi} |\u{Y^2 \psi}(\infty,\u v )| \gtrsim  | H_0^{\flat}[\phi]| \u{v}   + O(\ve \u v^\delta),\end{equation}
        \begin{equation}\label{eq:transient-nabla-Ricci}
            |\nabla_Y\u{\Ric}(Y,Y)(\infty, \u v)| \gtrsim |H_0^\flat[\phi]|^2 \u v + O(\ve^2 \u v^\delta).
        \end{equation}
\end{enumerate}
       
\end{prop}
\begin{proof} \textsc{Delayed decay}: We use \eqref{eq:Ypsi-representation-formula-t} to represent $\u{Y\psi}(\infty,\u v{}_2)$. Using \eqref{eq:varkappa-phi-horizon} and \eqref{eq:redshift-bound-sub} with $\u v{}_1 \leq \u v{}_2 = \beta$, we estimate the integrated term by
\begin{align*}
    \bigg|\int_{\u v{}_1}^{\beta}R_\flat (\u v')\u\kappa  \u\varkappa\u\phi (\infty,\u v')\,d\u v'\bigg| &\les  \int_{\u v{}_1}^{\beta} e^{ - \frac{4\pi}{\beta}(\beta -\u v') } \big( \beta^{-1/2} \ve \u v'^{-3/2 + 3\delta/4} + \ve^3 \u v'^{-3 + 3\delta/2}  \big) \,d\u v'\\ &  
     \les \ve \int_{\u v{}_1}^{\beta}   \u v'^{-2 + 3\delta/4}  \,d\u v'\les
           \ve \u v{}_1^{-1 + 3\delta/4}. 
\end{align*}
 In particular, this then shows
\[ 
\u{Y\psi} (\infty, \u v{}_1)  = \frac{1}{R_{\flat}( \u v{}_1)} e^{-4\pi} H_0^\flat[\phi] + O(\ve \u v{}_1^{-1+3\delta/4})
\] which together with \eqref{eq:redshift-bound-sub} shows \eqref{eq:transient-instability-1}  for $\u v \leq \beta$. For $\beta = \u v{}_1 \leq \u  v_2$, we estimate the integrated term by
\begin{align*}
    \bigg|\int_{\beta}^{\u v{}_2}R(\u v',\u v{}_2)\u\kappa \u\varkappa \u\phi (\infty,\u v')\,d\u v'\bigg| & \les  \int_{\beta}^{\u v{}_2} e^{ - \frac{4\pi}{\beta}(\u v{}_2 -\u v') } \big( \beta^{-1/2} \ve\u v'^{-3/2 + 3\delta/4} + \ve^3\u v'^{-3 + 3\delta/2}  \big)\, d\u v' \\
      & \les \ve \int_{\beta}^{\u v{}_2} e^{ - \frac{4\pi}{\beta}(\u v{}_2 -\u v') } \beta^{-1/2} \u v'^{-3/2 + 3\delta/4}  \,d\u v'\\
    & \les \ve \beta^{-1+3\delta/4} \left(\frac{ \u v{}_2 }{\beta}\right)^{-3/2+3\delta/4}  \les \ve \u v{}_2^{-1+3\delta/4}, 
\end{align*}
where we also used the estimate 
\[ \int_{1}^{y} e^{ x} x^{-b} dx \les_b e^y y^{-b}  \] for $y\geq 1$ and $b>0$. 
This yields \eqref{eq:transient-instability-1} for $\u v \geq \beta$. Now \eqref{eq:transient-Ricci-1} follows from \eqref{eq:RYY}. 

\textsc{Conditional growth}: We use \eqref{eq:Y2psi-formula} to represent $\u{Y^2\psi}(\infty,\u v)$. We decompose the integral term as
\begin{align*}
  \int_{1}^{\u v}R^2(\u v',\u v)\left(-\frac{2\u\kappa  Q^2}{\u r^4}\u{Y\psi}  +\u E\right)(\infty,\u v')\,d\u v' =I_1(\u v) + I_2(\u v). 
\end{align*}
Using the geometric estimates and \eqref{eq:transient-instability-1}, we first have 
\begin{align*}
    I_1 ( \u v) 
    & = - \int_{1}^{\u v} e^{ - \frac{8\pi}{\beta}(\u v -\u v') } \frac{2 Q^2}{r_+^4} \left( e^{-\frac{4\pi \u v'}{\beta}   } H_0^{\flat}[\phi]+ O(\ve \u v'^{-1 + \delta} ) \right) \,d\u v'  \\
     & = - \frac{2 Q^2}{r_+^4} H_0^{\flat}[\phi] \int_{1}^{\u v} e^{ - \frac{4\pi}{\beta}(2\u v - \u v' ) }  \,d\u v' + O\left( \ve \int_{1}^{\u v} e^{ - \frac{8\pi}{\beta}(\u v -\u v') }  \u v'^{-1 + \delta}    \,d\u v' \right)  \\
     & = - \frac{2 Q^2}{r_+^4} H_0^{\flat}[\phi]   \frac{\beta}{4\pi} \left(e^{ - \frac{4\pi}{\beta} \u v } -  e^{  \frac{4\pi}{\beta}( 1- 2\u v) }    \right) + O(  \ve \min \{\u v^{\delta}, \beta \u v^{-1+\delta}\}  ) 
      \\ & =- \frac{2 Q^2}{r_+^4} H_0^{\flat}[\phi]  \u v    \frac{\beta e^{ - \frac{4\pi}{\beta}\u v  } }{4\pi \u v } \left( 1- e^{ \frac{4\pi}{\beta}(1-\u v ) } \right)  + O(  \ve \min \{\u v^{\delta}, \beta \u v^{-1+\delta}\} ).
\end{align*}
 For the error term  $E(\infty, \u v')$, we use the geometric estimates and the estimate \eqref{eq:transient-instability-1} for $\u{Y\psi}$ to obtain
\begin{equation*}
    |E(\infty,\u v')|  \les \ve \u v'^{-1+\delta}  + |D'(r_+) \u{Y\psi}(\infty,\u v')| + |{\u v'}^{-2+\delta } \u{Y\psi}(\infty,\u v')| \les \ve \u v'^{-1+\delta} + \ve\beta^{-1}e^{-4\pi \u v / \beta} \les   \ve \u v'^{-1+\delta} .
\end{equation*}
Using this, we estimate $I_2$ as 
\begin{align*}
    |I_2(\u v)| & \les \int_{1}^{\u v} e^{ - \frac{8\pi}{\beta}(\u v -\u v') } | E(\infty,\u v')| \,d\u v'  \les \ve \int_{1}^{\u v} e^{ - \frac{8\pi}{\beta}(\u v -\u v') } \u v'^{-1+\delta} \,d\u v' \les \ve \min \{\u v^{\delta}, \beta \u v^{-1+\delta}\}.
\end{align*}
Combining the estimates for $I_1$ and $I_2$ completes the proof of \eqref{eq:aretakis-main-asy}.  Now \eqref{eq:main-Ricci-asy} follows from \eqref{eq:DYRYY}.

 Finally, \eqref{eq:transient-aretakisY^2psi}  holds since for $\u v \leq \beta$ we have \[    \frac{\beta e^{ - \frac{4\pi}{\beta}\u v  } }{4\pi \u v } \left( 1- e^{ \frac{4\pi}{\beta}(1-\u v ) } \right)  =    \frac{\beta e^{ - \frac{4\pi}{\beta}\u v  } }{4\pi \u v} \left( 1- e^{- \frac{4\pi}{\beta}\u v  } \right) + O(\u v^{-1})  \geq \frac{1}{4\pi} e^{-4\pi}  (1- e^{-4\pi}) + O(\u v^{-1}).\qedhere \]
\end{proof}

\subsubsection{The transient Aretakis charge map}

 We define the \emph{transient Aretakis charge map}  \index{H4@$\mathscr H_\flat$, transient Aretakis charge map} by 
\begin{align*}
        \mathscr H_\flat:\mathfrak M_\mathrm{sub}\cap \cyl(\ve_0,2)&\to \Bbb R\\
        \Psi &\mapsto H_0^\flat[\phi].
    \end{align*} 
     This map continuously extends the Aretakis charge map:
     
\begin{prop} \label{prop:H-is-continuous}
    The piecewise defined function 
    \begin{equation}\label{eq:piecewise-map}
         \begin{cases}
\mathscr H(\Psi)& \Psi\in \mathfrak M_\mathrm{stab}^{+1,-1},\\ 
\mathscr H_\flat(\Psi)& \mathfrak M_\mathrm{sub}\cap\cyl(\ve_0,2)
            \end{cases}
    \end{equation}
    lies in $C^{3\delta/4}_b\big(\mathfrak M_\mathrm{black}\cap\cyl(\ve_0,2)\big)$.
\end{prop}

We prove this proposition in the following two lemmas. First, we prove the H\"older estimate for $\mathscr H_\flat$ in $\mathfrak M_\mathrm{sub}\cap\cyl(\ve_0,2)$ and then verify continuity of the glued function across the threshold. The H\"older estimate then holds up to and including the threshold by a standard approximation argument. 

\begin{lem}\label{lem:H-trans-Holder}
    Set $\tilde{\mathscr H}_\flat\doteq \mathscr H_\flat\circ\check{\mathscr W}_\infty$. Then $\tilde{\mathscr H}_\flat \in C^{3\delta/4}_b(\mathring{\cyl}(\ve_0,1))$, namely, for any $\Upsilon_1,\Upsilon_2\in \mathring{\cyl}(\ve_0,1)$, it holds that 
    \begin{equation}\label{eq:transient-Holder}
        |\tilde{\mathscr H}_\flat(\Upsilon_1)-\tilde{\mathscr H}_\flat(\Upsilon_2)|\les \|\Upsilon_1-\Upsilon_2\|_\mathfrak{Z}^{3\delta/4}.
    \end{equation}
\end{lem}
\begin{proof}
    Let $\Upsilon_1=(x_1,\sigma_1)$ and $\Upsilon_2=(x_2,\sigma_2)$. We may assume without loss of generality that $0\le\sigma_1 \le \sigma_2$. Let $\Upsilon(z)= (1-z)\Upsilon_1+z\Upsilon_2$ for $z\in [0,1]$ and consider the solution arising from $\Psi(z)\doteq\check{\mathscr W}_\infty(\Upsilon(z))$. Note that $\|\lin\Psi\|_\mathfrak Z\les \|\Upsilon_1-\Upsilon_2\|_\mathfrak{Z}$.

    We can again use \eqref{eq:Hb-formula} to compute $\tilde{\mathscr H}_\flat(\Upsilon(z))$ and then differentiate in $z$ (here we are using that $\tilde{\mathscr U}$ is $C^1$ for asymptotically subextremal black holes by \cref{lem:U-C1-sub}) to obtain, analogously to \eqref{eq:lint-Ypsi},
\begin{multline*}
     e^{-4\pi}\frac{d}{dz}\tilde{\mathscr H}_\flat(\Upsilon(z))  = Y\psi(u_{\mathcal H^+},1) \lin R_\flat(1)+R_\flat(1)\lin{Y\psi}(u_{\mathcal H^+},1)-R_\flat(1)Y^2\psi(u_{\mathcal H^+},1)\lin{\mathscr U}
        \\ +\u\kappa\u\varkappa\u\phi(\infty,\beta)\lin\beta+\int_1^{\beta}  \big(\u\kappa \u\varkappa\u\phi\lin R_\flat+ R_\flat (\lin{\u\kappa\u\varkappa\u\phi})\big)\,d\u v'.
\end{multline*} Using \eqref{est:phi_bound} and \eqref{eq:varkappa-expansion-3}, we have $|\u\kappa\u\varkappa\u\phi(\infty,\u v)|\les \ve \u v^{-2+\delta/2}$. 
By \eqref{eq:lin-varkappa-lin-kappa} and \eqref{eq:lin-beta-2}, and \cref{lem:lin-phi-pw,lem:U-C1-sub,lem:subextremal-redshift}, we estimate
\begin{gather*}
  |\u\kappa\u\varkappa\u\phi(\infty,\beta)\lin\beta|\les  \ve \beta^{-1+\delta/2}\|\lin\Psi\|_\mathfrak{Z}+\ve\beta^{1+\delta/2}|\lin{\mathscr P}_\infty|, \\
   | \u\kappa \u\varkappa\u\phi(\infty,\u v')\lin R_\flat(\u v')|\les \ve \u v'^{-2+\delta/2}\|\lin\Psi\|_\mathfrak{Z}+\ve \beta \u v'^{-1+\delta/2}|\lin{\mathscr P}_\infty|+\ve \beta^{2-3\delta/2}\u v'^{-4+7\delta/2}|\lin{\mathscr P}_\infty|,\\
    |R_\flat(\u v') (\lin{\u\kappa\u\varkappa\u\phi})(\infty,\u v')|\les \u v'^{-2+\delta/2}\|\lin\Psi\|_\mathfrak{Z}+\beta^{-1}\u v'^{-1+\delta/2}\|\lin\Psi\|_\mathfrak{Z}+\beta^{2-3\delta/2} \u v'^{-2+3\delta/2}|\lin{\mathscr P}_\infty| + \u v'^{-1+\delta/2}\beta|\lin{\mathscr P}_\infty| 
\end{gather*} for $1\le \u v'\le\beta$. 
Using also \cref{lem:U-C1-sub}, we estimate
\begin{equation}\label{eq:dvHflat}
       \left|\frac{d}{dz}\tilde{\mathscr H}_\flat(\Upsilon(z))\right|\les \|\lin\Psi\|_\mathfrak{Z} + \beta^{2-3\delta/2}|\lin{\mathscr P}_\infty|.
\end{equation}
for $z\in[0,1]$. Now $\beta(\Psi(z)) \sim (1-(1-z)\sigma_1+z\sigma_2)^{-1/2}$ and $\lin{\mathscr P}_\infty = \sigma_2-\sigma_1$, so we have
\begin{equation*}
    \int_0^1  \beta^{2-3\delta/2}|\lin{\mathscr P}_\infty|\,dz \les (\sigma_2-\sigma_1)^{3\delta/4}.
\end{equation*}
Therefore, \eqref{eq:transient-Holder} follows from integrating \eqref{eq:dvHflat}.
\end{proof}

\begin{lem}\label{lem:H-boundary-limit}
For any $x\in B^\mathfrak{X}_{\ve_0}(x_0)$ and any sequence $\sigma_j$ with $|\sigma_j|<1$ and $\sigma_j\to \pm 1$ as $j\to\infty$, it holds that 
\begin{equation*}
        \lim_{j\to\infty}\tilde{\mathscr H}_\flat(x,\sigma_j) = \tilde{\mathscr H}(x,\pm 1). 
\end{equation*}
\end{lem}
\begin{proof} Let $0< \eta <1$. Let $\sigma_\infty=\pm 1$ and put $\Psi_j\doteq \check{\mathscr W}_\infty(x,\sigma_j)$. Then $\Psi_j\to\Psi_\infty$ in $\mathfrak Z$. Let $r_j$, $\phi_j$, etc., denote the quantities associated to $\mathcal S[\Psi_j]$. By the definitions and the triangle inequality,
\begin{multline*}
    |\tilde{\mathscr H}(x,\sigma_\infty)-\tilde{\mathscr H}_\flat(x,\sigma_j) | \le |H_0[\phi_\infty]-\u{Y\psi}{}_\infty(\infty,\eta^{-1})| \\ + |\u{Y\psi}{}_\infty(\infty,\eta^{-1})-\u{Y\psi}{}_j(\infty,\eta^{-1})|+|\u{Y\psi}{}_j(\infty,\eta^{-1})-e^{4\pi}\u{Y\psi}{}_j(\infty,\beta_j)|.
\end{multline*}
The first term is $\les \ve^3 \eta^{1-\delta}$ by \eqref{eq:Aretakis-main}. The third term is
\begin{equation*}
    \les |1-e^{-4\pi \eta^{-1}\beta^{-1}}||\u{Y\psi}{}_j(\infty,\beta_j)| + \ve \eta^{1+\delta} + \ve\beta_j^{-1+\delta}\les \ve \eta^{-1}\beta_j^{-1} + \ve\eta^{1+\delta}
\end{equation*}
for $j$ sufficiently large depending on $\eta$. Unpacking notation, the middle term is given by
\begin{equation*}
    Y\psi_\infty(\mathscr U(\Psi_\infty),\u{\mathfrak v}{}_\infty(\eta^{-1}))-Y\psi_j(\mathscr U(\Psi_j),\u{\mathfrak v}{}_j(\eta^{-1})).
\end{equation*}
Since $\mathscr U(\Psi_j)\to\mathscr U(\Psi_\infty)$ by \cref{thm:scaling} and $\u{\mathfrak v}{}_j(\eta^{-1})\to \u{\mathfrak v}{}_\infty(\eta^{-1})$ by \cref{lem:v-smoothness}, this expression is $o(1)$ as $j\to\infty$ for fixed $\eta>0$ by \cref{prop:smooth-dependence}. Therefore, 
\[  \limsup_{j\to\infty} |\tilde{\mathscr H}(x,\sigma_\infty)-\tilde{\mathscr H}_\flat(x,\sigma_j) |\les \ve\eta^{1-\delta}.\]
Since $\eta$ was arbitrary, this completes the proof.
\end{proof}

\subsubsection{Proof of \texorpdfstring{\cref{thm:instabilities}}{Theorem 5.19}, Part 3.}

\begin{proof}[Proof of \cref{thm:instabilities}, Part 3.]

  The estimates \eqref{eq:transient-sec5-1}--\eqref{eq:transient-sec5-4}  follow from \cref{prop:transient-instability-1}.     
  
For the second statement, we let $p\mapsto \Psi_p$ be a $C^1$ curve in $\cyl(\ve_0,2)$ with $\Psi_p \in \mathfrak M_\mathrm{sub}$ for $p<p_\ast$ and $\Psi_{p_\ast}\in \mathfrak M_\mathrm{stab}^{+1,-1}$. By \cref{prop:H-is-continuous} and \eqref{eq:T-as}, we have
\begin{equation}\label{eq:Holder-in-action}
      |\mathscr H_\flat(\Psi_p) - \mathscr H(\Psi_{p_\ast})| \lesssim \ve_0|p-p_\ast|^{3\delta/4} \lesssim \ve_0 (\sqrt c  \beta(\Psi_p))^{-3\delta/2},
\end{equation}
for $p<p_\ast$ sufficiently close to $p_\ast$ and where the  first implicit constant  depends  on the $C^{3\delta/4}$ Hölder norm of the map \eqref{eq:piecewise-map}. 
Thus, equations \eqref{eq:transient-curve-1} and \eqref{eq:transient-curve-3} hold true since 
\[ |e^{-4\pi v / \beta} (\sqrt c  \beta )^{-3\delta/2}| \les (\sqrt c v)^{-3\delta/2}.\] 
Combining \eqref{eq:transient-aretakisY^2psi}, \eqref{eq:transient-nabla-Ricci}, and \eqref{eq:Holder-in-action} we then also obtain \eqref{eq:transient-curve-2} and \eqref{eq:transient-curve-4}.
\end{proof}

\printindex

\newpage

\printbibliography[heading=bibintoc] 
\end{document}